%% file: lfi_cpv-June-2013-05-16.tex
\documentclass{JINST}
\usepackage{graphicx}
\usepackage{epsfig}
\DeclareGraphicsExtensions{.pdf,.eps,.ps,.png,.jpg,.mqps}
\usepackage{amssymb,amsmath}
\usepackage{longtable}
\usepackage{subfig}
\usepackage{multirow}
\usepackage{hhline}
\usepackage{afterpage}
\usepackage{rotating}
\usepackage{appendix}

\input{journaldefs}
\input{customdefs}

\input{00_title}

\begin{document}

%
%________________________________________________________________

\def\aap{A\&A}%
\def\aapr{A\&A~Rev.}%
\def\aaps{A\&AS}%
\def\Planck{\textit{Planck}}
\def\deg{$^\circ$}

% SYMBOLS REBA TUNING
%**********************
\def\Qone{Q_{1}} % Q1
\def\Qtwo{Q_{2}} % Q2
\def\Vone{V_{1}} % V1
\def\Vtwo{V_{2}} % V2
\def\GMFO{r_1}     
\def\GMFT{r_2}     
\def\Offset{\mathcal{O}}
\def\SecondQuant{S_\mathrm{q}}
\def\Qerr{\epsilon_{q}}
\def\Qerrdiff{\epsilon_{q,\mathrm{diff}}}
\def\Qerrsky{\epsilon_{q,\mathrm{sky}}}
\def\Qerrload{\epsilon_{q,\mathrm{ref}}}

%*****************************

\section{Introduction}
\label{sec:introduction}
  \input{01_introduction}

%________________________________________________________________

\section{Overview of the Planck-LFI Instrument}
\label{sec:overview_lfi}

  \input{02_overview_lfi}

% %________________________________________________________________

\section{Overview of Planck-LFI test campaign}
\label{sec:test_campaign}

  \input{03_test_campaign}

%________________________________________________________________

\section{In-flight testing of Planck-LFI}
\label{sec:lfi_tests}

  \input{04_lfi_tests}

%________________________________________________________________

\section{Conclusions}
\label{sec:conclusions}

  \input{06_conclusions}
\acknowledgments
    \Planck\ is a project of the European Space Agency with instruments funded by ESA member states, and with special contributions from Denmark and NASA (USA). The \Planck\ LFI project is developed by an International Consortium lead by Italy and involving Canada, Finland, Germany, Norway, Spain, Switzerland, UK, USA. The Italian contribution to \Planck\ is supported  by the Italian Space Agency (ASI). The work in this paper has been supported by in the framework of the ASI-E2 phase of the \Planck\ contract. The US \Planck\ Project is supported by the NASA Science Mission Directorate. In Finland, the \Planck\ LFI 70 GHz work was supported by the Finnish Funding Agency for Technology and Innovation (Tekes). 

During the ground tests and the full in-flight operations, the LFI instrument team is supported by the HFI team, in particular for what regards the cryogenic chain operations, by the ESA Mission Operation Centre, by the ESA Planck Science Office and by the ESA Planck Project. 

\clearpage

\appendix
\appendixpage
   \input{a04_groups_table}
   \input{a02_cryo01_sequence}
   \input{a01_spike_plots}
   \input{a03_drain_current}
   \input{a06_PH_SW_tuning_plots}
   \input{a05_test_details}
   \input{a06_tuning_plots}
   \input{a07_tuning_nonlin_plots}
   \input{a08_acronyms}

\newpage

\bibliographystyle{pippo}
\bibliography{prelaunch_earlypapers,custom}
\end{document}

%% file: journaldefs.tex
\def\aap{A\&A}%
\def\aapr{A\&A~Rev.}%
\def\aaps{A\&AS}%

%% file: customdefs.tex
\def\sp{\hspace{.2cm}}

%% file: 00_title.tex
\title{In-flight calibration and verification of the Planck-LFI instrument}

   \author{
        A.  Gregorio$^{a,b}$\thanks{Corresponding author},
        F.  Cuttaia$^c$,
        A.  Mennella$^d$,
        M.  Bersanelli$^d$,
        M.  Maris$^b$,        
        P.  Meinhold$^e$,	
        M.  Sandri$^c$, 
        L.  Terenzi$^c$,
        M.  Tomasi$^d$,        
        F.  Villa$^d$,                  
        M.  Frailis$^{b}$,
        G.  Morgante$^c$,
        D.  Pearson$^f$,
        A.  Zacchei$^{b}$,         
        P.  Battaglia$^d$,
	R.C. Butler$^c$,
        R.  Davis$^g$, 
        C.  Franceschet$^d$,
        E.  Franceschi$^c$, 
        S.  Galeotta$^{b}$,
        R.  Leonardi$^h$,
        S.  Lowe$^j$,
        N.  Mandolesi$^c$,
        F. Melot$^i$, 
        L. Mendes$^{h}$,
        P. Stassi$^i$,   
        L. Stringhetti$^c$,
        D. Tavagnacco$^{a,b}$,       
        A. Zonca$^{e}$,          
        A. Wilkinson$^g$,   
        P. Wilson$^f$,
        M. Charra$^k$, T. Maciaszek$^l$,
        S. Foley$^m$, C.J. Watson$^m$,
        M. Casale$^{h}$, R. Laureijs$^{h}$, J. Tauber$^{h}$, D. Texier$^{h}$, 
        M. Baker$^n$, L. Perez Cuevas$^n$, M. Krassenburg$^n$, P. Rihet$^o$ \\
    \llap{$^a$}Universit\`a degli Studi di Trieste, Dipartimento di Fisica, 
    Via Valerio 2, 34127 Trieste, Italy.\\
   \llap{$^b$}INAF-OATS Trieste, 
    Via Tiepolo 11, 34143 Trieste, Italy.\\
   \llap{$^c$}INAF-IASF Bologna, 
    Via P.Gobetti 101, 40129 Bologna, Italy.\\
   \llap{$^d$}Universit\`a degli Studi di Milano, Dipartimento di Fisica, 
   Via Celoria 16, 20133 Milano, Italy.\\
    \llap{$^e$}Department of Physics, University of California (UCSB), Santa Barbara, California, USA.\\ 
    \llap{$^f$}Jet Propulsion Laboratory (JPL), Pasadena, California, USA.\\
    \llap{$^g$}Jodrell Bank Observatory (JBO), Manchester, UK.\\
    \llap{$^h$}Planck Science Office (PSO), European Space Astronomy Centre, Madrid, Spain.\\ 
    \llap{$^i$}Laboratoire de Physique Subatomique et de Cosmologie (LPSC), UJF Grenoble 1, CNRS/IN2P3, INPG, Grenoble, France. \\
    \llap{$^j$}Las Cumbres Observatory Global Telescope (LCOGT).\\
    \llap{$^k$}Institut d’Astrophysique Spatiale (IAS), Universite Paris-Sud, Paris, France.\\
    \llap{$^l$}CNES Centre National d'\'Etudes Spatiales (CNES), France. \\
   \llap{$^m$}Mission Operations Centre (MOC), European Space Operation Centre, Darmstadt, Germany.\\
    \llap{$^n$}Planck Project, ESTEC, Keplerlaan 1, Noordwijk, The Netherlands.\\  
    \llap{$^o$}Thales Alenia Space, Boulevard du midi 100, 06150 Cannes, France.\\  
    Email: \email{Anna.Gregorio@ts.infn.it}
    }

   \abstract{
In this paper we discuss the Planck-LFI in-flight calibration campaign. After a brief overview of the ground test campaigns, we describe in detail the calibration and performance verification (CPV) phase, carried out in space during and just after the cool-down of LFI. We discuss in detail the functionality verification, the tuning of the front-end and warm electronics, the preliminary performance assessment and the thermal susceptibility tests. The logic, sequence, goals and results of the in-flight tests are discussed. All the calibration activities were successfully carried out and the instrument response was comparable to the one observed on ground. For some channels the in-flight tuning activity allowed us to improve significantly the noise performance.
    }

    \keywords{Instruments for CMB observations; Space instrumentation; Microwave radiometers; Instrument optimisation}

%% file: 01_introduction.tex
The Low Frequency Instrument (LFI) is an array of 22 coherent differential receivers in Ka, Q and V bands on board the European Space Agency \Planck\ satellite \cite{tauber2010, mandolesi2010}. The LFI shares the \Planck\ telescope focal plane with the High Frequency Instrument (HFI), a bolometric array in the 100-857~GHz range cooled to 0.1~K \cite{lamarre2010}. 
The \Planck\  full-sky measurements from the Lagrangian point L2 will provide cosmic variance- and foreground-limited measurements of the Cosmic Microwave Background (CMB) by scanning the sky in almost great circles with a 1.5~m dual reflector aplanatic telescope \cite{tauber2010a, 2002_villa_planck_telescope,2005_dupac_planck_scanning_strategy,2006_maris_planck_scanning_strategy}. 

After being successfully launched on May, 14$^{\rm th}$ 2009 with the infrared \textit{Herschel} satellite, \Planck\ was transferred to its final orbit around L2. A series of tests were performed during the first three months in the so-called \textit{calibration, performance and verification} (CPV) phase, aimed at verifying functionality, tuning of instrument parameters and assessing calibration and scientific performance. Since the start of nominal operations, \Planck\ has scanned the entire sky seven times and has started the eighth survey as we write. 
The measured LFI and HFI in-flight scientific performance meets all ground expectations making \Planck\ the most sensitive CMB space experiment to date \cite{mennella2011,planck2011-1.5}. 

In this paper we discuss the strategy adopted for the \Planck-LFI CPV campaign and provide detailed results of tuning, calibration, and verification tests that were key in meeting the challenging design performance. The cryogenic nature of the spacecraft and instrument required a rather complex operation scheme. This paper, therefore, may constitute a valuable source of information and experience for the execution of in-flight calibration campaigns in future CMB space missions.

After a brief overview of the LFI instrument (see Section~\ref{sec:overview_lfi}), in Section~\ref{sec:test_campaign} we give an overview of the LFI test campaigns, of the cooldown sequence, and of the CPV rationale. Then in Section~\ref{sec:lfi_tests}, the heart of this paper,  we discuss in detail the methodology and results of the main tests performed on LFI during the CPV phase. Summary and conclusions are provided in Section~\ref{sec:conclusions}.

%% file: 02_overview_lfi.tex
\subsection{Receiver schematics and signal model}
\label{sec_receiver_schematics_signal_model}

    The \Planck-LFI instrument is an array of 11 radiometric receivers in the Ka, Q and V bands, with centre frequencies close to 30, 44 and 70~GHz. The exact centre frequencies for each receiver are reported in \cite{mennella2011}; for simplicity, in this paper we will refer to the three channels using their nominal centre frequency. A detailed description of the LFI instrument is given in \cite{bersanelli2010}, and  references therein. Here we outline the main instrument features that are essential for a self-consistent discussion of the CPV tests. 

%     Best LFI noise performance is obtained with receivers based on High Electron Mobility Transistor (HEMT) amplifiers cryogenically cooled at 20~K by the Planck Sorption Cooler, a vibration-less hydrogen cooler providing more than 1~W of cooling power at 20~K. To optimise noise performance and cooling power the RF amplification is divided between a 20~K front-end unit and a $\sim$300~K back-end unit connected by composite waveguides~\cite{bersanelli2010}.

\begin{figure}
        \begin{center}
            \includegraphics[width=15cm]{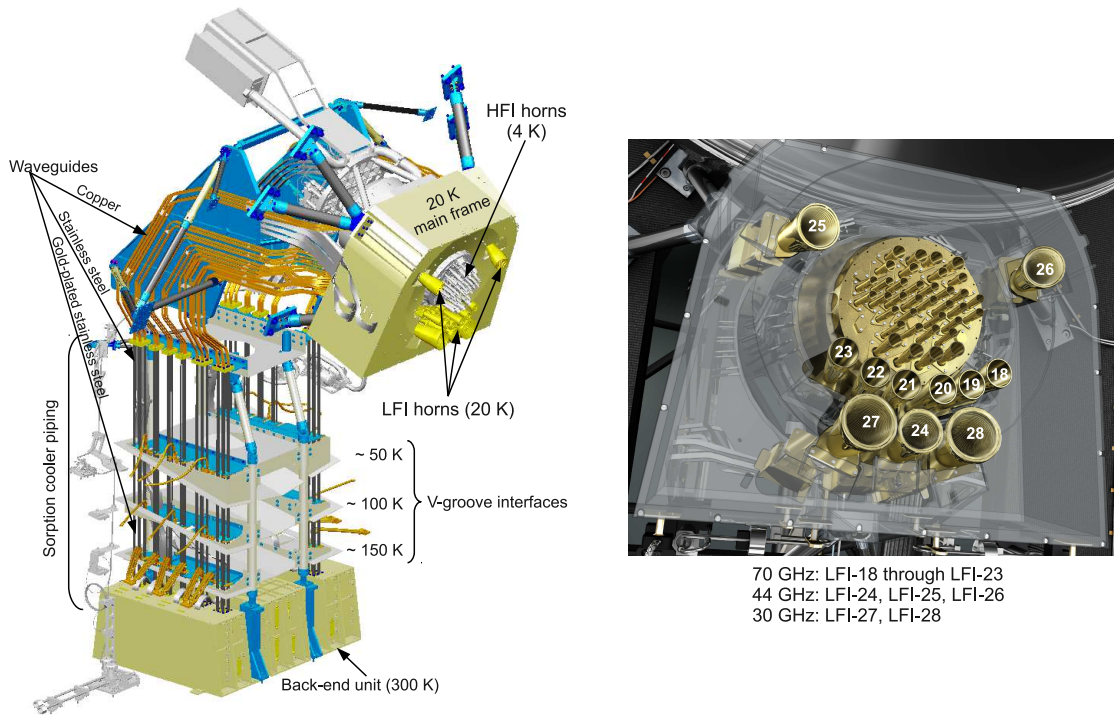}
        \end{center}
        \caption{Left panel: the LFI instrument with main thermal stages, focal plane, waveguides and sorption cooler piping highlighted. Right panel: labelling of feed horns on the LFI focal plane.}
        \label{fig_lfi_instrument}
    \end{figure}

\begin{figure}
        \begin{center}
            \includegraphics[width=15cm]{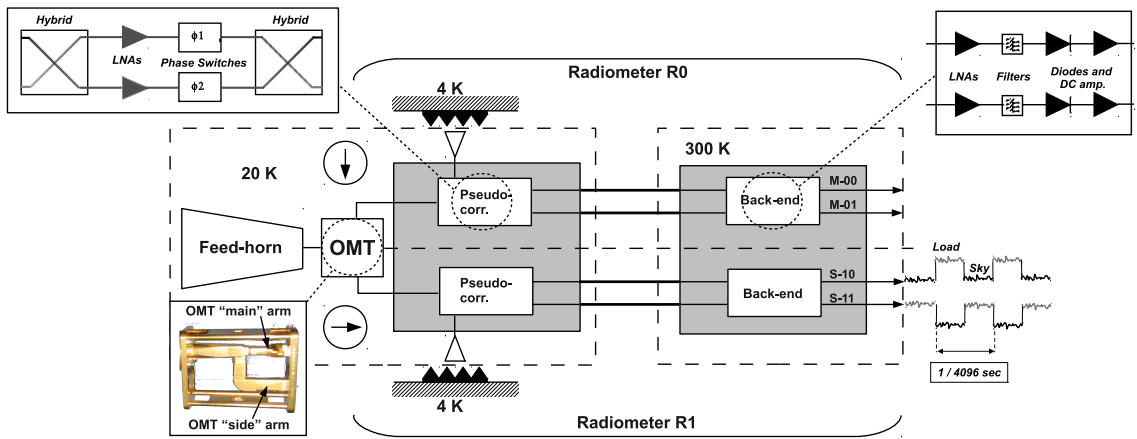}
        \end{center}
        \caption{A complete RCA from feed-horn to analog voltage output. The insets show the OMT, the details of the 20~K pseudo-correlator and of the back-end radio-frequency amplification, low-pass filtering, detection and DC amplification.}
        \label{fig_rca_schematic}
    \end{figure}

    The instrument (Figure~\ref{fig_lfi_instrument}) consists of a $20$~K focal plane unit (FPU) hosting the corrugated feed horns, the orthomode transducers (OMTs) and the receiver front-end modules (FEMs). Fourty four composite waveguides \cite{d'arcangelo2009a} are interfaced with three conical thermal shields and connect the front-end modules to the warm ($\sim 300$~K) back-end unit (BEU) containing a further radio frequency amplification stage, detector diodes and all the electronics for data acquisition and bias supply. Every radiometer chain assembly (RCA) consists of two radiometers, each feeding two diode detectors (Figure~\ref{fig_rca_schematic}), for a total of 44 detectors. The 11 RCAs are labelled by a numbers from 18 to 28 as outlined in the right panel of Figure~\ref{fig_lfi_instrument}.

    Figure~\ref{fig_rca_schematic} provides a more detailed description of each radiometric receiver. In each RCA, the two perpendicular linear polarisation components split by the OMT propagate through two independent pseudo-correlation differential radiometers, labelled as \texttt{M} or \texttt{S} depending on the arm of the OMT they are connected to (``Main'' or ``Side'', see lower-left inset of Figure~\ref{fig_rca_schematic}).
    In each radiometer the sky signal coming from the OMT output is continuously compared with a stable 4~K blackbody reference load mounted on the external shield of the HFI 4~K box \cite{valenziano2009}. After being summed by a first hybrid coupler, the two signals are amplified by $\sim 30$~dB, see upper-left inset of Figure~\ref{fig_rca_schematic}. The amplifiers were selected for best operation at low drain voltages and for gain and phase match between paired radiometer legs, which is crucial for good balance. Each amplifier is labelled with codes \texttt{1}, \texttt{2} so that the four outputs of the low noise amplifiers (LNAs) can be named with the sequence: \texttt{M1}, \texttt{M2} (radiometer \texttt{M}) and \texttt{S1}, \texttt{S2} (radiometer \texttt{S}). 
Tight mass and power constraints called for a simple design of the data acquisition electronics (DAE) box so that power bias lines were divided into five common-grounded power groups with no bias voltage readouts; only the total drain current flowing through the front-end amplifiers is measured and is available to the house-keeping telemetry\footnote{This design has important implications for front-end bias tuning, which depends critically on the satellite electrical and thermal configuration and was repeated at all integration stages, during on-ground and in-flight satellite tests.}.
A phase shift alternating between 0\deg and 180\deg at the frequency of 4~kHz is applied in one of the two amplification chains and then a second hybrid coupler separates back the sky and reference load components that are further amplified and detected in the warm BEU, with a voltage output ranging from $-$2.5~V to $+$2.5~V.
    Each radiometer has two output diodes which are labelled with binary codes \texttt{00}, \texttt{01} (radiometer \texttt{M}) and \texttt{10}, \texttt{11} (radiometer \texttt{S}), so that the four outputs of each radiometric chain can be named with the sequence: \texttt{M-00}, \texttt{M-01}, \texttt{S-10}, \texttt{S-11}.

    After detection, an analog circuit in the DAE box removes a programmable offset in order to obtain a nearly null DC output voltage and a programmable gain is applied to increase the signal dynamics and optimally exploit the analogue-to-digital converters (ADC) input range. After the ADC, data are digitally downsampled, requantised and compressed in the radiometer electronics box assembly (REBA) according to a scheme described in \cite{herreros2009,maris2009} before preparing telemetry packets. On ground, telemetry packets are converted to sky and reference load time ordered data after calibrating the analogue digital units (ADU) samples into volt considering the applied offset and gain factors.

    To first order, the mean differential power output for each of the four receiver diodes can be written as follows \cite{seiffert2002,mennella2003,bersanelli2010}:
    \begin{equation}
        P_{\rm out}^{\rm diode} = a\, G_{\rm tot}\,k\,\beta \left[ T_{\rm sky} + T_{\rm noise} - r\left(
            T_{\rm ref} + T_{\rm noise}\right) \right],
        \label{eq_p0}
    \end{equation}
    where $G_{\rm tot}$ is the total gain, $k$ is the Boltzmann constant, $\beta$ the receiver bandwidth and $a$ is the diode constant. $T_{\rm sky}$ and $T_{\rm ref}$ are the average sky and reference load antenna temperatures at the inputs of the first hybrid and $T_{\rm noise}$ is the receiver noise temperature.
    The gain modulation factor \cite{mennella2003,zacchei2011}, $r$, is defined by:
    \begin{equation}
        r = \frac{T_{\rm sky} + T_{\rm noise}}{T_{\rm ref} + T_{\rm noise}},
        \label{eq_r}
    \end{equation}
    and is used to balance (in software) the temperature offset between the sky and reference load signals and minimise the residual 1/$f$ noise in the differential datastream. This parameter is calculated from the average uncalibrated total power data using the relationship:
    \begin{equation}
        r = \langle V_{\rm sky} \rangle/ \langle V_{\rm ref}\rangle,
        \label{eq_r_v}
    \end{equation}
where $<V_{\rm sky}>$ and $<V_{\rm ref}>$ are the average sky and reference voltages calculated in a defined time range.
    The white noise spectral density at the output of each diode is essentially
    independent from the reference-load absolute temperature and is
    given by:
    \begin{equation}
        \Delta T_0^{\rm diode} = \frac{2\,(T_{\rm sky}+T_{\rm noise})}{\sqrt{\beta}}.
        \label{eq_deltat_diode_ideal}
    \end{equation}
\subsection{Phase switches commanding and configuration}
\label{sec_phase_switch_commanding}

\begin{figure}[thb!]
        \begin{center}
            \includegraphics[width=12cm]{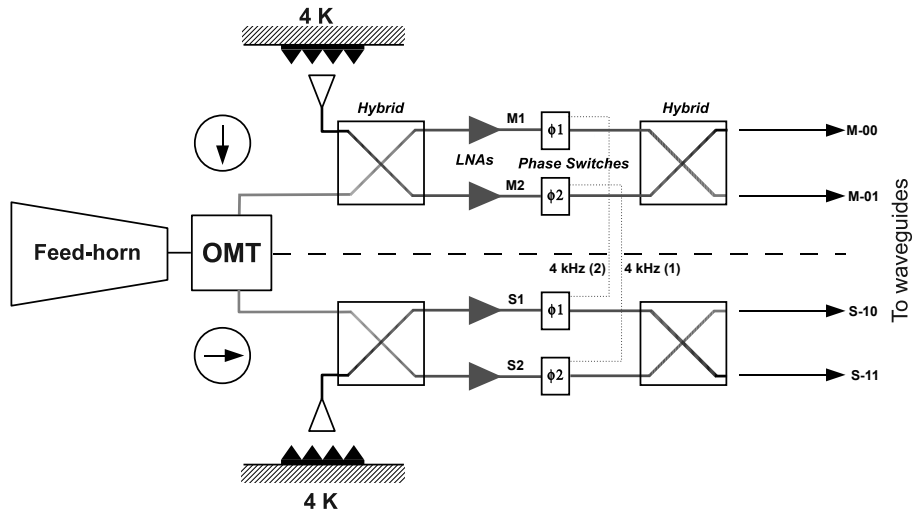}
            \caption{\label{fig_phase_switch_operation} Close-up of the two FEMs of an RCA. There are four phase switches, each one can be fixed in one of its two positions (labelled as 0, 1) or switch at 4~kHz between 0 and 1. Phase switches are clocked and biased by the DAE in pairs  (see text and Table 17 for further explanation).}
        \end{center}
    \end{figure}

    In Figure~\ref{fig_phase_switch_operation} we show a close-up of the two front end modules of an RCA with the four phase switches which are labelled according to the LNA they are coupled to (\texttt{M1}, \texttt{M2}, \texttt{S1}, \texttt{S2}). Each phase switch is characterised by two states: state 0 (no phase shift applied to the incoming wave) and state 1 (180$^\circ$ phase shift applied) and can either stay fixed in a state or switch at 4~kHz between the two states.

    Phase switches are clocked and biased by the DAE and their configuration can be programmed via telecommand. In order to simplify the instrument electronics, phase switches are configured and operated in pairs: by convention they are labelled \texttt{A/C} and \texttt{B/D}. This means that if phase switches \texttt{A/C} are switching at 4~kHz then \texttt{B/D} are fixed both in the same state (either 0 or 1). This simplification, required during the design phase to comply with mass and power budgets, comes at the price of loosing some setup redundancy. 

The correspondance between the phase switch labels (\texttt{A/C}, \texttt{B/D}) and the corresponding LNA names (\texttt{M1}, \texttt{M2}, \texttt{S1}, \texttt{S2}), and, as a consequence, the back end module (BEM) output diodes (\texttt{M-00}, \texttt{M-01}, \texttt{S-10}, \texttt{S-11}) is not the same for all RCAs. For the details the interested reader can refer to Table~\ref{tab_phsw_corr} in Appendix~\ref{app_PH_SW_plots}.

%% file: 03_test_campaign.tex
During its development, the LFI flight model was calibrated and tested at various integration levels from sub-systems~\cite{davis2009,artal2009,varis2009} to individual integrated receivers~\cite{villa2010} and the whole receiver array~\cite{mennella2010}. In every campaign we performed tests according to the following classification:
\begin{itemize}
    \item \textbf{Functionality tests}, performed to verify the instrument functionality.
    \item \textbf{Tuning tests}, to tune radiometer parameters (biases, DC electronics gain and offset, digital quantisation and compression) for optimal performance in flight-like thermal conditions.
    \item \textbf{Basic calibration and noise performance tests}, to characterise instrument performance (photometric calibration, isolation, linearity, noise and stability) in tuned conditions.
    \item \textbf{Susceptibility tests}, to characterise instrument susceptibility to thermal and electrical variations.
\end{itemize}

Where possible, the same tests were repeated in several test campaigns, in order to ensure enough redundancy and confidence in the instrument behaviour repeatability. A critical comparison, that is central to the subject of this work, is the one between the results of on-ground and in-flight test campaigns. A matrix showing the instrument parameters measured in the various test campaigns is provided in Table 1 of \cite{mennella2010}. In the next sections we briefly summarise the on-ground test activities and then provide an overview of the tests carried out during CPV.

\subsection{Ground tests}
\label{sec:ground_tests}
    The ground test campaign was developed in three main phases: cryogenic tests on the individual RCAs, cryogenic tests on the integrated receiver array (the so-called radiometer array assembly, RAA) and system-level tests after the integration of the LFI and HFI instruments onto the satellite. The first two phases were carried out at the Thales Alenia Space - Italia laboratories located in Vimodrone (Milano, Italy)\footnote{Receiver tests on 70 GHz RCAs were carried out in Finlad, at Yilinen laboratories.}, system level tests were conducted in a dedicated cryofacility at the Centre Spatiale de Li\'ege (CSL) located in Li\'ege (Belgium). 
    
    In Table~\ref{tab_temperatures_ground_tests} we list the temperature of the main cold thermal stages during ground tests compared to in-flight nominal values. These values show that system-level tests were conducted in conditions that were as much as possible flight-representative, while results obtained during RCA and RAA tests needed to be extrapolated to flight conditions to allow comparison. Details about the RCA test campaign are discussed in \cite{villa2010} while the RAA tests and the extrapolation methods are presented in \cite{mennella2010}.
    \begin{table}[h!]
        \begin{center}
            \caption{Temperatures of the main cold stages during the various ground test campaigns compared to in-flight nominal values.}
            \label{tab_temperatures_ground_tests}
            \begin{tabular}{l | c | c c c}
                \hline
                \hline
                Temperature     & \texttt{Nominal}        & \texttt{RCA tests}      & \texttt{RAA tests}         & \texttt{System-level} \\
                \hline
                Sky             & $\sim 3$\,K    & $\gtrsim 8$\,K & $\gtrsim 18.5$\,K & $\sim 4$\,K  \\
                Ref. load       & $\sim 4.5$\,K  & $\gtrsim 8$\,K & $\gtrsim 18.5$\,K & $\sim 4.5$\,K  \\
                Front-end unit  & $\sim 20$\,K   & $\sim 20$\,K   & $\sim 26$\,K      & $\sim 20$\,K \\
                \hline
            \end{tabular}
        \end{center}
    \end{table}

    Cryogenic system-level tests were split into three parts:
    \begin{itemize}
        \item \textbf{Thermal balance}, to validate the overall thermal mathematical model. The ground vacuum test equipment simulated the space environment.
        \item \textbf{System cryogenic test}, to check and optimise the satellite and instrument performance working at nominal temperatures.
        \item \textbf{Thermal cycling test}, to check the reliability of all electronics equipment and instruments to temperature variations\footnote{To allow the thermal control of the Service Vehicle Module (SVM), the warm part of the spacecraft was surrounded by a cryogenic shroud at a temperature lower than 100~K. The temperature of the shrouds was adjustable in order to allow the thermal cycling of the SVM during the tests. The temperature ranged between the two extreme values of 293~K to 373~K.}. 
    \end{itemize}

    During the various test campaigns the instrument was switched off and moved several times in a time period of about three years. A series of functional tests were always repeated at each location and also in flight, in order to verify the instrument functionality and the response repeatability. No failures or major problems have been identified due to transport and integration procedures.
    
\subsection{In-flight calibration, performance and verification}
\label{sec:cpv}
        
    The \Planck\ cryo-chain \cite{planck2011-1.3} is composed by three coolers: a 20\,K Hydrogen sorption cooler, passively pre-cooled by the 3$^{\rm rd}$ V-groove radiator, cools the LFI focal plane and pre-cools the HFI 4\,K cooler; a Stirling 4\,K cooler that cools the HFI box and feed-horns and provides a 4\,K blackbody reference signal to the LFI receivers; a 0.1\,K dilution cooler, which is pre-cooled by the 4\,K cooler and cools the HFI bolometer filters to $\sim$1.6\,K and the HFI bolometer detectors to $\sim$0.1\,K. The cooldown of the HFI 4\,K stage, in particular, was key during CPV for the LFI as it provided a variable input signal that was exploited during bias tuning (see Section~\ref{sec:lfi_LNAs_Tun}).

    The LFI CPV started on June, 4$^{\rm th}$ 2009 and lasted until August, 12$^{\rm th}$ when \Planck\ started scanning the sky in nominal mode. At the onset of CPV, the active cooling started when the radiating surfaces on the payload module reached their working temperatures ($\sim$50\,K on the 3$^{\rm rd}$ V-groove, and $\sim$40\,K on the reflectors) by passive cooling. This was achieved during the transfer phase. Nominal temperatures were achieved on July, 3$^{\rm rd}$ 2009, when the dilution cooler temperature reached 0.1\,K~\cite{planck2011-1.1,planck2011-1.3} (see Figure~\ref{fig_Funct_tests_schedule}).

    The CPV was carried out in four phases (see Table~\ref{tab:tests} for a summary of the overall CPV test campaign): (i) LFI switch on and basic functionality verification, (ii) tuning of front-end biases and back-end electronics, (iii) preliminary calibration tests and (iv) thermal tests. 

\begin{itemize}
	 \item {LFI switch on and basic functionality tests} aimed at verifying functionality either of warm  (DAE and REBA) and cold electronics. Functionality tests were performed just after the switch on and periodically during CPV to verify the correct instrument behaviour.

  \item {Tuning} tests were performed after the first functional tests to find and set the instrument parameters for optimal scientific performance. Front-end unit biases were tuned exploiting the varying input signal provided by the cooldown of the 4\,K cooler to calculate noise temperature and isolation for a large set of bias voltages. Afterwards, the DAE gain and offset parameters were tuned to optimise the DC voltage output at the ADC input range. Finally, optimal REBA parameters were sought to find the best trade-off between telemetry allocation and noise resolution in quantised data.

  \item {Calibration} tests provided a preliminary photometric calibration using the CMB dipole and an estimate of the main scientific instrument performance (white noise sensitivity, $1/f$ noise). 

  \item {Thermal} tests were carried out to characterize thermal and radiometric transfer functions that couple the radiometric response to temperature variation at different locations in the focal plane. This was done by acquiring scientific and housekeeping data with the sorption cooler temperature stabilisation assembly (TSA) turned off in order to increase the level of temperature fluctuations and amplify the effect.

\end{itemize}

    \begin{table*}
        \begin{center}
            \caption{List of CPV tests with objectives and references to relevant paper sections}
            \label{tab:tests}
            \begin{small}
            \begin{tabular}{l l l c}
        \hline
        \hline
        Phase & \texttt{Test}  & \texttt{Objective} & \texttt{Ref.} \\ \hline
        \textit{LFI switch on} & Switch on & Verify functionality of electronics units & \ref{sec:lfi_on_basic}	\\
\textit{Functionality} & \texttt{CRYO--01}& Switch on LFI front-end unit (FEU) and verify basic & \ref{sec:lfi_on_basic}	\\
        &                         & functionality of all front-end components &	\\
        & \texttt{CRYO--02}& Verify effectiveness of pseudo-correlation in reducing& \ref{sec:lfi_on_basic}	
        \\
        &                         & $1/f$ noise\\
        & Spike tests    & Characterise 1\,Hz spurious spikes  & \ref{sec:lfi_spike}	\\
        & Drain current test & Characterise LNAs I--V response & \ref{sec:lfi_drain}\\
        & Reference test & Set a functionality reference point & \ref{sec:lfi_ref} \\
        \hline
  {\textit{Tuning} } & Stability check & Verify instrument response stability prior to hypermatrix & \ref{sec:lfi_stab}	\\     
  &   Pre-tuning  & Constrain LNA bias space prior to hypermatrix tuning&  \ref{sec:lfi_LNAs_Tun}	\\
        & Phase switch tuning& Tune phase switch currents for optimal balance & \ref{sec:lfi_ps}\\
        & Hypermatrix tuning & Find the optimal LNA bias configuration by exploring & \ref{sec:lfi_LNAs_Tun}	\\
        &                         & the bias space $[V_{\rm g1}^{\rm LNA_1},V_{\rm g2}^{\rm LNA_2},V_{\rm g1}^{\rm LNA_1},V_{\rm g2}^{\rm LNA_2}]$.    &  \\
        &                         & The bias space defined during pre-tuning is scanned 4 &  \\
        &                         & times at different temperatures of the 4\,K stage. & \\
        &                         & For each bias quadruplet noise temperature and & \\
        &                         & isolation are calculated & \\
        & Tuning verification    & Verify optimal biases obtained from hypermatrix tuning & \ref{sec:lfi_ver}	\\
        & DAE tuning      & Optimise DC voltage output from each detector to the & \ref{sec:lfi_dae}	\\
        &                         & ADC input & 	\\
        & REBA tuning   & Optimise digital quantisation and compression  & \ref{sec:rebatuning}	\\
        \hline
        \multirow{1}{*}{\textit{Calibration}} & Noise properties & Preliminary assessment of noise performance and &        \ref{sec:calib} \\
        &                         & photometric calibration &\\
        \hline
        \multirow{1}{*}{\textit{Thermal}}     & Dyn. thermal model\dotfill & Assess FPU thermal damping & \ref{sec:thermal}	 	\\
        & Susceptibility & Assess radiometric thermal transfer functions & \ref{sec:thermal}	\\
        \hline
        \end{tabular}
        \end{small}
        \end{center}
    \end{table*}

        The need for stability over long periods is a key aspect of the Planck surveys and may affect how the instruments are tuned and operated. The large and varied number of activities during most of CPV did not allow to achieve survey-like stability in routine conditions. The first light survey was a period of 15 days at the end of CPV, from August 12$^{\rm th}$ to August 26$^{\rm th}$ 2009, where such stability could be achieved and constituted the transition from CPV into Routine phase. In this paper we will not discuss the first light survey.

    \subsection{CPV Operations}
    \label{sec:cpv_overview}

        Since the start of \Planck\ operations, during the daily tele-communication period (DTCP) the satellite is in contact with the ground station, operations are implemented and data can be analysed in ``realtime'' directly at the ESA mission operation centre (MOC). During the CPV, DTCP nominal duration was 5 hours during which telecommand stacks were loaded for test activities that were carried on out of visibility (``timeline'' mode). Data were retrieved at the beginning of the following DTCP and arrived at the instrument data processing centre about 5 hours later.

        LFI switch-on and the first functional tests (\texttt{CRYO-01}, \texttt{CRYO-02} and drain current verification, see Section~\ref{sec:functionality}) were performed in real-time in order to promptly monitor the instrument behaviour and stop the procedure in case of contingency. Although few additional short tests (e.g., REBA tuning) were added in realtime to speed up data analysis approach, the bulk of the CPV tests were carried out in timeline mode.

        During the CPV, \Planck\ scanned the sky at 1 revolution per minute repointing the spin axis by 1$^\circ$ each day: this way it was possible to efficiently integrate and remove the sky signal, which facilitates noise analysis. Moreover several tests were run before digital quantisation and compression optimisation. In these cases data were acquired unquantised and uncompressed with a sampling rate reduced to 16\,Hz for all channels in order to fit within the allocated telemetry bandwidth. At the end of CPV, \Planck\ entered its nominal scanning strategy and data acquisition mode which has been maintained throughout the subsequent operations.

%% file: 04_lfi_tests.tex
\subsection{Functionality tests}
\label{sec:functionality}

\input{04_Functionality}

    \subsubsection{LFI switch on and basic functionality}
    \label{sec:lfi_on_basic}
        
        \input{04_lfi_switch_on_and_basic_functional_tests}

    \subsubsection{Spurious 1\,Hz spikes}
    \label{sec:lfi_spike}
    
        \input{04_lfi_tests_spikes}

    \subsubsection{Drain current verification}
    \label{sec:lfi_drain}

        \input{04_lfi_tests_drain_current}

    \subsubsection{Stability check}
    \label{sec:lfi_stab}

       \input{04_lfi_tests_stability_check}

    \subsubsection{Reference test}
    \label{sec:lfi_ref}

        \input{04_lfi_tests_reference_test}

% % % % % % % % % % % % % % % % % % % % % % % % % % % % % % % % % 

\subsection{Front-end electronics tuning}
\label{sec:fetuning}

    \subsubsection{Phase switch tuning}
    \label{sec:lfi_ps}

     \input{04_lfi_tests_psw_tuning}

    \subsubsection{LNAs tuning}
    \label{sec:lfi_LNAs_Tun}

    \input{04_lfi_tests_LNAs_Tuning}

     \subsubsection{Tuning verification}
     	\label{sec:lfi_ver}
			 \input{04_lfi_tests_tuning_verif}

% % % % % % % % % % % % % % % % % % % % % % % % % % % % % % % % % 

\subsection{Back-end electronics tuning}
\label{sec:betuning}

    \subsubsection{DAE tuning}
    \label{sec:lfi_dae}

    \input{04_lfi_tests_dae_tuning}

    \subsubsection{REBA tuning}
    \label{sec:rebatuning}

    \input{04_lfi_tests_reba_tuning}

\subsection{Performance and calibration}
\label{sec:calib}

 \input{04_lfi_tests_performance_calibration}

\subsection{Thermal susceptibility}
\label{sec:thermal}

 \input{04_lfi_tests_thermal_susceptibility}

%% file: 04_Functionality.tex
\begin{figure}[h!]
    \begin{center}
        \includegraphics[width=\textwidth]{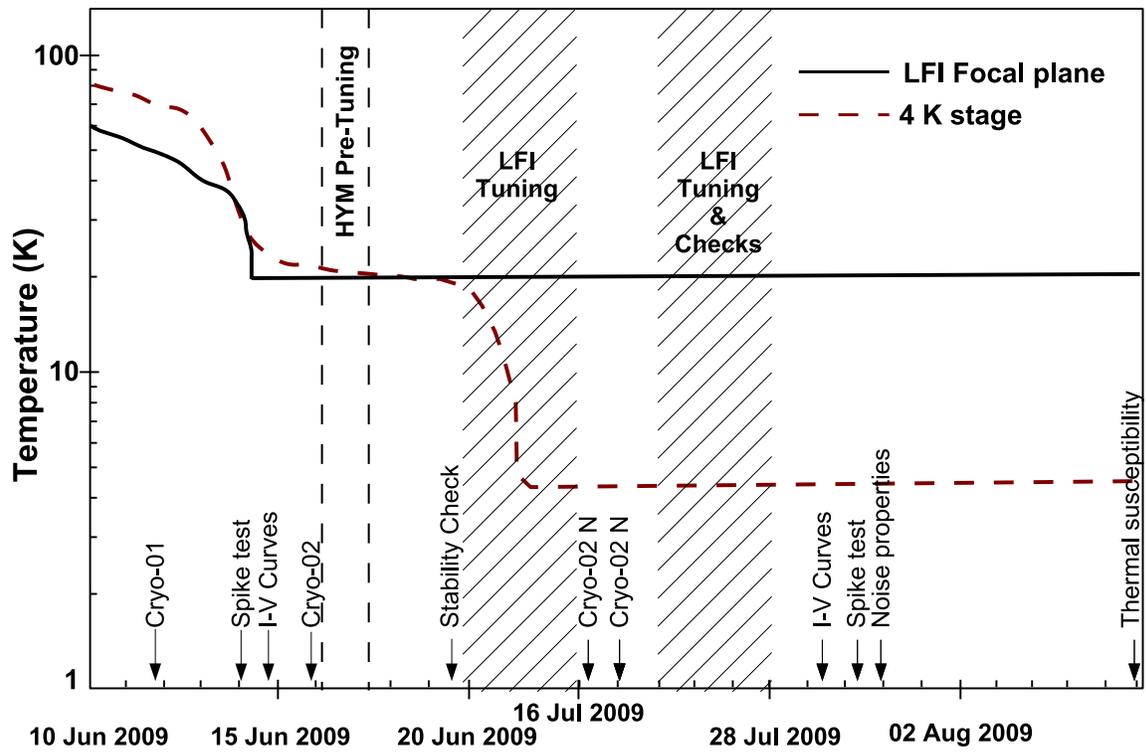}
    \end{center}
    \caption{Functional tests performed during CPV; the 4~K stage and the FPU Temperature are superimposed.}
    \label{fig_Funct_tests_schedule} 
\end{figure}

In this section we discuss the set of functional tests that were run immediately at switch on and several times during CPV when the instrument changed thermal configuration (see Figure~\ref{fig_Funct_tests_schedule}). Their main objectives were:
\begin{itemize}
    \item to verify communication between the analogue and digital electronics and between the instrument and the satellite,
    \item to check the electrical connection between the power supply units and the receiver active components (LNAs and phase switches),
    \item to verify the functionality of the receiver active components, 
    \item to assess the instrument stability in terms of voltage output and drain current,
    \item to measure the radiometer characteristic curves (drain current vs. voltage output) and compare them with ground measurements.
\end{itemize}

%% file: 04_lfi_switch_on_and_basic_functional_tests.tex
The LFI was switched on June, 4$^{\rm th}$ 2009. The procedure was carried out in two steps: during the first step the REBA was switched on and commissioned by verifying memory contents of its units (the Science and Data Processing Units), loading and starting the application software, and finally synchronising the REBA with spacecraft on board time. During the second step, the DAE was switched on and commissioned by verifying its main functionalities, synchronising the DAE clock with the spacecraft on board time, verifying its memory contents, and finally starting scientific data processing. At the end of this phase the LFI back-end unit was fully on.

\begin{figure}[h!]
   \begin{center}
        \includegraphics[width=10.0cm]{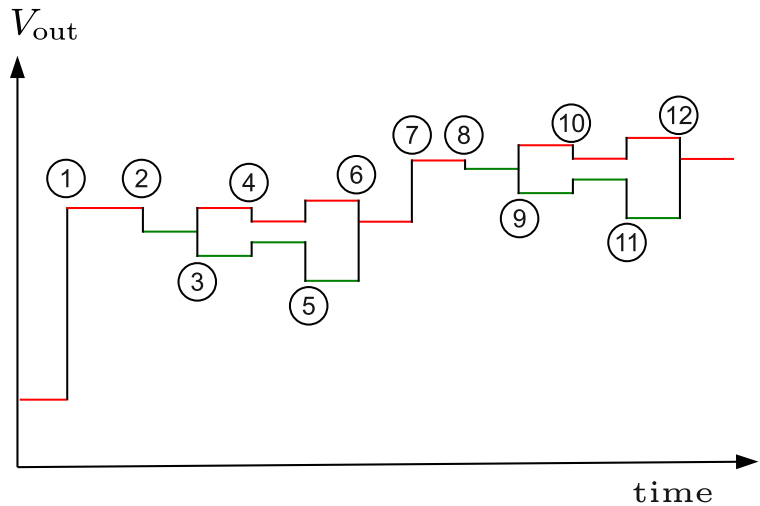}
        \caption[]{Expected voltage output pattern during the \texttt{CRYO-1} test, numbers correspond to the same reported Appendix~\ref{app_cryo01_sequence} in Table~\ref{tab_cryo01_sequence}. Red and green lines represent data labelled as ``sky'' or ``reference load'' samples. Steps showing just one line refer to the `4~kHz disabled' condition.}
        \label{fig_CRYO01_pattern}  
    \end{center}
\end{figure}

After commissioning, the LFI front-end modules were gradually biased, powering the LNAs and the corresponding phase switches one by one (\texttt{CRYO-01} test). The bias configurations were the same used on ground apart from \texttt{LFI18M1} and \texttt{LFI27M1}, that were biased with a lower drain voltage to reduce the risk of ADC saturation, and \texttt{LFI24M1}, for which the 
voltage was slightly changed  to avoid oscillations that were occasionally observed during ground tests. The sequence was chosen in order to minimise the electric cross-talk due to the common return resistance in the cryo harness (see Table~\ref{tab_power_groups_cryo01}, Appendix~\ref{app_power_groups_table}). The functionality was verified by checking that during the procedure the voltage output from each detector followed the expected pattern due to the sequential switching on of the FEM components. The details of the various steps are reported in Appendix~\ref{app_cryo01_sequence}, Table~\ref{tab_cryo01_sequence}, while the typical pattern followed is sketched in Figure~\ref{fig_CRYO01_pattern}.

\begin{figure}[h!]
    \begin{center}
        \includegraphics[width=11.4cm]{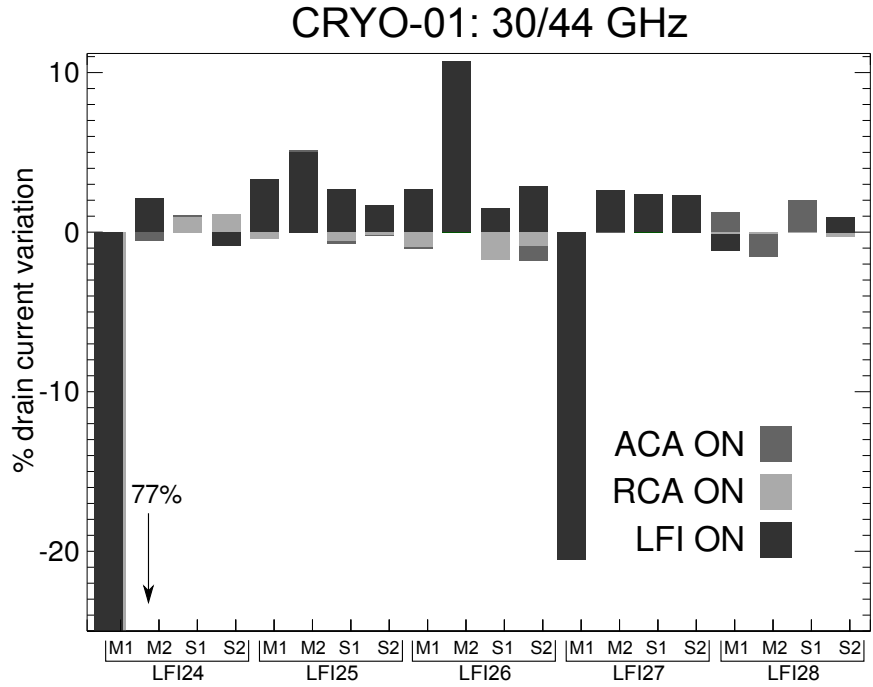} \\ \vspace{0.2cm}
 \hspace{-0.2cm}       \includegraphics[width=11.6cm]{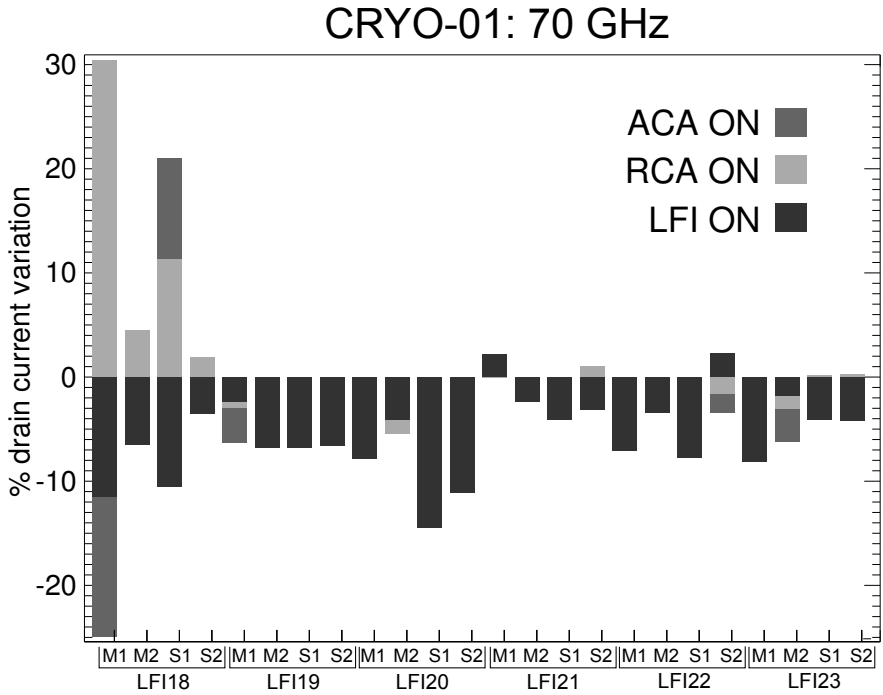}
        \caption{Comparison of drain currents measured during on-ground and in-flight \texttt{CRYO-1} tests. The percentage variation is defined by $(I_{\rm drain}^\mathrm{flight} - I_{\rm drain}^\mathrm{ground})/I_{\rm drain}^\mathrm{ground}$. The relative drain current variation is displayed for three different configurations: one amplifiers chain assembly (ACA) on, one RCA on, the whole LFI on. For each channel, the three data sets are stacked with different colors on the same bar. When the differences between the three steps are negligible, just one color is displayed.}
    \label{fig_CRYO01} 
    \end{center}
\end{figure}

We also compared the drain current measured in the various steps with the corresponding values measured on ground (see Figure~\ref{fig_CRYO01}). Although the focal plane temperature was different in the two cases (in the range [38.5\,K-35.8\,K] during \texttt{CRYO-01} performed on ground and in the range [36.1\,K- 20\,K] during the same in-flight test) drain current measurements were repeatable within 10\% in all cases, besides those channels for which the switch-on bias voltages was changed (\texttt{LFI24M1}, \texttt{LFI27M1},  \texttt{LFI18M1}).

After all the channels were nominally biased, we assessed instrument functionality and the effectiveness of the pseudo-correlation differential scheme in reducing 1/$f$ noise instabilities. To achieve this we checked that:
\begin{itemize}
   \item drain currents were comparable with those measured during ground tests ($\pm$5\%);
   \item 1/$f$ noise knee frequency\footnote{The knee frequency is defined as the frequency at which the $1/f$ and white noise contribute equally in power.}, $f_{\rm knee}$, was reduced to less than $\sim$1\,Hz after sky-reference load differencing\footnote{It is worth highlighting that this test was run in non nominal conditions (before tuning, with unstable 4\,K reference loads at $\sim$20\,K). Therefore we did not require the knee frequency to comply with scientific requirements.}.
\end{itemize}

The \texttt{CRYO-02} test consisted of a two-hours data acquisition split into four 30-minute steps in which the radiometers were operated in switching mode testing all the four possible phase switch configurations in series (see Table~\ref{tab_cryo02_test}).
\begin{table}[h!]
    \begin{center}
        \caption{\label{tab_cryo02_test} The four \texttt{CRYO-02} steps with the corresponding phase switch configurations; ``sw.'' indicates the 4\,kHz switching.}
        \vspace{.2cm}
        \begin{tabular}{c c c}
            \hline
            \hline
            Step n. & \texttt{A/C} & \texttt{B/D} \\
            \hline
                1   & sw. & 0 \\
                2   & sw. & 1 \\
                3   & 0 & sw. \\
                4   & 1 & sw.\\
            \hline
        \end{tabular}
    \end{center}
\end{table}

\begin{figure}[h!]
    \begin{center}
        \includegraphics[width=7.3cm]{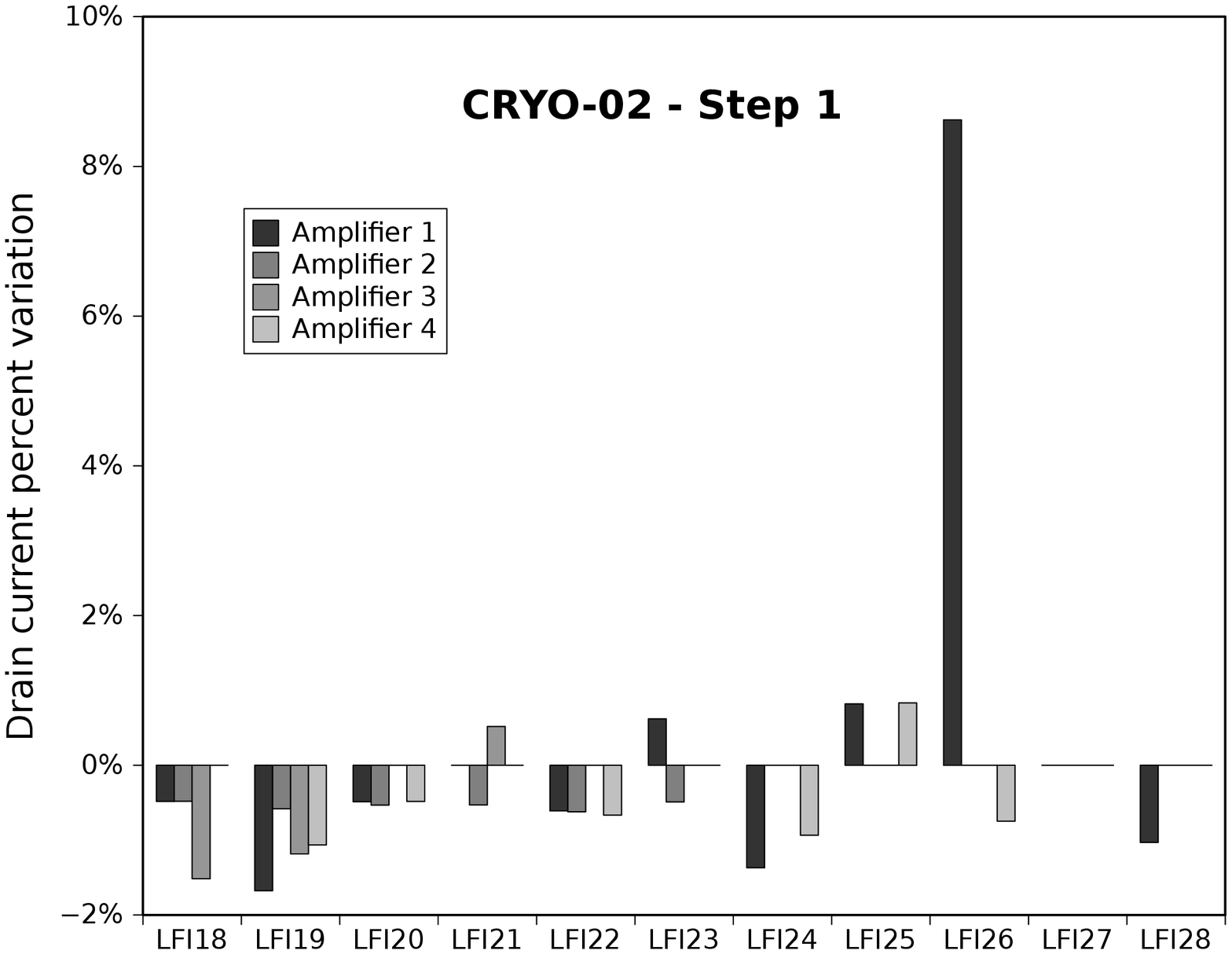}
        \includegraphics[width=7.3cm]{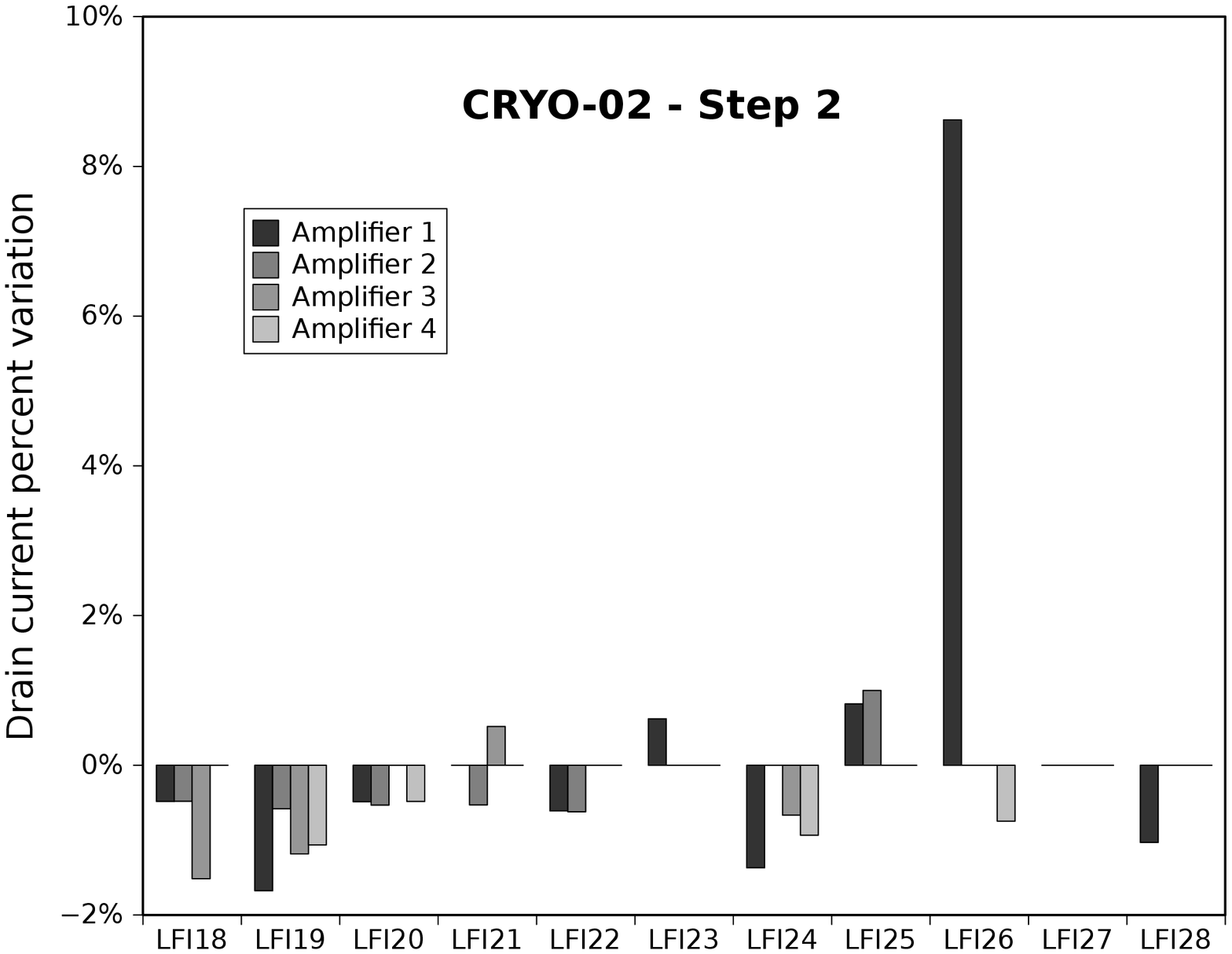}\\
        \includegraphics[width=7.3cm]{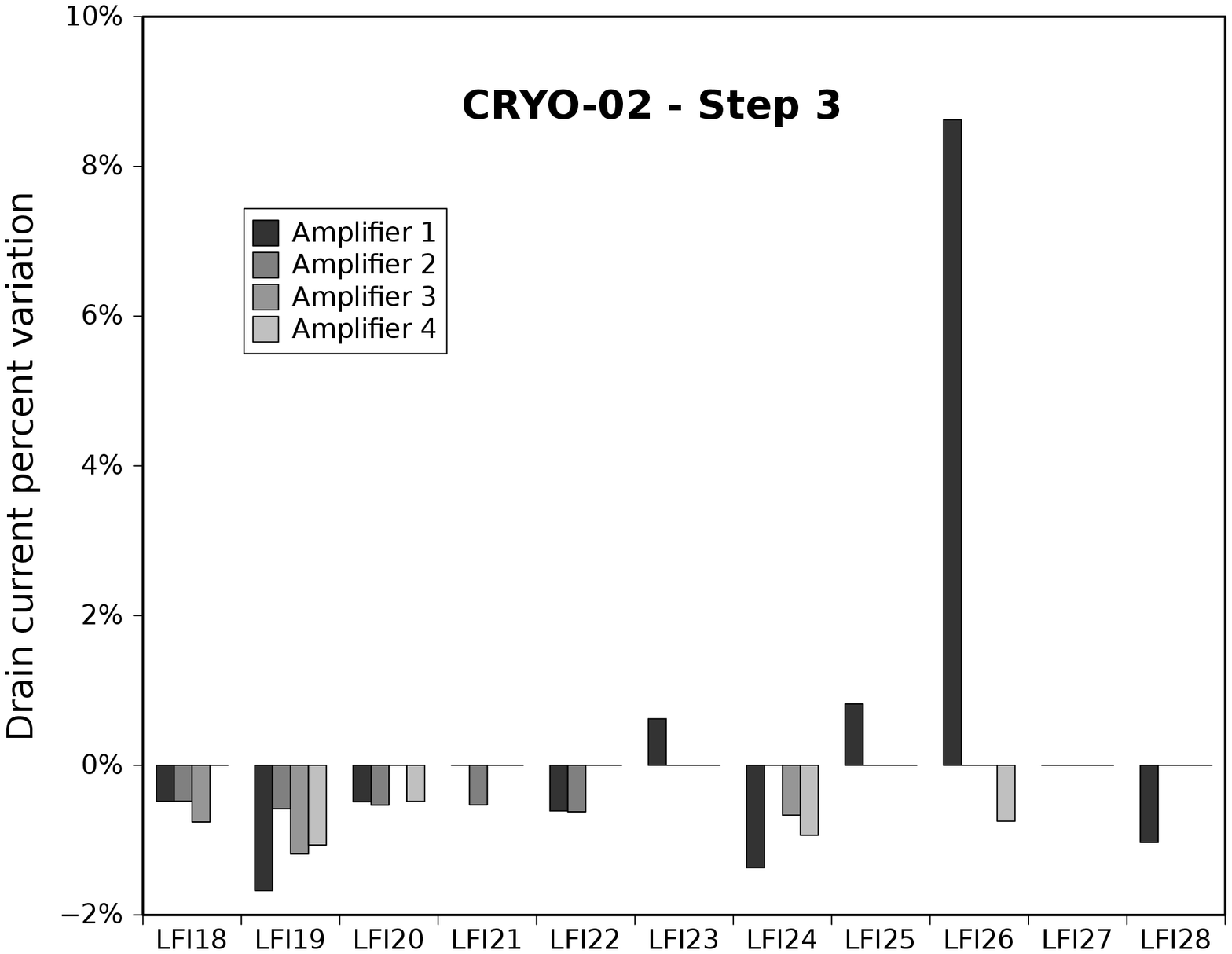}
        \includegraphics[width=7.3cm]{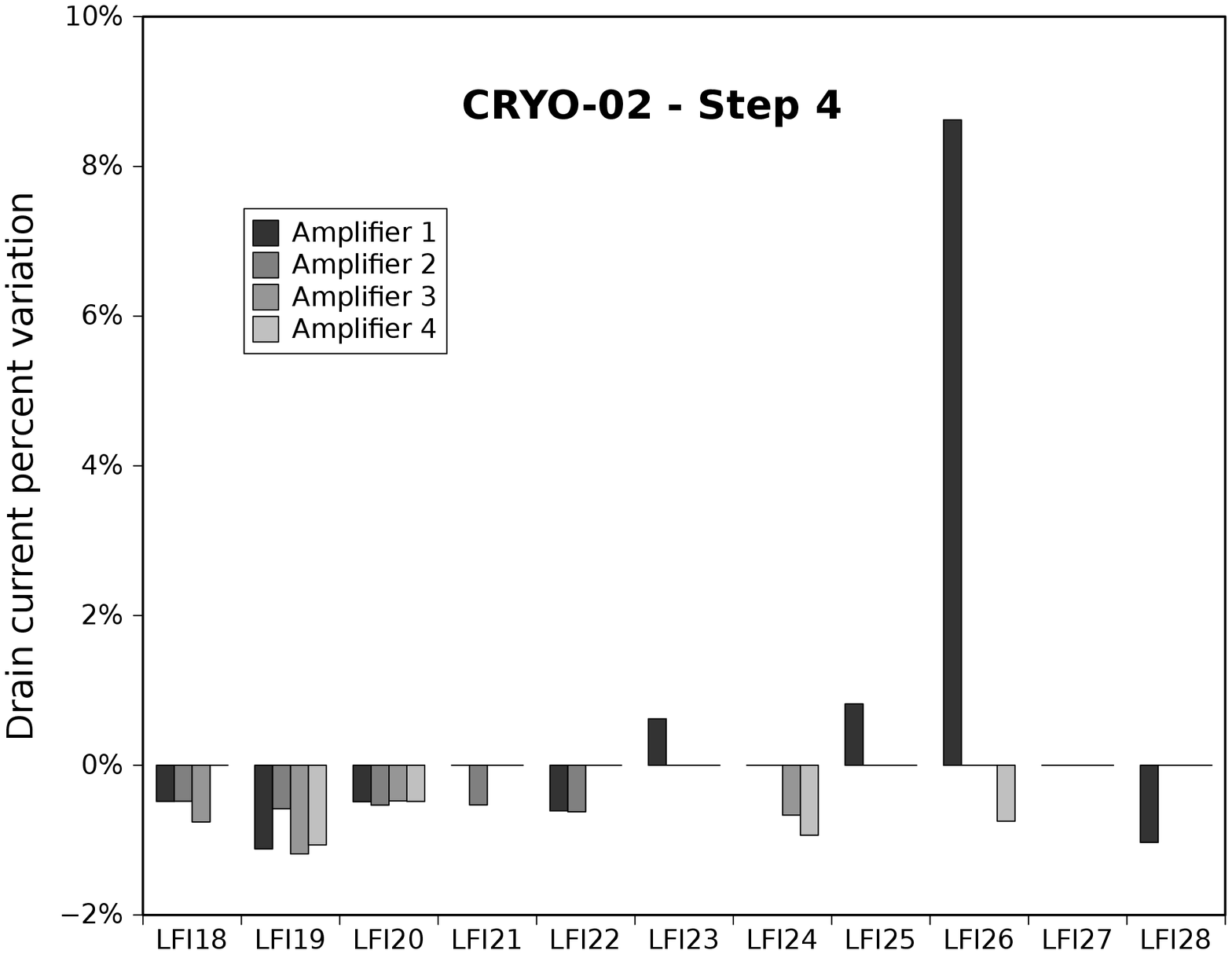}
    \end{center}
    \caption{Comparison of drain currents measured during \texttt{CRYO-02} tests performed in-flight and on-ground. The percentage variation is defined by $(I_{\rm drain}^\mathrm{flight} - I_{\rm drain}^\mathrm{ground})/I_{\rm drain}^\mathrm{ground}$. No bars are displayed when  the drain current variation was less than $0.1$\,mA}
    \label{fig_cryo02_idrain} 
\end{figure}

In Figure~\ref{fig_cryo02_idrain} we show a comparison of drain currents measured in-flight and on-ground \texttt{CRYO-02} test for the four phase switch configurations. Results show that all the drain currents have been found reproducible within $\pm$2\% apart from \texttt{LFI26M2} for which we recorded a 8.6\% ($\sim$1\,mA) higher drain current compared to ground measurements. Although the reason of this discrepancy has not been understood the LNA showed full functionality as proved by the drain current verification test (Section~\ref{sec:lfi_drain}).

\begin{figure}[h!]
    \begin{center}
        \includegraphics[width=7.2cm]{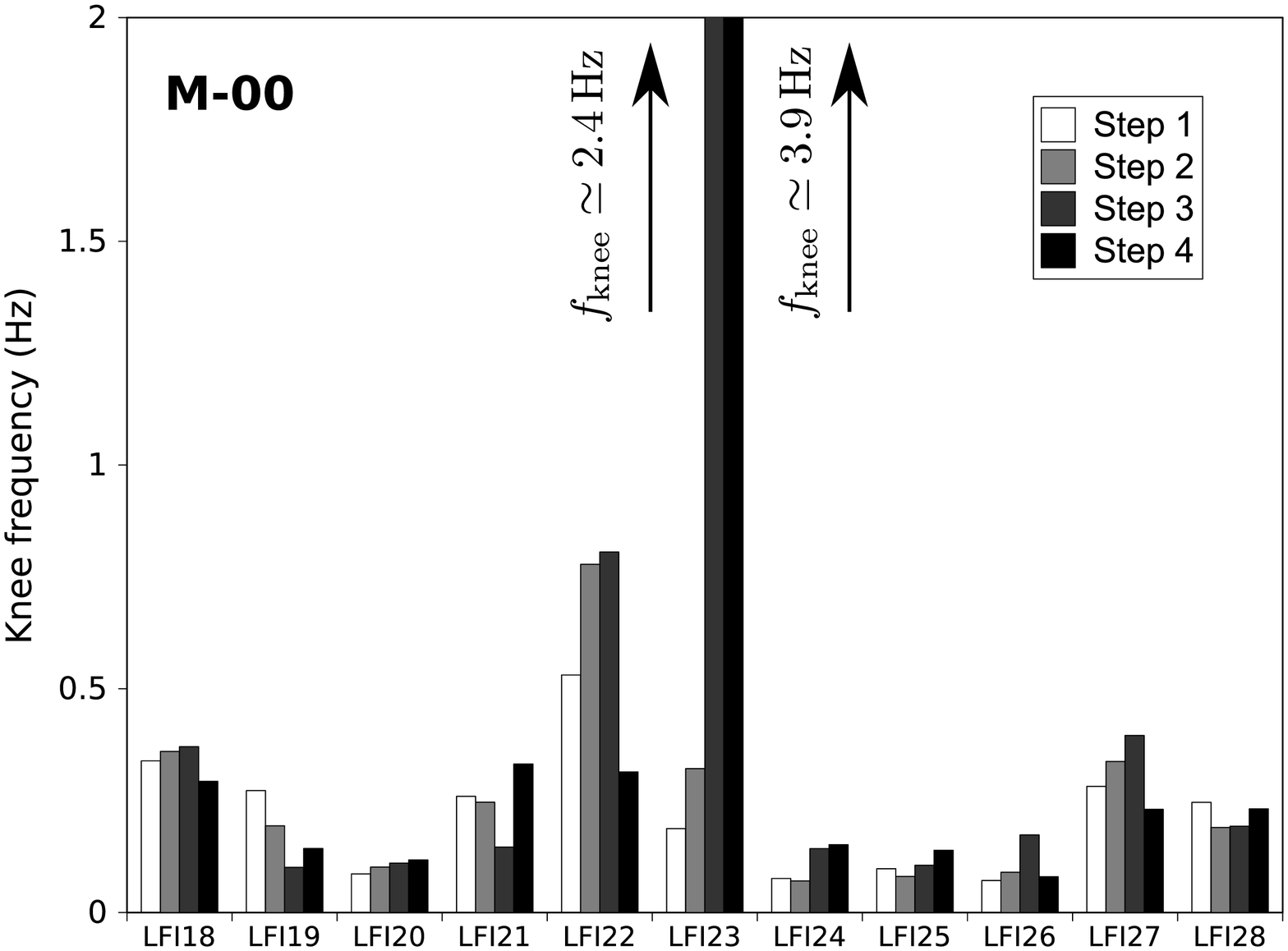}
        \includegraphics[width=7.2cm]{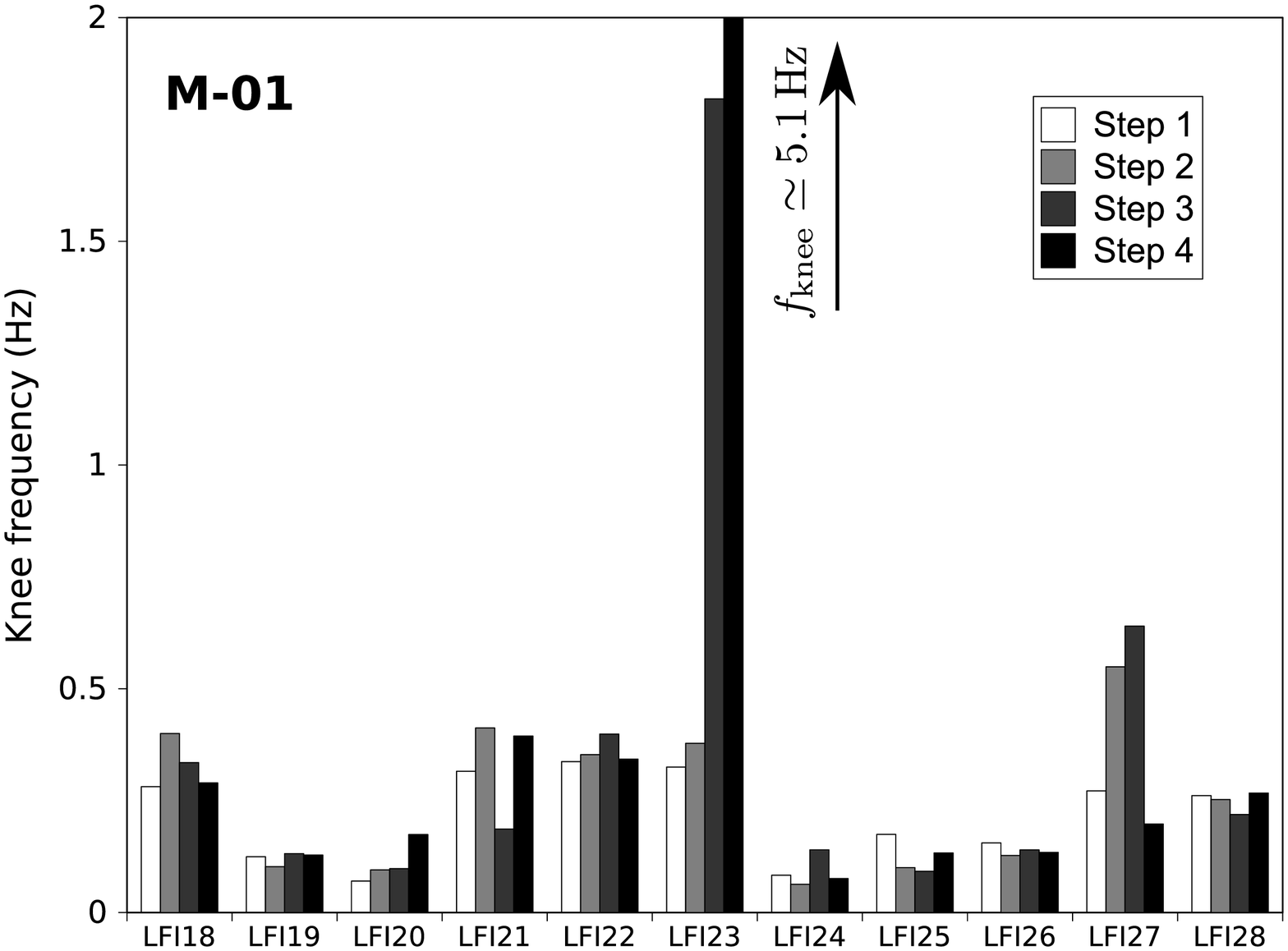}\\
        \includegraphics[width=7.2cm]{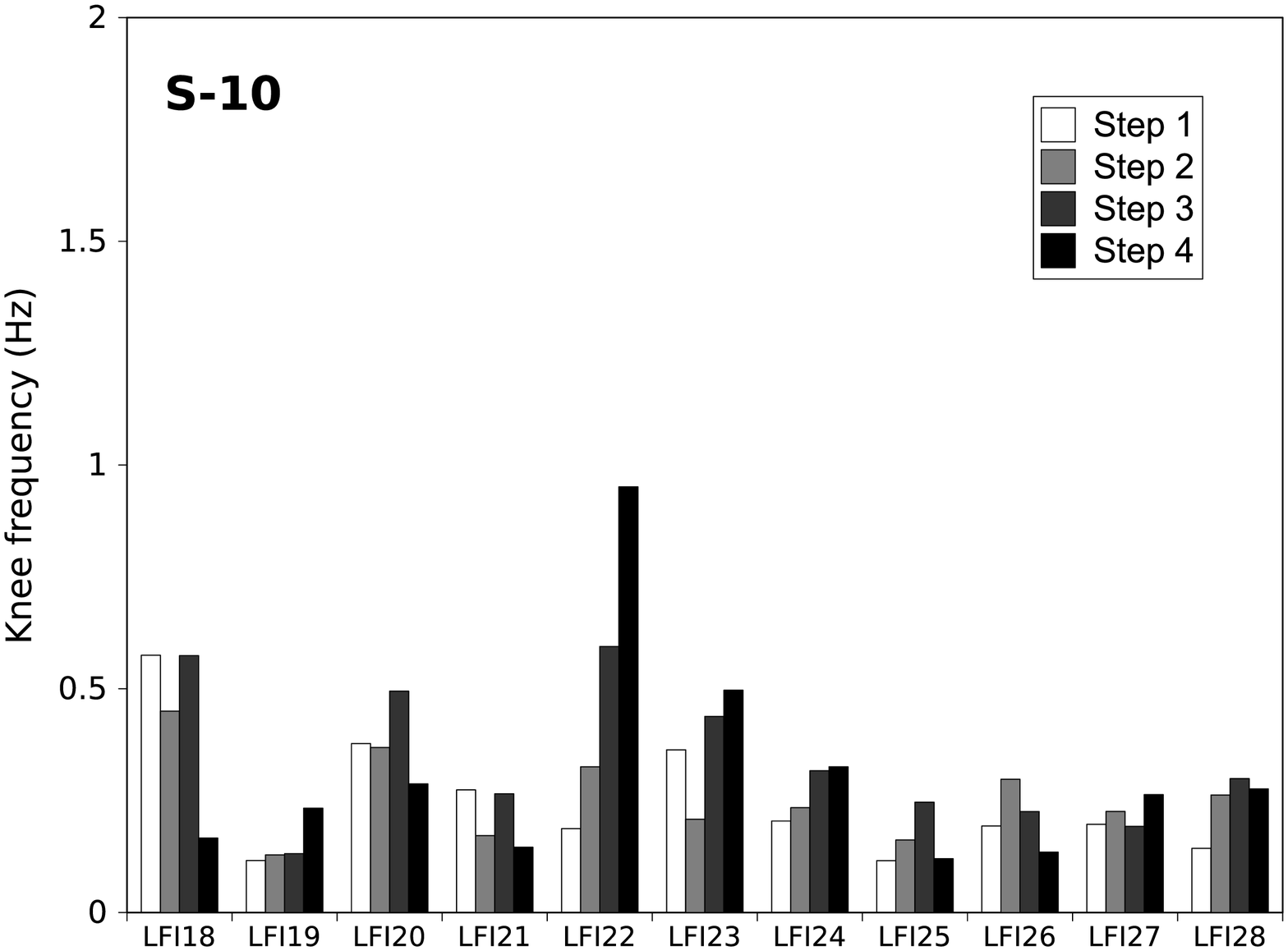}
        \includegraphics[width=7.2cm]{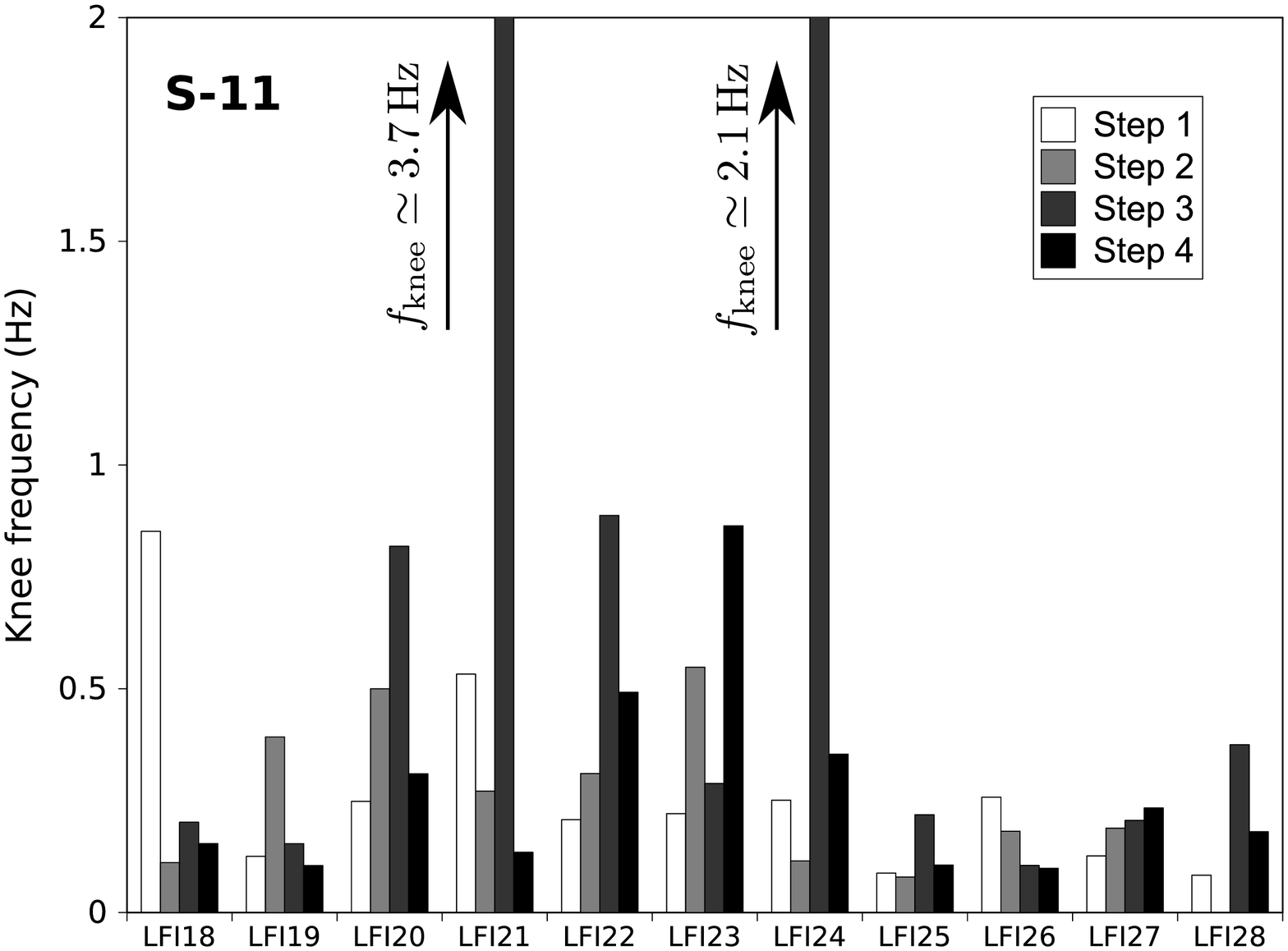}
    \end{center}
    \caption{Knee frequencies of differenced data measured during the \texttt{CRYO-02} test. For all channels apart from \texttt{LFI23M-00}, \texttt{LFI23M-01}, \texttt{LFI21S-11}, \texttt{LFI24S-11} knee frequencies were less than 1\,Hz and comparable in the four states of the phase switches.}
    \label{fig_cryo02_fk} 
\end{figure}

Figure~\ref{fig_cryo02_fk} summarises the calculated knee frequency for all 44 channels during the \texttt{CRYO-02} test for all the tested phase switch configurations. The knee frequency was essentially independent from the switch configuration and less than 1~Hz for all channels apart from the following cases:
\begin{itemize}
 \item \texttt{LFI23M-00} and \texttt{LFI23M-01} with \texttt{B/D} switching (steps 3 and 4),
 \item \texttt{LFI21S-11} and \texttt{LFI24S-11} with \texttt{B/D} switching and \texttt{A/C} set to 0 (step 3).
\end{itemize}
We now discuss these two cases separately.

\begin{figure}[h!]
    \begin{center}
        \includegraphics[width=14cm]{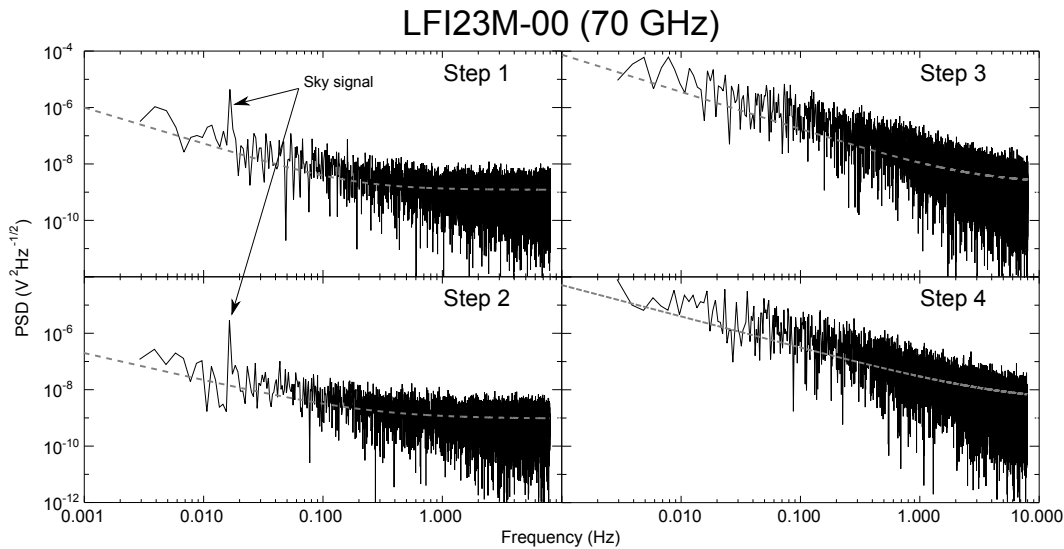}
    \end{center}
    \caption{PSD of differenced data from \texttt{LFI23M-00} detector during the \texttt{CRYO-02} test. The spike around 16~mHz in the the two panels on the left is determined by the sky signal (CMB dipole and galaxy).}
    \label{fig_cryo02_fk_2300}
\end{figure}

\paragraph{High knee frequency in \texttt{LFI23M-00} and \texttt{LFI23M-01}}
In Figure~\ref{fig_cryo02_fk_2300} we show the power spectral densities (PSDs) of differenced data streams from the \texttt{LFI23M-00} detector during the four steps of the \texttt{CRYO-02} test. The plots on the left indicate a nominal behaviour: a white noise plateau at high frequencies and a 1/$f$ tail below that does not prevent to detect the sky signal visible as a spike at 16\,mHz, the satellite spin frequency. The plots on the right, instead, (corresponding to steps 3 and 4, i.e. with \texttt{B/D} switching) indicate an anomalous behaviour: there is no white noise plateau and the spike at 16\,mHz caused by the sky signal (CMB dipole and galaxy) is not visible. The noise spectrum, in this case, was dominated by 1/$f$ noise. PSDs of \texttt{LFI23M-01} show the same trend and are not reported here for simplicity.

This anomaly was not new as was detected for the first time during the satellite ground tests and was confirmed in flight. The root cause of this extra noise was not fully understood but the anomaly could be avoided by keeping the \texttt{LFI23} RCA in the \texttt{A/C} switching configuration at the price of redundancy loss for the phase switch configurations of this RCA.

\begin{figure}[h!]
    \begin{center}
        \includegraphics[width=14cm]{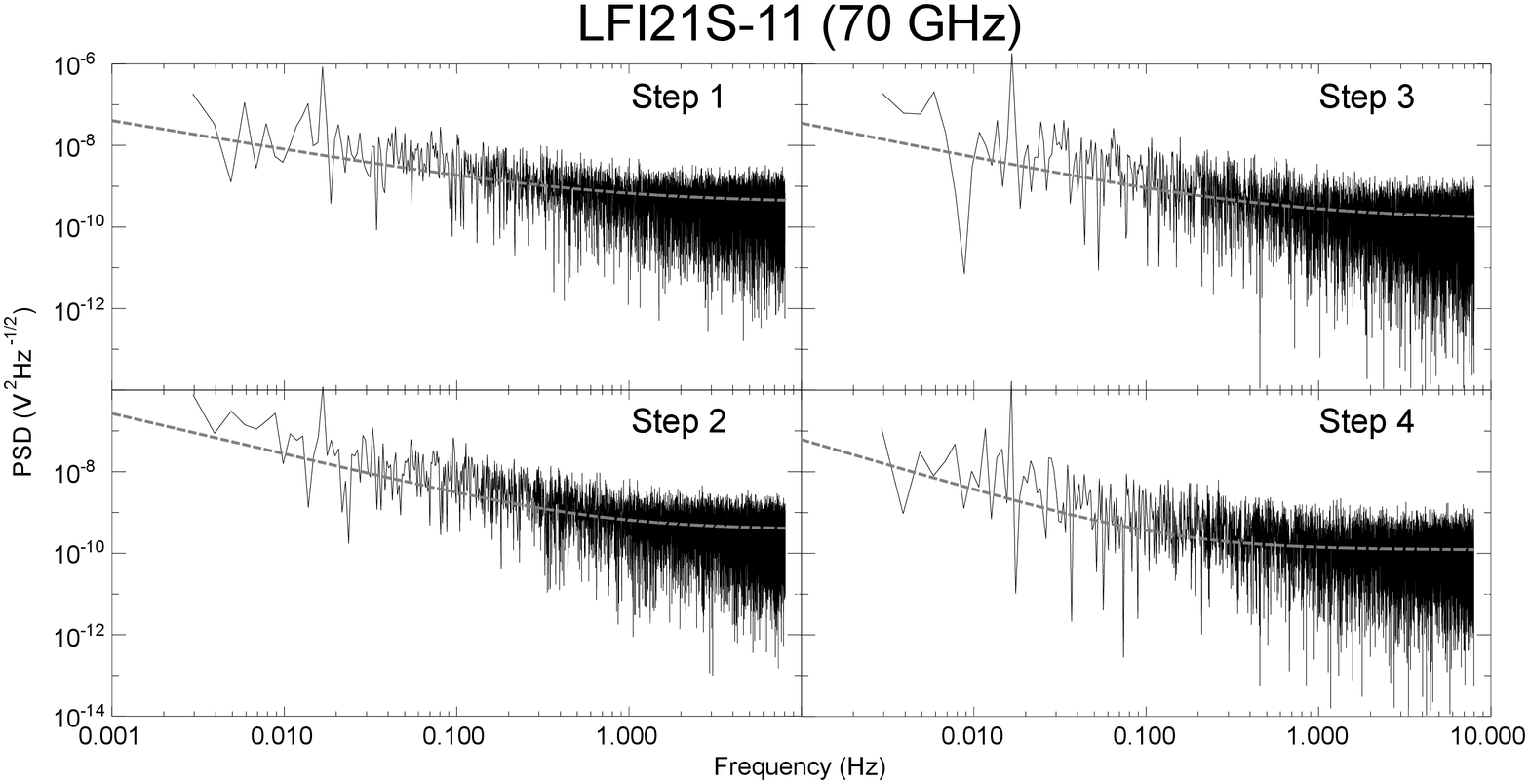}\\
        \vspace{0.5cm}
        \includegraphics[width=14cm]{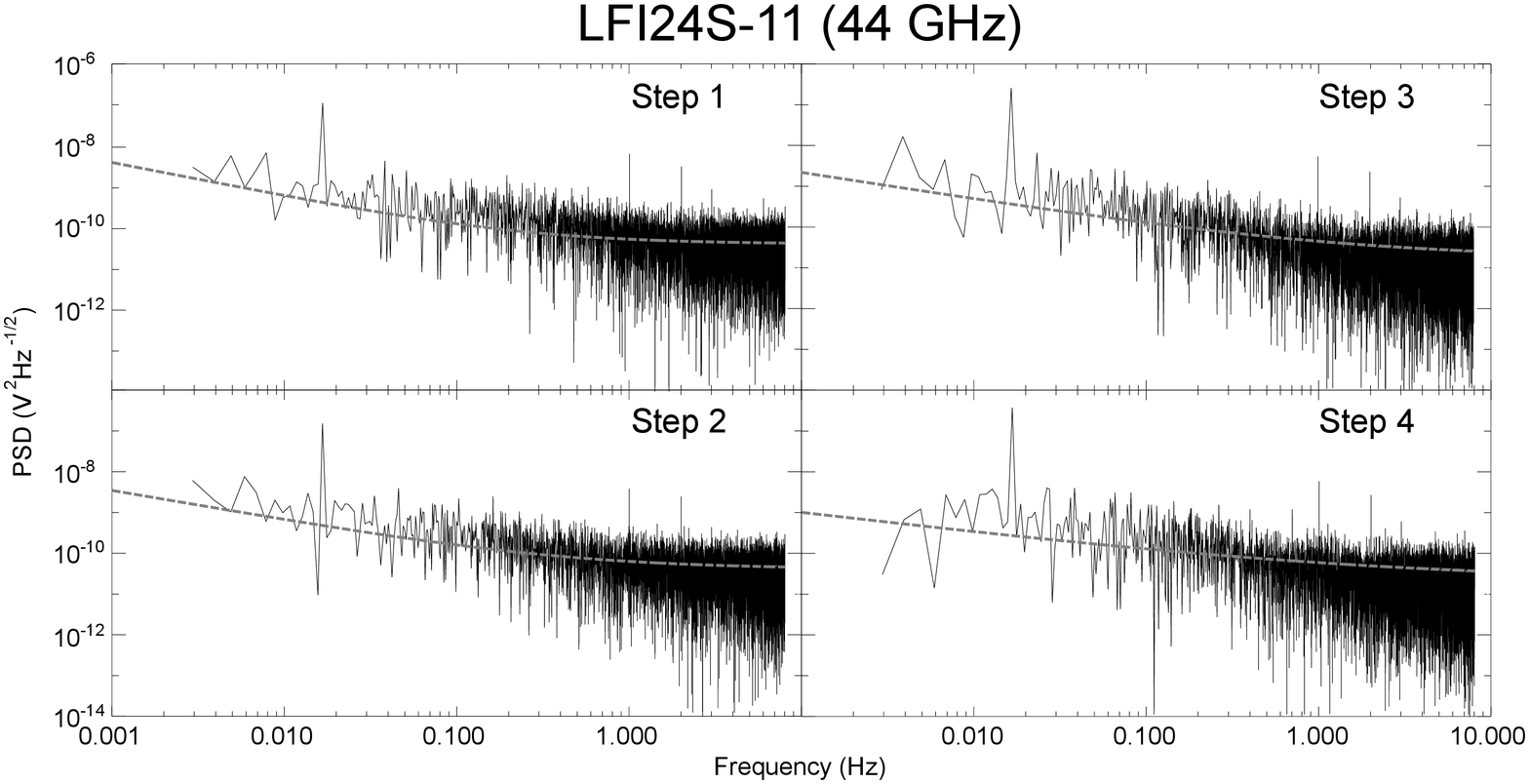}
    \end{center}
    \caption{PSD of differenced data from \texttt{LFI21S-11} and \texttt{LFI24S-11} detectors during the \texttt{CRYO-02} test. The behaviour was nominal in all the tested configurations.}
    \label{fig_cryo02_fk_21-24}
\end{figure}

\paragraph{High knee frequency in \texttt{LFI21S-11} and \texttt{LFI24S-11}}
In Figure~\ref{fig_cryo02_fk_21-24} we show the PSDs of differenced data streams from the \texttt{LFI21S-11} and \texttt{LFI24S-11} during the four steps of the \texttt{CRYO-02} test. In these two cases the four spectra are essentially comparable.
The high numerical value calculated for $f_{\rm knee}$ did not indicate, in this case, a real anomaly in the phase switch configuration.

%% file: 04_lfi_tests_spikes.tex
\begin{figure}
    \begin{center}
        \includegraphics[width=14cm]{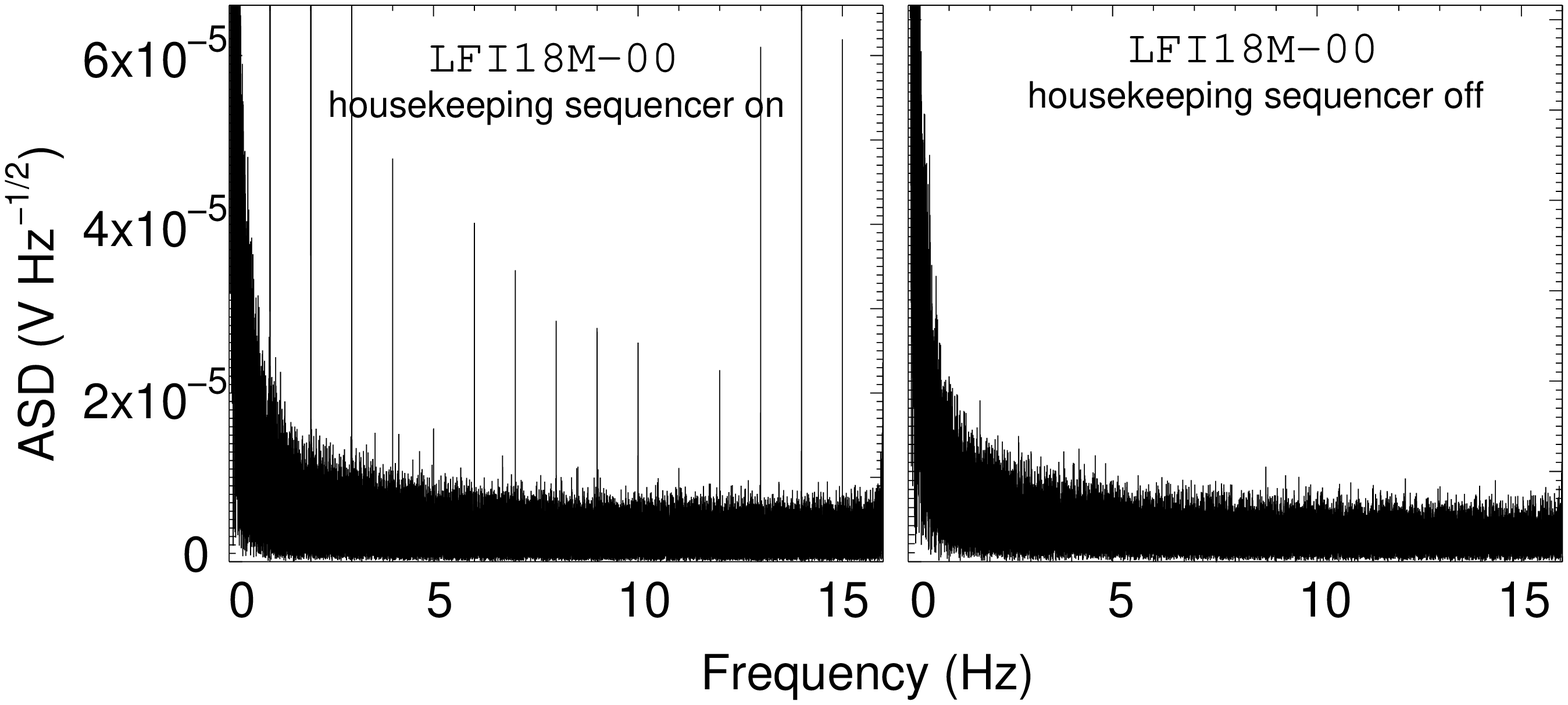}
    \end{center}
    \caption{Amplitude spectral density (ASD) of data from the \texttt{LFI18M-00} science channel with the housekeeping sequencer on (left panel) and off (right panel). In this test the radiometers were unbiased so that only the noise of the warm electronics was recorded. Spurious 1\,Hz spikes clearly disappeared when the housekeeping acquisition was turned off.}
    \label{fig_lfi1800_spikes_on_off}
\end{figure}

During instrument ground tests \cite{meinhold2009,mennella2010} an artifact was identified in the LFI data which is characterised as a set of extremely narrow spikes at 1\,Hz and higher harmonics. These artifacts are nearly identical in sky and reference samples, and are almost completely removed by the LFI differencing scheme. Extensive testing and analysis  identified the spikes as a disturbance on the science channels from the housekeeping data acquisition, which is performed by the DAE at 1~Hz sampling. An example is given in Figure~\ref{fig_lfi1800_spikes_on_off}. 

Frequency spikes in scientific data have been checked several times during ground and flight tests to verify their stability and their impact on the radiometric data. The test consisted in data acquisitions in several radiometer configurations:
\begin{itemize}
    \item Radiometers unbiased. In this case only the warm electronics noise was acquired. This was the most favourable condition to measure the spike amplitudes.
    \item Front-end modules unbiased and back-end modules in nominal bias conditions.
    \item All instrument in nominal bias conditions. In this case the data acquisition was repeated for all phase switch configurations to check for possible phase switch interactions.
\end{itemize}
Here we summarise results obtained in the first configuration, used to characterise spike stability, and in the third configuration, that highlights the impact of frequency spikes on nominal science data.

\begin{figure}
    \begin{center}
        \includegraphics[width=14cm]{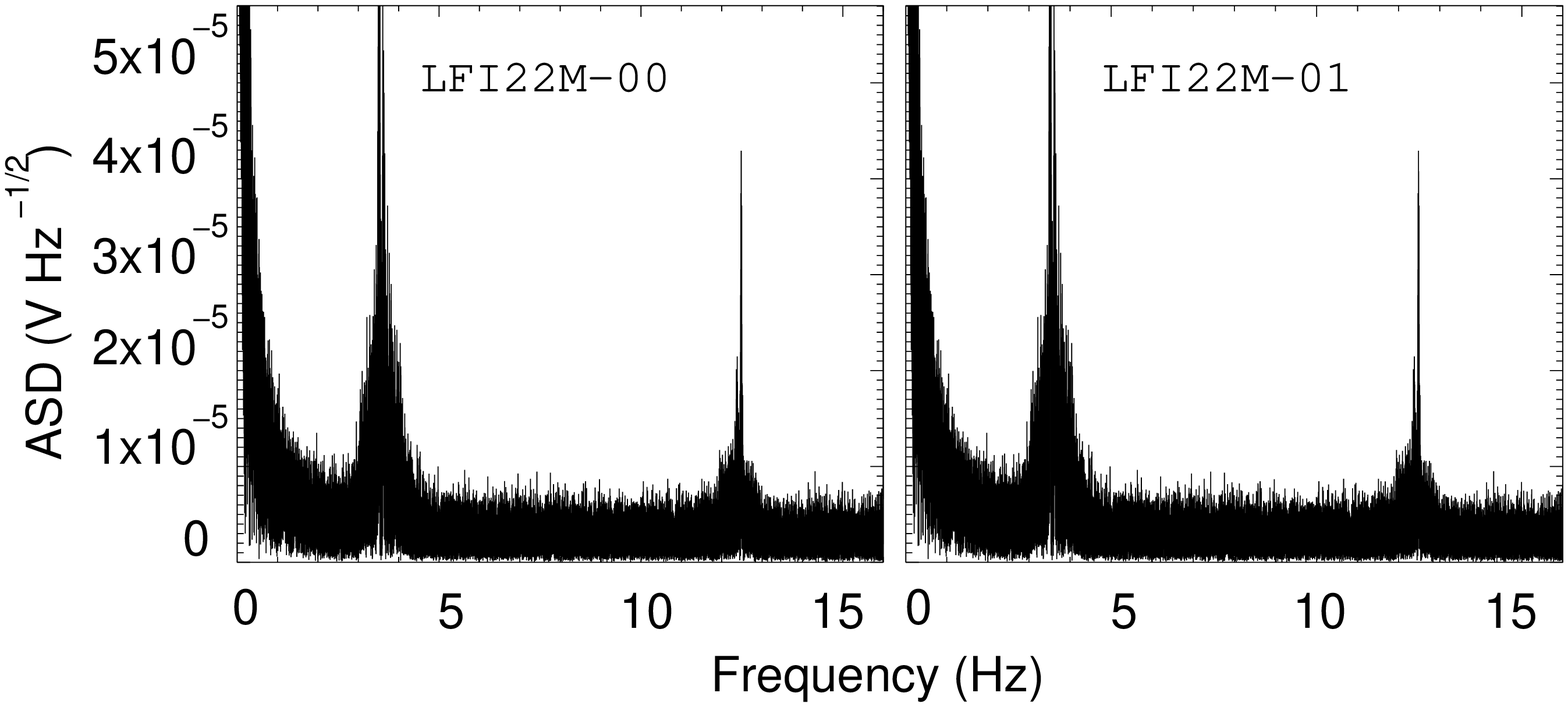}
    \end{center}
    \caption{ASD of data from the \texttt{LFI22M-00} and \texttt{LFI22M-01} science channels with the housekeeping sequencer off. Two broad spikes are present at frequencies of about 3\,Hz and 12\,Hz. These spikes do not depend on the status of the housekeeping sequencer and their cause is presently unknown. Also in this case the radiometers were unbiased.}
    \label{fig_spikes_22m}
\end{figure}

In Appendix~\ref{app_spike_plots} (Figure~\ref{fig_spikes}) we show the amplitude spectral density of 1\,Hz frequency spikes in all the LFI outputs measured during ground and flight tests with radiometers completely unbiased. These results show that the spike amplitude was very stable and reproducible among the various test campaigns. Two particular features can be noticed around 3.5\,Hz and  12.5\,Hz in the \texttt{LFI22M-00} and \texttt{LFI22M-01} plots. These correspond to two broad spikes that were detected in the electronics noise output of these two channels independently from the status of the housekeeping sequencer. Their cause is presently unknown. A detailed view of these frequency spikes is provided in Figure~\ref{fig_spikes_22m}.

Although spikes are below noise level in frequency domain, it was possible to obtain a template of the spurious signature in time domain exploiting its synchronous characteristics and binning several days of data in 1-second time windows. These templates have then been used to correct the 44\,GHz data, which are the more affected by this effect. Details of the spike removal pipeline are provided in \cite{zacchei2011} while the assessment of the residual on the scientific results is discussed in \cite{mennella2011}.

%% file: 04_lfi_tests_drain_current.tex
The receiver response to front-end bias changes was checked several times by means of the so-called \textit{drain current test}. During this test the two gate voltages, $V_{\rm g1}$ and $V_{\rm g2}$, were changed independently for each front-end LNA spanning a pre-defined set of values and, for each value, the average drain current was recorded. The signal unbalance between the sky (at $\sim 3$\,K) and the reference load (at $\gtrsim 20$\,K) temperatures was exploited to estimate the noise temperature according to the formula (``Y--factor method''):

\begin{equation}
    T_{\rm noise} = \frac{T_{\rm ref}-T_{\rm sky}}{Y-1},\, \mbox{with }\, Y=\frac{V_{\rm ref}}{V_{\rm sky}}.
    \label{eq_y_factor}
\end{equation}

The channels were grouped in pairs in order to minimise electric cross-talks caused by the cryo harness common ground return (see Appendix~\ref{app_power_groups_table}, Table~\ref{tab_IV_groups}). Two channels were tested at a time, all the other channels were kept to their default bias values. During CPV the drain current test was run three times: (i) before the bias pre-tuning (see Section~\ref{sec:lfi_LNAs_Tun}) to compare the response with on-ground test results, (ii) after the bias tuning phase (see Section~\ref{sec:lfi_LNAs_Tun}), to check for any changes that could have been induced by the several bias changes occurred during tuning, and (iii) at the very end of the tuning activities when new default biases were implemented.

\begin{figure}[h!]
    \begin{center}
        \includegraphics[width=15cm]{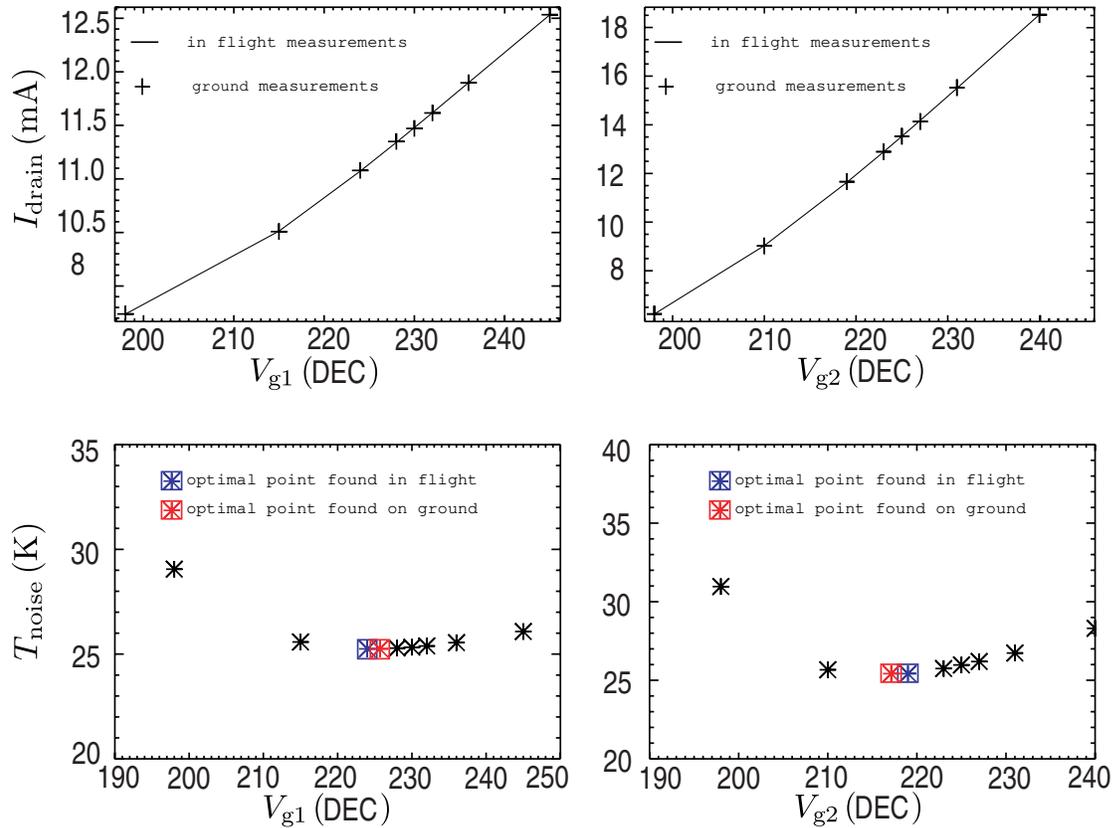}
    \end{center}
    \caption{Example of drain current test results for channel \texttt{LFI26M1}. Top panels:  drain currents (line for flight and crosses for ground data), bottom panels: noise temperatures measured in-flight, as a function of the input bias voltages. Minimum noise temperatures are highlighted in blue and red respectively. Note that the bias units are DEC units that are used to set the voltage in the DAE: see Table~4 for a rough conversion into physical units.}
    \label{IV_curves_260PLOT} 
\end{figure}

In Figure~\ref{IV_curves_260PLOT}, we show an example of the test results obtained for the channel \texttt{LFI26M1} during the first run. The top panels show drain currents as a function of the two input gate bias voltages, while the bottom panels show  noise temperature variations as a function of $V_{\rm g1}$ and $V_{\rm g2}$. 
Figure~\ref{IV_curves_260PLOT} shows an excellent match between ground and flight measurements  providing a further verification of the functionality of front-end LNAs after launch and cooldown.

Notice that gate voltages are displayed using the adimensional code (from 0 to 255) used to set the voltage in the DAE rather than the voltage in physical units. The reason is that the actual voltage reaching the FEMs depends on the harness resistive drop which depends on the satellite thermal configuration. The DAE code, therefore, was the most robust way to refer to the LNA bias voltages. The interested reader can roughly convert the DAE code into the actual delivered voltage (neglecting the harness resistance drop) using a linear calibration with the values detailed in Table~\ref{tab_dae_calibration}. It is important to underline that the calibration table actually used to convert DAE codes to physical units is channel dependent, but it is not provided here for simplicity.
\begin{table}[h!]
    \begin{center}
        \caption{Values for a zero-order calibration of DAE bias codes into physical units.}
        \label{tab_dae_calibration}
        \begin{tabular}{l  c c c   c c c }
        \hline         \hline
          & \multicolumn{3}{c}{\texttt{DAE code = 0}} & \multicolumn{3}{c}{\texttt{DAE code = 255}} \\
          & $V_{\rm drain}$ & $V_{\rm g1}$ & $V_{\rm g2}$ & $V_{\rm drain}$ & $V_{\rm g1}$ & $V_{\rm g2}$\\
        \hline
            \multicolumn{1}{l}{30 GHz} & 0.0\,V & -4.0\,V & -4.0\,V & 1.4\,V & 2.0\,V & 2.0\,V \\
            \multicolumn{1}{l}{44 GHz}  & 0.0\,V & -4.0\,V & -4.0\,V & 1.4\,V & 2.0\,V & 2.0\,V \\
            \multicolumn{1}{l}{70 GHz}  & 0.0\,V &  0.0\,V &  0.0\,V & 1.0\,V & 2.0\,V & 2.0\,V \\
        \hline
        \end{tabular}
    \end{center}
\end{table}

\begin{figure}[h!]
    \begin{center}
        \includegraphics[width=7.5cm]{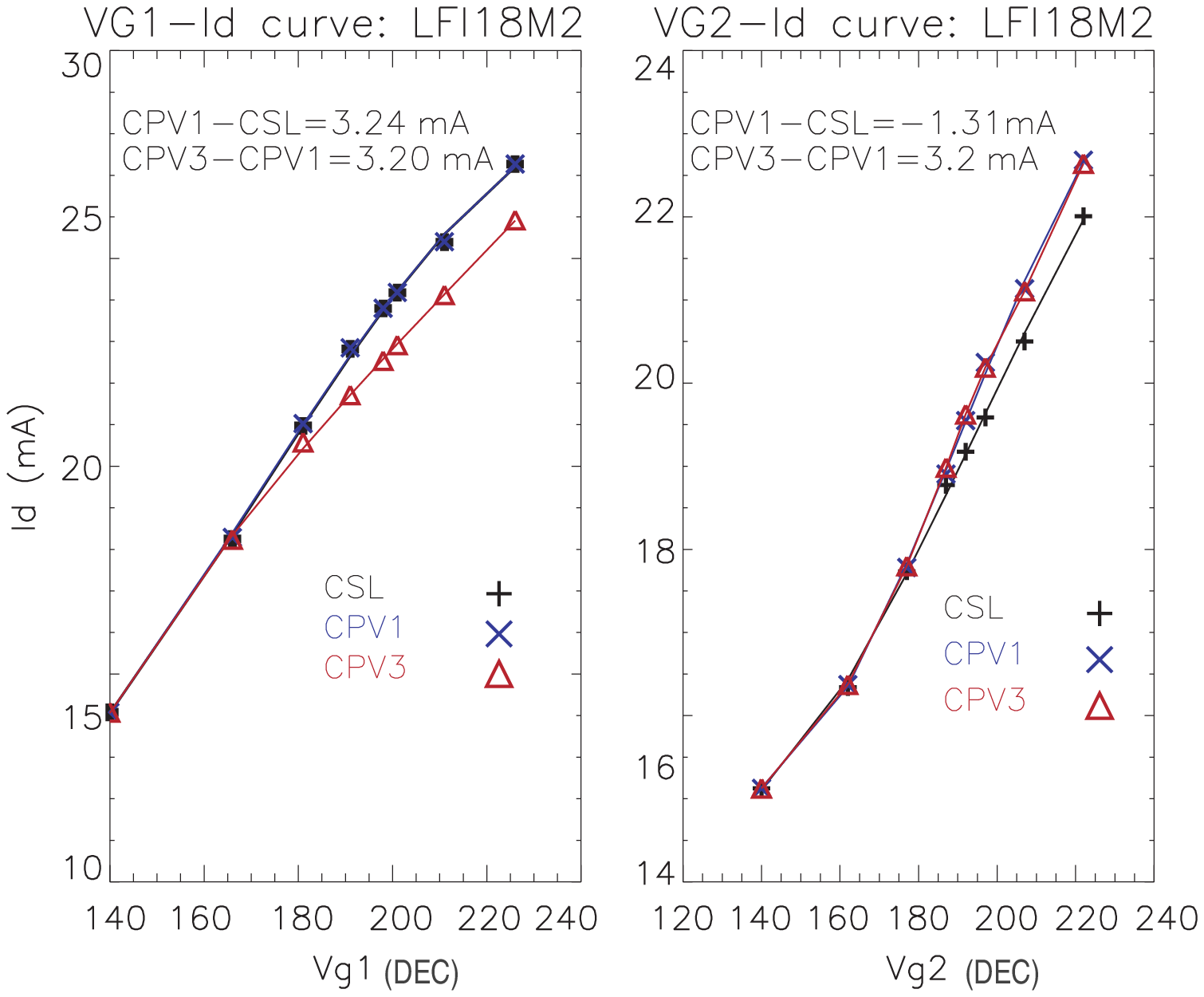} 
           \includegraphics[width=7.5cm]{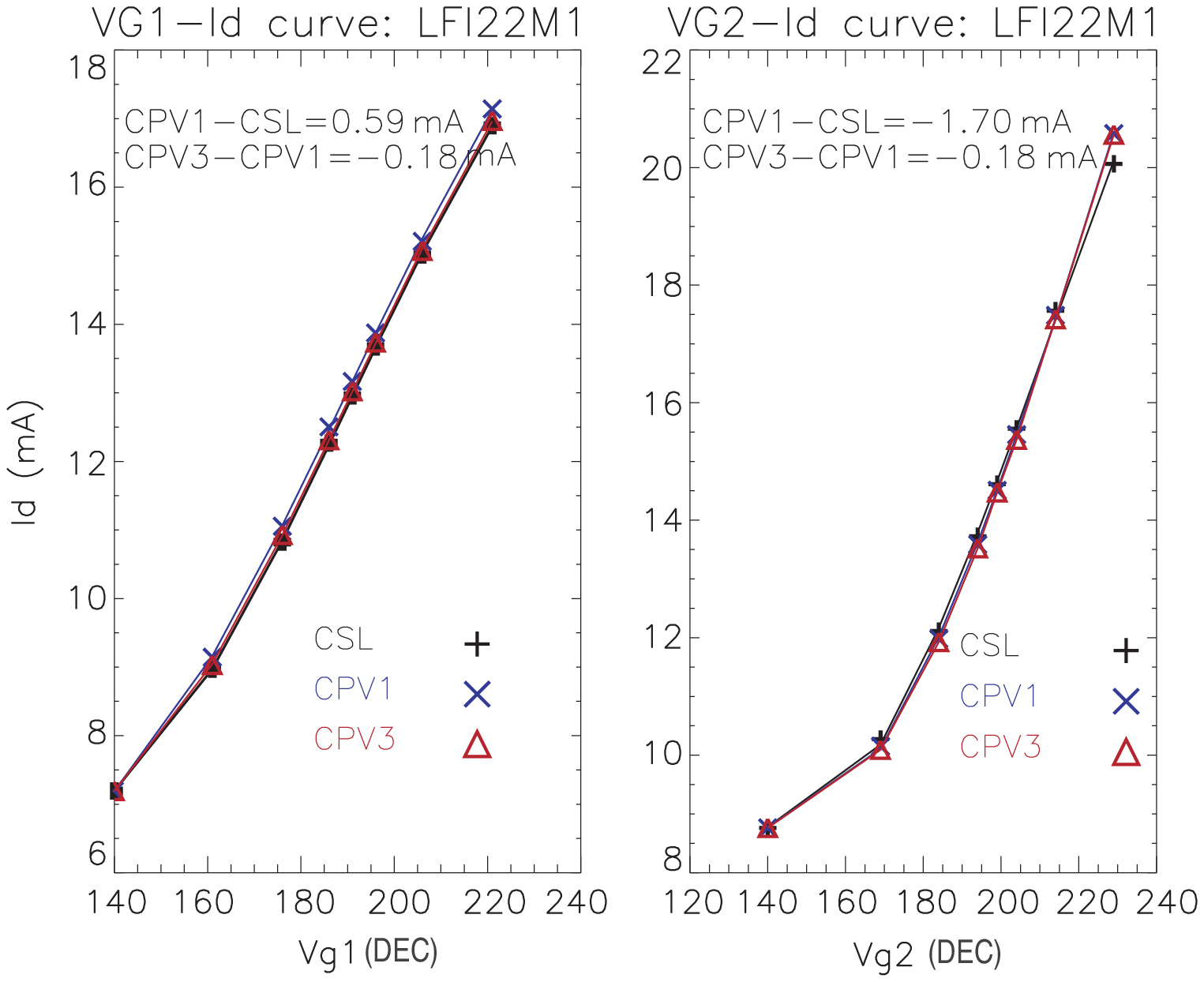}\\ 
        \includegraphics[width=7.5cm]{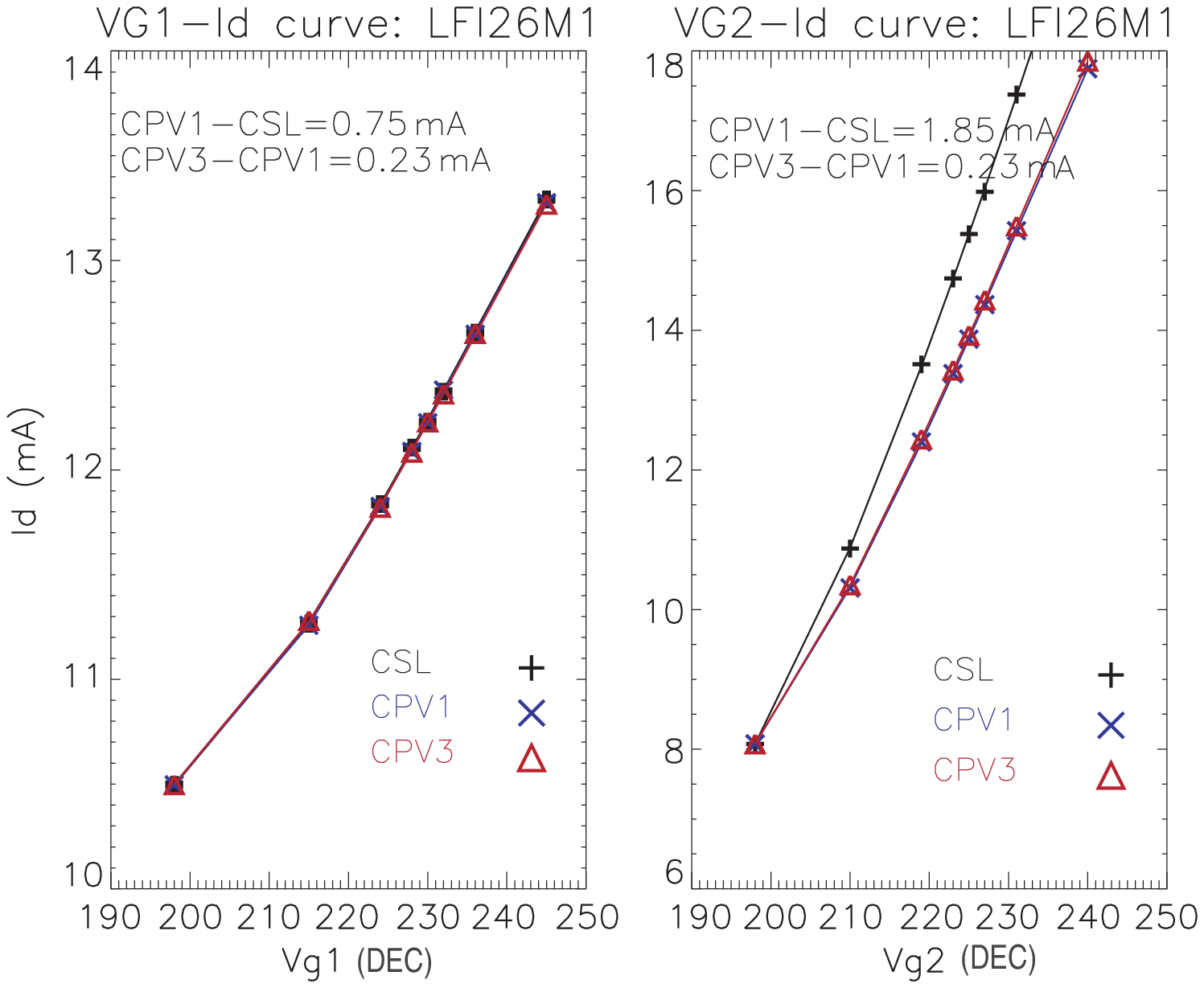} 
           \includegraphics[width=7.5cm]{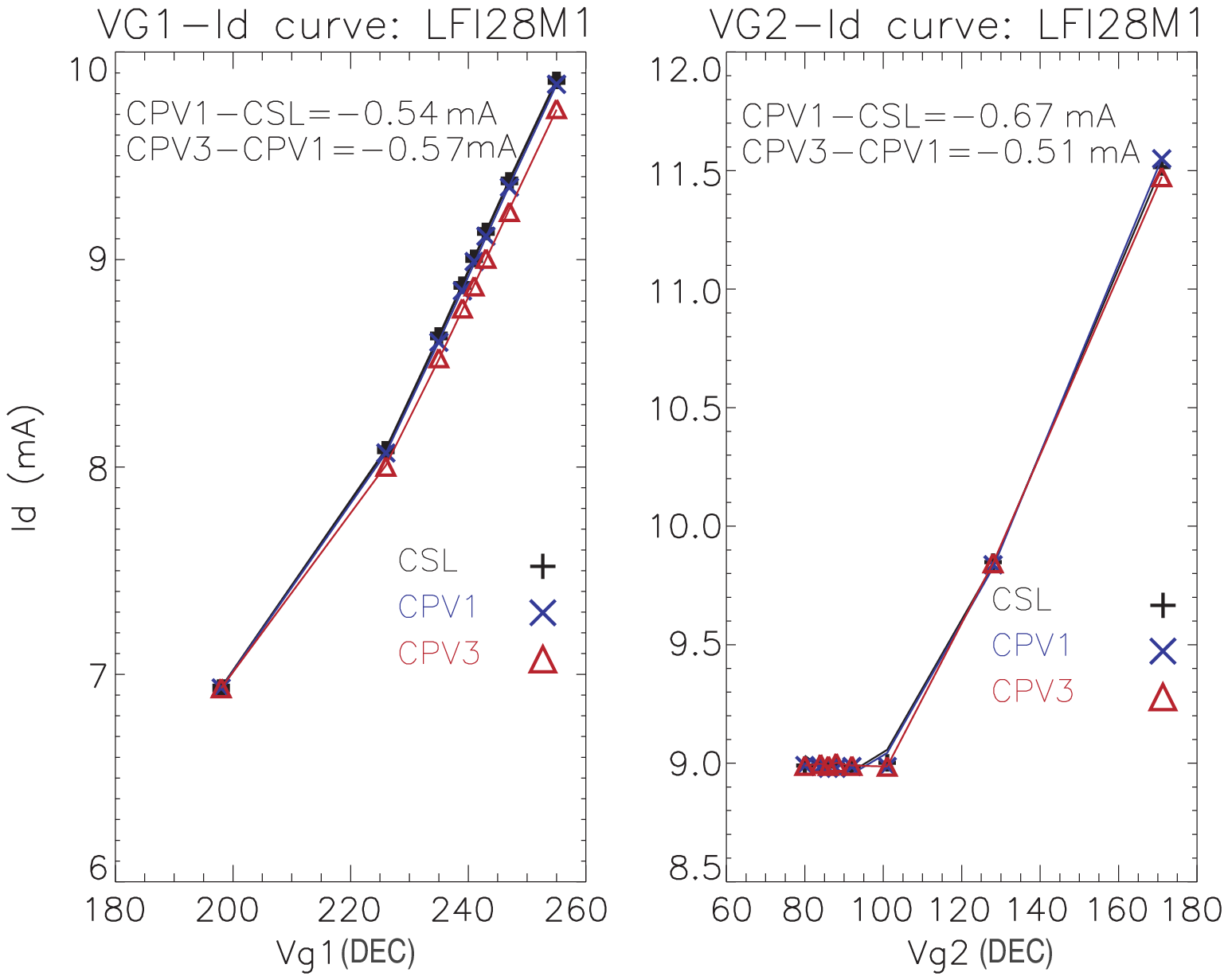}\\ 
    \end{center}
    \caption{I-V curves: for each channel (\texttt{LFI18M2},  \texttt{LFI22M1},  \texttt{LFI26M1},  \texttt{LFI28M1}) we show two plots representing the drain current behavior respectively versus $V_{\rm g1}$ and $V_{\rm g2}$. Black crosses refer to results from CSL system level tests, red triangles to the 3$^{\rm rd}$ one, blue crosses to 1$^{\rm st}$ run; solid lines represents the corresponding polinomial fits. Curves are offset to facilitate the comparison, the offset amount is reported on the internal labels. Note that the bias units are DEC units that are used to set the voltage in the DAE: see Table~4 for a rough conversion into physical units.}
    \label{IV3_representative_plots} 
\end{figure}

In Figure~\ref{IV3_representative_plots}, we show a comparison of results coming from the same test at system level in CSL and during the CPV runs first and third, for different channels. To facilitate the comparison, the 1$^{\rm st}$ run and CSL curves were offset on the Y-axis as reported on the label (CPV1-CSL and CPV3-CPV1). 
In  most cases the matching is very good, despite the different bias defaults; in a few cases,  mostly corresponding to channels characterized by a large difference in their bias setup in the two frames, or in the bias setup of other channels belonging to the same power group, we found larger differences. 
Plots for all channels are given in Appendix~\ref{app_draincurr_plots}.

%% file: 04_lfi_tests_stability_check.tex
During this test the instrument was left undisturbed for 12 hours in nominal, pre-tuned bias configuration, with the objective to verify the instrument readiness for the bias tuning activity. The instrument stability was verified by checking the following observables: (i) drain currents, (ii) output voltages, (iii)  knee frequency of differenced data. 

It is important to notice that during this test the temperature of the 4\,K stage was still around 20\,K with no stability optimisation. This resulted in a knee frequency which was much higher than the required 50\,mHz for optimal scientific performance. Therefore in this test we checked that the knee frequency calculated from differenced datastreams were much less than those calculated from undifferenced data and, in general, less than $\sim$1\,Hz.

\begin{figure}
    \begin{center}
        \includegraphics[width=13.3cm]{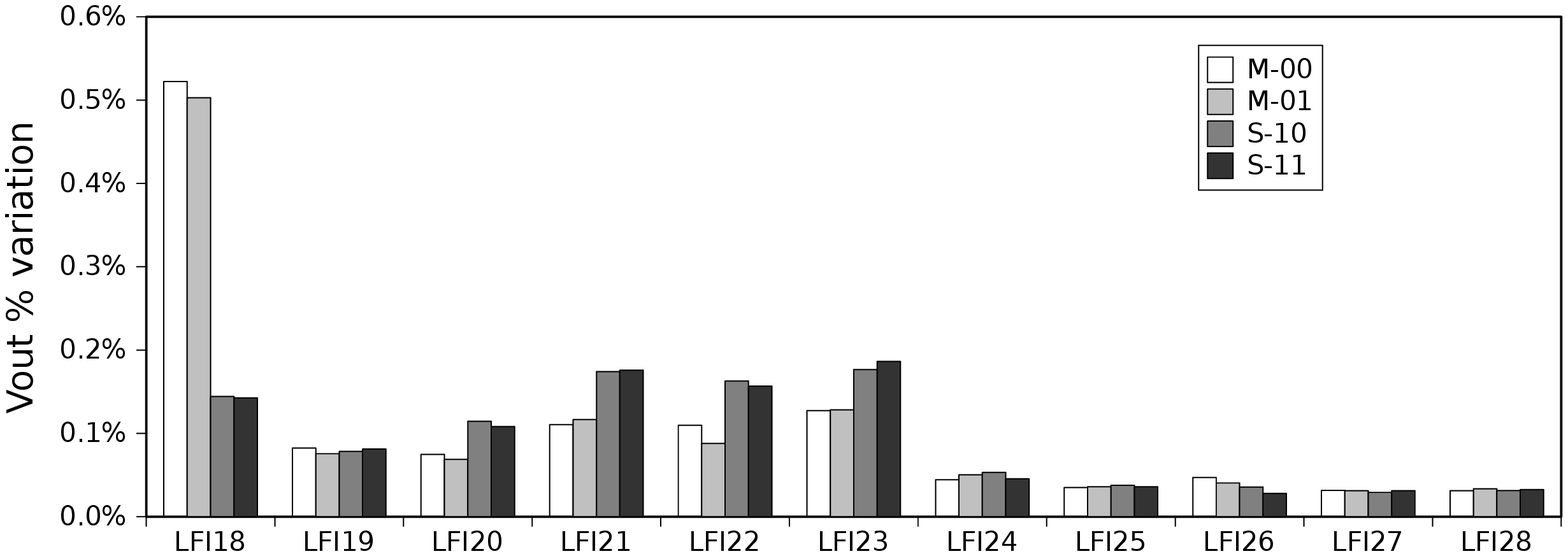}\\
        \includegraphics[width=13.3cm]{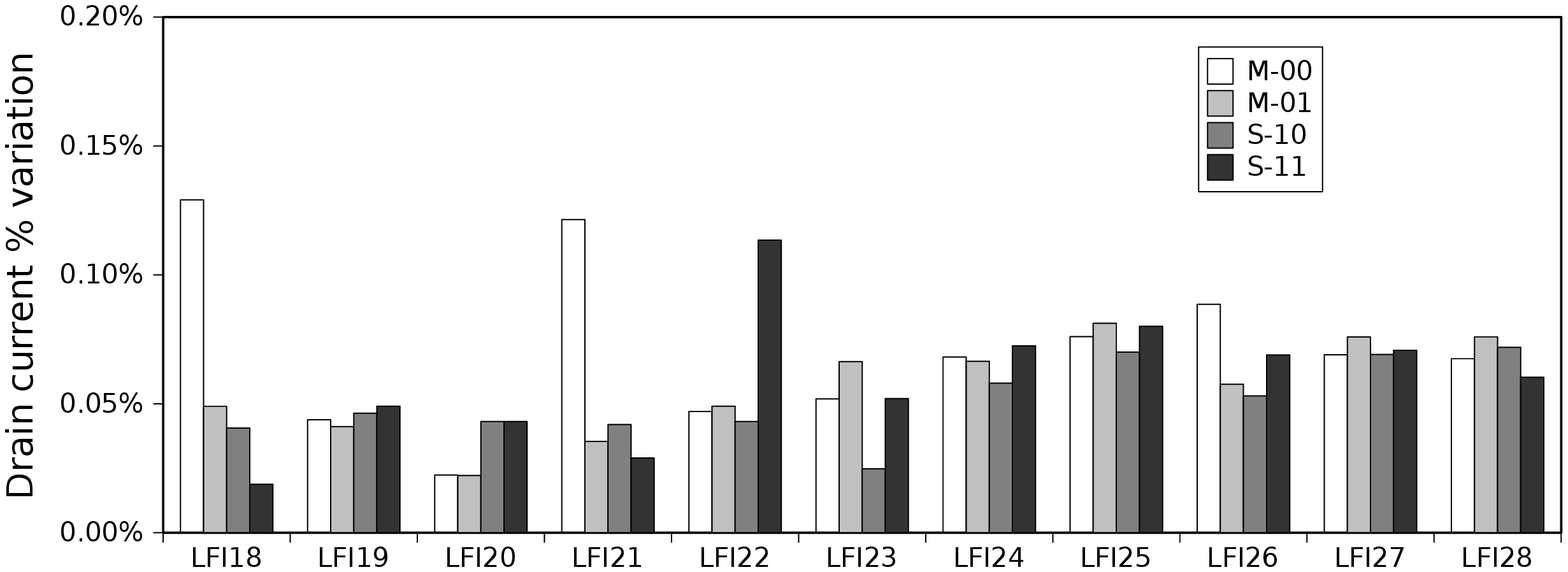}\\
        \includegraphics[width=13.3cm]{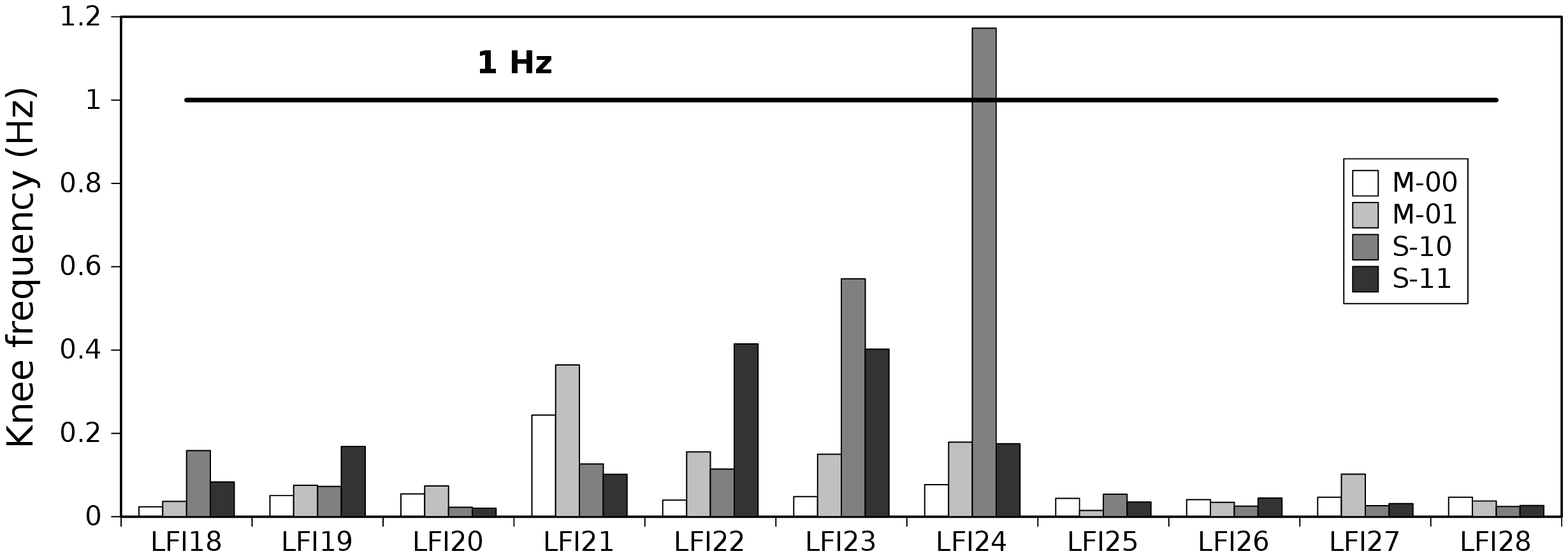}
    \end{center}
    \caption{Main stability indicators measured during the stability test. Top panel: sky output voltage \% variation. Middle panel: drain current \% variation. Bottom panel: $1/f$ knee frequencies of differenced data.}
    \label{fig_stability_plots}
\end{figure}

\begin{figure}
    \begin{center}
        \includegraphics[width=13cm]{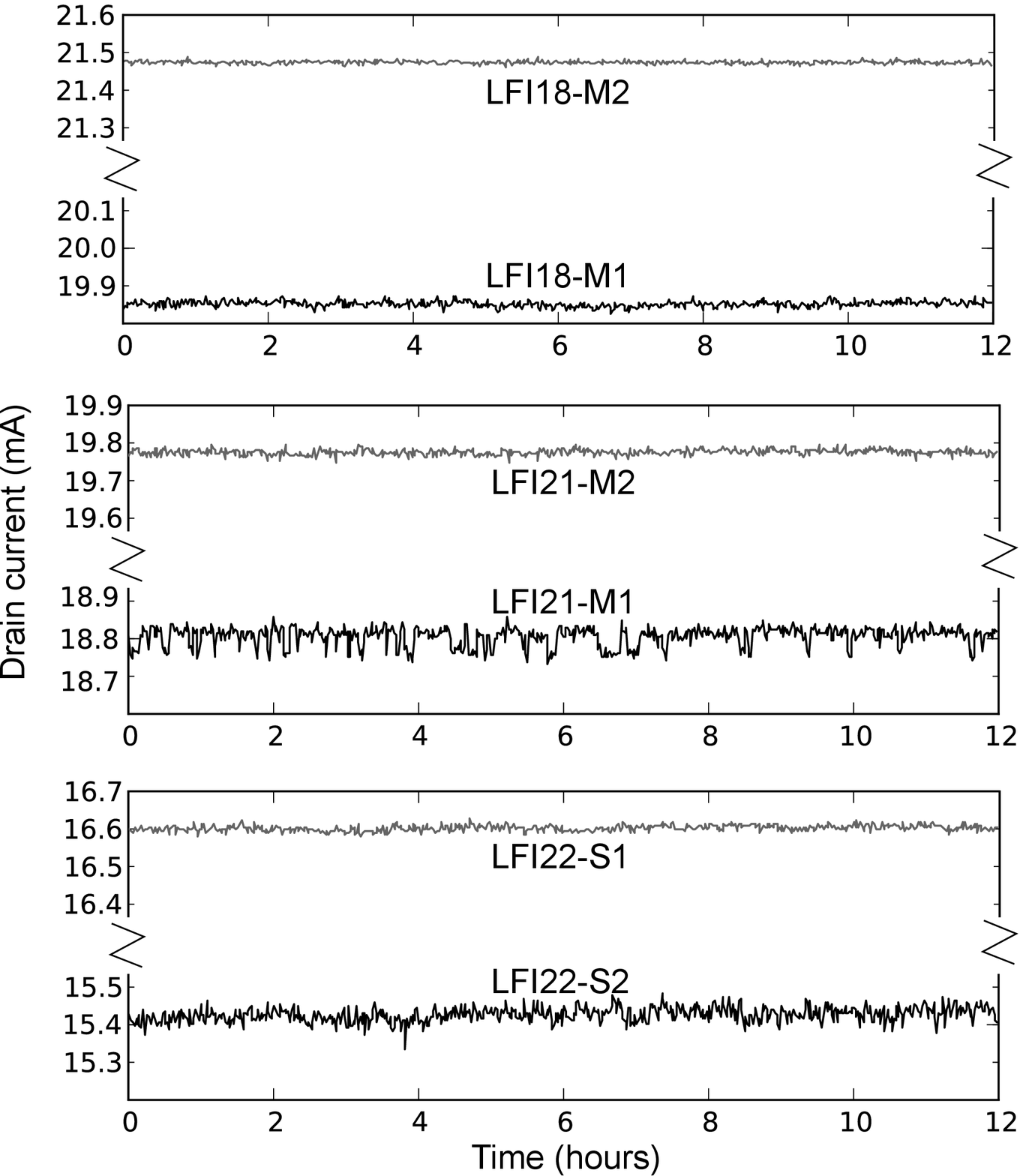}     
    \end{center}
    \caption{Drain currents during the stability test. Instabilities in \texttt{LFI18M1}, \texttt{LFI21M1} and \texttt{LFI22S2} are highlighted by comparison with drain currents from their twin amplifiers.}
    \label{fig_stability_idrains}
\end{figure}

In the top and middle panels of Figure~\ref{fig_stability_plots} we show the percentage variation of (sky) voltage output and drain currents during the stability test. These results show that variations in the sky voltage output were less than 0.2\%, with the exception of \texttt{LFI18M-00} and \texttt{LFI18M-01} showing a stability of 0.5\%, a typical figure for these two channels. Drain currents were stable within 0.1\% with three cases slightly exceeding this figure: \texttt{LFI18M1}, \texttt{LFI21M1} and \texttt{LFI22S2}. The drain current behaviour for these three devices is plotted in Figure~\ref{fig_stability_idrains} compared to the drain current measured for the twin amplifier. The higher unstability level of these three amplifiers was already known since ground tests and has never represented a problem.

The bottom panel in Figure~\ref{fig_stability_plots} shows the $1/f$ knee frequency of differenced data for all channels.

%% file: 04_lfi_tests_reference_test.tex
Close to the end of the test Campaign, a functionality assessment was performed with a double objective: 
\begin{itemize}
	\item to verify that no damages, or any other evident changes, were occurred in the radiometers during the test campaign;
	\item to characterize the LNAs and phase switches response with the LFI in nominal conditions. This test was thought as a tool to check for any possible anomalies in one or several channels during the nominal survey,  without disturbing the acquisition of all the other channels.
		\end{itemize}
The test was run in two parts. The first part (\texttt{RT-1}) is identical to the \texttt{CRYO-02}, with radiometers powered with exactly the same \texttt{CRYO-02} biases but with different thermal boundaries (during the first  \texttt{CRYO-02} the LFI focal plane was almost at its nominal temperature but the 4~K stage was still at about 21~K). The second part (\texttt{RT-2}) followed a procedure very similar to the \texttt{CRYO-01} (see  Appendix~\ref{app_cryo01_sequence}) but with the tuned LNAs biases. 
The two steps were executed about ten days apart due to scheduling requirements but this did not affect the general result.

The objective of \texttt{RT-1} was to  check the functionality of radiometers in all the possible phase switch and 4~kHz combinations after the hypermatrix tuning: LNAs drain currents and noise properties (1/$f$, white noise, spikes) were compared with those measured during \texttt{CRYO-02}. The 
sky scientific outputs were compared as well (reference outputs were not compared due to the different thermal conditions of the 4~K stage during the two tests).

\begin{figure}[h!]
    \begin{center}  
\includegraphics [angle=90,width=10.25cm ]{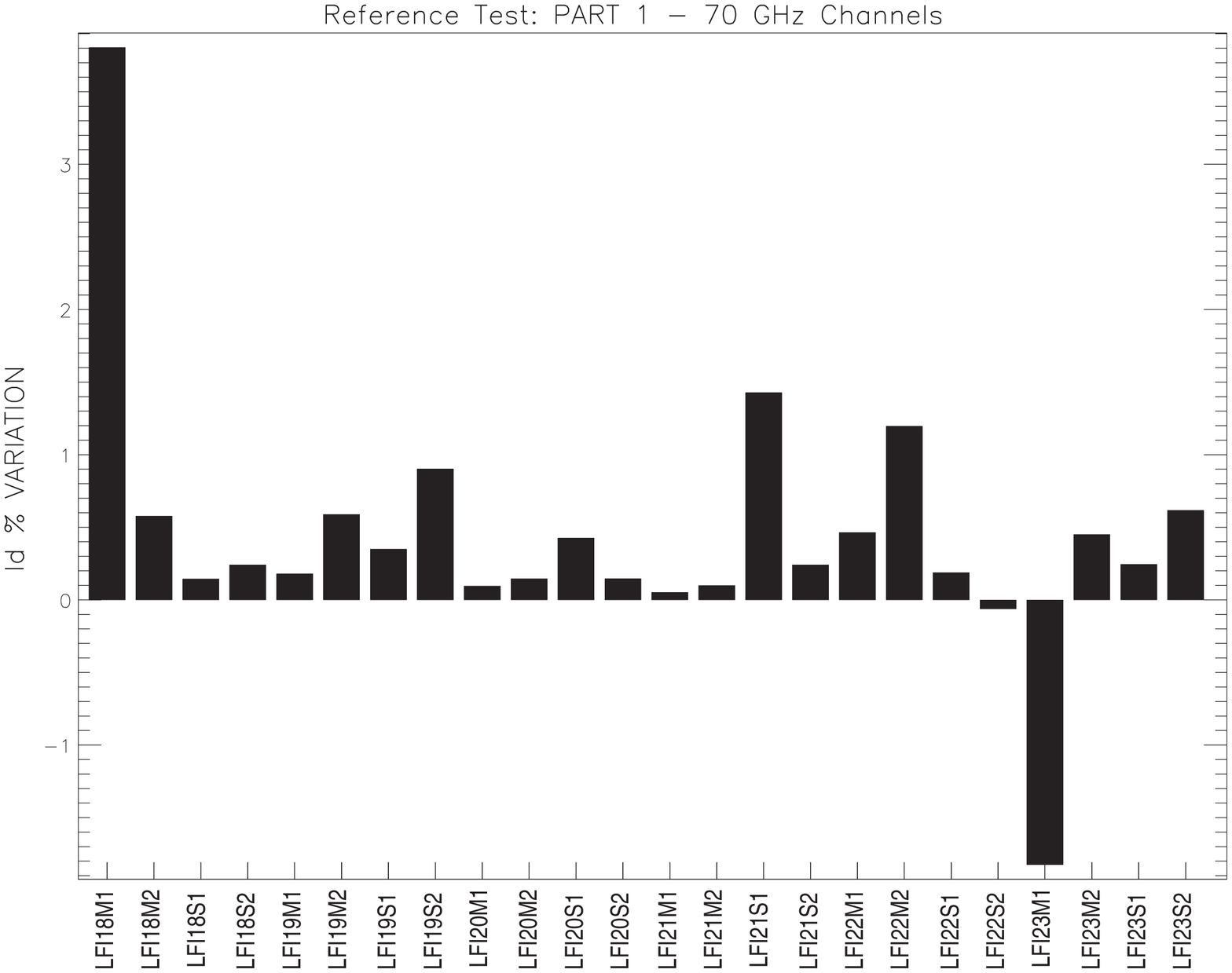}%[width=10cm] \\
\vspace{0.4cm} \hspace{0.2cm}
        \includegraphics [angle=90,width=10.25cm ]{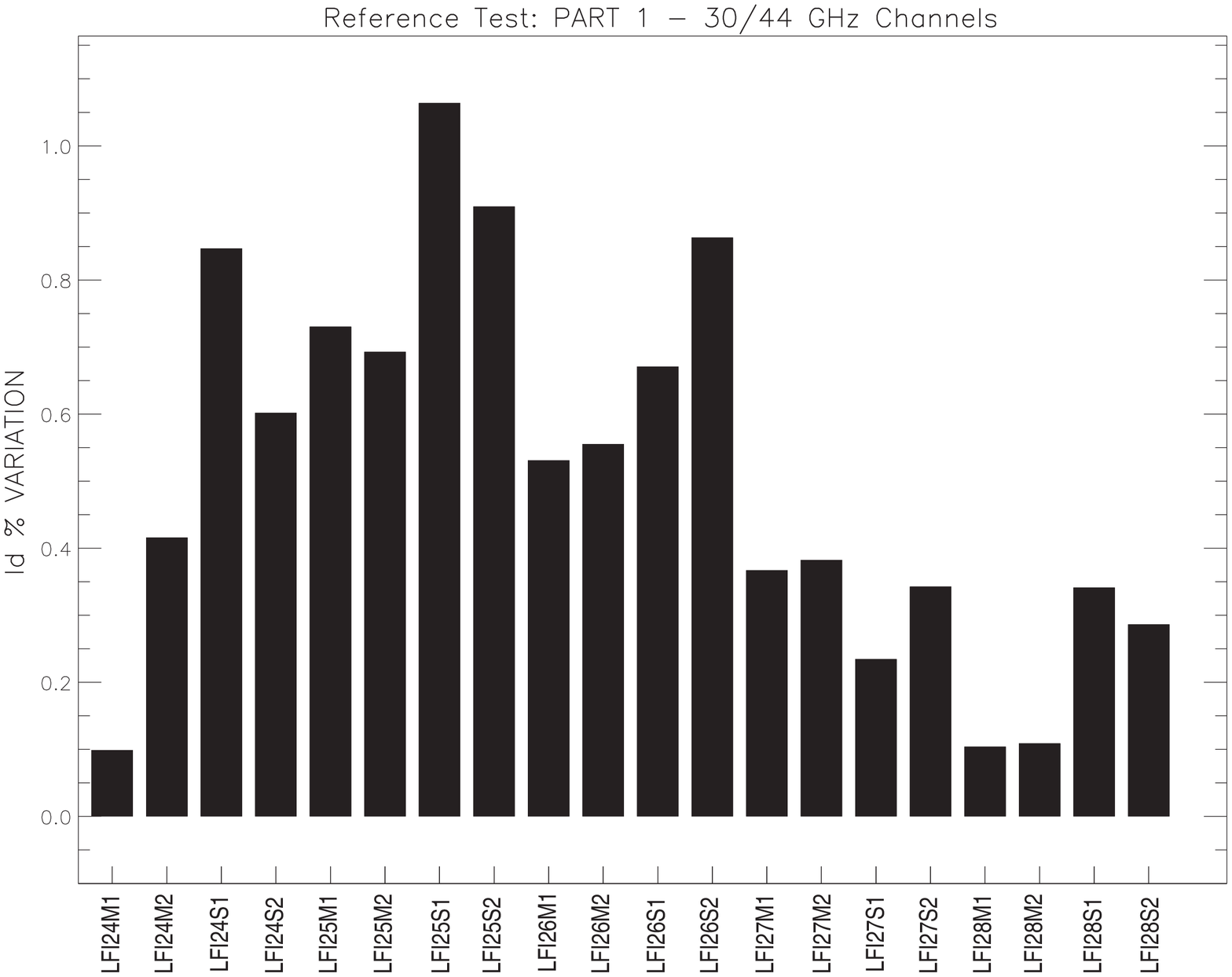}%[width=10cm]
    \end{center}
    \caption{Reference Test Part 1: Drain Current comparison with \texttt{CRYO-02} data.}% For each channel, outputs are indicated as \texttt{0, 1, 2, 3}, referring to LNAs respectively labelled as: \texttt{M1, M2, S1, S2}.)  }
    \label{Plot_REF_TEST_P1_Id} 
\end{figure}

\begin{figure}[h!]
    \begin{center}
        \includegraphics [angle=90,width=10.95cm ]{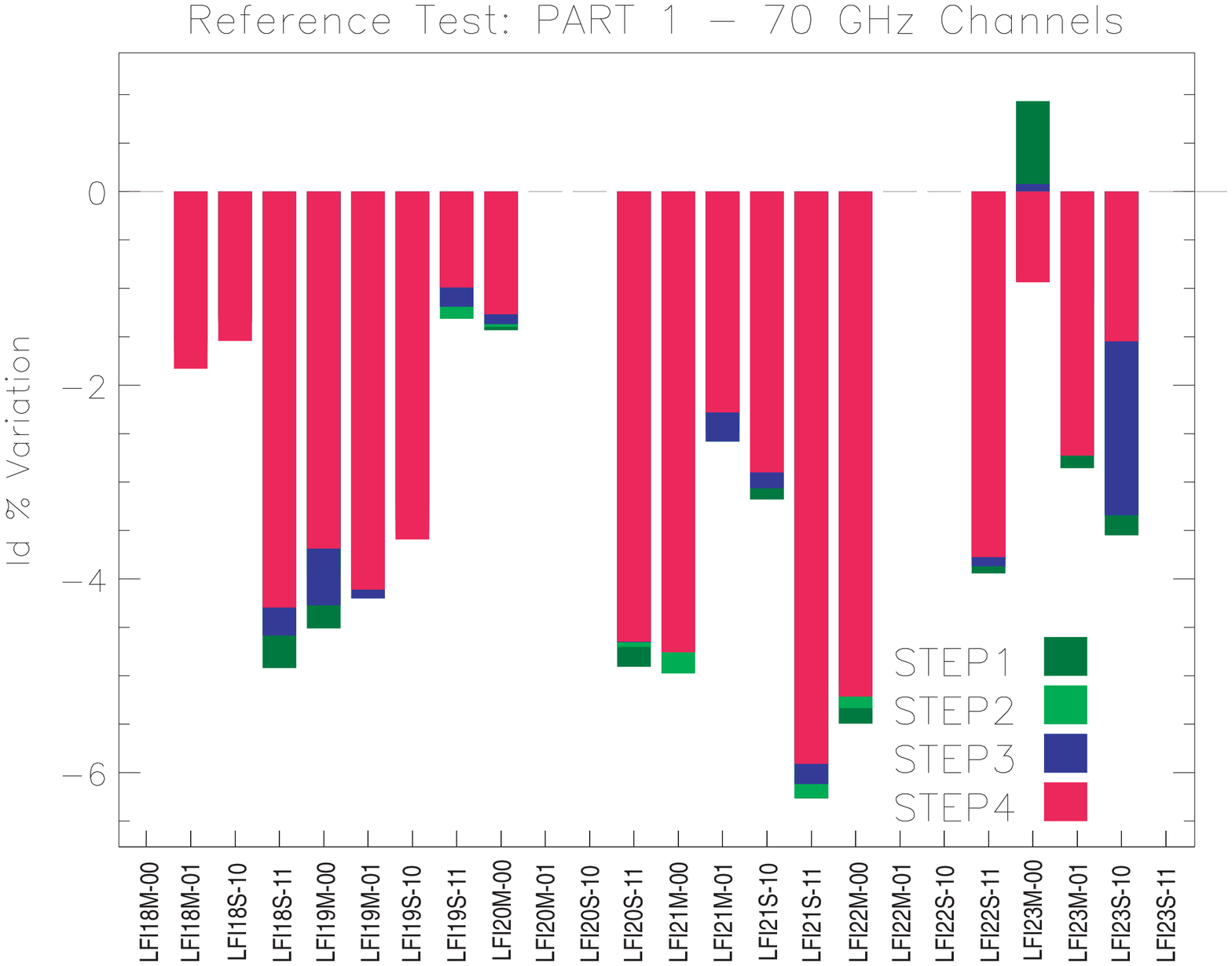}%[width=10cm] \\
\vspace{0.4cm} \hspace{0.2cm}
        \includegraphics [angle=90,width=10.25cm ]{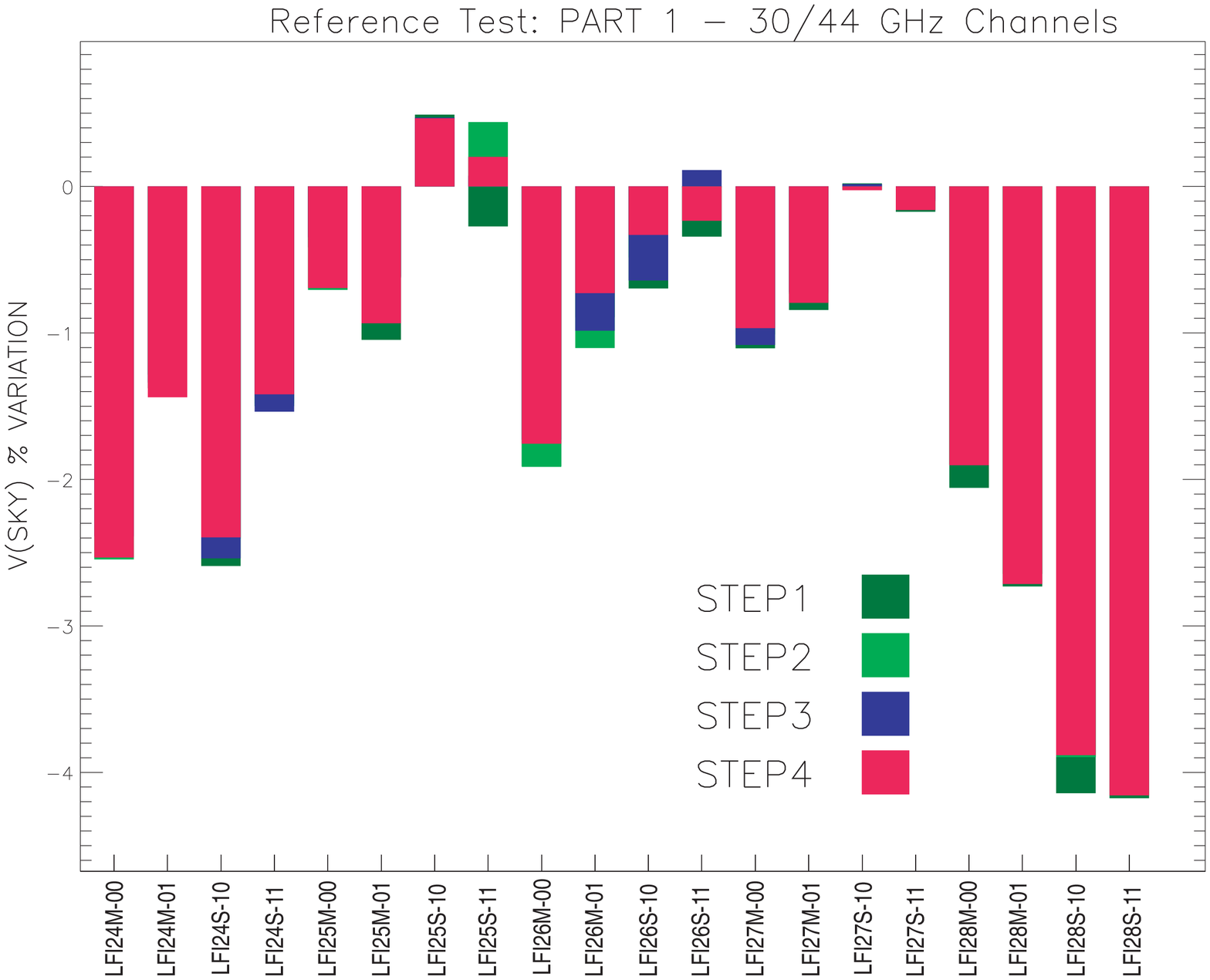}%[width=10cm]
    \end{center}
    \caption{Reference Test Part 1: Output voltage (sky) with  \texttt{CRYO-02} data sets: the four steps are stacked on the same bar for each channel. When differences among the four steps are negligible, just one color is displayed on each bar. For three radiometers (\texttt{LFI20M-00}, \texttt{LFI20M-01}, \texttt{LFI22M-00}, \texttt{LFI22M-01}, \texttt{LFI23S-10}, \texttt{LFI23S-11}) the comparison is missing because of an ADC saturation caused by a wrong settings in the DAE biases. }
    \label{Plot_REF_TEST_Vout} 
\end{figure}

A strong drift in the Voltage output characterized the whole test; this was probably due to variation in LNA power dissipation once \texttt{CRYO-02} biases were set. As during the first \texttt{CRYO-02}, some instabilities were observed in the drain currents of a few radiometers (\texttt{LFI21S1} and \texttt{LFI22S2}).\\
Drain current comparison (see Figure~\ref{Plot_REF_TEST_P1_Id}) with previous \texttt{CRYO-02} showed the excellent consistency of the two tests: differences are all below 5\%, on average below 2\% with the exception of \texttt{LFI18M1}, already known for being characterized by a larger instability. \\
The sky-voltage comparison (Figure~\ref{Plot_REF_TEST_Vout}) confirmed the agreement: differences are below 1\% in  most cases, around 5\% in the remaining confirming the good radiometer isolation.

The second part of the reference test, \texttt{RT-2}, %was aimed at testing each radiometer separately, both in LNAs and phase switches functionality, minimizing the perturbation of all the other channels.\\
confirmed the full functionality of radiometers as all the devices responded as expected. In detail we checked:
\begin{itemize}
\item the functionality of V$_{\rm g1}$ and V$_{\rm g2}$ amplification stages;
\item phase switch diodes ability to exchange signals (sky/ref) when the polarization status is inverted and to separate signals when I$_1$ or I$_2$ phase switch bias currents are lowered or increased;
\item the 4 kHz switching;
\item DAE ability to tag signals (sky/ref) correctly when the P/S status or the switching 4~kHz circuit are inverted;
\item the functionality of the Back-End warm LNAs.
\end{itemize}

This test also provided  a quantitative reference point for any comparison during the mission. 

%% file: 04_lfi_tests_psw_tuning.tex
The objective of this test was to find the optimal bias currents to the front-end phase switches that balance the wave amplitude in the two phase switch states. Because an optimal balance affects the receiver isolation, the test was performed before the tuning of the front-end modules amplifiers. The test was repeated after the front-end amplifier tuning to show that the optimal phase switch biases were independent from the amplifier bias values. As planned the test was performed only on 30 and 44~GHz RCAs by varying the current supply (I$_1$ and I$_2$) in order to reach the optimal balance (this choice was driven by the switch time response that in the 70 GHz devices was longer compared to the 30 and 44~GHz and further increased when the diode currents were lowered: the option to setting phase switch biases to maximum allowed values was preferred to phase balancing \cite{cuttaia2009,varis2009}). When this condition is reached, the sky-reference output difference is very close as possible to zero when the amplifier in the opposite leg is set to zero bias. In fact, when 
an ACA is switched off, 
the output at each diode in the two switch states is (Seiffert et al. 2002 page 1187):
               \begin{eqnarray}
            V_{\rm even}^{\rm bem} \propto {1\over \sqrt{2}} \sqrt{A_2} e^{i\phi_2} G_{\rm FEM} \left ( N_{\rm FEM}  + {S_{\rm sky} + S_{\rm ref} \over \sqrt{2} }\right ) \\
            V_{\rm odd}^{\rm bem} \propto {1\over \sqrt{2}} \sqrt{A_1} e^{i\phi_1} G_{\rm FEM} \left ( N_{\rm FEM}  + {S_{\rm sky} + S_{\rm ref} \over \sqrt{2} }\right ),
            \label{eq:psw_tuning_output}
        \end{eqnarray}

where ${\phi_{1,2}}$ represent the phase shifts in the two switch states (in LFI baseline ${\phi_1}  = 0$ and ${\phi_2} = \pi$); $A_{1,2}$ represent the fraction of the signal amplitudes  transmitted after the phase switch in the two switch states (for lossless switches $A_{1,2} = 1$); $N_{\rm FEM}$ is the Front-End LNA noise .
In the case of a tuned phase switch, the difference between the two voltages of the two states, $\delta$ = $V_{\rm even}^{\rm bem}$-$V_{\rm odd}^{\rm bem}$, is close to 0.
\begin{figure}
	\centering
		\includegraphics[width=0.50\textwidth]{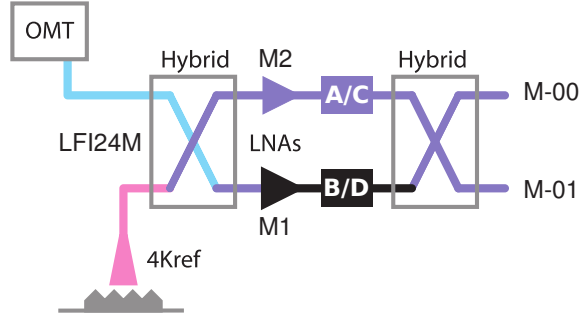}
		 \caption{\label{fig:pippo1} Radiometer scheme showing the phase switch tuning. The figure shows the case of the LFI24M when one amplifier is switched off. Under this condition the output of both hybrid output arms experience the same signal unless the phase switches is not well balanced. Reference and sky signals, disentangled at the 2$\rm ^{nd}$ hybrid output, are expected to be close under balanced conditions. The figure is a shot of a 1/8000~s.}
\end{figure}
        
For each pair  I$_1$ and I$_2$ we measured the following quantities: 
\begin{itemize}
	\item 	$V_{1}^{\rm even}$ and $V_{1}^{\rm odd}$: average over the time window respectively of the even and odd samples of the signal at first detector;
	\item $\Delta V_{\rm 1}$: difference at the first detector defined as $\Delta V_{\rm 1} = V_{1}^{\rm even} - V_{1}^{\rm odd}$;
	\item  $V_{2}^{\rm even}$ and $V_{2}^{\rm odd}$: average over the time window respectively of the even and odd samples of the signal at second detector;
	\item  $\Delta V_{\rm 2}$: difference at the second detector defined as $\Delta V_{\rm 2} = V_{2}^{\rm even} - V_{2}^{\rm odd}$;
	\item  $\Delta V$: figure of merit parameter defined as  $\Delta V = \sqrt{\Delta V_{1}^{\rm 2} +\Delta V_{2}^{\rm 2}}$.
\end{itemize}

For each phase switch a matrix of  I$_1$ and I$_2$ values was applied, the ranges were set on the basis of the ground test results.  The matrix approach, already verified on ground \cite{cuttaia2009}, was selected as the best option to evaluate the best balance condition for phase switches by computing the minimum value of $\Delta V$. 
Notice that phase switch currents are displayed using the adimensional code (from 0 to 255) used to set the bias in the DAE rather than the current in physical units. At first order the interested reader can convert the DAE code into the actual delivered current using a linear calibration with the values detailed in Table~\ref{tab_dae_calibration_i}. It is important to underline that the calibration table actually used to convert DAE codes to physical units is channel dependent, but it is not provided here for simplicity.
\begin{table}[h!]
    \begin{center}
        \caption{Values for a linear calibration of DAE bias codes into physical units, for all LFI channels.}
        \label{tab_dae_calibration_i}
        \begin{tabular}{ c  c }
        \hline  \hline
           \multicolumn{1}{c}{\texttt{DAE code = 0}} & \multicolumn{1}{c}{\texttt{DAE code = 255}} \\
           I$_{\rm 1}$, I$_{\rm 2} $ & I$_{\rm 1}$, I$_{\rm 2}$\\
        \hline
0.0~mA & 1.0~mA \\
        \hline
        \end{tabular}
    \end{center}
\end{table}

\begin{figure}
	\begin{center} 
            \includegraphics[width=7.1cm]{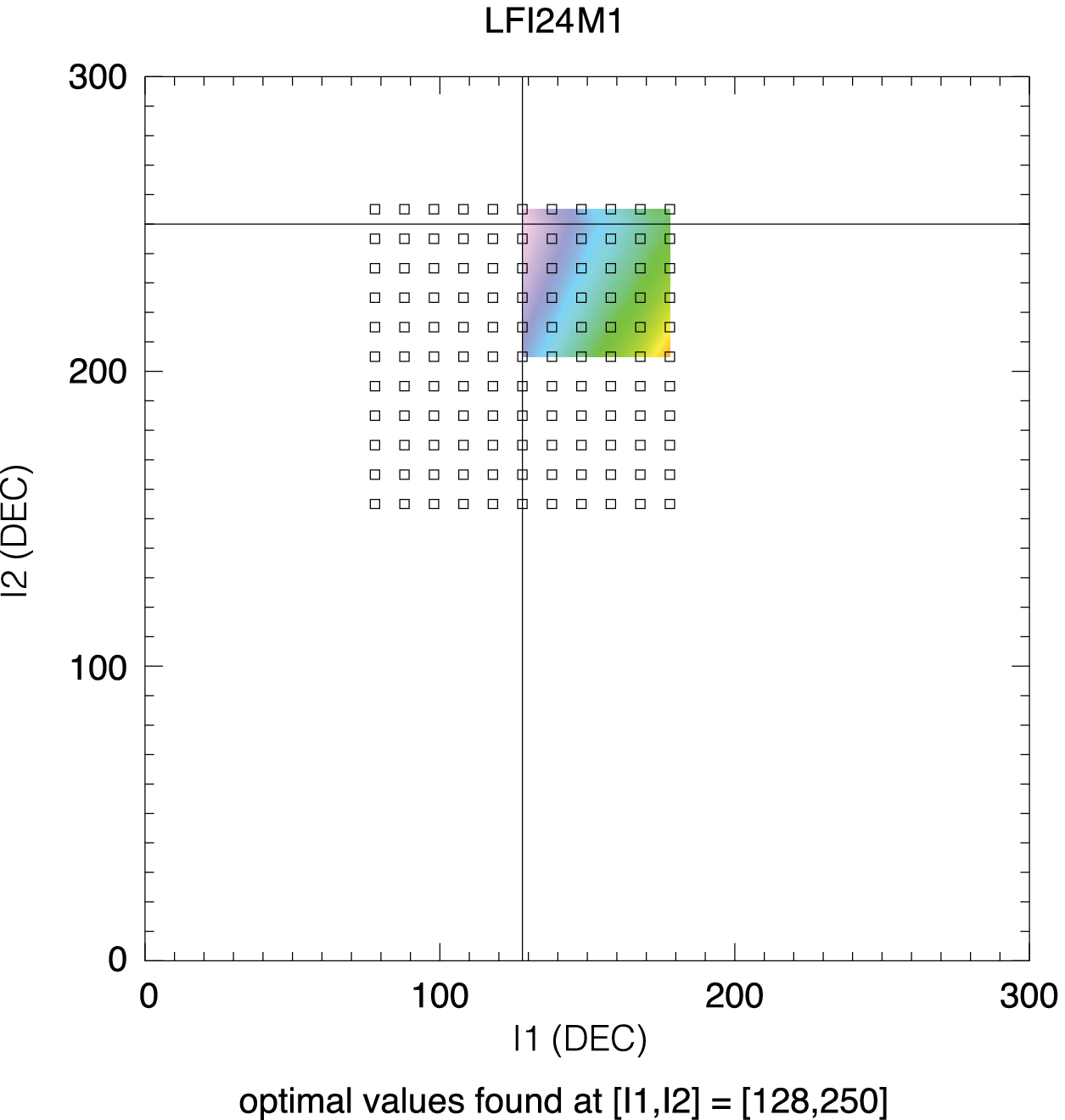}
            \includegraphics[width=7.5cm]{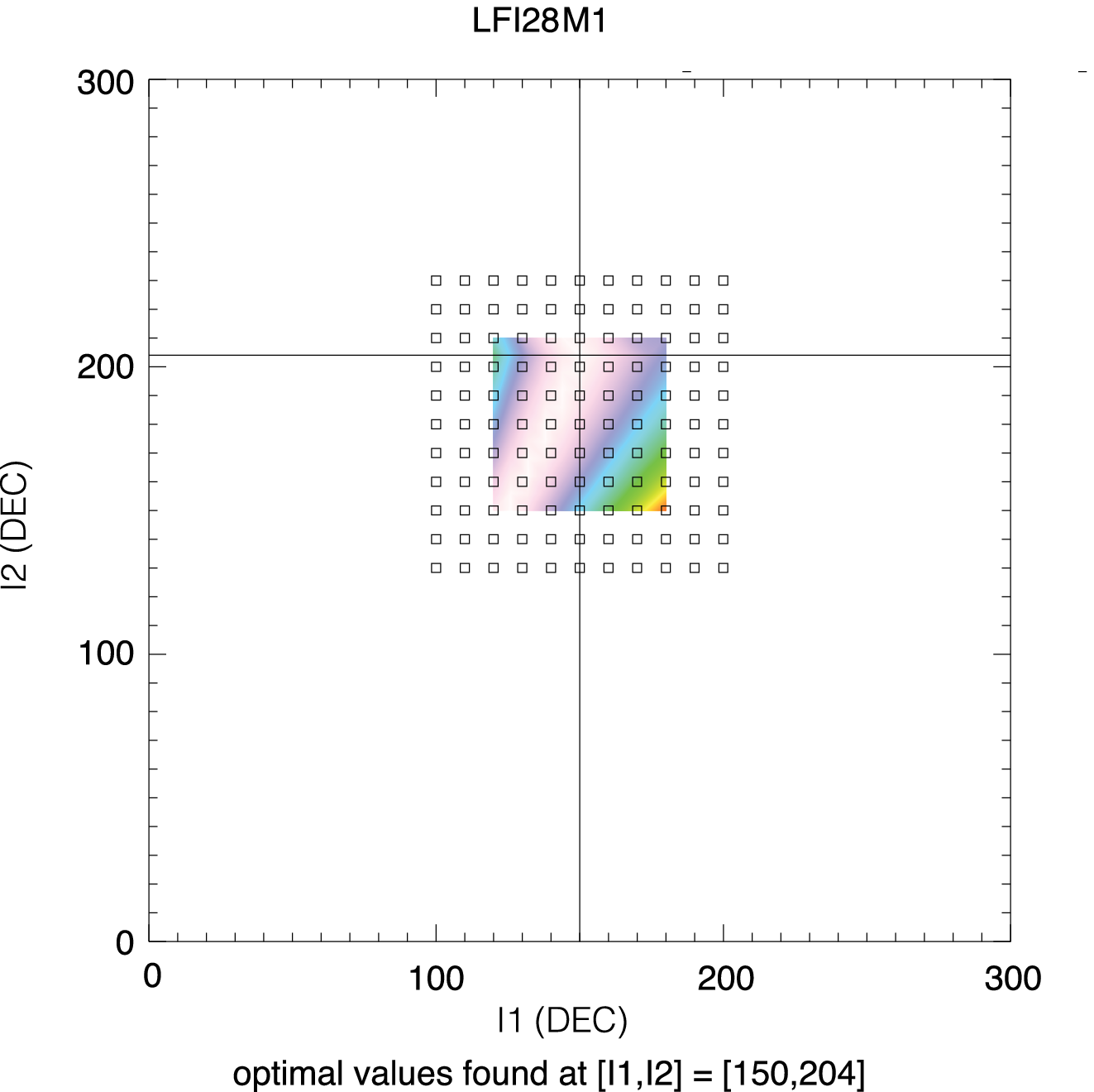}
		 \caption{\label{fig_PHSW_tun_grid} CSL system level tests phase switch tuning maps for the two channels \texttt{LFI24M1} (left panel) and \texttt{LFI28M1} (right panel) compared to the bias grid sampled during the same test performed in CPV. I$_1$ bias lies on x-axis, I$_2$ bias on y-axis. The bright zone identifies small unbalances. The map represents the figure of merit $\Delta V$ as defined in the text. Open squares represent the  points explored during CPV. Note that the bias units are DEC units that are used to set the bias in the DAE: see Table~5 for a conversion into physical units.}
   	\end{center} 
              \end{figure}

Two examples of the analysis performed are shown in Figure~\ref{fig_PHSW_tun_det}, while the behavior of all the 30~GHz and 44~GHz channels is displayed in  Appendix~\ref{app_PH_SW_plots}, Figure~\ref{fig_PHSW_tun} in terms of relative percentage balance maps $\delta = 100\cdot  \Delta V / \left ( V_{\rm ave}\right )$, in colored contour plots for the whole set of applied I$_1$ and I$_2$ combinations.\\ 
\begin{figure}
	\begin{center} 
            \includegraphics[width=7.3cm]{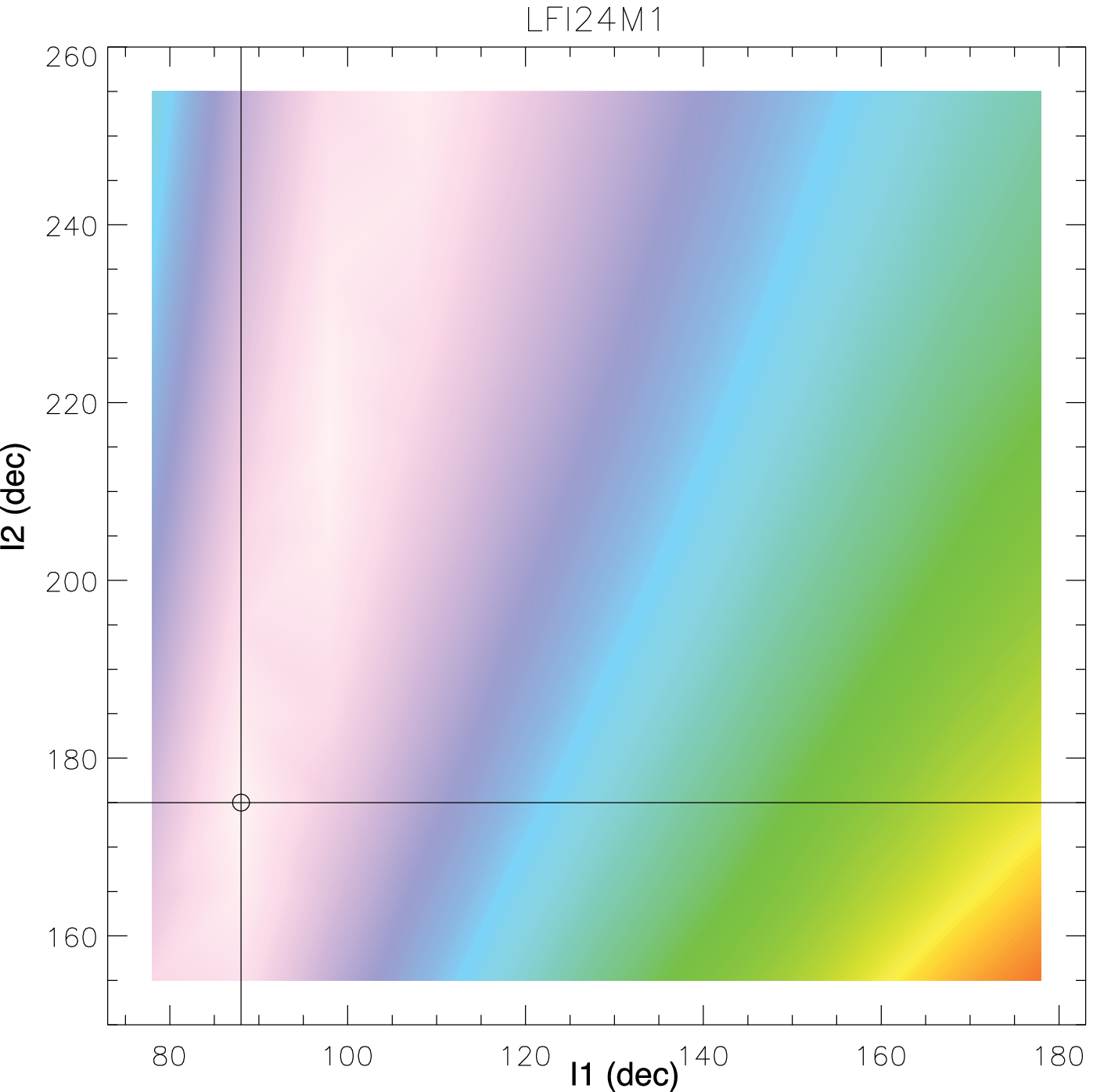}
           \hspace{0.1cm}             
            \includegraphics[width=7.3cm]{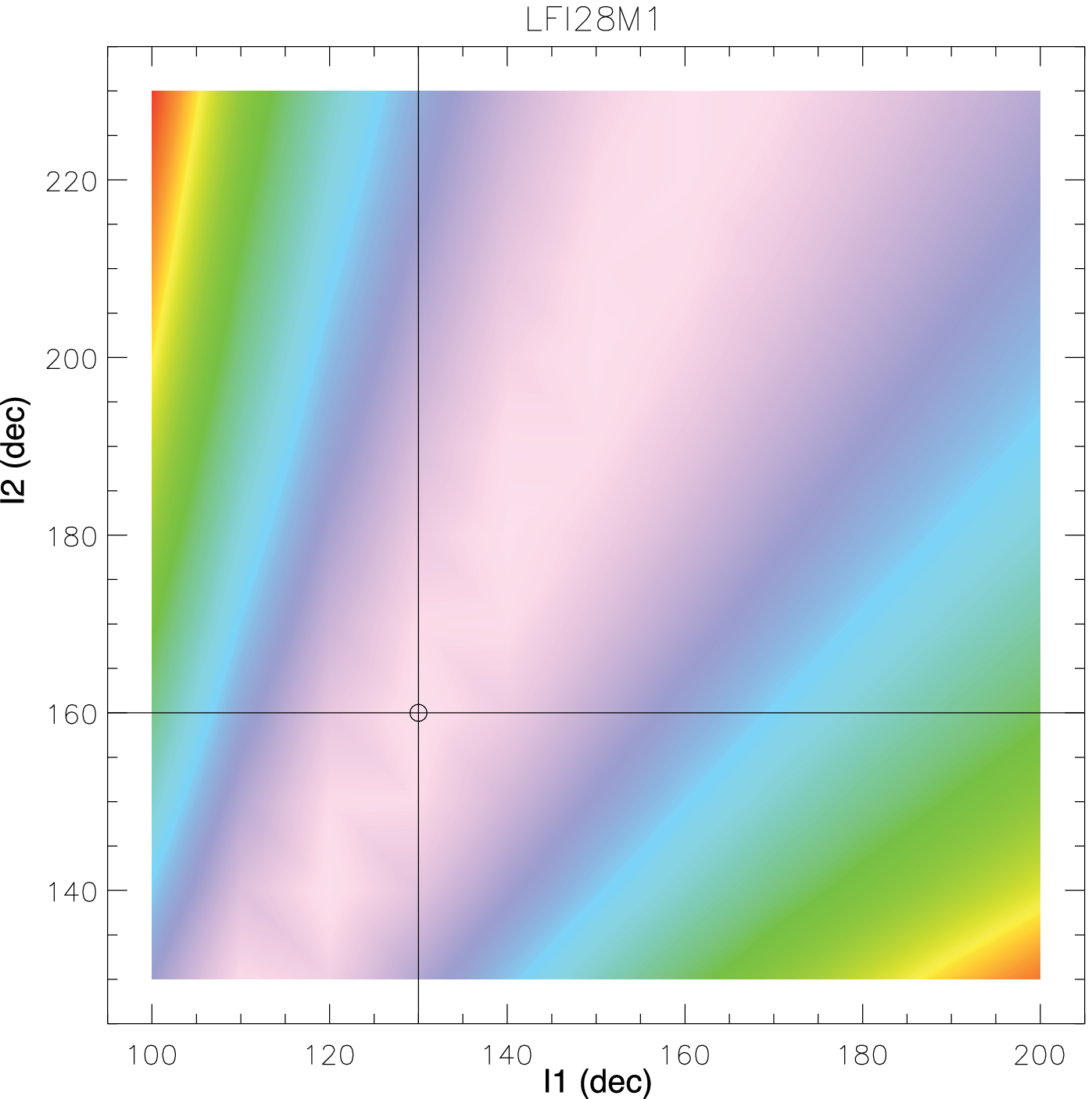}
		 \caption{\label{fig_PHSW_tun_det} Phase switch tuning maps for the two channels \texttt{LFI24M1} (left panel) and \texttt{LFI28M1} (right panel). I$_1$ bias lies on x-axis, I$_2$ bias on y-axis. The bright zone identifies small unbalances. The map represents the figure of merit, $\Delta V$, as defined in the text. Note that the bias units are DEC units that are used to set the bias in the DAE: see Table~5 for a conversion into physical units.}
  	\end{center} 
              \end{figure}
\indent Table~\ref{tab_phsw_Tun_results} in Appendix~\ref{app_detail_tests}, reports the values found from the automatic routine, the optimized values and the flight default values applied during the mission. It should be noted that in some cases the best tuning condition was clearly in a region outside the grid scanned with phase switch tuning performed during system level tests  in CSL. This is the case of \texttt{LFI24M1}, \texttt{LFI25M2}, \texttt{LFI25S1}, \texttt{LFI26M1}, \texttt{LFI26S2}, \texttt{LFI27M1}, and marginally of  \texttt{LFI24M2}, \texttt{LFI24S2}, \texttt{LFI27M2}, \texttt{LFI27S1}, \texttt{LFI27S2}, \texttt{LFI28S2}. Two examples of resampled grids  compared to the original  regions explored during phase switch tuning at system level are reported in Figure~\ref{fig_PHSW_tun_grid}.

The optimal I$_1$ and I$_2$ pairs derived automatically from the routine were checked one by one and eventually optimized on the basis of the following criteria:
\begin{itemize}
\item Comparison with cryogenic ground tests at CSL \cite{cuttaia2009};
\item Balancing of I$_1$ and I$_2$ privileging biases providing high phase switch currents (guaranteeing small losses). 
\end{itemize}
According to the last two items, results were hand refined for the following ACAs: \texttt{LFI24M1}, \texttt{LFI25M2}, \texttt{LFI27M2}, \texttt{LFI27S2}, \texttt{LFI28M2}. 

\paragraph{Phase switch tuning verification}
This test was aimed at verifying that the phase switch tuning results are independent (or marginally dependent) on the bias configuration of the LNAs. In fact the phase switch tuning was performed before the LNAs hypermatrix tuning, with the LNAs in pre-tuned conditions (with ground test CPV biases). The phase switch tuning verification was instead run after the hypermatrix tuning, with the LNAs in ``Tuned conditions''. \\
However two events  during the test compromised the result: an electric oscillation occurred in the \texttt{RCA24} (caused by an error in the \textit{soft switch-on} procedure\footnote{During on-ground tests \texttt{RCA24} and \texttt{RCA28} showed non nominal behaviours if switched on from a zero bias configuration. Dedicated switch-on procedures named \textit{soft switch-on} procedure were adopted for both of them.}) and the signal saturation on the \texttt{RCA27} (caused by a wrong bias set at DAE level, amplifying the signal outside the allowed range). Because of these, the full purpose of this test was not reached. Actually, even if at first order the phase switch tuning test results were confirmed, any second order effects due to the phase switch interaction with the different LNAs bias remained unexplored.

%% file: 04_lfi_tests_LNAs_Tuning.tex
Optimal front-end biases were determined during CPV by exploring a large set of voltage values and assessing, for each of them, the radiometer performance in terms of noise temperature and isolation. Tuning was performed in two phases: a \textit{pre-tuning} phase to constrain the bias space around best-performance regions and a \textit{proper tuning} phase during which the bias regions identified during pre-tuning were finely sampled in order to converge towards the optimal values.

During both phases the bias space was sampled according to a so-called \textit{hypermatrix} (\texttt{HYM}) strategy in which several bias quadruplets, $[(V_{\rm g1},V_{\rm g2})_{\rm LNA-1},(V_{\rm g1},V_{\rm g2})_{\rm LNA-2}]$, were varied simultaneously for each radiometer. This strategy increased the sampled parameter space with respect to ground tuning tests \cite{cuttaia2009}, where the bias space was sampled independently for the two ACAs over a two-dimensional space. 

In both phases the test was performed simultaneously on several RCAs which were grouped according to a scheme that minimised the bias cross talk effects along the harness lines. The details of the grouping scheme are reported in Appendix~\ref{app_power_groups_table}, Table~\ref{tab_LNAs_Tuning_groups}. ACAs not under test were set at their default biases (CSL biases) with the only exception of the phase switch biases that during the \textit{proper tuning} phase were set to the optimal values resulting from the phase switch tuning (see Section~\ref{sec:lfi_ps}). 

\begin{figure}[htb!]
    \begin{center}
      \includegraphics[width=7.7 cm]{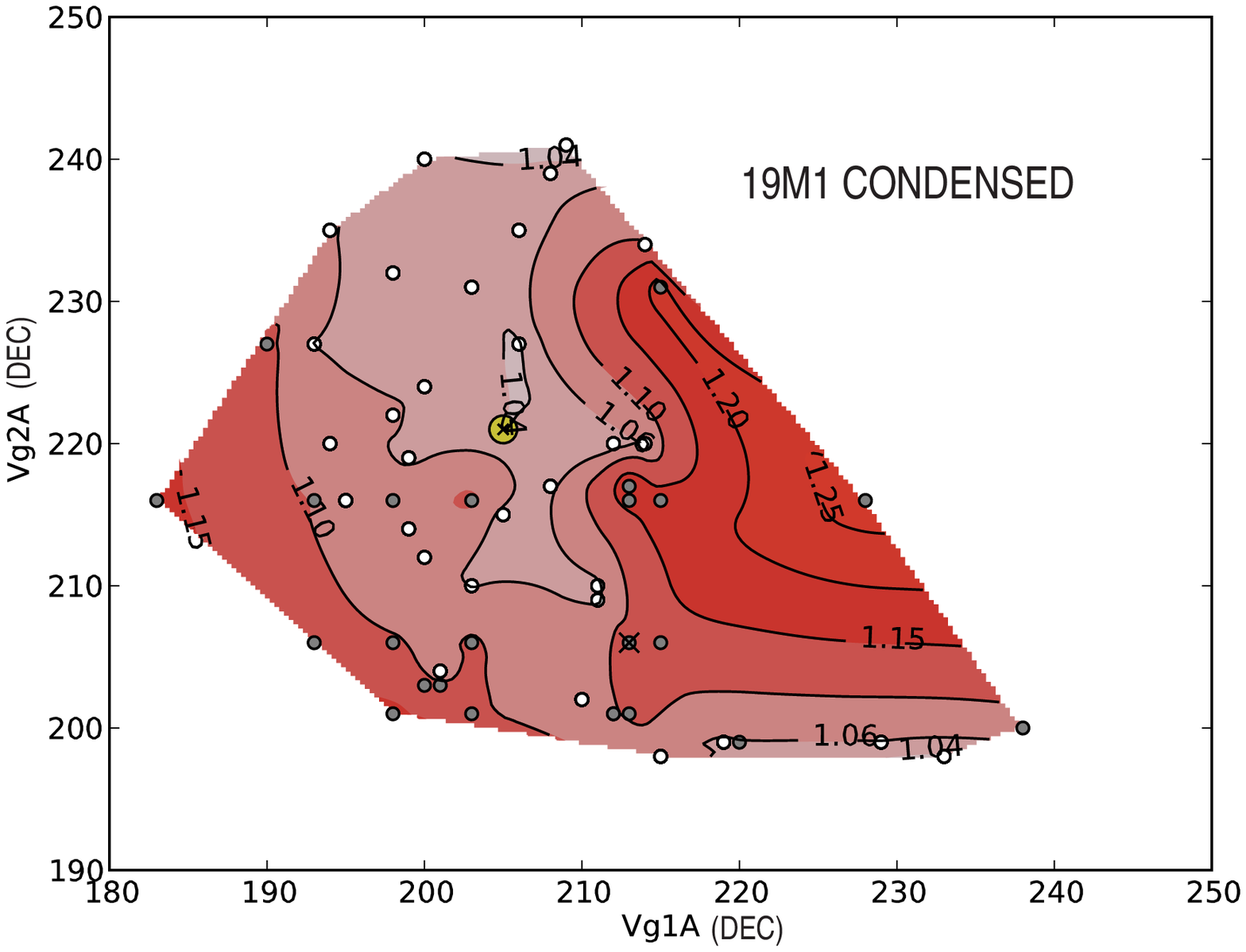} \hspace{-0.5cm}
      \includegraphics[width=7.7 cm]{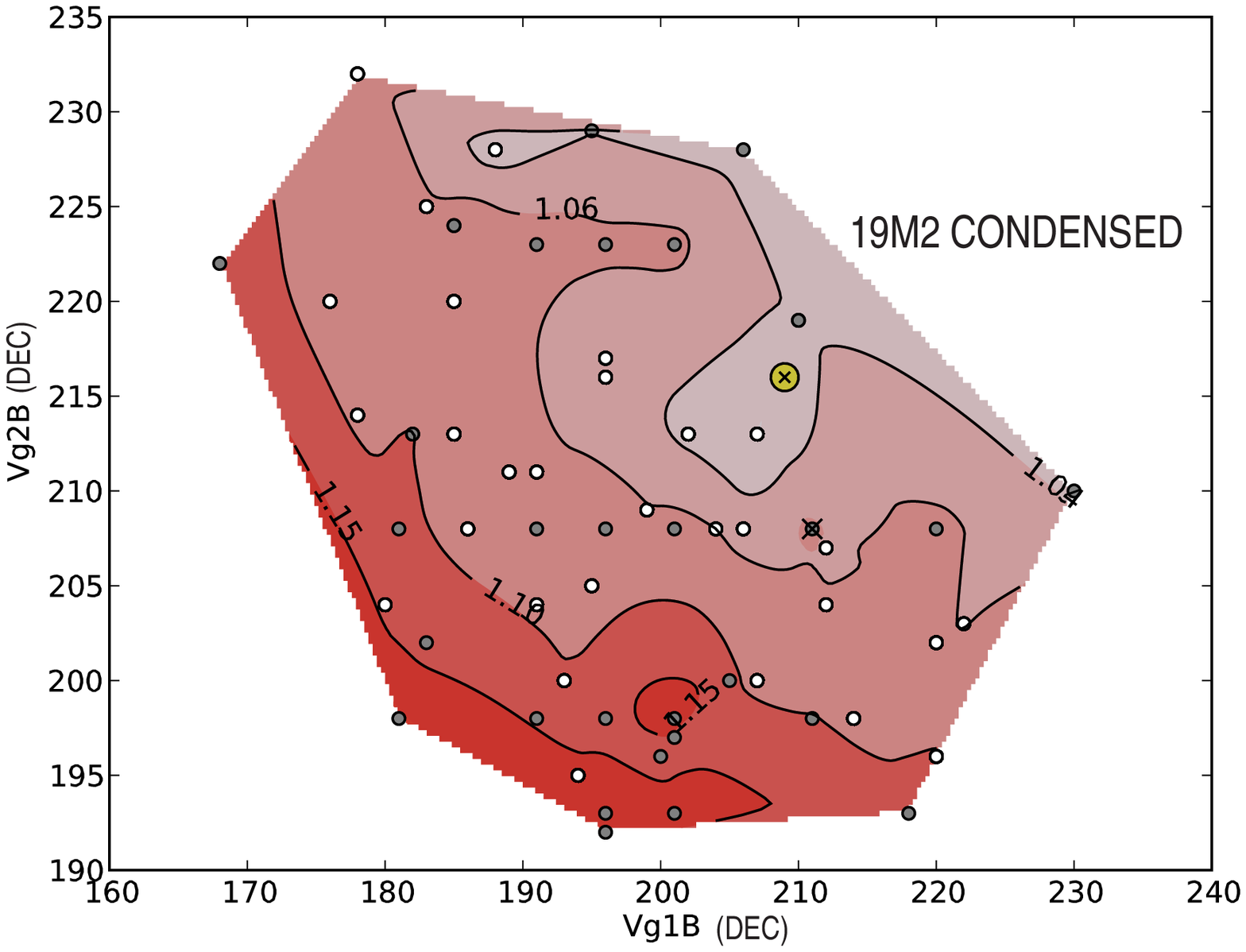}\\
      \includegraphics[width=7.7 cm]{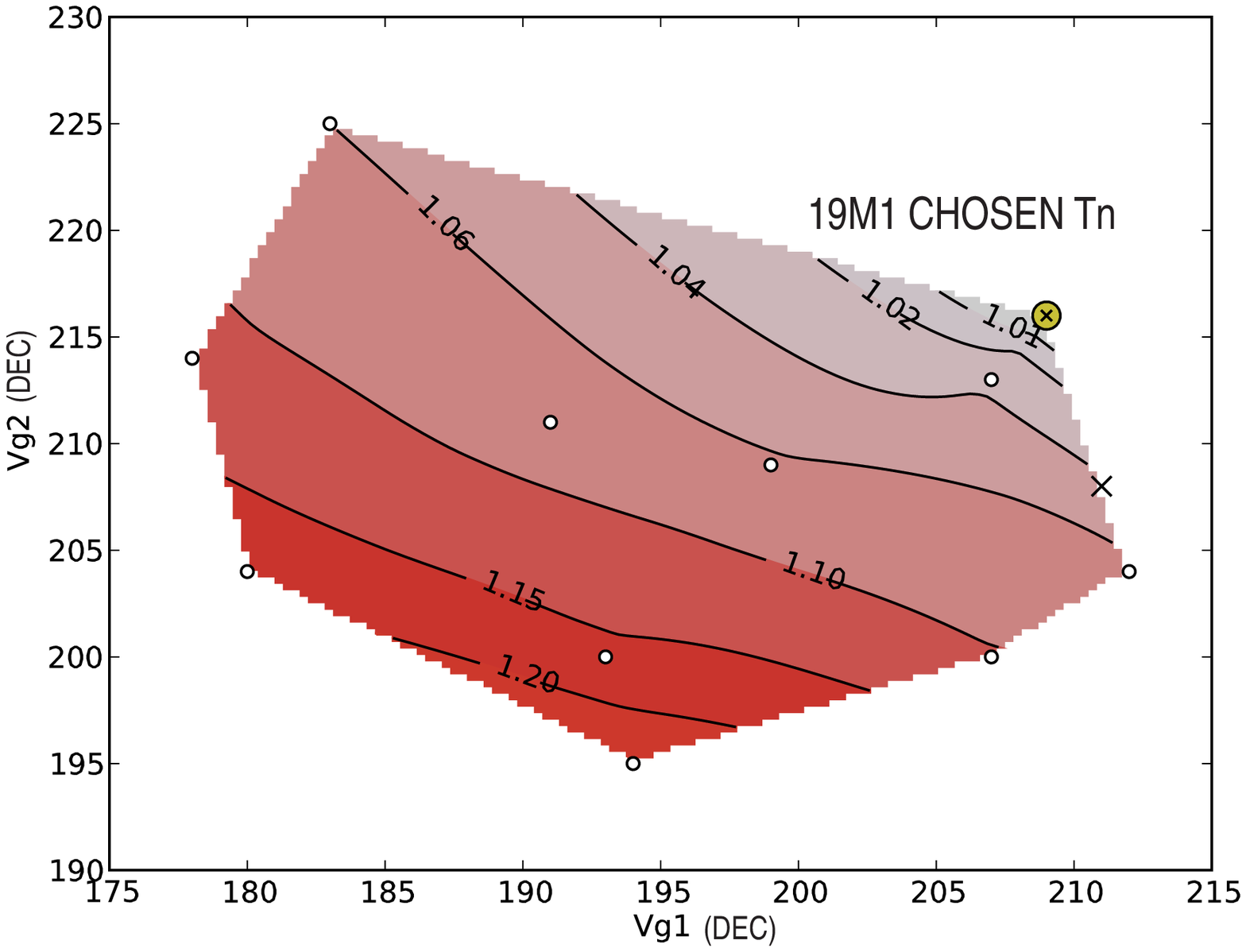} \hspace{-0.5cm}
      \includegraphics[width=7.7 cm]{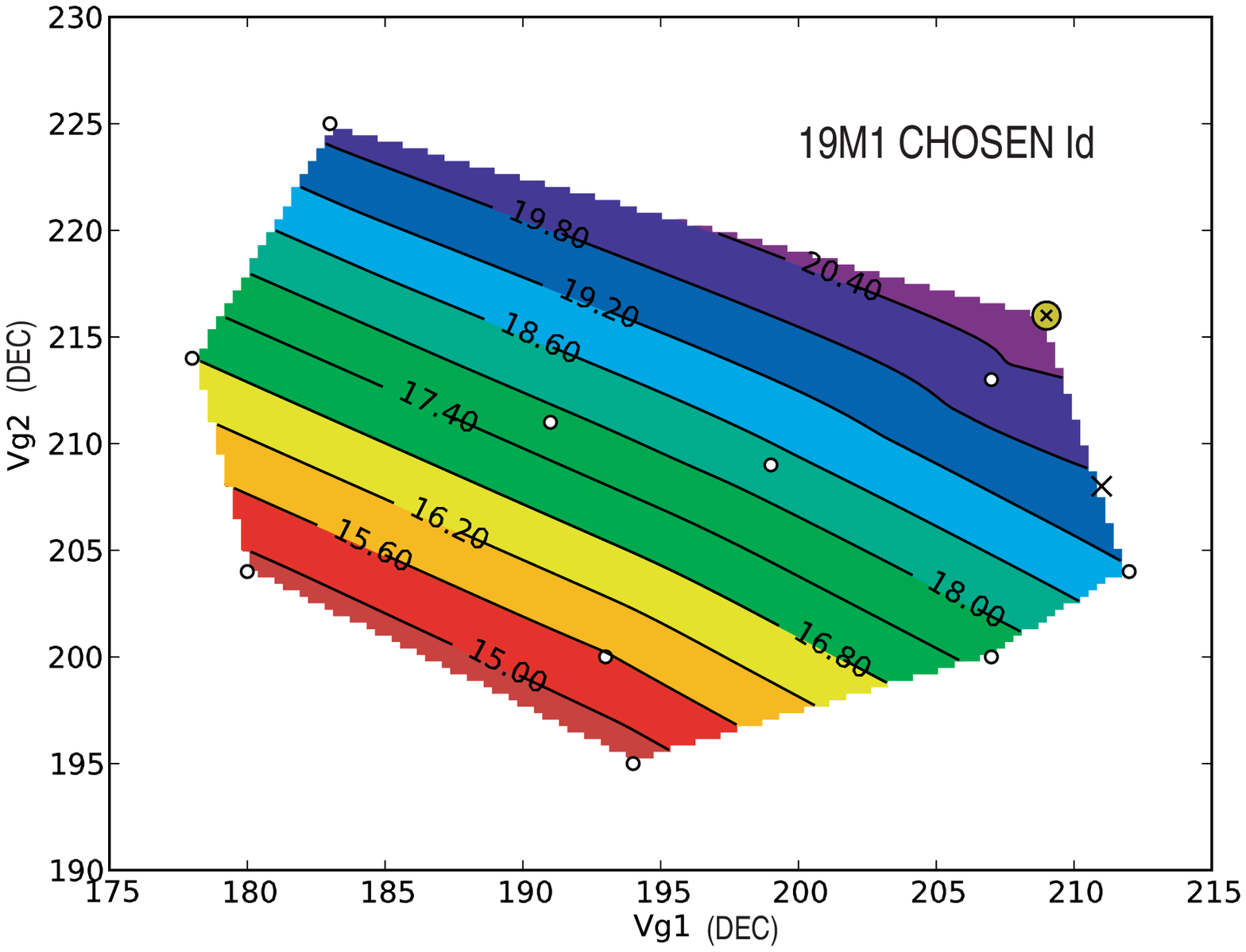}
    \end{center}
    \caption{Condensed bias maps for \texttt{LFI19M} as a function of $V_{\rm g1}$, on x-axis, and $V_{\rm g2}$, on y-axis. Top panels, noise temperature maps: each point represents the condensed properties of bias quadruplets in the bias space of the \texttt{M1} amplifier (left panel) and of the \texttt{M2} amplifier (right panel). Colour scale ranges from red (highest noise temperatures) to light-grey (lowest noise temperatures), normalized to the lowest noise temperature. Yellow circles correspond to the optimal bias quadruplets. Crosses correspond to results from CSL system level tests. Contours represents level of constant noise temperature.
Bottom panels: noise temperature map obtained when the \texttt{M1} bias corresponding to the best condensed noise temperature are fixed (left panel), the colour scale and contour meaning is the same as for the top panels; drain current map corresponding to the noise temperature map shown on the left (right panel), the colour map, from red to blue, represents increasing drain current values, contours represents level of constant drain current. Refer to the text for further explanations. }
    \label{fig_tuning_maps} 
\end{figure}

A dedicated code was developed to analyse the tuning data and navigate the four-dimensional bias space. Data display was performed using the concept of \textit{condensed bias maps} of radiometer noise temperatures, isolation and drain currents: see Figure~\ref{fig_tuning_maps} for a few examples. These are contour plots in the $V_{\rm g1},V_{\rm g2}$ space for the LNAs of a given radiometer. Each point in the plot is the average of the best 20\% noise temperature values determined by the quadruplets sharing that particular $V_{\rm g1},V_{\rm g2}$ pair. The interface allows to show, for each point, the quadruplet yielding best performance among all the possible combinations. The same approach was used to map isolation and drain currents. 

Hypermatrix analysis was further refined by applying a non linear model of detector response \cite{mennella2009-2} and other corrections accounting for gain variations due to thermal instabilities along the four steps (see \ref{par:nonlinear} section). As expected, with respect to the  standard approach, the hypermatrix method did not produce major changes in mapping the optimal quadruplets. 

At the end of the \textit{proper tuning}, a verification test (see Section~\ref{sec:lfi_ver}) was performed to directly calculate the white noise for each bias quadruplet. Probably due to the too short integration time, the power of this test was weaker than expected: despite of the overall consistency with the \texttt{HYM} tuning results, this test was not able to provide any extra improvement to the bias optimization process. 

\paragraph{Pre-tuning.}
\label{par:pretuning}

The aim of this first phase was to determine, for each radiometer, a relatively small bias region where best performance was expected. Pre-tuning was performed by scanning a large bias volume determined using results from ground tests combined with a drain current model applied to predict the expected drain current for each bias configuration. This test lasted 27 hours. 

During the test the temperature of the 4\,K cooler was still above 20\,K and we exploited the large unbalance between the sky and reference load signal to estimate the noise temperature for each diode according to the equation:
\begin{equation}
	T_{\rm noise} = \frac{T_{\rm ref}-Y^{\rm *} \cdot T_{\rm sky}}{Y^{\rm *}-1},
	\label{eq_tnoise_yfactor}
\end{equation}

where $Y^{\rm *} = V_{\rm ref} / V_{\rm sky}$, and $V_{\rm sky}$ and $V_{\rm ref}$ are the sky and reference load voltage outputs measured by each diode thanks to the 4\,kHz phase switch. This is a modification of the standard approach based on the Y--factor (see Section~\ref{eq_y_factor}) calculated on the temperature variations of the target load between two values \textit{high} and \textit{low} \cite{cuttaia2009}. Although pre-tuning scheme intrinsically prevented to calculate isolation, we measured drain currents to assess the gain balance of each amplifier pair in order to exclude bias voltage configurations leading to a too large drain current imbalance.

\begin{figure}[htb!]
 \begin{center}
        \includegraphics[width=7.4cm]{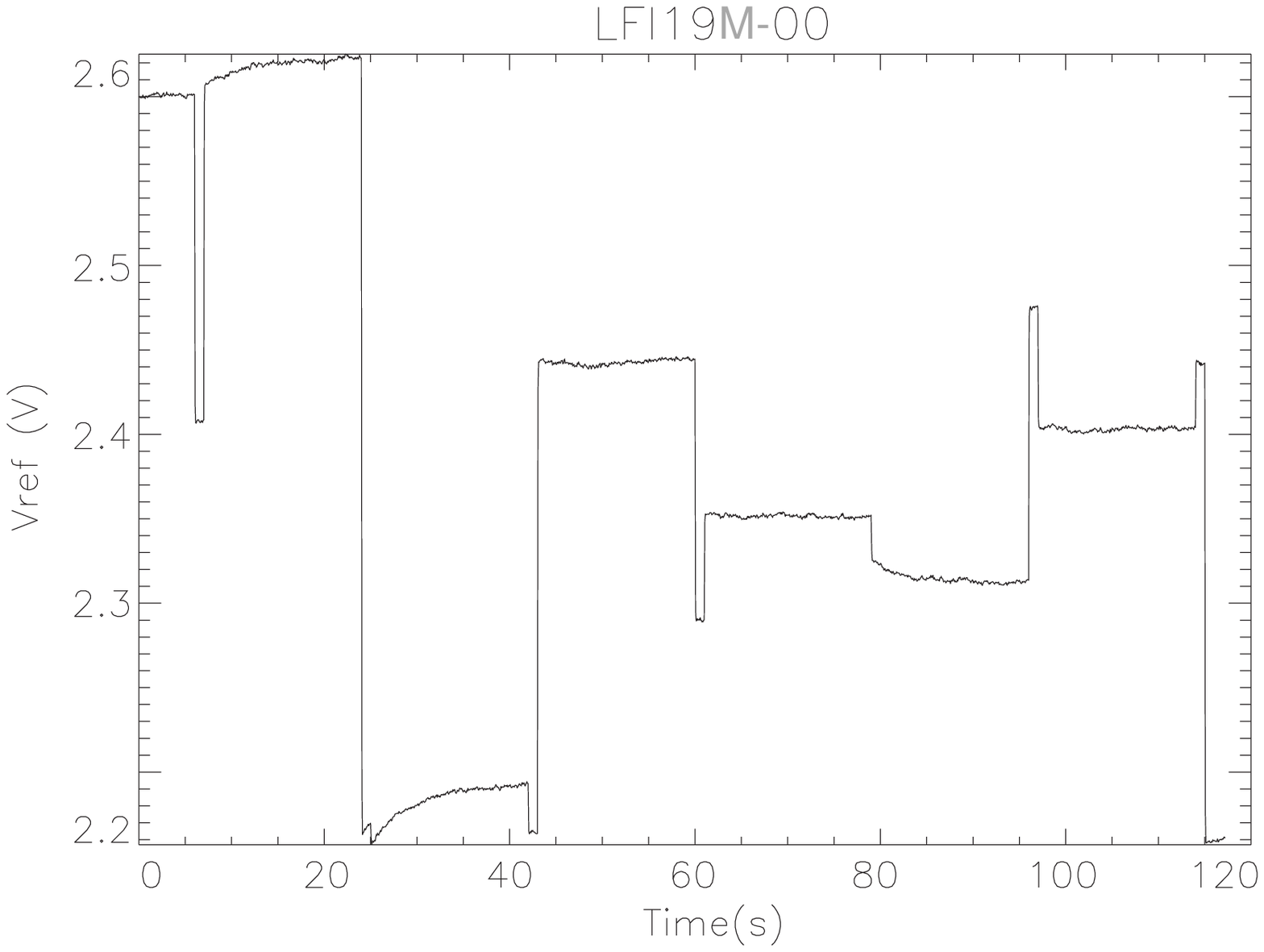} \\
        \includegraphics[width=7.4cm]{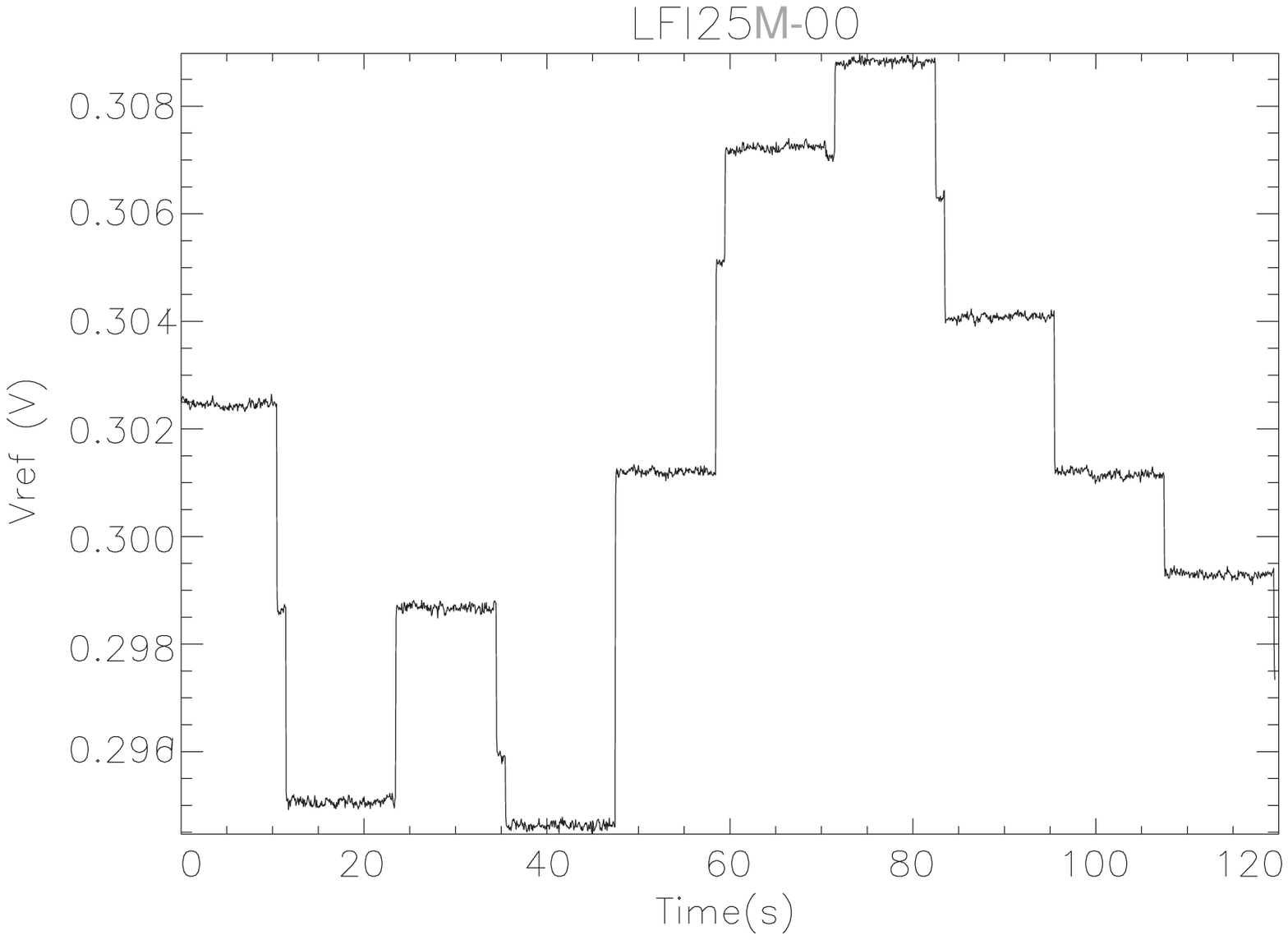} 
        \includegraphics[width=7.4cm]{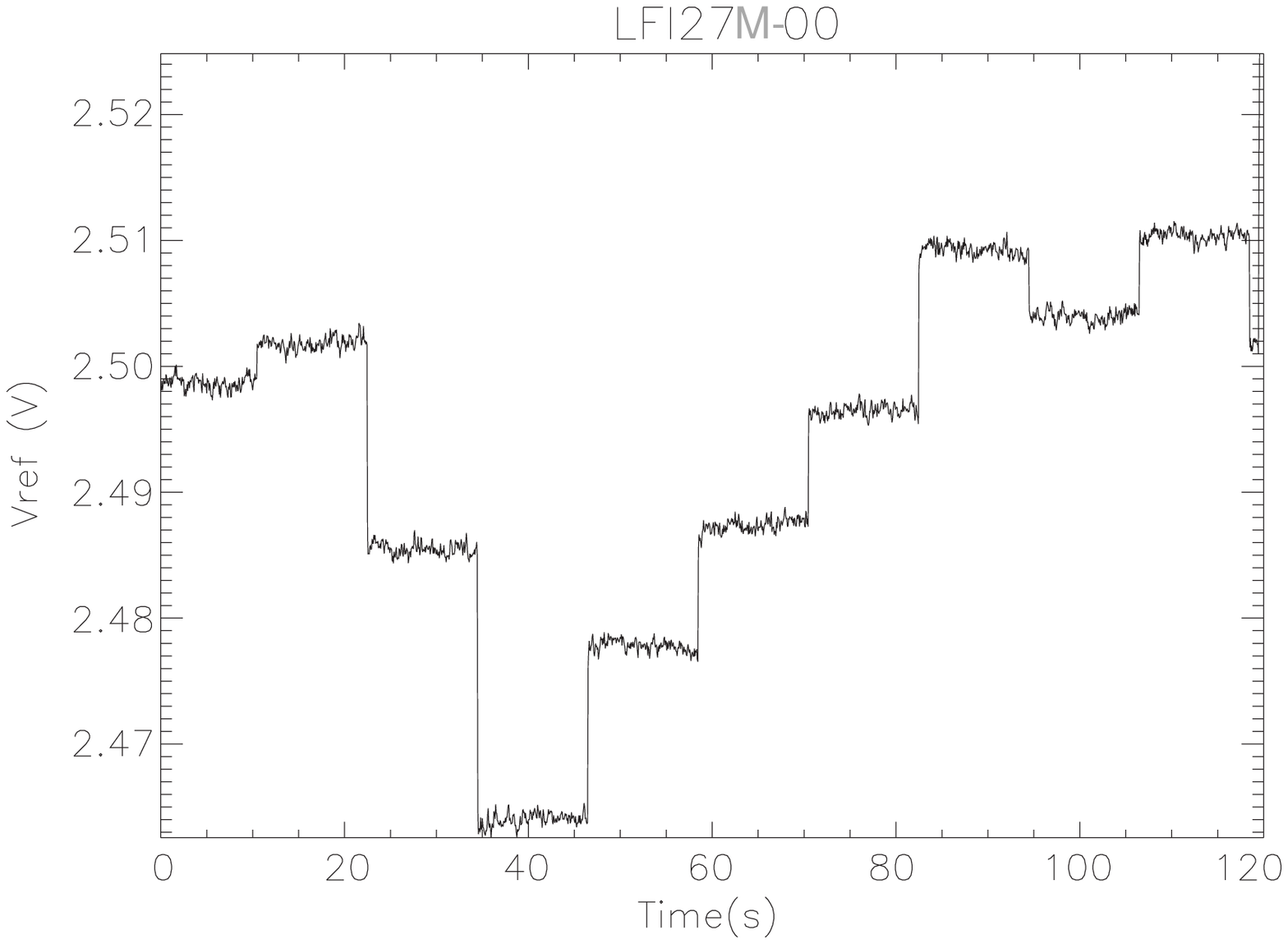}
    \end{center}
		\caption{Three examples of the voltage output step caused by a bias change. Top panel: \texttt{LFI19M-00}, bottom panels: \texttt{LFI25M-00} (left), \texttt{LFI27M-00} (right). Each panel covers a time window of 120 seconds. Voltage outputs of 30 and 44\,GHz channels relaxed faster than 70\,GHz channels. In all cases only the last three seconds of data were used in the analysis.}
    \label{fig_relaxation_times} 
\end{figure}

The integration time for each bias was a compromise between the test duration constraints and the necessary relaxation time after each bias change. We chose 15 seconds for the 70\,GHz channels and 9 seconds for 30 and 44\,GHz channels which relax faster (see Figure~\ref{fig_relaxation_times}). For each bias step, only the last three seconds of data were effectively used in the analysis. 

The accuracy in the noise temperature measurements depends on the levels and stability of sky and reference load signals ($T_{\rm sky}, T_{\rm ref}, \sigma_{T_{\rm sky}}, \sigma_{T_{\rm ref}}$) and on the levels and stability of the output voltages ($V_{\rm sky}, V_{\rm ref}, \sigma_{V_{\rm sky}}, \sigma_{V_{\rm ref}}$). In particular we have that:
\begin{equation}
	\sigma_{T_{\rm noise}} = \left[ 
	\left(\frac{\partial T_{\rm noise}}{\partial T_{\rm sky}}\right)^2\sigma_{T_{\rm sky}}^2 +
	\left(\frac{\partial T_{\rm noise}}{\partial T_{\rm ref}}\right)^2\sigma_{T_{\rm ref}}^2 +
	\left(\frac{\partial T_{\rm noise}}{\partial V_{\rm sky}}\right)^2\sigma_{V_{\rm sky}}^2 +
	\left(\frac{\partial T_{\rm noise}}{\partial V_{\rm ref}}\right)^2\sigma_{V_{\rm ref}}^2
	\right]^{1/2}.
	\label{fig_accuracy_noise_temperature}
\end{equation}

If we consider $V_{\rm sky}\sim$ 1\,V, $V_{\rm ref}\sim$ 2\,V, $\sigma_{V_{\rm sky}}\sim\sigma_{V_{\rm ref}}\sim 10$\,mV\footnote{Typical values of  the \texttt{LFI18M} radiometer, the one characterised by the highest voltage output and lowest stability},  $T_{\rm sky}\sim 3$\,K, $\sigma_{T_{\rm sky}}\sim 3.5$\,mK (the maximum peak-to-peak CMB dipole amplitude), we see that an accuracy of 1\% in the noise temperature requires the reference load temperature to be stable at the level of $\lesssim 600$\,mK during the useful integration time for each bias configuration. 

\begin{figure}[htb!]
    \begin{center}
        \includegraphics[width=7.4cm]{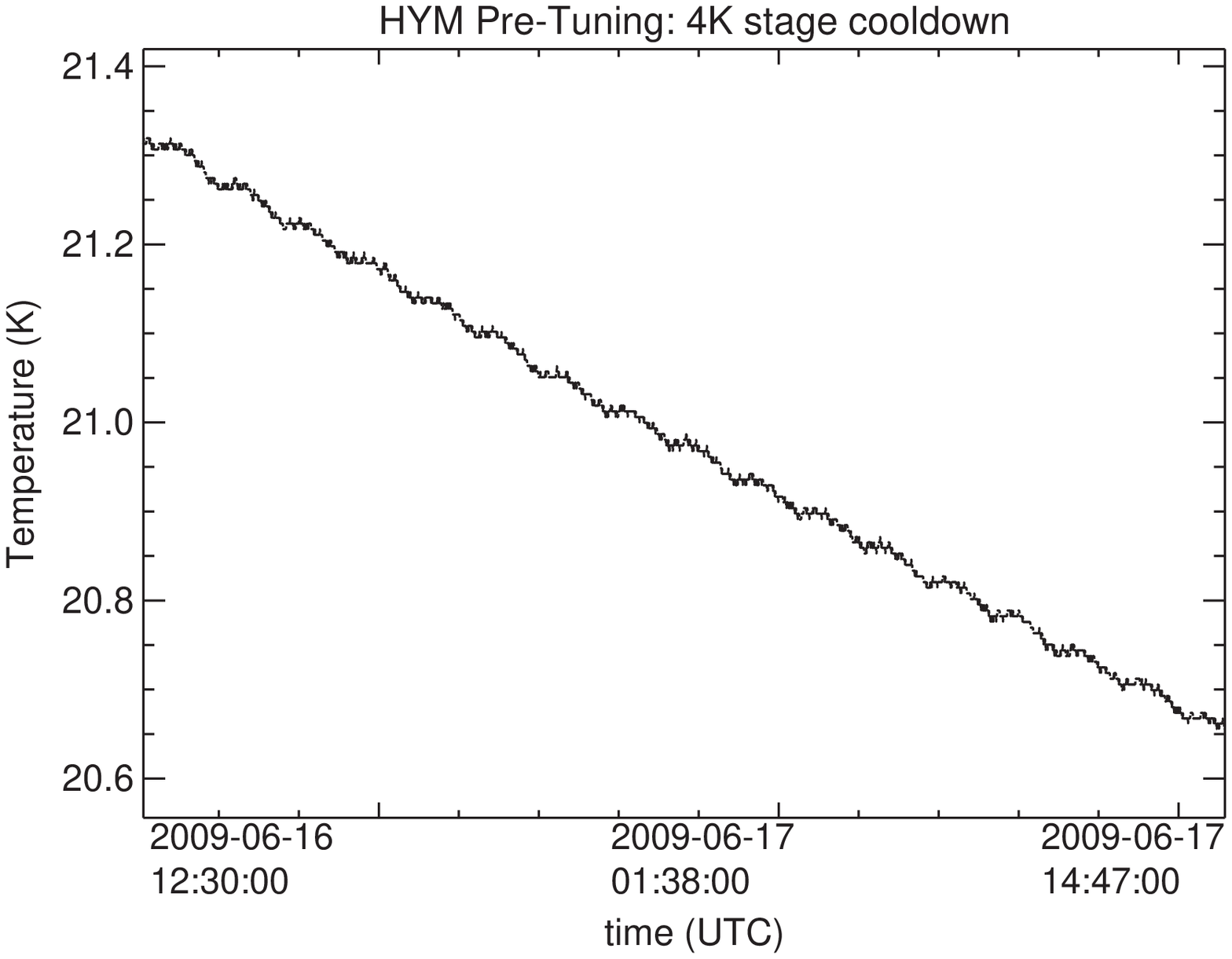}
        \includegraphics[width=7.4cm]{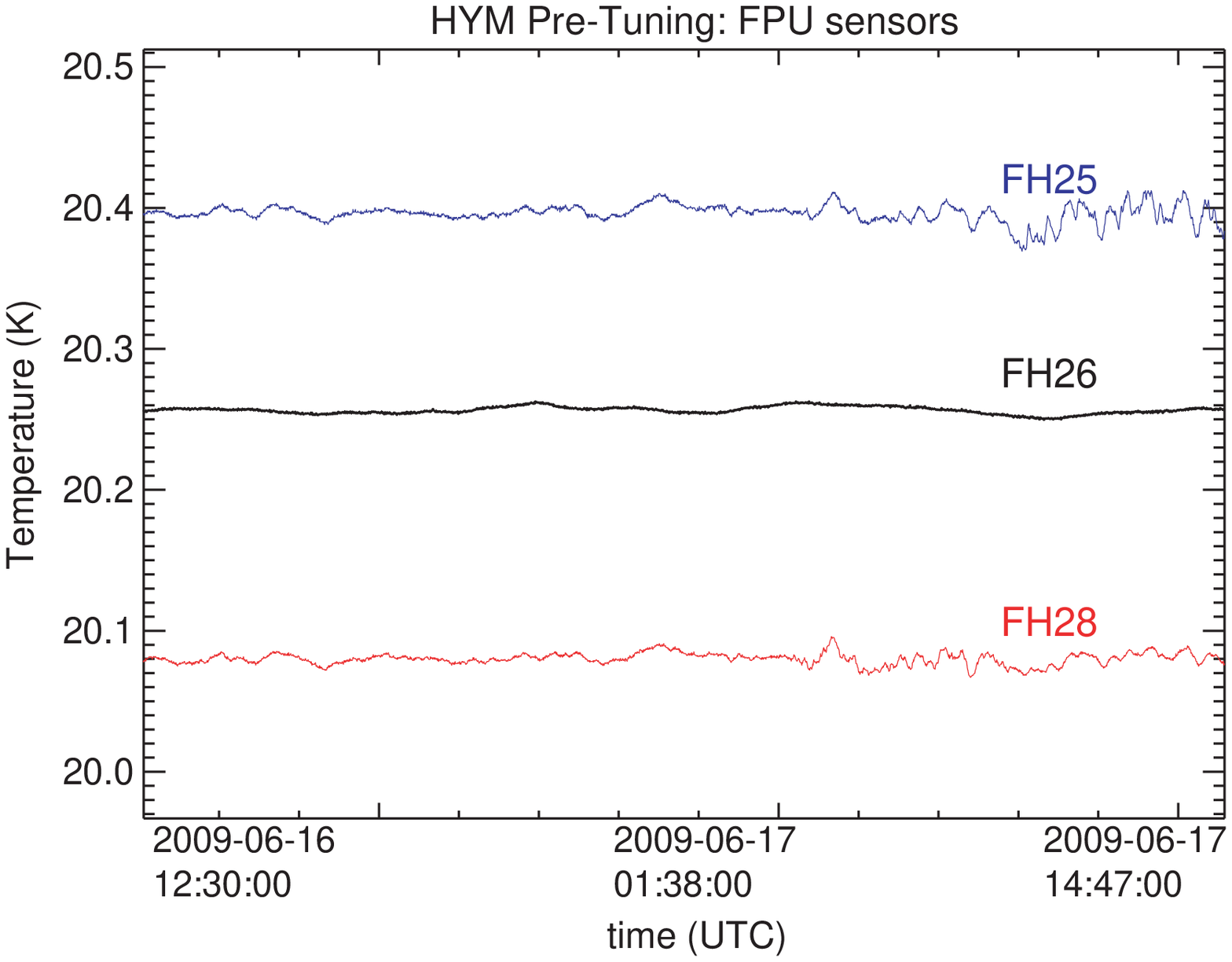} \\
        \includegraphics[width=7.4cm]{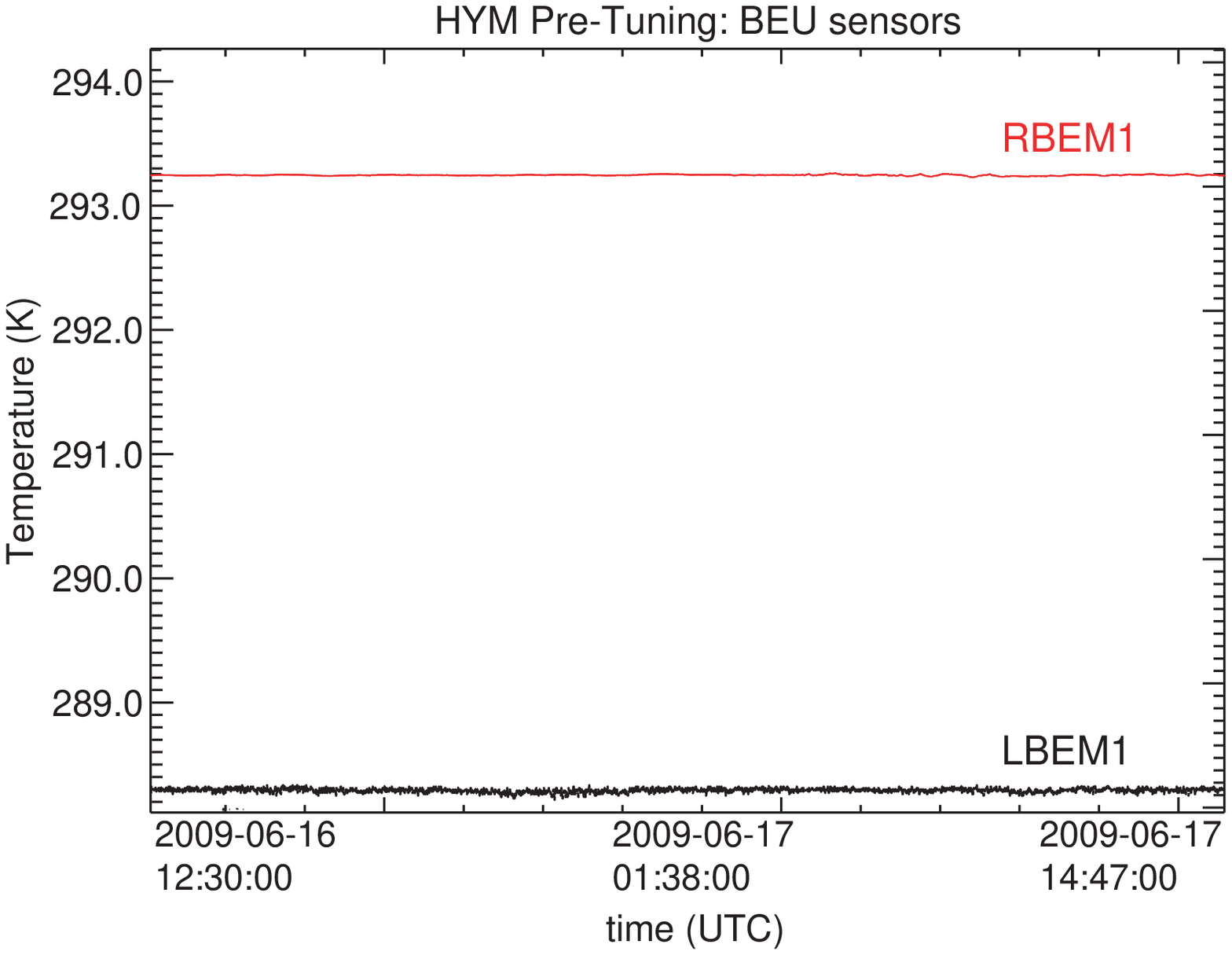}
    \end{center}
    \caption[]{Top panels: thermal behaviour of the 4~K sensor (on the left, \texttt{HD028260}, sampled at 1~Hz, placed in the bottom part of the 4~K thermal shield), Front-End Unit sensors (on the right, \texttt{FH25}, \texttt{FH26}, \texttt{FH28}), bottom panel: Back-End Unit sensors (\texttt{RBEM-1} and \texttt{LBEM-1}), during the hypermatrix pre-tuning; the Front-End and Back-End sensors position is displayed in Figure~\ref{Fit_THF}.}
    \label{plot_LNAs_Pre_Tuning_thermal}     
\end{figure}

Figure~\ref{plot_LNAs_Pre_Tuning_thermal} shows the behaviour of the main thermal stages during the test. The 4\,K cooler temperature slowly drifted at a rate of about 26\,mK/hour with temperature fluctuations on shorter time-scales of few tens of mK, which allowed to determine the noise temperature with an accuracy better than 1\%. 

\paragraph{Proper tuning.}
\label{par:propertuning}

\begin{figure}[htb!]
    \begin{center}
     \includegraphics[width=15 cm]{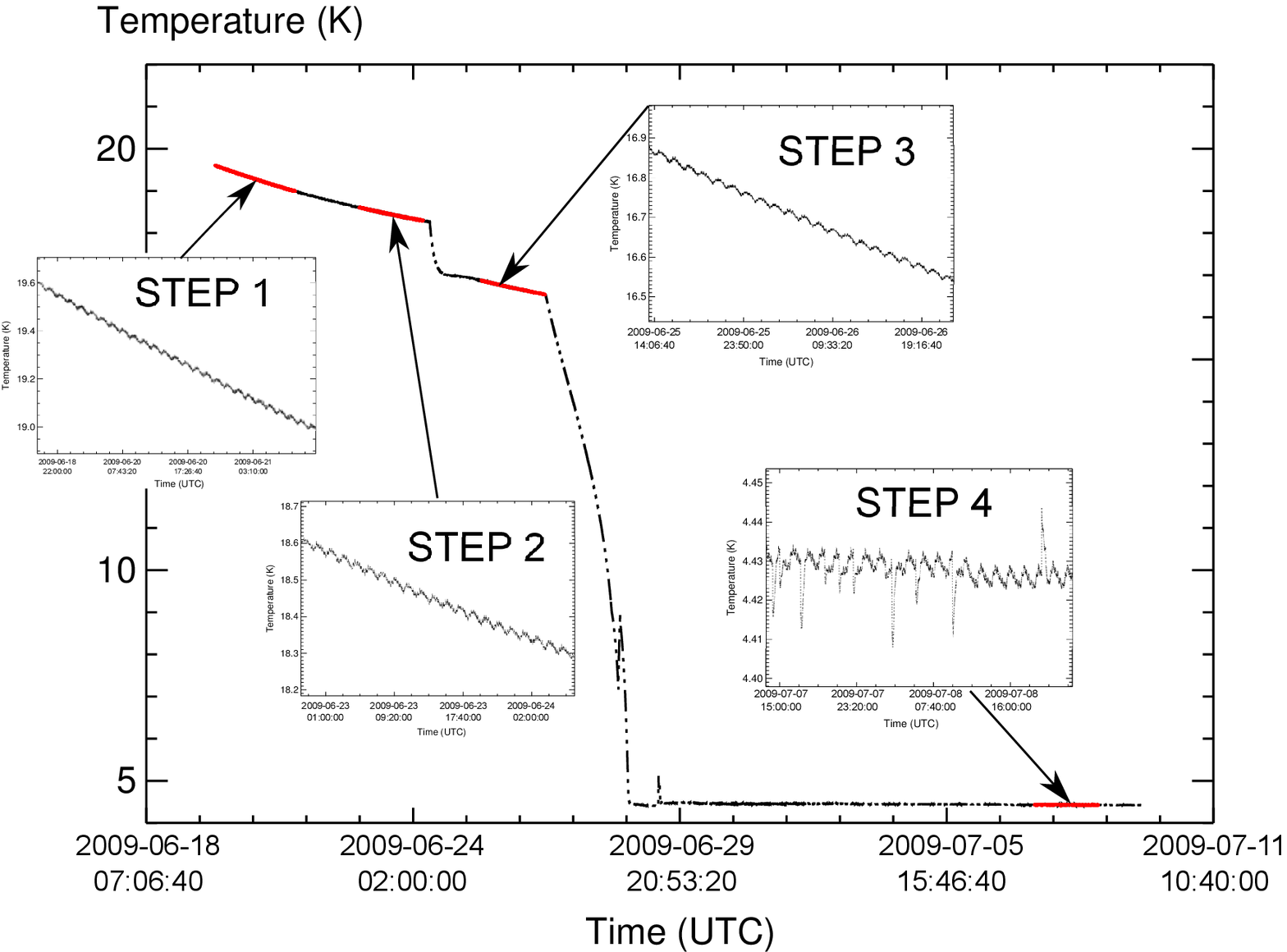}
     \end{center}
    \caption{Temperature profile of the 4\,K stage during the proper tuning phase. The four insets show the details of the temperature variations during each phase.}
    \label{Tuning_4K_cooldown_profile} 
\end{figure}

The test started on June 19$^{\rm th}$ at 19:00:00 UTC and was successfully completed on July 9$^{\rm th}$ at 23:42:32 UTC. It was made of four steps, each corresponding to a different 4~K stage temperature. In Figure~\ref{Tuning_4K_cooldown_profile} we show a plot of the temperature profile of the 4\,K stage temperature versus time. Table~\ref{tab_Tun_4K_rate_summary} reports, for each step, the temperature value and drift compared with the test requirements.
The r.m.s. temperature variations of the front-end and back-end stages were less than 15~mK and 70~mK, respectively, which did not represent a problem for the tuning analysis.

\begin{table}[htb!]
    \begin{center}
        \caption{\label{tab_Tun_4K_rate_summary}Average temperatures of the 4\,K stage and  cooldown speed achieved during the four tuning steps. The average speed is calculated along the whole duration of each step while the maximum speed is calculated during one hour. The last two columns report the required temperature and maximum cooldown rate.}
        \vspace{.1cm}
        \begin{tabular}{c|c c |c c|c c}
	\hline
 	\hline   
		STEP	&	\texttt{\small $T_{\rm start}$}	&	\texttt{\small $T_{\rm stop}$}	&	\texttt{\small Av. Speed}	&	\texttt{\small Max Speed} &	 \texttt{\small Req. $T$} &	\texttt{\small Req. max. speed}	\\
			&	(K)	&	(K)	&	(mK/h) 	&(mK/h) 	  &	(K) 	&	(mK/h) \\
				\hline 
		1	&	19.60	&	19.11	&	22	&	27	&	23.0 $\pm$1.0	&	40	\\
		2	&	18.61	&	18.29	&	20	&	30	&	18.0 $\pm$1.0	&	15	\\
		3	&	16.88	&	16.55	&	15	&	25	&	15.0 $\pm$1.0	&	40	\\
		4	&	\,\,4.75&	\,\,4.70&	<2	&	<5	&	\,\,\,5.0 $\pm$ 0.5&	15	\\
	\hline
	\end{tabular}
    \end{center}
\end{table}

For each radiometer, 748 bias quadruplets were explored through the pre-tuning: the resulting noise temperature maps were analysed to identify the new (smaller) bias regions identifying the bias quadruplets to be tested during the proper tuning phase. A baseline set of 22$\times$22=484 quadruplets was selected for each radiometer according to the following rules: 
\begin{itemize}
  \item select bias regions around noise temperature minima measured during pre-tuning; 
  \item select bias regions outside pre-tuning boundaries if there were indications of possible low noise temperatures;
  \item discard bias quadruplets characterised by drain current imbalance above 25\% as this could indicate gain imbalance and poor isolation.
\end{itemize}

\begin{figure}[htb!]
    \begin{center}
\vspace{0.7cm}
     \includegraphics[width=15.0 cm]{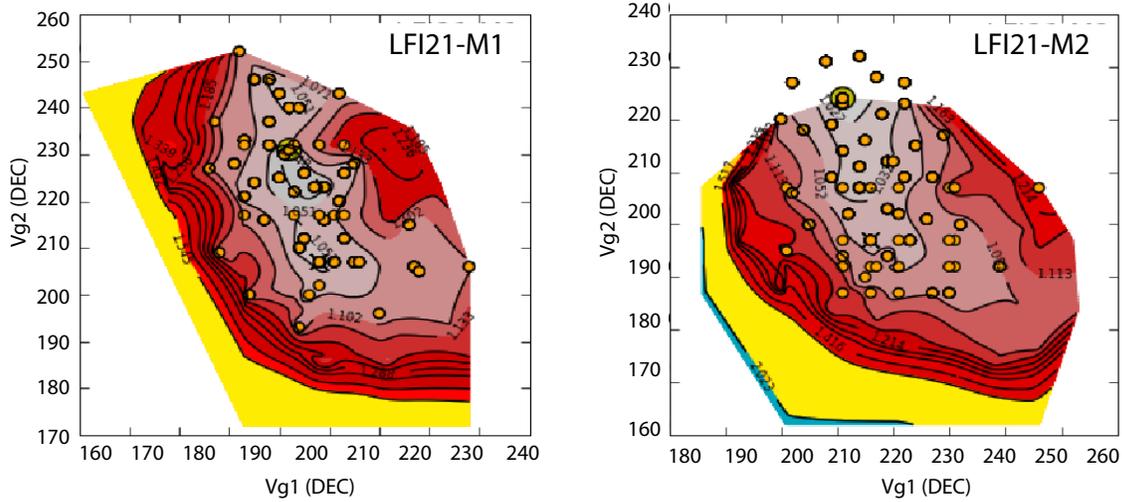}
    \end{center}
    \caption{Resampled maps. Noise temperature is displayed for channels \texttt{LFI21-M1} (left panel) and \texttt{LFI21-M2} (right panel) as a function of $V_{\rm g1}$, on x-axis, and $V_{\rm g2}$, on y-axis. Colour scale ranges from light blue (highest noise temperatures), through yellow and red to light-grey (lowest noise temperatures), normalized to the lowest noise temperature. Contours represents level of constant noise temperature. Orange bullets in the left and right panels of each channel correspond to the 22 + 22 hand-chosen points, originating the 22$\times$22=484 quadruplets limiting the selected bias sub-region to be resampled. In some cases (as for \texttt{LFI21-M1}), the selected region was contained in the original map; in other cases (as for \texttt{LFI21-M2}), where there was an evidence for a possible improvement, it was chosen out of the boundaries. Note that the bias units are DEC units that are used to set the voltage in the DAE: see Table~4 for a rough conversion into physical units.}
    \label{fig_tuning_resampled_maps} 
\end{figure}

In Figure~\ref{fig_tuning_resampled_maps} we show an example of the noise temperature maps with the chosen configurations for the two amplifiers of radiometers \texttt{LFI21M} and \texttt{LFI24S}. In the left panel we see that some of the chosen configurations were selected outside the region tested during the pre-tuning, as the minimum noise temperature was close to the region boundary.

\noindent This baseline set was enlarged adding the following configurations:
\begin{itemize}
  \item the 50 quadruplets characterised by the lowest noise temperatures during ground tests \cite{cuttaia2009};
  \item the 6 optimal bias points found during ground test campaigns and from the drain current verification test described in Section~\ref{sec:lfi_drain}. This allowed a comparison between results taken in the same instrument configuration;
  \item the 75 quadruplets characterised by the lowest noise temperatures from the pre-tuning analysis;
  \item 10 points were left available for the procedure optimization. In order to reduce the drift after each bias change, the bias quadruplets were ordered according to the square root sum of the four biases. In correspondence of the ten largest bias steps, the integration time was doubled by setting twice these 10 points. 
\end{itemize}
 
Noise Temperature, Isolation  and Gain balance were measured for each of the 615 independent gate voltage quadruplets of each radiometer. 

After having tested all these 615 $V_{\rm gate}$ quadruplets, for each channel we also varied the drain voltage, $V_{\rm drain}$, moving along a subset of bias configurations. For each radiometer we varied three values for $V_{\rm drain,1}$ and three for $V_{\rm drain,2}$ for a subset of 15 $V_{\rm g}$ bias quadruplets testing $3\times3\times15$ combinations. The 15 gate voltage quadruplets were chosen among the best results from pre-tuning. This  strategy was aimed at an additional fine bias tuning as previously verified during on ground tests performed on Flight Spare Units \cite{cuttaia2009}. 

To minimize the signal drift due to the bias changes, the duration of each bias step was increased to 20 seconds for the 70~GHz and to 15 seconds for the 30 and 44~GHz channels. The analysis was based only on the last 3 seconds of each step in order to minimize the signal transients.

Each step took about 33 hours; however, the fourth step was completed only after 13 operational days after the end of the 3$\rm ^{rd}$ step (see Figure~\ref{Tuning_4K_cooldown_profile}). This delay was partly foreseen in the schedule to comply with HFI activities carried out in the meanwhile, and partly due to 4~K temperature instabilities that required extra investigations and a better stabilization. 

\paragraph{Thermal Setup.}
\label{par:thermalsetup}

Different sensors were used to monitor the 4~K reference Loads, depending on the step and temperature range considered:\\
- for 1$^{\rm st}$, 2$^{\rm nd}$ and 3$^{\rm rd}$ steps, the sensor \texttt{HD028260}, placed in the bottom part of the 4~K thermal shield, was used for all the loads as it was the only one provided with calibration curves for temperature above 7~K.\\
- for the 4$^{\rm th}$ step, several sensors were available to track the 30/44~GHz and the 70~GHz loads.  The \texttt{Ther-PID4N} sensor (placed on the 4~K focal plate) was used to track the 70~GHz loads; the \texttt{Ther-4KL1} (placed in the middle of the 4~K shield, near \texttt{LFI25}), to track the 30 and 44~GHz loads, assuming a cylindric symmetry of the heat transfer. Data were provided by the HFI Team sampled at 180~Hz and re-sampled at 2~Hz.

\begin{figure}[htb!]
    \begin{center}
       \includegraphics[width=7.5 cm]{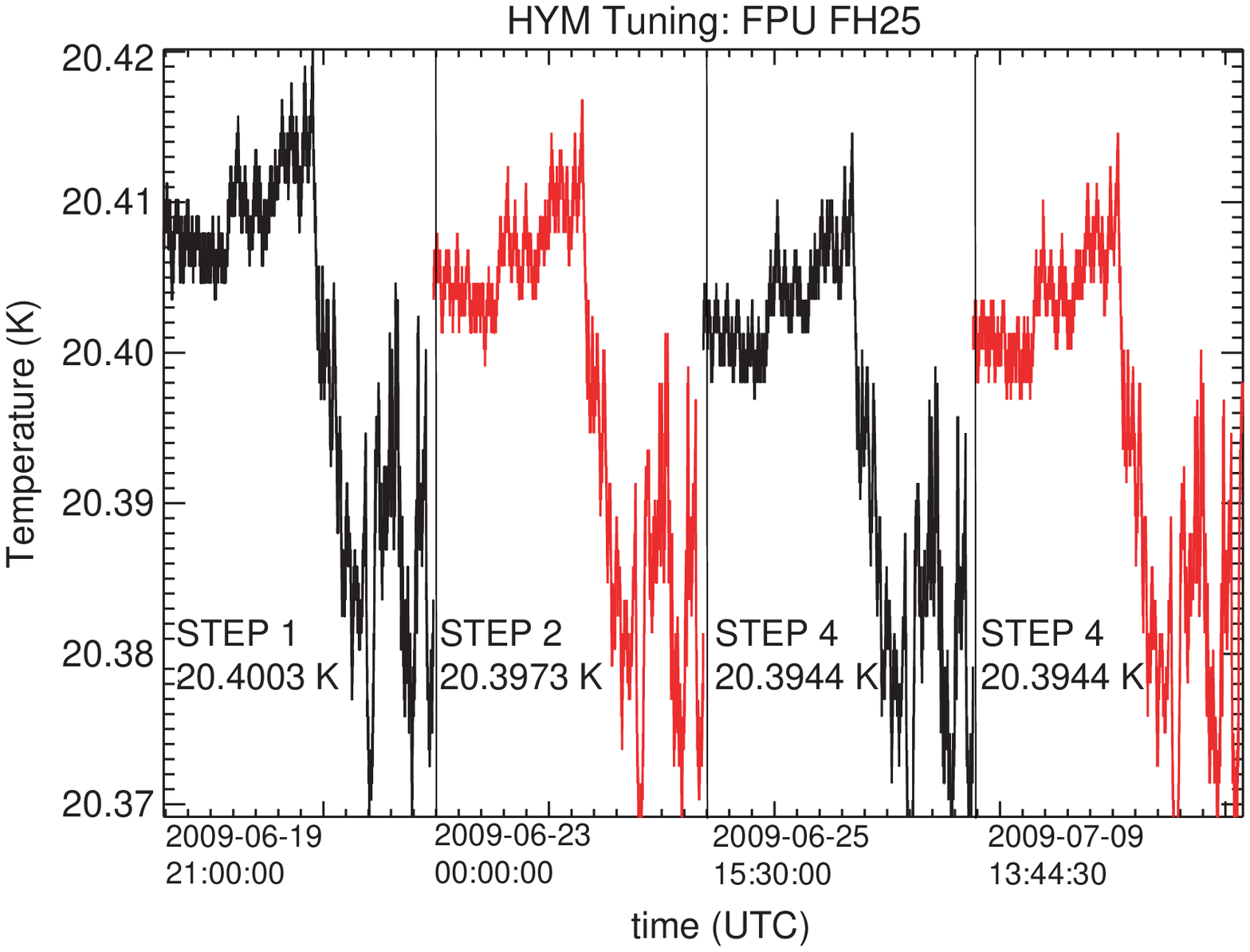}
       \includegraphics[width=7.5 cm]{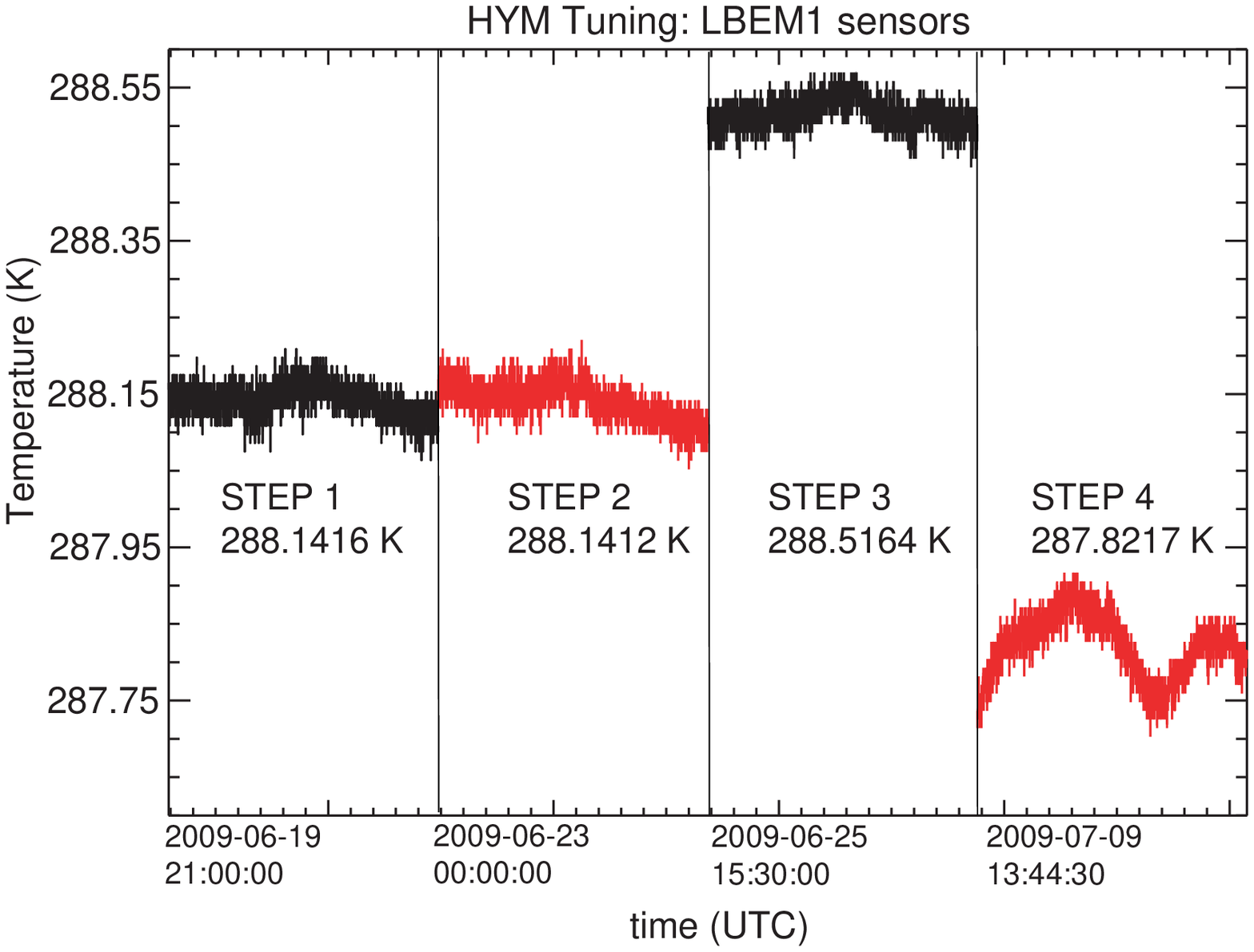}\\
     \end{center}
    \caption[]{Example showing the FPU (left panel \texttt{FH25} sensor) and BEU (right panel \texttt{LBEM1} sensor) thermal stability during the four nominal steps of the hypermatrix tuning. The sensors position is shown in Figure~\ref{Fit_THF}. The shape of each curve in the FPU sensors was determined by the changes in power dissipation when changing the bias quadruplets: the four steps exhibited similar thermal conditions. Regarding the BEM curves, the absolute temperature was strongly dependent on the step considered due to the HFI activities performed at 4~K cooler level to keep the required stability during the cooldown and to the transponder \texttt{ON/OFF} effect.}
    \label{HYM_Tuning_4steps_FPU_sensors} 
\end{figure}

As it can be seen in Figure~\ref{HYM_Tuning_4steps_FPU_sensors}, the FPU temperature remained very stable for the whole duration of the tuning tests, preventing changes in Noise and in the Gain at level of the cold LNAs. The Back-End Unit instead suffered some instabilities. They had mostly three causes: (i) the daily perturbations due to the Transponder switch \texttt{ON/OFF} \footnote{The spacecraft telecommunication subsystem consists of a redundant set of transponders using X-Band for the uplink, and X-Band for the downlink. Depending on the mission phase, the transponder was routed via RF switches to different antennas: the transponder switch \texttt{ON/OFF}, during DTCP phases, caused a warm-up/down modulation in the Back-End unit.}, (ii) the long term drift of the Back End, (iii) perturbation induced in the warm power box of radiometers caused by bias changes. Temperature changes at BEU level induced gain changes in the warm LNAs and offset changes in the warm diodes mimicking true signal changes. These effects (see {\em non linear solution} 
section) were taken into account in the data analysis.

An overview of the thermal conditions (FEU and BEU sensors) during the four hypermatrix tuning steps is given in Appendix~\ref{app_detail_tests}, Table~\ref{tab_Tun_thermal_sensors}. 

\paragraph{Selection of optimal bias values.} 
\label{par:optimalbias}

\begin{figure}[htb!]
    \begin{center}
       \includegraphics[width=7.9 cm]{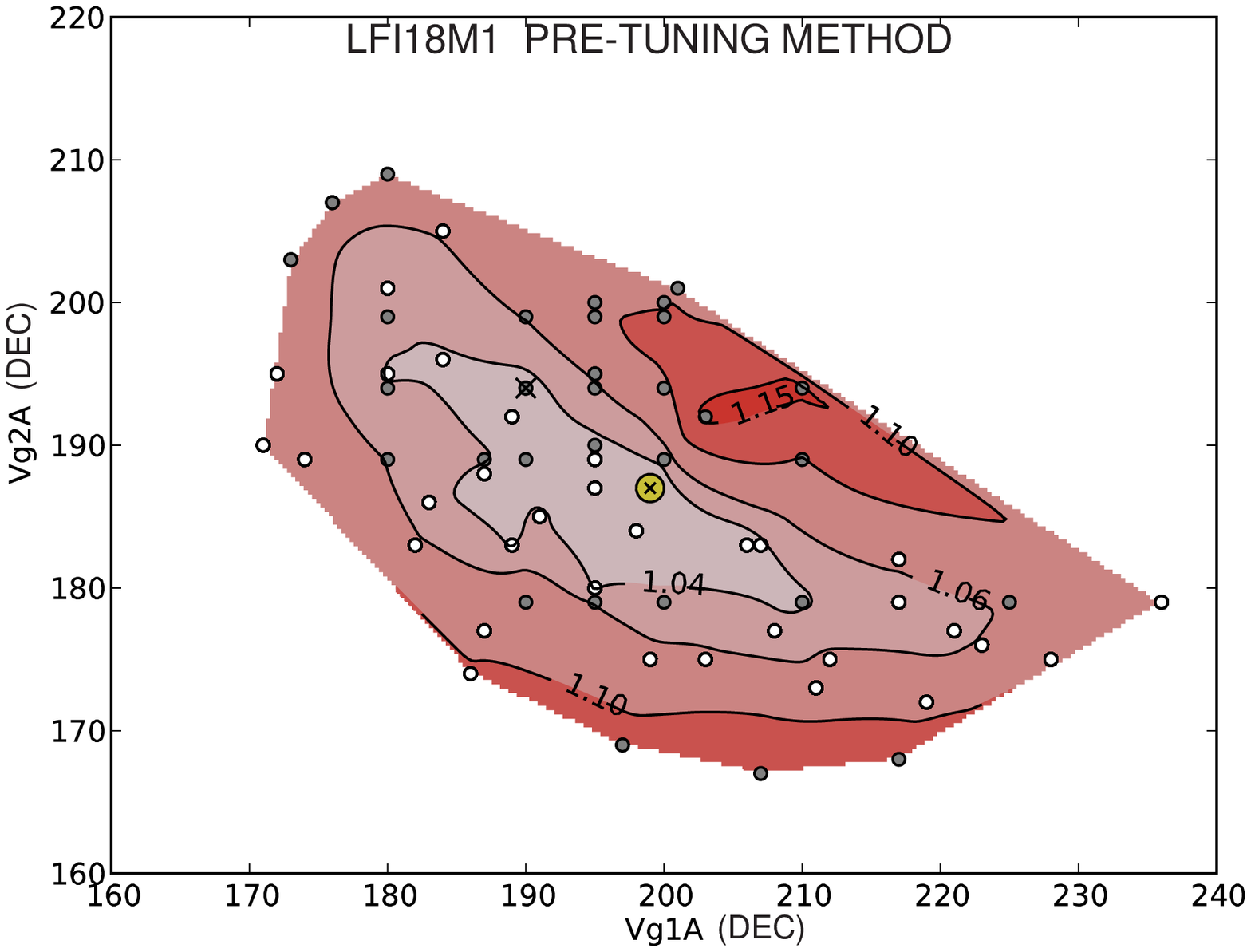}
       \hspace{-0.9cm}           
       \includegraphics[width=7.9 cm]{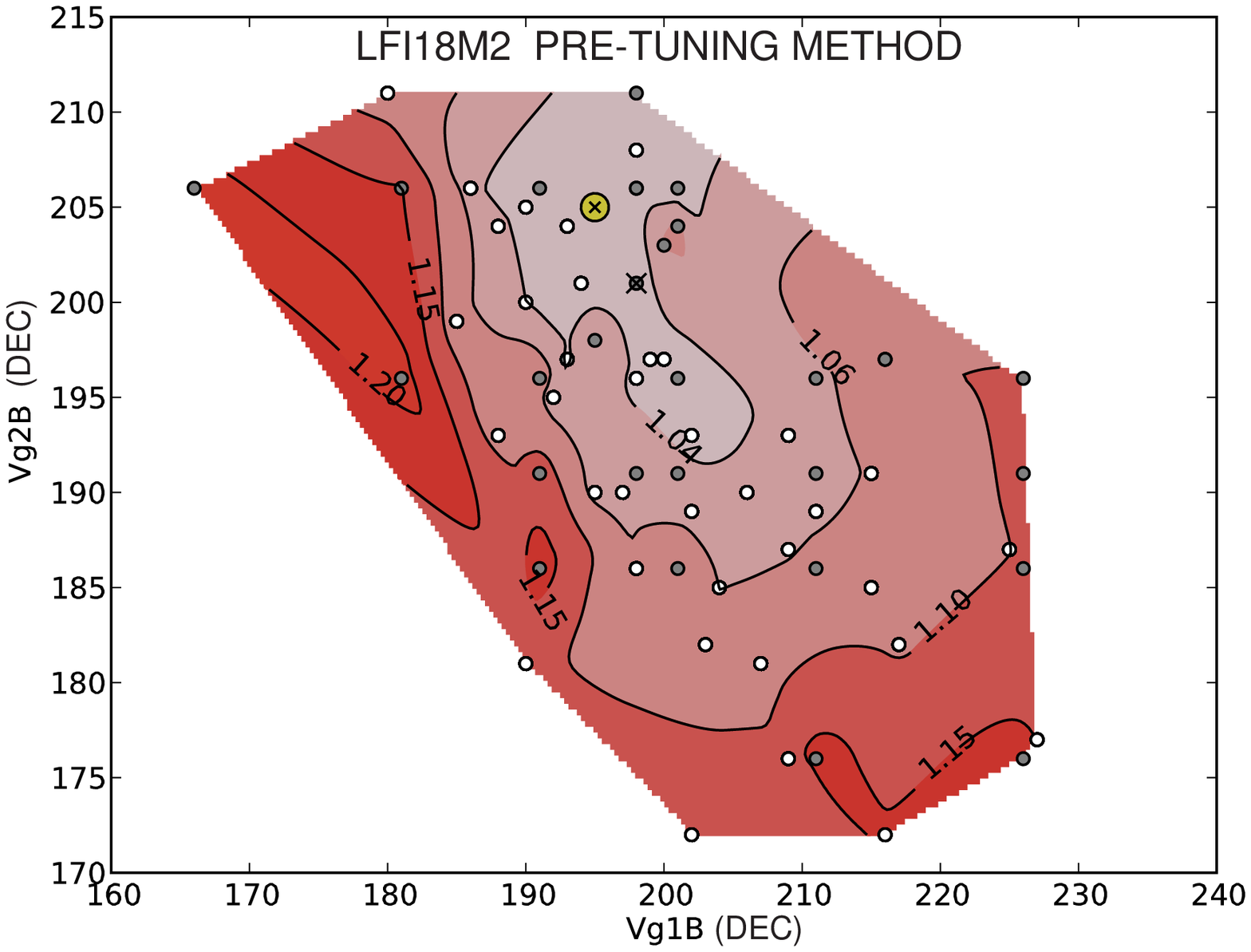} \\  
       \includegraphics[width=7.9 cm]{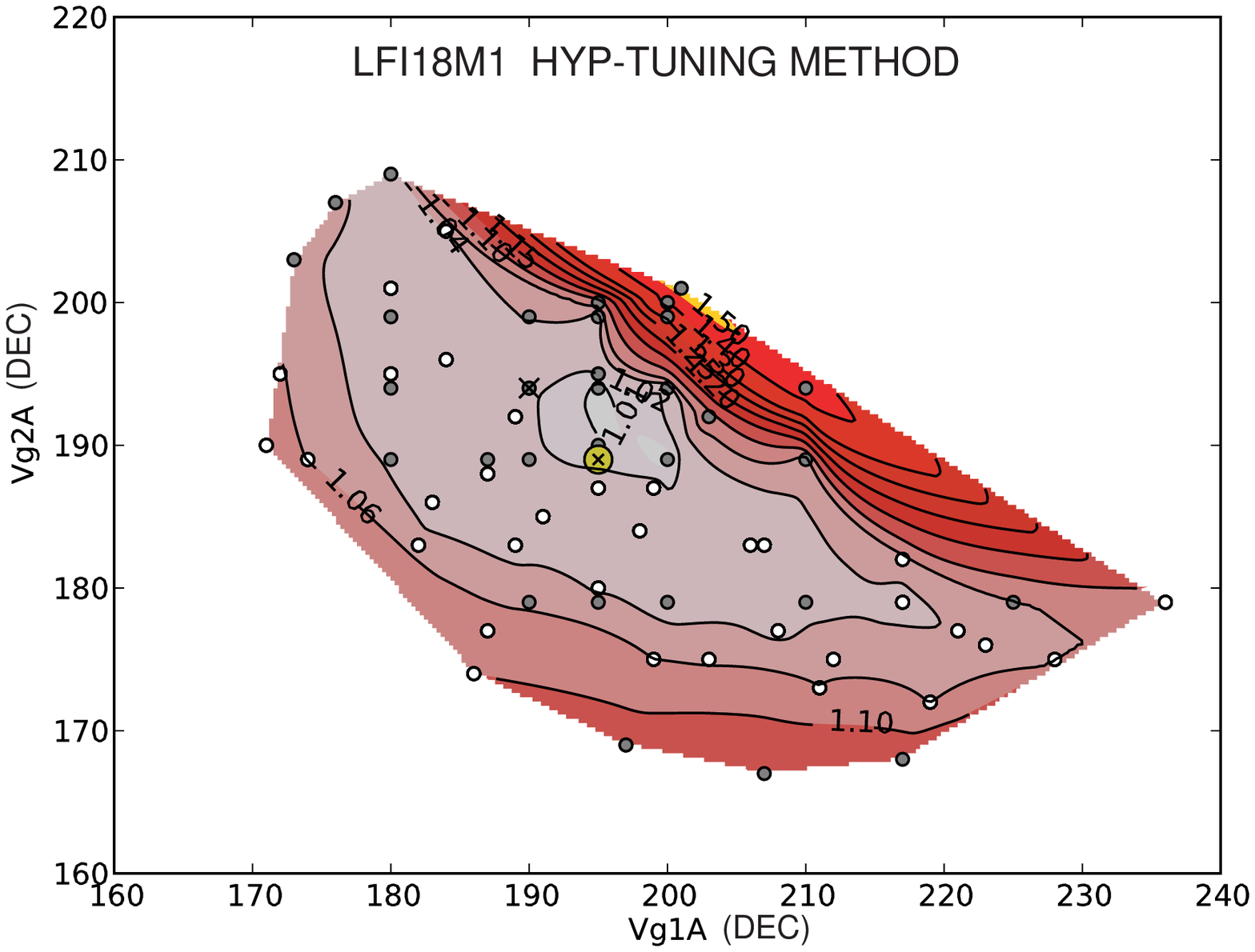}
       \hspace{-0.9cm}           
       \includegraphics[width=7.9 cm]{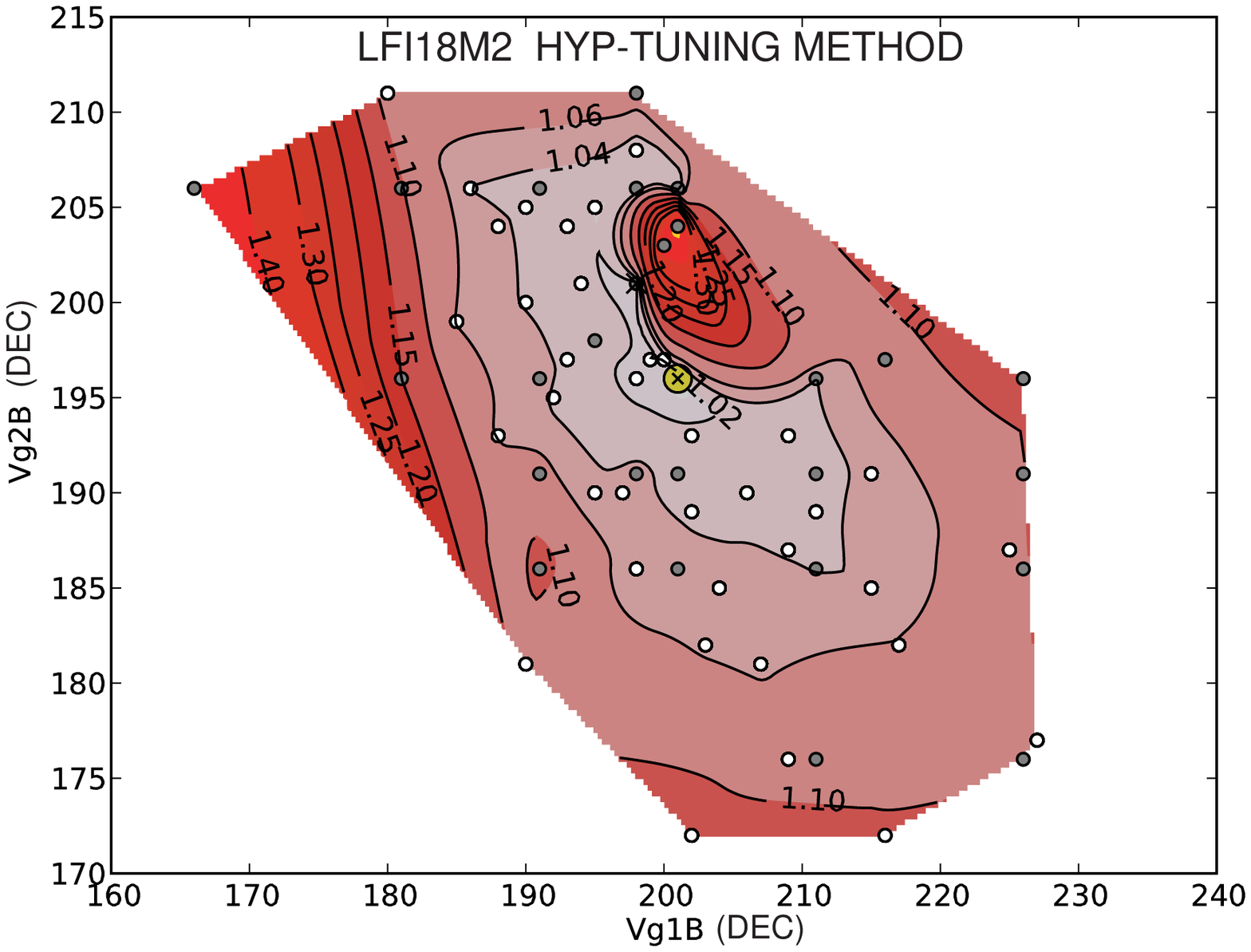} \\   
      \vspace{-0.3cm}          
    \end{center}
    \caption{Noise temperature maps for \texttt{LFI18M} as a function of $V_{\rm g1}$, on x-axis, and $V_{\rm g2}$, on y-axis.  Comparison between results obtained from pre-tuning approach (based on $Y^{\rm *}$ applied to the 4$^{\rm th}$ step of \texttt{HYM)}, top panels, and from the proper \texttt{HYM} approach (based on \texttt{Y-factor}  applied to 1$^{\rm st}$ and 4$^{\rm th}$ steps), bottom panels. The condensed noise maps refer to \texttt{M1} LNA on the left, and to \texttt{M2} LNA on the right. Colour scale ranges from red (highest noise temperatures) to light-grey (lowest noise temperatures), normalized to the lowest noise temperature. Contours represents level of constant noise temperature. Even if $Y^{\rm *}$ is here calculated from conditions of small thermal unbalance (the input difference between sky and reference was less than 1.7~K in the fourth step), results are comparable. Grey dots correspond to bias pairs common to CSL matrix tuning and to CPV hypermatrix tuning; black crosses correspond to the best performance pairs measured at  system level (CSL); yellow circles with black crosses correspond to the best performance pairs measured at CPV level. Note that the bias units are DEC units that are used to set the voltage in the DAE: see Table~4 for a rough conversion into physical units.} 
    \label{plots_LFI18-M_condensed_Tn}
\end{figure}

Noise temperature and isolation maps were analysed to find, for each radiometer, the bias configuration yielding optimal performance. Special care was taken to avoid sharp noise temperature minima and bias values with poor isolation and/or unbalanced drain currents. 
The  \texttt{HYM} proper tuning provided in general improved performance with respect to the optimal bias quadruplets resulting from CSL tuning and in most cases the best performing quadruplets were found well inside the four-dimensional sub-regions indicated by the \texttt{HYM} Pre-Tuning. This confirmed the good capability of this technique to provide a resonable guess of the overall performance. Examples of noise temperature maps are provided in Figure~\ref{plots_LFI18-M_condensed_Tn}. 

\begin{figure}[htb!]
    \begin{center} 
\includegraphics[width=7.45 cm]{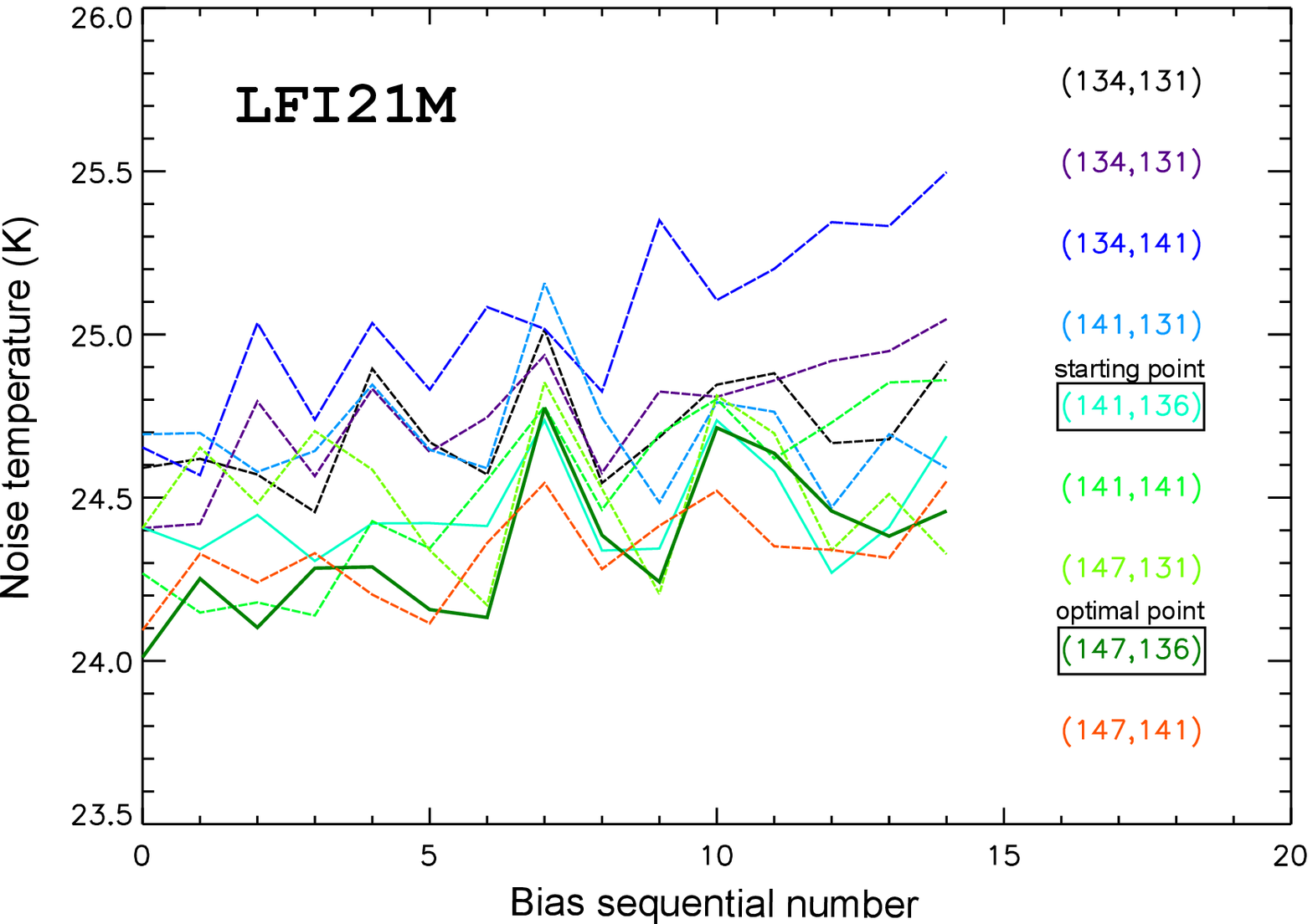}  \hspace{0.05cm} 
       \includegraphics[width=7.33 cm]{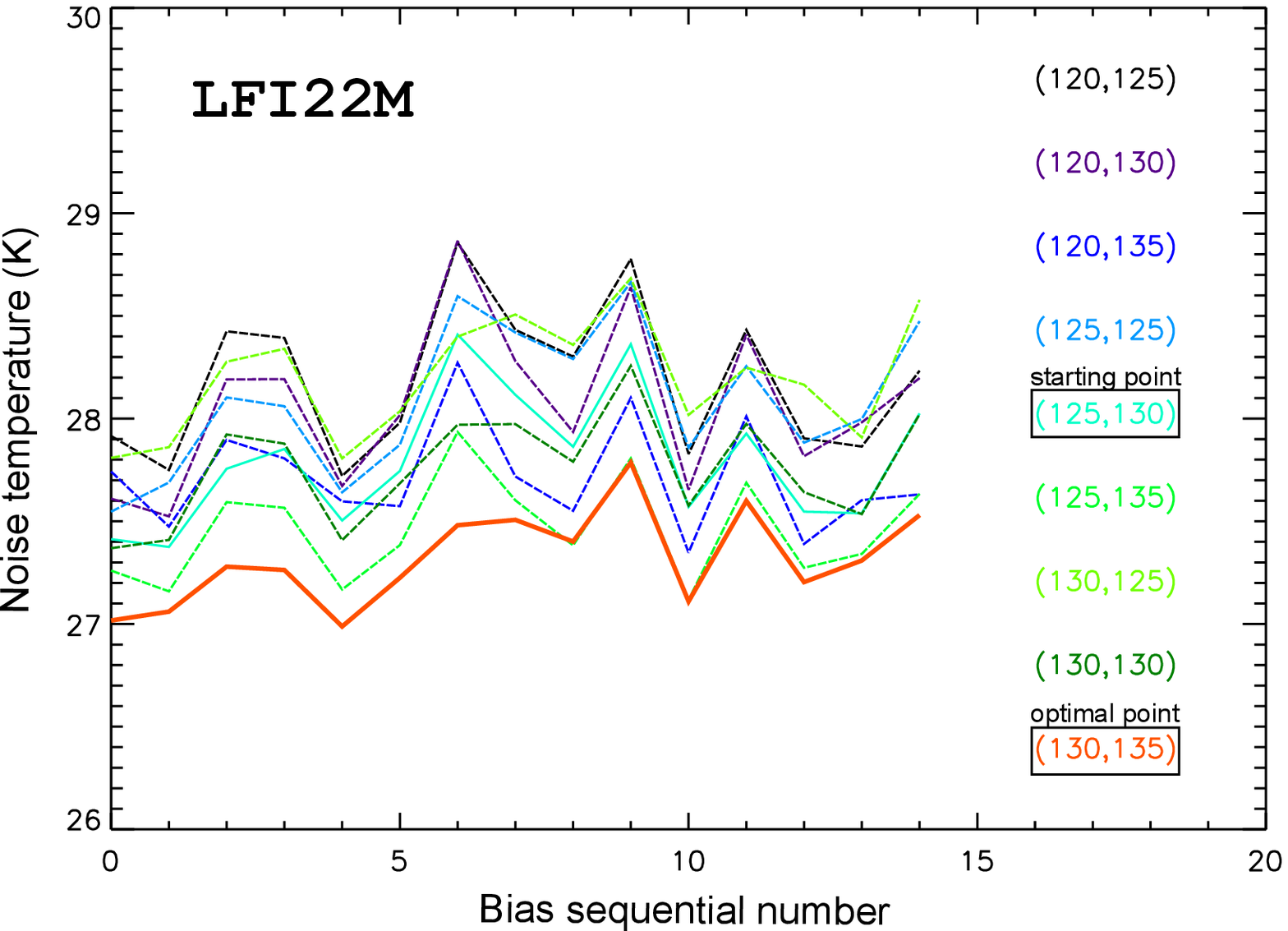}\\
       \vspace{0.5cm} \hspace{0.08cm}
       \includegraphics[width=7.28 cm]{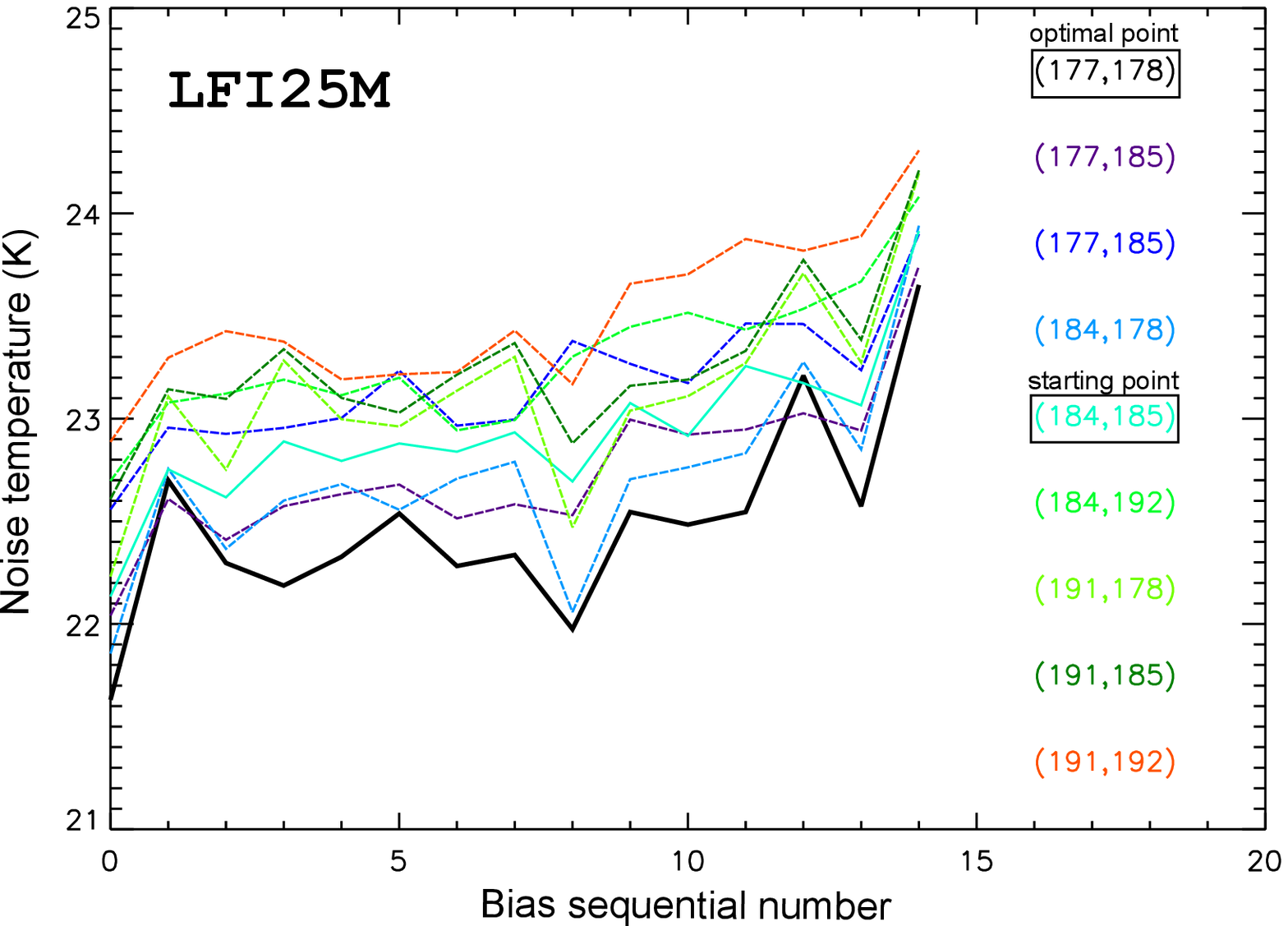}
       \includegraphics[width=7.42 cm]{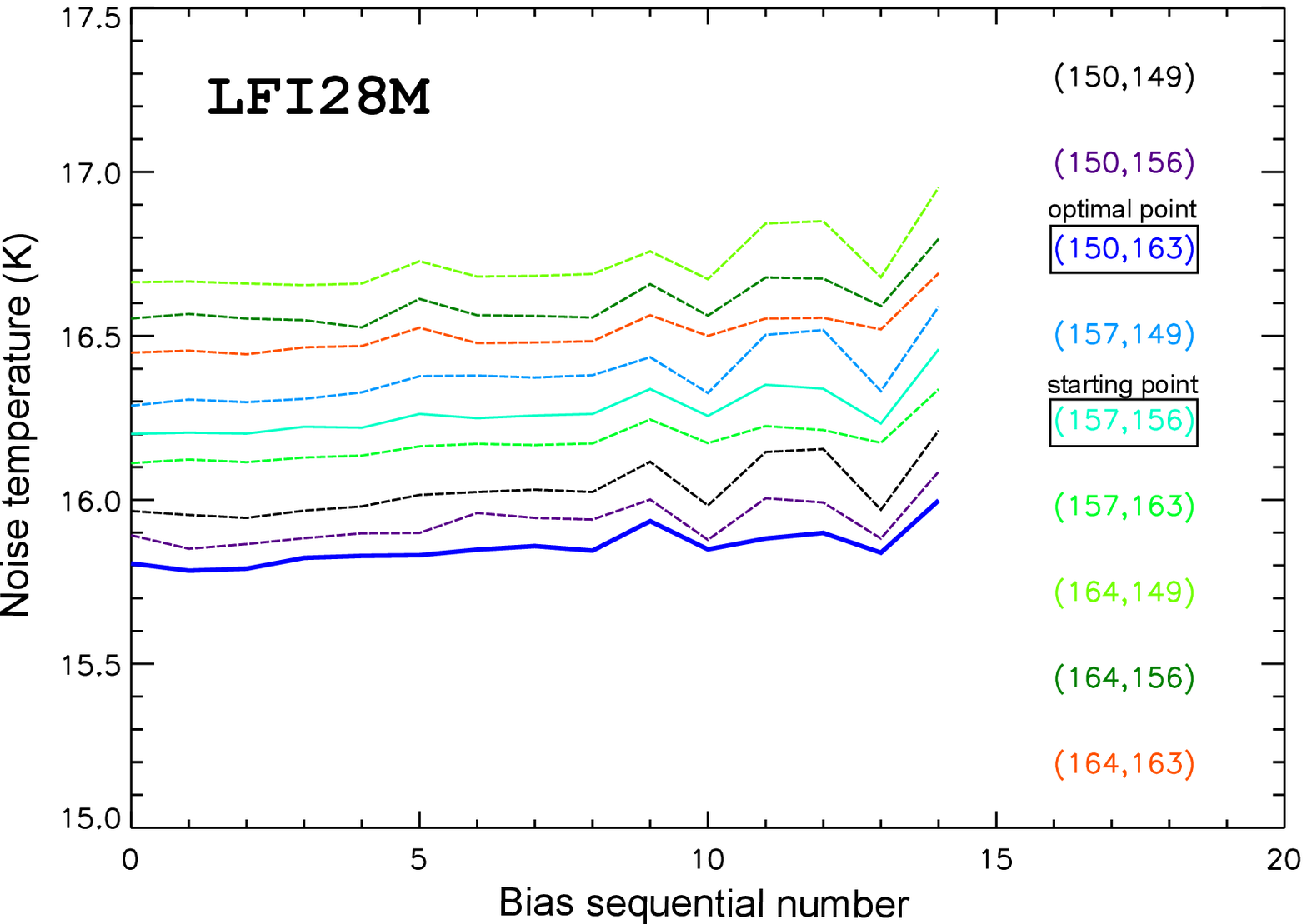}\\
     \end{center}
    \caption{Drain voltage tuning plots: the four plots correspond to the four cases where drain current tuning yielded improved performance. On the abscissa axis we report a sequential number corresponding to each of the 15 $V_{\rm g}$ bias quadruplets under test, on the vertical axis the noise temperature (K). Each curve corresponds to a $V_{\rm drain}$  bias pair, as reported in the legend. Solid thin curves show the noise temperature for the starting point, solid thick curves show the noise temperature for the optimal point.}
    \label{plots_HYMT_VD_CURVES} 
\end{figure}

\indent In most cases tuning the drain biases did not produce % results compliant wih the stringent criteria defined. Improvements were sometimes not perfectly 
clear improvements, as the shape of the curves was not simple as desired. % or the bias quadruplet associated with the improved performance did not have a counterpart along the four $V_{\rm g}$ steps. 
Drain biases were changed with respect to the ground test default values only when noise temperature improvement was larger than 0.5\,K, without degrading Isolation. This happened only in four cases: \texttt{LFI21M}, \texttt{LFI22M}, \texttt{LFI25M}, and \texttt{LFI28M}). The $V_{\rm drain}$ curves for these four radiometers are shown in Figure~\ref{plots_HYMT_VD_CURVES}.

\paragraph{Comparison to system level tests results.}
\label{par:comparisonSLT}

\begin{figure}[thb!]
    \begin{center}
       \includegraphics[width=8.9 cm]{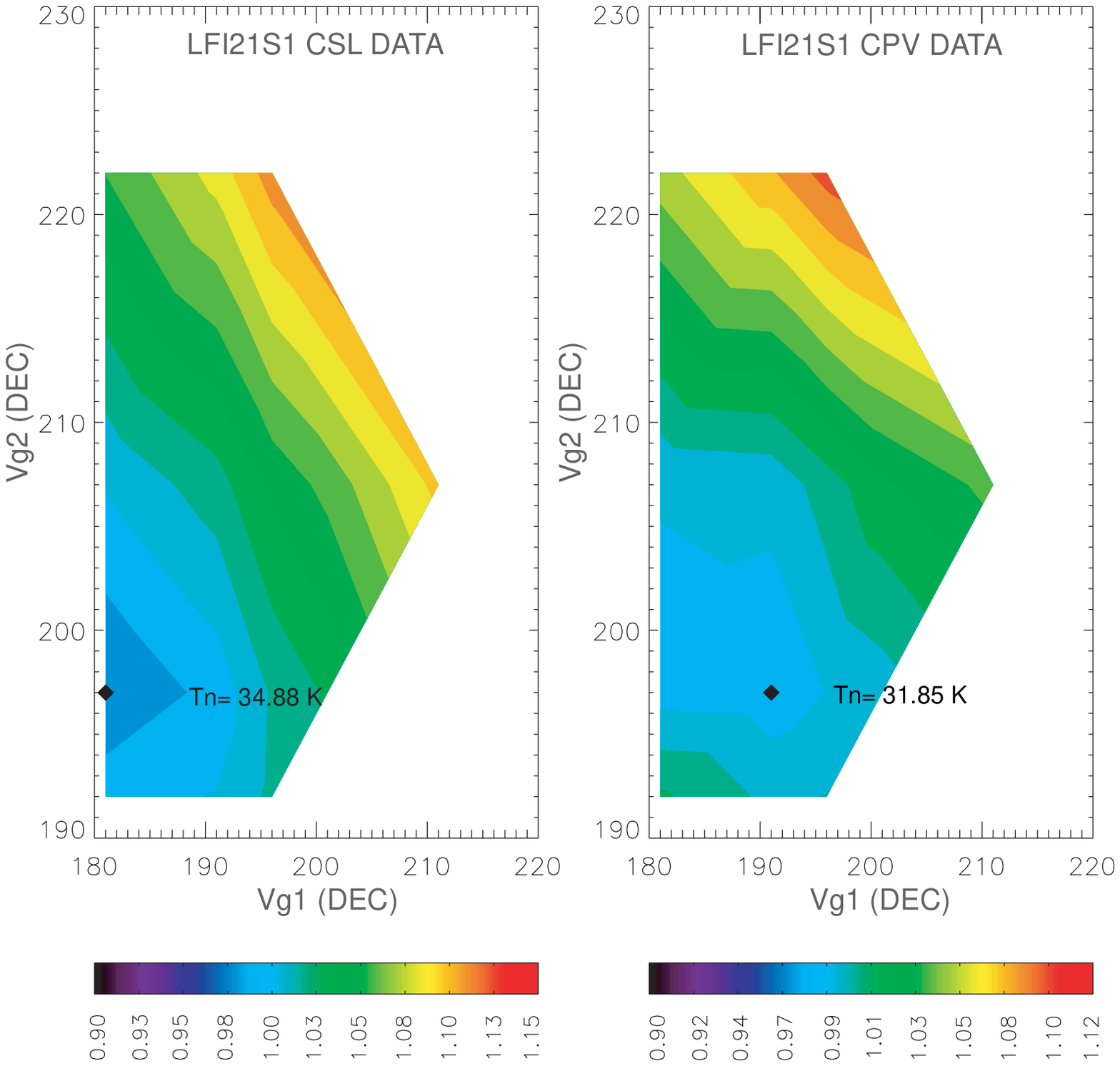} \\
\vspace{0.6cm}
       \includegraphics[width=8.9 cm]{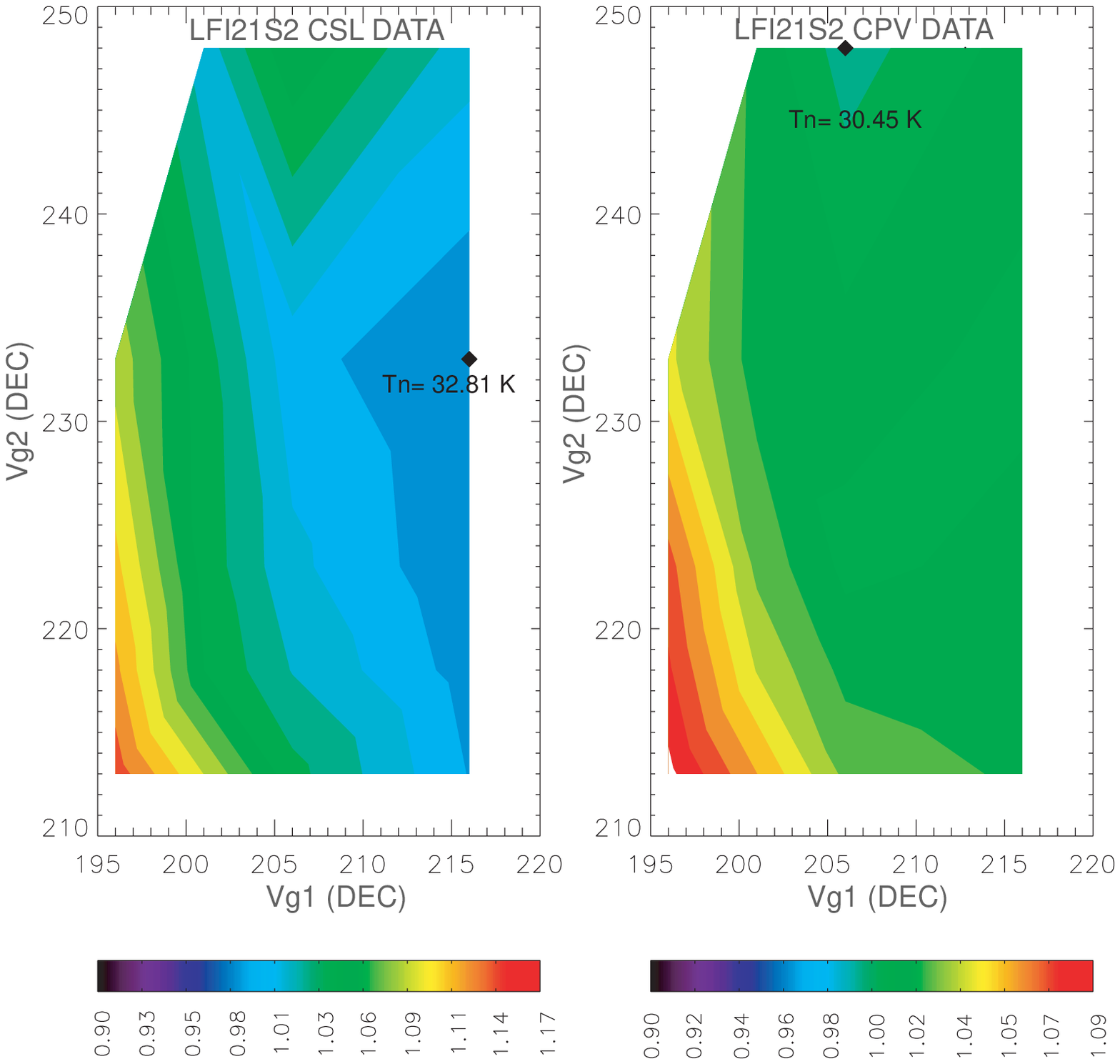}
     \end{center}
    \caption{\texttt{LFI21S1} (top panels) and \texttt{LFI21S2} (bottom panels) maps are compared in the same bias regions covered by the common quadruplets explored both in CSL system level tests (left panel) and in CPV (right panel): the qualitative agreement is very good, showing the consistency of \texttt{HYM} and the simpler Matrix philosophy used during ground tests. The two dimensional bias spaces shown represent only two slices of the whole four dimensional bias space mapped by hypermatrix tuning. The number of samples depends on the channel and determined the resolution of each map. Note that the bias units are DEC units that are used to set the voltage in the DAE: see Table~4 for a rough conversion into physical units.}
    \label{plot_tuning_CSL_CPV_compar_21S} 
\end{figure}

Direct comparison to results from CSL system level tests (see Figure~\ref{plot_tuning_CSL_CPV_compar_21S}) was possible only over a reduced bias range, covering all the quadruplets common both to CPV and CPV matrices. %Because of the  sequencial choices of \texttt{HYM} to the \texttt{Pretuning}, in some cases only a few quadruplets were available. 
In many cases maps were well defined and showed a good agreement between the two cases, confirming the validity of the two methods and showing that LFI radiometers did not suffer major changes (complete results are given in Appendix~\ref{app_HYM_plots}).

In general the resulting biases were very close to ground measured ones, with average variation in $V_{\rm g1}$ and $V_{\rm g2}$ of the order of 1\% and few cases where in-flight optimal biases differed from the ground ones by 10-20\%. We therefore generally maintained the ground configuration apart from a few cases in which the new configuration provided a noise temperature improvement better than 5\% and/or an isolation improvement better than 50\% without degrading the 1/$f$ noise performance. Indeed the short integration times for each bias did not allow to evaluate the 1/$f$ noise knee frequency which required at least 1 hour data and was evaluated when tuning was completed.

\begin{figure}[htb!]
  \begin{center}
      \includegraphics[width=10cm]{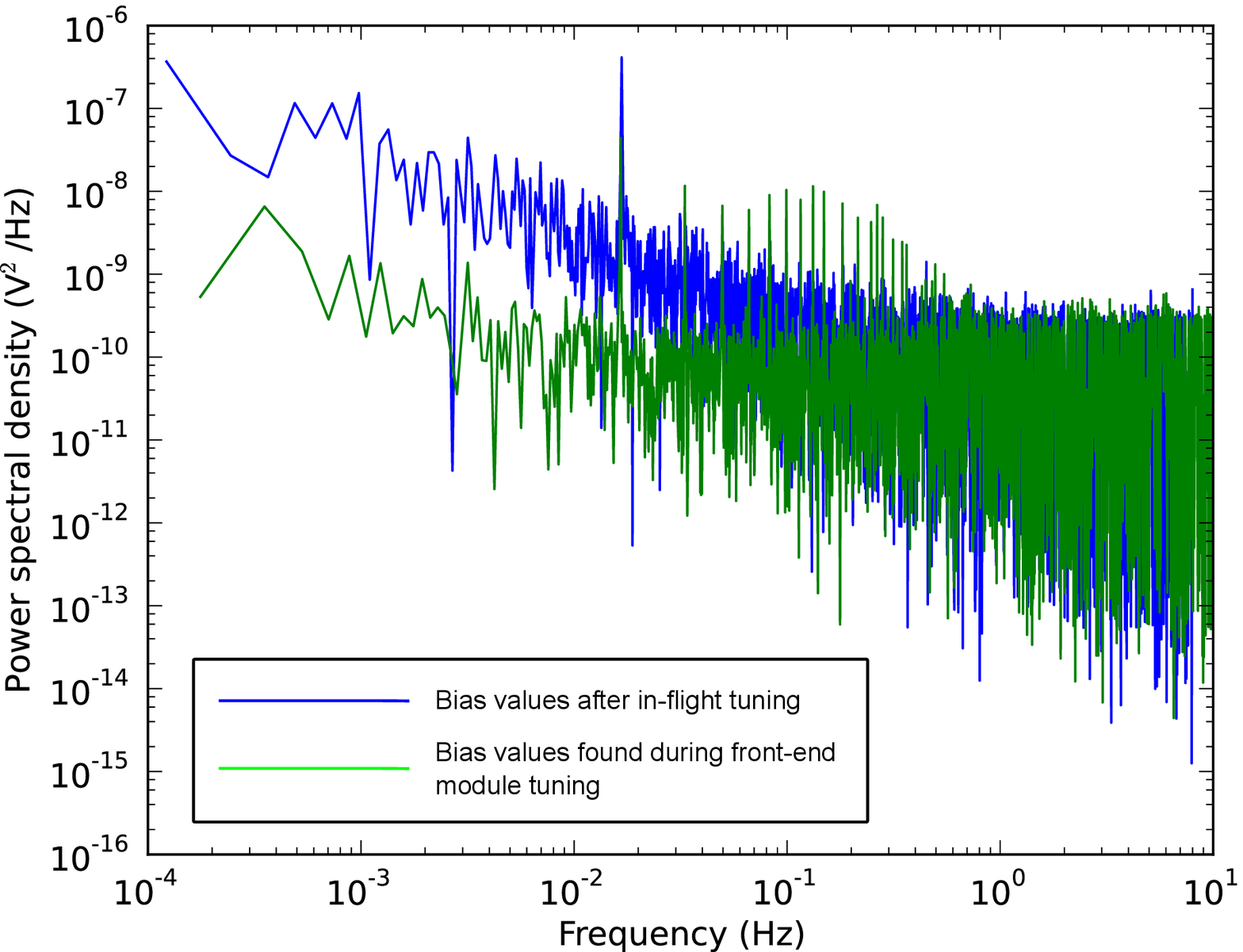}
      \caption{\label{fig_LFI25_1_f_drift} Power spectral density of data from \texttt{LFI25M-00} detector with two different bias sets. Blue: optimal biases determined after flight tuning tests. Green: optimal biases determined during front-end module tuning before integration.}                            
  \end{center}
\end{figure}

An unexpectedly high level of $1/f$ noise fluctuations was observed for the 44\,GHz RCAs, using either the new flight bias settings or the old ground ones. To test this behaviour we performed several hours of data acquisition with different bias configurations. We observed this instability, both with the in-flight and on-ground optimal biases, only with a particular phase switch configuration of 70\,GHz \texttt{LFI23}. We also observed that this instability disappeared, independently from the \texttt{LFI23} phase switch configuration, when the radiometers were biased with the values resulting from the optimisation of the individual front-end modules before instrument integration (see Figure~\ref{fig_LFI25_1_f_drift}). This interaction between RCAs belonging to different frequency channels was unexpected and the root cause was never fully understood. The most likely explanation was a parasitic oscillation triggered by unexpected cross-talk in the warm electronics. 

\begin{figure}[htb!]
  \begin{center}
  \includegraphics[height=4.8cm]{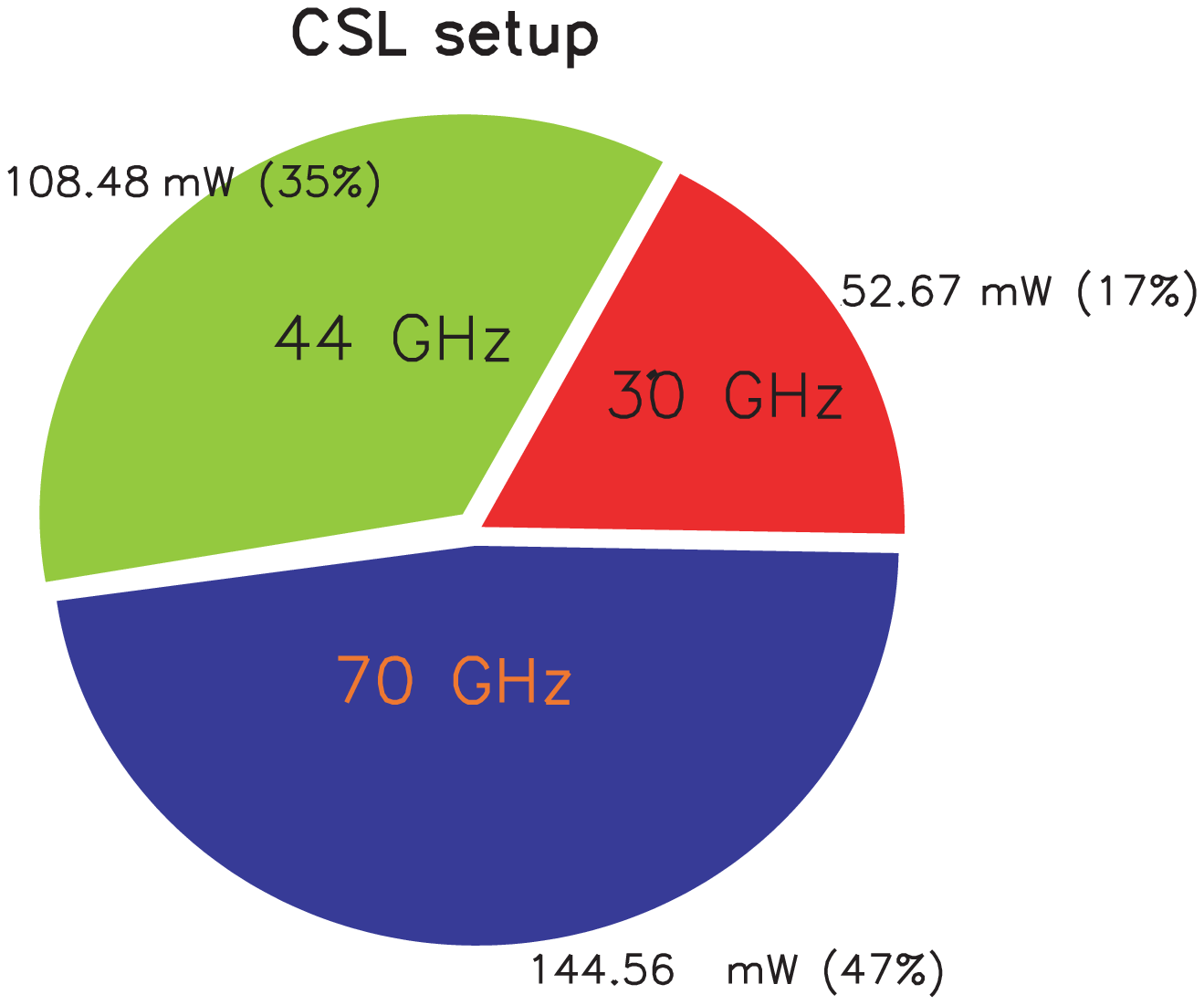}
  \includegraphics[height=4.8cm]{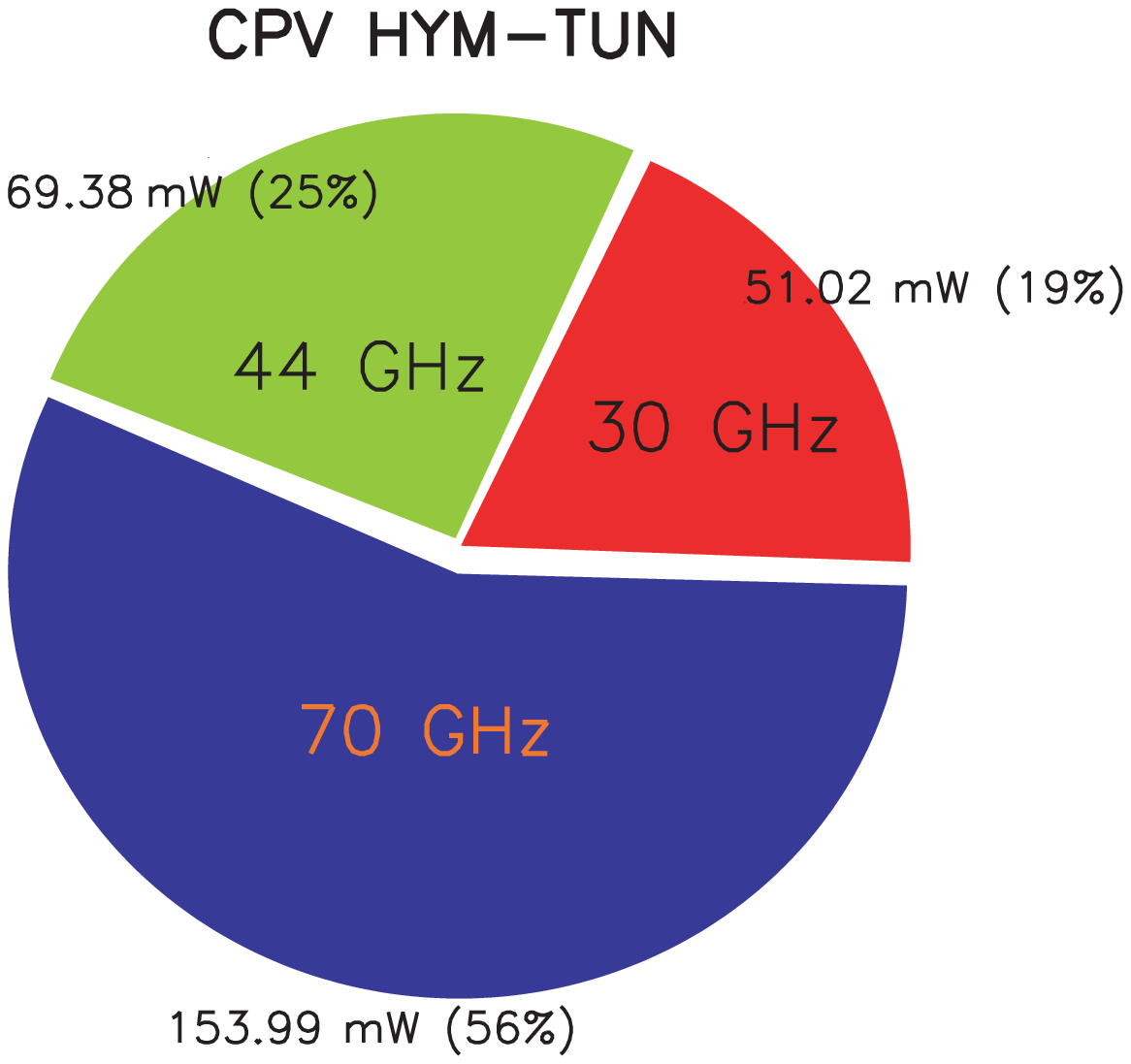}  \\
\vspace{0.5cm}
  \includegraphics[height=4.8cm]{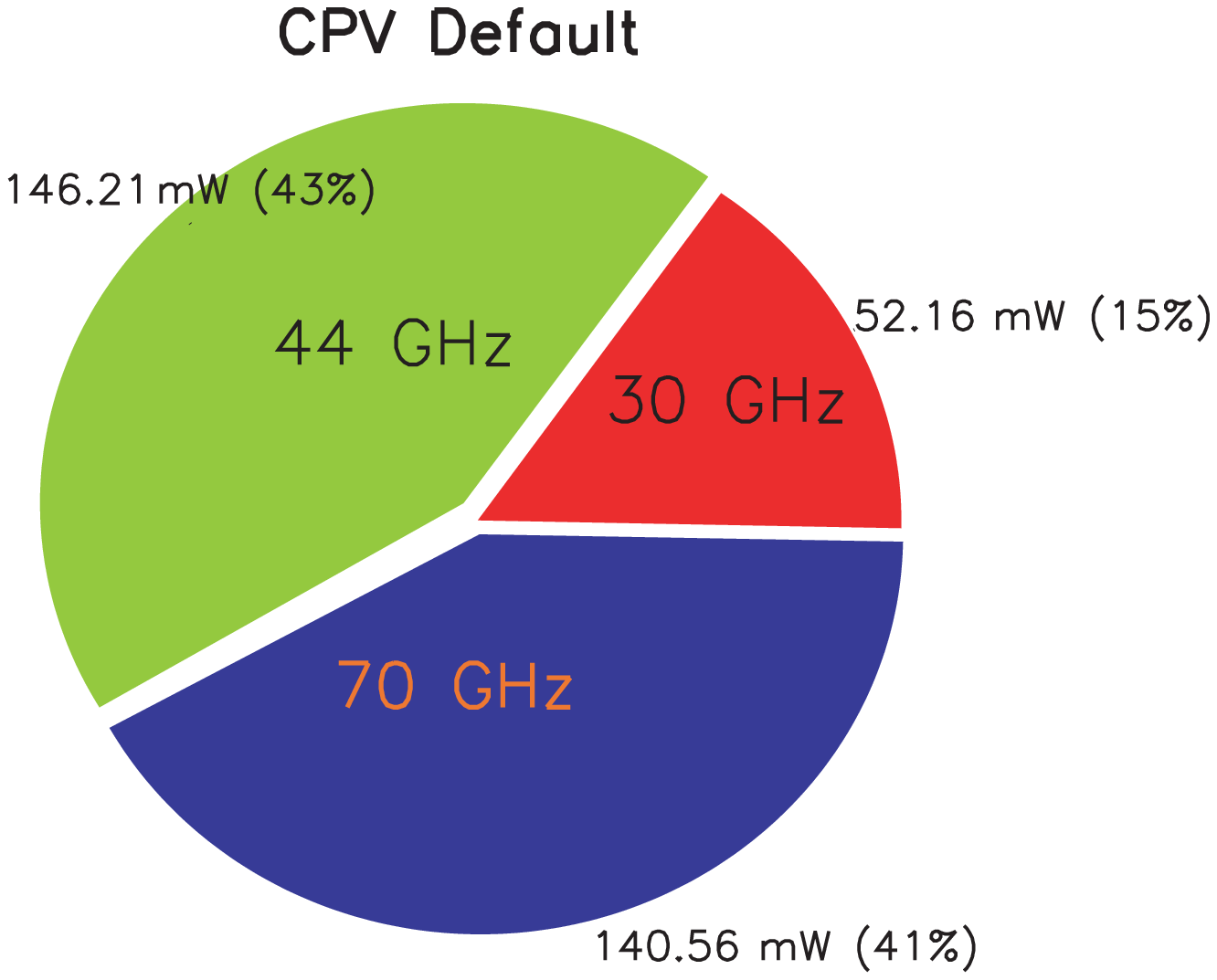}
  \caption{Power consumption distribution among the channels grouped by frequency. The pie charts refer to: the system level tests bias (CSL, top left panel); optimal bias resulting from hypermatrix tuning (top right panel); bias default set at the end of the CPV phase (bottom panel). The corresponding total LFI FPU power consumption was 305.71~mW (CSL), 274.39~mW (hypermatrix) to be compared with the flight conditions of 338.93~mW. }
  \label{fig_power_consumption_cakes}
    \end{center}
\end{figure}

For this reason this latter final bias setting was chosen for all the 44\,GHz radiometers \cite{davis2009}. This setting was characterised by a slightly higher power consumption ($\sim 40$\,mW) but similar noise and isolation performance. The power consumption distribution in the two bias frames (optimal bias resulting from hypermatrix tuning and bias default set at the end of the CPV phase) is shown in Figure~\ref{fig_power_consumption_cakes}.

\begin{figure}[htb!]
  \begin{center}
  \includegraphics[width=10cm]{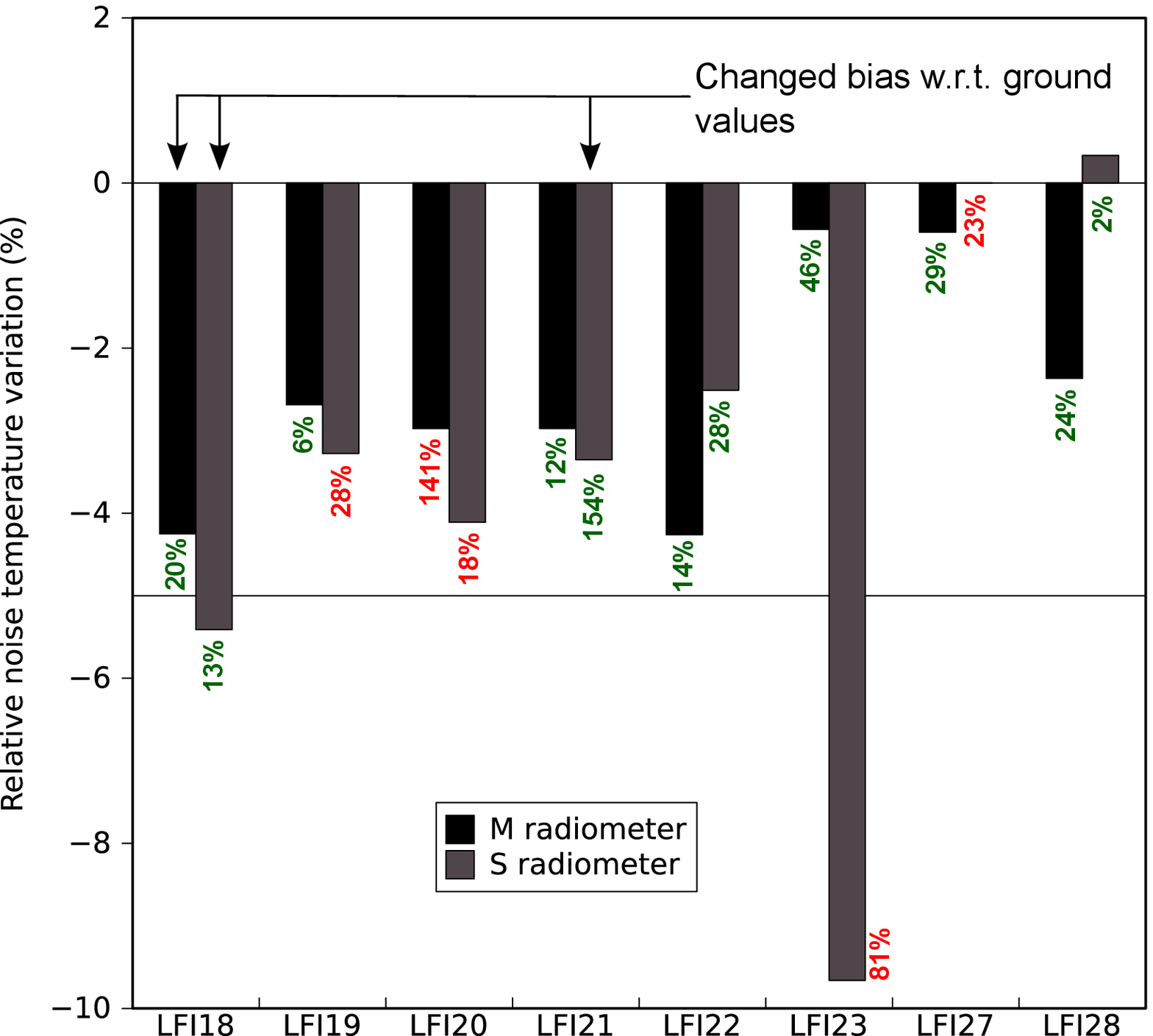}
  \caption{Relative variation in noise temperature and isolation between the flight and ground bias configurations for the 30 and 70\,GHz receivers. Negative variation in noise temperature means a smaller noise temperature measured in flight. Green numbers indicate a better isolation measured in flight and red numbers vice-versa. The horizontal line represents the 5\% noise temperature improvement threshold.}
  \label{fig_noise_temperature_iso}
  \end{center}
\end{figure}

In Figure~\ref{fig_noise_temperature_iso} we show the relative variation in noise temperature and isolation for the 30 and 70\,GHz receivers between the in-flight and on-ground optimal configurations. From the figure we see that in two cases we had a clear indication of a performance improvement that led us to choose the flight bias settings. These were: \texttt{LFI18S}, for which we obtained a noise temperature reduction larger than 5\% and \texttt{LFI21S} for which the noise temperature reduction was limited but the isolation  improved by more than a factor of 2 (in dB). Eventually we decided to apply the flight bias setting also to \texttt{LFI18M} which showed a 20\% isolation improvement and was close to the 5\% threshold we set on noise temperature. For all the remaining receivers we maintained the optimal biases found during satellite on-ground tests. Table~\ref{tab_final_bias_settings} summarises the bias settings chosen for the front-end amplifiers at the end of CPV. Concerning the 44~GHz phase switch biases, coherently with the choices done at LNAs level, we decided to apply the unit level test I$_1$, I$_2$ values (about 0.8~mA for both, coming from RF-bench optimization). Although the bias region covering these values was not mapped during the CPV (the CPV phase switch tuning focused on a very limited region were the best performance were expected), nevertheless the extrapolated performance from ground data showed that, even if not optimal, this setup was acceptable (Section~\ref{sec:lfi_ps}). At 30~GHz we applied the flight biases that were coincident with ground bias. At 70~GHz, the phase switch setup remained fixed to the default applied since the ground level tests (about 1~mA).\\ These LNAs and phase switch biases have never been changed since the start of nominal operations.

\begin{table}[htb!]
  \begin{center}
    \caption{Origin of final bias settings. \texttt{M} and \texttt{S} refer to (V$_{\rm g1}$, V$_{\rm g2}$) corresponding to  \texttt{Main} and \texttt{Side} radiometers. \texttt{PH/SW} identifies the LNAs phase switch bias setup (I$\rm_1$, I$\rm_2$) applied to each channel.
``Flight'' label corresponds to values determined during in-flight tests, ``Ground'' to satellite-level ground tests, ``FEM'' to module-level ground tests, ``MAX'' to maximum current value allowed from specifications and ``Gr./Fl.'' when Flight and Ground results are coincident.}
    \label{tab_final_bias_settings}
    \begin{tabular}{l l l l}
      \hline
      \hline
	& \multicolumn{2}{c}{R{\sc adiometer}} &\\
	\hhline{~--}
	\multicolumn{1}{c}{RCA} & \multicolumn{1}{c}{\texttt{M}}& \multicolumn{1}{c}{\texttt{S}} & \multicolumn{1}{c}{\texttt{PH/SW}}\\
	\hline
	\texttt{LFI18}& Flight  &  Flight &  MAX \\
	\texttt{LFI19}& Ground&  Ground&  MAX\\
	\texttt{LFI20}& Ground&  Ground&  MAX\\
	\texttt{LFI21}& Ground&  Flight&  MAX\\
	\texttt{LFI22}& Ground&  Ground&  MAX\\
	\texttt{LFI23}& Ground&  Ground&  MAX\\
\noalign{\vskip 8pt}
	\texttt{LFI24}& FEM&  FEM & FEM\\
	\texttt{LFI25}& FEM&  FEM & FEM\\
	\texttt{LFI26}& FEM&  FEM & FEM\\
\noalign{\vskip 8pt}
	\texttt{LFI27}& Ground&  Ground &  Gr./Fl.\\
	\texttt{LFI28}& Gr./Fl.&  Ground&  Gr./Fl.\\
\hline
      \end{tabular}
\vskip 8pt
\end{center}
\end{table}

\paragraph{Non linear solution.}
\label{par:nonlinear}

The best performance quadruplets were chosen basing the analysis on the \emph{Y-factor} provided by the linear analysis of the four temperature steps. From the CSL matrix tuning it was already known that the relative improvements in the bias space were only slightly modified by the non linear response of radiometers, only 44~GHz channels showing non linearity effects. 
Although, as described in the \ref{par:thermalsetup} section, the poor accuracy of the thermometers used during the tuning reduced the sensitivity of the method, nevertheless we applied the non linear model of the Gain \cite{mennella2009-2} to compare with results from the same analysis performed at system level and unit level, and to results from the linear approach. The analysis was applied only to 30~GHz and 44~GHz channels: in fact 70~GHz channels were already known from RCA and  system level  tests to show a linear response. This analysis was complicated by other effects (for instance the BEU thermal fluctuations, see Figure~\ref{HYM_Tuning_4steps_FPU_sensors}) contributing to masking or generating spurious non linear responses.

Both the linear and the non linear solutions were corrected for the BEU thermal drift along the four steps  causing gain changes in the Back-End amplifiers. This correction was applied by imposing the unicity of the sky signal during the four steps of the tuning, neglecting the small variations due to the moving sky and the second order variations due to non perfect Isolation of the LNAs. 
During the four steps, the Front End temperature remained quite constant, for any given bias quadruplet; the FPU temperature pattern along each step could be ascribed to the different power dissipation of the LNAs, when operated through different bias quadruplets. However, the pattern of instabilities looked repeatable along the four steps (see Figure~\ref{HYM_Tuning_4steps_FPU_sensors}), allowing to consider constant  the gain of Front End amplifiers   for each quadruplet. \\
Results confirmed that non linear effects and BEU thermal drift influence only weakly the choice of the optimal biases, apart for those channels (namely the two 30~GHz) showing a very flat response due to the criteria previously adopted to restrict the bias hypermatrix: for these channels only small changes are registered in the noise Temperature when surfing the bias map. For each channel, complete results showing the best bias quadruplets corresponding to the three analysis applied (Linearity, Gain Model, Gain Model considering the Back-End drift correction) are presented in Appendix~\ref{app_detail_tests}, Table~\ref{tab_HYM_nonlin_Tun_vs_CSL}. 

Plots resulting from the non linear Gain-Model analysis corrected for the Thermal Drift of the Back End amplifiers are fully presented in Appendix~\ref{app_HYM_nonlin_plots}. It is worth noting that the correction applied  improved w.r.t. the technique described above used in CPV. In fact, data were here corrected on the basis of thermal transfer functions calculated along 40 nominal survey days from August 22$\rm ^{nd}$ to September 30$\rm ^{th}$, 2009. 

Tables reporting the comparison in terms of performance and non linear coefficients among several bias configurations (corresponding to unit level tuning, RCA tuning, CSL system level  matrix tuning, CPV default) applied in CPV are reported in Appendix~\ref{app_detail_tests},  Tables~\ref{tab_HYM_Tun_summary_24} to~\ref{tab_HYM_Tun_summary_28}.

%% file: 04_lfi_tests_tuning_verif.tex
Tuning of the LFI has been accomplished at several stages of
integration, with different procedures. During these procedures, the
system noise temperature (T$_{\rm sys}$) and isolation were used as the
figures of merit for optimising performance, since they can be
estimated with high signal to noise in a short period of time. In
fact, for the LFI receivers, the calibrated noise and 1/$f$
characteristics are the true indicators of scientific performance. In
principle, calibrated white noise can be derived directly from the
system temperature and noise effective bandwidth, but in practice
there are noise contributions which make it
hard to be sure that white noise predicted by T$_{\rm sys}$ and bandwidth
will be achieved. With a receiver topology as complex as LFI, it is
even possible that optimising T$_{\rm sys}$ and isolation may cause us to
miss the actual optimum white noise bias point. With this in mind we
developed the following verification test based on the hypermatrix
tuning:
\begin{itemize}
\item Set LFI for nominal operations (DAE gain and offset tuned to
  allow measurement of the true radiometer white noise);

\item Acquire data (30 seconds) at each of the nominal hypermatrix
  tuning bias points, in the same manner as was done for the
  hypermatrix tuning. This samples the LFI bias space around the
  points most likely to yield good performance;

\item Change the 4~K load temperature by a known amount. This change is
  provided by the HFI team using the PID controller of the 4~K stage;

\item Again acquire data at all the hypermatrix bias points;

\item White noise is estimated from each 30 second period, and then
  calibrated using the corresponding data from the known temperature
  step of the 4~K load.
\end{itemize}

The results of the white noise hypermatrix are consistent with the normal hypermatrix tuning but provide no extra power to optimise the bias. This result was unexpected. Based on both analytic and monte carlo calculations before flight we estimated only a few percent error in the calibrated white noise, which would have been sufficient.

A possible explanation is in the settling time for the receivers after bias changes. In analyzing this test we allowed a few seconds settling time, but it is possible that the noise characteristics are not stable in this short a time. Given the time limitations on the test, we could not consider larger integration times  per bias point, which might have given more stable and discriminatory results.

%% file: 04_lfi_tests_dae_tuning.tex
The 44 analog voltage outputs from the radiometer back-end modules are digitized in the Digital Acquisition Electronics (DAE) box by 44 14-bit ADCs. The signal $x$ is processed according to the following formula:
\begin{equation}
\label{eq:DAEProcessing}
x\,\textrm{[V]} 
\quad\rightarrow\quad (x - V_0) \times G_\textrm{DAE} + b_0 \equiv y\,\text{ [ADU]},
\end{equation}
where $V_0$ and $G_\textrm{DAE}$ are the programmable offset and gain respectively, the latter being responsible for the conversion from Volt to ADU, that is 14-bit integers. The quantity $b_0$ is a 14-bit number that depends on the value of $G_\textrm{DAE}$. Once the sequence of $y$ is acquired on ground, the \Planck-LFI pipeline applies the inverse transformation
\begin{equation}
\label{eq:InvDAEProcessing}
x = \frac{y - b_0}{G_\textrm{DAE}} + V_0
\end{equation}
to reconstruct the signal in Volt.

Each of the 44 ADC converters can be individually programmed. The calibration of the ADCs consists in finding 44 pair of values $(V_0, G_\textrm{DAE})$ satisfying the three following requirements: (i) the voltage $x - V_0$ before the amplification stage must be between $-2.5\,\textrm{V}$ and %greater than $0.0\,\textrm{V}$ and smaller than 
$2.5\,\textrm{V}$, as this is the operating range of the converter; (ii) the value of $G_\textrm{DAE}$ should be as large as possible, in order to reduce the quantization noise in the output $y$; (iii) the value of $y$ should not require more than 14 bit (i.e. it must be in the range $[0, 2^{14} - 1]$), in order to prevent overflows.

We were able to set the value of $V_0$ and $G_\textrm{DAE}$ by sending telecommands to the DAE which encoded a pair of integer numbers for each ADC. Each integer number matched some specific value for $V_0$ and $G_\textrm{DAE}$, so that there were 256 different values available for $V_0$ and 10 for $G_\textrm{DAE}$. The matches between the integer parameters used in the telecommand and the actual values of $V_0$ and $G_\textrm{DAE}$ were established on ground with a dedicated calibration.
\begin{figure}[tbf]
  \centering
  \includegraphics[width=0.8\textwidth]{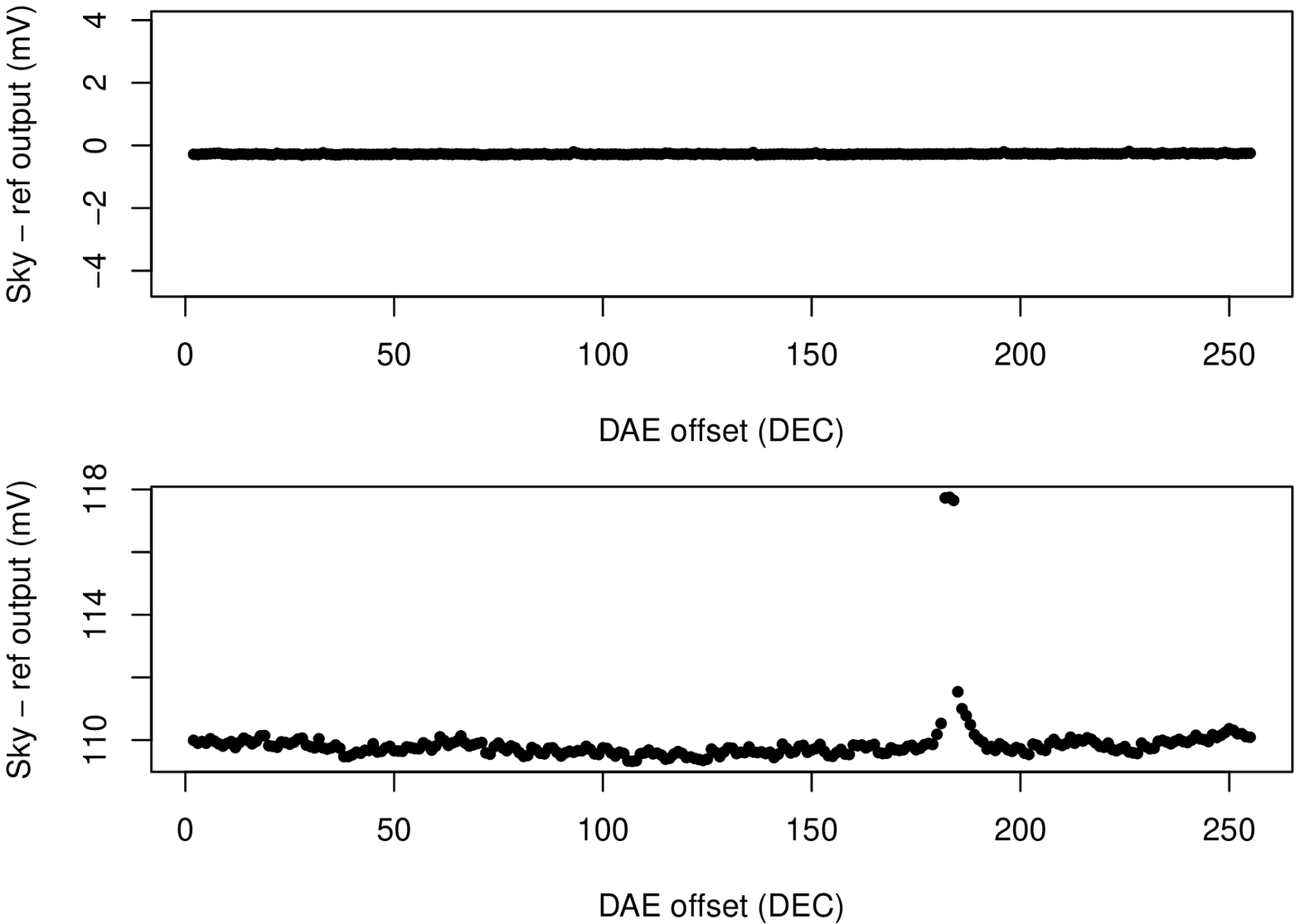}
  \caption{\label{fig:DAEDiffExample} Top panel: plot of the quantity $x_\textrm{sky} - x_\textrm{ref}$ (reconstructed from $y_\textrm{sky}$ and $y_\textrm{ref}$ following eq.~\protect\ref{eq:InvDAEProcessing}) as a function of the applied offset $V_0$, using data acquired at room temperature (sky and reference loads both at 300\,K). The balance between the two loads implies that $x_\textrm{sky} - x_\textrm{ref}$ is roughly zero. The average value does not change while varying $V_0$: this proves that we are correctly reconstructing the voltage $x$ from $y$. Bottom panel: when sky and reference load temperatures are unbalanced (sky at 30\,K, load at 22\,K), the average is no longer zero as expected, but an unexpected spike at $V_0 \sim 185\,\textrm{DEC}$ reveals the presence of the additional offset $\Delta$ in Eq. \protect\eqref{eq:DAENonIdeality}.}
\end{figure}

\begin{figure}[tbf]
  \centering
  \includegraphics[width=0.8\textwidth]{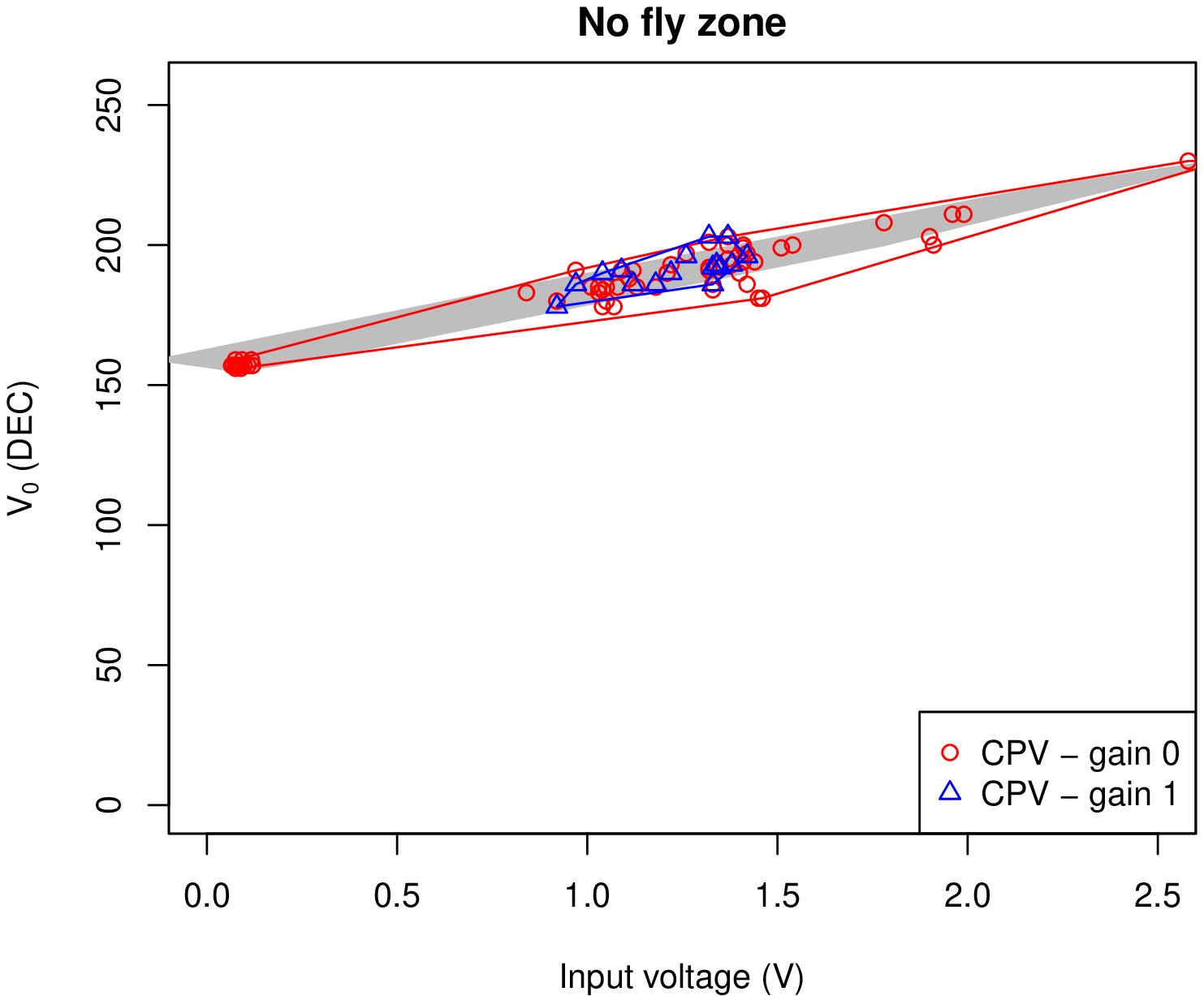}
  \caption{\label{fig:noFlyZoneCPV} The red and blue points show for which values of the input voltage $V$ entering the ADC the DAE offset $V_0$ in eq.~\protect\eqref{eq:DAEProcessing} introduces an error in the measure. The cloud of points profiles a region in the $V \times V_0$ plane called ``no-fly zone''. Red points refer to gain state $G_\textrm{DAE} = 1$, blue to $G_\textrm{DAE} = 2$ (higher gains are not included as the ADCs saturated before reaching the no-fly zone). The gray region is the no-fly zone measured during the 2006 ground tests (see Figure~18 in \cite{cuttaia2009}). Note the good overlap with the red/blue points.}
\end{figure}

\paragraph{The ``no-fly'' zone.} 

During the LFI ground tests we discovered that for some values of $V_0$ (the so-called ``no-fly zone'') there was a spurious offset $\Delta$ at the output of the ADC:
\begin{equation}
\label{eq:DAENonIdeality}
x \quad\rightarrow\quad (x - V_0) \times G_\textrm{DAE} + b_0 + \Delta(x, V_0),
\end{equation}
where the notation $\Delta(x, V_0)$ stresses that such offset depends both on the input voltage $x$ and the offset $V_0$ applied to the DAE (see also \cite{cuttaia2009}). Figure~\ref{fig:DAEDiffExample} explains this effect by plotting the value of the difference in the voltages $x_\textrm{sky} - x_\textrm{ref}$ reconstructed on ground using eq.~\eqref{eq:InvDAEProcessing}, which is not constant as expected when varying the value of $V_0$ (taking the difference emphasizes the effect over other systematic effects, most notably slow thermal drifts occurred during the tests).

The presence of $\Delta$ is potentially dangerous for the LFI radiometers, as they are feeding the ADC with pairs of voltages $x_\textrm{sky}, x_\textrm{ref}$ measuring the temperature of two mismatched loads (the sky at 2.7\,K and the reference load at 4.5\,K): therefore, picking a value of $V_0$ for which $\Delta \not= 0$ would lead to a difference $\Delta_\textrm{sky} - \Delta_\textrm{ref}$ that is not cancelled by the differential strategy.

We empirically found a region in the $x \times V_0$ plane producing nonzero $\Delta$: such values $V_0^\textrm{no-fly}$ depend on the input voltage $x$. For all those detectors showing enough low noise we were able to clearly identify \emph{two} pairs $(x, V_0^\textrm{no-fly})$ corresponding respectively to sky signal $x_\textrm{sky}$, and to reference signal $x_\textrm{ref}$. Figure~\ref{fig:noFlyZoneCPV} compares CPV results (in the $x \times V_0$ plane) to those obtained during on ground tests (2006). The consistency of results from the two tests \footnote{$V_0^\textrm{no-fly}$ points are all contained in the shaded region, called \emph{no-fly zone}}, proved the stability with time of this effect and its independence from the environmental conditions (laboratory/space) and from the applied gain $G_\textrm{DAE}$. 

Even if we did not understand the real cause of this spurious offset $\Delta$, we were able to produce a ``safe'' calibration for LFI by picking $V_0$ values not falling into the no-fly zone.

\paragraph{Calibration methodology.} 

\begin{figure}[f]
	\centering
	\includegraphics[width=0.75\textwidth]{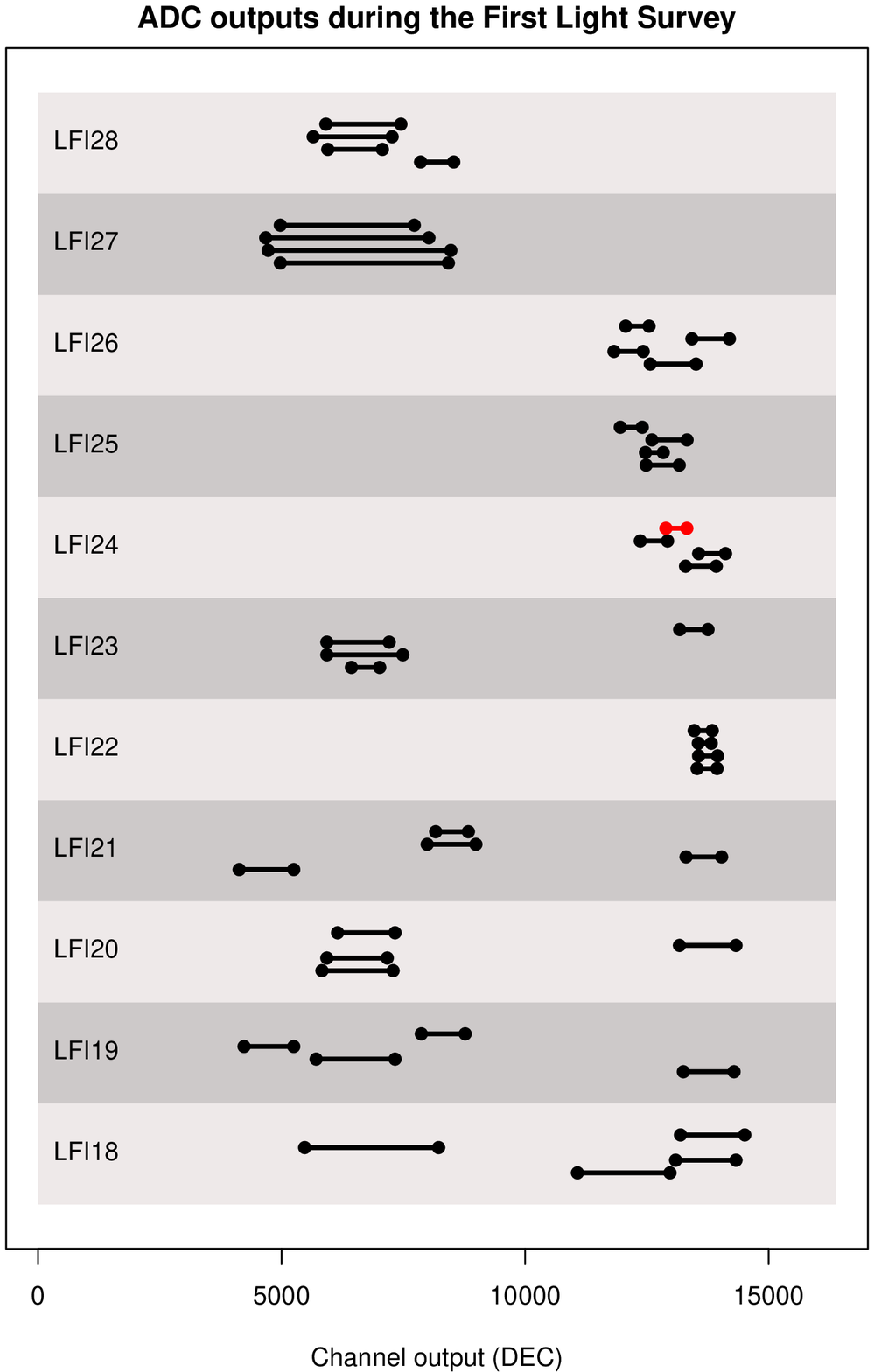}
	\caption{\label{fig:ADCRangesDuringFLS} Range of variations of
          the scientific output (i.e. $y_\textrm{sky/ref}$ in
          eq.~\protect\ref{eq:DAEProcessing}) for the 44 LFI channels.
          For each RCA the four diodes are (from the lowest to the
          topmost): M-00, M-01, S-10, S-11. The time considered for
          the production of the plot is the so-called ``First Light
          Survey'', which lasted 15 days. Since an unexpected
          gain change provoked a saturation in channel
          \texttt{LFI24S-11} (marked with red) during the first days
          of the first light survey, in this case we discarded the first three days
          of the first light survey in computing the range.}
\end{figure}

The calibration of the DAE was divided into two parts:
\begin{enumerate}
\item \emph{Calibration}. Data were acquired while exercising the 44 ADC converters in each possible configuration; such data were used to determine the best values for $V_0$ and $G_\textrm{DAE}$ for each converter.

 This phase required the REBA to acquire 30 seconds data in \texttt{AVR1} uncompressed mode (see Section~\ref{sec:rebatuning}) for each of the possible gain and offset states of the DAE ($10 \times 255 = 2\,550$ states). The negligible cross-talk between the 44 ADCs allowed to  change their state in parallel. The full test required about 21 hours ($2\,550 \times 30\,\,\textrm{s}$).

\item \emph{Verification}. Performed after having uploaded the best
  configuration in the DAE. It simply required to acquire data for a
  few minutes  to verify the consistency of output from each ADC with
  analytical predictions. Figure~\ref{fig:ADCRangesDuringFLS} shows
  the ranges of variation of the voltages during the First Light
  Survey.
\end{enumerate}

\afterpage{\clearpage}

\begin{figure}[tbf]
	\centering
	\includegraphics[width=\textwidth]{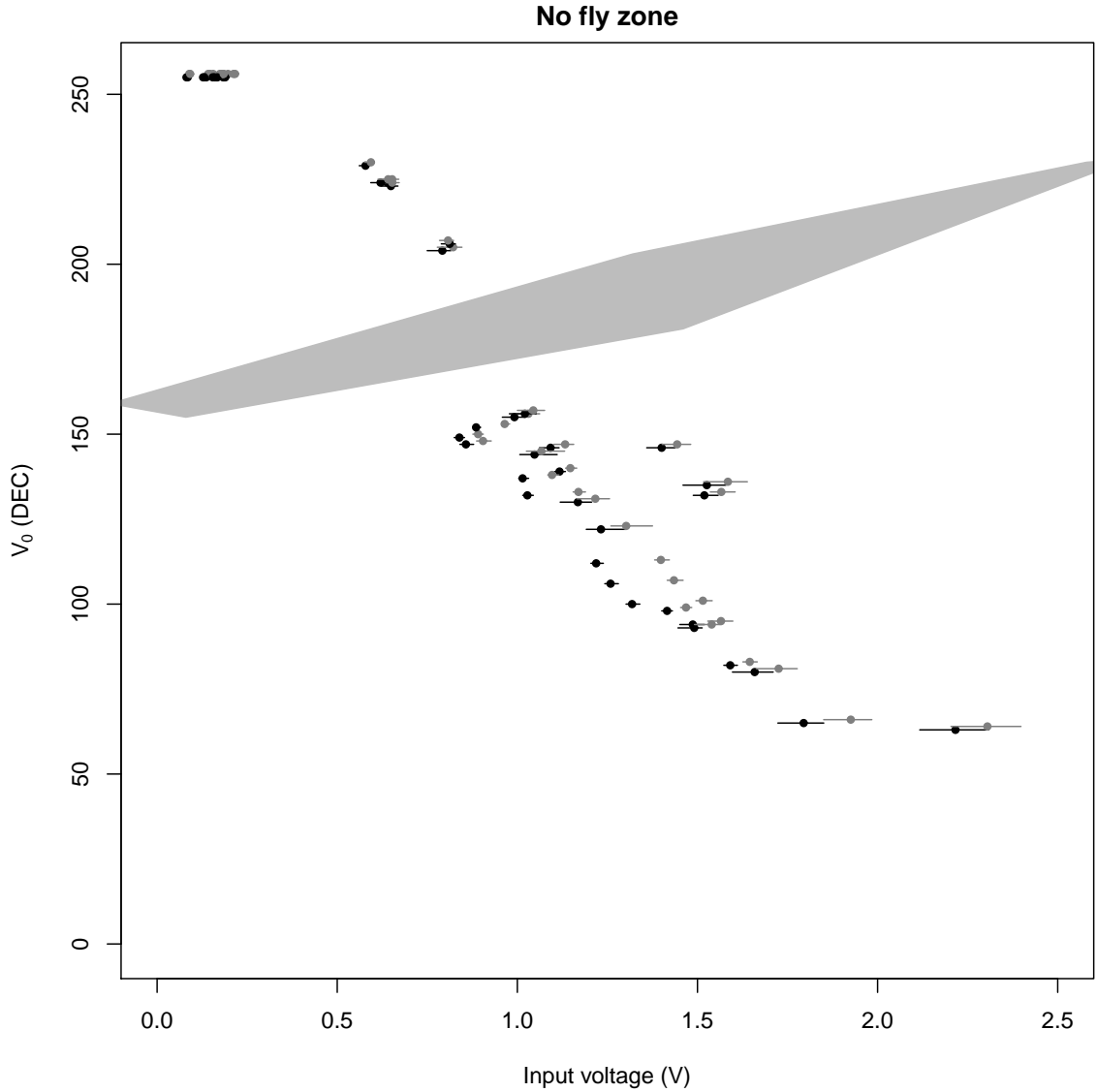}
	\caption{\label{fig:DAECalibrationResults} The values of the
          DAE offset $V_0$ (eq.~\protect\ref{eq:DAEProcessing})
          resulting from the DAE calibration adopted for the nominal
          survey were chosen well out of the no-fly zone (gray
          polygon). $V_0$ is plotted against the average value of the
          sky (black points) and reference (gray points) signals
          $x_\textrm{sky}$ and $x_\textrm{ref}$ ($44 \times 2 = 88$
          points). The horizontal bars show the typical range of
          variation of the output during the first 2 years of data
          acquisition. For clarity, the gray points have been shifted
          horizontally by +1.}
\end{figure}
\begin{figure}[tbf]
	\centering
	\includegraphics[width=0.8\textwidth]{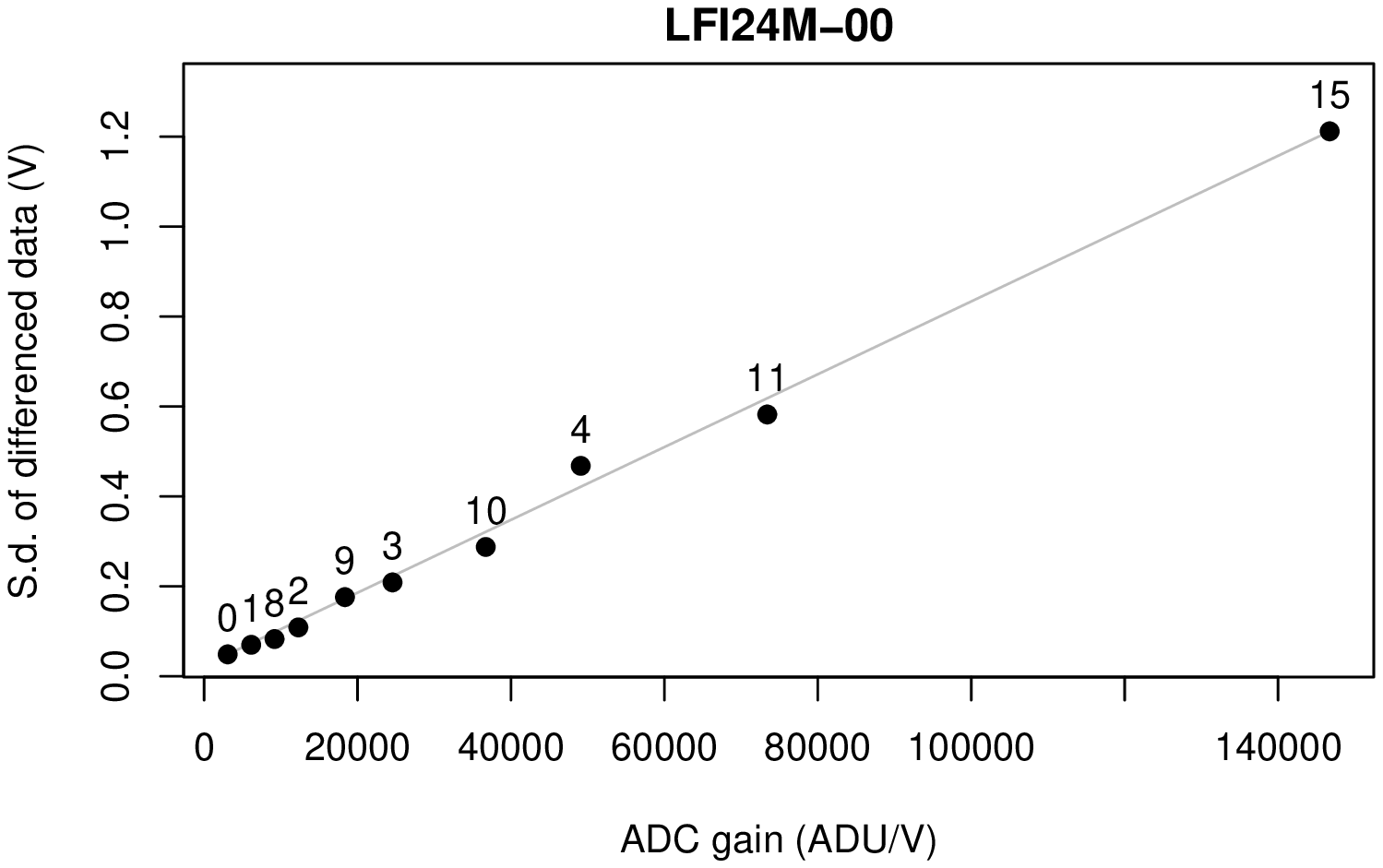}
	\includegraphics[width=0.8\textwidth]{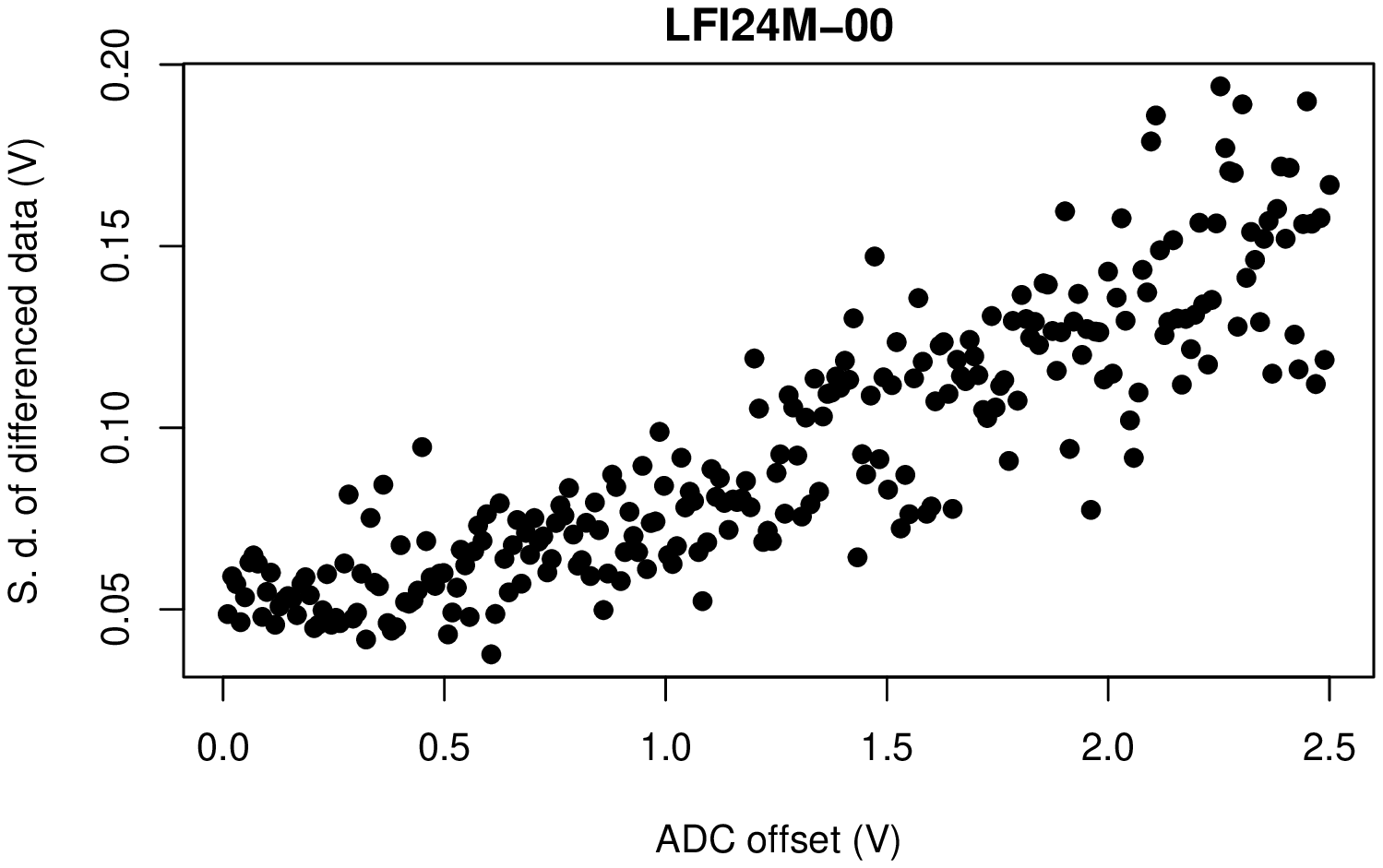}
	\caption{\label{fig:AduRmsVsDaeGainAndOfs} Top panel: the
          measured standard deviation (s.d.) of the digital data
          produced by the ADC of the channel \texttt{LFI24M-00}
          (44\,GHz) during the DAE calibration test as a function of
          the applied gain. We chose to show \texttt{LFI24M-00} as it
          has a low intrinsic output ($\sim$ 0.4\,V) and therefore did
          not saturate even when applying the maximum gain
          $G_\textrm{DAE} = 15$. Note the expected linear correlation
          between the r.m.s. of the data and the gain. Botton panel:
          Unlike 30 and 70\,GHz channels, 44\,GHz channels like
          \texttt{LFI24M-00} show an increase in the r.m.s. level of
          their differenced output which is correlated with the value
          of $V_0$ ($G_\textrm{DAE}$ was kept constant at 0\,DEC, the
          lowest gain state). For further details, see text.}
\end{figure}

\noindent The results of the DAE calibration can be schematized as follows:
\begin{itemize}
\item For each detection line, the set of pairs $(G_\textrm{DAE}, V_0)$ for which the ADC saturates were measured and are consistent with our analytical models.

\item The increase in the r.m.s. of the signal is well correlated with the ADC gain (see e.g.  \texttt{LFI24M-00} in the top plot in Figure~\ref{fig:AduRmsVsDaeGainAndOfs}), as it was expected (note we were able to check this property only for those channels with a voltage output small enough to exercise all the gain states without saturating the ADC).

\item The r.m.s. did not change while varying the ADC offset $V_0$, except for the 44\,GHz channels (see the bottom plot in Figure~\ref{fig:AduRmsVsDaeGainAndOfs}). These channels are characterized by an unusually low output ($\sim$0.1\,V), which therefore allows to detect the noise of the ADC itself. This was hardly a problem, as the main purpose of $V_0$ in this context was to make the signal entering the ADC close to zero: for the 44\,GHz channels the small signal allowed us to pick up $V_0 = 0\,\mathrm{V}$, thus avoiding the problem.

\item The best configuration $(G_\textrm{DAE}, V_0)$ was chosen for each ADC in order to maximize the gain while keeping the ADC far from saturating. The ADCs were proven to be within the allowed limits during the period called ``First Light Survey'' (corresponding to the Operational Days 91--105 corresponding to August 12$^{\rm th}$ to 26$^{\rm th}$, 2009). Also, the value for $V_0$ was always chosen to be far enough from the no-fly zone (see Figure~\ref{fig:DAECalibrationResults}).

\item We had the proof that the no-fly zone effect is remarkably stable, as from the analysis of the CPV tests we found the very same results of the RAA test campaign (which was performed three years before). Also, since for the first time we were able to study this effect for every value of the gain $g$, we finally had the proof that the shape of the no-fly zone does not depend on $G_\textrm{DAE}$. See Figure~\ref{fig:noFlyZoneCPV}.
\end{itemize}

%% file: 04_lfi_tests_reba_tuning.tex
The LFI Signal Processing Unit is a module of the REBA whose purpose, among others,
is to downsample and compress the scientific data acquired 
by the radiometers and digitized by the DAE at a sampling rate of $\approx 8$~kHz,
 in order to fit the spacecraft telemetry bandwidth. Preprocessing includes a downsampling stage and a 
tunable lossy compression scheme which allows to achieve high compression ratios 
($C_\mathrm{r}\sim 2.4$) at the expense of increasing the noise level. 
Details on the full preprocessing and compression stage, as well as its 
optimization and characterization are given in 
\cite{maris2009}.% A full account of inflight optimization, monitoring and performances of REBA Signal Processing Unit  is in preparation, here a short account is given. 
Since the preprocessing plus compression is tunable by defining a set of 4 parameters for each detection line ($\Offset$, $\SecondQuant$, $r_1$ and $r_2$),\footnote{ $r_1$ and $r_2$ are the two parameters transforming the sky and reference load data streams $V_{\rm sky}$, $V_{\rm ref}$ into two differential data streams $V_i$, according to the formula:
        \begin{eqnarray}
            V_i &=& V_{\rm sky} - r_i\, V_{\rm ref} \nonumber
           \label{eq:reba_differencing_2}
        \end{eqnarray}

The two differential datastreams are hence quantised according to the formula:
        \begin{equation}
            Q_i= {\rm round}\left[(V_i + \Offset)\times \SecondQuant\right], i=1,2,
            \label{eq:quantisation}
        \end{equation}
        where $\Offset$ and $\SecondQuant$ are an offset and a quantisation factor.} the objective of the REBA tuning was to find the set of 44 quadruplets that allowed to fit the total bandwidth within the requirements while keeping the digitization noise and other possible distortions as low as possible \cite{maris2009}.

The calibration and verification of the REBA compressor exploits the ability of the REBA to acquire each LFI detector in two modes at the same time, that is in \texttt{AVR1} mode (the ground station receives the data \emph{after} they are downsampled but \emph{before} they enter the compression stage) or \texttt{COM5} mode (ground receives the data \emph{after} the compression stage). Due to inflight bandwidth constrains, it is not possible to make all the 44 detectors work in \texttt{AVR1} mode, in the nominal configuration of LFI all the detectors work in \texttt{COM5} mode while in turn one detector is sampled also in \texttt{AVR1} for half an hour realizing a round--robin scheme whith a 22~hours cycle.

Here are the stages of the REBA optimization done during the CPV: 
\begin{enumerate}
  \item The nominal round--robin scheme was switched off.
  \item Each of the 44 detectors was kept in stable conditions and data were acquired in 
\texttt{AVR1} mode for 45 minutes. To save time, 11 detectors were acquired at the same time, so 
that the overall duration of the data acquisition phase was only three hours.

  \item Using the software suite and the
optimization methods reported in \cite{maris2009}, 
for each detector statistics of downsampled signals was derived and used to asses
a first analytical pre--optimization.

  \item \label{itm:REBACalibration} 
We further optimized the 
compression configurations on the uncompressed data acquired 
exploring a regularly gridded subset of each of 4-dimensional parameter space
around the pre optimized parameter sets.

  \item To determine the best configuration, we used the following criteria: 
(1) the compression ratio $C_\mathrm{r}$ must be not smaller than the requirement, which is 2.4; 
(2) the quantization error on a weighted averages of 
the sky and reference signals\footnote{During ground tests we favoured 
configurations which produced the smallest errors in the 
\emph{differenced} signal $V_\mathrm{sky} - r V_\mathrm{ref}$, see \cite{maris2009}. 
However, we found that putting the constraint on a weighted average 
of the sky and reference signals would produce significantly 
smaller noise in total-power data (which is useful for better estimating the gain modulation 
factor $r$) at the only expense of a slight increase in the r.m.s. of the difference 
$V_\mathrm{sky} - r V_\mathrm{ref}$.} 
measured by the detector 
must be the lowest one. Both the $C_\mathrm{r}$ and the quantization errors are measured on 
the real data by using a software which reproduces both the onboard production
of \texttt{COM5} data out of the
\texttt{AVR1} data and the subsequent decompression and decoding at the Ground Segment.

  \item Once the best configuration was found, the procedure repeated from step 
\ref{itm:REBACalibration} by centering the grid in the parameter space around this configuration 
and by shrinking it 10 times along each of the four dimensions. 
This produced a new ``best'' configuration. We compared the 
compression ratios, quantization levels and required bandwidths of the two ``best'' 
configurations in order to make the final choice. 

  \item After the REBA was programmed with the quadruplets found in the previous step, 
the instrument was put in nominal mode and started with the nominal round--robin 
for a 22 hours cycle.

  \item The 22 hours of data were analysed %by {\tt OCA2K}
in order to check if the differences between the \texttt{AVR1} and \texttt{COM5} datastreams for each radiometers were in 
agreement with the analysis done in step \ref{itm:REBACalibration}.

\end{enumerate}

The results of the test were satisfactory. All the 44 channels were able to compress 
the data so that the overall compression ratio was $C_{\mathrm{r}} \approx 2.4$, corresponding 
to the requirement. The quantization induced by the compressor accounted for a 
$\sim 5\%$ increase in the r.m.s. level, which is considered to be acceptable.

%% file: 04_lfi_tests_performance_calibration.tex
After the instrument was fully tuned its noise performance was assessed during a dedicated test that was split into two parts: 

\begin{enumerate} 
  \item a data acquisition performed with the radiometers unswitched, where the noise spectra were compared for all four phase switch configurations. The objective of this test was to assess whether any phase switch configuration was preferable from the point of 1/$f$ noise;
  \item a data acquisition with the radiometers nominally switching in order to characterise the full noise properties in both switching configurations (\texttt{A/C} or \texttt{B/D}), with the twin phase switch in its nominal position. Photometric calibration was performed using the CMB dipole allowing a preliminary assessment of the in-flight white noise instrument sensitivity.
\end{enumerate}

The instrument thermal configuration was nominal. The front-end unit temperature ranged from 19.6\,K to 20.6\,K with a stability of $\pm 5$~mK, the 4\,K stage was at $4.37\pm0.002$~K and the back-end unit temperature ranged from 287.7~K to 309.8~K with a stability of $\pm 0.1$~K.

\subsubsection{Noise properties with radiometers unswitched}
\label{sec_noise_properties_unswitched}

This test consisted in four acquisitions of 20 minutes each, in order to test all the phase switch configurations. The only exception was the \texttt{LFI23} receiver for which the \texttt{B/D} phase switches were always kept in the \texttt{0} position to avoid the anomalous 1/$f$ noise described in Section~\ref{sec:lfi_on_basic} (Figure~\ref{fig_cryo02_fk_2300}). In all cases the 4\,kHz switching was set to zero for both \texttt{A/C} and \texttt{B/D}. The complete set of phase switch configurations tested is summarised in Table~\ref{tab_unswitched_configurations}.
\begin{table}[h!]
  \caption{Phase switch configurations during the four acquisitions with the radiometers unswitched. The configurations relative to \texttt{LFI23} are marked in boldface because for that receiver the \texttt{B/D} switches were always kept in the \texttt{0} position to avoid the anomalous noise level triggered by the \texttt{B/D=1} configuration (see Section~\protect \ref{sec:lfi_on_basic}, Figure~\protect \ref{fig_cryo02_fk_2300}).}
  \label{tab_unswitched_configurations}
  \begin{center}
    \begin{tabular}{c c c c c c c}
	  \hhline{===~===}
		  &\texttt{Configuration 1}	&	&	&	&\texttt{Configuration 2}	&\\
	  \hhline{---~---}
	  RCA	&\texttt{A/C} pos	&\texttt{B/D} pos	&	&RCA	&\texttt{A/C} pos	&\texttt{B/D} pos\\
	  \texttt{LFI18}	&1	&0	&	&\texttt{LFI18}	&0	&0\\
	  \texttt{LFI19}	&1	&0	&	&\texttt{LFI19}	&0	&0\\
	  \texttt{LFI20}	&1	&0	&	&\texttt{LFI20}	&0	&0\\
	  \texttt{LFI21}	&1	&0	&	&\texttt{LFI21}	&0	&0\\
	  \texttt{LFI22}	&1	&0	&	&\texttt{LFI22}	&0	&0\\
	  \textbf{\texttt{LFI23}}	&\textbf{1}	&\textbf{0}	&	&\textbf{\texttt{LFI23}}	&\textbf{0}	&\textbf{0}\\
	  \texttt{LFI24}	&1	&0	&	&\texttt{LFI24}	&0	&0\\
	  \texttt{LFI25}	&1	&0	&	&\texttt{LFI25}	&0	&0\\
	  \texttt{LFI26}	&1	&0	&	&\texttt{LFI26}	&0	&0\\
	  \texttt{LFI27}	&1	&0	&	&\texttt{LFI27}	&0	&0\\
	  \texttt{LFI28}	&1	&0	&	&\texttt{LFI28}	&0	&0\\
	  \hhline{---~---}
		  &	&	&	&	&	&\\
	  \hhline{===~===}
		  &\texttt{Configuration 3}	&	&	&	&\texttt{Configuration 4}	&\\
	  \hhline{---~---}
	  RCA	&\texttt{A/C} pos	&\texttt{B/D} pos	&	&RCA	&\texttt{A/C} pos	&\texttt{B/D} pos\\
	  \texttt{LFI18}	&0	&1	&	&\texttt{LFI18}	&1	&1\\
	  \texttt{LFI19}	&0	&1	&	&\texttt{LFI19}	&1	&1\\
	  \texttt{LFI20}	&0	&1	&	&\texttt{LFI20}	&1	&1\\
	  \texttt{LFI21}	&0	&1	&	&\texttt{LFI21}	&1	&1\\
	  \texttt{LFI22}	&0	&1	&	&\texttt{LFI22}	&1	&1\\
	  \textbf{\texttt{LFI23}}	&\textbf{0}	&\textbf{0}	&	&\textbf{\texttt{LFI23}}	&\textbf{1}	&\textbf{0}\\
	  \texttt{LFI24}	&0	&1	&	&\texttt{LFI24}	&1	&1\\
	  \texttt{LFI25}	&0	&1	&	&\texttt{LFI25}	&1	&1\\
	  \texttt{LFI26}	&0	&1	&	&\texttt{LFI26}	&1	&1\\
	  \texttt{LFI27}	&0	&1	&	&\texttt{LFI27}	&1	&1\\
	  \texttt{LFI28}	&0	&1	&	&\texttt{LFI28}	&1	&1\\
	  \hhline{---~---}
    \end{tabular}
  \end{center}
\end{table}

\begin{figure}[h!]
    \begin{center}
        \includegraphics[width=14.5cm]{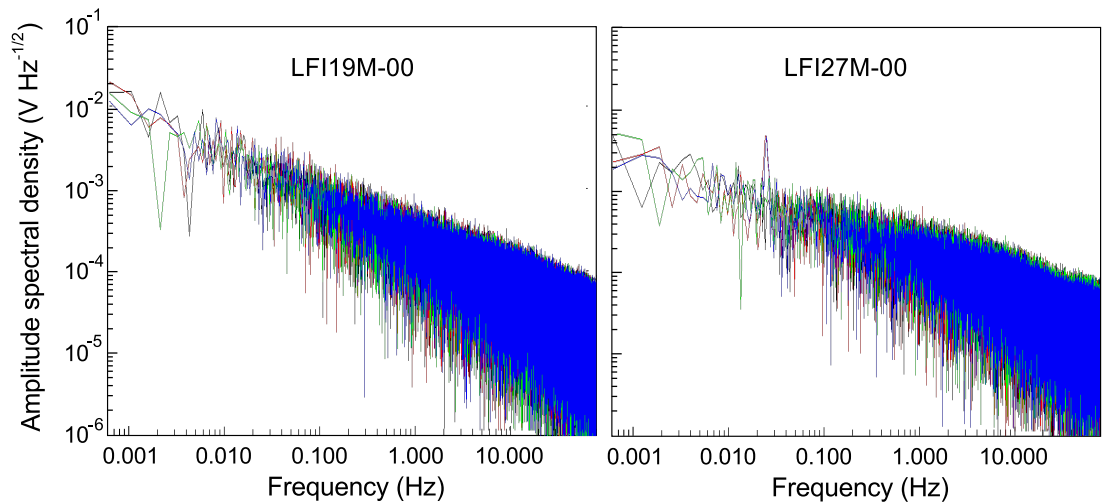}
    \end{center}
    \caption{Amplitude spectral density plots from \texttt{LFI19M-00} and \texttt{LFI27M-00} detectors data acquired with radiometers unswitched. Each plot contains spectra from all four possible phase switch configurations (red, green, blue and black lines). Spectra overlap, showing that the four configurations are equivalent in terms of noise contribution. Spectra from all the other detectors are equivalent to those presented here and are not shown for simplicity.}
    \label{fig_unswitched} 
\end{figure}

In Figure~\ref{fig_unswitched} we show two examples of noise amplitude spectral density plots relative to detectors \texttt{LFI19M-00} and \texttt{LFI27M-00}. In each plot we have overplotted spectra from all four configurations with different colours (black for configuration 1, red for configuration 2, blue for configuration 3 and green for configuration 4). The four spectra essentially overlap showing that any extra noise introduced by the phase switch is equivalent in the four possible switch states.

The remaining forty-two plots not shown in this paper reproduce the same behaviour of the two examples presented here. Therefore we have not included them for simplicity.

\subsubsection{Noise properties with radiometers switched}
\label{sec_noise_properties_switched}

This test consisted in two 12-hour acquisitions in nominal bias and thermal conditions in order to check for 1/$f$ noise differences determined by the switching configuration. In Table~\ref{tab_switched_configurations} we list the two switching configurations tested. Notice that the switching configuration of \texttt{LFI23} was not changed because of the anomalous noise caused by the configuration with \texttt{B/D} switching. Therefore \texttt{LFI23} did not change configuration in the two acquisitions and could be used as a repeatability verification channel.
  \begin{table}[h!]
    \caption{Phase switch configurations in the two data acquisitions for noise characterisation. The receiver \texttt{LFI23}, which did not change configuration, is highlighted in boldface.}
    \label{tab_switched_configurations}
    \begin{center}
      \begin{small}
	\begin{tabular}{c c c c c c c c c}
	  \hhline{====~====}
		 \multicolumn{4}{c}{\texttt{Configuration 1}}	&	& \multicolumn{4}{c}{\texttt{Configuration 2}}	\\
	  \hhline{----~----}
	  RCA	&4\,kHz sw.		&Fixed sw. & pos.	&	&RCA	&4\,kHz sw.	&Fixed sw. & pos.\\
	  \hhline{----~----}
	  \texttt{LFI18}	&\texttt{B/D}	&\texttt{A/C} & 1	&	&\texttt{LFI18}	&\texttt{A/C}	&\texttt{B/D} & 0\\
	  \texttt{LFI19}	&\texttt{B/D}	&\texttt{A/C} & 1	&	&\texttt{LFI19}	&\texttt{A/C}	&\texttt{B/D} & 0\\
	  \texttt{LFI20}	&\texttt{B/D}	&\texttt{A/C} & 1	&	&\texttt{LFI20}	&\texttt{A/C}	&\texttt{B/D} & 0\\
	  \texttt{LFI21}	&\texttt{B/D}	&\texttt{A/C} & 1	&	&\texttt{LFI21}	&\texttt{A/C}	&\texttt{B/D} & 0\\
	  \texttt{LFI22}	&\texttt{B/D}	&\texttt{A/C} & 1	&	&\texttt{LFI22}	&\texttt{A/C}	&\texttt{B/D}& 0\\
	  \textbf{\texttt{LFI23}}	&\textbf{\texttt{A/C}}	&\textbf{\texttt{B/D}} & \textbf{\texttt{0}}	&	&\textbf{\texttt{LFI23}}	&\textbf{\texttt{A/C}}	&\textbf{\texttt{B/D}} & \textbf{\texttt{0}}\\
	  \texttt{LFI24}	&\texttt{B/D}	&\texttt{A/C} & 0	&	&\texttt{LFI24}	&\texttt{A/C}	&\texttt{B/D} & 0\\
	  \texttt{LFI25}	&\texttt{B/D}	&\texttt{A/C} & 0	&	&\texttt{LFI25}	&\texttt{A/C}	&\texttt{B/D} & 0\\
	  \texttt{LFI26}	&\texttt{B/D}	&\texttt{A/C} & 0	&	&\texttt{LFI26}	&\texttt{A/C}	&\texttt{B/D} & 0\\
	  \texttt{LFI27}	&\texttt{B/D}	&\texttt{A/C} & 0	&	&\texttt{LFI27}	&\texttt{A/C}	&\texttt{B/D} & 0\\
	  \texttt{LFI28}	&\texttt{B/D}	&\texttt{A/C} & 1	&	&\texttt{LFI28}	&\texttt{A/C}	&\texttt{B/D} & 0\\
	  \hhline{----~----}
	\end{tabular}
      \end{small}
    \end{center}
  \end{table}  

\begin{figure}[h!]
	\begin{center}
	    \includegraphics[width=12cm]{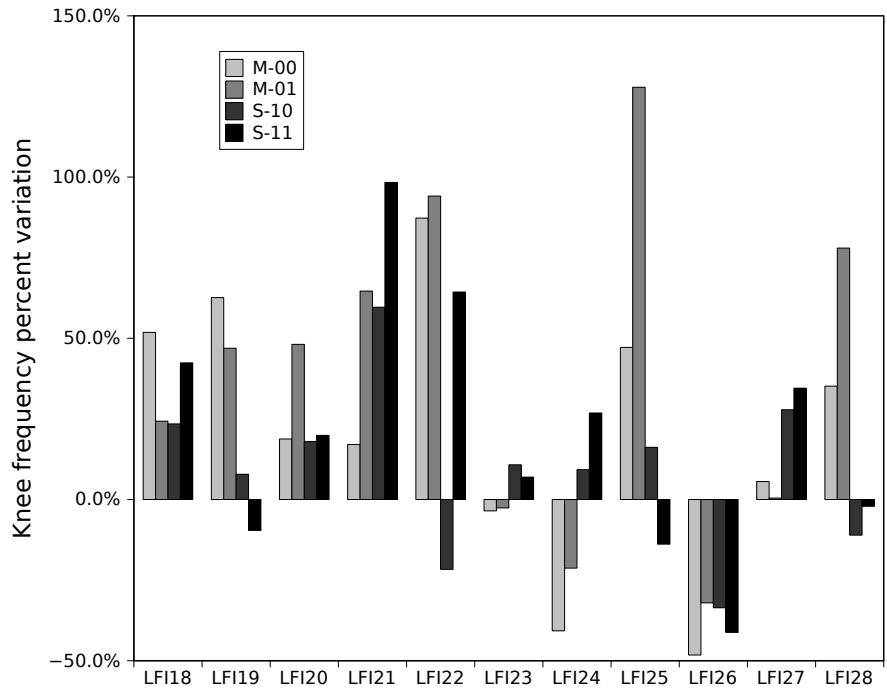}
	\end{center}
	\caption{Percent variation in knee frequency between the two phase switch configurations ($\delta = (f_{\rm k}^\mathrm{Conf\,2} - f_{\rm k}^\mathrm{Conf\,1})/f_{\rm k}^\mathrm{Conf1}$). The variation in \texttt{LFI23} can be considered as a reference for the repeatability of this comparison, as the phase switch configuration was not changed for this receiver.}
	\label{fig_fk_switch} 
    \end{figure}

  \paragraph{Differences in 1/$f$ noise in the two switching configurations.}
    In Figure~\ref{fig_fk_switch} we show the percent difference in 1/$f$ knee frequency for the two tested configurations. Positive numbers indicate that knee frequencies in configuration 2 are larger and vice versa. If we consider a 10\% level of repeatability as indicated by \texttt{LFI23} we see that, in general, in configuration 1 the noise was more stable (smaller knee frequencies). The only noticeable exceptions were represented by \texttt{LFI26} and \texttt{LFI24M} for which the displayed knee frequency was smaller in configuration 2.
    
    Because configuration 1 was more extensively tested during the whole test campaign and the 1/$f$ level of \texttt{LFI26} in this configuration was still compatible with the \texttt{MADAM} map-making code~\cite{zacchei2011,poutanen2004}, we eventually preferred configuration 1 as the default switching configuration for all receivers.
  
    In Table~\ref{tab_fk_switch} we list the measured knee frequencies in mHz for all 44 LFI detectors in the default switching configuration. For a more detailed discussion about the 1/$f$ noise performance measured during nominal operations see~\cite{mennella2011}.
    \begin{table}[h!]
      \begin{center}
	\caption{Knee frequencies in mHz measured in flight during CPV.}
	\label{tab_fk_switch}
      	\begin{tabular}{l c c c c}
      	\hline
      	\hline
	  	&\texttt{M-00}	&\texttt{M-01}	&  \texttt{S-10}	&\texttt{S-11} \\
	  	\hline
		LFI18	&\sp 52	&\sp 58	&\sp 43	&\sp 50\\
		LFI19	&\sp 47	&\sp 49	&\sp 79	&\sp 86\\
		LFI20	&\sp 32	&\sp 28	&\sp 45	&\sp 43\\
		LFI21	&\sp 56	&\sp 45	&\sp 45	&\sp 30\\
		LFI22	&\sp 51	&\sp 44	&\sp 81	&\sp 43\\
		LFI23	&104	&\sp 76	&\sp 58	&\sp 60\\
		LFI24	&\sp 65	&\sp 48	&\sp 60	&\sp 48\\
		LFI25	&\sp 29	&\sp 25	&\sp 39	&\sp 45\\
		LFI26	&105	&\sp 71	&117	&111\\
		LFI27	&105	&119	&\sp 83	&\sp 67\\
		LFI28	&\sp 79	&\sp 64	&\sp 62	&\sp 61\\
		\hline
      	\end{tabular}
      \end{center}
    \end{table}

  \paragraph{Photometric calibration.}

    The CMB Doppler dipole anisotropy induced by Earth and spacecraft motions with respect to the CMB rest frame was used as a calibrator to convert the voltage output into antenna temperature units. The nominal calibration procedure and accuracy has been discussed in detail in \cite{mennella2011,zacchei2011} and will not be repeated here. It is worth mentioning that during CPV tests a simplified calibration pipeline was used, in which a 20$^\circ$ galactic cut was applied to the data to remove the galactic signal, but no iterative algorithm was implemented to estimate and remove the CMB anisotropy. 
    
\begin{figure}[h!]
	\begin{center}
	    \includegraphics[width=10cm]{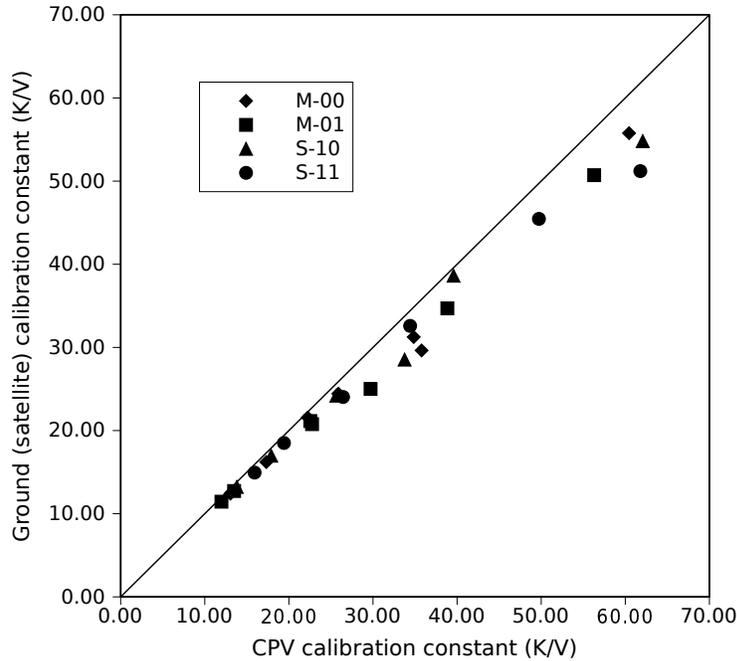}
	\end{center}
	\caption{Photometric calibration constants measured in flight during CPV ($x$-axis) and on ground during satellite tests ($y$-axis).}
	\label{fig_calibration_constants} 
    \end{figure}

\begin{figure}[h!]
    \begin{center}
        \includegraphics[width=15cm]{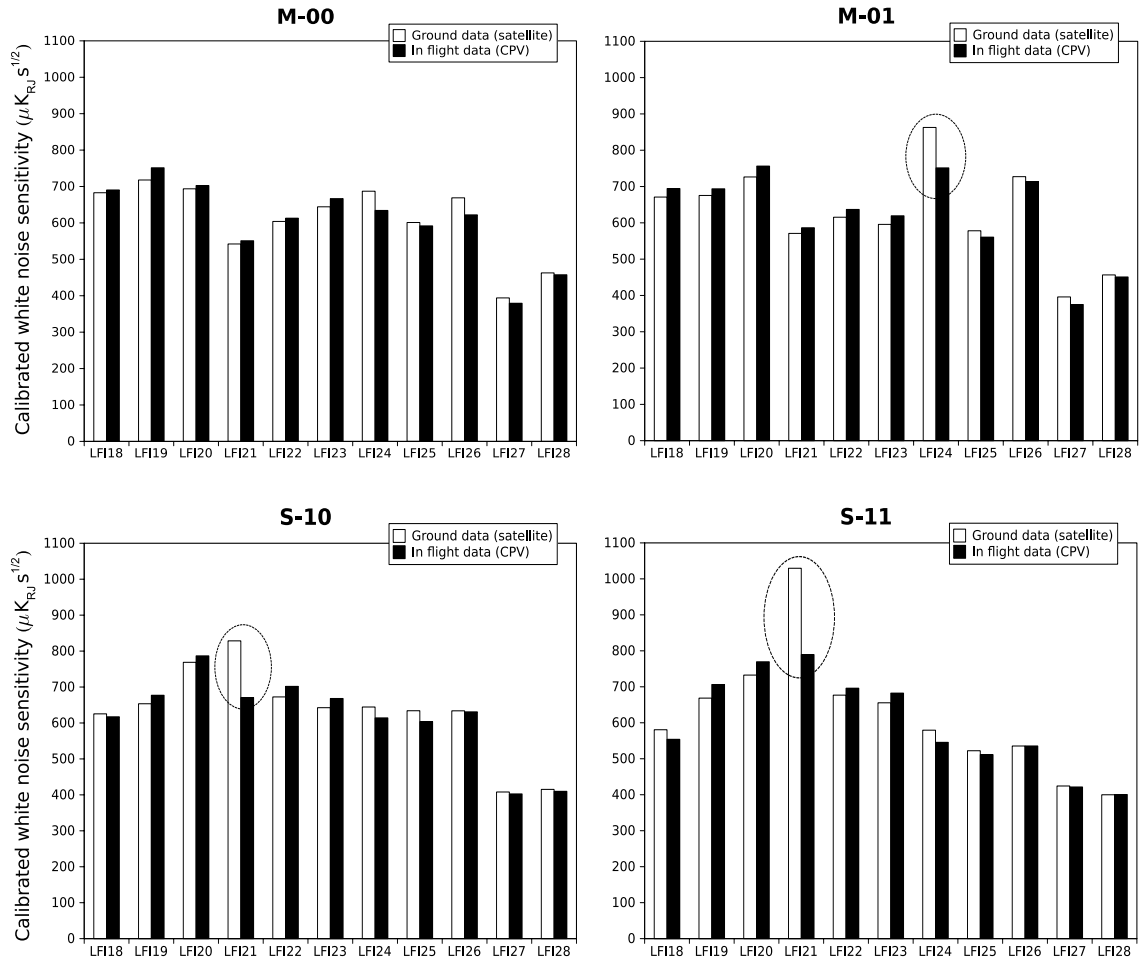}
    \end{center}
    \caption{White noise sensitivities in antenna temperature measured on ground (satellite) and in flight (CPV). Sensitivity increased for \texttt{LFI21S-10}, \texttt{LFI21S-11} and \texttt{LFI24M-01} (circled bars) thanks to in-flight improved bias tuning.}
    \label{fig_white_noise} 
\end{figure}

    In Figure~\ref{fig_calibration_constants} we show the calibration constants as measured in flight during CPV ($x$-axis) compared to those measured on ground during satellite tests ($y$-axis) where we used temperature variations of a 4\,K black-body load that illuminated the focal plane and simulated the sky cold load. In the figure we report the comparison only for the receivers that were biased with the same voltages (see Table~\ref{tab_final_bias_settings}).  The figure shows a broad agreement between the two calibration constant sets with systematically larger values measured in flight, with differences ranging from $\sim 2\%$ to $\sim 17\%$ depending on the receiver. Because differences in calibrated white noise were found to be less than 5\% (see Figure~\ref{fig_white_noise}) we believe that differences in calibration constant values indicated different values of the receiver gains probably due to slightly different bias voltages at the cold amplifiers compared to the ground test environment.

\paragraph{White noise performance.}

Figure~\ref{fig_white_noise} shows the white noise sensitivity in antenna temperature units measured in flight during CPV compared with measurements performed on ground during satellite tests. The comparison show a remarkable agreement apart from three outstanding cases: \texttt{LFI21S-10}, \texttt{LFI21S-11} and \texttt{LFI24M-00}. For these three channels the white noise sensitivity improved by 24\%, 30\% and 15\%, respectively, due to improved bias tuning. For all remaining channels the agreement was better than 10\% with a median discrepancy of about 2\%.

This showed that we were able to generally match the ground-measured instrument sensitivity performance \cite{mennella2010} while improving on a small subset of channels. More detailed discussion on the stability of the sensitivity performance during the first year of operations is provided in \cite{mennella2011}.

%% file: 04_lfi_tests_thermal_susceptibility.tex
The effect of temperature fluctuations on the LFI radiometers is originated in the \Planck\ cold end interface of the hydrogen sorption cooler  to the instrument focal plane, the LVHX2.
The temperature is actively controlled through a dedicated stage, the TSA, providing a first reduction of the effect. The thermal mass of the focal plane strongly contribute to reduce residual fluctuations.

The physical temperature fluctuations propagated at the front end modules cause a correlated fluctuation in the radiometer signal degrading the quality of scientific data. The accurate characterization of this effect is crucial for possibly removing it from raw data by exploiting the housekeeping information of thermal sensors. 
The propagation of the temperature oscillations through the focal plane and the instrument response to thermal changes were characterized through two main tests performed during CSL system level tests and verified in the latest part of the CPV (instruments already tuned, in nominal configuration):
the thermal dynamic response \cite{tomasi2009} and the radiometers thermal susceptibility \cite{terenzi2009b}. 

\subsubsection{Dynamic thermal verification}

\begin{figure}[htb!]
      \begin{center}    
            \includegraphics[width=12.5cm]{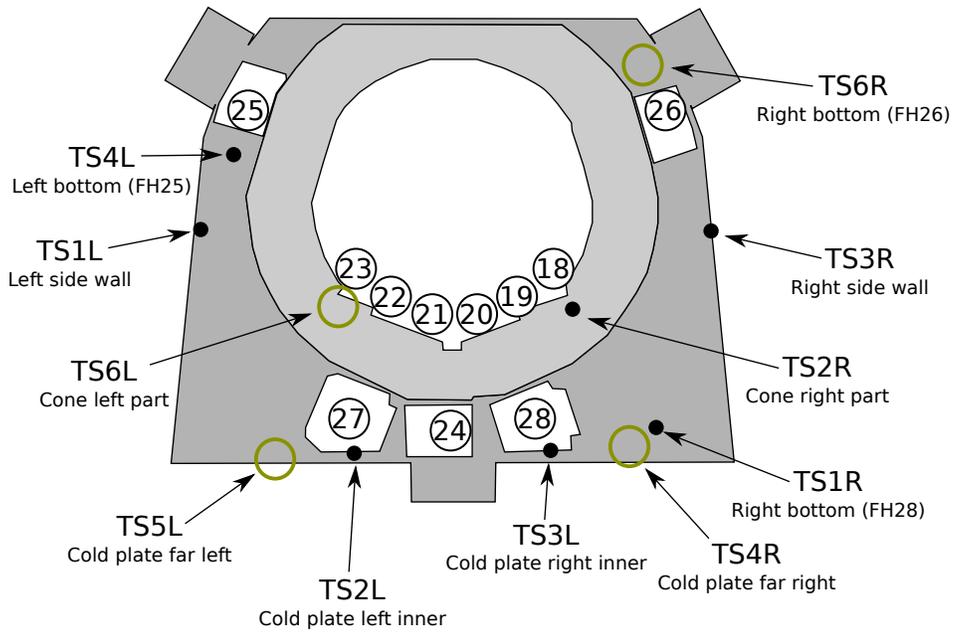} 
            \caption{Schema of the focal plane sensor locations. Sensor TS5R {\em FH28 Flange}, not included here, is located on the flange of LFI28 horn.}
            \label{LFI_sens}
      \end{center}  		
		\end{figure}

\begin{figure}[htb!]
      \begin{center}
   \includegraphics[width=12.5cm]{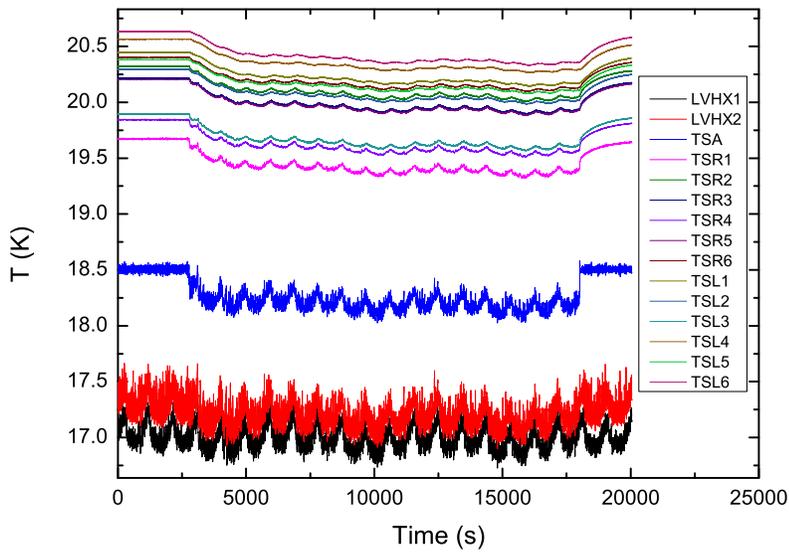}
            \caption{Dynamic behaviour of the sensors when TSA control is switched off; fluctuations at 1~mHz are reduced more than 40\%.}
            \label{TH_Dyn}
      \end{center} 
		\end{figure}

This test was aimed at measuring the dynamic thermal behaviour of the LFI Focal Plane; it started on July 29$^{\rm th}$ at 6:00 UTC with a duration of four hours. 
In order to amplify the effect and to get a more accurate measurement, the active control from the TSA was switched off. The resulting increased fluctuations, propagating at the cooler frequencies, were used to evaluate transfer functions between the TSA stage and the FPU sensors (Figures~\ref{LFI_sens} and \ref{TH_Dyn}). The analysis, conducted following the two methods reported in \cite{tomasi2009}, produced damping factors of 2--5 at about 1~mHz .

The source of fluctuations was characterized by two typical periods of the sorption cooler during the final CPV phase: (i) the single bed cycle time, 940~s, (ii) the complete cooler period, six times larger, 5640~s.

The first analysis method consisted in performing a Discrete Fourier Transfom (DFT) of the sensor timestreams.
The second method consisted in fitting the sensor timestreams with a double sinusoidal function:
	${\rm T(t) = T_0 + A1\cdot sin(2\pi\nu_1 t+ \phi_1) + A2\cdot sin(2\pi\nu_2 t+ \phi_2)}$
where ${\rm \nu_1}$ = 1/940 Hz and ${\rm \nu_2}$ = 1/5640~Hz. In both methods transfer functions were evaluated by the ratio of the resulting amplitudes and by the phase differences at the considered frequencies.

\begin{figure}[h!]
        \begin{center}
            \includegraphics[width=11.2cm]{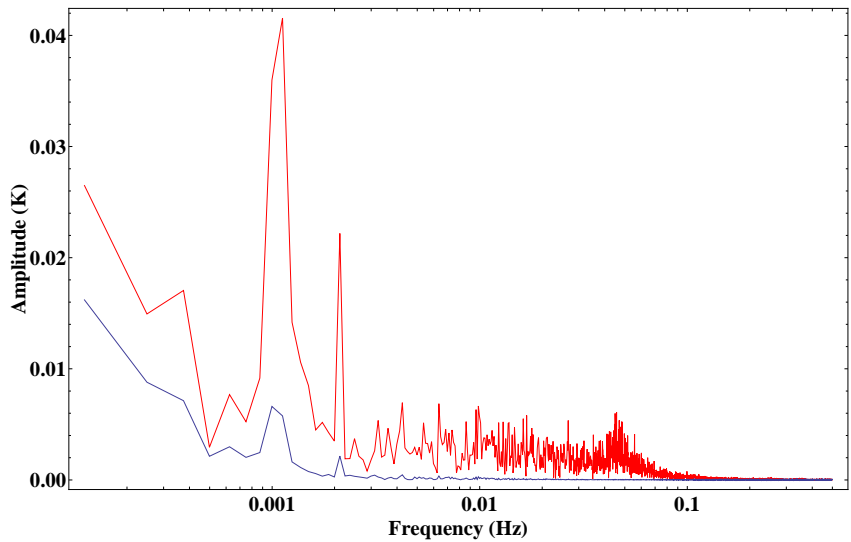}
            \caption{Fourier transform comparison of TSA sensor (red line) with one of the sensors in the left side of the focal plane, TS2L (blue line). }
            \label{t_fits}
        \end{center}            
		\end{figure}

\begin{figure}[h!]
                \begin{center}
            \includegraphics[width=7.0cm]{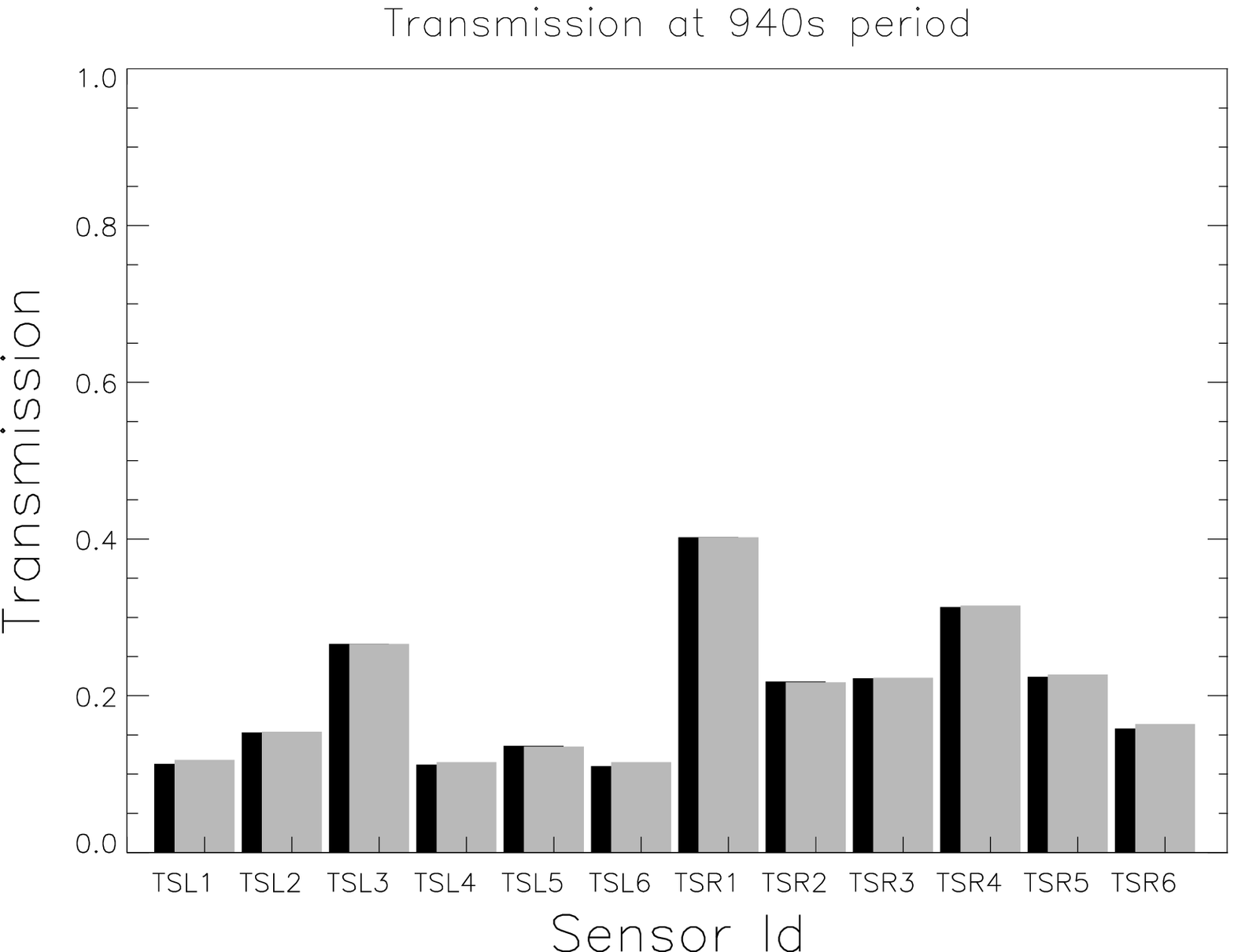} 
                        \hspace{0.5cm}
            \includegraphics[width=7.0cm]{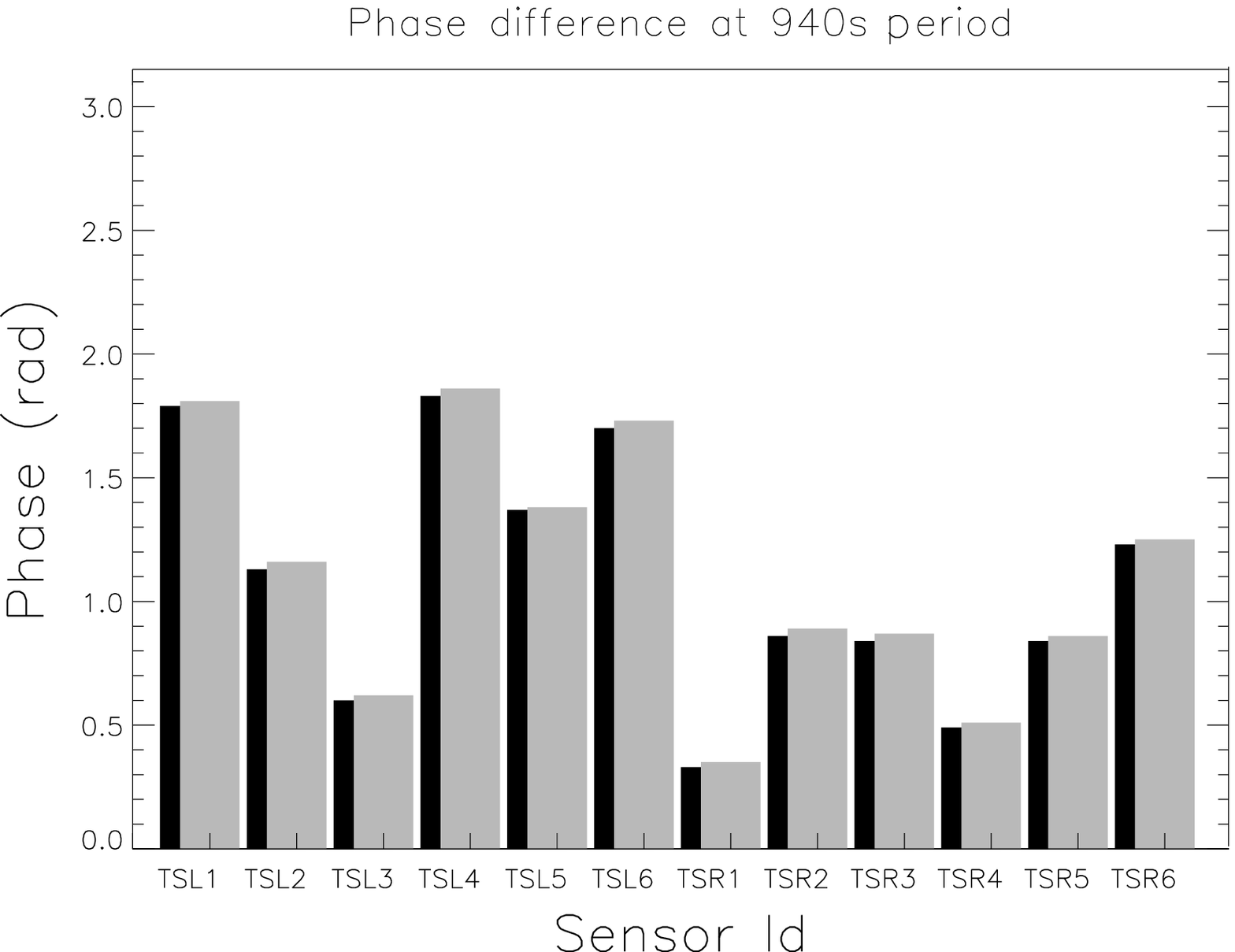}\\
                        \vspace{0.2cm}
            \includegraphics[width=7.0cm]{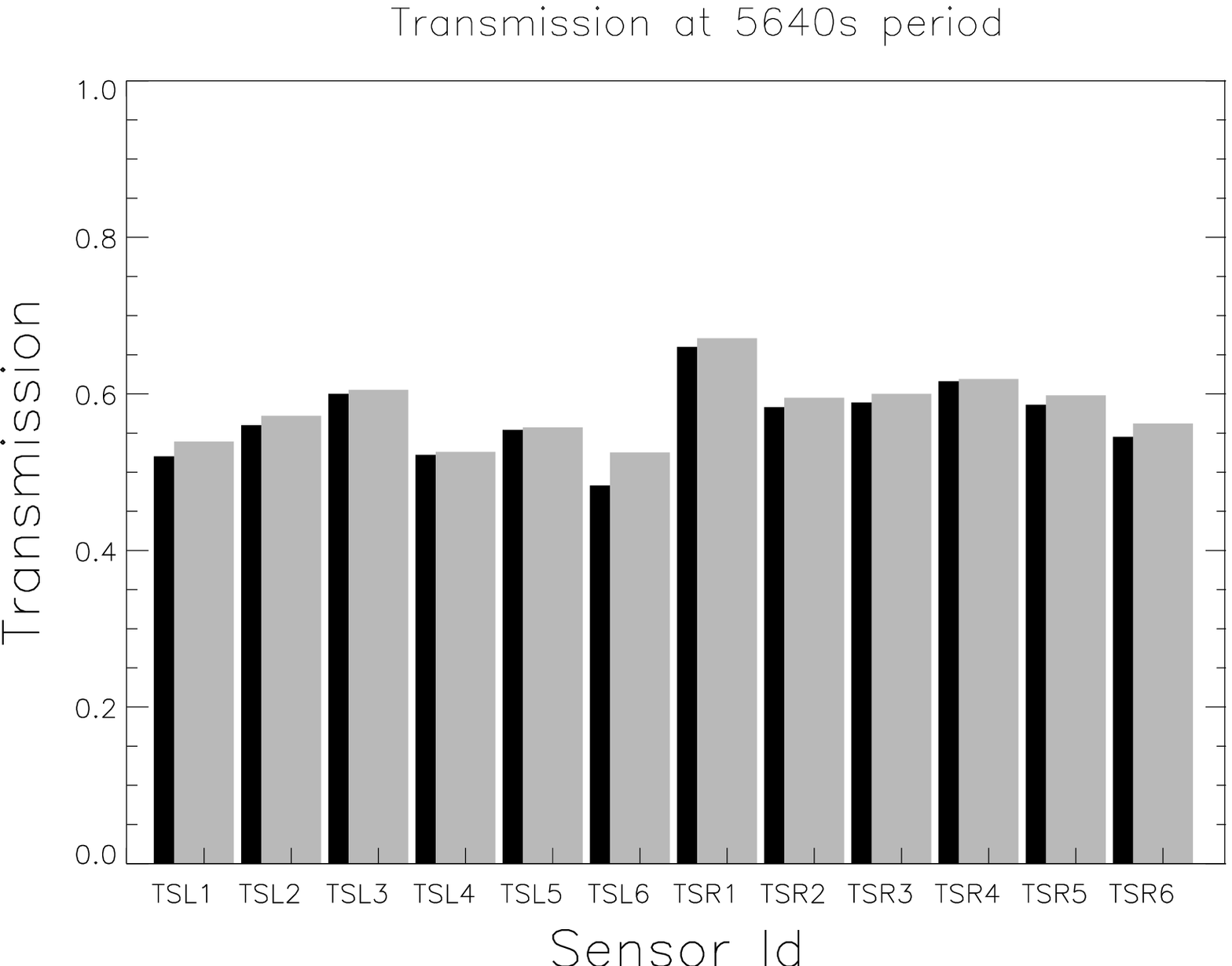}
                        \hspace{0.5cm}
            \includegraphics[width=7.0cm]{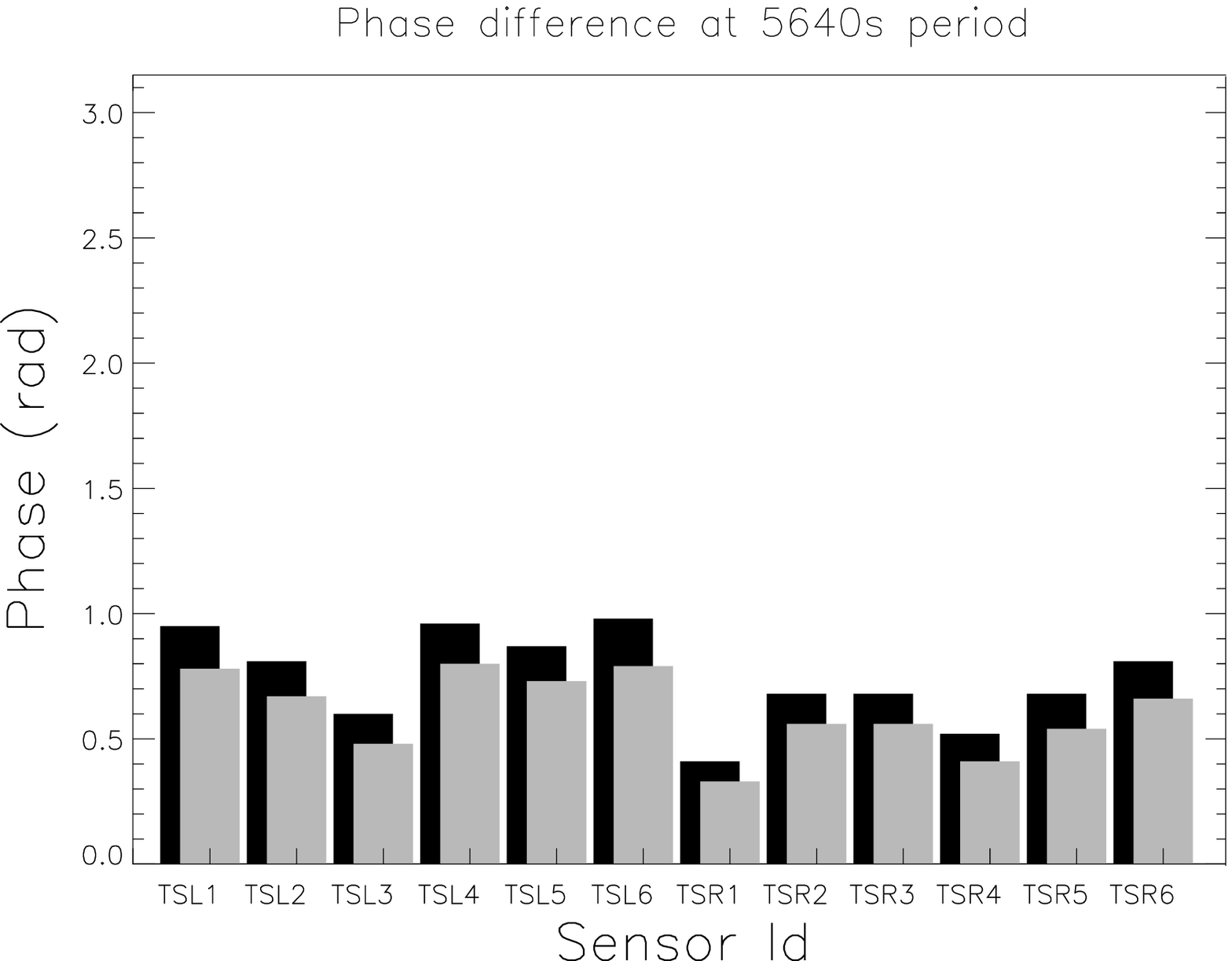}\\
\vspace{0.5cm}
            \caption{Comparison between CPV (black bars) and CSL (grey bars) measurements of amplitude (left panels) and phase (right panels) of the transfer functions for the single bed fluctuation period of 940~s (top panels) and for the whole cooler fluctuation period of 5640~s (bottom panels).}
            \label{CSL_Comp}
                \end{center}	
		\end{figure}

Figure~\ref{t_fits} shows the time stream available for the analysis in the frequency domain: this method did not allow to efficiently sample the lowest frequency peak and a  comparison between the two methods is possible only for the 940~s component.
Results from both methods are in agreement, as shown in the Table~\ref{dyn_res} reporting data from the sinusoidal fit together with relative differences between the two methods.
Sorting the sensors by the transfer function amplitudes in descending order (second column of the table), the route of the propagation of temperature fluctuations through the focal plane sensors (shown in Figure~\ref{LFI_sens}) was reproduced as expected: the largest amplitudes are in the sensor closest to the right bottom corner interface with the working cooler and they decrease in the direction left upwards.  
The measured values in flight are compared to what measured during the ground test at satellite level in Figure~\ref{CSL_Comp}, showing good agreement. The 20\% difference in the phase measured at the time scale of 5640~s is coherent with a larger uncertainty in the fit for a dataset including just about 1.5 times the fluctuation period (about two hours of data were used). 

\begin{table}
	\caption{Transmission amplitudes and phases of fluctuations at the two main cooler frequencies during the transient test. Results are taken from the double sinusoidal fit method (see text and references). Typical fit parameters uncertainties are at 1\% level. In the last two columns relative differences between the results of the two methods at 1~mHz are reported.}
 \begin{tabular}{l c c c c c c}
 \hline \hline
	& \multicolumn{2}{c}{\texttt{\small Results 1.064 mHz}}	& \multicolumn{2}{c}{\texttt{\small Results 0.177 mHz}} & \multicolumn{2}{c}{\texttt{\small Compare $\sim$1 mHz}} \\
Sensor ID	&  Ampl. & Phase (rad)	&  Ampl. 	& Phase (rad) & rel. ${\rm \Delta}$ Amp (\%) & rel. ${\rm \Delta \phi}$ (\%)\\
 \hline
TSL1	& 0.113	& 1.79	& 0.520	& 0.95 & 0 & 11\\
TSL2	& 0.153	& 1.13	& 0.560	& 0.81 & 9 &  7\\
TSL3	& 0.266	& 0.60	& 0.600	& 0.60 & 5 &  0\\
TSL4	& 0.112	& 1.83	& 0.522	& 0.96 & 1 & 11\\
TSL5	& 0.136	& 1.37	& 0.554	& 0.87 & 8 &  9\\
TSL6	& 0.110	& 1.70	& 0.483	& 0.98 & 2 & 12\\
TSR1	& 0.402	& 0.33	& 0.660	& 0.41 & 2 &  3\\
TSR2	& 0.218	& 0.86	& 0.583	& 0.68 & 6 &  2\\
TSR3	& 0.222	& 0.84	& 0.589	& 0.68 & 6 &  1\\
TSR4	& 0.313	& 0.49	& 0.616	& 0.52 & 4 &  2\\
TSR5	& 0.224	& 0.84	& 0.586	& 0.68 & 6 &  1\\
TSR6	& 0.158	& 1.23	& 0.545	& 0.81 & 8 &  7\\
\hline
\end{tabular}
\label{dyn_res}
\end{table}

\subsubsection{Thermal susceptibility}

\begin{figure}[htb!]
            \begin{center}
            \includegraphics[width=13.0cm]{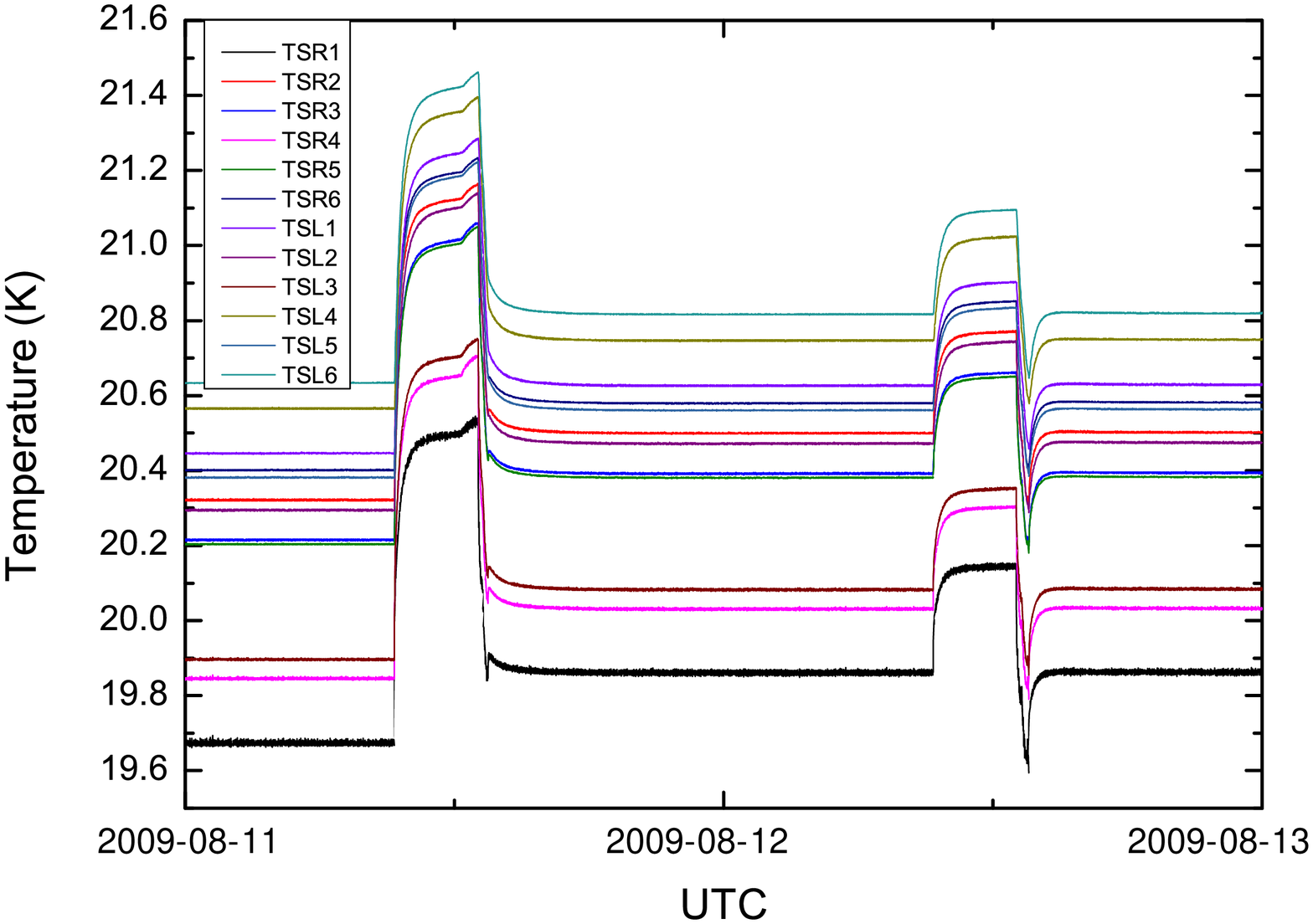}  \\ \vspace{-1.5cm}
            \hspace{-1.cm}  
            \includegraphics[width=13.3cm]{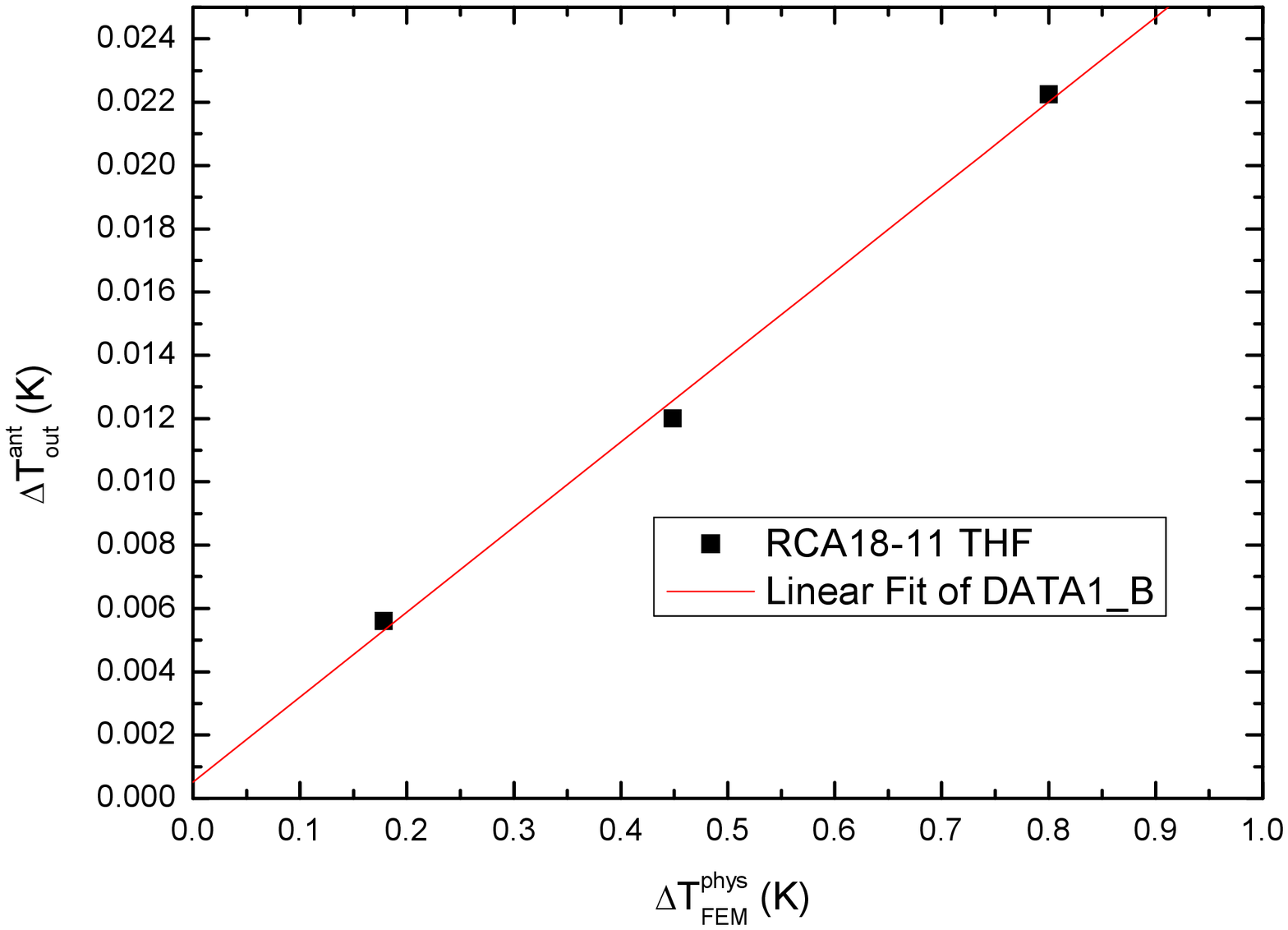}
            \caption{Top panel: overall view of the LFI sensors in the final part of the susceptibility test. Bottom panel: the slope of the fitted line is a measure of the signal variation as a function of changing focal plane temperature.}
            \label{Fit_THF}
                       \end{center} 
		\end{figure}

Fluctuations of the focal plane temperature would cause variations of important parameters (mainly the low noise amplifier gains and noise temperatures), impacting the radiometer output signal, as detailed in \cite{terenzi2009b}.
The response of the LFI radiometers to thermal fluctuations was estimated by inducing discrete temperature steps on the focal plane through TSA setpoint changes. The test started on August 11$^{\rm th}$ at 9:15 UTC. 
The setpoint was changed over four values (Figure~\ref{Fit_THF}, left) and after a stabilization of at least two hours, the measured receivers output was characterized as a function of each temperature variation of about 0.3~K. Due to reduced duration of the DTCP, the temperature steps were performed during two consecutive visibility periods. At the beginning of the first DTCP, a major temperature step up ($\Delta T \approx +0.8$~K) was performed while the TSA setpoint was decreased of a step downwards just before the loss of visibility with the spacecraft. in order to guarantee that the heat load on the sorption cooler interface was at the lowest possible level during the long period of 20 hours with no direct monitoring.

The slope of the resulting ${\rm T_{ant}}$ vs ${\rm T_{phys}}$ plot is the measured response of the receivers to a change in the temperature (Figure~\ref{Fit_THF}, bottom).
Results, reported in the Table~\ref{THF_res}, confirmed that physical temperature fluctuations in the main frame are furtherly reduced when convolved with the radiometer thermal susceptibility coefficients: the derived output fluctuations, measured in antenna temperature, were actually reduced by an extra factor of 10 to 200 (according to the channel considered), of the same order of ground test results \cite{terenzi2009b}. This corresponds to reduce the mean peak-to-peak amplitudes of  fluctuations measured by high resolution sensors, of the order of 4~mK in steady condition, of at least one order of magnitude in the output timestream.

\begin{figure}[h!]
                \begin{center}  
            \hspace{-0.5cm}            
            \includegraphics[width=8.32cm]{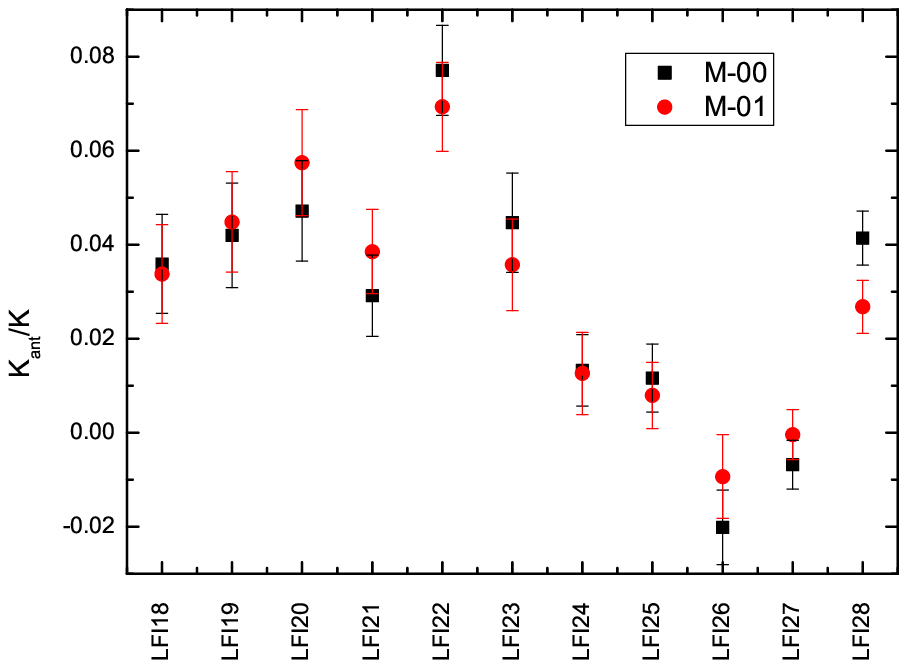} 
            \hspace{-1.3cm}            
            \includegraphics[width=8.32cm]{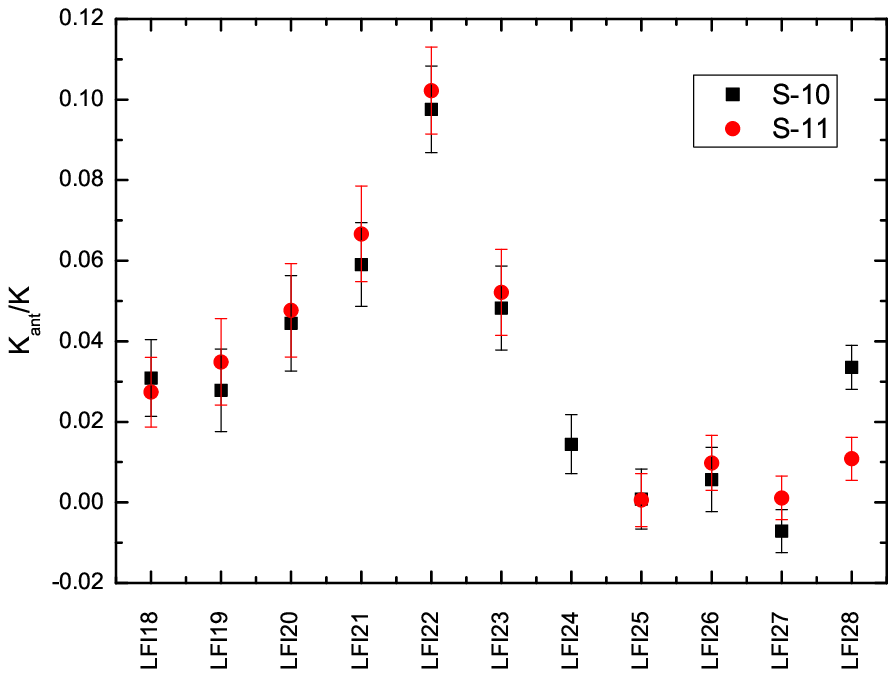}\\
            \caption{Results of the susceptibility test for the RCAs \texttt{main} arm (left panel) and \texttt{side} arm (right panel) radiometers.}
            \label{Results_THF}
                \end{center}
		\end{figure}

Figure~\ref{Results_THF} shows the same results in two separate plots, one for each radiometer \texttt{Main} and \texttt{Side}. Values from coupled detectors are in general in line with the above considerations: the main parameters affected by the temperature changes are noise and gain of the amplifier chain, which is shared by the two coupled diodes of each radiometer. \texttt{LFI28} showed a different behaviour: the calibrated susceptibilities measured for the two coupled diodes are non consistent within the error bars. It is worth noting that \texttt{LFI28} shows very peculiar features in terms of high $r$  (see eq.~\ref{eq_r_v}) factors and asymmetry between the paired detectors: even if the above inconsistency can not be univocally related to these features, it is likely that they are contributing to generate it. 

\begin{table}
	\caption{Results of thermal susceptibility test. Units are K$_{\rm ant}$/K$_{\rm phys}$. Together with the radiometers, the closest sensor used as reference for the physical temperature is indicated. Results for the detector \texttt{LFI24S-11} are missing because it suffered a period of signal saturation during the test.}
 \begin{tabular}{l c c c c}
\hline \hline
Radiometer (sensor)	& \texttt{M-00} & \texttt{M-01} & \texttt{S-10} & \texttt{S-11}\\
\hline
LFI18 (TS2R)	& 0.040	$\pm$ 0.010 & 0.030	$\pm$0.010	&  0.031	$\pm$ 0.009 & 0.027 $\pm$ 0.009\\
LFI19 (TS2R)	& 0.040	$\pm$ 0.010 & 0.050	$\pm$0.010	&  0.030	$\pm$ 0.010 & 0.040 $\pm$ 0.010\\
LFI20 (TS2R)	& 0.050	$\pm$ 0.010 & 0.060	$\pm$0.010	&  0.040	$\pm$ 0.010 & 0.050 $\pm$ 0.010\\
LFI21 (TS6L)	& 0.029	$\pm$ 0.009 & 0.039	$\pm$0.009	&  0.060	$\pm$ 0.010 & 0.070 $\pm$ 0.010\\
LFI22 (TS6L)	& 0.080	$\pm$ 0.010 & 0.069	$\pm$0.009	&  0.100	$\pm$ 0.010 & 0.100 $\pm$ 0.010\\
LFI23 (TS6L)	& 0.050	$\pm$ 0.010 & 0.040	$\pm$0.010	&  0.050	$\pm$ 0.010 & 0.050 $\pm$ 0.010\\
LFI24 (TS2L)	& 0.013	$\pm$ 0.008 & 0.013	$\pm$0.009	&  0.014	$\pm$ 0.007 & --    \\
LFI25 (TS4L)	& 0.012	$\pm$ 0.007 & 0.008	$\pm$0.007	&  0.001	$\pm$ 0.007 & 0.001 $\pm$ 0.007\\
LFI26 (TS6R)	& -0.020$\pm$ 0.008 & -0.009$\pm$	0.009 &  0.006	$\pm$ 0.008 & 0.010 $\pm$ 0.007\\
LFI27 (TS2L)	& -0.007$\pm$ 0.005 & 0.000	$\pm$0.005	&  -0.007	$\pm$ 0.005 & 0.001 $\pm$ 0.005\\
LFI28 (TS5R)	& 0.041	$\pm$ 0.006 & 0.027	$\pm$0.006	&  0.034	$\pm$ 0.005 & 0.011 $\pm$ 0.005\\
\hline
\end{tabular}
\label{THF_res}
\end{table}

\clearpage

%% file: 06_conclusions.tex
We have summarized the overall CPV plan of \Planck-LFI, designed to ensure proper in-flight verification and tuning of all the key element of the instrument. The plan was designed taking into account the constraints and the opportunities provided by the gradual cool-down of the various interfaces of the instrument assembly after launch. 
The functionality of the key LFI subsystems, as well as of the instrument as a whole, was repeatedly verified in-flight in a wide range of conditions, including and expanding over those covered by ground tests. Spurious 1-Hz spikes and I-V response of the LNAs were measured for all radiometers and compared with results from ground tests. The results show good repeatability of the instrument behaviour.

A rather sophisticated tuning procedure was operated on the main LFI subsystems to maximise the instrument scientific performance. We exploited the cool-down of the 4K stage to calculate noise temperature and isolation in a large parameter space of LNA biases (the so-called hypermatrix tuning) and identified optimal parameter sets. We also tuned  phase switch currents to reach optimal balance. We tuned the DAE gain and offset to optimise the voltage input at the ADC for each LFI channel. Finally, we set the REBA parameters to optimise the tradeoff between digital quantisation and compression efficiency.

As soon as the instrument was operational, and well before the beginning of the first light survey, we used the CMB dipole signal to routinely perform a first-order photometric calibration, which allowed us to monitor the main instrument performance (white noise, 1/f noise) during CPV. As expected, the photometric calibration and noise parameters were somewhat different from those measured on the ground. However, accounting for the somewhat different thermal and electrical interfaces, the agreement is very good. 

A crucial aspect in the understanding of the LFI behaviour is the sensitivity of the radiometers response to changes in the thermal interfaces. In particular, we have extensively measured the transfer functions coupling the radiometric response to temperature variations in the focal plane by switching off the stabilisation PID of the 20~K stage. This measurement provides a key informaiton for analysis of systematic effects. While thermal tests were carried out also on-ground, the stability of the space environment in L2 allowed us to obtain much more accurate results in the CPV campaign.

Overall the CPV campaign has shown excellent consistency with the results of ground tests, and in general it has yielded improved accuracy. All the calibration activities were performed successfully and yielded very satisfactory results leading to a slightly improved LFI performance compared to ground tests. 

%% file: a04_groups_table.tex
\section{Channel grouping tables}
\label{app_power_groups_table}

\begin{table}[h!]
    \begin{center}
        \caption{\label{tab_power_groups_cryo01} Grouping of LFI channels in power groups: channels belonging to each power group share the same power box. This grouping has been used during \texttt{CRYO-01} test. }
        \vspace{.2cm}
        \begin{tabular}{c c }
            \hline
            \hline
            Power Group & \texttt{Channels}  \\
            \hline
                1   & LFI 18, LFI 26 \\
                2   & LFI 19, LFI 20, LFI 28\\
                3   & LFI 21, LFI 22, LFI 24, LFI 27\\
                4   & LFI 23, LFI 25\\
            \hline
        \end{tabular}
    \end{center}
\end{table}

 \begin{table}[h!]
     \begin{center}
         \caption{\label{tab_IV_groups}  Grouping of LFI Channels used during the drain current test. }
         \vspace{.2cm}
         \begin{tabular}{c c }
             \hline
             \hline
             Power Group & \texttt{Channels}  \\
             \hline
                 1   & LFI 18, LFI 21 \\
                 2   & LFI 19, LFI 22\\
                 3   & LFI 20, LFI 23\\
                 4   & LFI 25, LFI 24\\
                 5   & LFI 26, LFI 27\\
                 6   & LFI 28\\
             \hline
         \end{tabular}
     \end{center}
 \end{table}

\begin{table}[h!]
    \begin{center}
        \caption{\label{tab_LNAs_Tuning_groups} Grouping of LFI channels used during LNAs hypermatrix pre-tuning and proper tuning.}
        \vspace{.2cm}
        \begin{tabular}{c c }
            \hline
            \hline
            Power Group & \texttt{Channels}  \\
            \hline
                1   & LFI 18, LFI 19, LFI 22 \\
                2   & LFI 20, LFI 21, LFI 23\\
                3   & LFI 24, LFI 26, LFI 28\\
                4   & LFI 25, LFI 27\\
            \hline
        \end{tabular}
    \end{center}
\end{table}

\clearpage

%% file: a02_cryo01_sequence.tex
\section{Sequence of steps in CRYO-01 functionality tests}
\label{app_cryo01_sequence}

    \begin{table}[h!]
        \begin{center}
            \caption{Sequence of steps performed during the \texttt{CRYO-01} test.}
            \label{tab_cryo01_sequence}
            \begin{small}
            \begin{tabular}{ l  p{6cm}  | p{6cm} }
                \hline \hline
                \small
                \textbf{Step} &\texttt{Description}    &\texttt{Expected response}   \\
                \hline
                1       &The first ACA is biased with LNAs and phase switch biases ($V_{\rm g1}$, $V_{\rm g2}$, $V_{\rm d}$, $I_1$, $I_2$)  
                        &The signal rises on the pair of output diodes connected to the ACA$^\dagger$     \\
                \hline
                2       &The phase switch polarisation is changed       
                        &The signal switches from sky to reference load or vice versa \\
                \hline
                3       &The 4 kHz switching is activated       
                        &From each diode two output streams appear with different voltage outputs. \\
                \hline
                4       &The phase switch bias current $I_1$ is lowered    
                        &The separation between the two datastreams changes (either they converge or they diverge) \\
                \hline
                5       &The phase switch bias current $I_1$ is returned to its nominal value and $I_2$ is lowered    
                        &The separation changes in the opposite direction          \\
                \hline
                6       &$I_2$ is returned to its nominal value, the 4 kHz switching is turned off and the phase switch state is returned to 0     
                        &The signal returns as after step 1$^\ddagger$        \\
                \hline
                7       &The second ACA is biased with LNAs and phase switch biases ($V_{\rm g1}$, $V_{\rm g2}$, $V_{\rm d}$, $I_1$, $I_2$) 
                        &The output voltage increases in both diodes$^\dagger$  \\
                \hline
                8       &The phase switch polarisation is changed       
                        &The signal switches from sky to reference load or vice versa \\
                \hline
                9       &The 4 kHz switching is activated       
                        &From each diode two output streams appear with different voltage outputs. \\
                \hline
                10      &The phase switch bias current $I_1$ is lowered    
                        &The separation between the two datastreams changes (either they converge or they diverge) \\
                \hline
                11      &The phase switch bias current $I_1$ is returned to its nominal value and $I_2$ is lowered    
                        &The separation changes in the opposite direction         \\
                \hline
                12      &$I_2$ is returned to its nominal value, the 4 kHz switching is turned off and the phase switch state is returned to 0     
                        &The signal returns as after step 1$^\ddagger$        \\
               
                \hline
            \end{tabular}
            \end{small}
            
        \end{center}
        \begin{small}
        \noindent $^\dagger$ Drain current is recorded and compared to ground test CSL values.  \\
        \noindent $^\ddagger$ Here the output voltage can be different because of drifts caused by the thermal environment and amplifier stabilisation.  
        \end{small}
        
    \end{table}

\clearpage

%% file: a01_spike_plots.tex
\section{Plots of 1~Hz frequency spikes}
\label{app_spike_plots}

\begin{figure}[htb]
        \begin{center}
            \includegraphics[width=14cm]{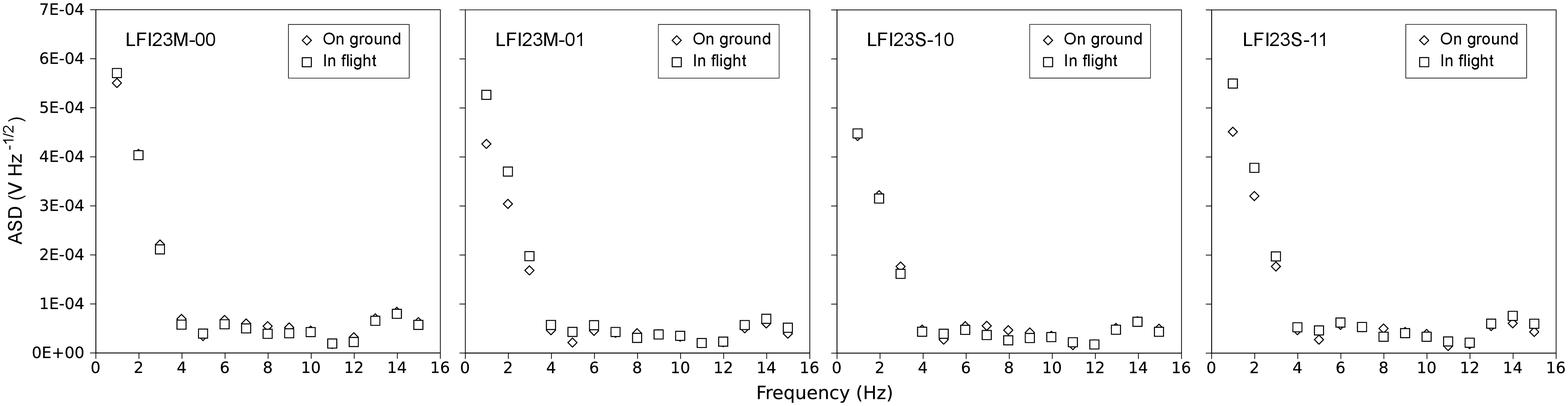}\\
            \includegraphics[width=14cm]{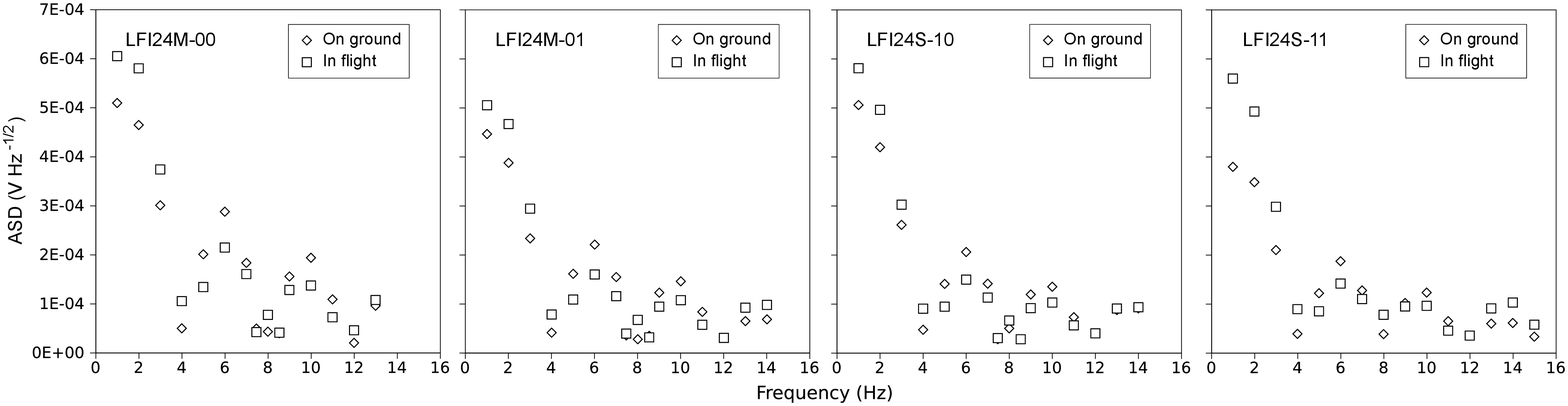}\\
            \includegraphics[width=14cm]{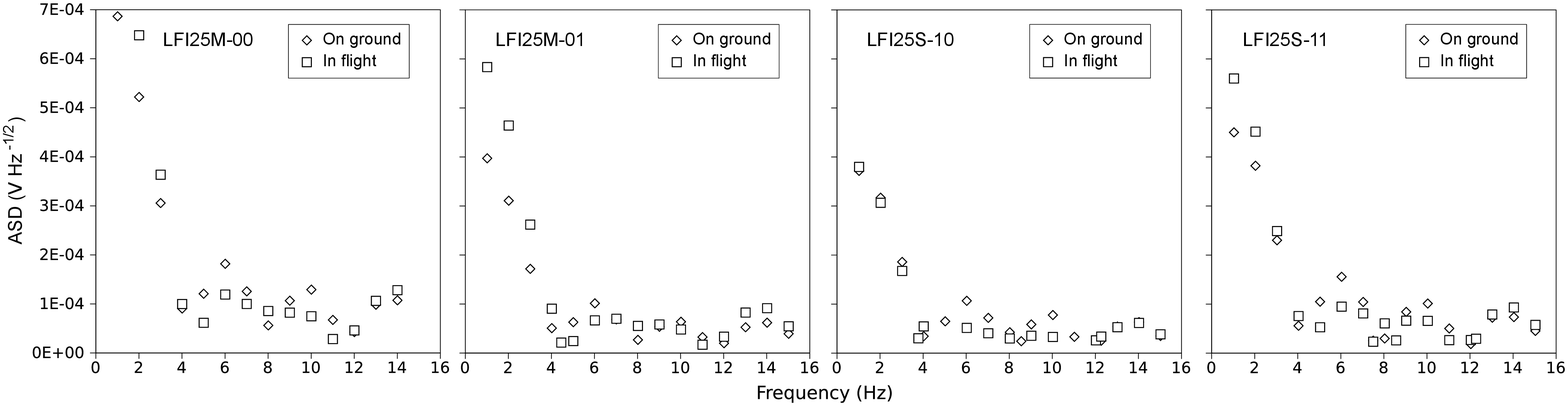}\\
            \includegraphics[width=14cm]{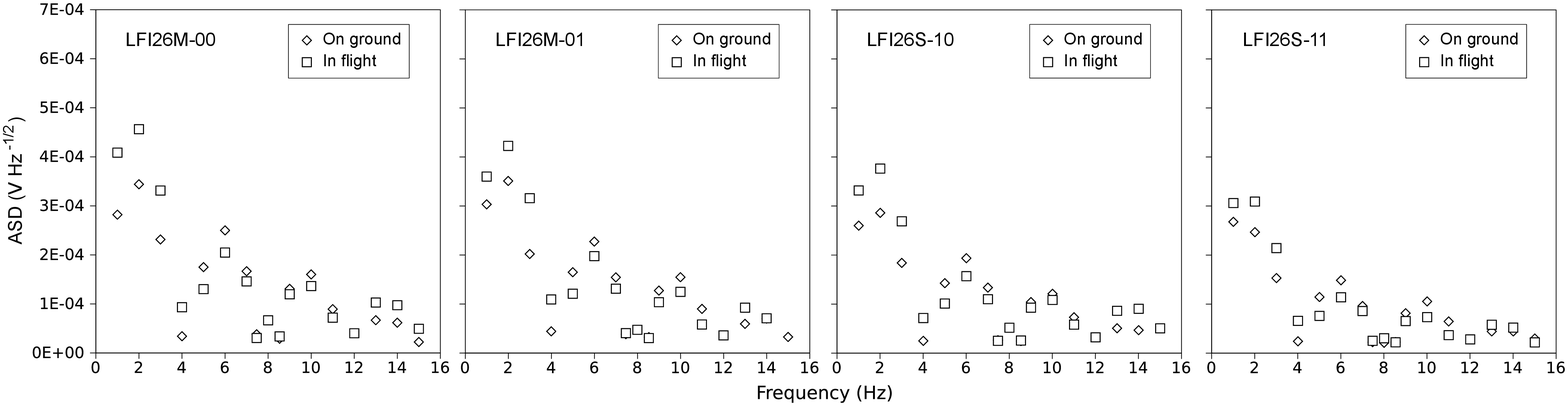}\\
            \includegraphics[width=14cm]{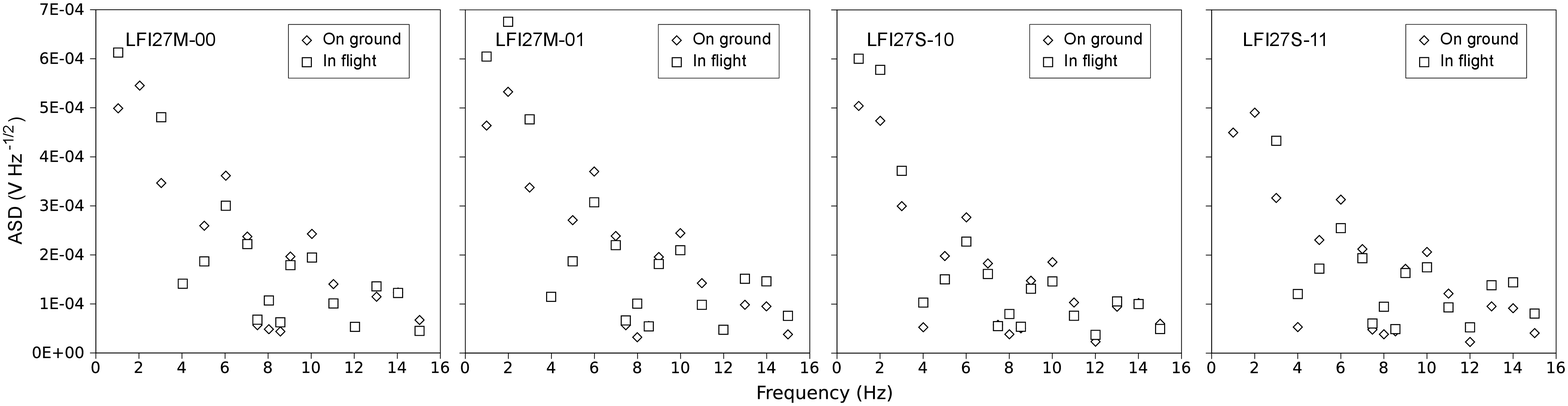}\\
            \includegraphics[width=14cm]{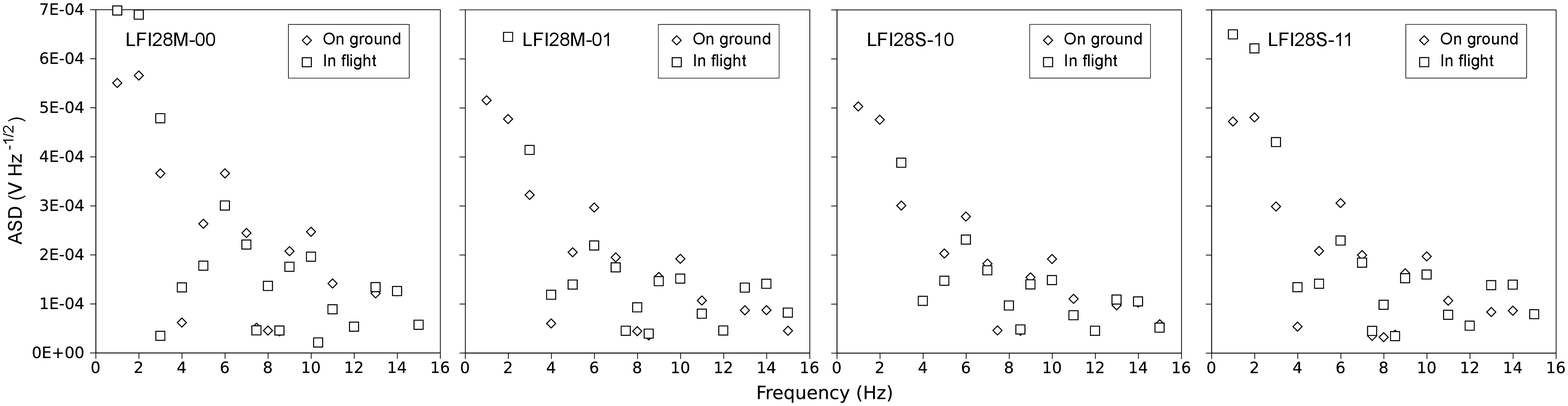}\\
        \end{center}
        \caption{ASD of 1~Hz frequency spikes detected in the warm electronics noise and measured on-ground and in-flight with the radiometers left unbiased. All 44 output channels are shown. Two anomalous spikes were detected in the two outputs \texttt{LFI22M-00} and \texttt{LFI22M-01}.}
        \label{fig_spikes}
    \end{figure}

    In Figure~\ref{fig_spikes} we show the amplitude spectral density of 1~Hz frequency spikes in all the LFI outputs measured during on-ground and in-flight tests with radiometers completely unbiased.
    \begin{figure}[htb]
        \begin{center}
            \includegraphics[width=14cm]{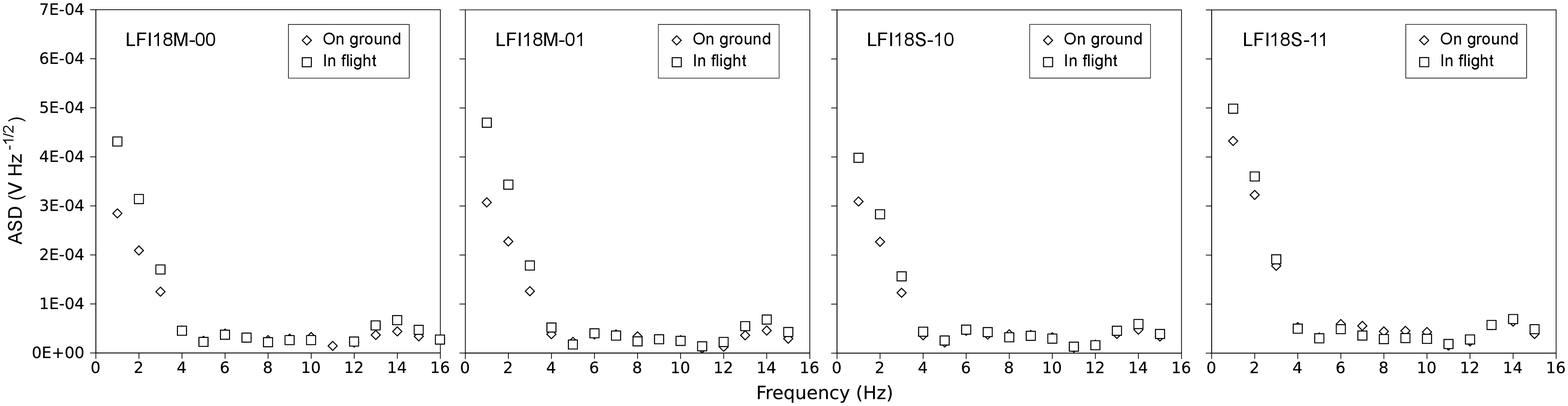}\\
            \includegraphics[width=14cm]{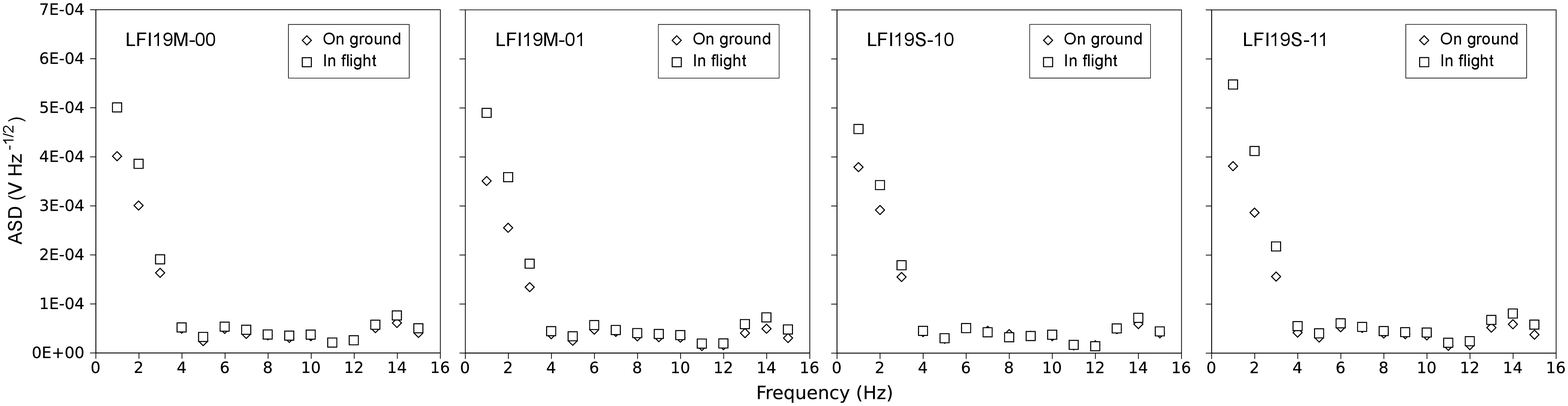}\\
            \includegraphics[width=14cm]{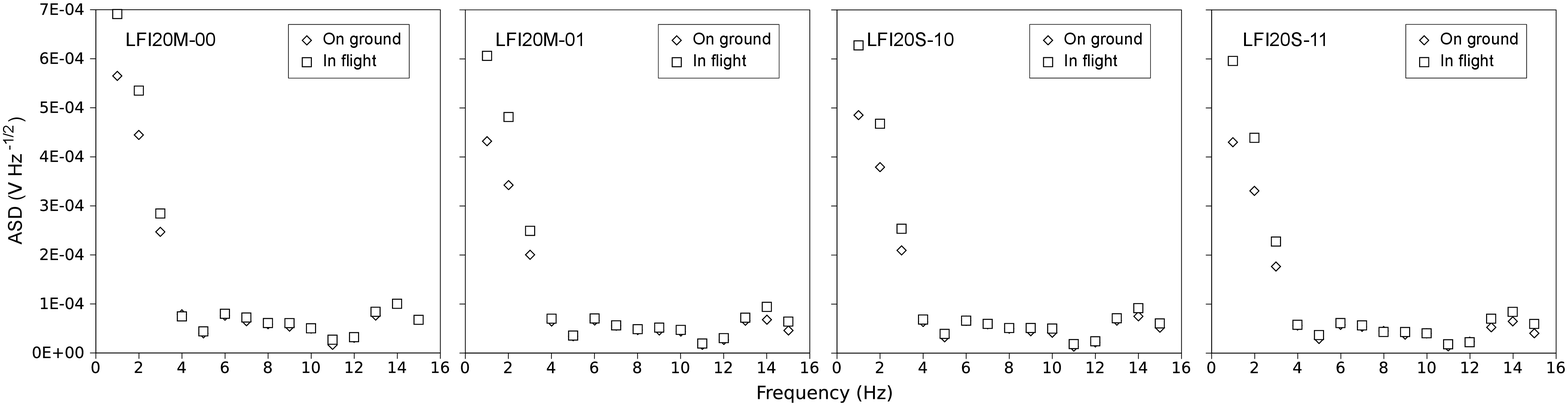}\\
            \includegraphics[width=14cm]{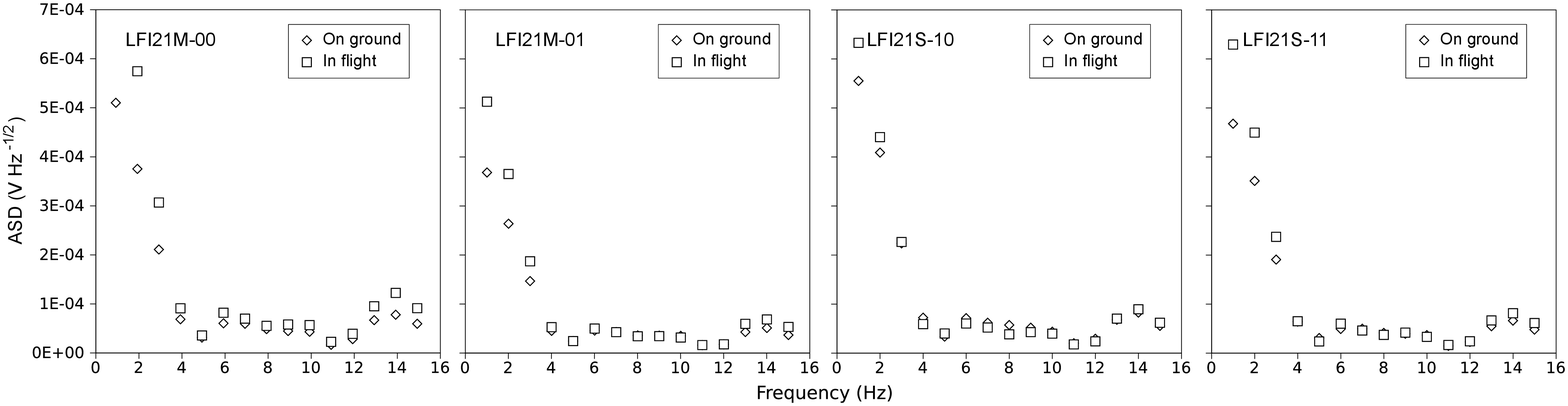}\\
            \includegraphics[width=14cm]{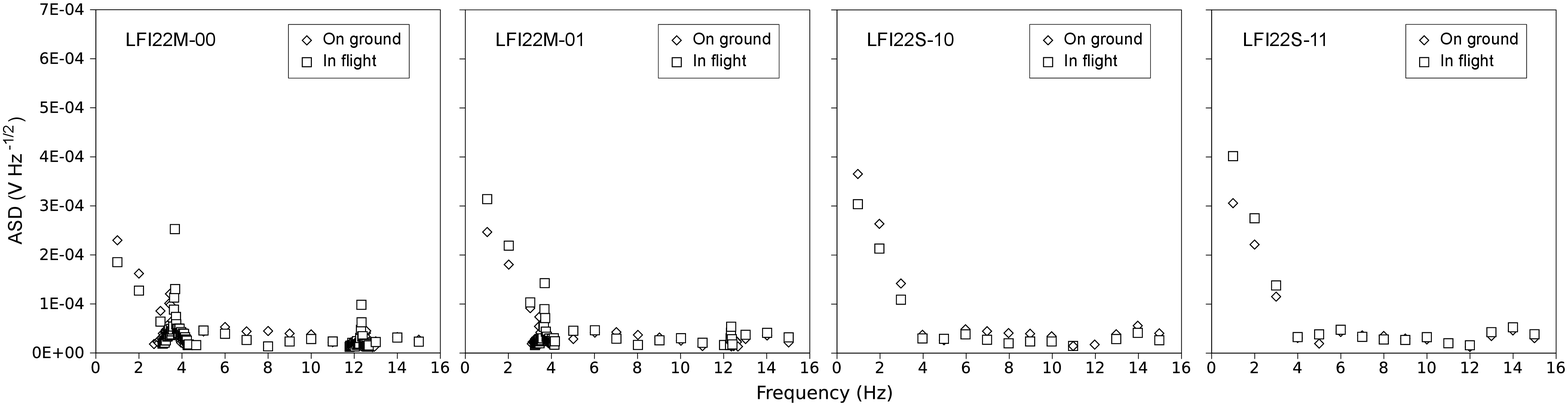}\\
        \end{center}
    \end{figure}

    \clearpage 

%% file: a03_drain_current.tex
\section{Plots of drain current}
\label{app_draincurr_plots}

\begin{figure}[htb]
        \begin{center} 
            \textbf{LFI-28}\\ \vspace{0.1cm}
            \includegraphics[width=7.0cm]{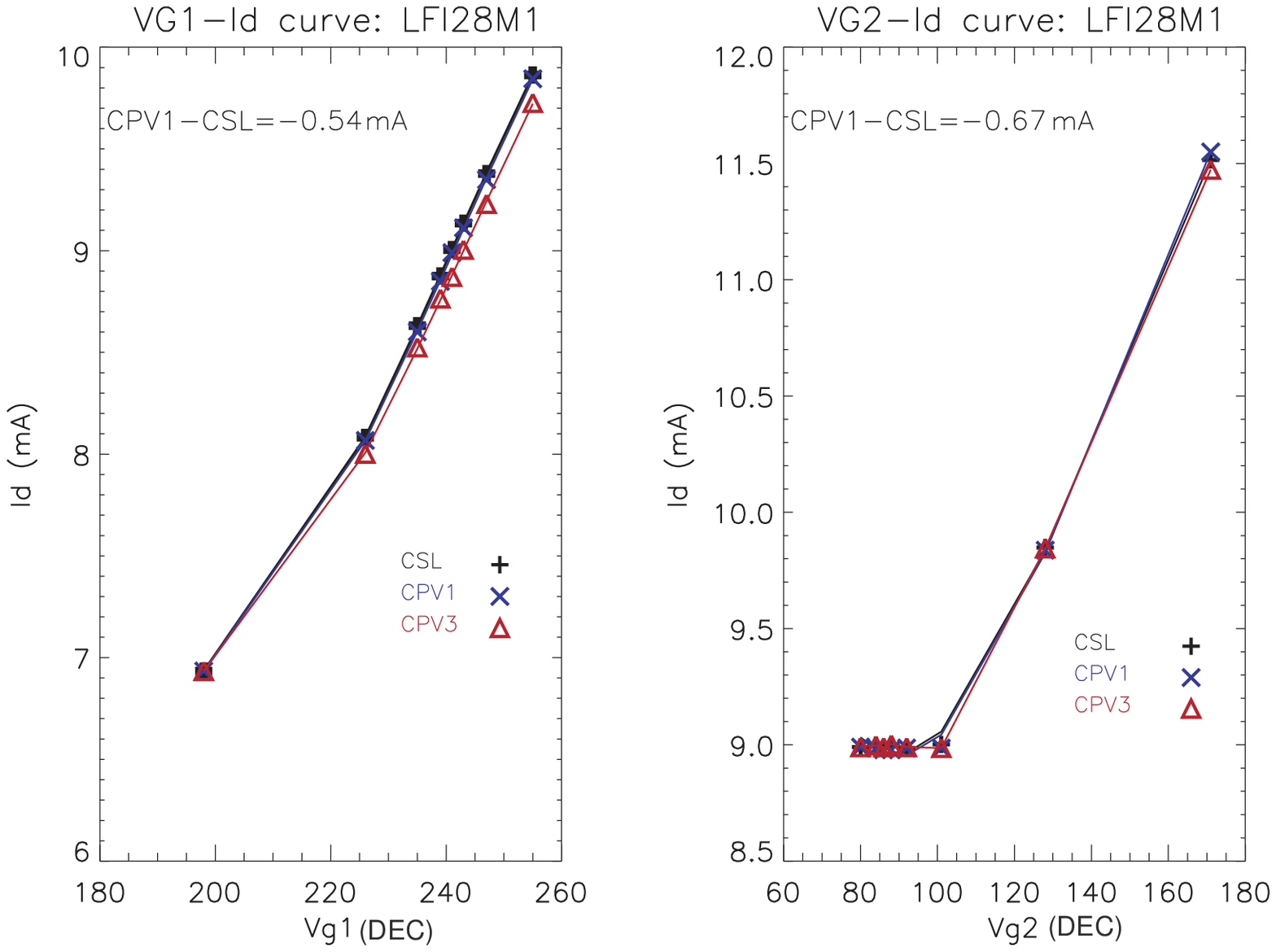}
            \includegraphics[width=7.0cm]{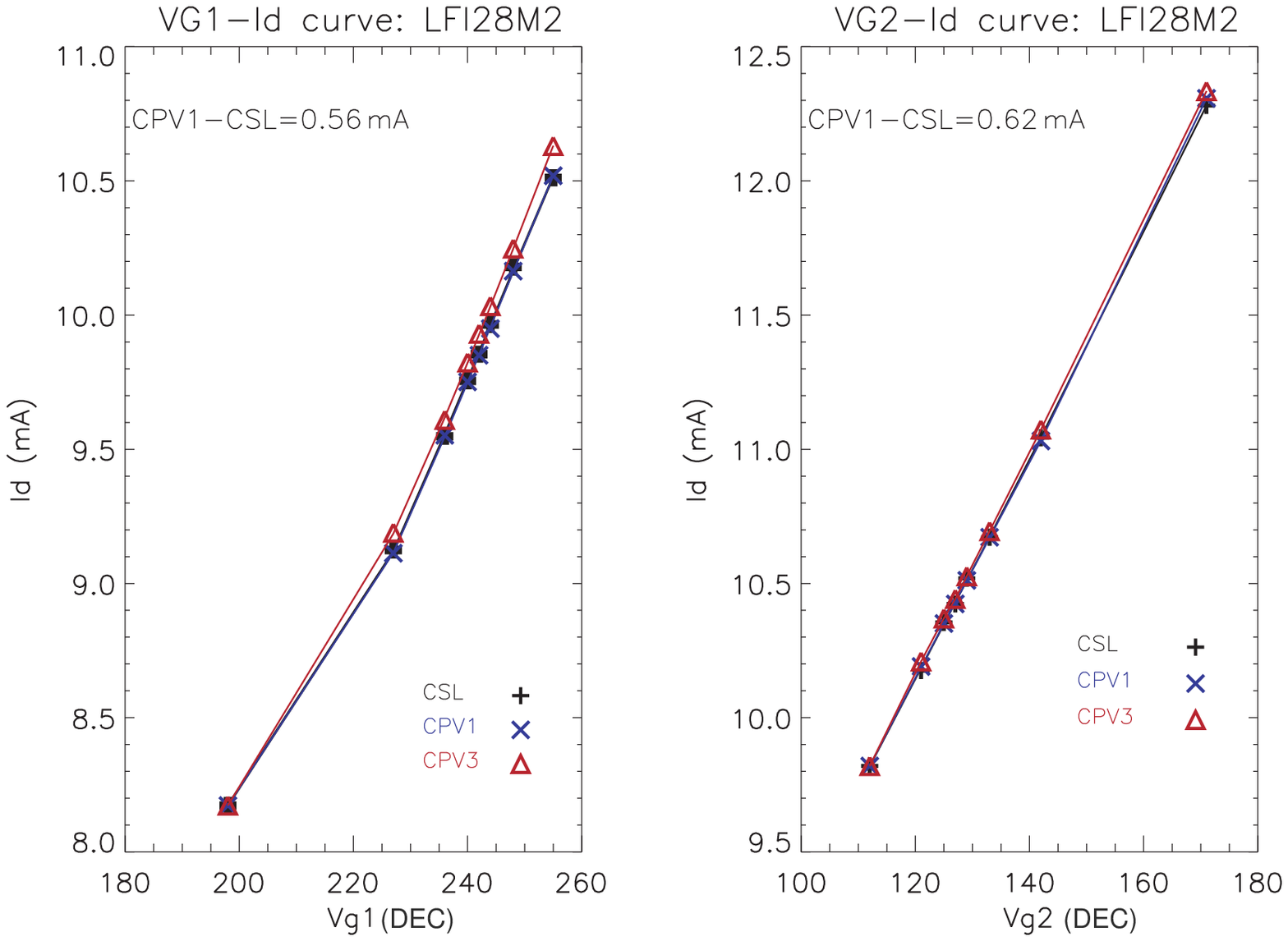}\\
            \includegraphics[width=7.0cm]{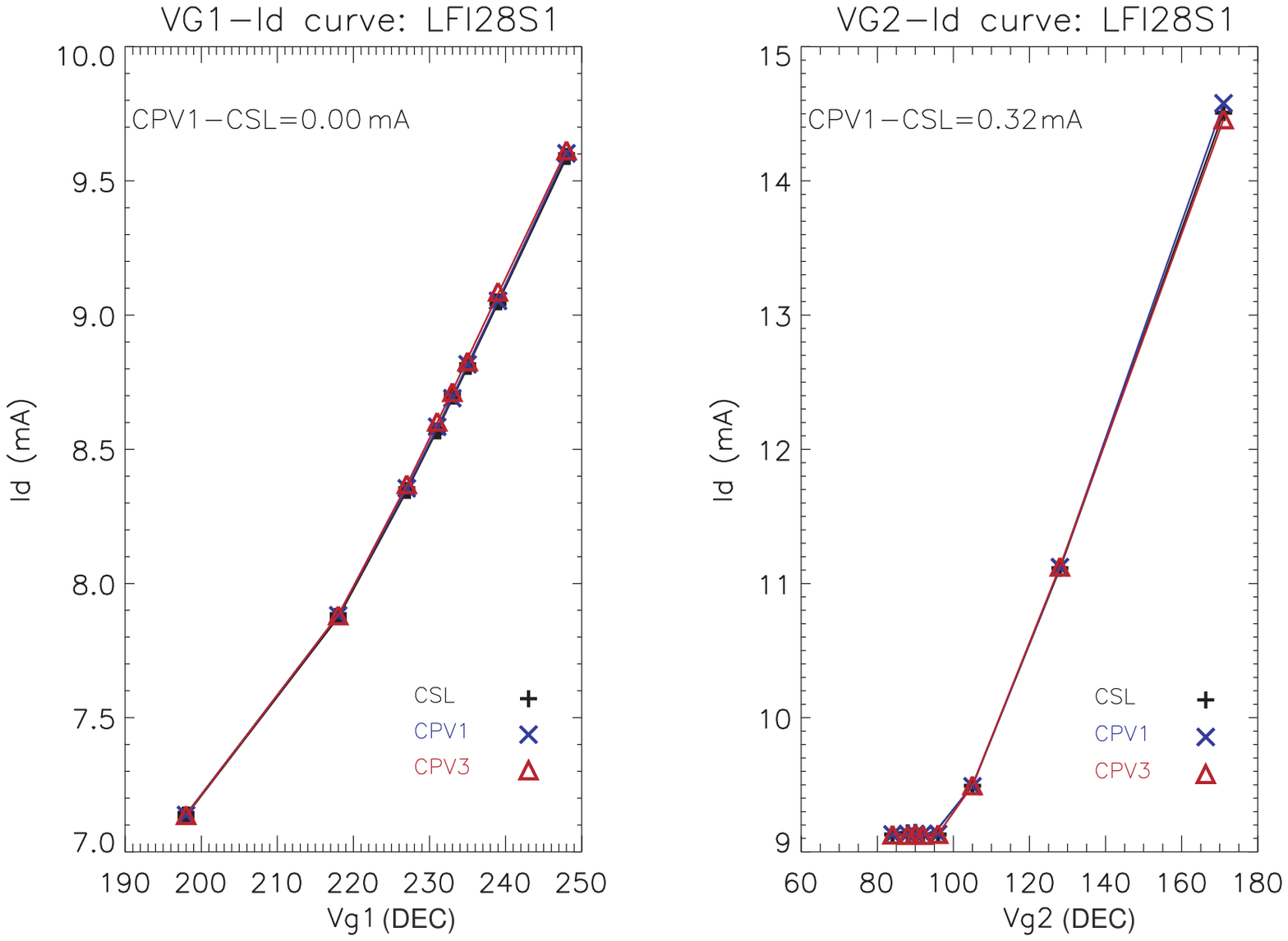}
            \includegraphics[width=7.0cm]{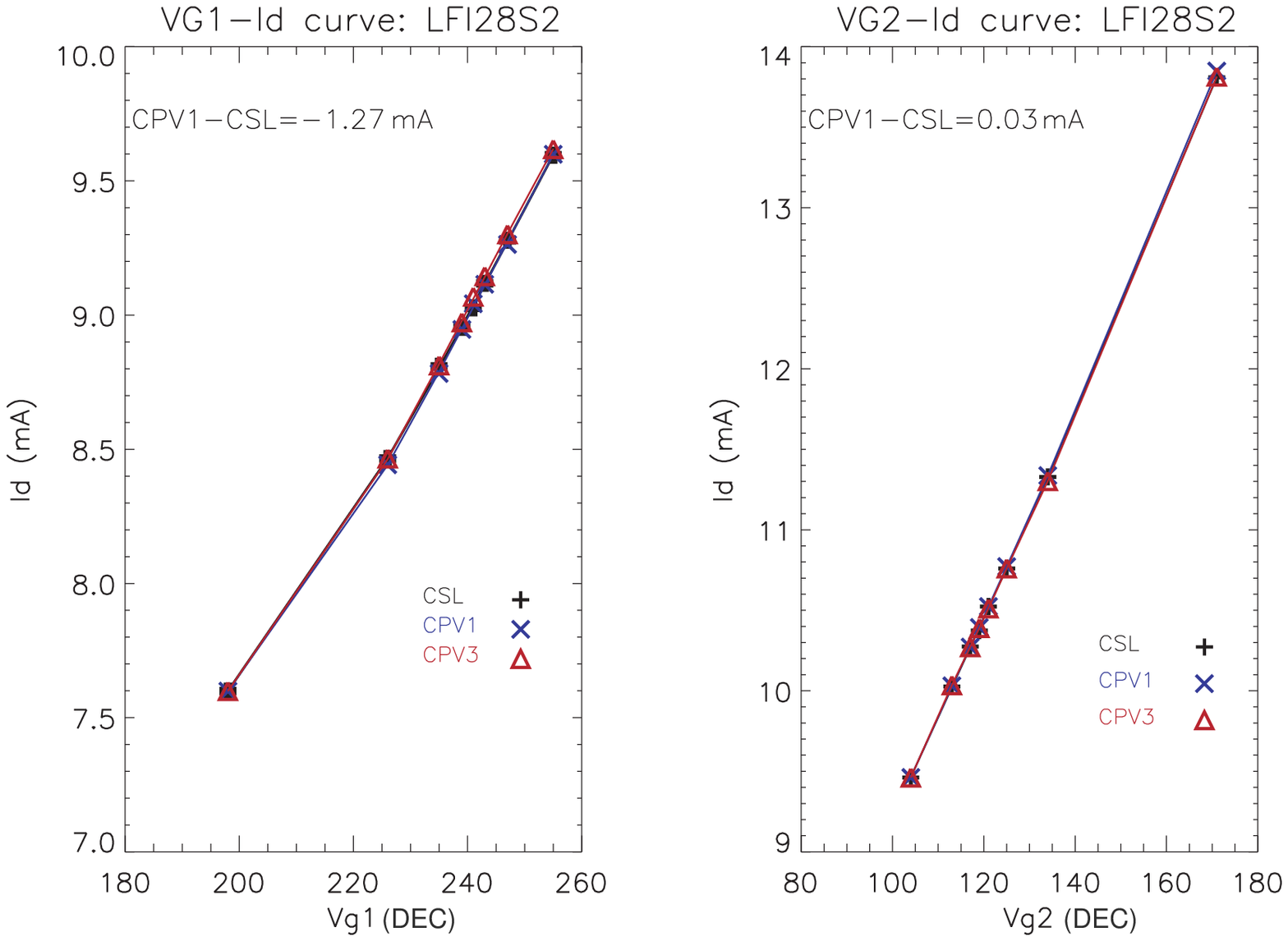}\\
        \end{center}
        \caption{I-V curves: for each channel $V_{\rm g1}$ and  $V_{\rm g2}$ plots are both shown. Black ``plus sign'' symbols refer to system level tests results in CSL; blue ``crosses'' to CPV 1$^{\rm st}$ run, red ``triangles'' to the 3$^{\rm rd}$ run. Solid lines represent the corresponding polynomial fits. Curves from the CSL system level tests and CPV 1$^{\rm st}$ run are offset to facilitate the comparison; the offset terms are print on top left of each picture. Note that the bias units are DEC units that are used to set the voltage in the DAE: see Table~4 for a rough conversion into physical units.}
        \label{fig_iv_test}
    \end{figure}

In Figure~\ref{fig_iv_test} we show for all the LFI channels the I-V curves (for each diode); black lines (``plus sign'' symbols) refer to system level tests data measured in CSL, blue lines (``cross'' symbols) refer to the $1^{\rm st}$ and red lines (``triangle'' symbols) to the $3^{\rm rd}$ run performed during CPV. 
    \begin{figure}[htb]
        \begin{center}
        		\textbf{LFI-18}\\ \vspace{0.1cm}
            \includegraphics[width=7.0cm]{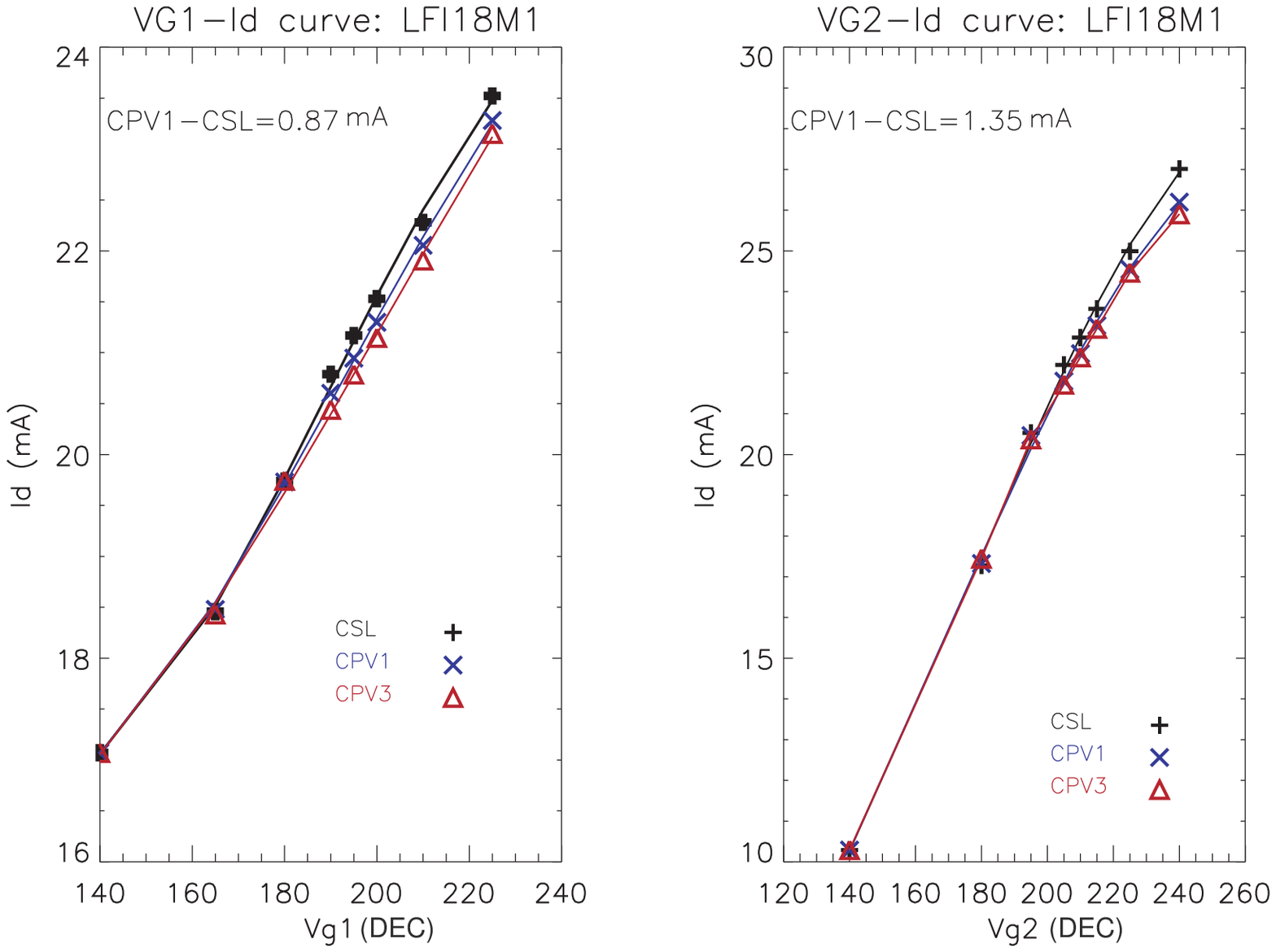}
            \includegraphics[width=7.0cm]{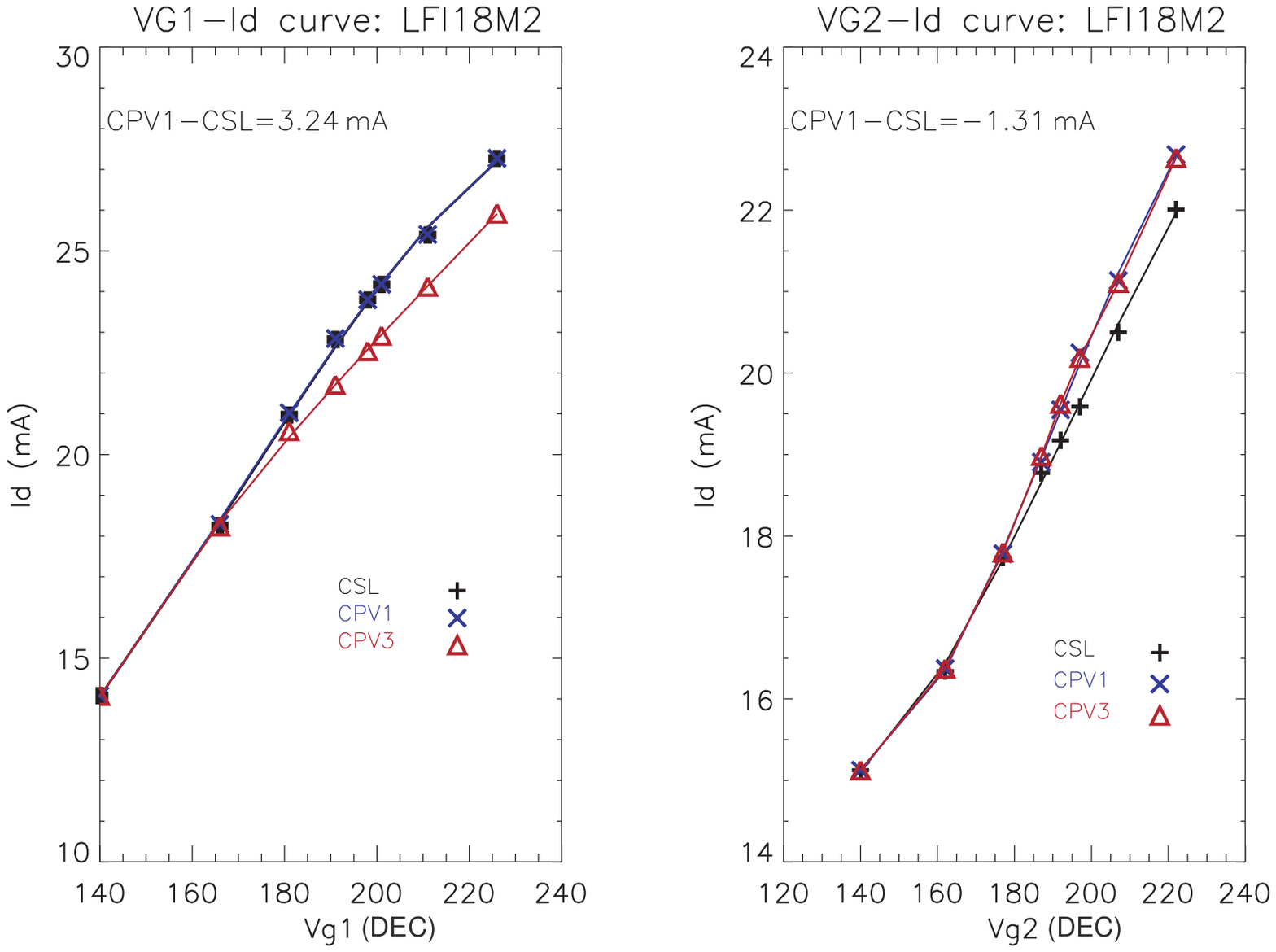}\\
            \includegraphics[width=7.0cm]{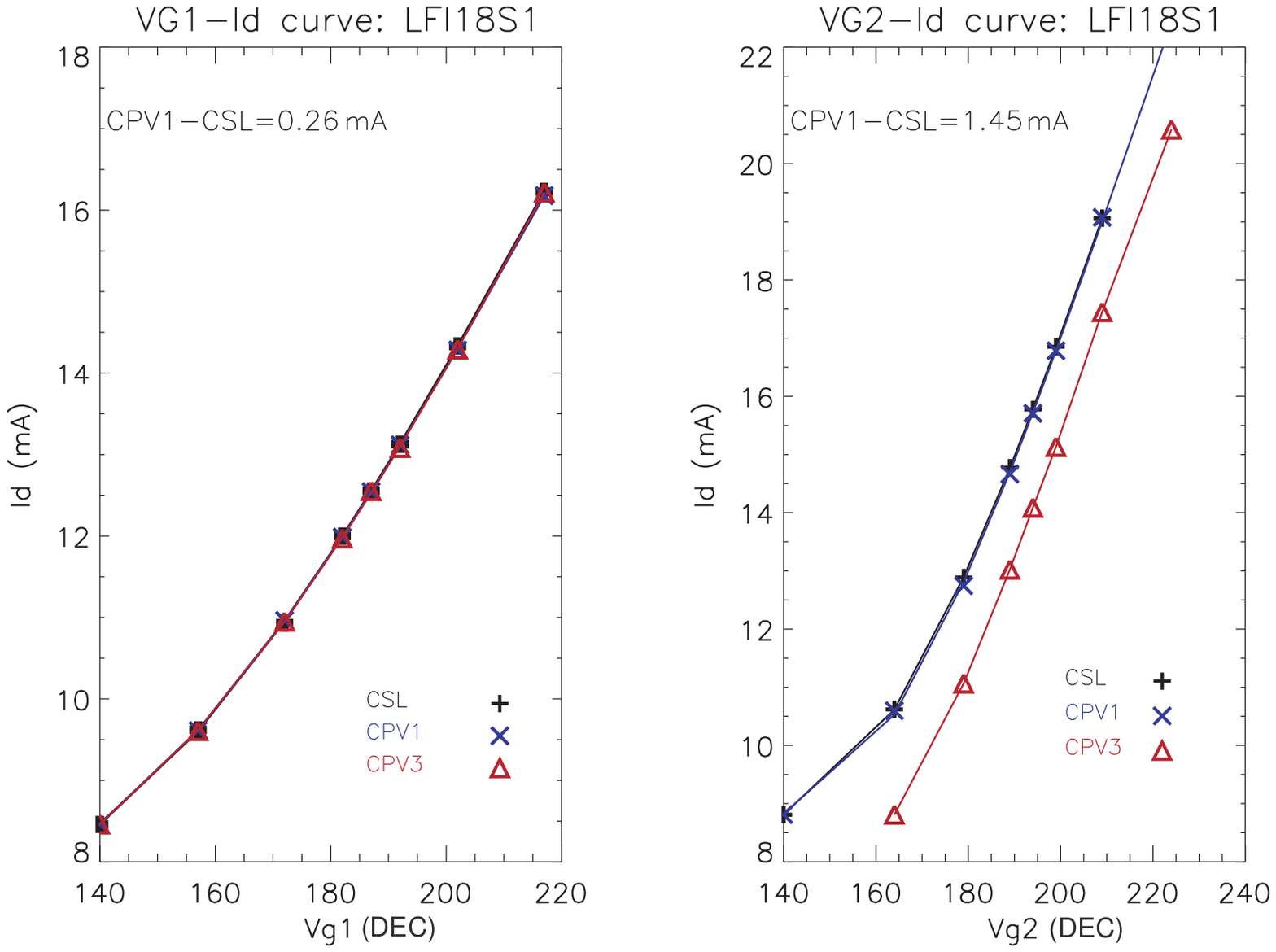}
            \includegraphics[width=7.0cm]{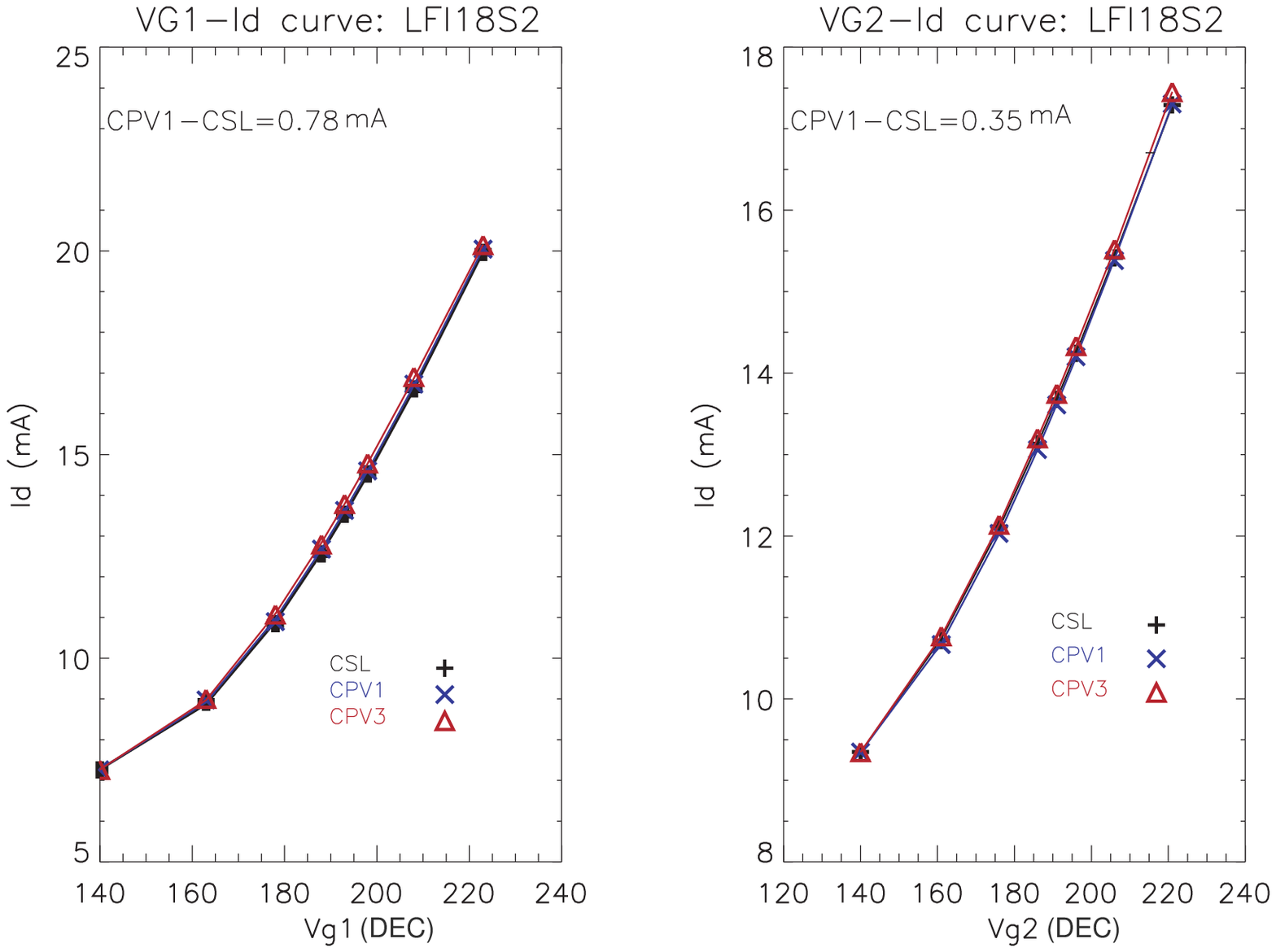}\\
            \textbf{LFI-19}\\ \vspace{0.1cm}
            \includegraphics[width=7.0cm]{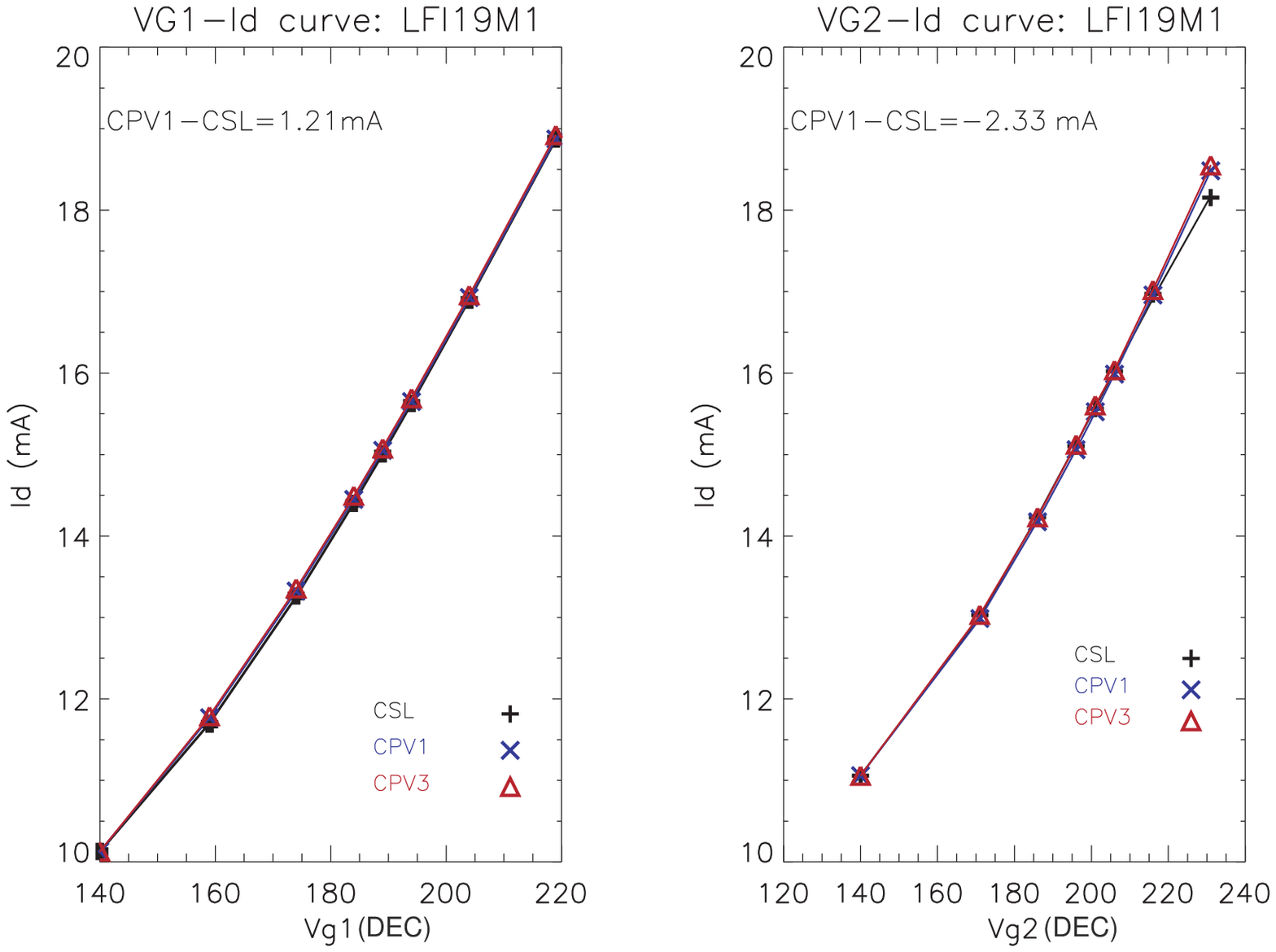}
            \includegraphics[width=7.0cm]{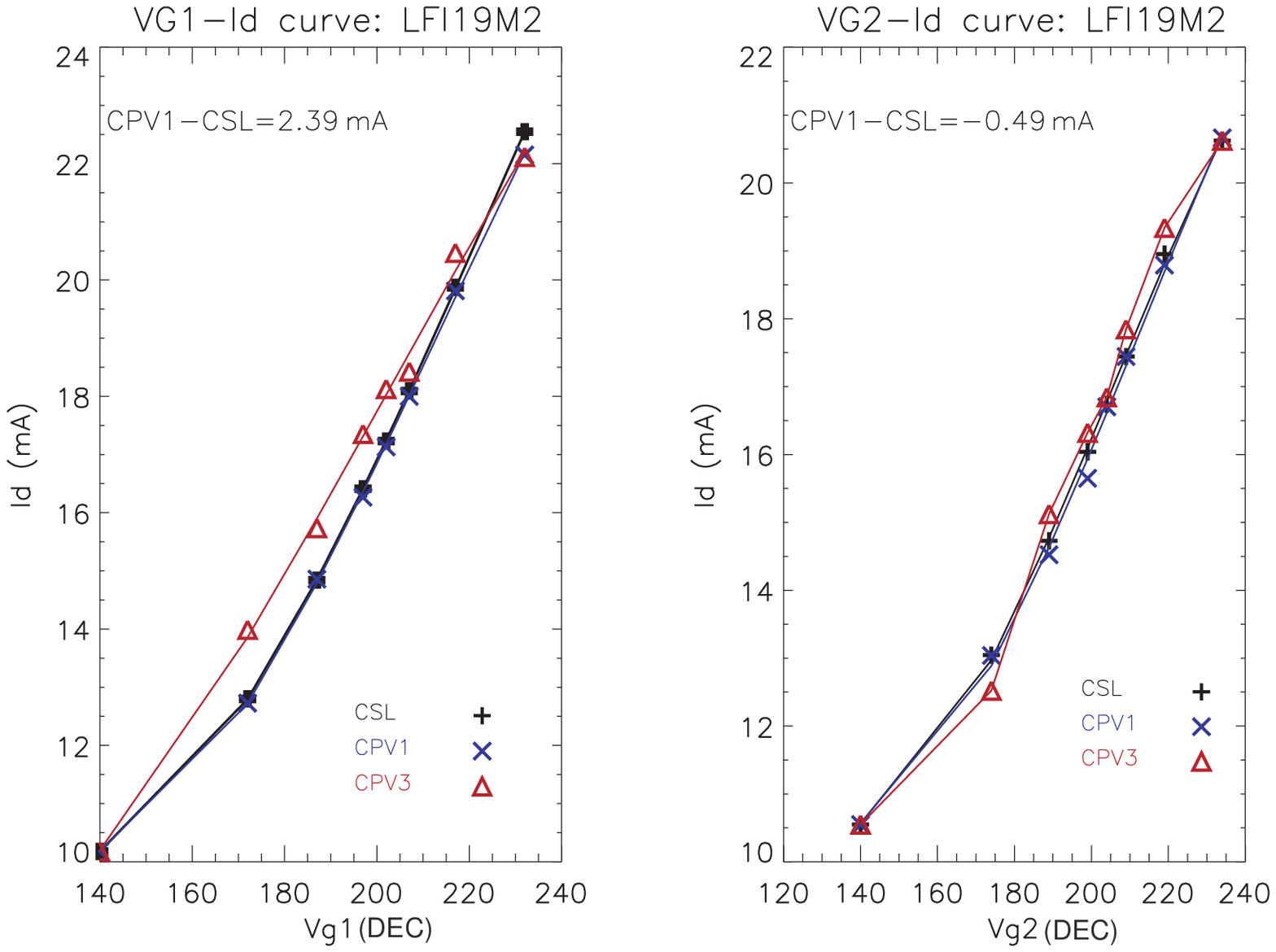}\\
            \includegraphics[width=7.0cm]{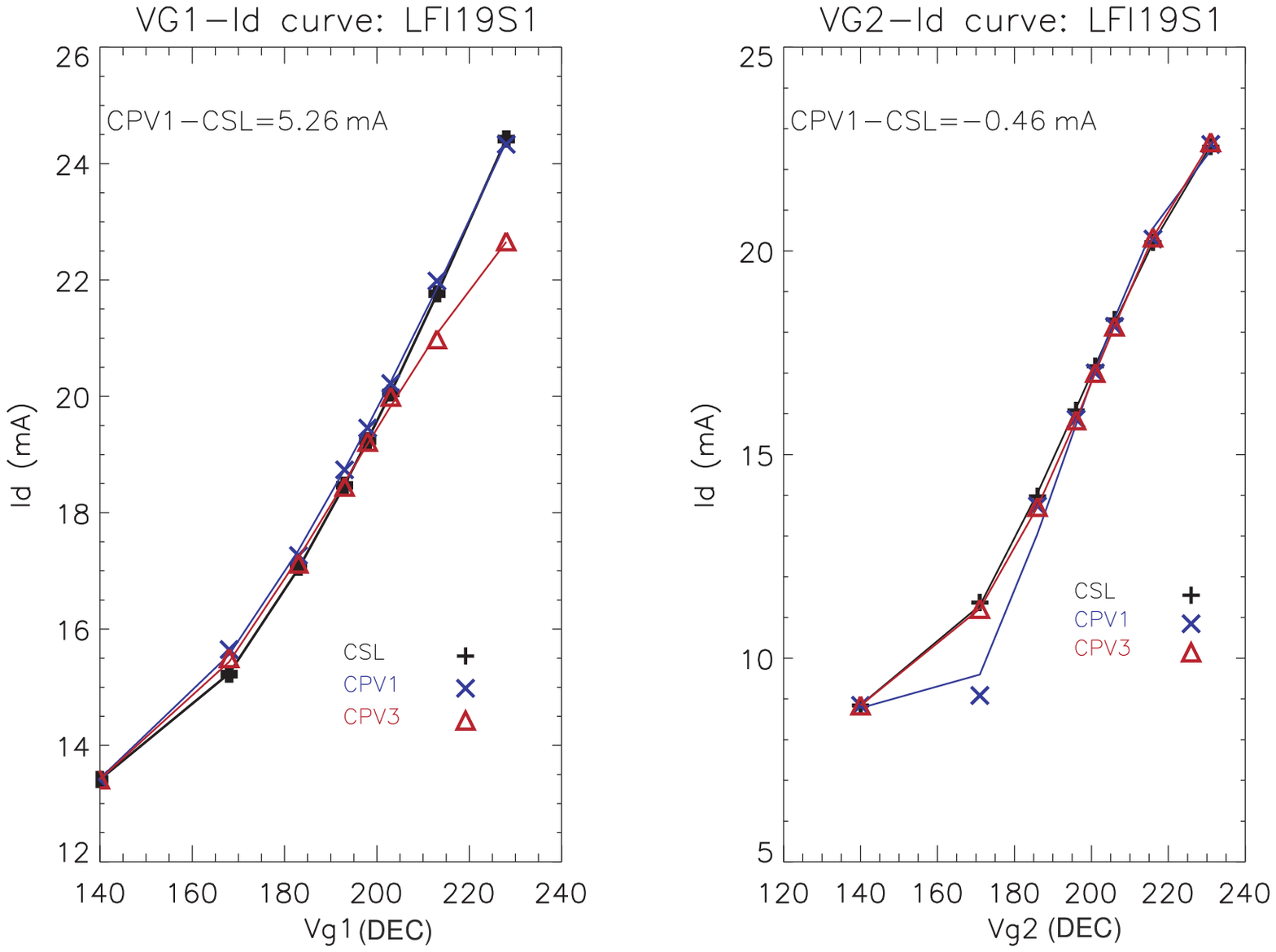}
            \includegraphics[width=7.0cm]{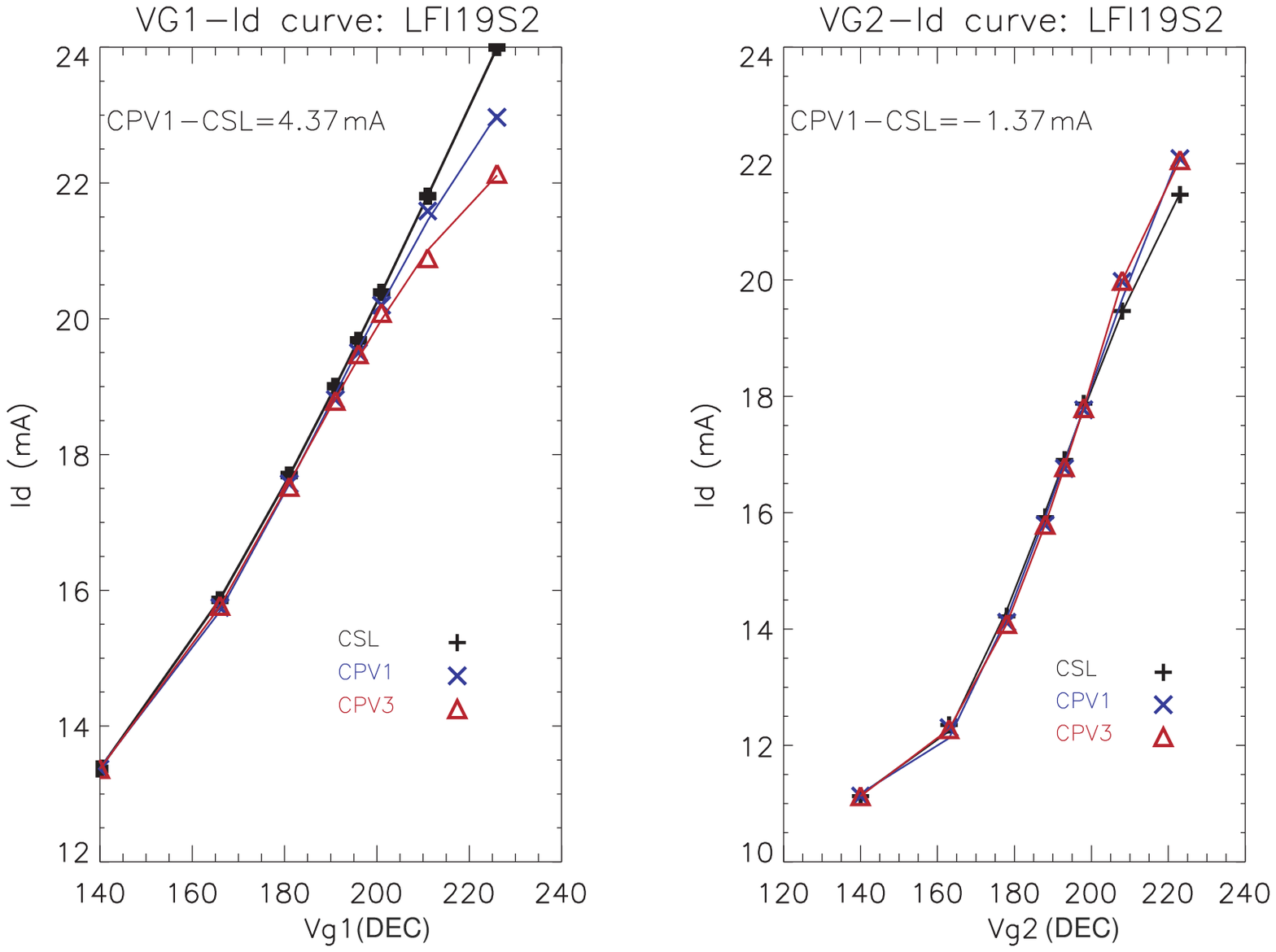}\\
        \end{center}
    \end{figure}

     \begin{figure}[htb]
        \begin{center}         
            \textbf{LFI-20}\\ \vspace{0.1cm}
            \includegraphics[width=7.0cm]{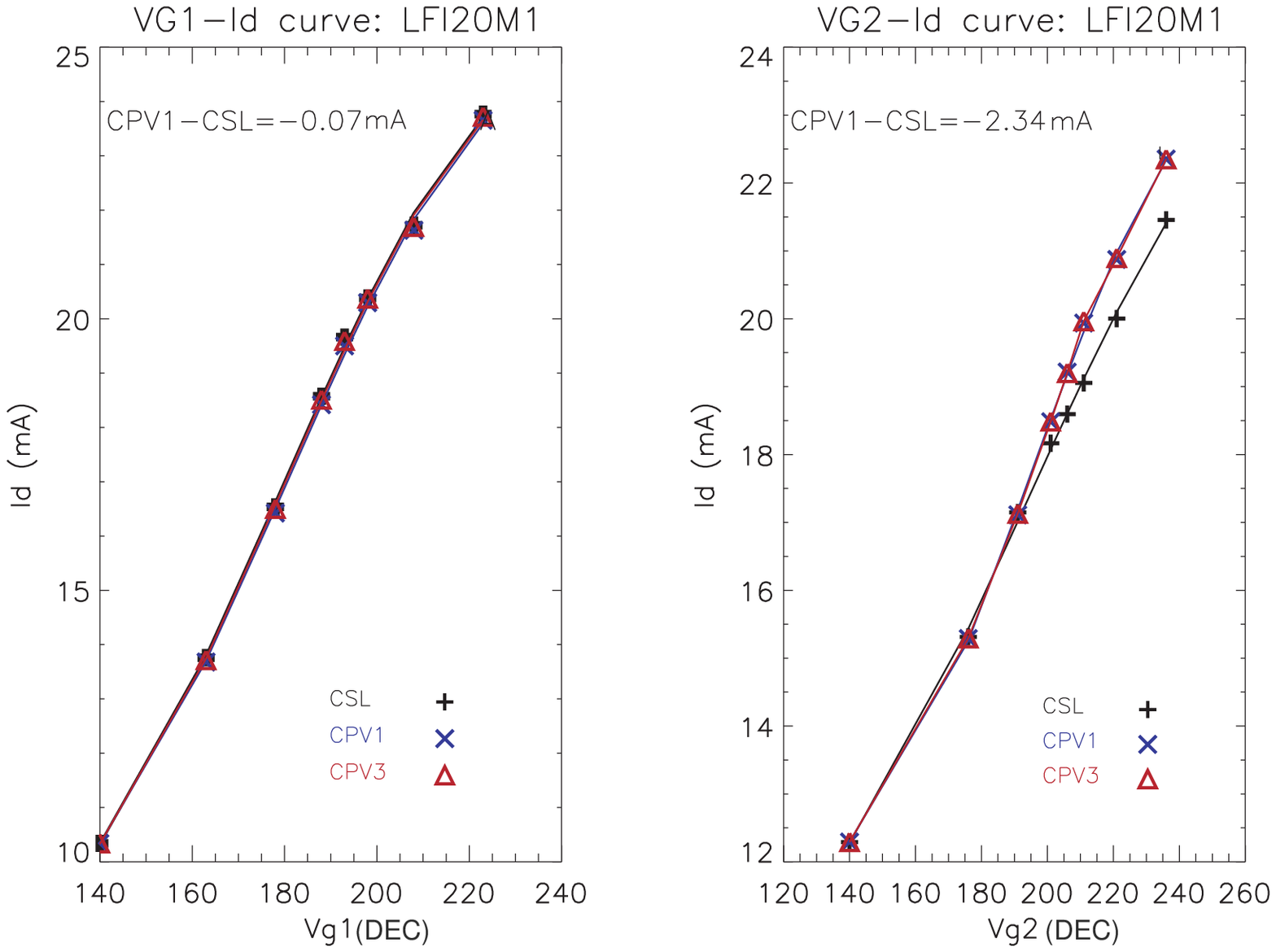}
            \includegraphics[width=7.0cm]{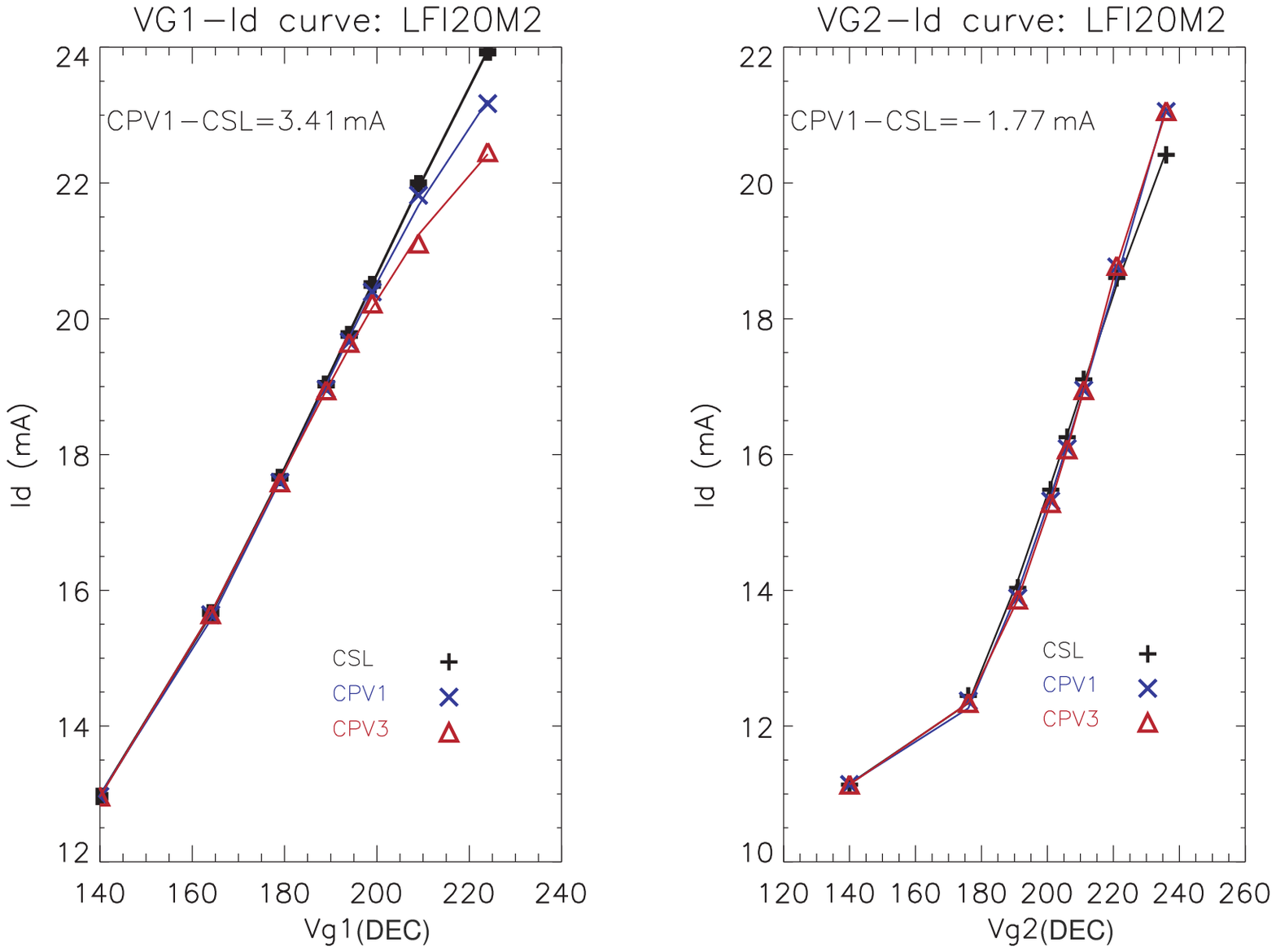}\\
            \includegraphics[width=7.0cm]{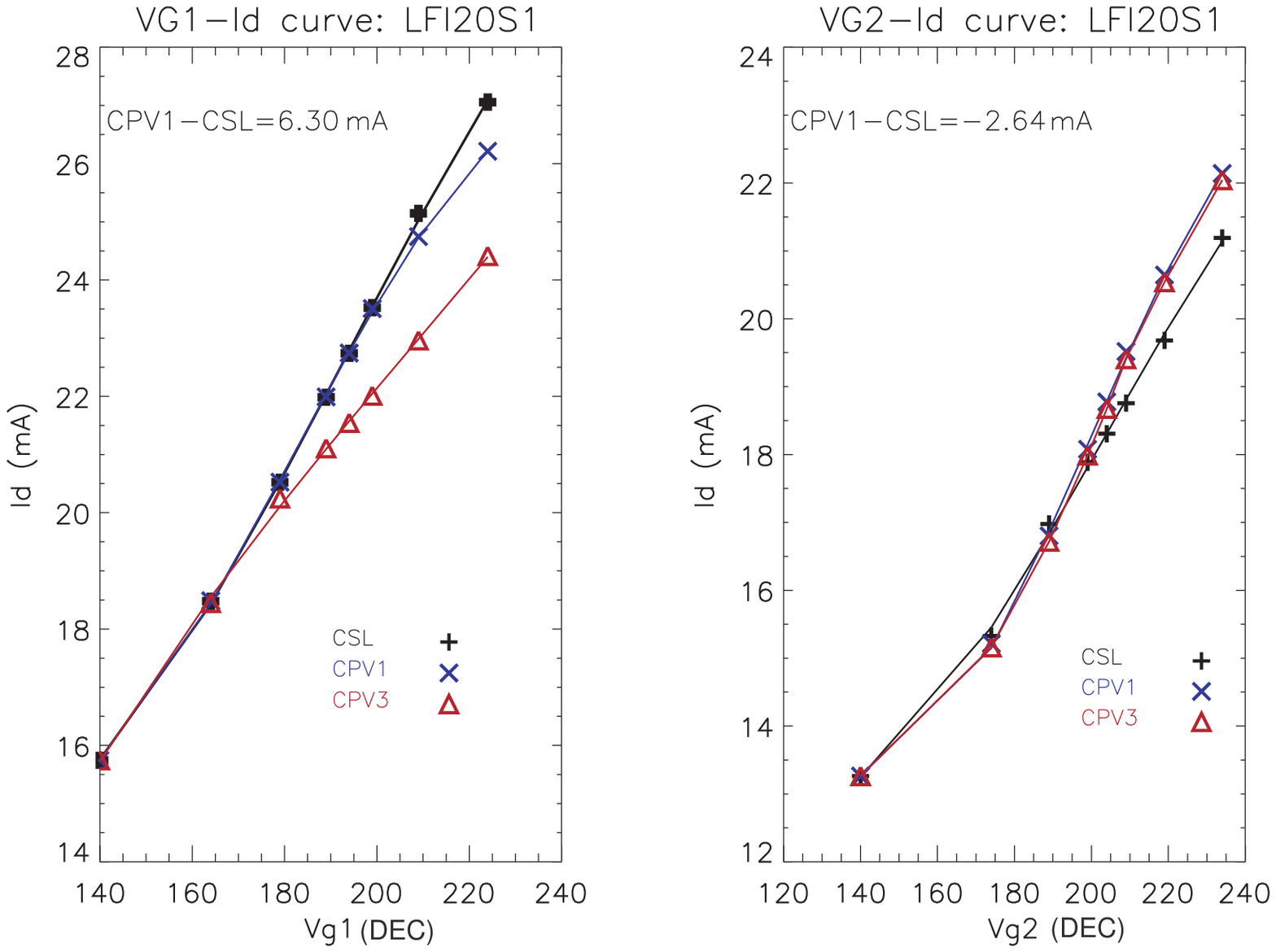}
            \includegraphics[width=7.0cm]{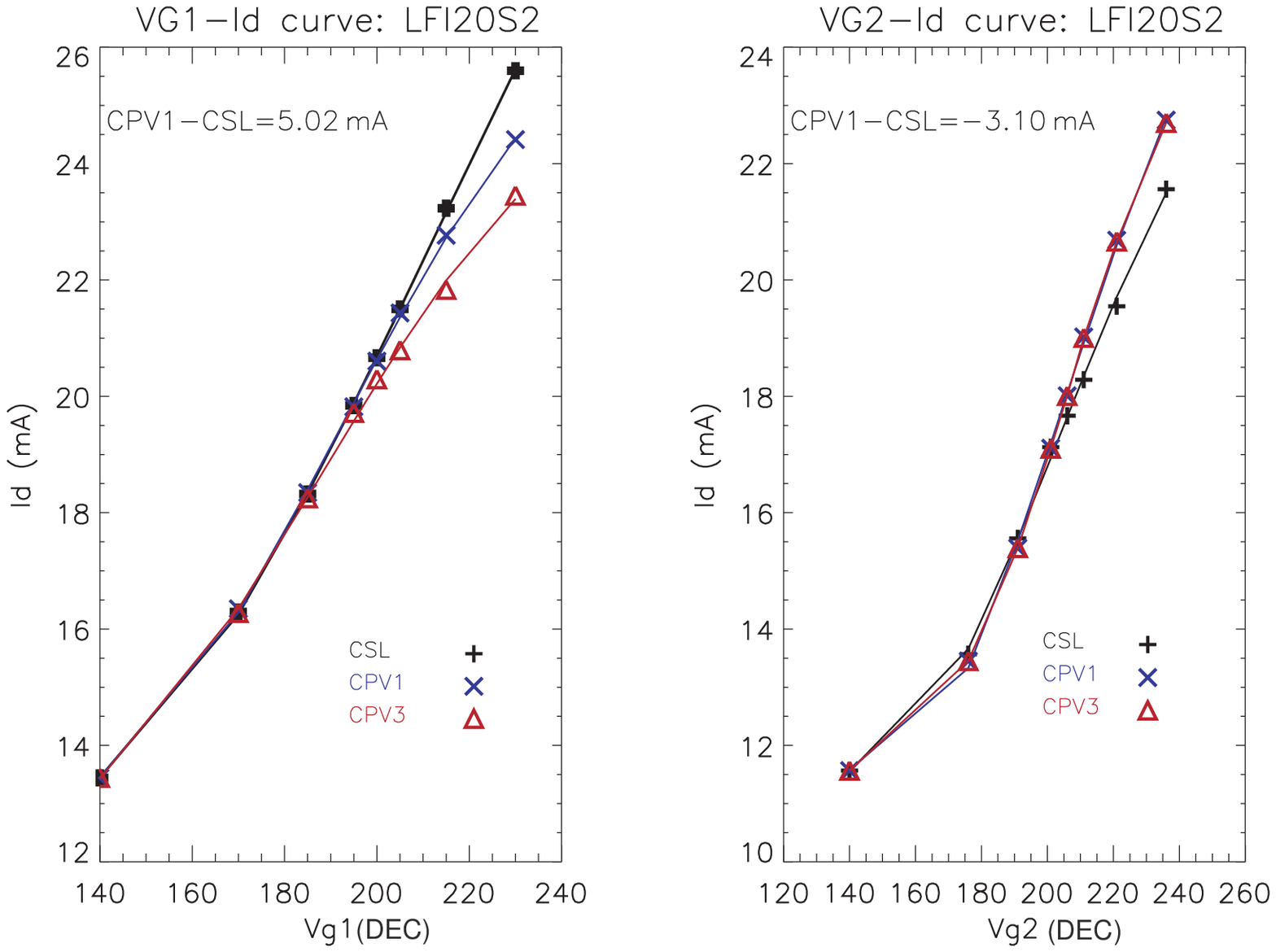}\\
            \textbf{LFI-21}\\ \vspace{0.1cm}
            \includegraphics[width=7.0cm]{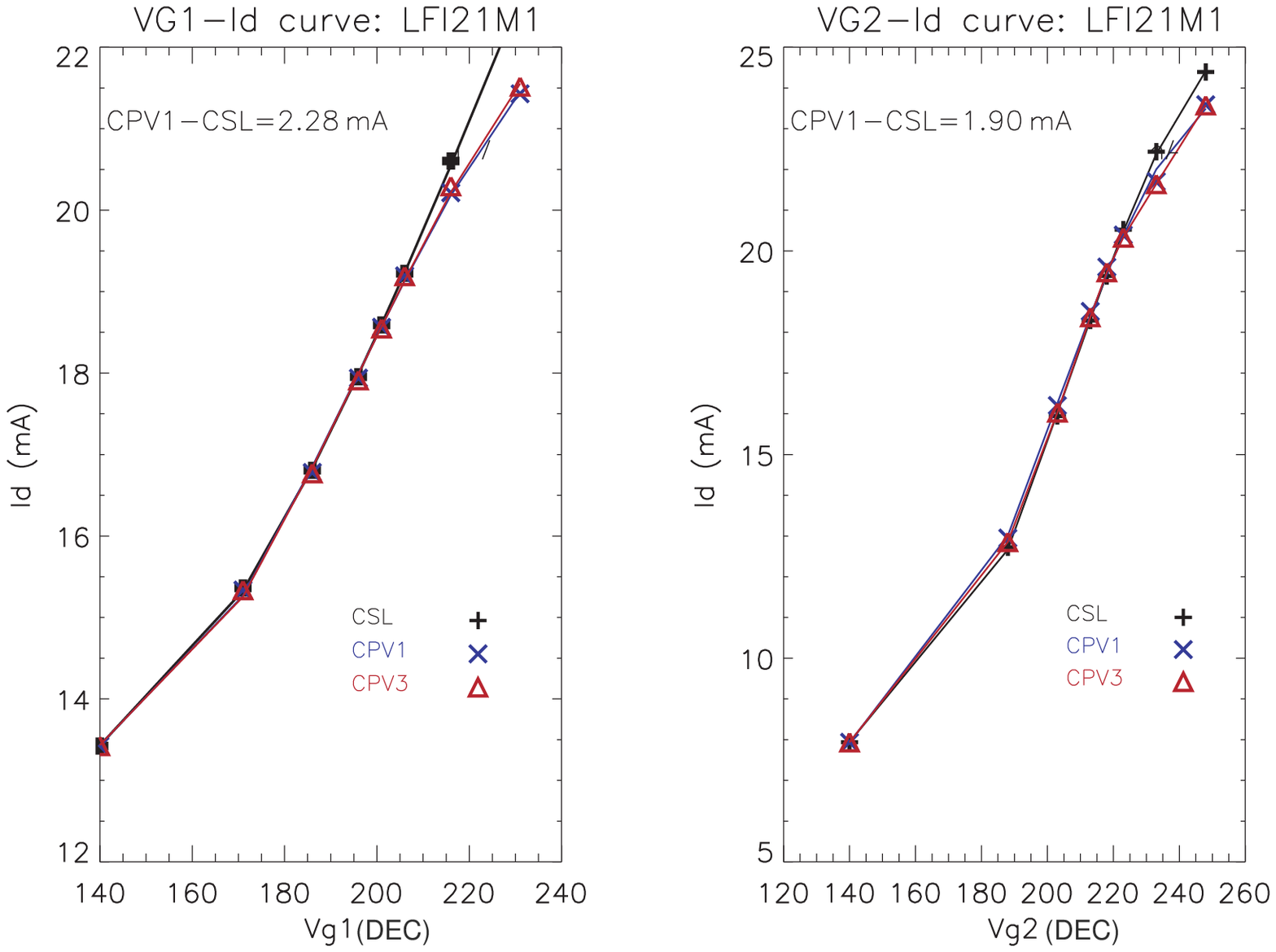}
            \includegraphics[width=7.0cm]{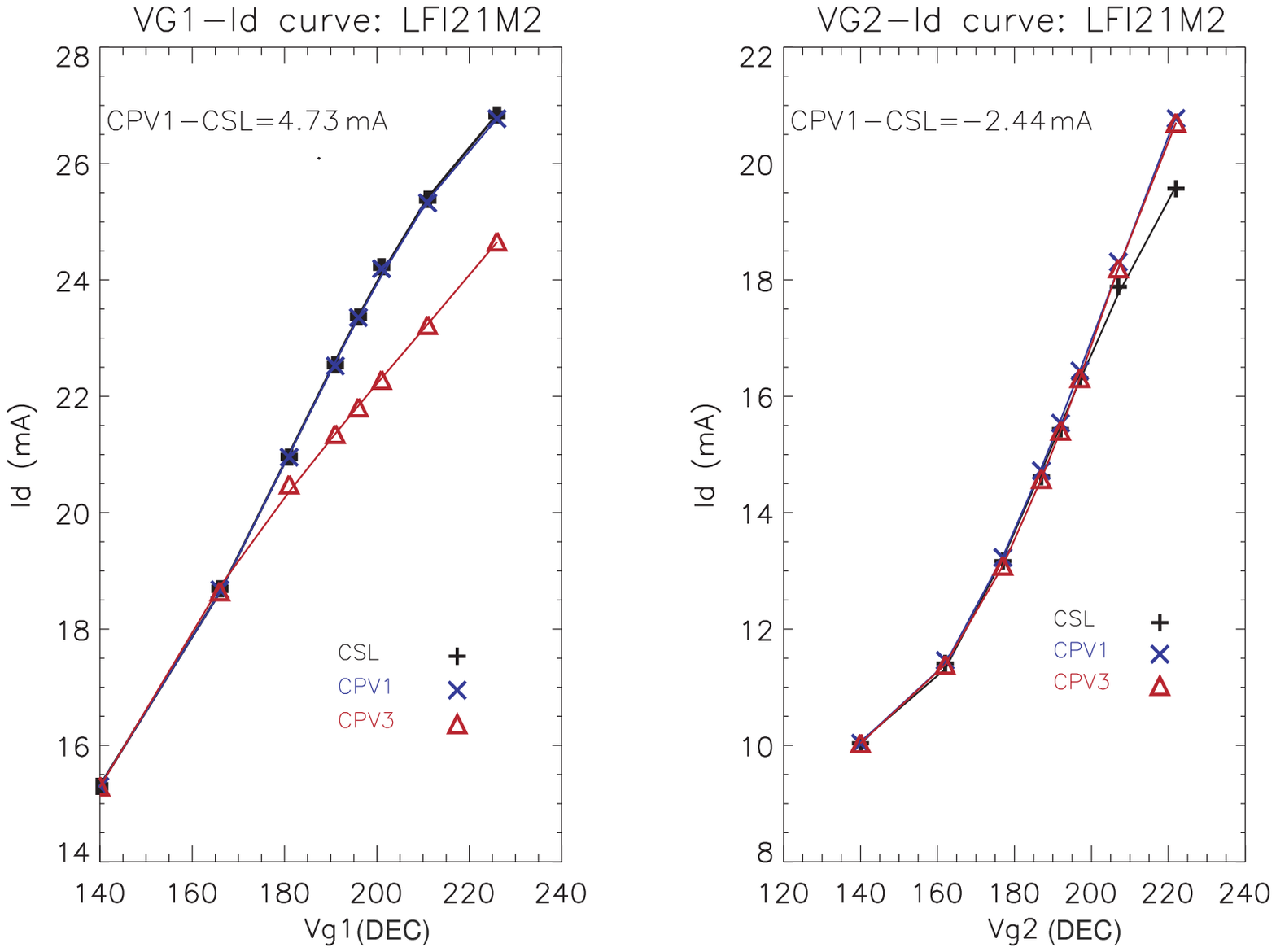}\\
            \includegraphics[width=7.0cm]{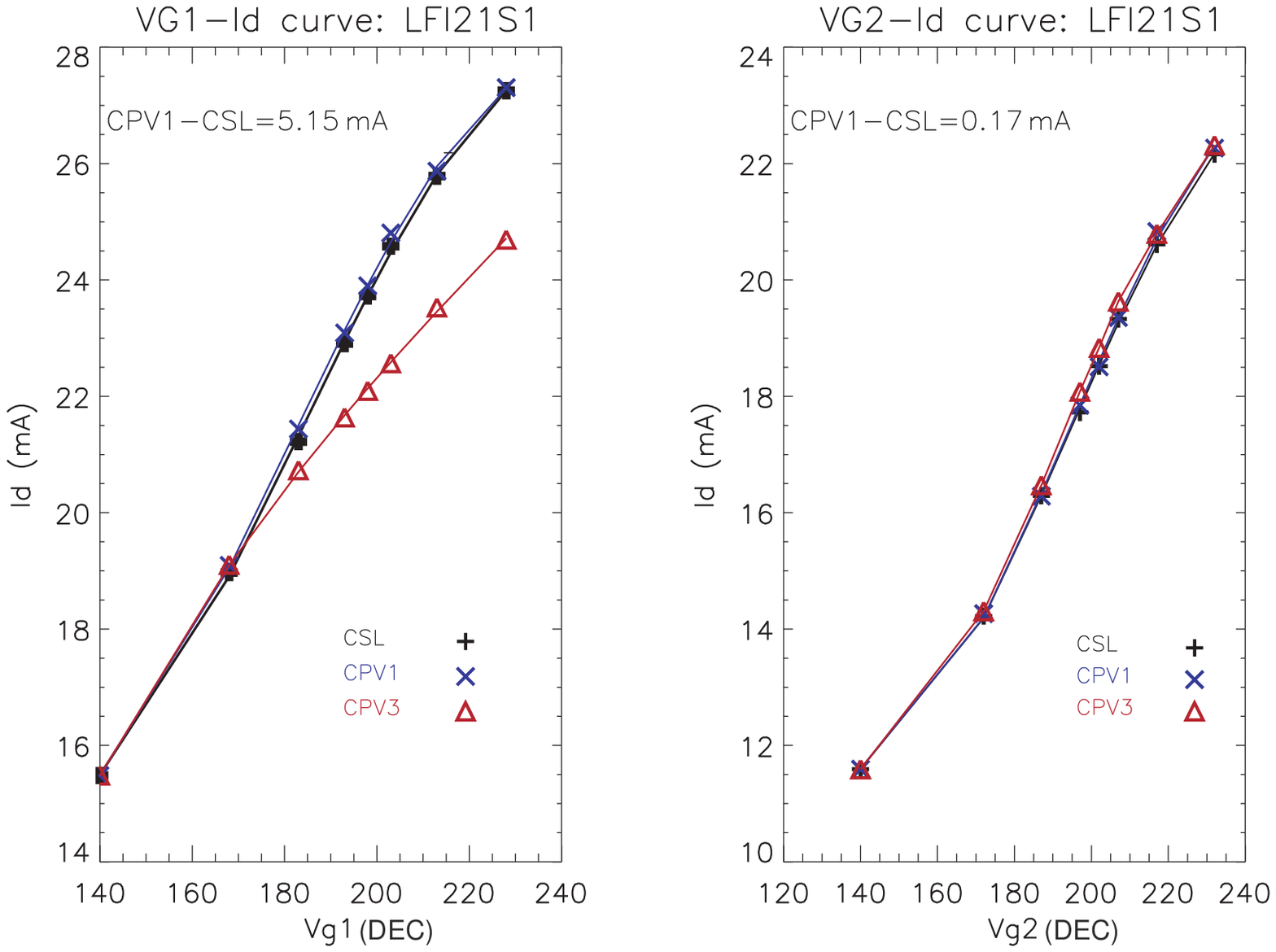}
            \includegraphics[width=7.0cm]{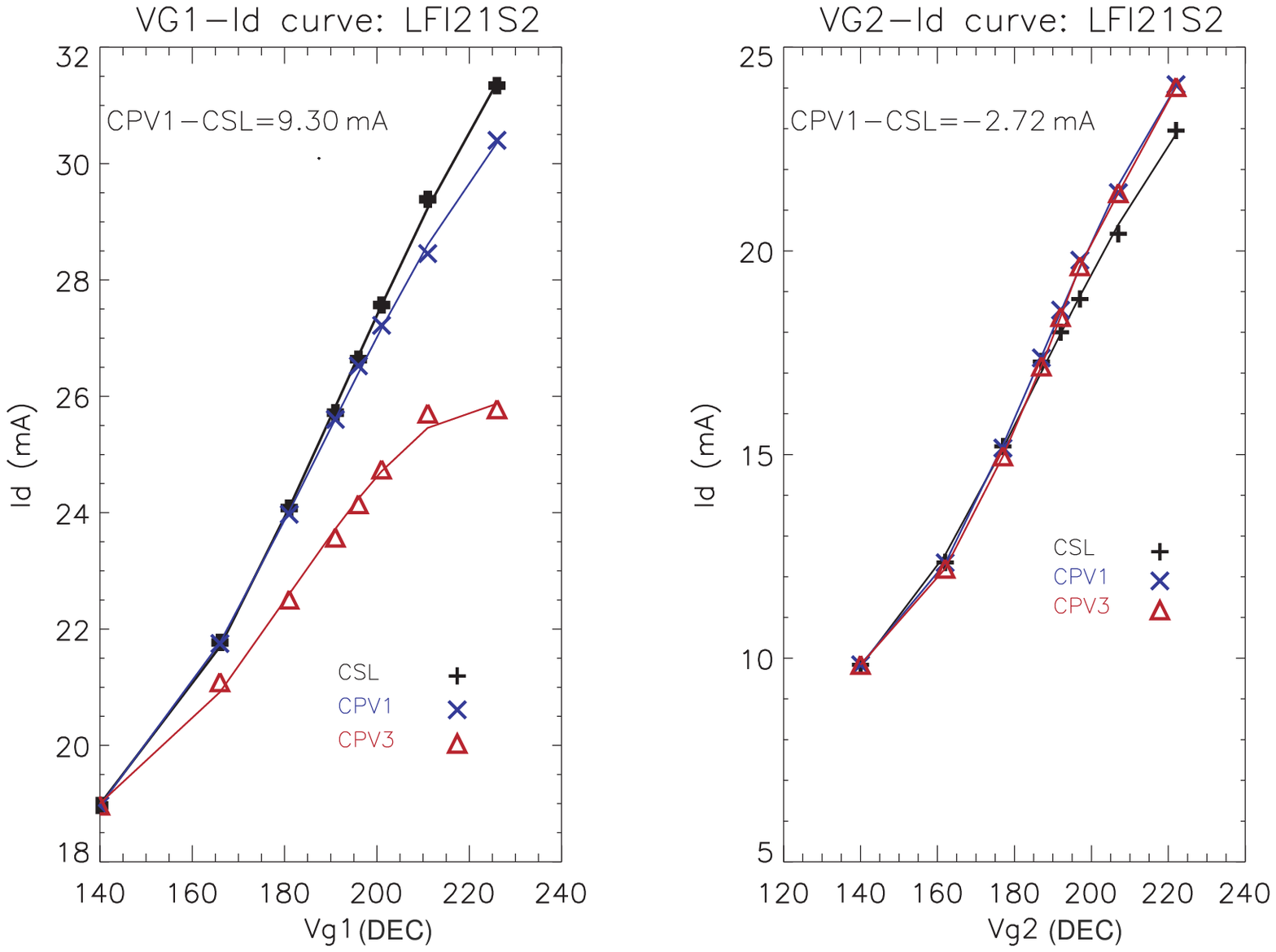}\\
        \end{center}
    \end{figure}
          \begin{figure}[htb]
        \begin{center}   
            \textbf{LFI-22}\\ \vspace{0.1cm}
            \includegraphics[width=7.0cm]{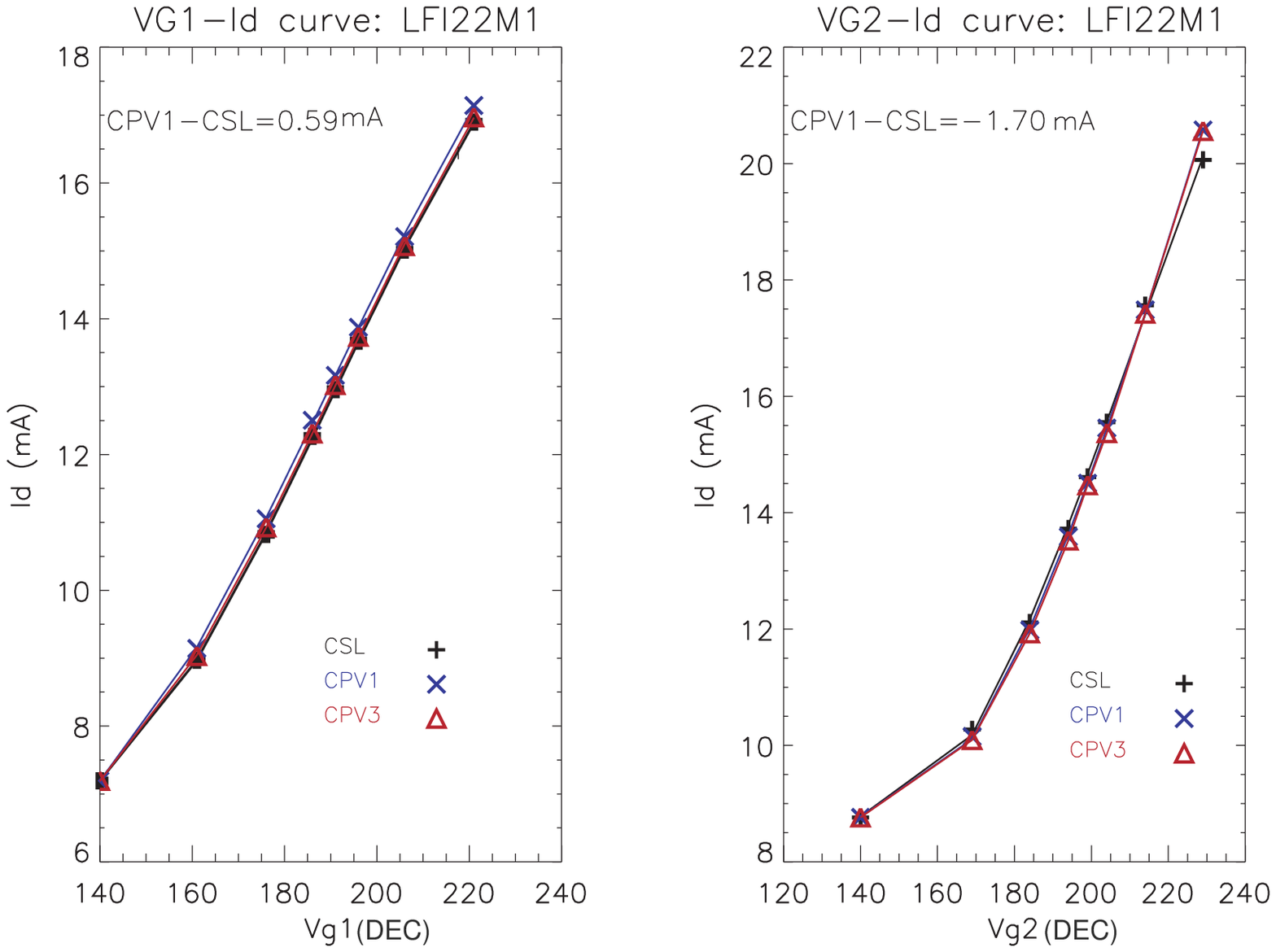}
            \includegraphics[width=7.0cm]{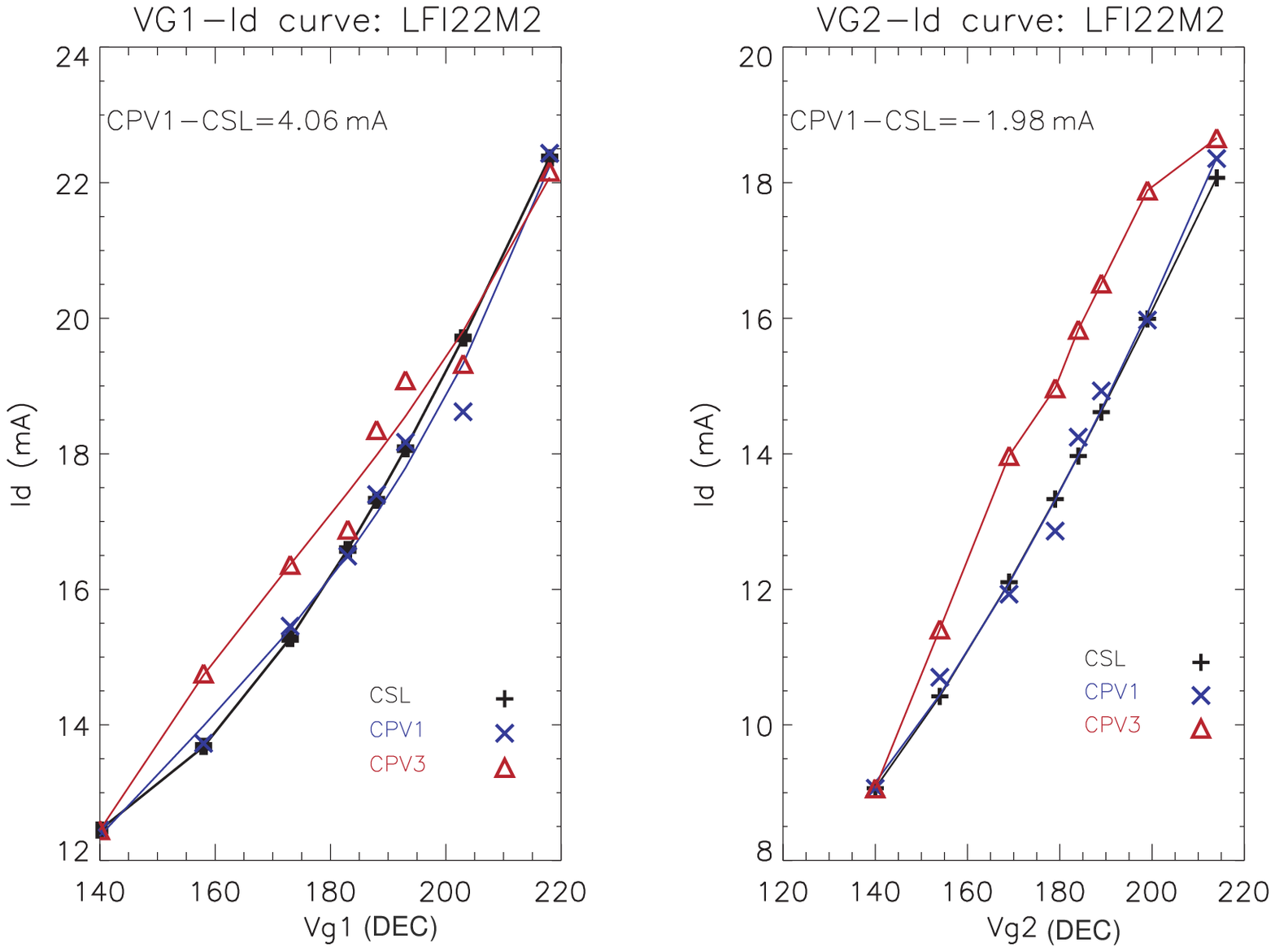}\\
            \includegraphics[width=7.0cm]{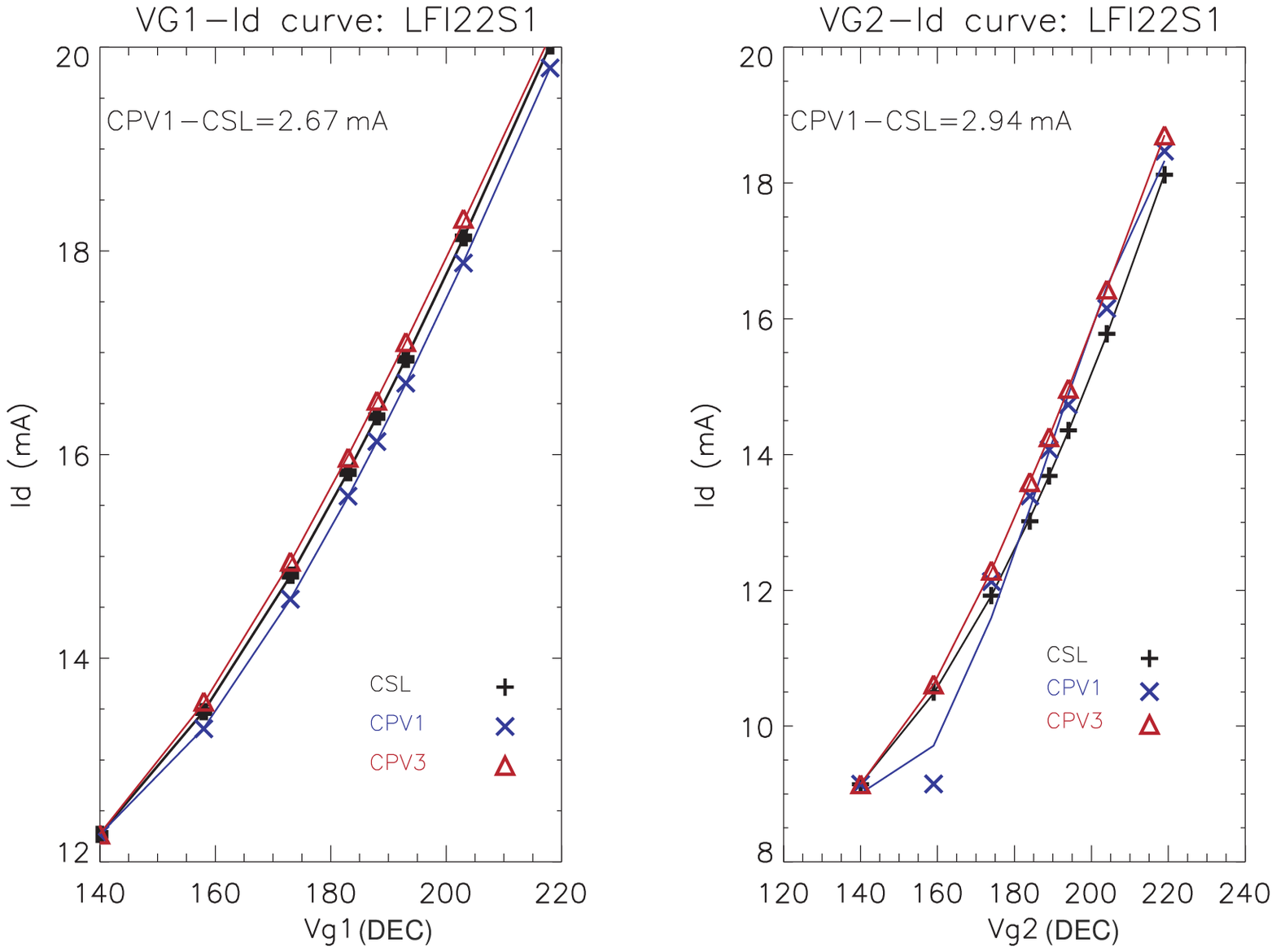}
            \includegraphics[width=7.0cm]{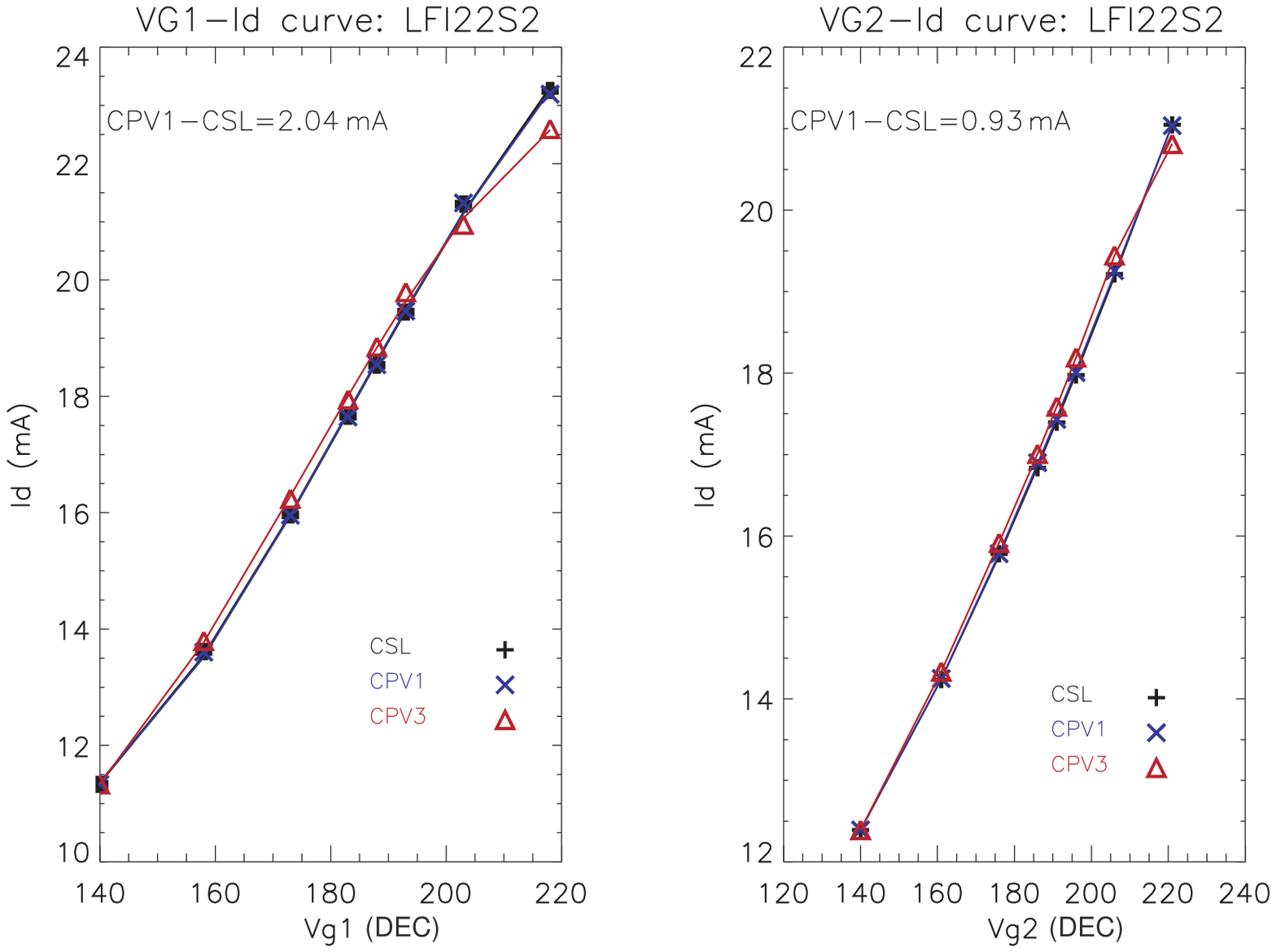}\\
            \textbf{LFI-23}\\ \vspace{0.1cm}
            \includegraphics[width=7.0cm]{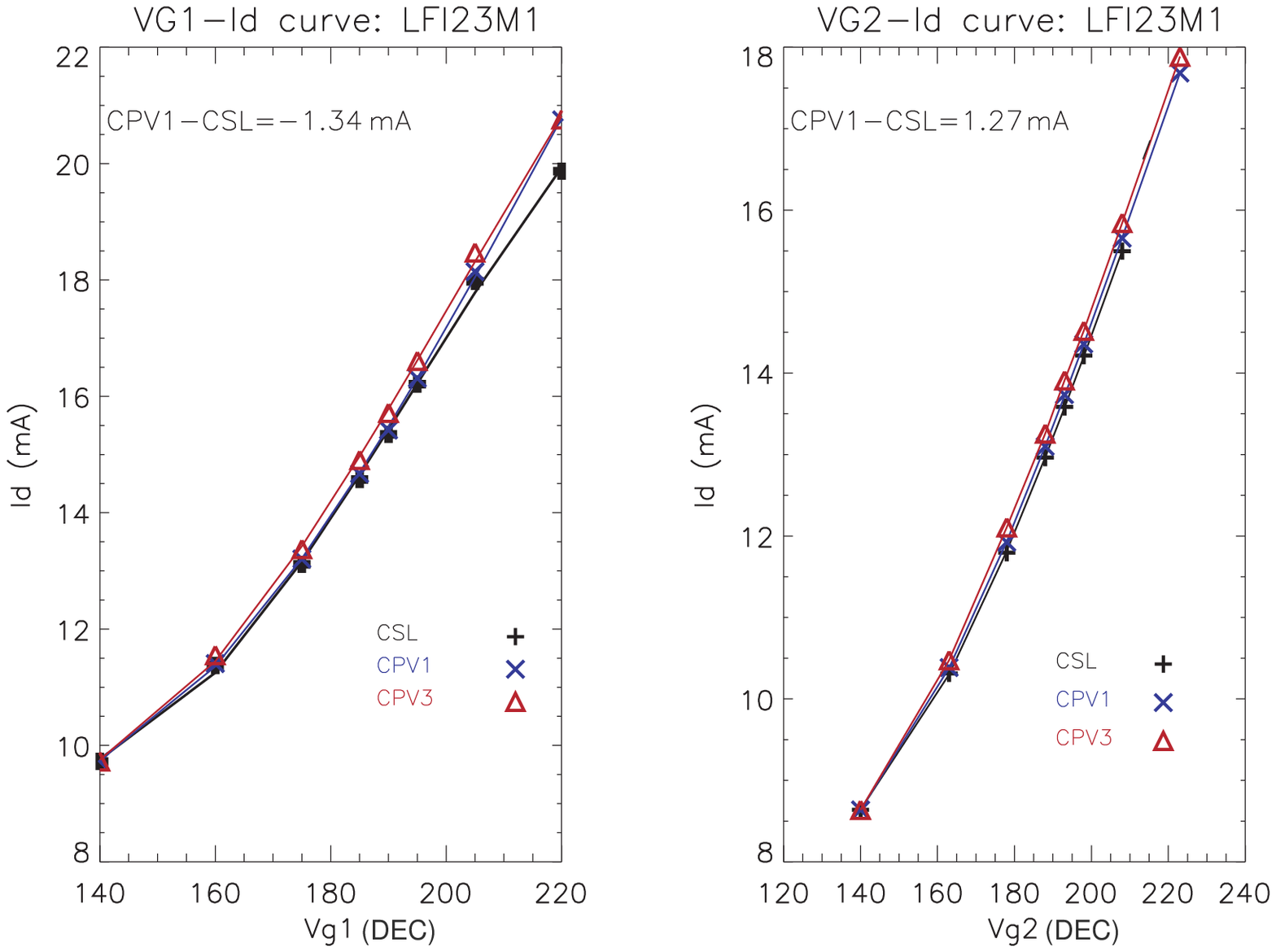}
            \includegraphics[width=7.0cm]{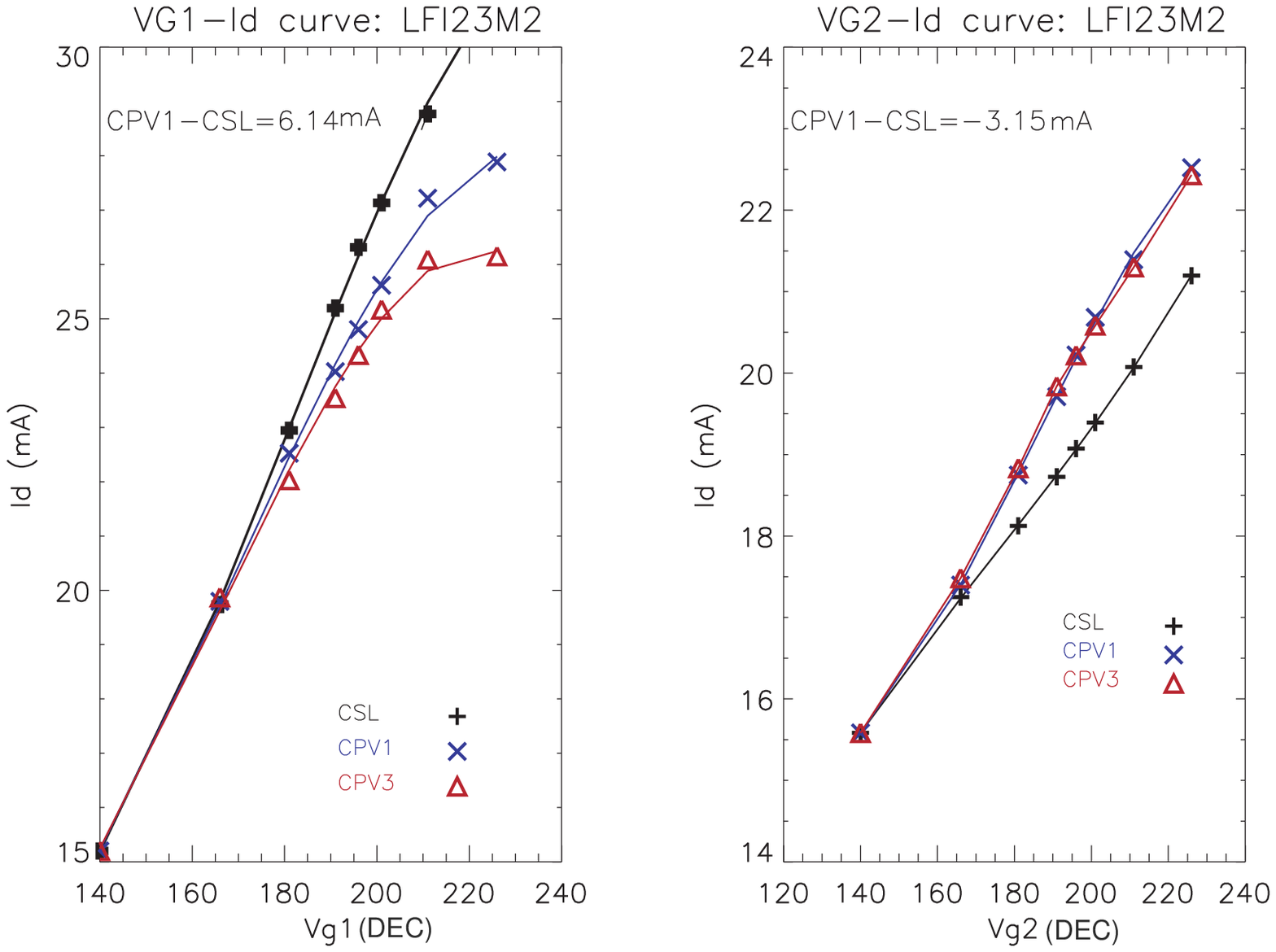}\\
            \includegraphics[width=7.0cm]{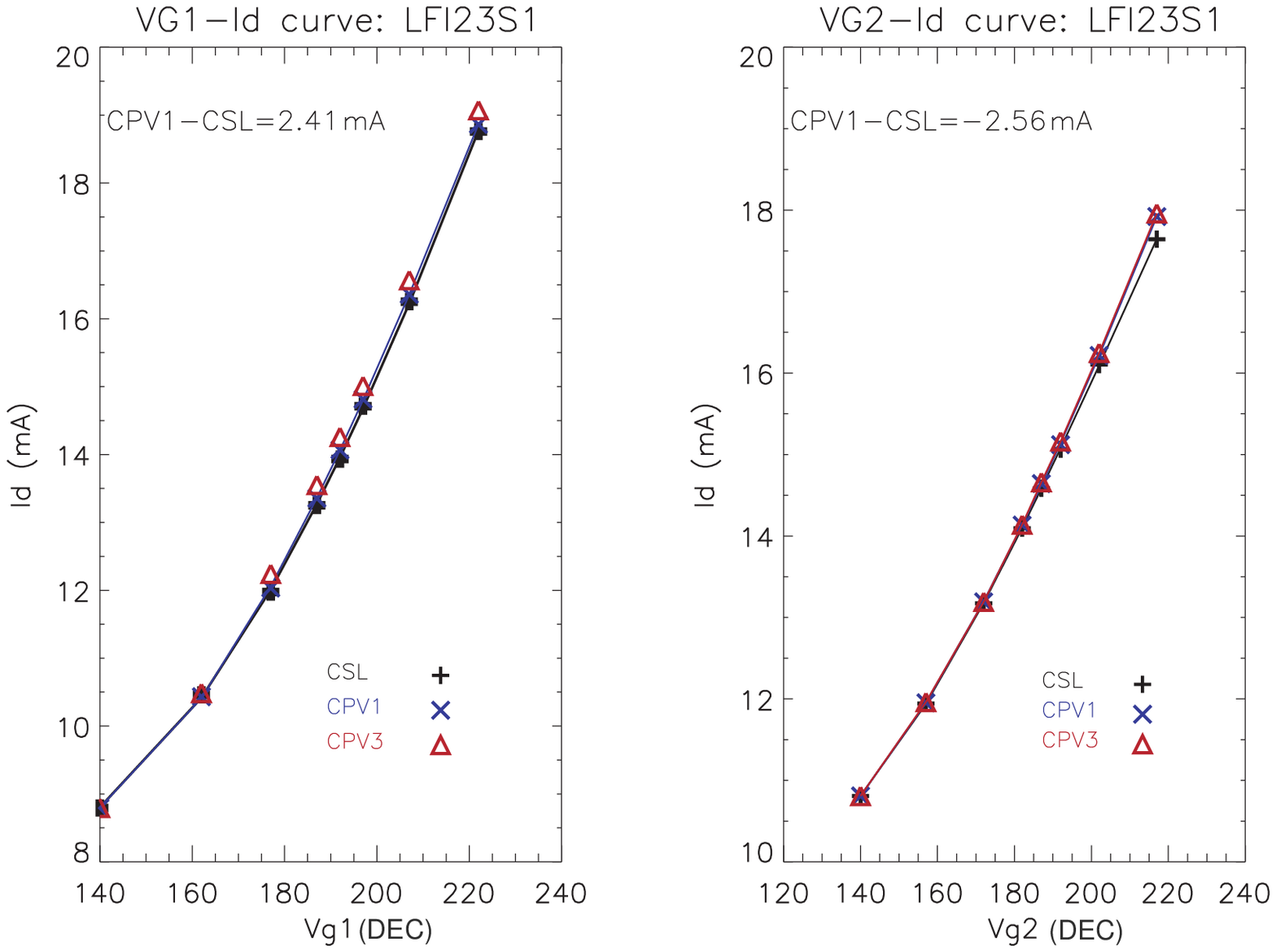}
            \includegraphics[width=7.0cm]{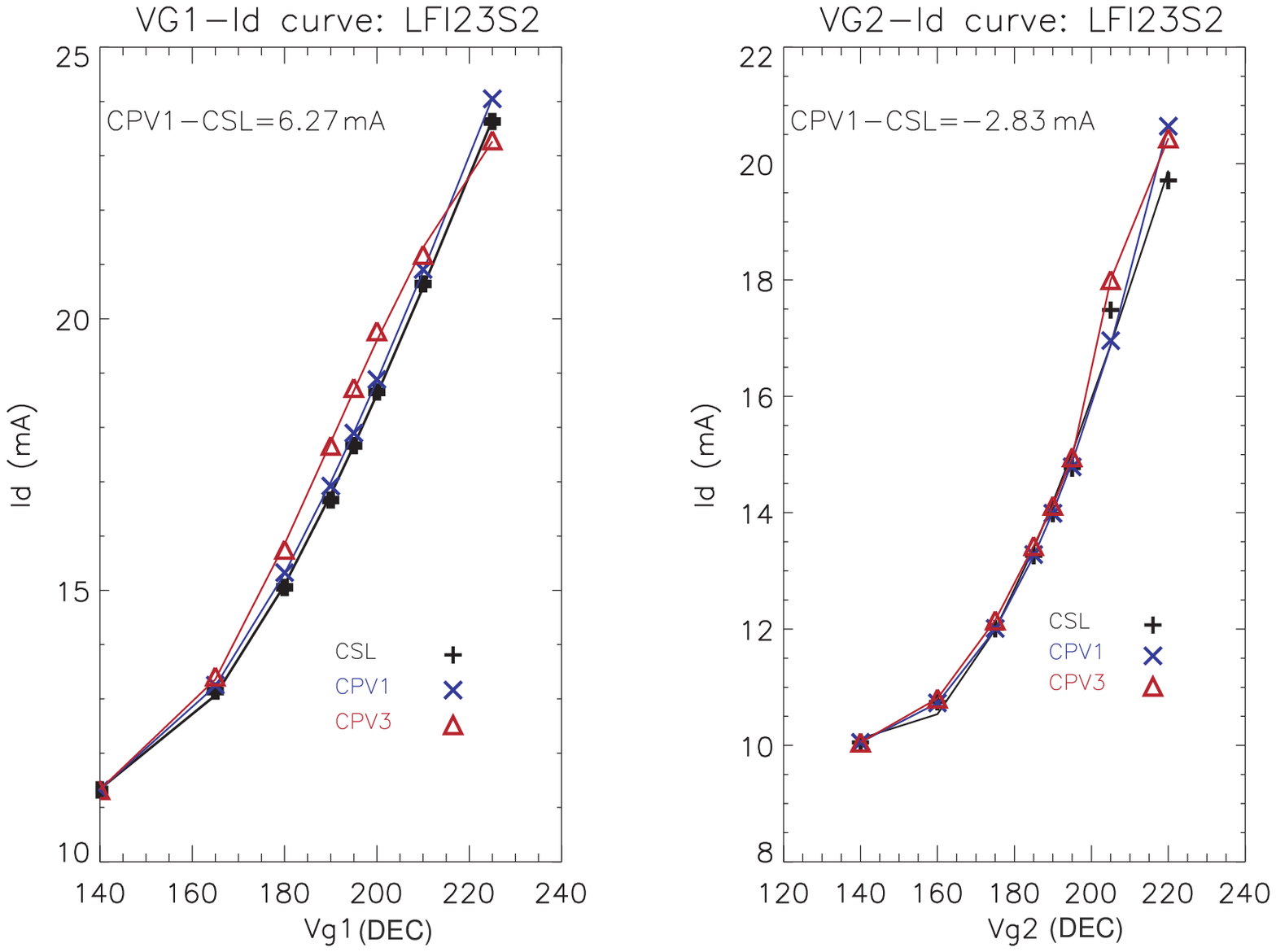}\\
        \end{center}
    \end{figure}

    \begin{figure}[htb]
        \begin{center}
           	\textbf{LFI-24}\\  \vspace{0.1cm}
            \includegraphics[width=7.0cm]{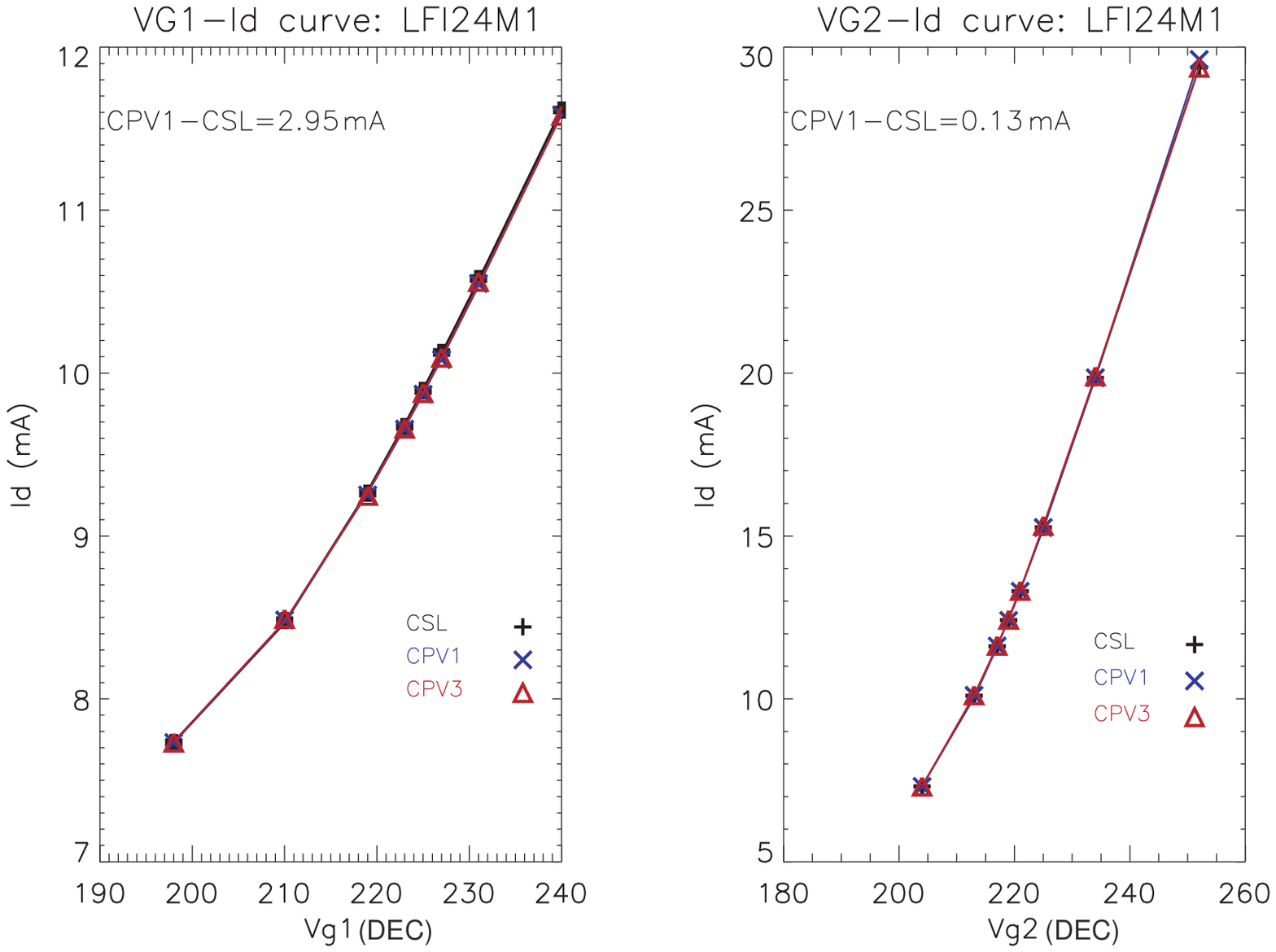}
            \includegraphics[width=7.0cm]{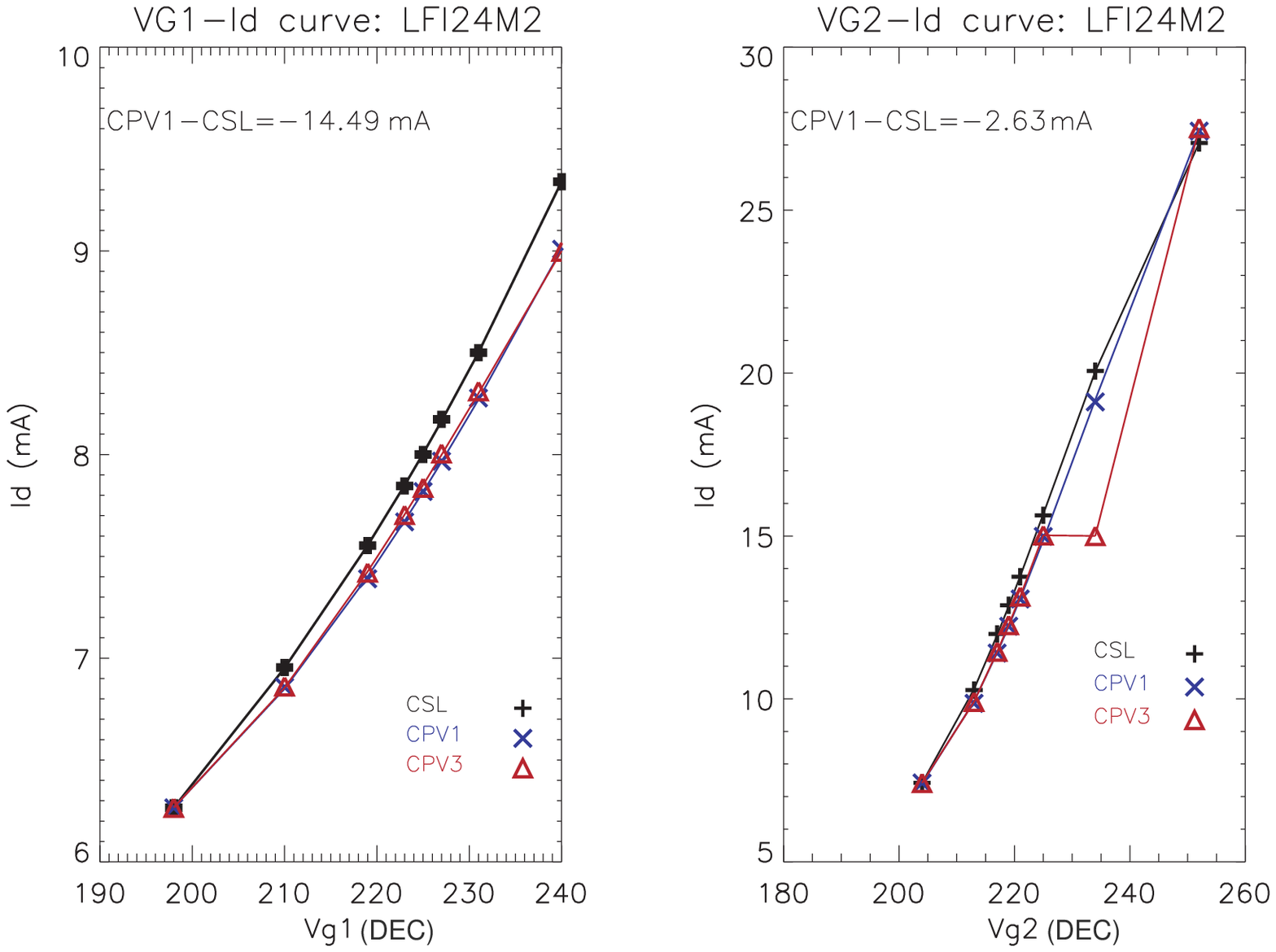}\\
            \includegraphics[width=7.0cm]{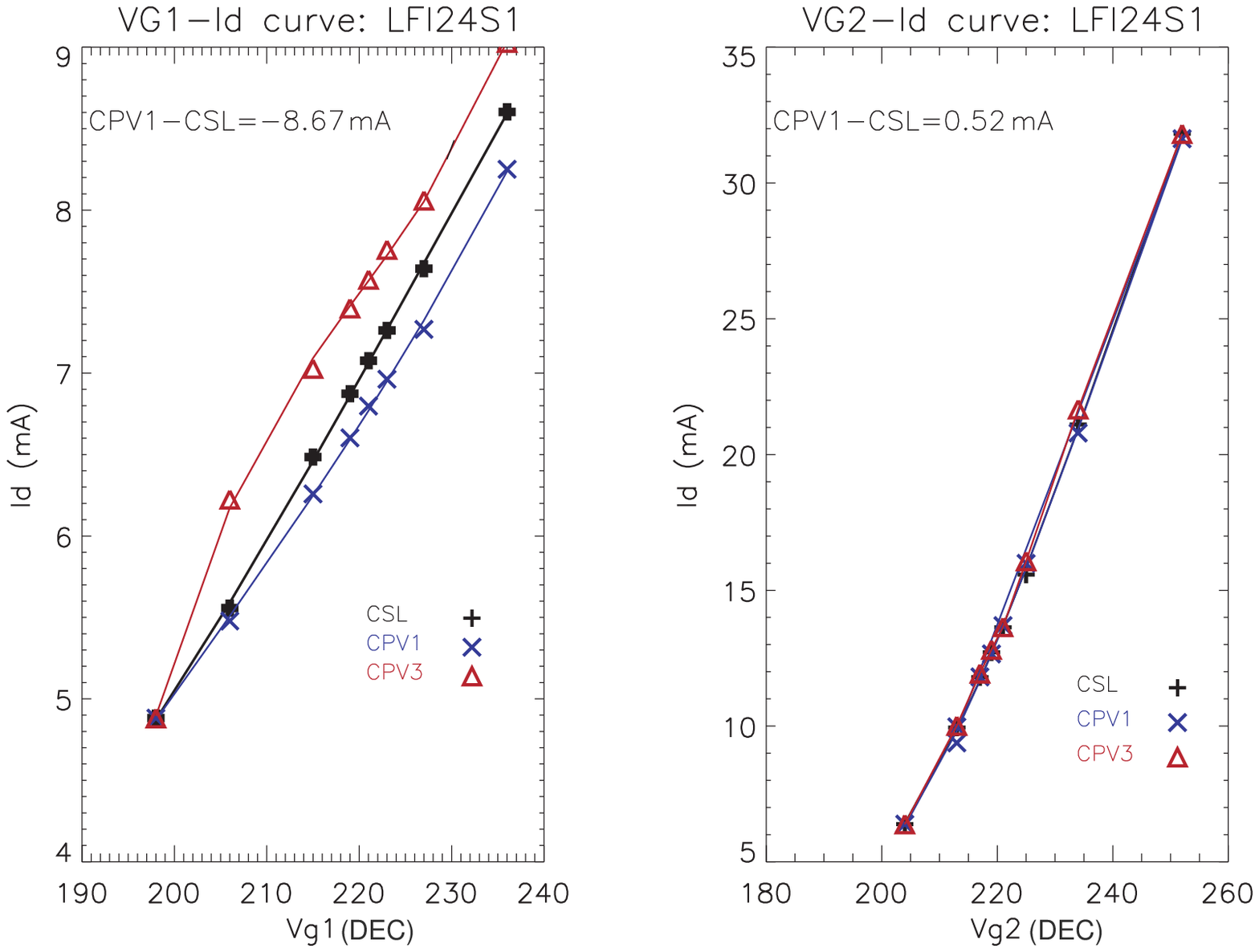}
            \includegraphics[width=7.0cm]{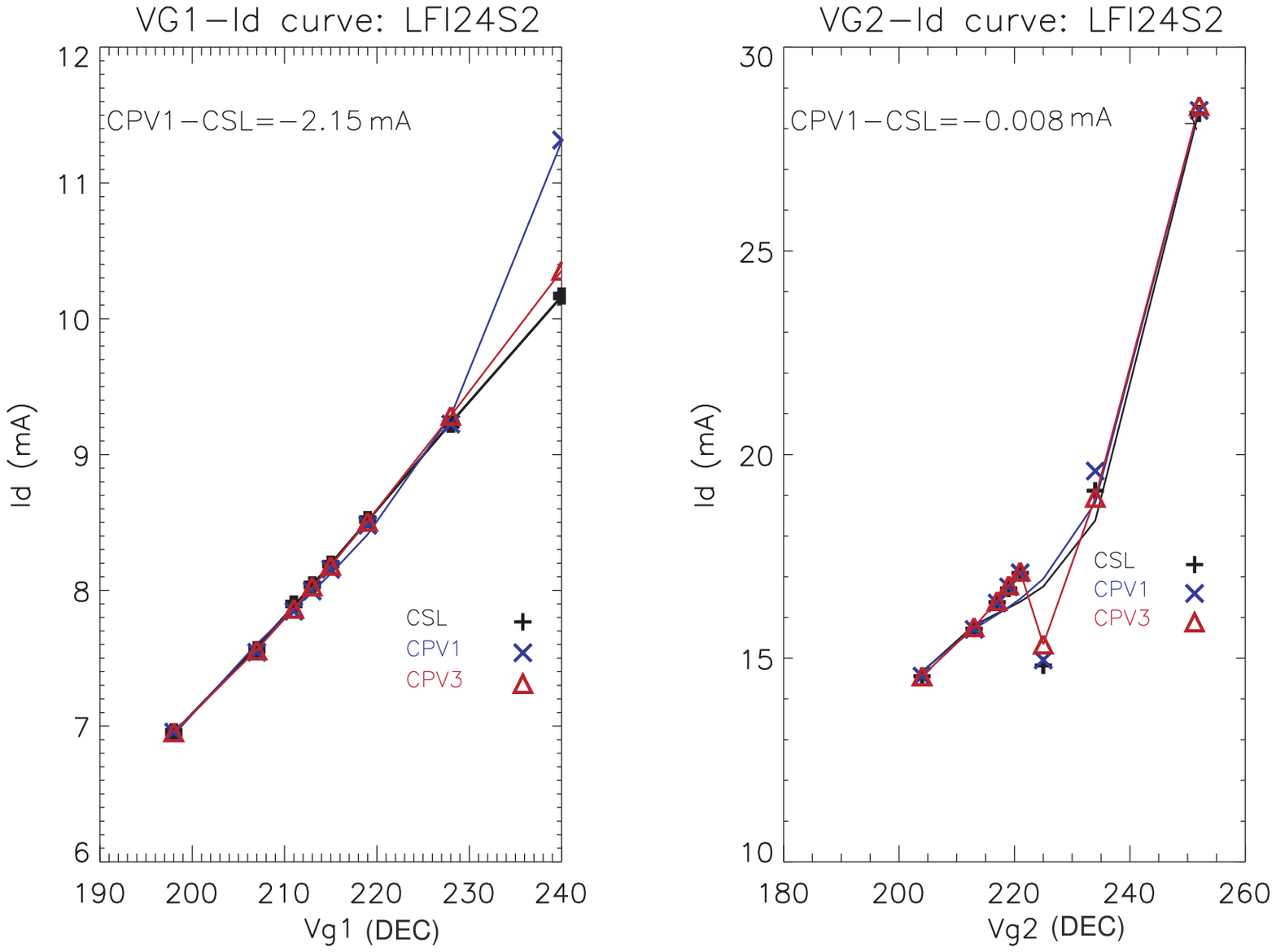}\\
            \textbf{LFI-25}\\ \vspace{0.1cm}
            \includegraphics[width=7.0cm]{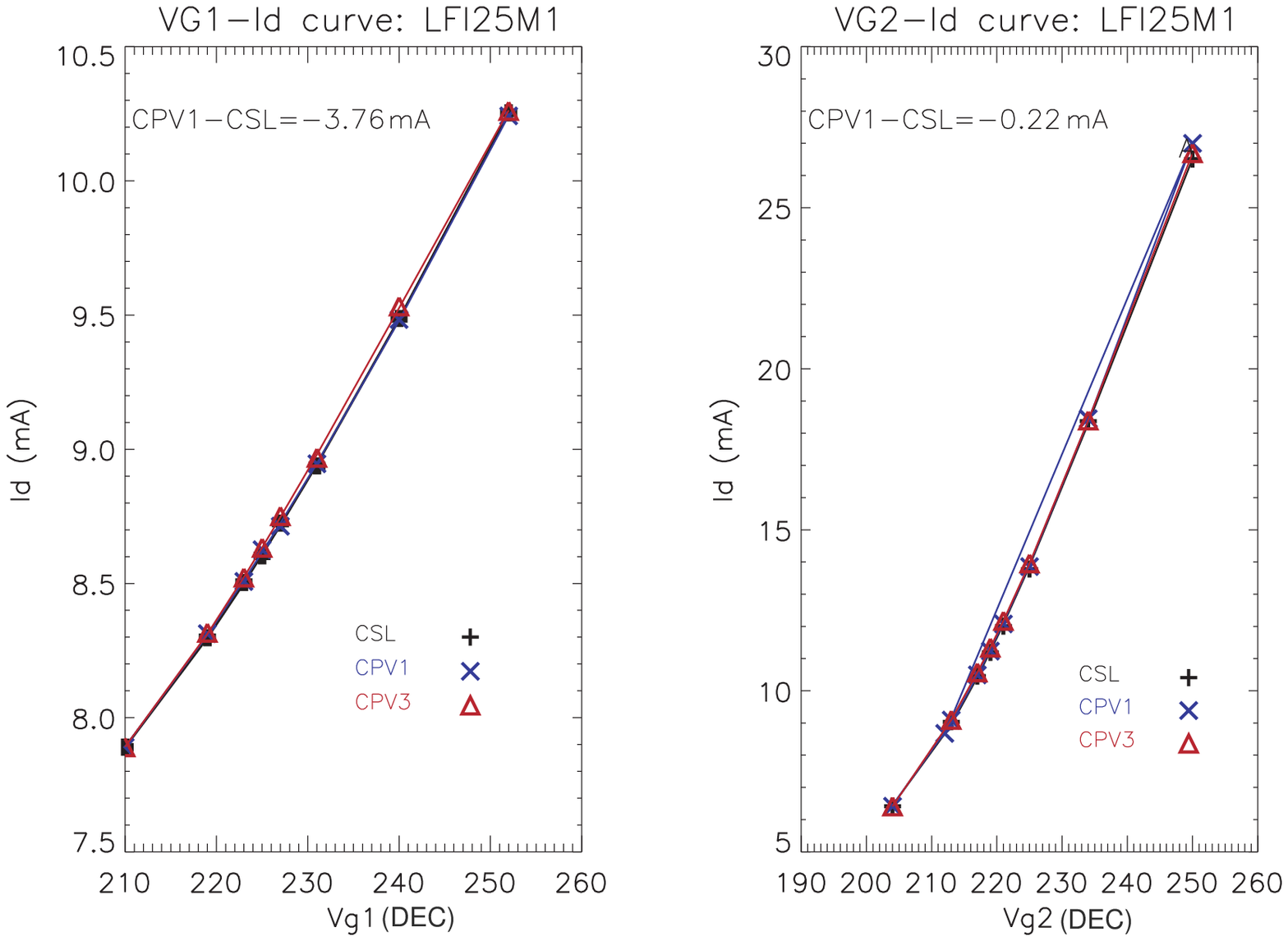}
            \includegraphics[width=7.0cm]{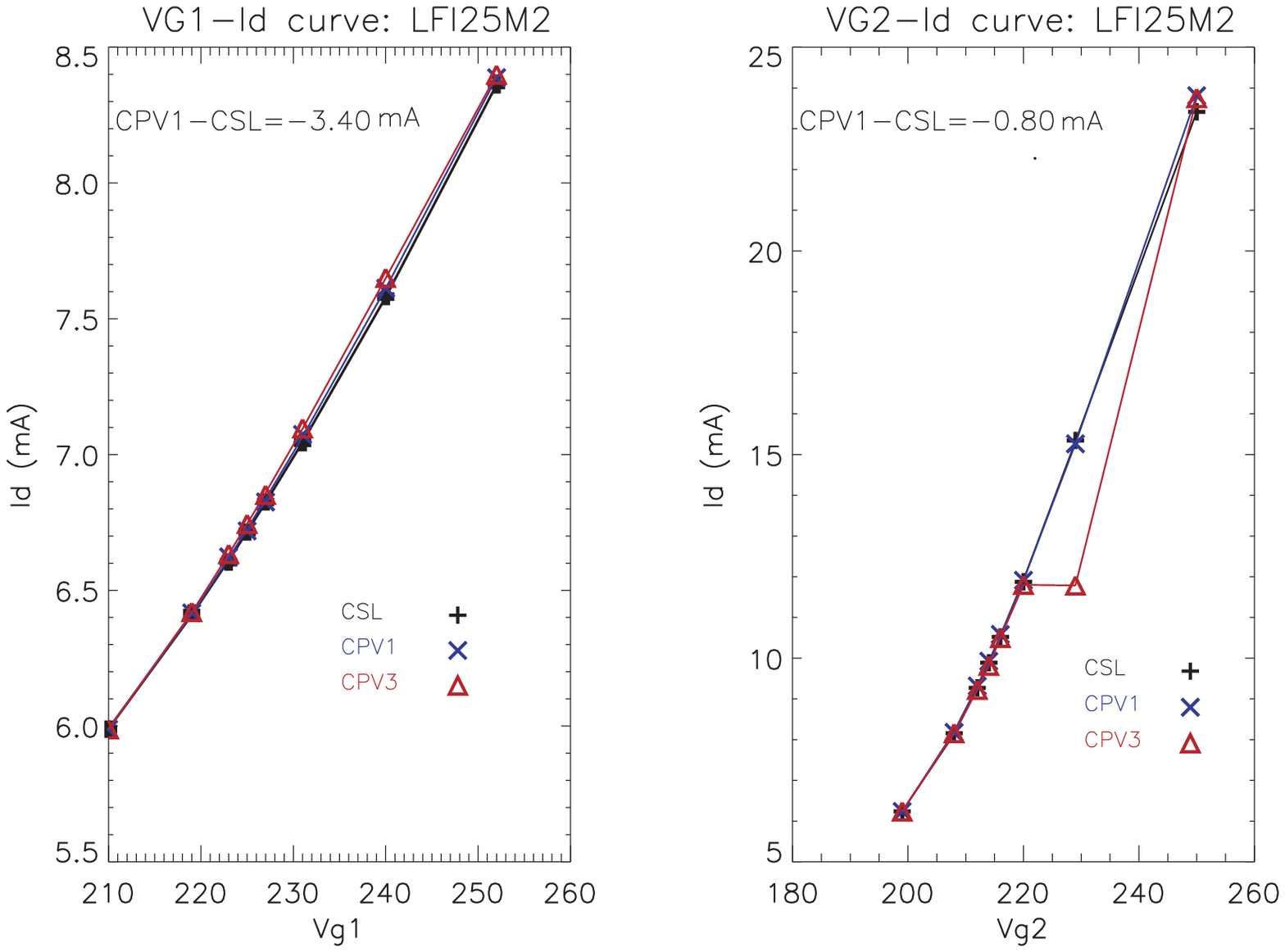}\\
            \includegraphics[width=7.0cm]{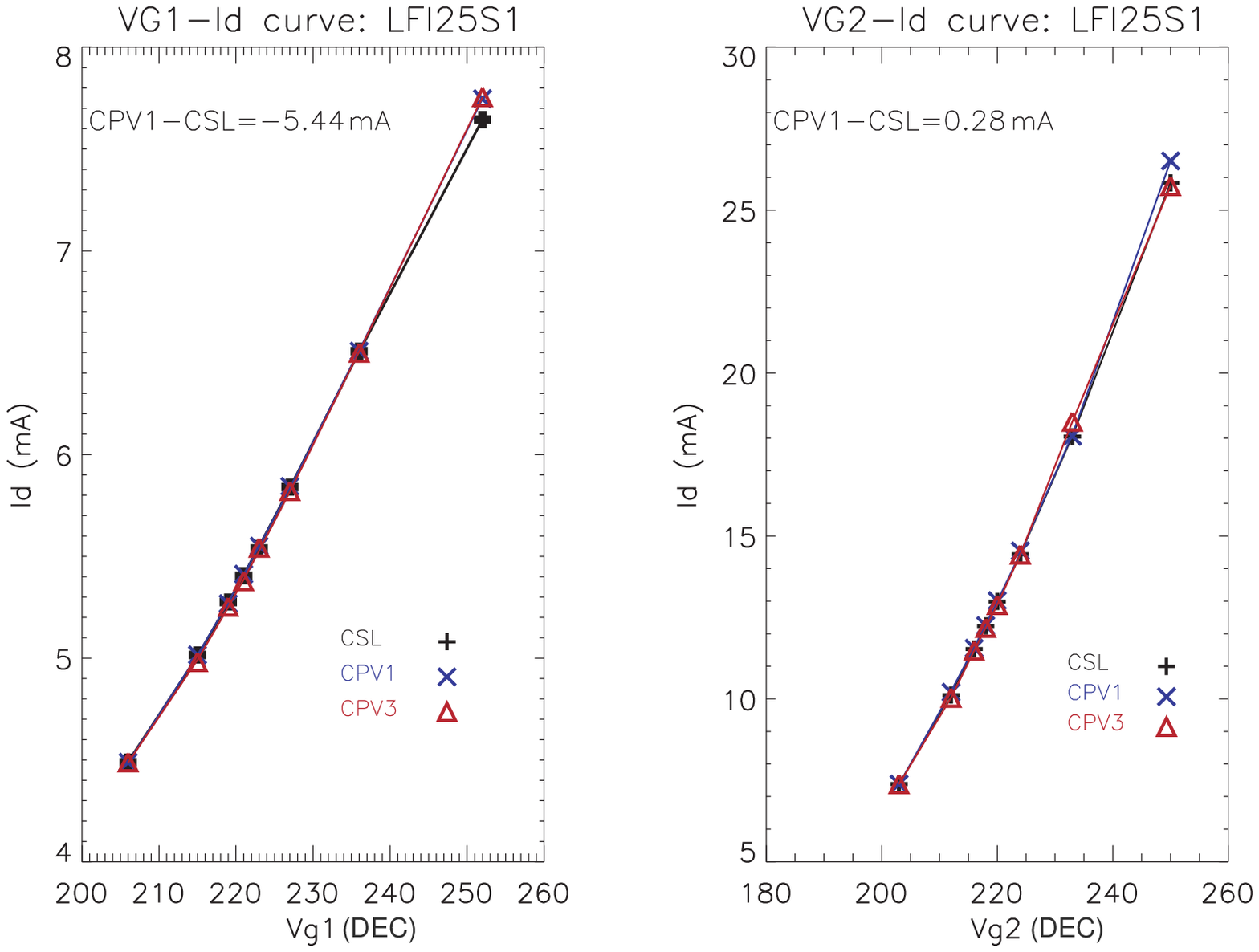}
            \includegraphics[width=7.0cm]{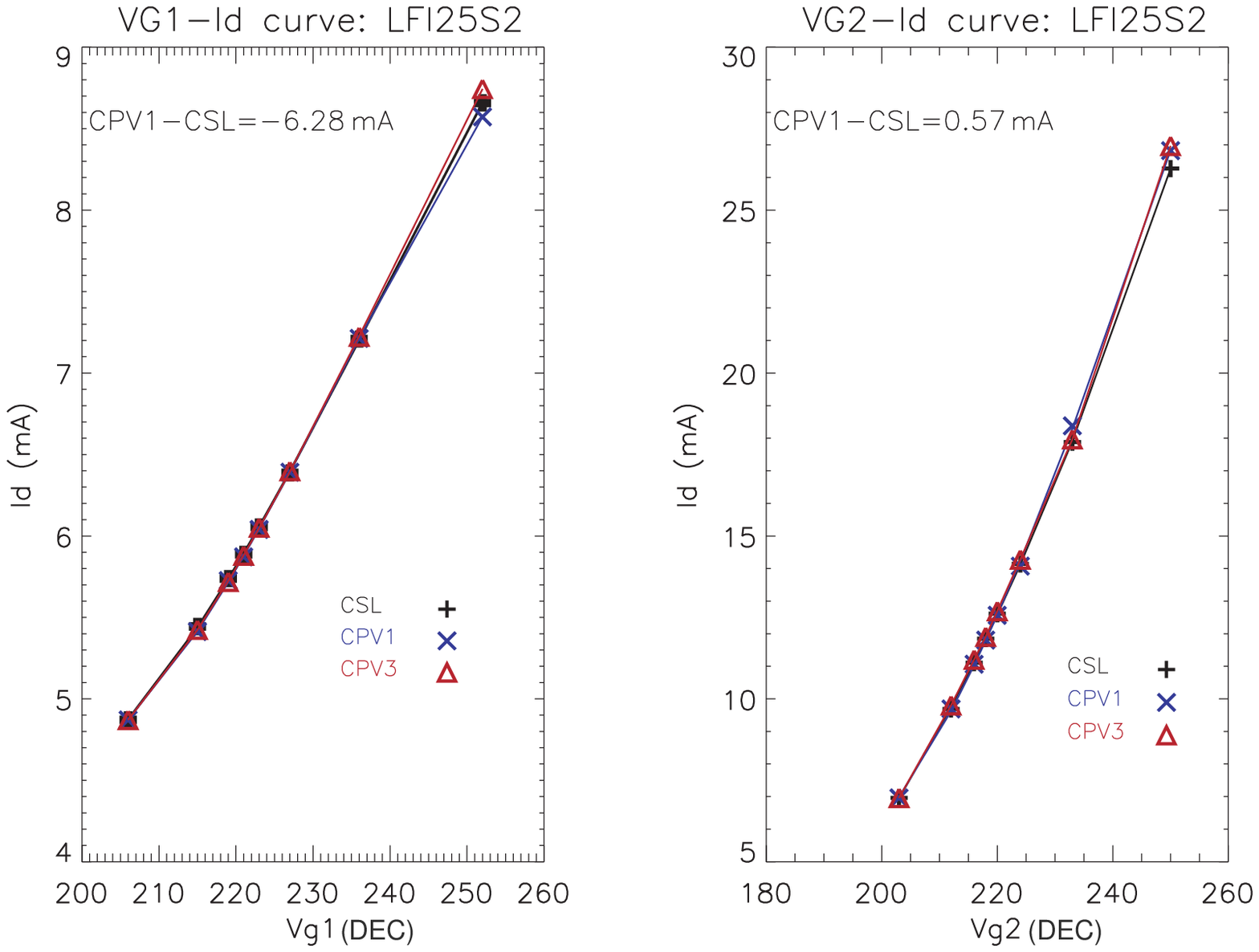}\\
        \end{center}
    \end{figure}
         \begin{figure}[htb]
        \begin{center} 
            \textbf{LFI-26}\\ \vspace{0.1cm}
            \includegraphics[width=7.0cm]{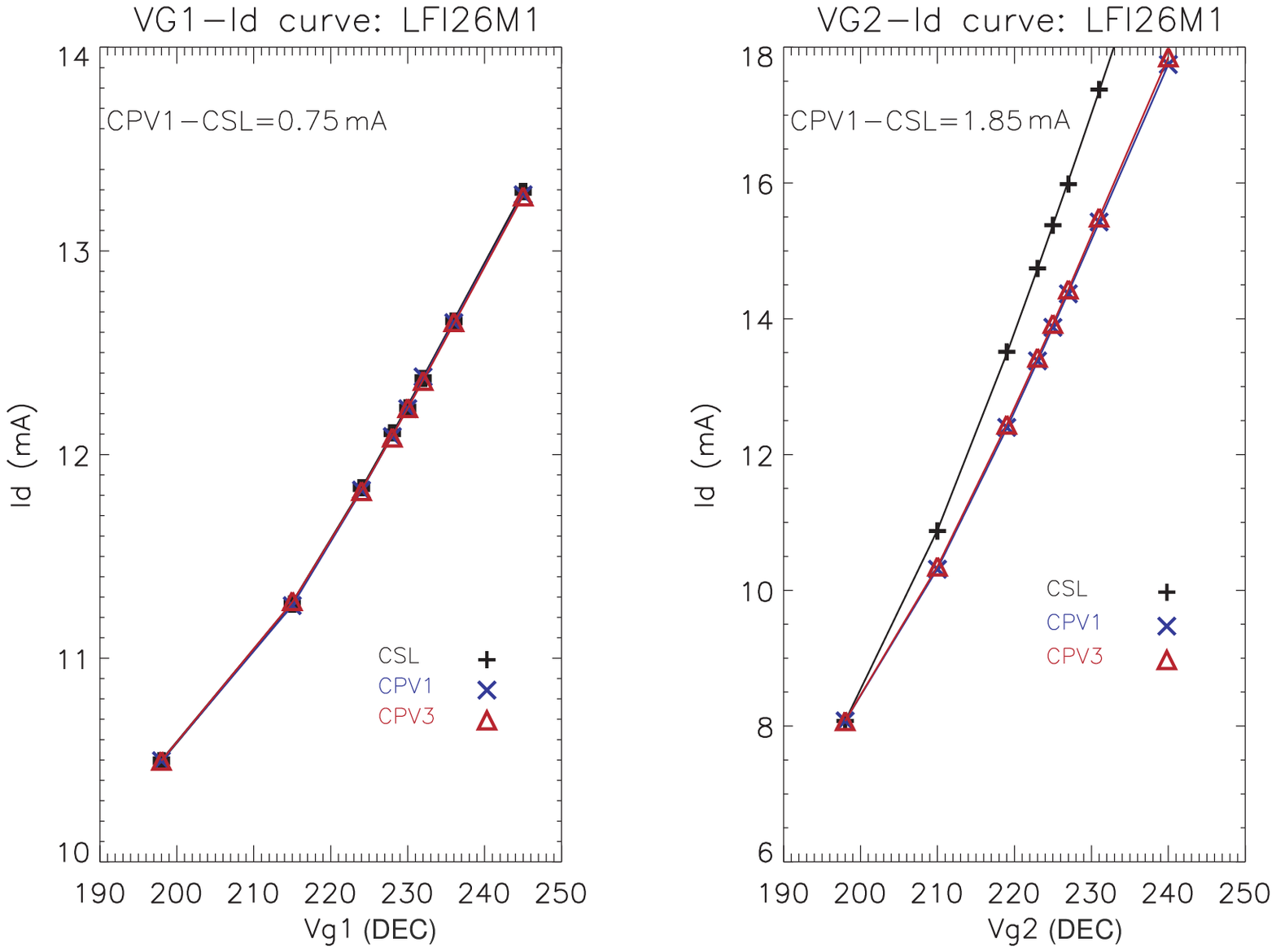}
            \includegraphics[width=7.0cm]{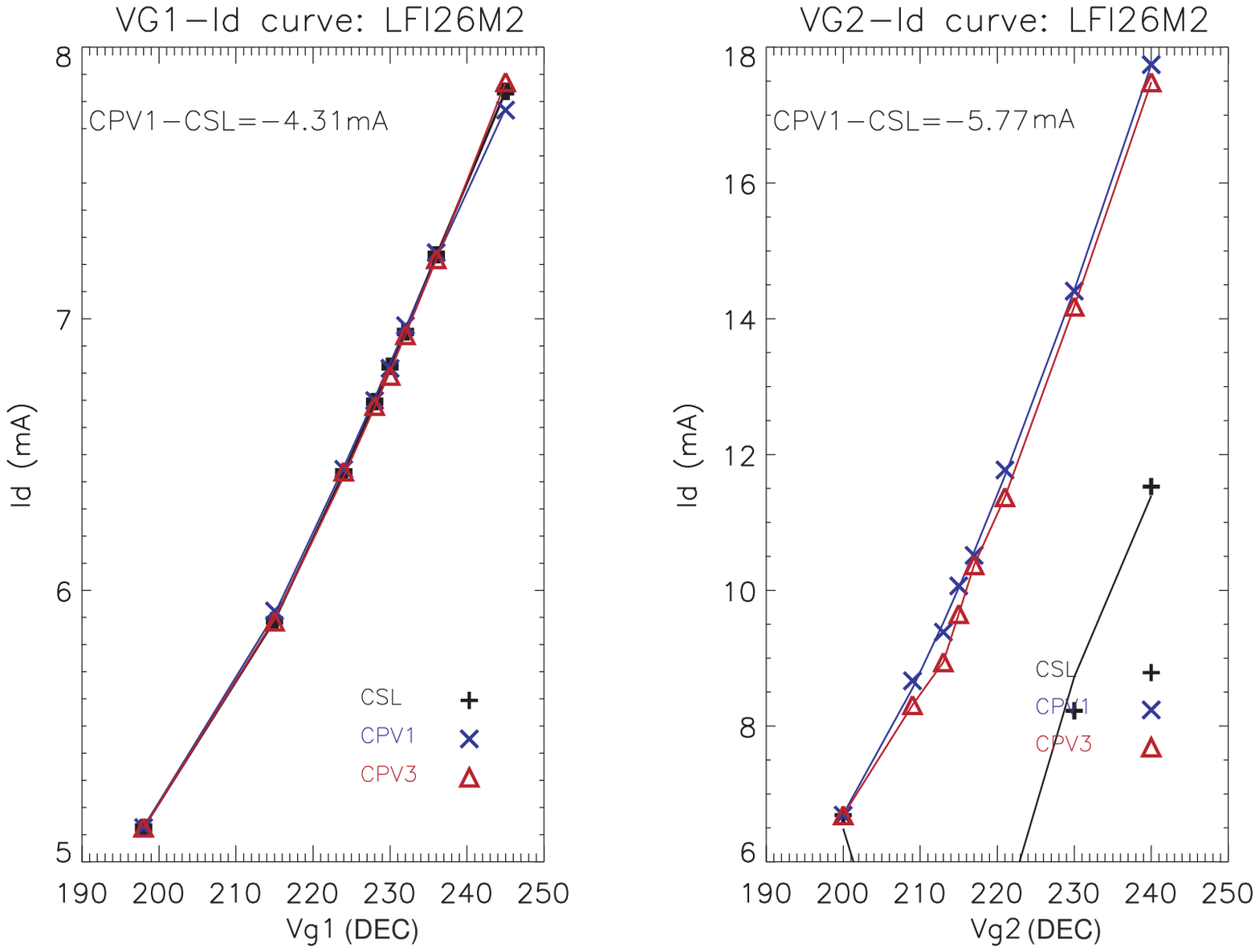}\\
            \includegraphics[width=7.0cm]{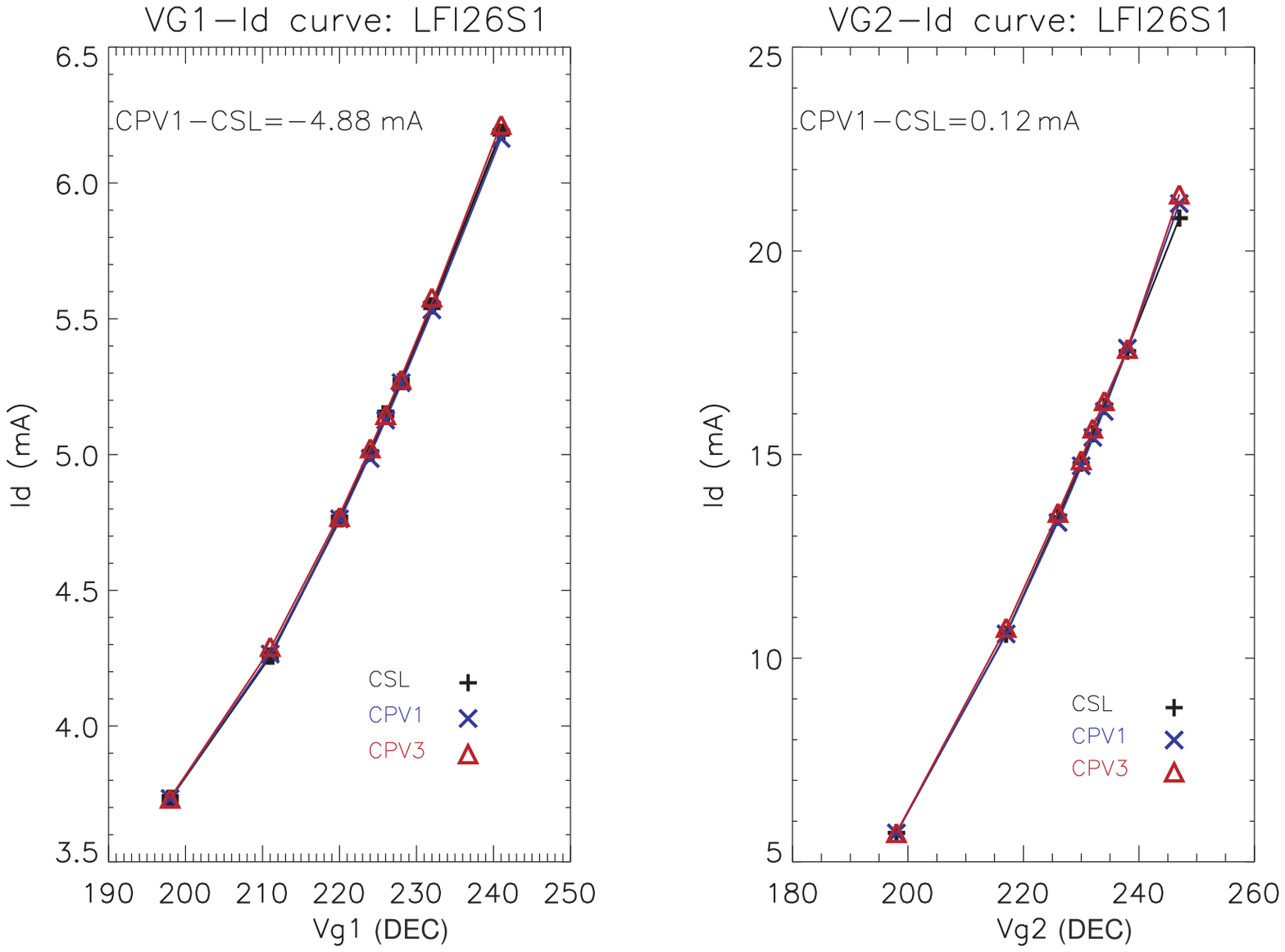}
            \includegraphics[width=7.0cm]{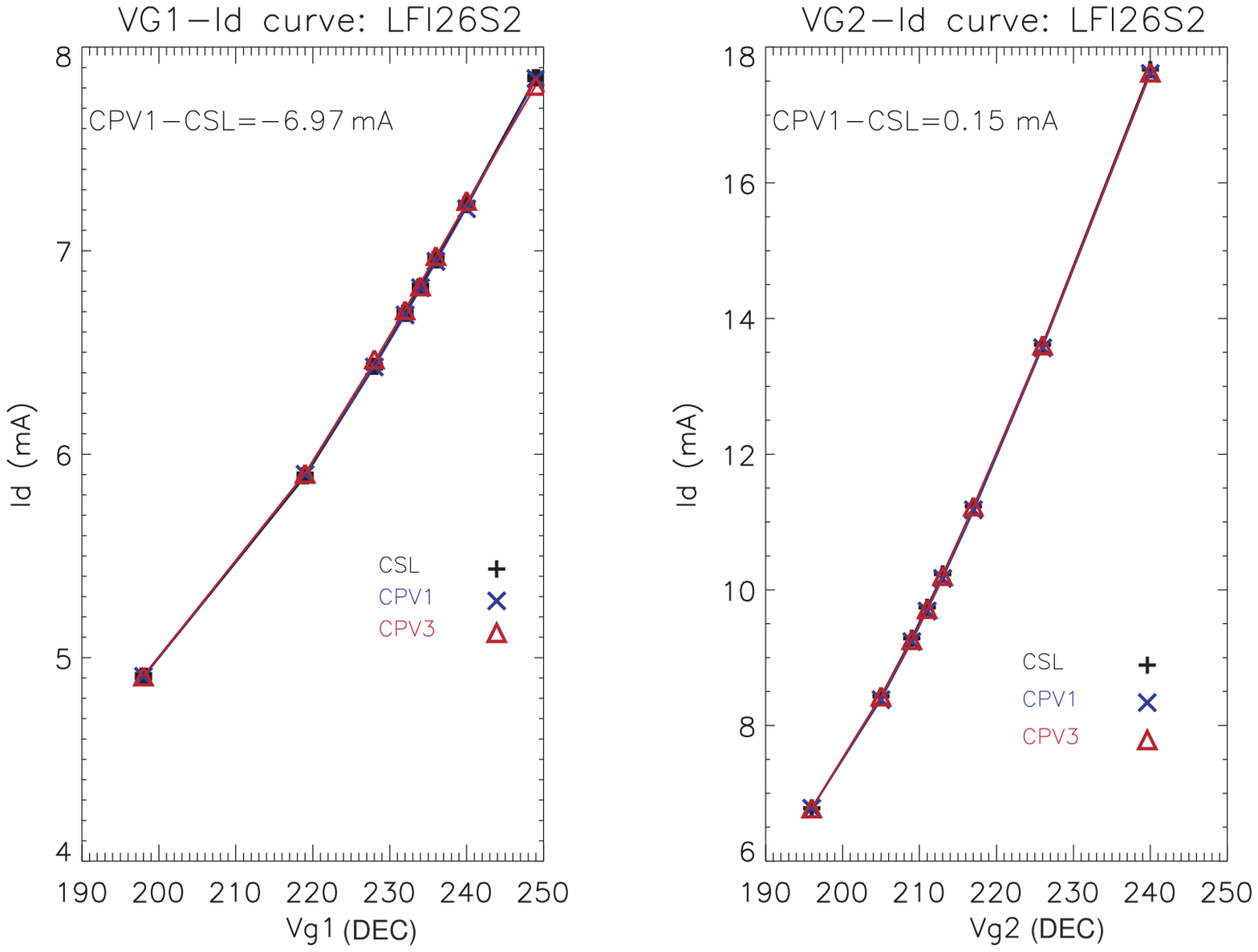}\\
            \clearpage
            \textbf{LFI-27}\\ \vspace{0.1cm}
            \includegraphics[width=7.0cm]{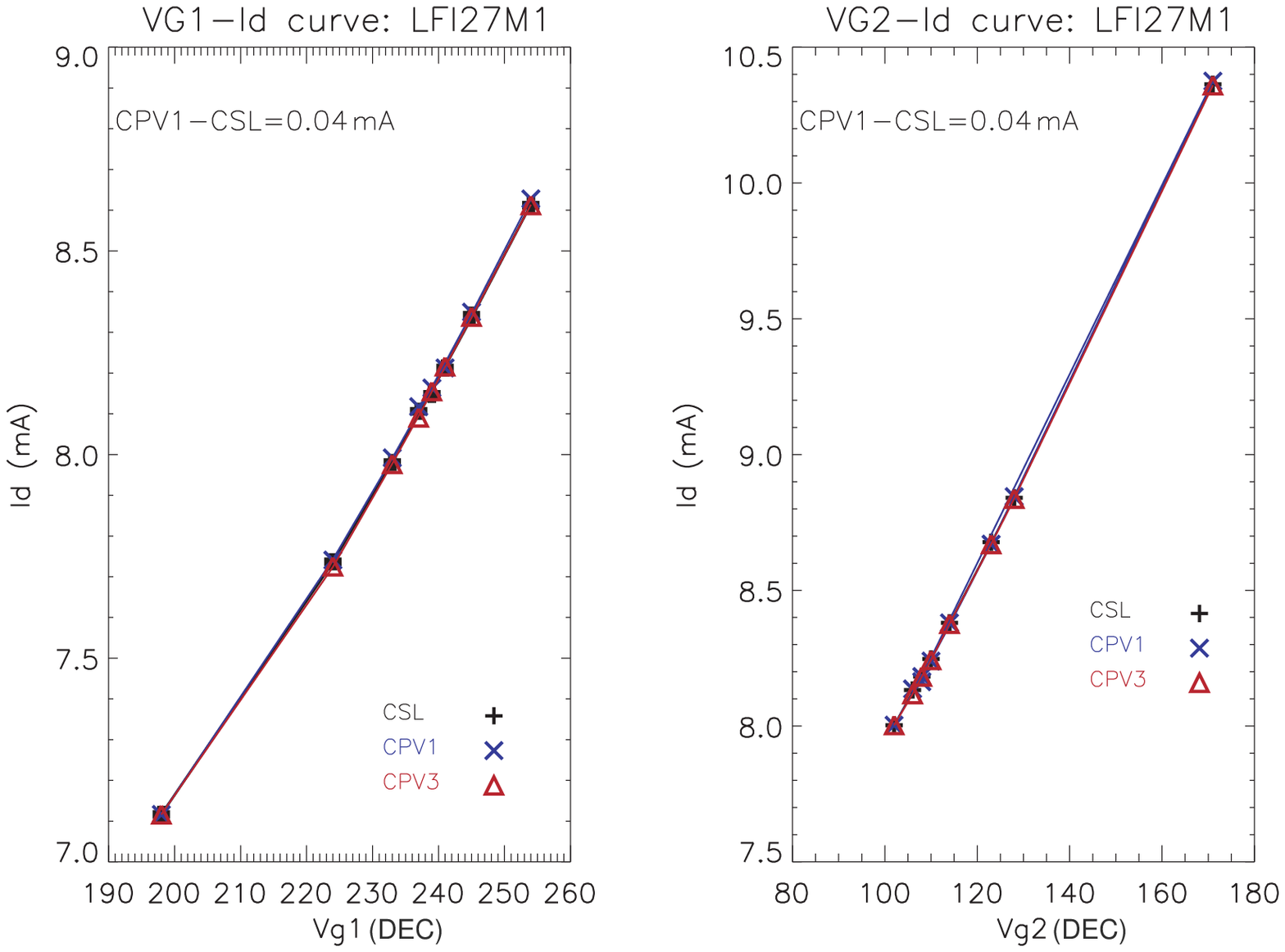}
            \includegraphics[width=7.0cm]{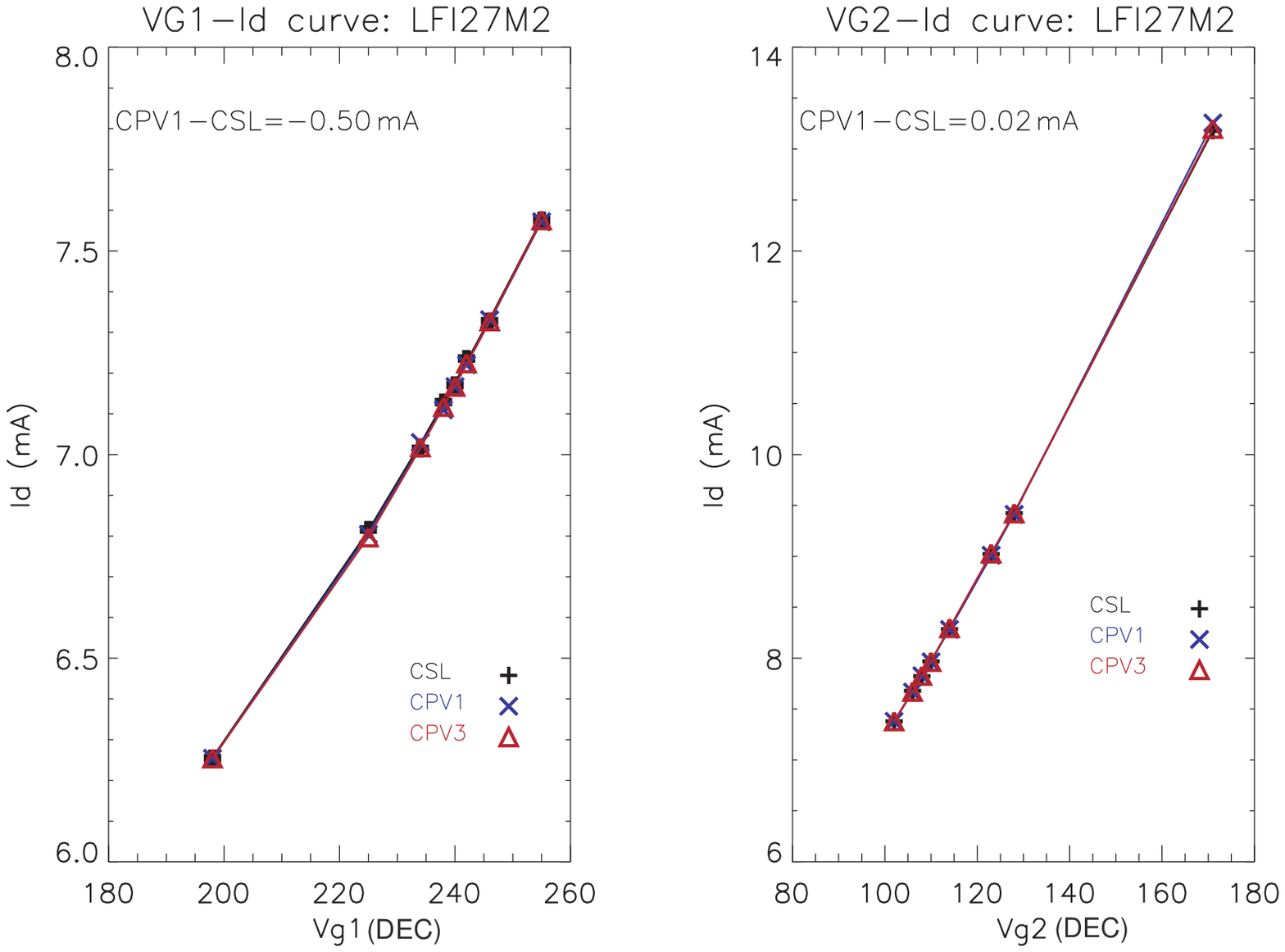}\\
            \includegraphics[width=7.0cm]{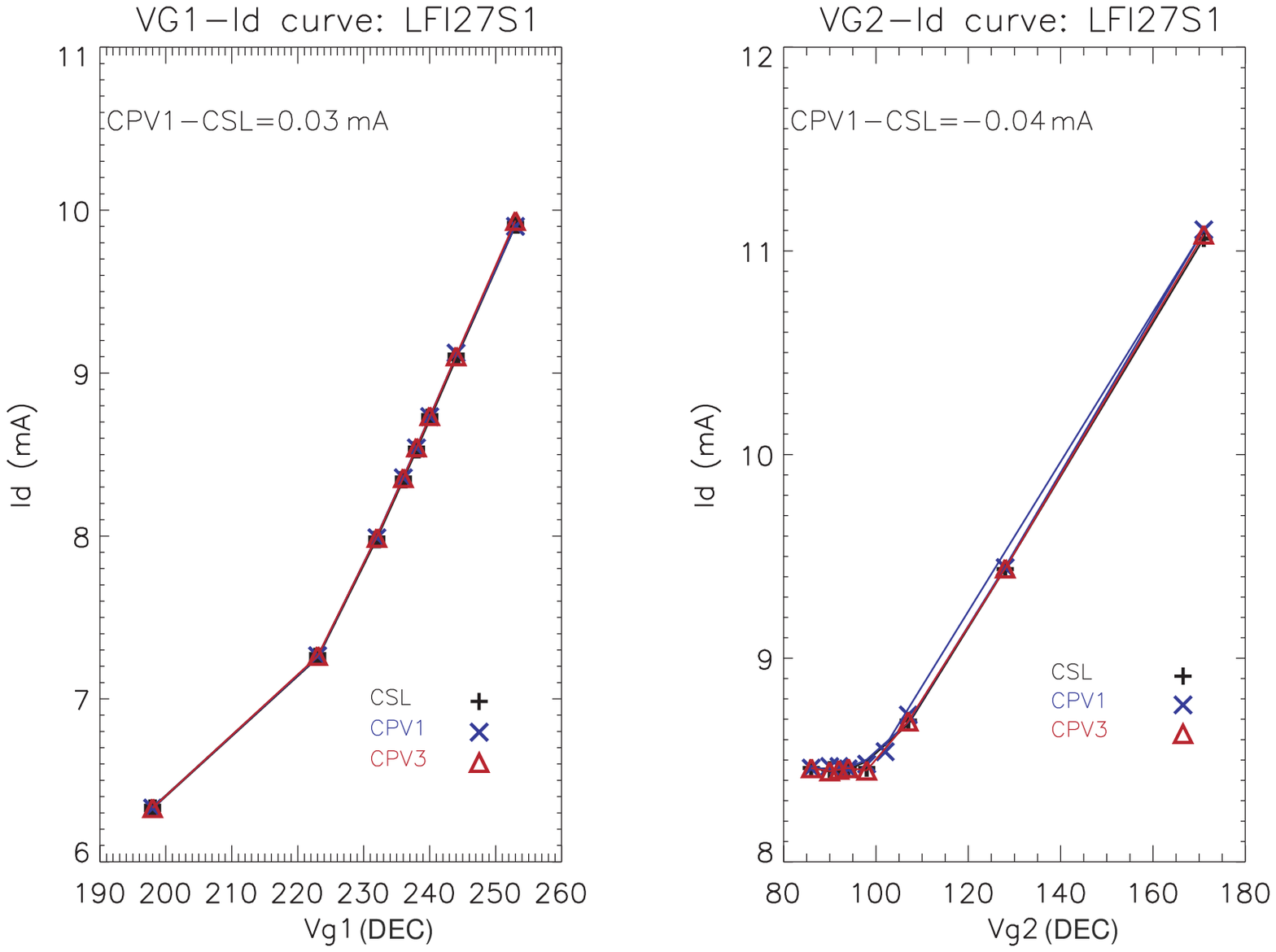}
            \includegraphics[width=7.0cm]{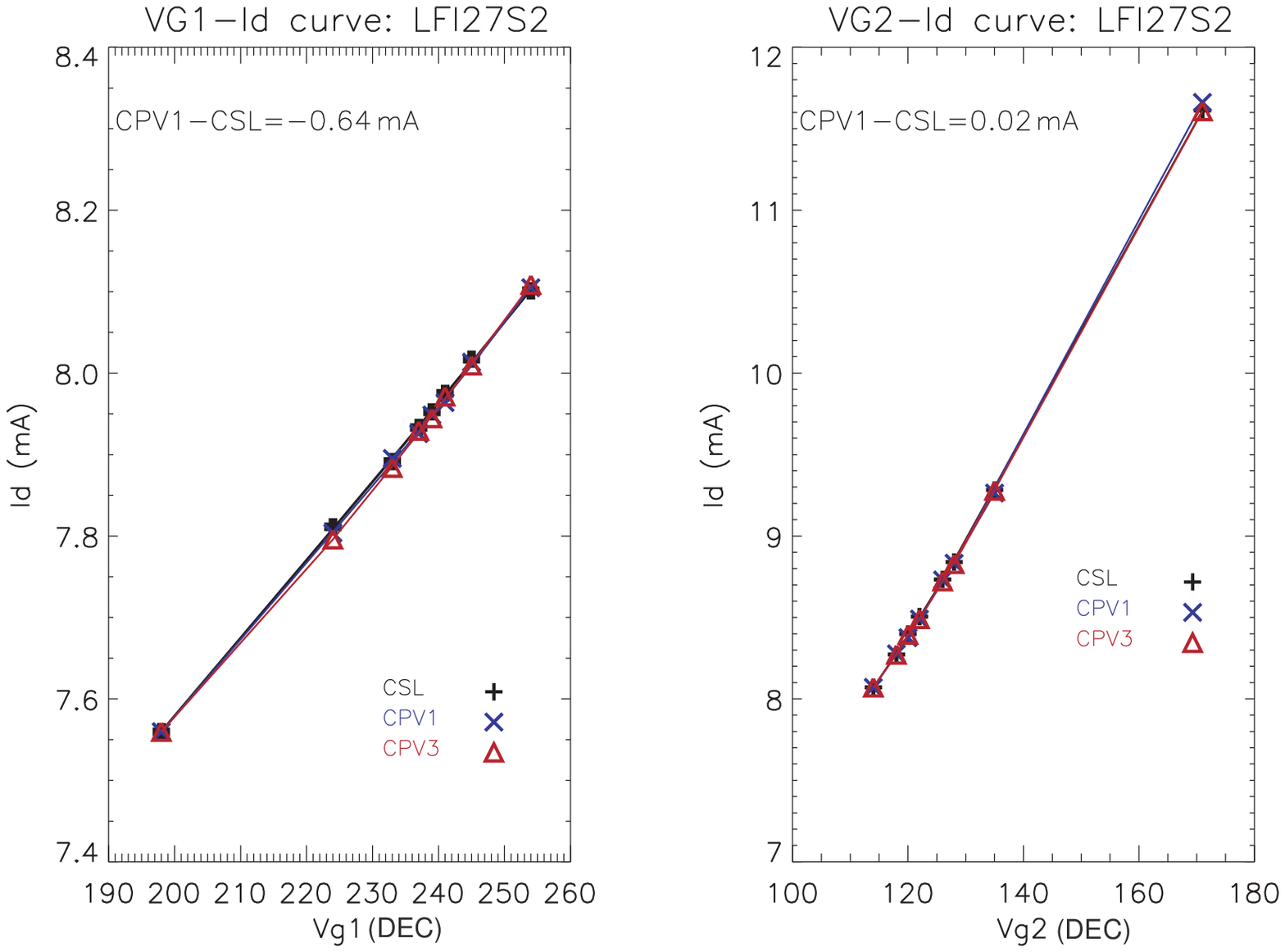}\\
        \end{center}
    \end{figure}
\clearpage

%% file: a06_PH_SW_tuning_plots.tex
\section{Phase switch labelling and tuning maps }
\label{app_PH_SW_plots}

    In Table~\ref{tab_phsw_corr} we show the correspondance between  LNA names (\texttt{M1}, \texttt{M2}, \texttt{S1}, \texttt{S2}) and the corresponding phase switch pairs (\texttt{A/C}, \texttt{B/D}) and BEM output diodes (\texttt{M-00}, \texttt{M-01}, \texttt{S-10}, \texttt{S-11}) for all RCAs.
\begin{table}[h!]
    \begin{center}
        \caption{Phase switch correspondance to LNAs and BEM diodes.}
        \label{tab_phsw_corr} \small
        \begin{tabular}{l  c  c c } 
 \hline  \hline   
\multirow{2}{*}{RCA}	&	\multicolumn{1}{c}{\texttt{Amplifier}}	&	\multicolumn{1}{c}{\texttt{Phase switch state}	}&	\multirow{2}{*}{\texttt{Detector}}	\\ 
&	 \multicolumn{1}{c}{\texttt{\& Phase switch}	}&	\multicolumn{1}{c}{\texttt{\& 4 kHz}}	& \\ \hline
%RCA	&		\texttt{Amplifier \& Phase switch}	&	\texttt{ \& 4 kHz}	&	\texttt{Detector}	\\ \hline
\multirow{4}{*}{LFI18}	&	S2	&	A/C	&	\multirow{2}{*}{S-10; S-11}	\\
	&	S1	&	B/D	&		\\  
	&	M1	&	A/C	&	\multirow{2}{*}{M-00; M-01}	\\
	&	M2	&	B/D	&		\\         \hline
\multirow{4}{*}{LFI19}	&	S2	&	A/C	&	\multirow{2}{*}{S-10; S-11}	\\
	&	S1	&	B/D	&		\\  
	&	M1	&	A/C	&	\multirow{2}{*}{M-00; M-01}	\\
	&	M2	&	B/D	&		\\         \hline
\multirow{4}{*}{LFI20}	&	S2	&	A/C	&	\multirow{2}{*}{S-10; S-11}	\\
	&	S1	&	B/D	&		\\  
	&	M1	&	A/C	&	\multirow{2}{*}{M-00; M-01}	\\
	&	M2	&	B/D	&		\\         \hline
\multirow{4}{*}{LFI21}	&	S2	&	A/C	&	\multirow{2}{*}{S-10; S-11}	\\
	&	S1	&	B/D	&		\\  
	&	M1	&	A/C	&	\multirow{2}{*}{M-00; M-01}	\\
	&	M2	&	B/D	&		\\         \hline
\multirow{4}{*}{LFI22}	&	S2	&	A/C	&	\multirow{2}{*}{S-10; S-11}	\\
	&	S1	&	B/D	&		\\  
	&	M1	&	A/C	&	\multirow{2}{*}{M-00; M-01}	\\
	&	M2	&	B/D	&		\\         \hline
\multirow{4}{*}{LFI23}	&	S2	&	A/C	&	\multirow{2}{*}{S-10; S-11}	\\
	&	S1	&	B/D	&		\\  
	&	M1	&	A/C	&	\multirow{2}{*}{M-00; M-01}	\\
	&	M2	&	B/D	&		\\         \hline
\multirow{4}{*}{LFI24}	&	M2	&	A/C	&	\multirow{2}{*}{M-00; M-01}	\\
	&	M1	&	B/D	&		\\  
	&	S2	&	A/C	&	\multirow{2}{*}{S-10; S-11}	\\
	&	S1	&	B/D	&		\\         \hline
\multirow{4}{*}{LFI25}	&	M1	&	A/C	&	\multirow{2}{*}{M-00; M-01}	\\
	&	M2	&	B/D	&		\\  
	&	S1	&	A/C	&	\multirow{2}{*}{S-10; S-11}	\\
	&	S2	&	B/D	&		\\         \hline
\multirow{4}{*}{LFI26}	&	M2	&	A/C	&	\multirow{2}{*}{M-00; M-01}	\\
	&	M1	&	B/D	&		\\  
	&	S2	&	A/C	&	\multirow{2}{*}{S-10; S-11}	\\
	&	S1	&	B/D	&		\\         \hline
\multirow{4}{*}{LFI27}	&	M1	&	A/C	&	\multirow{2}{*}{M-00; M-01}	\\
	&	M2	&	B/D	&		\\  
	&	S1	&	A/C	&	\multirow{2}{*}{S-10; S-11}	\\
	&	S2	&	B/D	&		\\         \hline
\multirow{4}{*}{LFI28}	&	M1	&	A/C	&	\multirow{2}{*}{M-00; M-01}	\\
	&	M2	&	B/D	&		\\  
	&	S1	&	A/C	&	\multirow{2}{*}{S-10; S-11}	\\
	&	S2	&	B/D	&		\\         \hline
        \end{tabular}
    \end{center}
\end{table}

\begin{figure}[htb]
        \begin{center}
           	\textbf{LFI-24}\\
            \includegraphics[width=3.5cm]{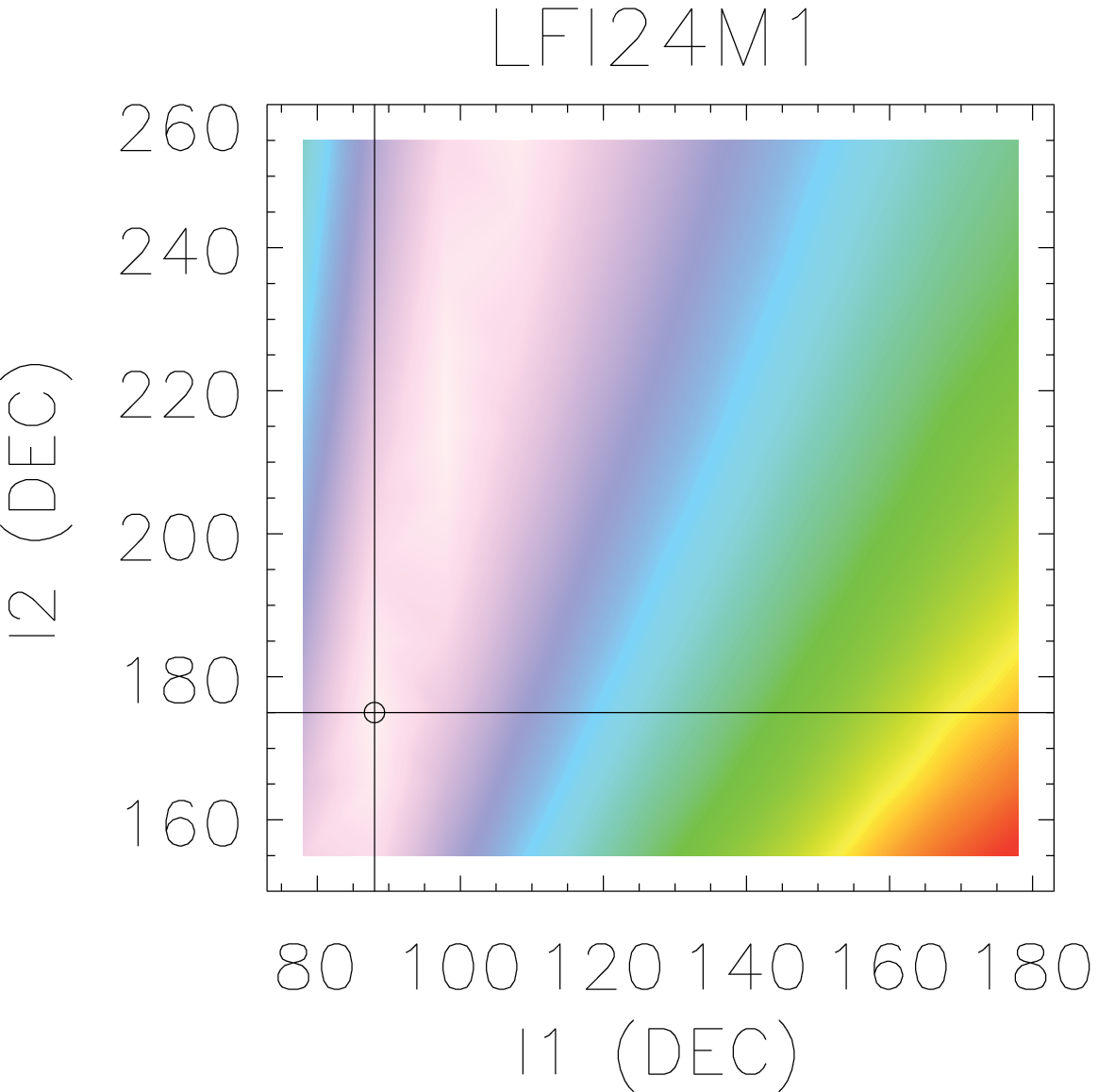}
            \includegraphics[width=3.5cm]{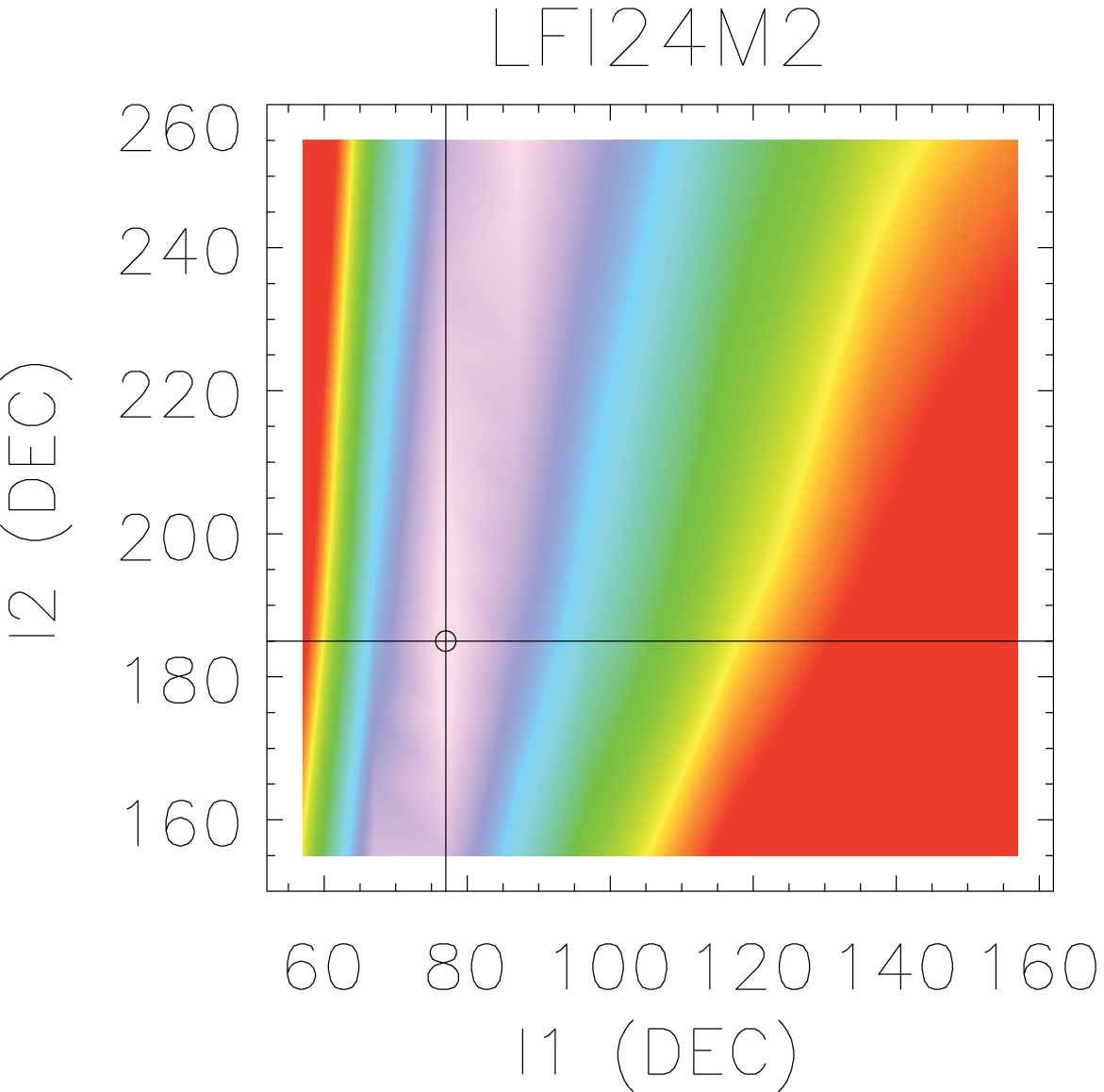}
            \includegraphics[width=3.5cm]{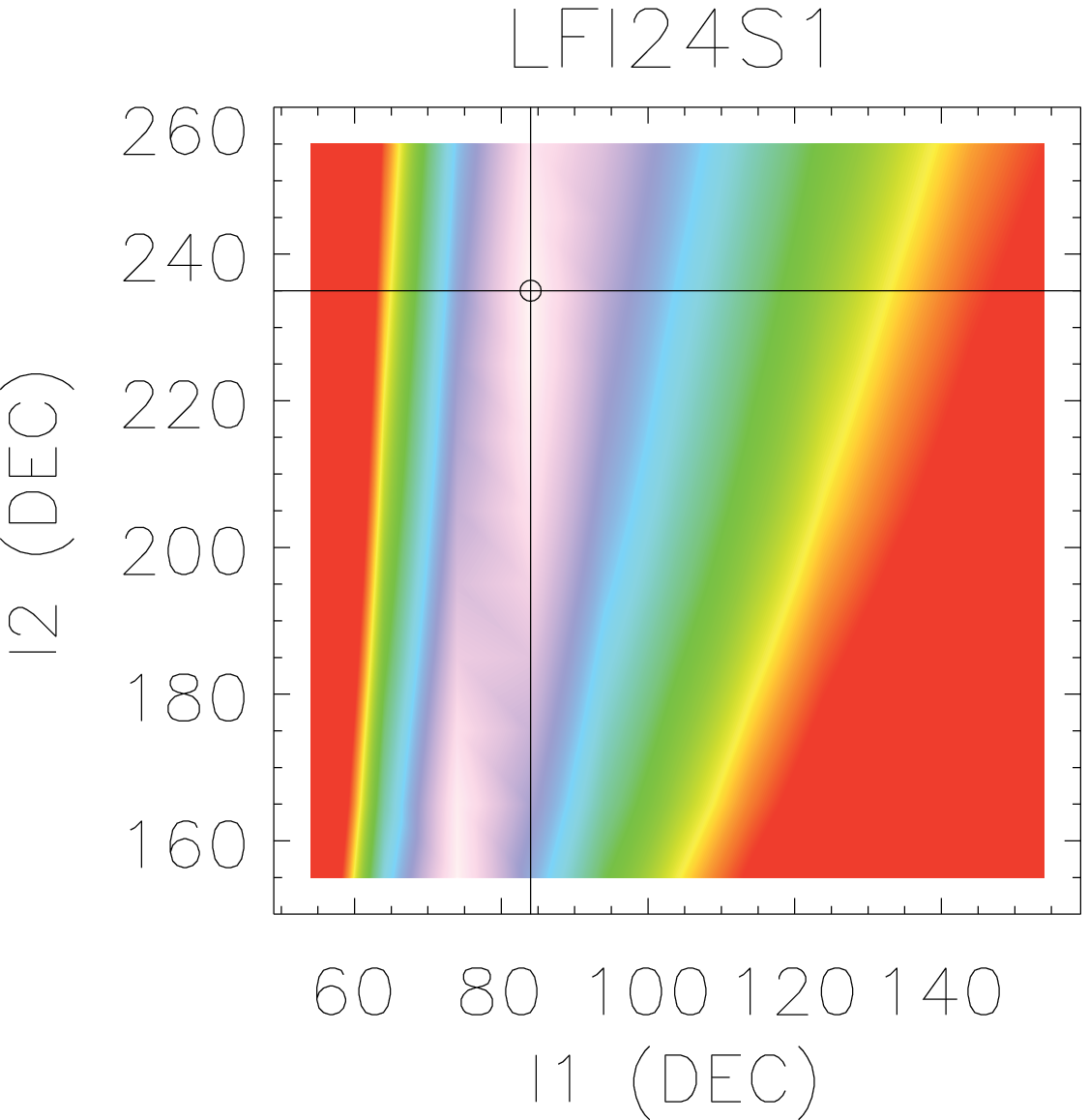}
            \includegraphics[width=3.5cm]{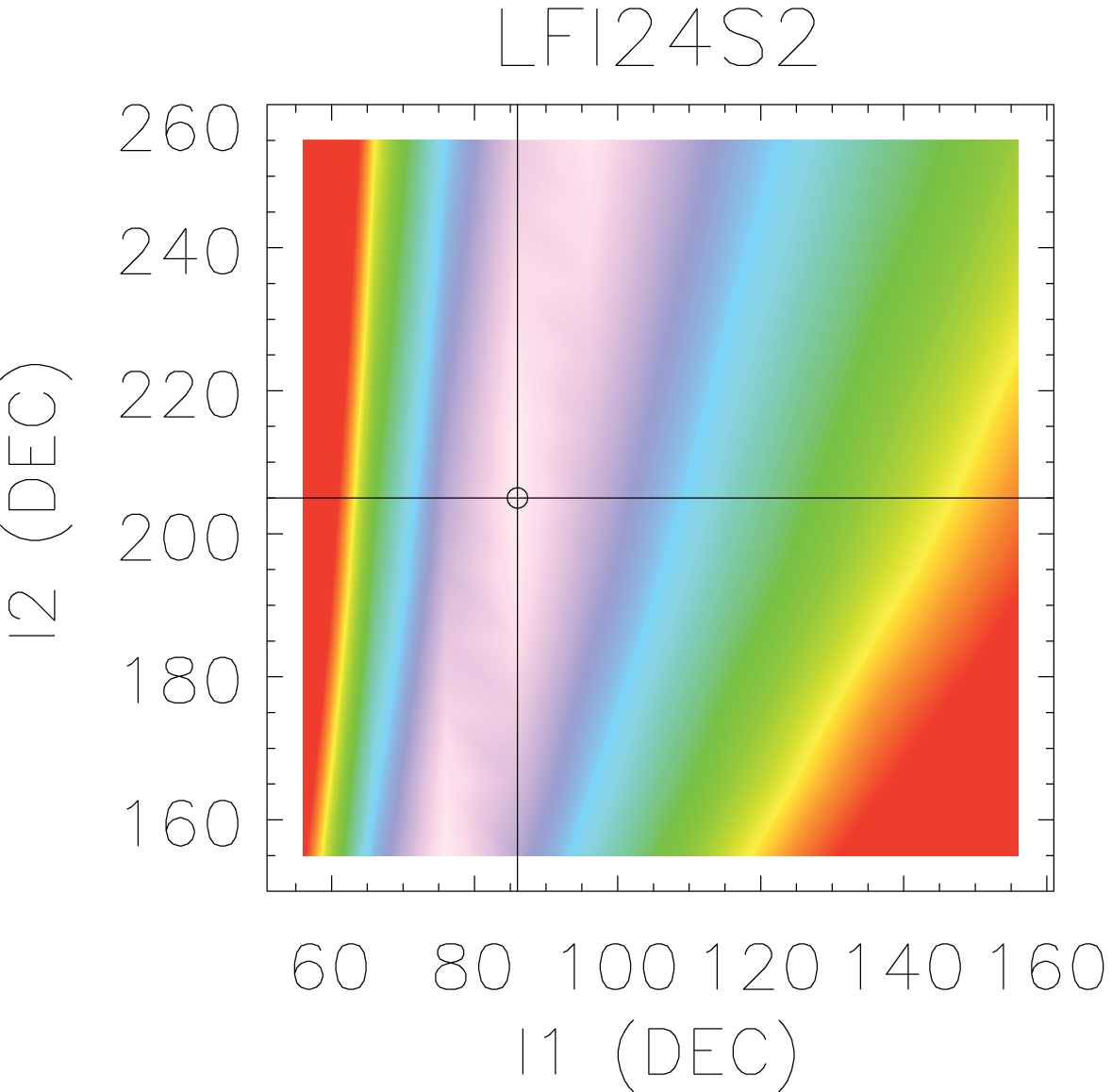}\\
            \textbf{LFI-25}\\
            \includegraphics[width=3.5cm]{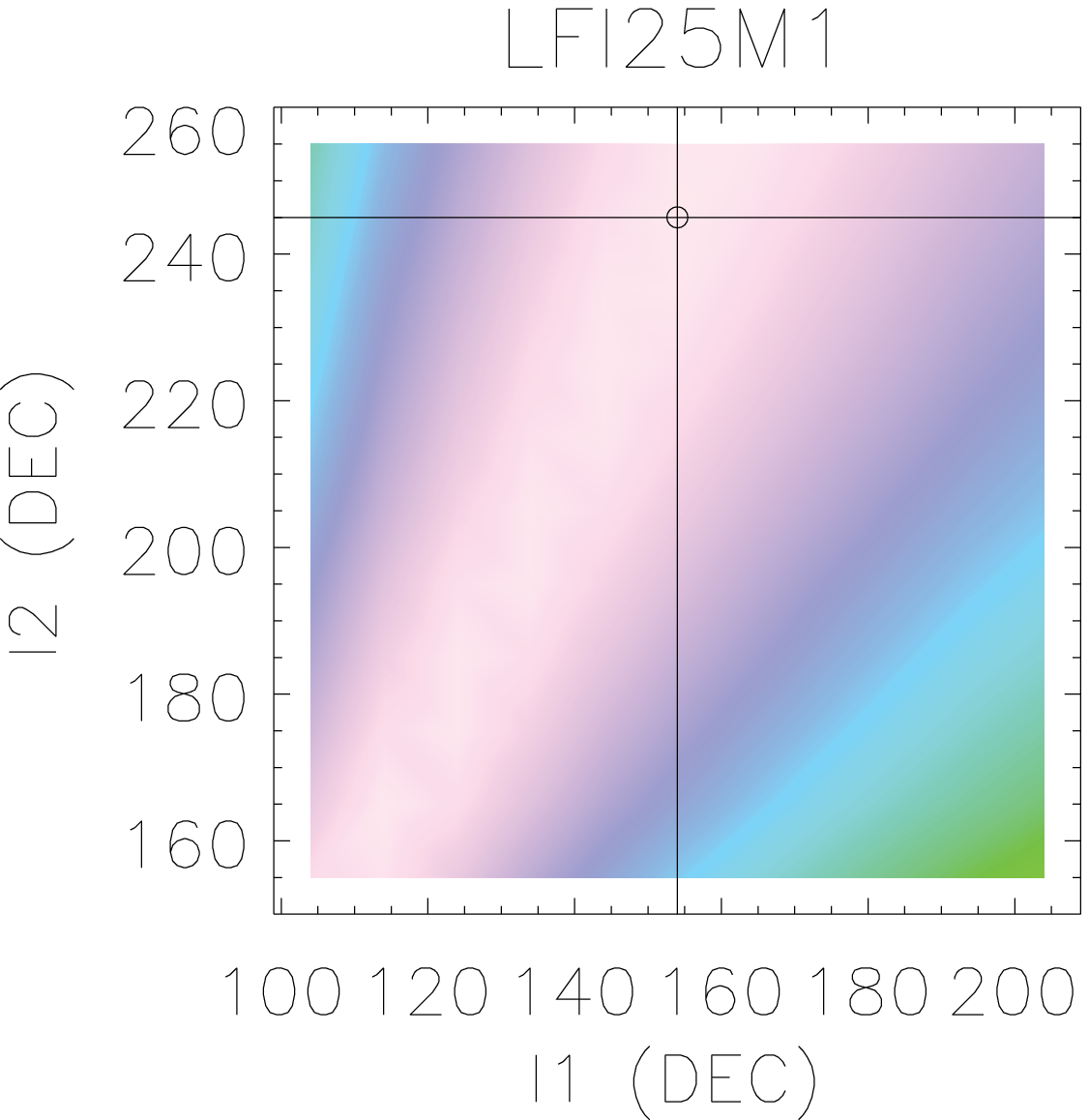}
            \includegraphics[width=3.5cm]{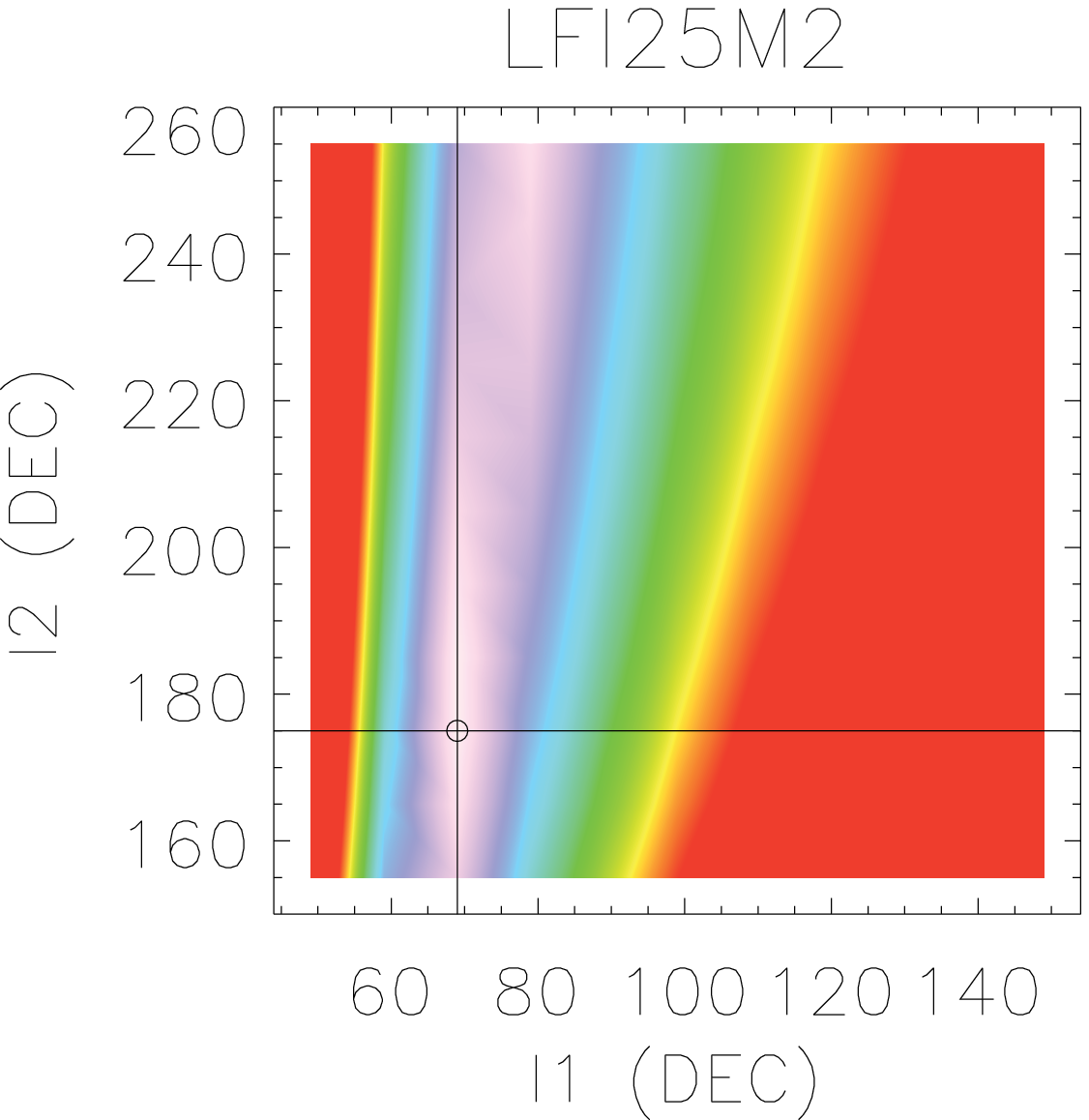}
            \includegraphics[width=3.5cm]{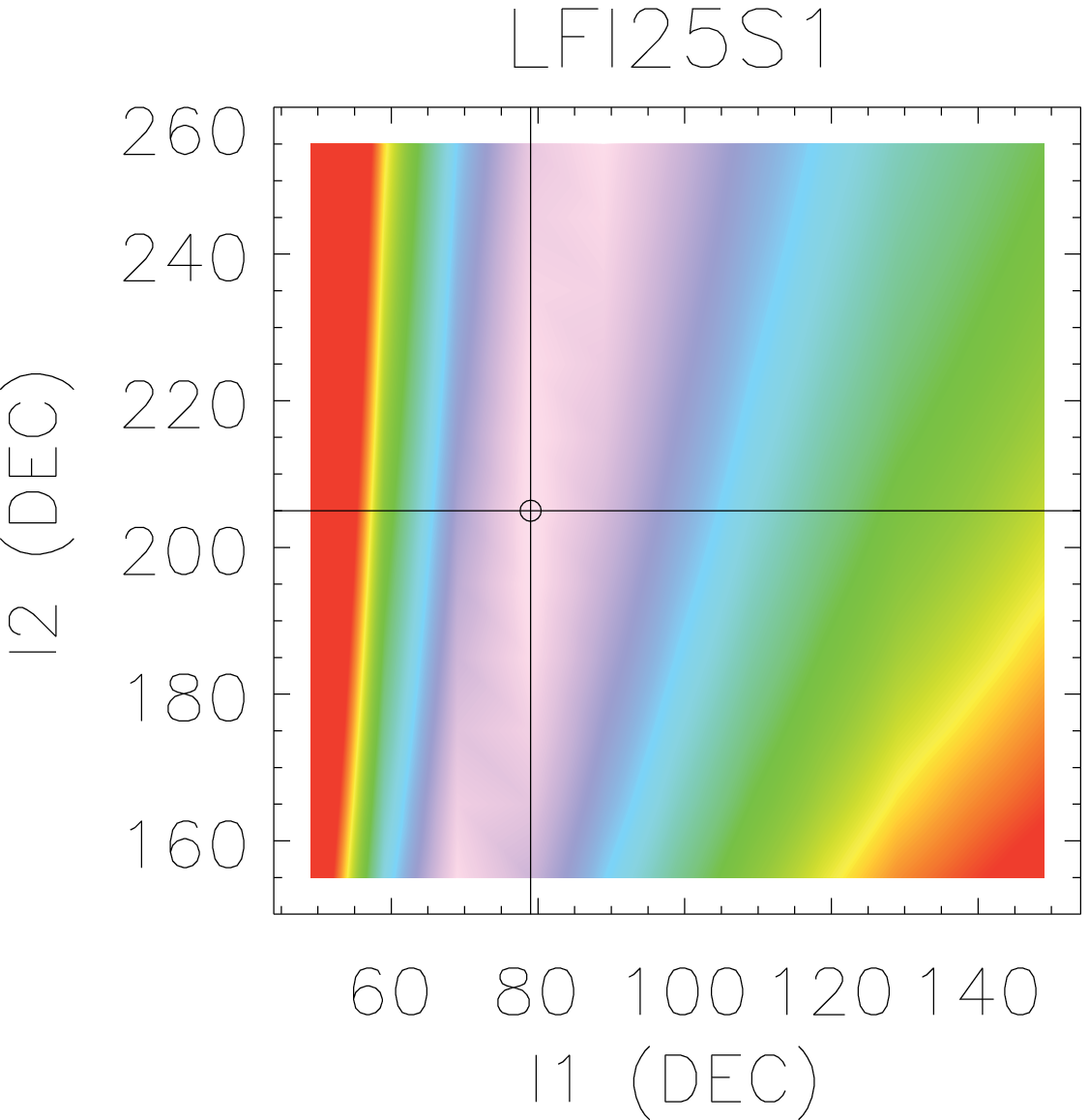}
            \includegraphics[width=3.5cm]{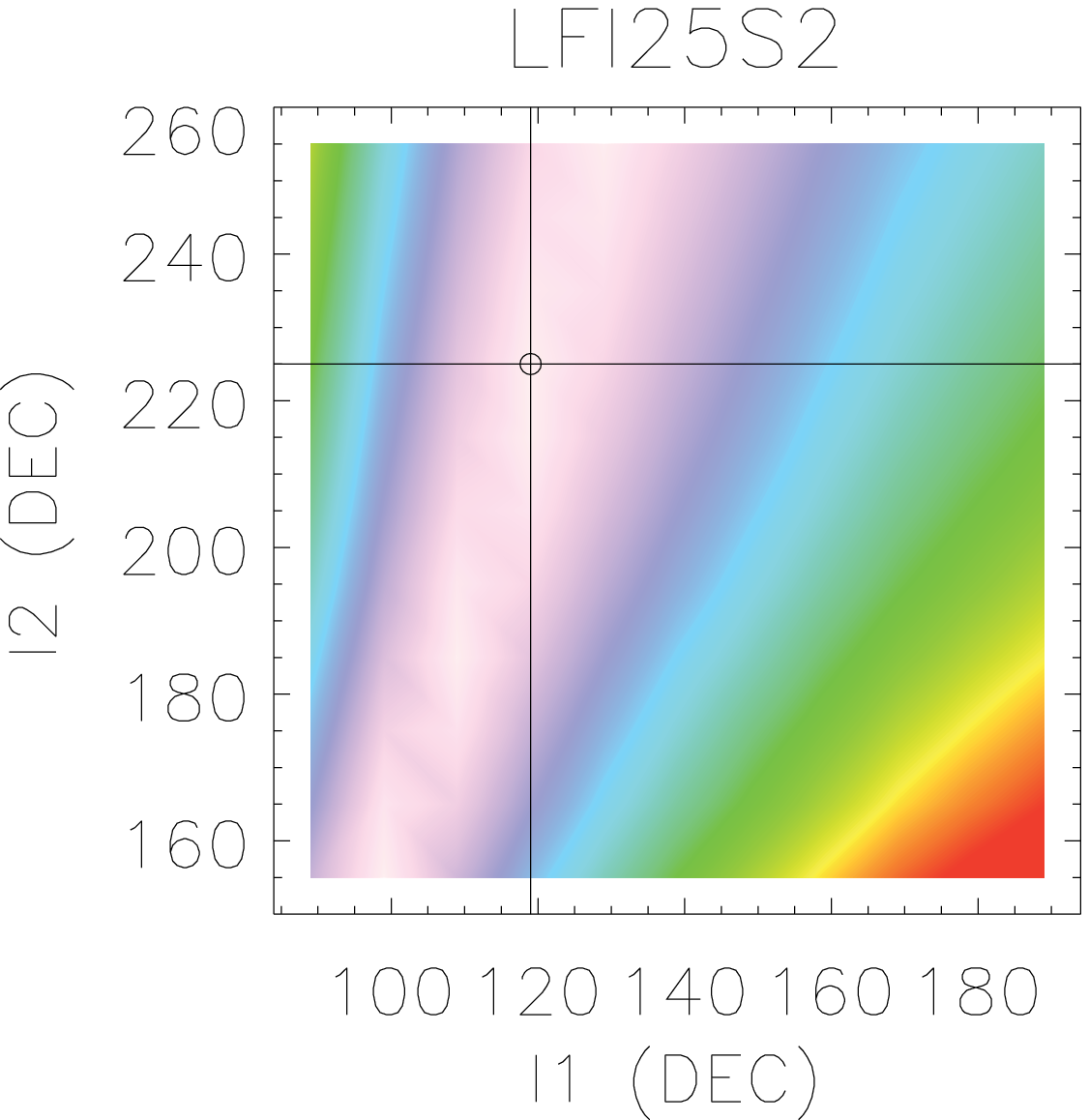}\\
            \textbf{LFI-26}\\
            \includegraphics[width=3.5cm]{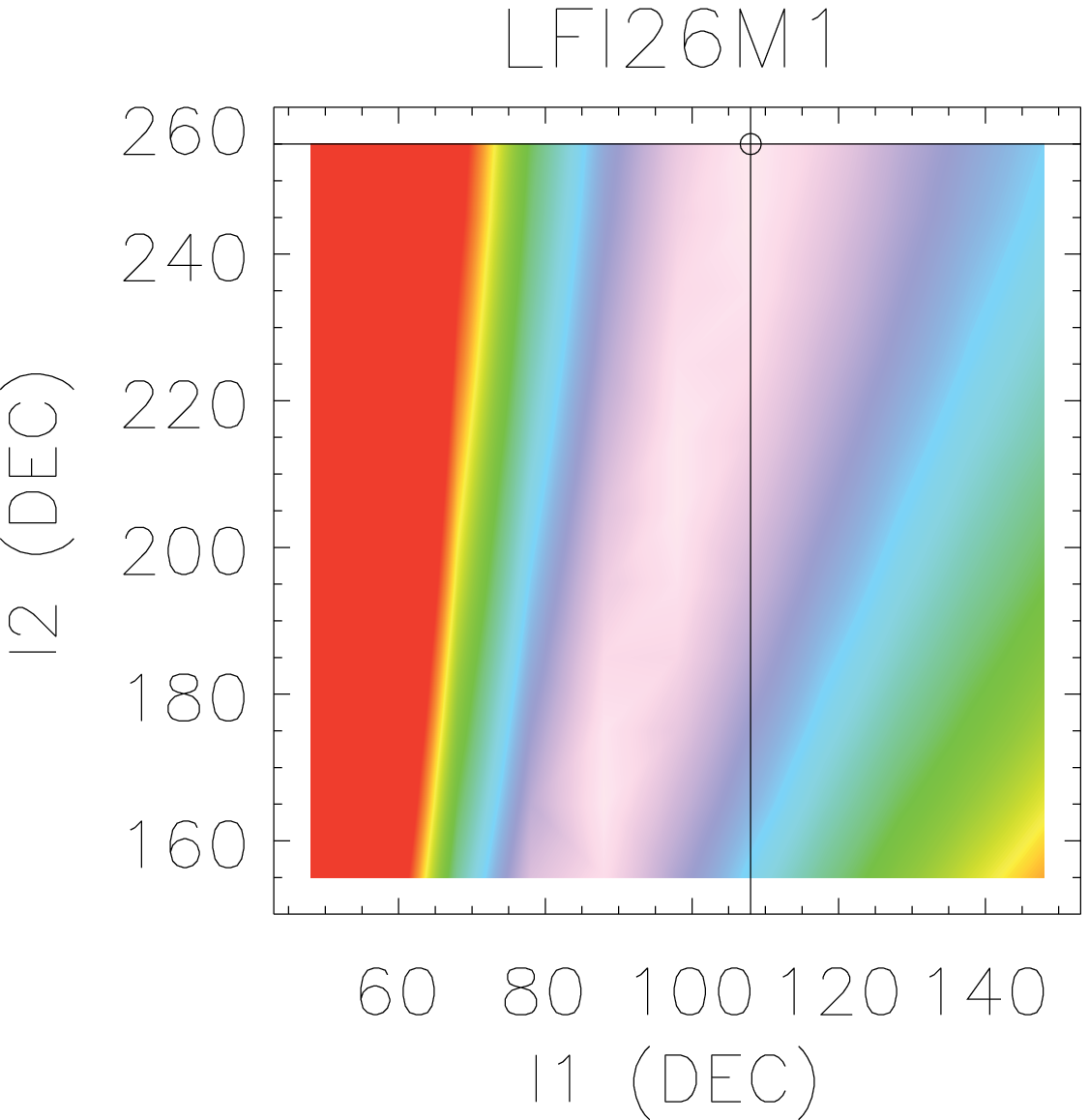}
            \includegraphics[width=3.5cm]{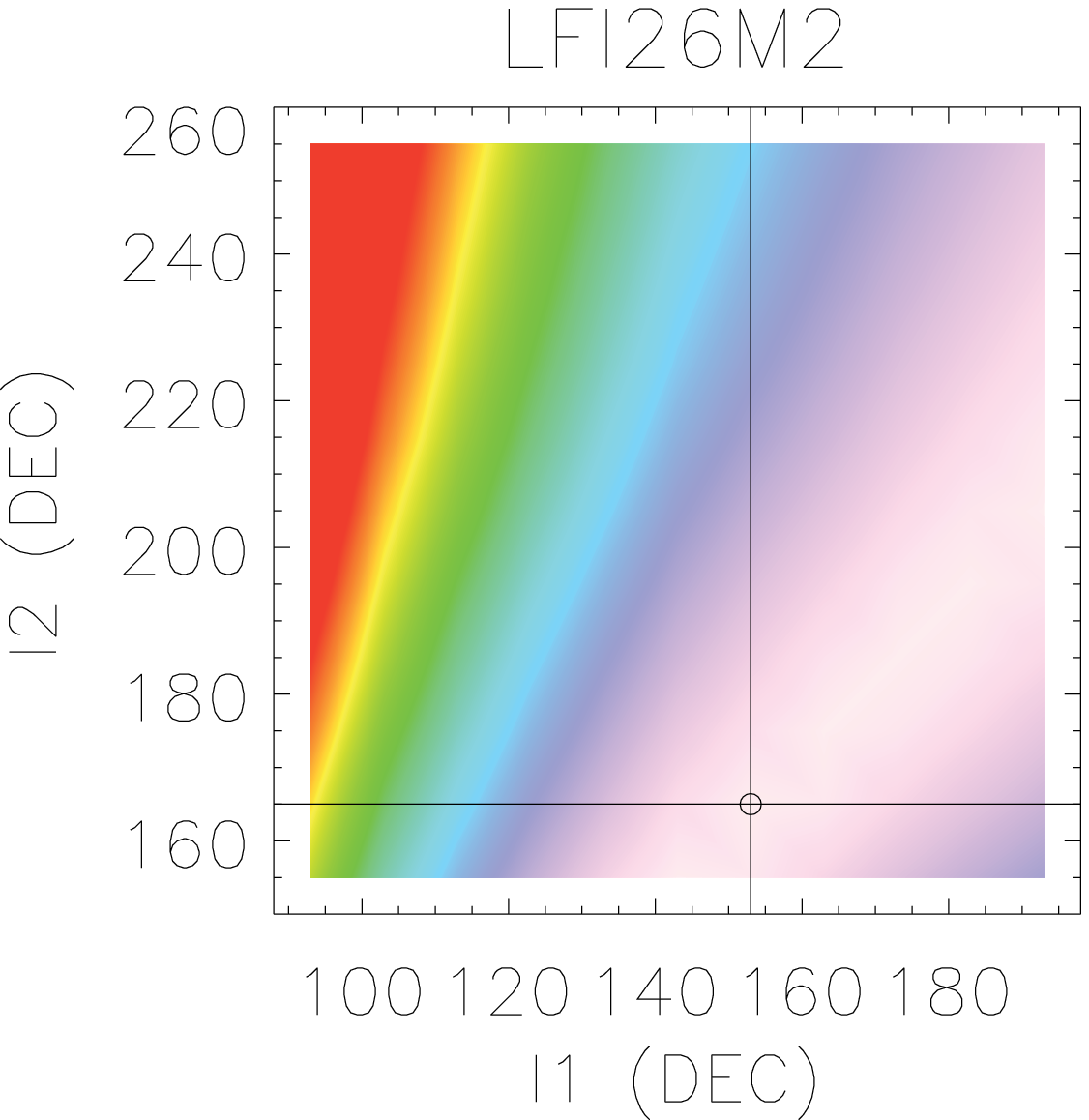}
            \includegraphics[width=3.5cm]{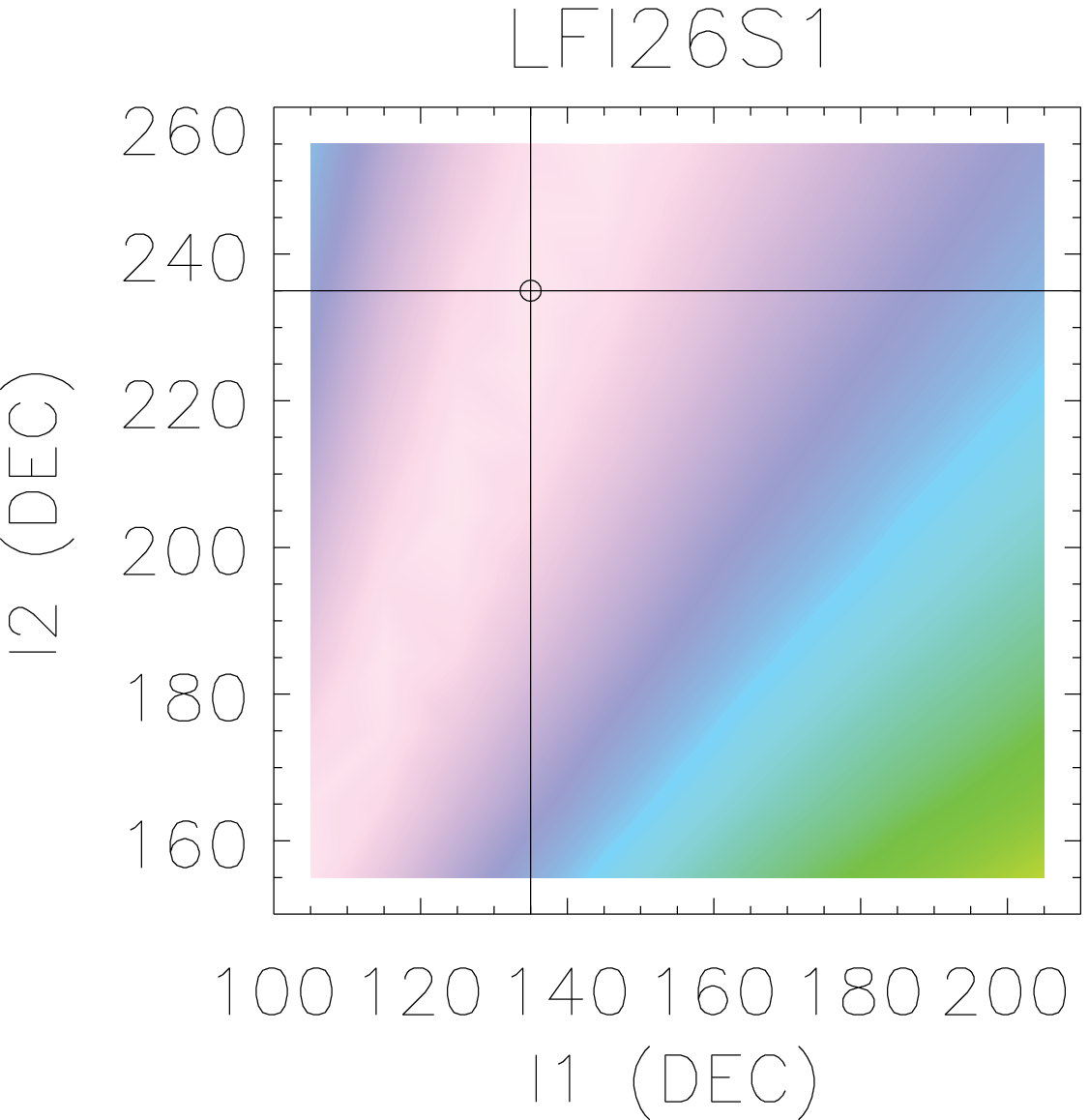}
            \includegraphics[width=3.5cm]{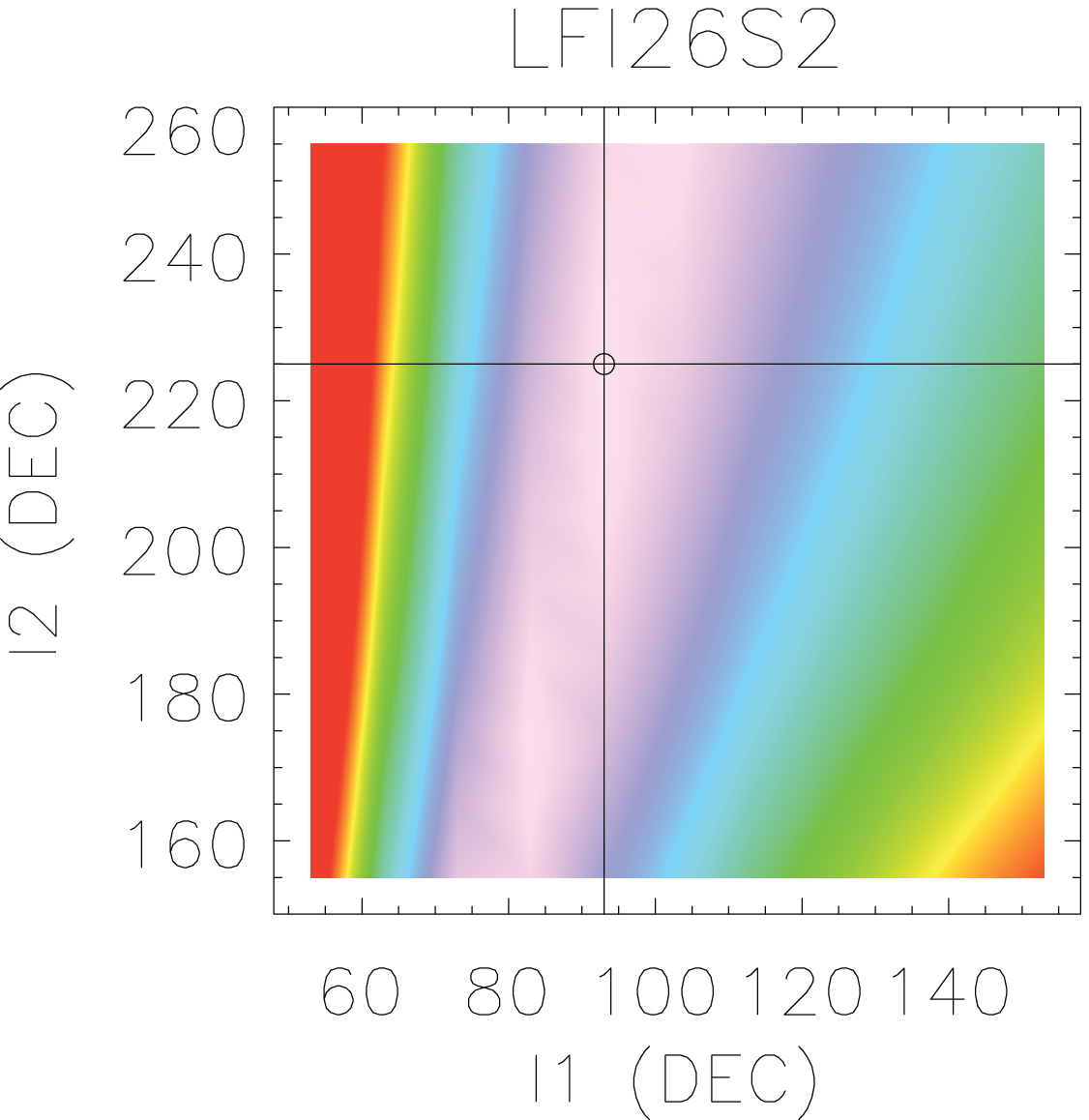}\\
            \textbf{LFI-27}\\
            \includegraphics[width=3.5cm]{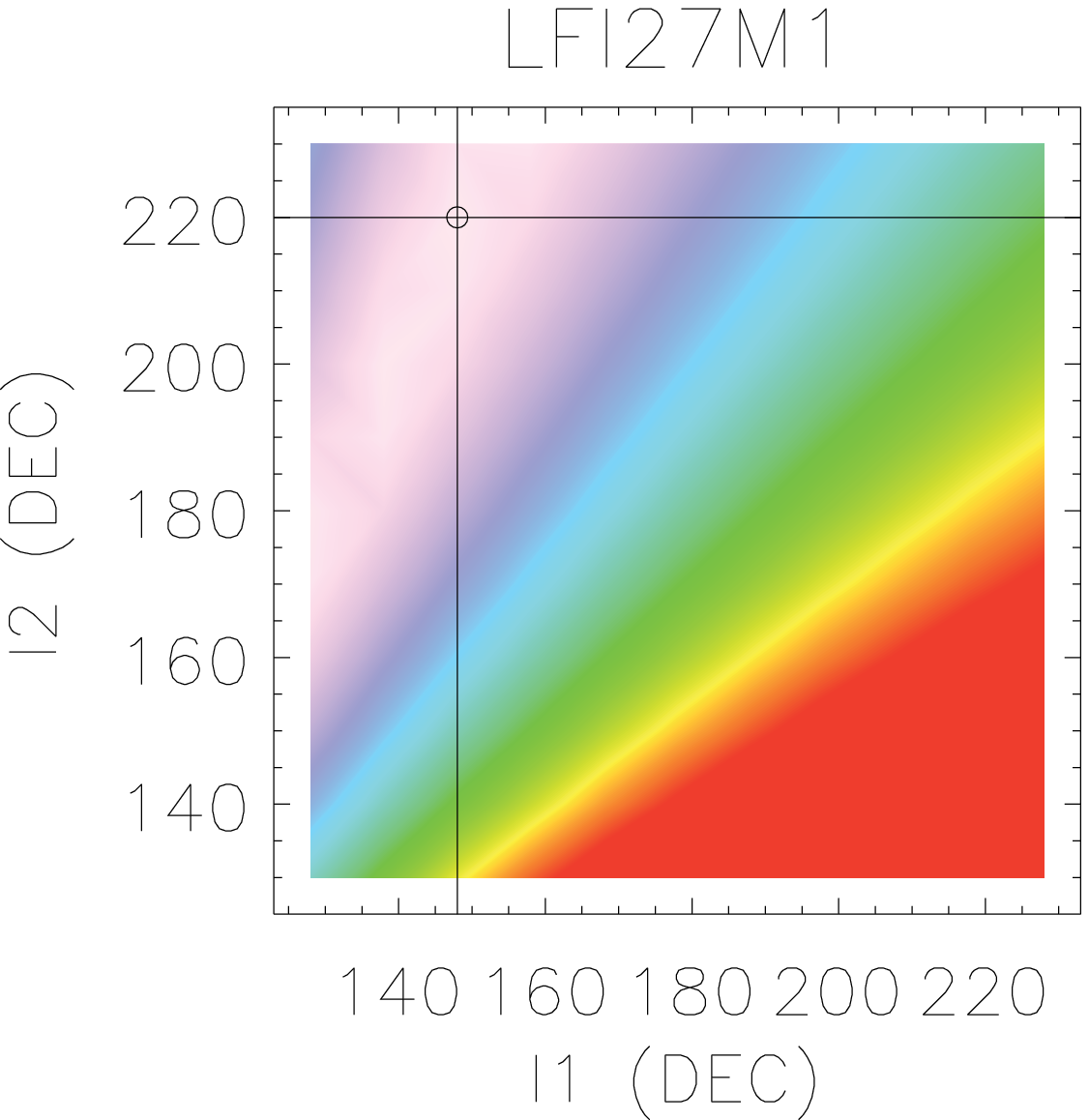}
            \includegraphics[width=3.5cm]{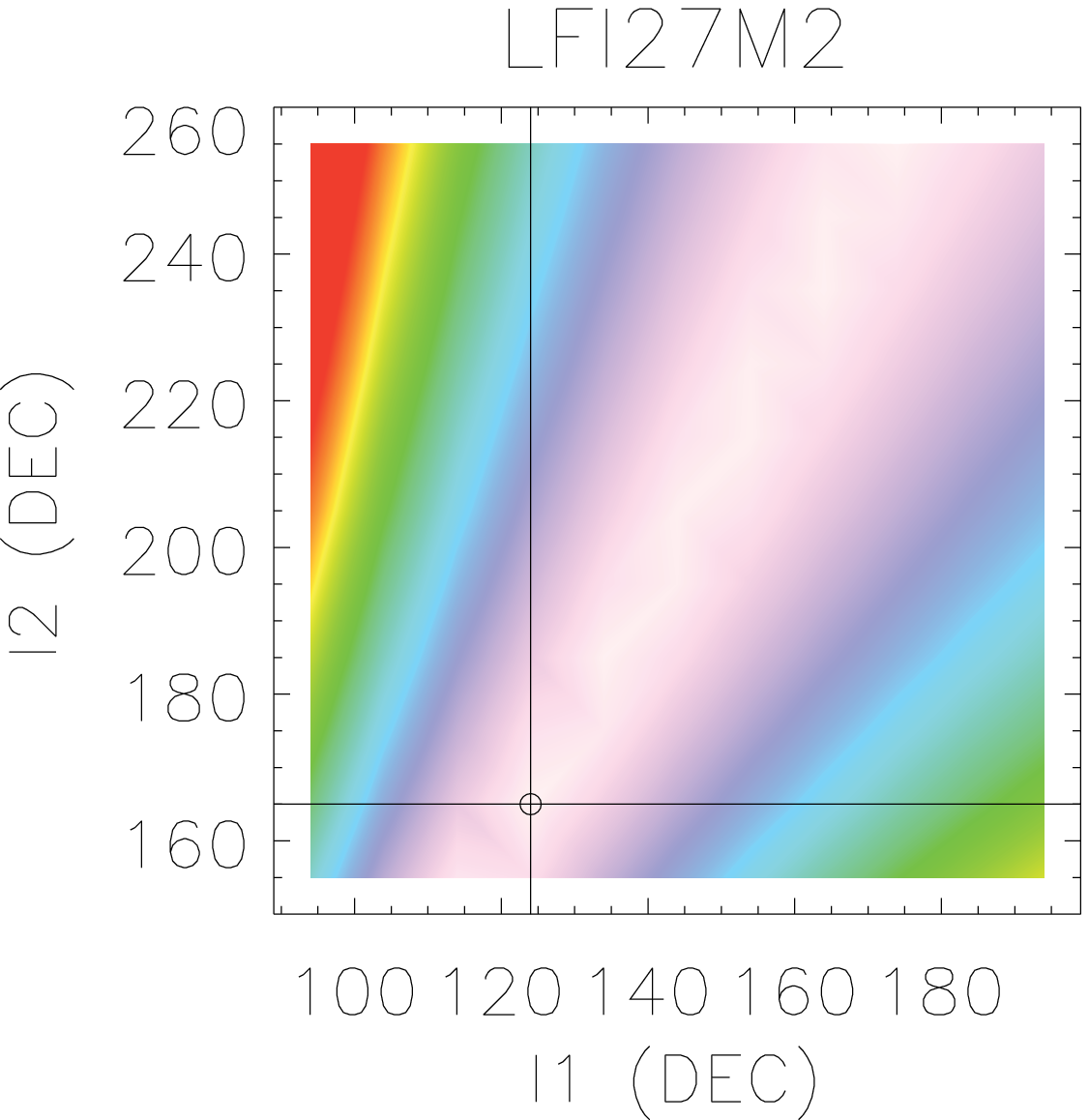}
            \includegraphics[width=3.5cm]{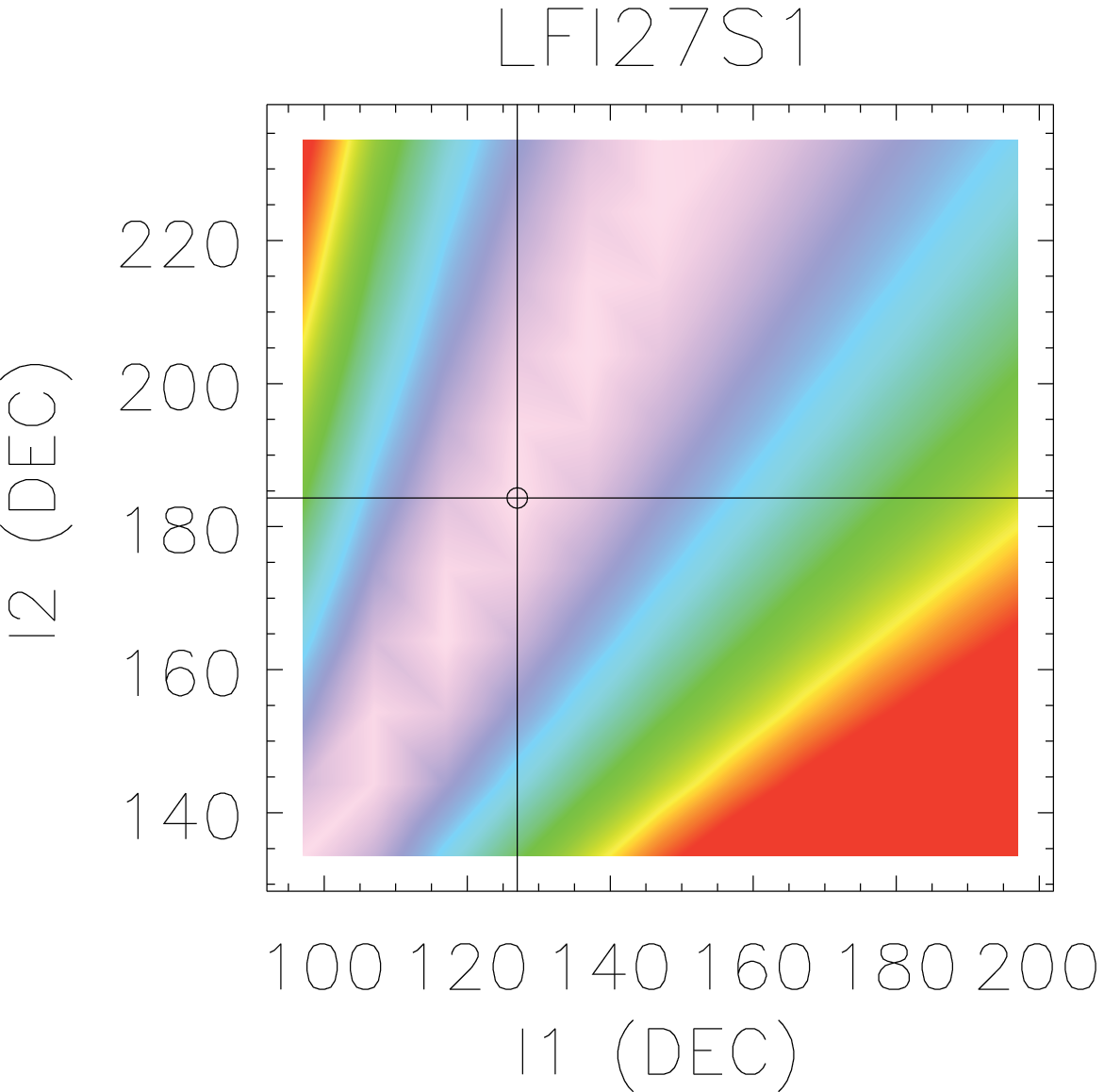}
            \includegraphics[width=3.5cm]{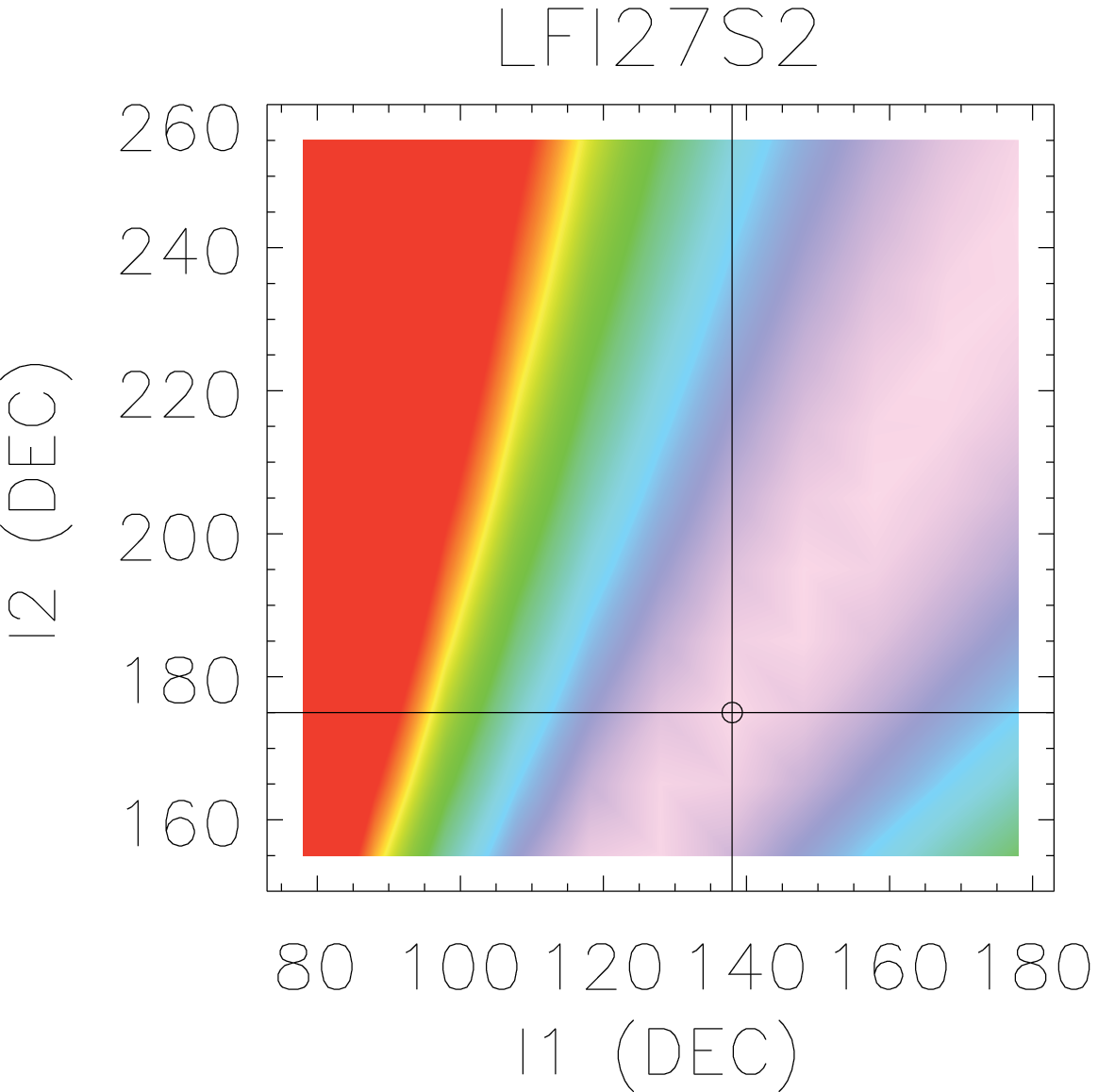}\\
            \textbf{LFI-28}\\
            \includegraphics[width=3.5cm]{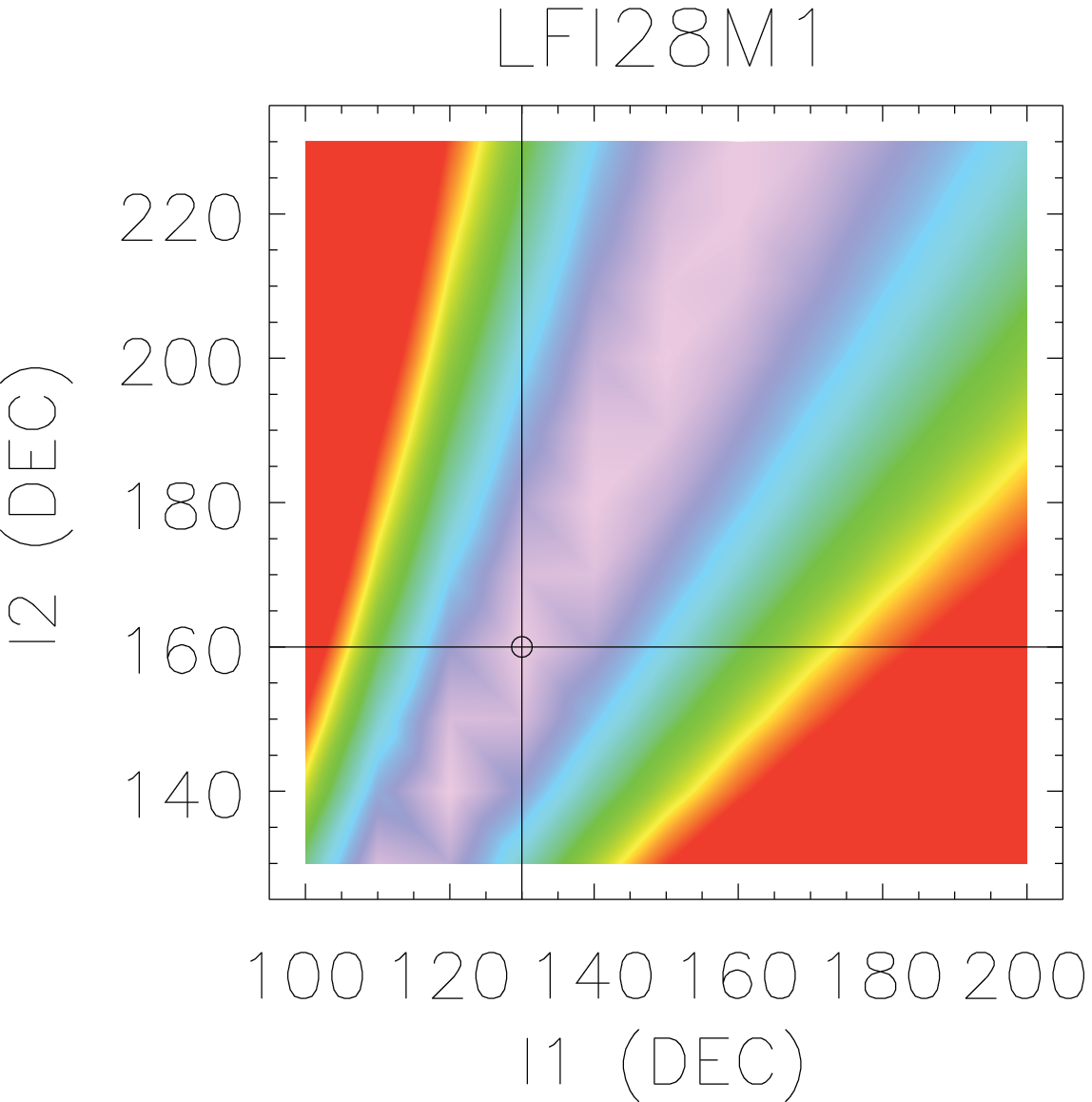}
            \includegraphics[width=3.5cm]{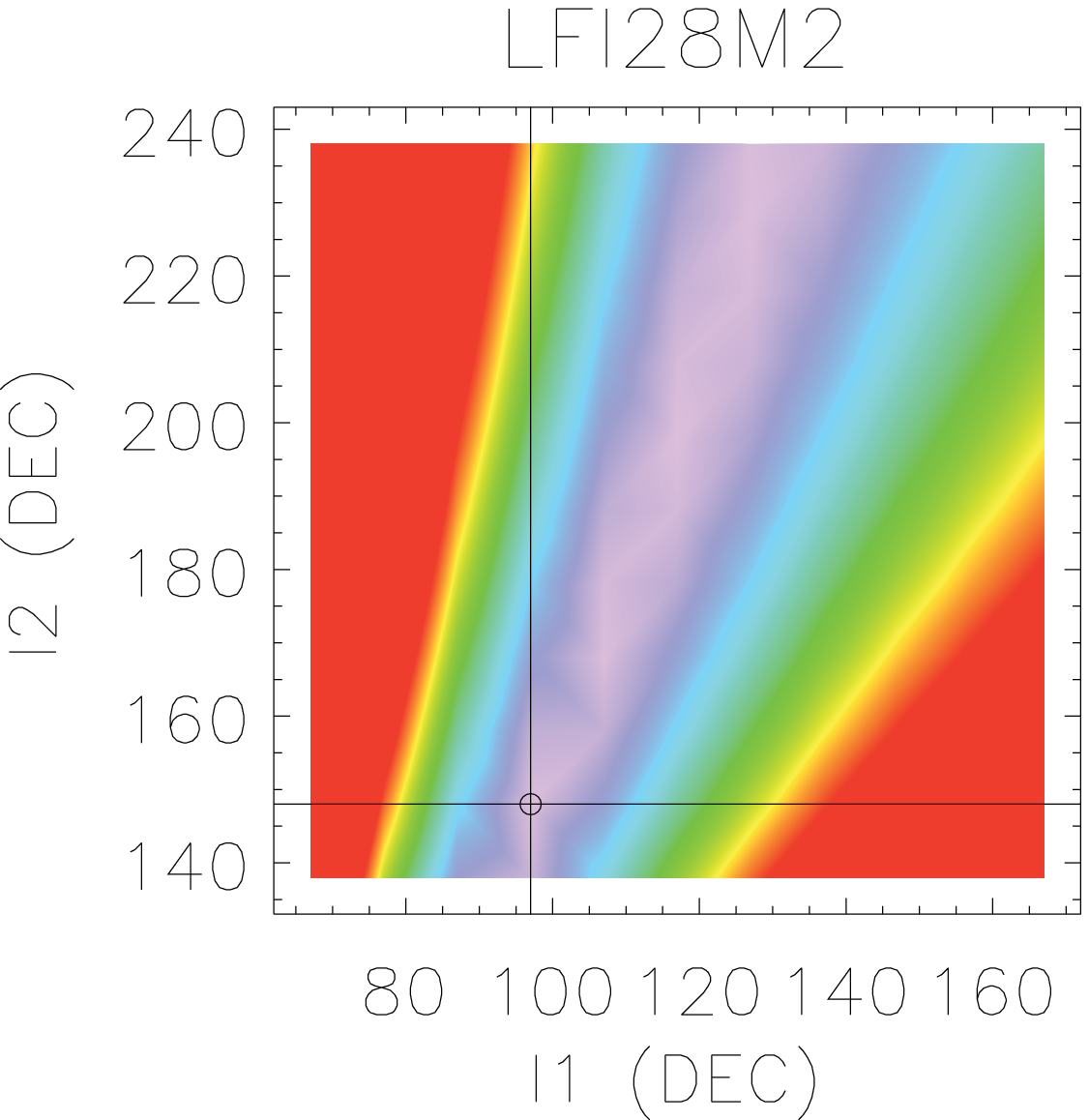}
            \includegraphics[width=3.5cm]{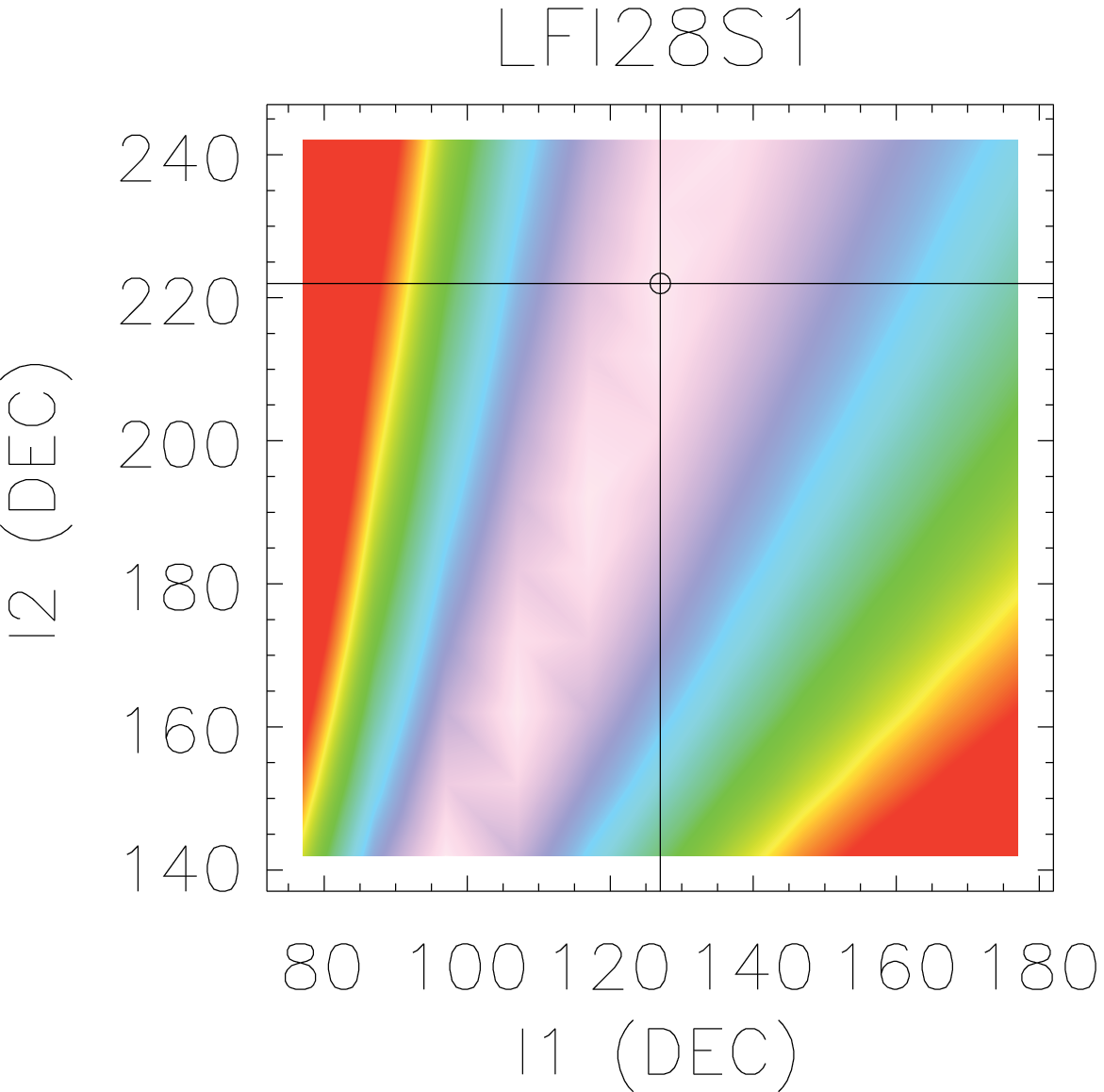}
            \includegraphics[width=3.5cm]{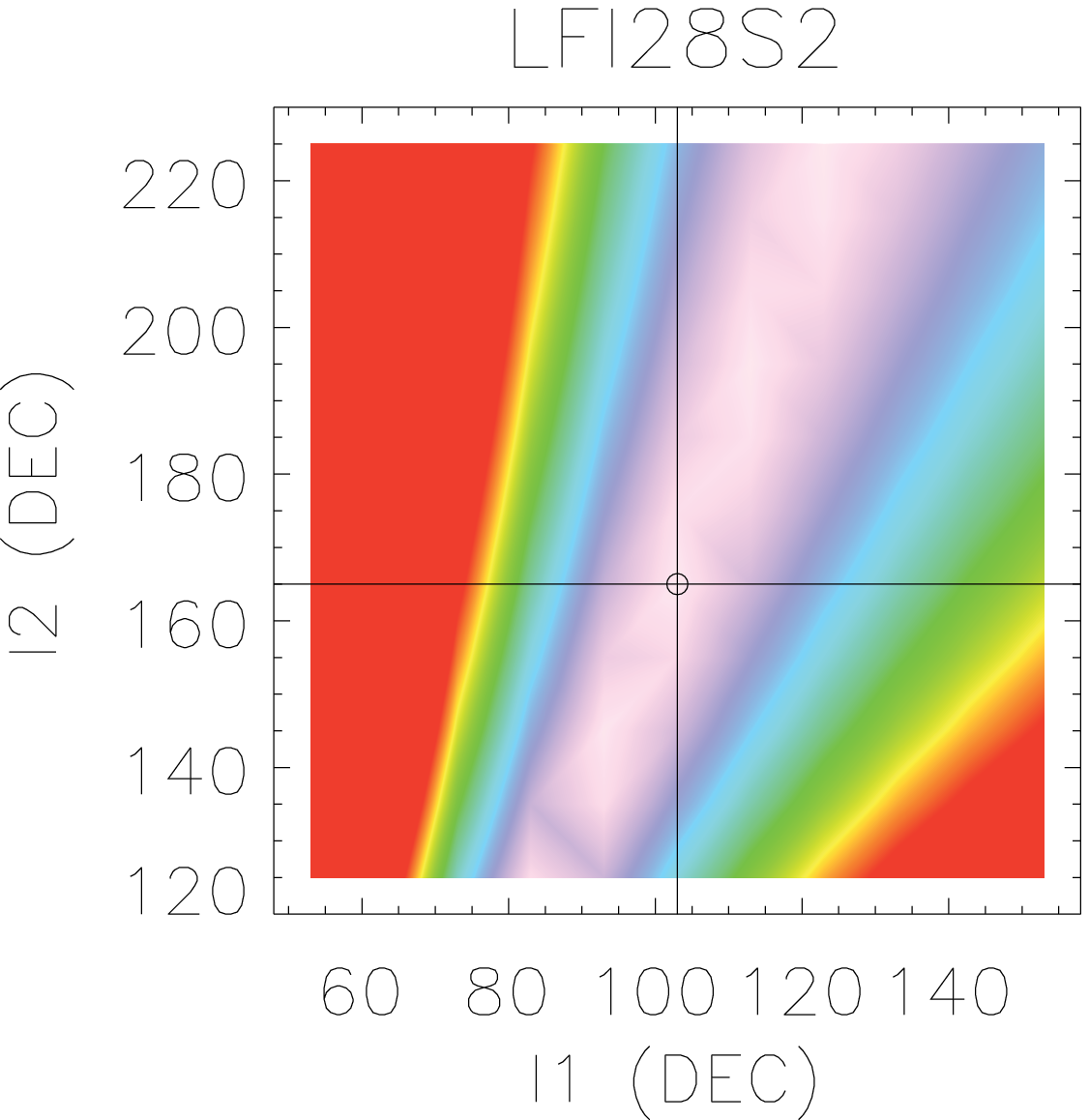}\\           
\vspace{-.5cm}
        \end{center}
        \caption{Phase switch tuning maps. For each channel they are represented from left to right panels the detectors: \texttt{M1}, \texttt{M2}, \texttt{S1}, \texttt{S2}. I$_1$ bias lies on x-axis, I$_2$ bias on y-axis. The color scale shows smaller umbalance in bright and larger unbalance in dark. Values are normalized to the average signal output. The intercept of the two lines indicates the optimal bias selected. Note that the bias units are DEC units that are used to set the bias in the DAE: see Table~5 for a conversion into physical units. }
        \label{fig_PHSW_tun}
    \end{figure}

    In Figure~\ref{fig_PHSW_tun} we show for the  30~GHz and 44~GHz LFI  channels the condensed noise temperature maps corresponding to the linear hypermatrix tuning analysis: these maps were used as input to select the optimal bias.
    \clearpage 

%% file: a05_test_details.tex
\section{Collection of tables with detailed test results}
\label{app_detail_tests}

\begin{table}[h!]
    \begin{center}
        \caption{\label{tab_phsw_Tun_results} Phase switch tuning bias: optimal bias (O{\sc pt}) compared to best values (B{\sc est}) from automatic procedure and to system level test results (C{\sc sl}). Values are given in the decimal units that are used to set the bias in the DAE: see Table 5 for a conversion into physical units. }
        \vspace{.05cm}
        \begin{tabular}{l |c c|c c|c c}
 \hline \hline  \small    

& \multicolumn{2}{c}{O{\sc pt}}& \multicolumn{2}{c}{B{\sc est}}& \multicolumn{2}{c}{C{\sc SL}} \\	   
			&	I$_1$ &	I$_2$ &	I$_1$ &	I$_2$ &	I$_1$ &	I$_2$	\\
			 	\hline      
\texttt{LFI24M1	}	&	98	&	215	&	88	&	175	&	128	&	250	\\
\texttt{LFI24M2	}	&	77	&	185	&	77	&	185	&	77	&	205	\\
\texttt{LFI24S1	}	&	84	&	235	&	84	&	235	&	84	&	235	\\
\texttt{LFI24S2	}	&	86	&	205	&	86	&	205	&	86	&	215	\\
\texttt{LFI25M1	}	&	154	&	245	&	154	&	245	&	174	&	235	\\
\texttt{LFI25M2	}	&	79	&	255	&	69	&	175	&	89	&	250	\\
\texttt{LFI25S1	}	&	79	&	205	&	79	&	205	&	89	&	250	\\
\texttt{LFI25S2	}	&	119	&	225	&	119	&	225	&	119	&	225	\\
\texttt{LFI26M1	}	&	108	&	255	&	108	&	255	&	98	&	245	\\
\texttt{LFI26M2	}	&	153	&	165	&	153	&	165	&	153	&	210	\\
\texttt{LFI26S1	}	&	135	&	235	&	135	&	235	&	135	&	230	\\
\texttt{LFI26S2	}	&	93	&	225	&	93	&	225	&	93	&	230	\\
\texttt{LFI27M1	}	&	148	&	220	&	148	&	220	&	178	&	180	\\
\texttt{LFI27M2	}	&	145	&	205	&	124	&	165	&	144	&	214	\\
\texttt{LFI27S1	}	&	127	&	184	&	127	&	184	&	117	&	154	\\
\texttt{LFI27S2	}	&	148	&	195	&	138	&	175	&	128	&	200	\\
\texttt{LFI28M1	}	&	130	&	160	&	130	&	160	&	150	&	204	\\
\texttt{LFI28M2	}	&	127	&	228	&	97	&	148	&	105	&	164	\\
\texttt{LFI28S1	}	&	127	&	222	&	127	&	222	&	111	&	168	\\
\texttt{LFI28S2	}	&	103	&	165	&	103	&	165	&	93	&	155	\\
 	\hline
\end{tabular}
    \end{center}
\end{table}

\begin{table}[h!]
    \begin{center}
        \caption{\label{tab_Tun_Timeline} Hypermatrix tuning timeline in operational days (OD) starting from the Launch date. }
        \vspace{.1cm}
        \begin{tabular}{c c  c  c  c}
 \hline
 	\hline \small    
STEP	&	\texttt{OD Start}	&	\texttt{OD Stop}	&	\texttt{Time Start (UTC)}	&	\texttt{Time Stop (UTC)}	\\
 	\hline 
1	&	37	&	38	&	2009-06-19 19:00:00&	2009-06-21 04:12:32\\
2	&	40	&	41	&	2009-06-22 22:00:00&	2009-06-24 07:12:32\\
3	&	43	&	44	&	2009-06-25 13:30:00&	2009-06-26 22:42:32\\
4	&	55	&	56	&	2009-07-07 13:30:00&	2009-07-08 22:42:32\\
	\hline
		\end{tabular}
  			  \end{center}
\end{table}

\begin{table}[h!]
    \begin{center}
        \caption{\label{tab_Tun_thermal_sensors} Average values and sigma measured from FPU and BEU sensors during the four steps of the hypermatrix tuning.}
        \vspace{.1cm} \small    
        \begin{tabular}{l c |c c|c c|c c|c c}
 \hline
 	\hline           
    	&	 &	\texttt{STEP 1}&	&	\texttt{STEP 2}& &	\texttt{STEP 3}& &	\texttt{STEP 4}& \\
    		\hline    

NAME	&	PARAM	&	T	&	sigma	&	T	&	sigma	&	T	&	sigma	&	T	&	sigma	\\
\hline
\multicolumn{10}{c}{FPU sensors (units are K)} \\ \hline
CONE R	&	LM302332	&	20.211	&	0.008	&	20.209	&	0.008	&	20.207	&	0.008	&	20.209	&	0.008	\\
CONE L&	LM206332	&	20.488	&	0.011	&	20.486	&	0.011	&	20.483	&	0.011	&	20.485	&	0.011	\\
FH28 Fl&	LM305332	&	20.084	&	0.008	&	20.083	&	0.008	&	20.081	&	0.008	&	20.083	&	0.008	\\
FH26 R	&	LM306332	&	20.260	&	0.011	&	20.258	&	0.011	&	20.255	&	0.011	&	20.256	&	0.011	\\
FH25 L	&	LM204332	&	20.400	&	0.012	&	20.397	&	0.012	&	20.394	&	0.012	&	20.394	&	0.012	\\
\hline
\multicolumn{10}{c}{BEU sensors (units are K)} \\ \hline
LBEM1	&	LM207332	&	288.141	&	0.022	&	288.141	&	0.024	&	288.516	&	0.019	&	287.822	&	0.041	\\
LBEM2	&	LM208332	&	290.640	&	0.019	&	290.641	&	0.024	&	291.027	&	0.018	&	290.308	&	0.044	\\
RFEM1	&	LM309332	&	292.605	&	0.051	&	292.567	&	0.051	&	292.955	&	0.053	&	292.256	&	0.08	\\
RFEM2	&	LM310332	&	288.819	&	0.011	&	288.825	&	0.014	&	289.154	&	0.013	&	288.434	&	0.045	\\
LFEM1	&	LM209332	&	292.481	&	0.057	&	292.499	&	0.062	&	292.875	&	0.057	&	292.208	&	0.067	\\
LFEM2	&	LM210332	&	288.129	&	0.017	&	288.133	&	0.022	&	288.509	&	0.017	&	287.810	&	0.041	\\
RBEM2	&	LM308332	&	288.366	&	0.011	&	288.366	&	0.015	&	288.695	&	0.012	&	287.963	&	0.042	\\
RBEM1	&	LM307332	&	290.891	&	0.010	&	290.902	&	0.013	&	291.246	&	0.012	&	290.519	&	0.043	\\
\hline
\end{tabular}
    \end{center}
\end{table}

\begin{table}[h!]
    \begin{center}
        \caption{\label{tab_Tun_compare_30_44} Average Noise Temperature (K) and Isolation from 30~GHz and 44~GHz channels hypermatrix tuning: comparison between \texttt{CSL} and  \texttt{CPV} results.}
        \vspace{.05cm}  \small       
        \begin{tabular}{l|c c|c c|c c|c c|c c}
 \hline
 	\hline         
	&	\texttt{24M}	&	\texttt{24S}	&	\texttt{25M}	&	\texttt{25S}	&	\texttt{26M}	&	\texttt{26S}	&	\texttt{27M}	&	\texttt{27S}	&	\texttt{28M}	&	\texttt{28S}	\\
\hline
\multicolumn{11}{c}{Noise Temperature (K)} \\  
\texttt{CSL}&	21.8	&	23.3	&	24.3	&	22.8	&	26.9	&	23.05	&	16.8	&	18.2	&	16.2	&	15.5	\\
\texttt{CPV}	&	20.3	&	21.2	&	21.7	&	20.0	&	25.0	&	21.0	&	16.7	&	18.2	&	15.8	&	15.5	\\
\hline
\multicolumn{11}{c}{Isolation} \\  
\texttt{CSL}	&	0.011	&	0.0285	&	0.016	&	0.002	&	0.044	&	0.012	&	0.0085	&	0.004	&	0.029	&	0.044	\\
\texttt{CPV}	&	0.011	&	0.0275	&	0.015	&	0.001	&	0.028	&	0.011	&	0.0065	&	0.005	&	0.023	&	0.043	\\
\hline
\end{tabular}
    \end{center}
\end{table}

\begin{table}[h!]
    \begin{center}
        \caption{\label{tab_Tun_compare_70} Average Noise Temperature (K) and Isolation from 70~GHz channels hypermatrix tuning: comparison between  \texttt{CSL} and \texttt{CPV} results.}
        \vspace{.05cm} \small    
        \begin{tabular}{l|c c|c c|c c|c c|c c|c c}
 \hline
 	\hline             
&	\texttt{18M}	&	\texttt{18S}	&	\texttt{19M}	&	\texttt{19S}	&	\texttt{20M}	&	\texttt{20S}	&	\texttt{21M}	&	\texttt{21S}	&	 \texttt{22M}	&	\texttt{22S}	&	\texttt{23M}	&	\texttt{23S}	\\
	 	\hline
\multicolumn{13}{c}{Noise Temperature (K)} \\  
\texttt{CSL}	&	28.1	&	27.6	&	29.6	&	26.3	&	32.3	&	29.0	&	32.5	&	24.0	&	28.6	&	27.8	&	31.7	&	30.5	\\
\texttt{CPV}	&	26.6	&	26.5	&	28.6	&	25.6	&	31.0	&	28.2	&	30.1	&	23.3	&	27.9	&	26.6	&	28.8	&	28.4	\\
 	\hline
\multicolumn{13}{c}{Isolation} \\  
\texttt{CSL}	&	N.A.	&	N.A.	&	0.026	&	N.A.	&	0.027	&	0.0187	&	0.167	&	0.0235	&	0.038	&	-0.001	&	0.010	&	0.025	\\
\texttt{CPV}	&	0.035	&	0.029	&	0.027	&	0.041	&	0.032	&	0.0145	&	0.021	&	0.0215	&	0.028	&	-0.001	&	0.025	&	0.015	\\
 	\hline
\end{tabular}
    \end{center}
\end{table}

\begin{table}[h!]
    \begin{center}
        \caption{\label{tab_VD_Tun_compare} Optimal average noise temperature T$\rm _{n}$ (K) and Isolation (Isol) resulting from different tuning phases: optimal bias from ground tests (\texttt{CSL}), from CPV  \texttt{HYM} using only gate voltages V$\rm _g$ (\texttt{TUN}) and from the full \texttt{HYM} tests, including the drain voltage V$\rm _d$ part (\texttt{SET}). Results are given in terms of V$\rm _d$ pairs and V$_{\rm g \, 1,2}$ quadruplets (in decimal values): Noise Temperature (K) and Isolation (dB) averaged over the two coupled diodes, drain currents (mA) of the two coupled ACAs (Id$\rm _{1,2}$). Voltage units are decimal: conversion in volts can be roughly obtained referring to Table~4. }
        \vspace{.05cm} \small 
        \begin{tabular}{l|c|c|c|c|c|c c}
 \hline
 	\hline               
 &\texttt{Phase} & \texttt{(V$\rm _{d,1}$,V$\rm _{d,2}$)} & \texttt{(V$\rm _{g\,1,1}$,V$\rm _{g\,2,1}$),(V$\rm _{g\,1,2}$,V$\rm _{g\,2,2}$)} & \texttt{T$\rm _{n}$} & \texttt{Isol} & \texttt{Id$\rm _1$} & \texttt{Id$\rm _2$}\\
\hline															
\multirow{3}{*}{21M}	&	\texttt{CSL}& (141, 136) & (198, 207) , (196, 197) &24.05	&	-16.40	&	18.42	&	19.56	\\
	&	\texttt{TUN}	&	=	&	(192, 240) , (194, 232)	&	24.69	&	-19.20	&	21.92	&	25.54	\\
	&	\texttt{SET}	&	 (147, 136)	&	 (192, 231) , (191, 224) 	&	23.30	&	-16.70	&	21.54	&	23.87	\\
\hline															
\multirow{3}{*}{22M	}&	\texttt{CSL}&	(125, 130)	&	(203, 194) , (178, 176)	&	28.14	&	-19.80	&	14.16	&	14.94	\\
	&	\texttt{TUN}	&	=	&	(218, 210) , (188, 188)	&	28.06	&	-20.20	&	18.49	&	18.10	\\
	&	\texttt{SET}	&	 (130, 135)	&	 (208, 218) , (188, 188 )	&	26.64	&	-23.50	&	18.87	&	18.53	\\
\hline															
\multirow{3}{*}{23S}	&	\texttt{CSL}&	(118, 122)	&	(181, 211) , (190, 208)	&	31.89	&	-19.20	&	20.83	&	15.16	\\
	&	\texttt{TUN}	&	=	&	(184, 213) , (198, 213)	&	30.31	&	-15.51	&	21.34	&	17.09	\\
	&	\texttt{SET}	&	 (123, 127)	&	 (180, 222) , (198, 213)	&	28.78	&	-16.00	&	21.77	&	17.74	\\
\hline															
\multirow{3}{*}{25M}	&	\texttt{CSL}&	(184, 185)	&	(227, 212) , (219, 212)	&	24.32	&	-17.72	&	9.4	&	10.04	\\
	&	\texttt{TUN}	&	=	&	(223, 203) , (211, 200)	&	23.28	&	-19.00	&	6.65	&	6.78	\\
	&	\texttt{SET}	&	 (177, 178)	&	 (231, 203) , (218, 200)	&	21.75	&	-18.20	&	6.28	&	6.28	\\
\hline															
\multirow{3}{*}{28M}	&	\texttt{CSL}&	(157, 156)	&	(243, 101) , (240, 112)	&	16.30	&	-15.50	&	9.62	&	9.15	\\
	&	\texttt{TUN}	&	=	&	(243, 101) , (240, 112)	&	16.30	&	-15.50	&	9.62	&	9.15	\\
	&	\texttt{SET}	&	 (150, 163)	&	(243, 101) , ( 240, 112)	&	15.87	&	-16.40	&	8.99	&	9.77	\\
 	\hline
\end{tabular}
    \end{center}
\end{table}

 \begin{sidewaystable}[htdp]																																			
  \begin{center}																											
    \caption{\texttt{LFI-24} Tuning: Gain-Model results$^\dagger$. Results from several sets of biases tested during the CPV are compared among each other and to ground test results (\texttt{RCA paper} \cite{villa2010}). 
For each radiometer (separately \texttt{Main} and \texttt{Side}) and for each detector (labelled with 1, 2) the gain model parameters are shown on the right side of the table: noise temperatures $T\rm _n$ (given in K), the non-linear coefficients $b$ (adimensional) and the Gains $G$ (expressed as V/K). Corresponding values corrected for the BEU thermal drift are reported on the right side  (G.M. correction). 
The bias settings considered are optimal biases from: RCA test campaign (\texttt{RCA}), unit level test campaign (\texttt{FEM}), system level test campaign (\texttt{CSL}), hypermatrix tuning (\texttt{CPV}). 	} 
																		
    \label{tab_HYM_Tun_summary_24}																											\small  
    \begin{tabular}{l l l l l l l | l l l l l l}																												
      \hline																											
      \hline								  																			
	& \multicolumn{12}{c}{M{\sc ain Radiometer }} \\																										
	& \multicolumn{6}{c}{G{\sc ain model}}& \multicolumn{6}{c}{G.M.{\sc correction}} \\																										
	\hhline{~------------}																										
	\multicolumn{1}{c}{SOURCE} & \multicolumn{1}{c}{$T\rm _{n,1}$}& \multicolumn{1}{c}{$b_1$} & \multicolumn{1}{c}{$G_1$} & \multicolumn{1}{c}{$T\rm _{n,2}$}& \multicolumn{1}{c}{$b_2$} & \multicolumn{1}{c}{$G_2$} & \multicolumn{1}{c}{$T\rm _{n,1}$}& \multicolumn{1}{c}{$b_1$} & \multicolumn{1}{c}{$G_1$} & \multicolumn{1}{c}{$T\rm _{n,2}$}& \multicolumn{1}{c}{$b_2$} & \multicolumn{1}{c}{$G_2$}\\																										
	\hline																										
	\texttt{RCA Paper}	&	N.A.	&	N.A.	&	N.A.	&	N.A.	&	N.A.	&	N.A.	&	15.5	&	1.794	&	0.0048	&	15.3	&	1.486	&	0.0044	\\	
	\texttt{RCA}	&	24.3	&	-0.34	&	0.0031	&	23.1	&	-0.57	&	0.0031	&	17.0	&	1.46	&	0.0050	&	16.7&	0.99	&	0.0051	\\	
	\texttt{FEM}	&	24.1	&	-0.29	&	0.0032	&	23.0	&	-0.57	&	0.0032	&	16.7	&	1.51	&	0.0051	&	16.6 	&	1.03	&	0.0051	\\	
	\texttt{CSL}	&	23.7	&	-0.53	&	0.0028	&	25.1	&	-0.41	&	0.0037	&	16.3	&	1.58	&	0.0045	&	16.3&	0.99	&	0.0044	\\	
	\texttt{CPV}	&	27.3	&	-2.99	&	0.0016	&	27.3	&	-1.53	&	0.0026	&	15.4	&	2.21	&	0.0032	&	15.0&	1.86	&	0.0032	\\	

\hline
\hline	
	& \multicolumn{12}{c}{S{\sc ide Radiometer }} \\																										
	& \multicolumn{6}{c}{G{\sc ain model}}& \multicolumn{6}{c}{G.M.{\sc correction}} \\																										
	\hhline{~------------}																									
	\multicolumn{1}{c}{SOURCE} & \multicolumn{1}{c}{$T\rm _{n,1}$}& \multicolumn{1}{c}{$b_1$} & \multicolumn{1}{c}{$G_1$} & \multicolumn{1}{c}{$T\rm _{n,2}$}& \multicolumn{1}{c}{$b_2$} & \multicolumn{1}{c}{$G_2$} & \multicolumn{1}{c}{$T\rm _{n,1}$}& \multicolumn{1}{c}{$b_1$} & \multicolumn{1}{c}{$G_1$} & \multicolumn{1}{c}{$T\rm _{n,2}$}& \multicolumn{1}{c}{$b_2$} & \multicolumn{1}{c}{$G_2$}\\		
				
	\hline																										
	\texttt{RCA PAPER}	&	N.A.	&	N.A.	&	N.A.	&	N.A.	&	N.A.	&	N.A.	&	15.8	&	1.44	&	0.0062	&	15.8	&	1.45	&	0.0062	\\	
	\texttt{RCA}	&	24.3	&	-0.34	&	0.0031	&	25.5	&	-0.26	&	0.0046	&	17.2	&	0.88	&	0.0072	&	17.4	&	0.95	&	0.0079	\\	
	\texttt{FEM}	&	24.1	&	-0.29	&	0.0032	&	23.0	&	0.16	&	0.0088	&	17.4	&	0.87	&	0.0073	&	18.1	&	0.82	&	0.0076	\\	
	\texttt{CSL}	&	25.1	&	-0.69 	&	0.0037	&	22.9	&	-0.93	&	0.0028	&	16.0	&	1.11	&	0.0066	&	16.1	&	1.19	&	0.0072	\\	
	\texttt{CPV}	&	25.2	&	-1.37	&	0.0026	&	25.0	&	-2.55	&	0.0015	&	15.4	&	1.33	&	0.0049	&	15.6	&	1.41	&	0.0053	\\	

\hline
      \end{tabular}																											
\end{center}																		 
\begin{small}
        \noindent $^\dagger$ A gain model was developed based on the results of Daywitt (1989), modified for the LFI \cite{mennella2009-2}. The total power output voltage $V_{\rm out}$ can be written as:
\begin{equation}
V_{\rm out} = \frac{G}{ 1+ b \cdot G \cdot ( T_{\rm in}+T_{\rm n} ) } \cdot (T_{\rm in}+T_{\rm n})
\end{equation}
where $T_{\rm in}$ is $T_{\rm sky}$ or $T_{\rm ref}$, $T_{\rm n}$ the noise temperature (K), $G$ the total gain (V/K) in the case of a linear radiometer, $b$ the linear coefficient.  \\ 
        \end{small}
  		
\end{sidewaystable}
 
 \begin{sidewaystable}[htdp]															 															
  \begin{center}																											
    \caption{\texttt{LFI-25} Tuning: Gain-Model results. Results from several sets of biases tested during the CPV are compared among each other and to ground test results (\texttt{RCA paper} \cite{villa2010}). 
For each radiometer (separately \texttt{Main} and \texttt{Side}) and for each detector (labelled with 1, 2) the gain model parameters are shown on the right side of the table: noise temperatures $T\rm _n$ (given in K), the non-linear coefficients $b$ (adimensional) and the Gains $G$ (expressed as V/K). Corresponding values corrected for the BEU thermal drift are reported on the right side  (G.M. correction). 
The bias settings considered are optimal biases from: RCA test campaign (\texttt{RCA}), unit level test campaign (\texttt{FEM}), system level test campaign (\texttt{CSL}), hypermatrix tuning (\texttt{CPV}). 	}
				
    \label{tab_HYM_Tun_summary_25}																											\small    
    \begin{tabular}{l l l l l l l | l l l l l l}																											
      \hline																											
      \hline																											
	& \multicolumn{12}{c}{M{\sc ain Radiometer }} \\																										
	& \multicolumn{6}{c}{G{\sc ain model}}& \multicolumn{6}{c}{G.M.{\sc correction}} \\			 
	\hhline{~------------}																										
	\multicolumn{1}{c}{SOURCE} & \multicolumn{1}{c}{$T\rm _{n,1}$}& \multicolumn{1}{c}{$b_1$} & \multicolumn{1}{c}{$G_1$} & \multicolumn{1}{c}{$T\rm _{n,2}$}& \multicolumn{1}{c}{$b_2$} & \multicolumn{1}{c}{$G_2$} & \multicolumn{1}{c}{$T\rm _{n,1}$}& \multicolumn{1}{c}{$b_1$} & \multicolumn{1}{c}{$G_1$} & \multicolumn{1}{c}{$T\rm _{n,2}$}& \multicolumn{1}{c}{$b_2$} & \multicolumn{1}{c}{$G_2$}\\																											
	\hline																										
	\texttt{RCA PAPER}	&	N.A.	&	N.A.	&	N.A.	&	N.A.	&	N.A.	&	N.A.	&	17.5	&	1.22	&	0.0086	&	17.9	&	1.17	&	0.0085	\\	
	\texttt{RCA}	&	22.6	&	0.33	&	0.0073	&	23.7	&	0.09	&	0.0073	&	14.8	&	1.25	&	0.0124	&	16.2	&	1.03	&	0.0104	\\	
	\texttt{FEM}	&	21.9	&	0.49	&	0.0087	&	23.3	&	0.22	&	0.0087	&	14.9	&	1.21	&	0.0140	&	16.2&	1.01	&	0.0118	\\
	\texttt{CSL}	&	22.3	&	0.34	&	0.0073	&	23.4	&	0.09	&	0.0073	&	15.1	&	1.21	&	0.0118	&	16.5	&	0.99	&	0.0099	\\	
	\texttt{CPV}	&	20.7	&	0.39	&	0.0054	&	21.6	&	0.16	&	0.0054	&	12.3	&	1.89	&	0.0101	&	13.2	&	1.73	&	0.0088	\\	
	
      \hline		
      \hline		
	& \multicolumn{12}{c}{S{\sc ide Radiometer }} \\																									
	& \multicolumn{6}{c}{G{\sc ain model}}& \multicolumn{6}{c}{G.M.{\sc correction}} \\			 
	\hhline{~------------}																										
	\multicolumn{1}{c}{SOURCE} & \multicolumn{1}{c}{$T\rm _{n,1}$}& \multicolumn{1}{c}{$b_1$} & \multicolumn{1}{c}{$G_1$} & \multicolumn{1}{c}{$T\rm _{n,2}$}& \multicolumn{1}{c}{$b_2$} & \multicolumn{1}{c}{$G_2$} & \multicolumn{1}{c}{$T\rm _{n,1}$}& \multicolumn{1}{c}{$b_1$} & \multicolumn{1}{c}{$G_1$} & \multicolumn{1}{c}{$T\rm _{n,2}$}& \multicolumn{1}{c}{$b_2$} & \multicolumn{1}{c}{$G_2$}\\																											
	\hline																										
	\texttt{RCA PAPER}	&	N.A.	&	N.A.	&	N.A.	&	N.A.	&	N.A.	&	N.A.	&	18.6	&	0.80	&	0.0079	&	18.4	&	1.01	&	0.0071	\\	
	\texttt{RCA}	&	23.5	&	0.05	&	0.0081	&	23.0	&	0.15	&	0.0081	&	16.9	&	0.71	&	0.0125	&	15.3	&	1.05	&	0.0120	\\	
	\texttt{FEM}	&	22.3	&	0.19	&	0.0088	&	23.0	&	0.16	&	0.0088	&	15.9	&	0.83	&	0.0135	&	15.4 &	1.05	&	0.0119	\\	
	\texttt{CSL}	&	22.4	&	0.07	&	0.0072	&	N.A.	&	N.A.	&	N.A.	&	16.0	&	0.83	&	0.0110	&	14.9	&	1.13	&	0.0102	\\	
	\texttt{CPV}	&	20.8	&	-0.19	&	0.0034	&	N.A.	&	N.A.	&	N.A.	&	10.7	&	2.67	&	0.0074	&	10.1	&	3.16	&	0.0068	\\	

\hline
      \end{tabular}																											
\end{center}																		 											
\end{sidewaystable}

\begin{sidewaystable}[htdp]															 														
  \begin{center}																											
    \caption{\texttt{LFI-26} Tuning: Gain-Model results. Results from several sets of biases tested during the CPV are compared among each other and to ground test results (\texttt{RCA paper} \cite{villa2010}). 
For each radiometer (separately \texttt{Main} and \texttt{Side}) and for each detector (labelled with 1, 2) the gain model parameters are shown on the right side of the table: noise temperatures $T\rm _n$ (given in K), the non-linear coefficients $b$ (adimensional) and the Gains $G$ (expressed as V/K). Corresponding values corrected for the BEU thermal drift are reported on the right side  (G.M. correction). 
The bias settings considered are optimal biases from: RCA test campaign (\texttt{RCA}), unit level test campaign (\texttt{FEM}), system level test campaign (\texttt{CSL}), hypermatrix tuning (\texttt{CPV}). }
																
    \label{tab_HYM_Tun_summary_26}																											
\small  
    \begin{tabular}{l l l l l l l | l l l l l l}																												
      \hline																											
      \hline										  																	
	& \multicolumn{12}{c}{M{\sc ain Radiometer }} \\																										
	& \multicolumn{6}{c}{G{\sc ain model}}& \multicolumn{6}{c}{G.M.{\sc correction}} \\																										
	\hhline{~------------}																										
	\multicolumn{1}{c}{SOURCE} & \multicolumn{1}{c}{$T\rm _{n,1}$}& \multicolumn{1}{c}{$b_1$} & \multicolumn{1}{c}{$G_1$} & \multicolumn{1}{c}{$T\rm _{n,2}$}& \multicolumn{1}{c}{$b_2$} & \multicolumn{1}{c}{$G_2$} & \multicolumn{1}{c}{$T\rm _{n,1}$}& \multicolumn{1}{c}{$b_1$} & \multicolumn{1}{c}{$G_1$} & \multicolumn{1}{c}{$T\rm _{n,2}$}& \multicolumn{1}{c}{$b_2$} & \multicolumn{1}{c}{$G_2$}\\																										
	\hline																										
	\texttt{RCA PAPER}	&	N.A.	&	N.A.	&	N.A.	&	N.A.	&	N.A.	&	N.A.	&	18.4	&	1.09	&	0.0052	&	17.4	&	1.42	&	0.0067	\\	
	\texttt{RCA}	&	27.8&	-0.29	&	0.0046	&	25.5	&	0.65	&	0.0046	&	19.7	&	0.84	&	0.0073	&	14.8	&	1.73	&	0.0143	\\	
	\texttt{FEM}	&	26.9	&	-0.15	&	0.0046	&	25.3	&	0.66	&	0.0046	&	19.5	&	0.91	&	0.0071	&	15.3	&	1.73	&	0.0133	\\	
	\texttt{CSL}	&	27.7	&	-0.56	&	0.0034	&	26.3	&	0.41	&	0.0034	&	18.6	&	1.12	&	0.0059	&	14.6	&	2.02	&	0.0108	\\	
	\texttt{CPV}	&	26.3	&	-0.68	&	0.0026	&	23.4	&	0.68	&	0.0026	&	15.5	&	2.16	&	0.0052	&	10.7	&	3.28	&	0.0102	\\	

      \hline		
      \hline	
	& \multicolumn{12}{c}{S{\sc ide Radiometer }} \\																								
	& \multicolumn{6}{c}{G{\sc ain model}}& \multicolumn{6}{c}{G.M.{\sc correction}} \\			 
	\hhline{~------------}																										
	\multicolumn{1}{c}{SOURCE} & \multicolumn{1}{c}{$T\rm _{n,1}$}& \multicolumn{1}{c}{$b_1$} & \multicolumn{1}{c}{$G_1$} & \multicolumn{1}{c}{$T\rm _{n,2}$}& \multicolumn{1}{c}{$b_2$} & \multicolumn{1}{c}{$G_2$} & \multicolumn{1}{c}{$T\rm _{n,1}$}& \multicolumn{1}{c}{$b_1$} & \multicolumn{1}{c}{$G_1$} & \multicolumn{1}{c}{$T\rm _{n,2}$}& \multicolumn{1}{c}{$b_2$} & \multicolumn{1}{c}{$G_2$}\\																										
	\hline																										
	\texttt{RCA PAPER}	&	N.A.	&	N.A.	&	N.A.	&	N.A.	&	N.A.	&	N.A.	&	16.8	&	0.94	&	0.0075	&	16.5	&	1.22	&	0.0082	\\	
	\texttt{RCA}	&	22.3	&	-0.13	&	0.0056	&	23.7	&	0.083	&	0.0056	&	14.5	&	1.07	&	0.0095	&	16.2	&	1.08	&	0.0097	\\	
	\texttt{FEM}	&	22.7	&	-0.14 &	0.0060	&	24.3	&	0.05	&	0.0060	&	14.9&	0.94	&	0.0101	&	16.6&	0.98	&	0.0103	\\
	\texttt{CSL}	&	22.6	&	-0.16	&	0.0059	&	24.8	&	-0.0470	&	0.0059	&	15.2	&	0.89	&	0.0097	&	17.1	&	0.90	&	0.0098	\\	
	\texttt{CPV}	&	21.9	&	-0.53	&	0.0028	&	23.2	&	-0.38	&	0.0028	&	9.9	&	3.24	&	0.0070	&	11.8	&	2.72	&	0.0067	\\	

\hline
      \end{tabular}																											
\end{center}																		 													
\end{sidewaystable}
 
\begin{sidewaystable}[htdp]															 															
  \begin{center}																											
    \caption{\texttt{LFI-27} Tuning: Gain-Model results. Results from several sets of biases tested during the CPV are compared among each other and to ground test results (\texttt{RCA paper} \cite{villa2010}). 
For each radiometer (separately \texttt{Main} and \texttt{Side}) and for each detector (labelled with 1, 2) the gain model parameters are shown on the right side of the table: noise temperatures $T\rm _n$ (given in K), the non-linear coefficients $b$ (adimensional) and the Gains $G$ (expressed as V/K). Corresponding values corrected for the BEU thermal drift are reported on the right side  (G.M. correction). 
The bias settings considered are optimal biases from: RCA test campaign (\texttt{RCA}), unit level test campaign (\texttt{FEM}), system level test campaign (\texttt{CSL}), hypermatrix tuning (\texttt{CPV}). 	}
																
    \label{tab_HYM_Tun_summary_27}																					

\small  						
    \begin{tabular}{l l l l l l l | l l l l l l}																												
      \hline																											
      \hline						
																					
	& \multicolumn{12}{c}{M{\sc ain Radiometer }} \\																										
	& \multicolumn{6}{c}{G{\sc ain model}} & \multicolumn{6}{c}{G.M.{\sc correction}} \\			 	
	\hhline{~------------}																										
	\multicolumn{1}{c}{SOURCE} & \multicolumn{1}{c}{$T\rm _{n,1}$}& \multicolumn{1}{c}{$b_1$} & \multicolumn{1}{c}{$G_1$} & \multicolumn{1}{c}{$T\rm _{n,2}$}& \multicolumn{1}{c}{$b_2$} & \multicolumn{1}{c}{$G_2$} & \multicolumn{1}{c}{$T\rm _{n,1}$}& \multicolumn{1}{c}{$b_1$} & \multicolumn{1}{c}{$G_1$} & \multicolumn{1}{c}{$T\rm _{n,2}$}& \multicolumn{1}{c}{$b_2$} & \multicolumn{1}{c}{$G_2$}\\																											
\hline																										
\texttt{RCA PAPER}	&	N.A.	&	N.A.	&	N.A.	&	N.A.	&	N.A.	&	N.A.	&	12.1	&	0.12	&	0.0723	&	11.9	&	0.12	&	0.0774	\\	
\texttt{RCA}	&	15.4	&	0.03	&	0.0776	&	16.1	&	0.02	&	0.0786	&	13.7	&	0.06	&	0.0934	&	13.7	&	0.06	&	0.1015	\\	
\texttt{FEM}	&	16.0	&	0.02	&	0.0774	&	16.2	&	0.02	&	0.0774	&	13.8	&	0.06	&	0.0911	&	13.8 &	0.0564	&	0.0992	\\
\texttt{CSL}	&	15.4	&	0.03	&	0.0765	&	15.5	&	0.03	&	0.0765	&	13.2	&	0.07	&	0.0904	&	13.2	&	0.0672	&	0.0992	\\	
\texttt{CPV}	&	15.4	&	0.03	&	0.0776	&	16.6	&	0.02	&	0.0675	&	13.2	&	0.07	&	0.0927	&	13.3	&	0.06	&	0.0998	\\	
	
      \hline																											
      \hline	
	& \multicolumn{12}{c}{S{\sc ide Radiometer }} \\																									
& \multicolumn{6}{c}{G{\sc ain model}} & \multicolumn{6}{c}{G.M.{\sc correction}} \\				 
	\hhline{~------------}																									
	\multicolumn{1}{c}{SOURCE} & \multicolumn{1}{c}{$T\rm _{n,1}$}& \multicolumn{1}{c}{$b_1$} & \multicolumn{1}{c}{$G_1$} & \multicolumn{1}{c}{$T\rm _{n,2}$}& \multicolumn{1}{c}{$b_2$} & \multicolumn{1}{c}{$G_2$} & \multicolumn{1}{c}{$T\rm _{n,1}$}& \multicolumn{1}{c}{$b_1$} & \multicolumn{1}{c}{$G_1$} & \multicolumn{1}{c}{$T\rm _{n,2}$}& \multicolumn{1}{c}{$b_2$} & \multicolumn{1}{c}{$G_2$}\\																												
\hline																										
\texttt{RCA PAPER}	&	N.A.	&	N.A.	&	N.A.	&	N.A.	&	N.A.	&	N.A.	&	13.0	&	0.13	&	0.0664	&	12.5	&	0.14	&	0.0562	\\	
\texttt{RCA}	&	18.7	&	0.01	&	0.0687	&	16.8	&	0.0132	&	0.0687	&	15.6	&	0.06	&	0.0855	&	14.3	&	0.06	&	0.0733	\\	
\texttt{FEM}	&	18.8	&	0.01	&	0.0660	&	16.9	&	0.01	&	0.0660	&	15.9	&	0.06	&	0.0808	&	14.5 &	0.06	&	0.0699	\\	
\texttt{CSL}	&	18.4	&	0.02	&	0.0674	&	16.7	&	0.01	&	0.0674	&	15.5	&	0.06	&	0.0828	&	14.3&	0.06	&	0.0704	\\	
\texttt{CPV}	&	18.2	&	0.02	&	0.0675	&	16.6	&	0.02	&	0.0675	&	15.4	&	0.07	&	0.0829	&	14.2	&	0.07	&	0.0701	\\	

\hline
      \end{tabular}																											
\end{center}																		 														
\end{sidewaystable}
 
\begin{sidewaystable}[htdp]															 																
  \begin{center}																											
    \caption{\texttt{LFI-28} Tuning: Gain-Model results. Results from several sets of biases tested during the CPV are compared among each other and to ground test results (\texttt{RCA paper} \cite{villa2010}). 
For each radiometer (separately \texttt{Main} and \texttt{Side}) and for each detector (labelled with 1, 2) the gain model parameters are shown on the right side of the table: noise temperatures $T\rm _n$ (given in K), the non-linear coefficients $b$ (adimensional) and the Gains $G$ (expressed as V/K). Corresponding values corrected for the BEU thermal drift are reported on the right side  (G.M. correction). 
The bias settings considered are optimal biases from: RCA test campaign (\texttt{RCA}), unit level test campaign (\texttt{FEM}), system level test campaign (\texttt{CSL}), hypermatrix tuning (\texttt{CPV}). 	}																		
    \label{tab_HYM_Tun_summary_28}
\small
    \begin{tabular}{l l l l l l l | l l l l l l}			 
      \hline																											
      \hline						
    																					
	& \multicolumn{12}{c}{M{\sc ain Radiometer }} \\																										
	& \multicolumn{6}{c}{G{\sc ain model}} & \multicolumn{6}{c}{G.M.{\sc correction}} \\			 	
	\hhline{~------------}																										
	\multicolumn{1}{c}{SOURCE} & \multicolumn{1}{c}{$T\rm _{n,1}$}& \multicolumn{1}{c}{$b_1$} & \multicolumn{1}{c}{$G_1$} & \multicolumn{1}{c}{$T\rm _{n,2}$}& \multicolumn{1}{c}{$b_2$} & \multicolumn{1}{c}{$G_2$} & \multicolumn{1}{c}{$T\rm _{n,1}$}& \multicolumn{1}{c}{$b_1$} & \multicolumn{1}{c}{$G_1$} & \multicolumn{1}{c}{$T\rm _{n,2}$}& \multicolumn{1}{c}{$b_2$} & \multicolumn{1}{c}{$G_2$}\\																											
\hline																							
\texttt{RCA PAPER}	&	N.A.	&	N.A.	&	N.A.	&	N.A.	&	N.A.	&	N.A.	&	10.6	&	0.19	&	0.0621	&	10.3	&	0.16	&	0.0839	\\
\texttt{RCA}	&	14.9	&	0.03	&	0.0647	&	14.2	&	0.04	&	0.0647	&	12.7	&	0.08	&	0.0771	&	12.4	&	0.0735	&	0.1028	\\
\texttt{FEM}	&	15.7	&	0.02	&	0.0598	&	15.2	&	0.03	&	0.0598	&	13.2	&	0.07	&	0.0728	&	13.0	&	0.07	&	0.0958	\\
\texttt{CSL}	&	15.7	&	0.01	&	0.0597	&	14.7	&	0.03	&	0.0597	&	13.0	&	0.07	&	0.0735	&	12.6	&	0.07	&	0.0984	\\
\texttt{CPV}	&	15.7	&	0.01	&	0.0597	&	14.7&	0.03	&	0.0597	&	13.0	&	0.07	&	0.0735	&	12.6	&	0.07	&	0.0984	\\

      \hline																											
      \hline	
	& \multicolumn{12}{c}{S{\sc ide Radiometer }} \\																									
& \multicolumn{6}{c}{G{\sc ain model}} & \multicolumn{6}{c}{G.M.{\sc correction}} \\				 
	\hhline{~------------}																								
	\multicolumn{1}{c}{SOURCE} & \multicolumn{1}{c}{$T\rm _{n,1}$}& \multicolumn{1}{c}{$b_1$} & \multicolumn{1}{c}{$G_1$} & \multicolumn{1}{c}{$T\rm _{n,2}$}& \multicolumn{1}{c}{$b_2$} & \multicolumn{1}{c}{$G_2$} & \multicolumn{1}{c}{$T\rm _{n,1}$}& \multicolumn{1}{c}{$b_1$} & \multicolumn{1}{c}{$G_1$} & \multicolumn{1}{c}{$T\rm _{n,2}$}& \multicolumn{1}{c}{$b_2$} & \multicolumn{1}{c}{$G_2$}\\																											
\hline																									
\texttt{RCA PAPER}	&	N.A.	&	N.A.	&	N.A.	&	N.A.	&	N.A.	&	N.A.	&	9.9	&	0.19	&	0.0607	&	9.8	&	0.20	&	0.0518	\\
\texttt{RCA}	&	15.8	&	0.01	&	0.0573	&	14.8&	0.01	&	0.0573	&	13.3	&	0.07	&	0.0692	&	9.9	&	0.14	&	0.0792	\\
\texttt{FEM}	&	16.0	&	0.01	&	0.0566	&	14.9	&	0.01	&	0.0566	&	13.5&	0.07	&	0.0689	&	9.7	&	0.15	&	0.0815	\\
\texttt{CSL}	&	15.9	&	0.01	&	0.0564	&	14.7&	0.03	&	0.0597	&	13.3	&	0.07	&	0.0686	&	9.7	&	0.14	&	0.0804	\\
\texttt{CPV}	&	15.7	&	0.01	&	0.0540	&	15.7	&	0.01	&	0.0540	&	13.1	&	0.07	&	0.0663	&	9.4	&	0.16	&	0.0782	\\
 
\hline
      \end{tabular}																											
\end{center}																		 												
\end{sidewaystable}
 
\begin{table}[h!]
    \begin{center}
        \caption{\label{tab_HYM_nonlin_Tun_vs_CSL} Optimal biases resulting from \texttt{HYM}  Maps when the Linear Model (\texttt{LIN}), the Gain Model (\texttt{G.M.}) and the Gain Model corrected (labelled as \texttt{G.M.+C}) for the BEU thermal drift  are applied. Voltage units are decimal values: conversion in Volts can be roughly obtained referring to Table 4. }
        \vspace{.05cm} \small    
        \begin{tabular}{l|c|c|c}
 \hline
 	\hline                    

&			\texttt{LIN}	&	\texttt{G.M.}	&	\texttt{G.M.+C}	\\
\texttt{RCA}	&	$V\rm _{g1}$	,	$V\rm _{g2}$	&	$V\rm _{g1}$	,	$V\rm _{g2}$	&	$V\rm _{g1}$,	$V\rm _{g2}$ \\
 	\hline  
\texttt{24M2} 	&	227	,	204	&	230	,	211	&	227	,	204	\\
\texttt{24M1} 	&	219	,	204	&	228	,	208	&	215	,	219	\\
\texttt{24S2} 	&	225	,	208	&	219	,	213	&	220	,	218	\\
\texttt{24S1} 	&	218	,	207	&	219	,	219	&	210	,	223	\\
\texttt{25M1} 	&	231	,	203	&	208	,	202	&	237	,	203	\\
\texttt{25M2} 	&	218	,	200	&	222	,	223	&	218	,	200	\\
\texttt{25S1} 	&	231	,	196	&	231	,	191	&	231	,	191	\\
\texttt{25S2} 	&	223	,	199	&	238	,	193	&	232	,	194	\\
\texttt{26M2} 	&	226	,	200	&	248	,	207	&	232	,	209	\\
\texttt{26M1} 	&	247	,	203	&	238	,	195	&	233	,	207	\\
\texttt{26S2} 	&	240	,	197	&	225	,	201	&	225	,	201	\\
\texttt{26S1} 	&	227	,	194	&	231	,	200	&	231	,	200\\
\texttt{27M1} 	&	242	,	97	&	243	,	109	&	240	,	108	\\
\texttt{27M2} 	&	255	,	96	&	244	,	70	&	253	,	77\\
\texttt{27S1} 	&	235	,	86	&	243	,	98	&	238	,	86\\
\texttt{27S2} 	&	248	,	113	&	246	,	94	&	255	,	106	\\
\texttt{28M1} 	&	243	,	101	&	240	,	101	&	240	,	101\\
\texttt{28M2} 	&	240	,	112	&	246	,	152	&	237	,	156\\
\texttt{28S1} 	&	235	,	81	&	235	,	88	&	234	,	104	\\
\texttt{28S2} 	&	249	,	90	&	248	,	121	&	254	,	116\\
 	\hline
\end{tabular}
    \end{center}
\end{table}
 
\begin{table}[h!]
    \begin{center}
        \caption{\label{tab_DAE_def_Cons} DAE Default: Front-End Bias and  Power Consumption.}
        \vspace{.05cm}       \small  
        \begin{tabular}{l|c c|c c|c c|c}
 \hline
 	\hline               

	&	\texttt{V$\rm _{g1}$ (V)}	&	\texttt{V$\rm _{g2}$ (V)}	&	\texttt{V$\rm _{d}$ (V)}	&	\texttt{I$\rm _1$ (mA)}	&	\texttt{I$\rm _2$ (mA)} 	&	\texttt{I$\rm _d$ (mA)}	&	\texttt{Power (mW)}\\
	 	\hline  
\texttt{18M1}	&	1.44	&	1.40	&	0.35	&	0.99	&	0.99	&	13.11	&	4.65\\
\texttt{18M2}	&	1.47	&	1.49	&	0.35	&	0.99	&	0.99	&	14.43	&	5.00\\
\texttt{18S1}	&	1.13	&	1.60	&	0.39	&	0.99	&	0.99	&	15.92	&	6.22\\
\texttt{18S2}	&	1.61	&	1.34	&	0.29	&	0.99	&	0.99	&	18.45	&	5.29\\
\texttt{19M1}	&	1.58	&	1.52	&	0.32	&	0.99	&	0.99	&	18.19	&	5.82\\
\texttt{19M2}	&	1.56	&	1.54	&	0.32	&	0.99	&	0.99	&	19.92	&	6.40\\
\texttt{19S1}	&	1.59	&	1.54	&	0.31	&	0.99	&	0.99	&	17.90	&	5.47\\
\texttt{19S2}	&	1.51	&	1.60	&	0.33	&	0.99	&	0.99	&	17.02	&	5.58\\
\texttt{20M1}	&	1.54	&	1.62	&	0.30	&	0.99	&	0.99	&	20.63	&	6.15\\
\texttt{20M2}	&	1.59	&	1.64	&	0.32	&	1.00	&	1.00	&	20.63	&	6.64\\
\texttt{20S1}	&	1.47	&	1.64	&	0.35	&	0.99	&	0.99	&	18.82	&	6.56\\
\texttt{20S2}	&	1.38	&	1.48	&	0.33	&	0.99	&	0.99	&	18.62	&	6.14\\
\texttt{21M1}	&	1.47	&	1.54	&	0.39	&	0.99	&	0.99	&	18.49	&	7.25\\
\texttt{21M2}	&	1.45	&	1.46	&	0.37	&	0.99	&	0.99	&	19.43	&	7.18\\
\texttt{21S1}	&	1.25	&	1.65	&	0.37	&	0.99	&	0.99	&	19.11	&	7.08\\
\texttt{21S2}	&	1.52	&	1.82	&	0.34	&	0.99	&	0.99	&	22.04	&	7.56\\
\texttt{22M1}	&	1.50	&	1.43	&	0.35	&	0.99	&	0.99	&	14.07	&	4.89\\
\texttt{22M2}	&	1.31	&	1.29	&	0.36	&	0.99	&	0.99	&	14.88	&	5.43\\
\texttt{22S1}	&	1.51	&	1.39	&	0.35	&	0.99	&	0.99	&	16.35	&	5.74\\
\texttt{22S2}	&	1.53	&	1.51	&	0.36	&	1.00	&	1.00	&	15.10	&	5.48\\
\texttt{23M1}	&	1.53	&	1.42	&	0.32	&	0.99	&	0.99	&	14.71	&	4.71\\
\texttt{23M2}	&	1.56	&	1.44	&	0.32	&	0.99	&	0.99	&	14.20	&	4.53\\
\texttt{23S1}	&	1.33	&	1.56	&	0.29	&	0.99	&	0.99	&	20.57	&	5.96\\
\texttt{23S2}	&	1.40	&	1.54	&	0.33	&	0.99	&	0.99	&	14.79	&	4.85\\
\texttt{24M1}	&	1.19	&	1.19	&	0.91	&	0.80	&	0.80	&	16.53	&	14.98\\
\texttt{24M2}	&	1.19	&	1.19	&	0.88	&	0.80	&	0.80	&	15.00	&	13.18\\
\texttt{24S1}	&	1.19	&	1.19	&	0.73	&	0.80	&	0.80	&	15.58	&	11.33\\
\texttt{24S2}	&	1.19	&	1.19	&	0.73	&	0.80	&	0.80	&	16.46	&	12.00\\
\texttt{25M1}	&	1.19	&	1.19	&	0.88	&	0.80	&	0.80	&	14.37	&	12.59\\
\texttt{25M2}	&	1.19	&	1.19	&	0.89	&	0.80	&	0.80	&	14.87	&	13.17\\
\texttt{25S1}	&	1.19	&	1.19	&	0.79	&	0.80	&	0.80	&	14.81	&	11.64\\
\texttt{25S2}	&	1.19	&	1.19	&	0.78	&	0.80	&	0.80	&	14.67	&	11.38\\
\texttt{26M1}	&	1.19	&	1.19	&	0.84	&	0.80	&	0.80	&	12.85	&	10.79\\
\texttt{26M2}	&	1.19	&	1.19	&	0.84	&	0.80	&	0.80	&	14.72	&	12.37\\
\texttt{26S1}	&	1.19	&	1.19	&	0.83	&	0.80	&	0.80	&	13.40	&	11.19\\
\texttt{26S2}	&	1.19	&	1.19	&	0.85	&	0.80	&	0.80	&	13.72	&	11.59\\
\texttt{27M1}	&	1.54	&	-1.56	&	0.75	&	0.58	&	0.86	&	8.12	&	6.05\\
\texttt{27M2}	&	1.64	&	-1.98	&	0.75	&	0.57	&	0.80	&	7.22	&	5.45\\
\texttt{27S1}	&	1.47	&	-1.70	&	0.75	&	0.50	&	0.72	&	8.44	&	6.33\\
\texttt{27S2}	&	1.68	&	-1.42	&	0.75	&	0.58	&	0.76	&	7.99	&	5.96\\
\texttt{28M1}	&	1.61	&	-1.73	&	0.74	&	0.51	&	0.62	&	9.63	&	7.11\\
\texttt{28M2}	&	1.54	&	-1.47	&	0.73	&	0.50	&	0.89	&	9.19	&	6.75\\
\texttt{28S1}	&	1.53	&	-2.12	&	0.74	&	0.50	&	0.86	&	9.11	&	6.74\\
\texttt{28S2}	&	1.65	&	-1.26	&	0.74	&	0.40	&	0.64	&	10.49	&	7.77\\

 	\hline
\end{tabular}
    \end{center}
\end{table}
 
\clearpage

%% file: a06_tuning_plots.tex
\section{Hypermatrix tuning noise maps}
\label{app_HYM_plots}

    In the next figures we show the condensed noise temperature maps (used as input to select the optimal biases) corresponding to the linear hypermatrix tuning analysis. \\
Each plot shows the hypermatrix tuning noise temperature map corresponding to $V\rm _{g2}$ biases (y-axis) versus  $V\rm _{g1}$ biases (x-axis). For each channel  they are represented from left to right: first row: \texttt{M1}, \texttt{M2}; second row: \texttt{S1}, \texttt{S2}.  Lower noise temperatures are shown in grey, higher in red (normalized to the lowest noise temperature). The best condensed noise temperature is enhanced by a yellow circle; system level test default by a black cross.\\
\clearpage
   \begin{figure}[htb]

        \begin{center}
                		\textbf{LFI-18}\\
    \vskip 0.5 cm  
  	    \includegraphics[width=6.9cm]{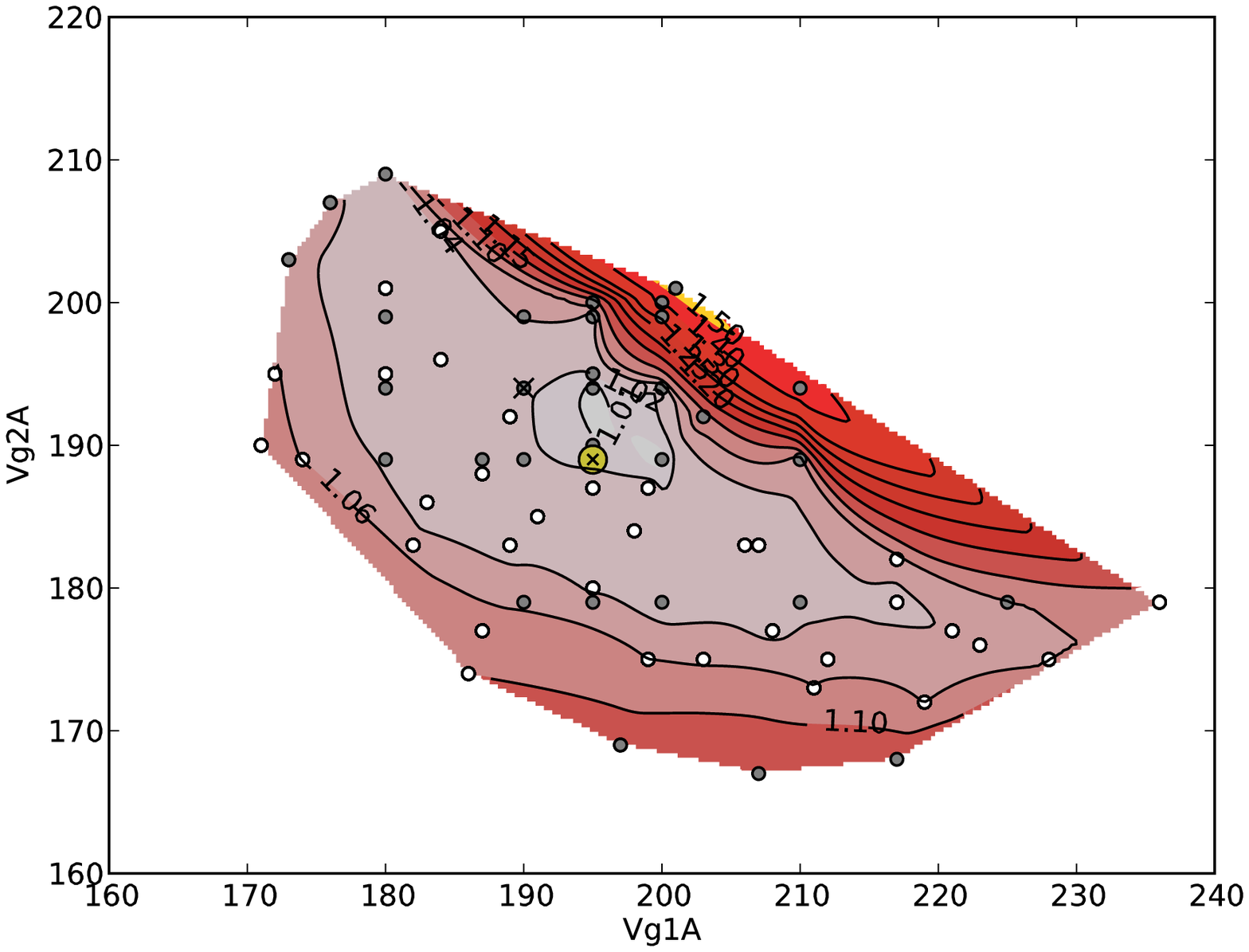} 
            \includegraphics[width=6.9cm]{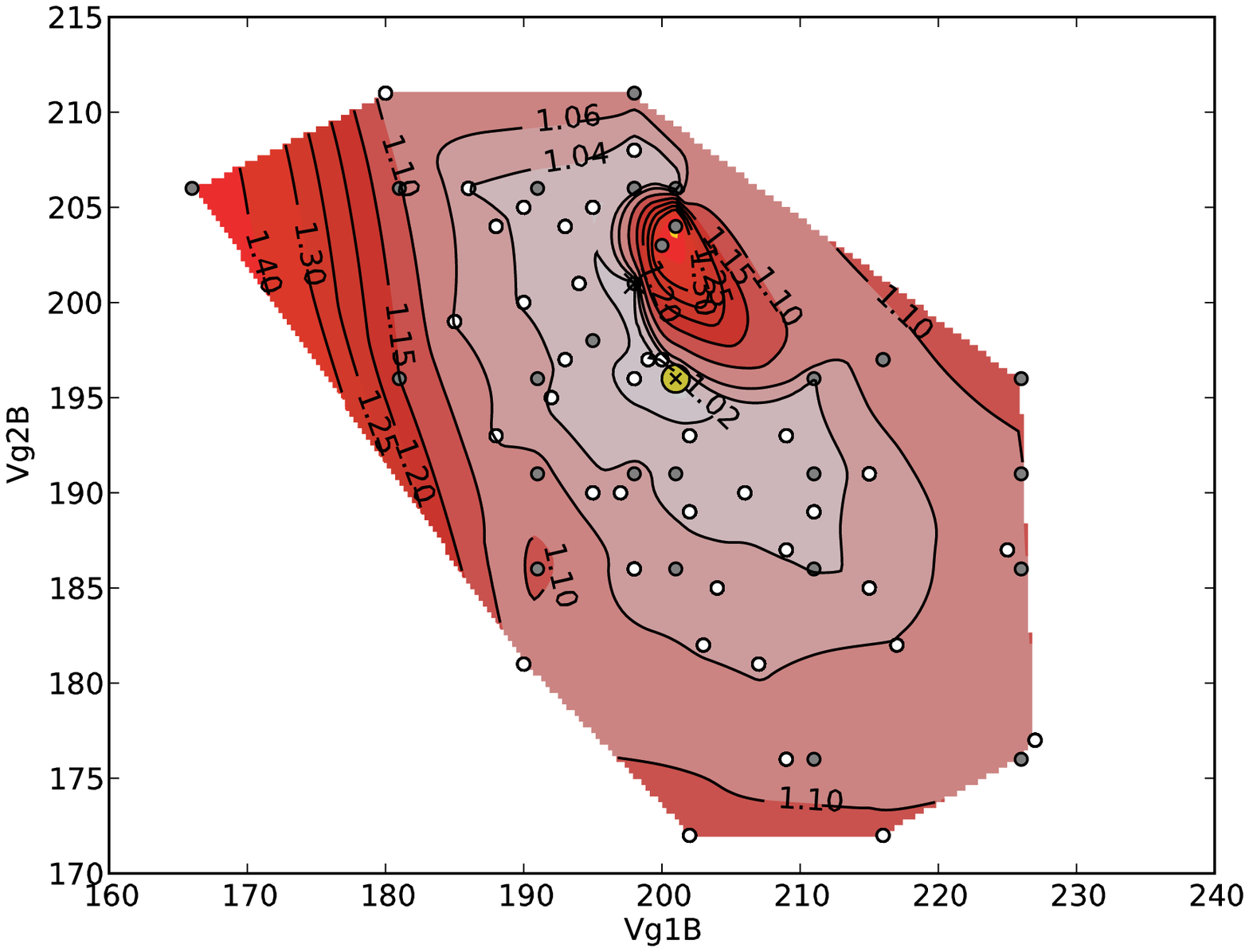}\\
            \includegraphics[width=6.9cm]{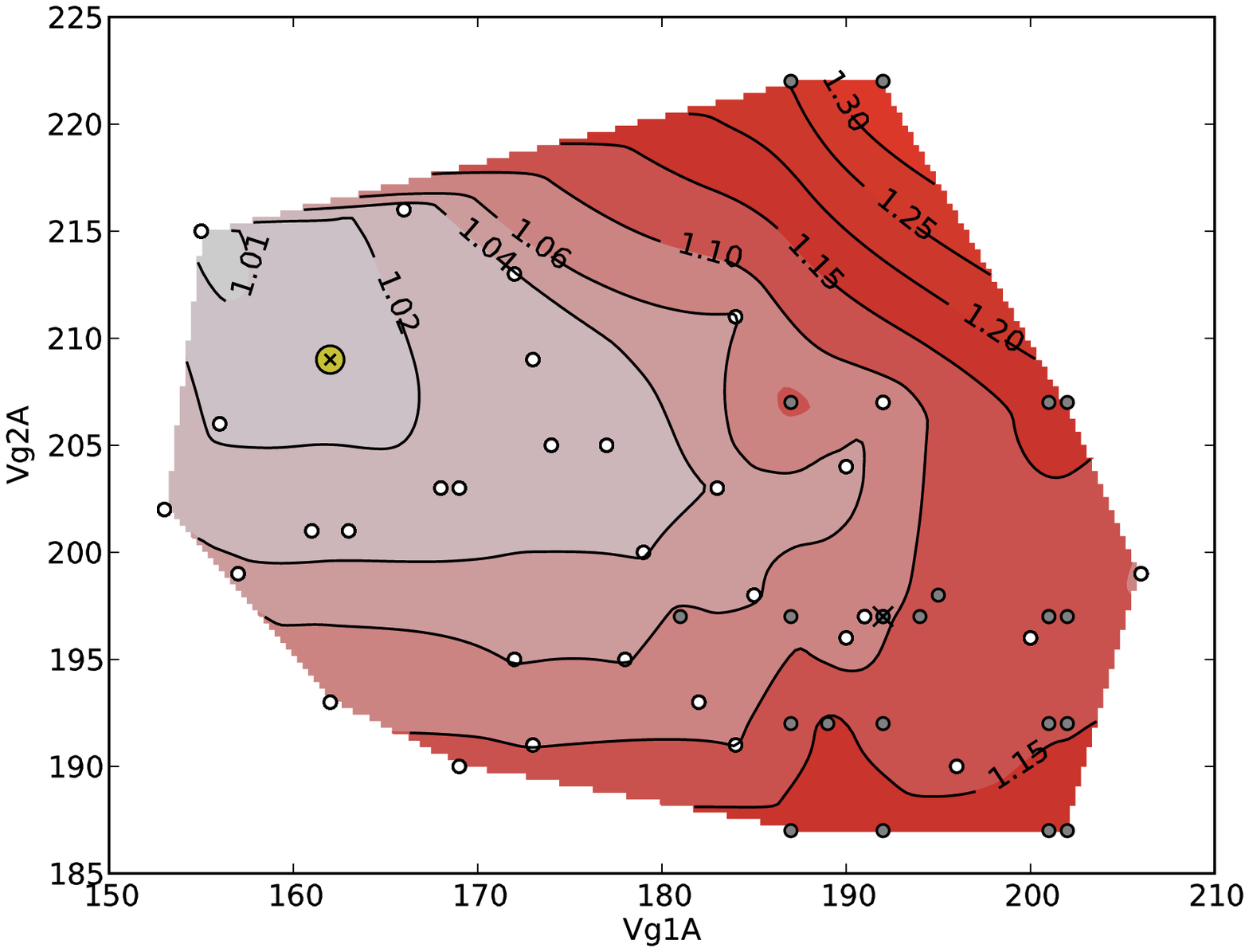}
            \includegraphics[width=6.9cm]{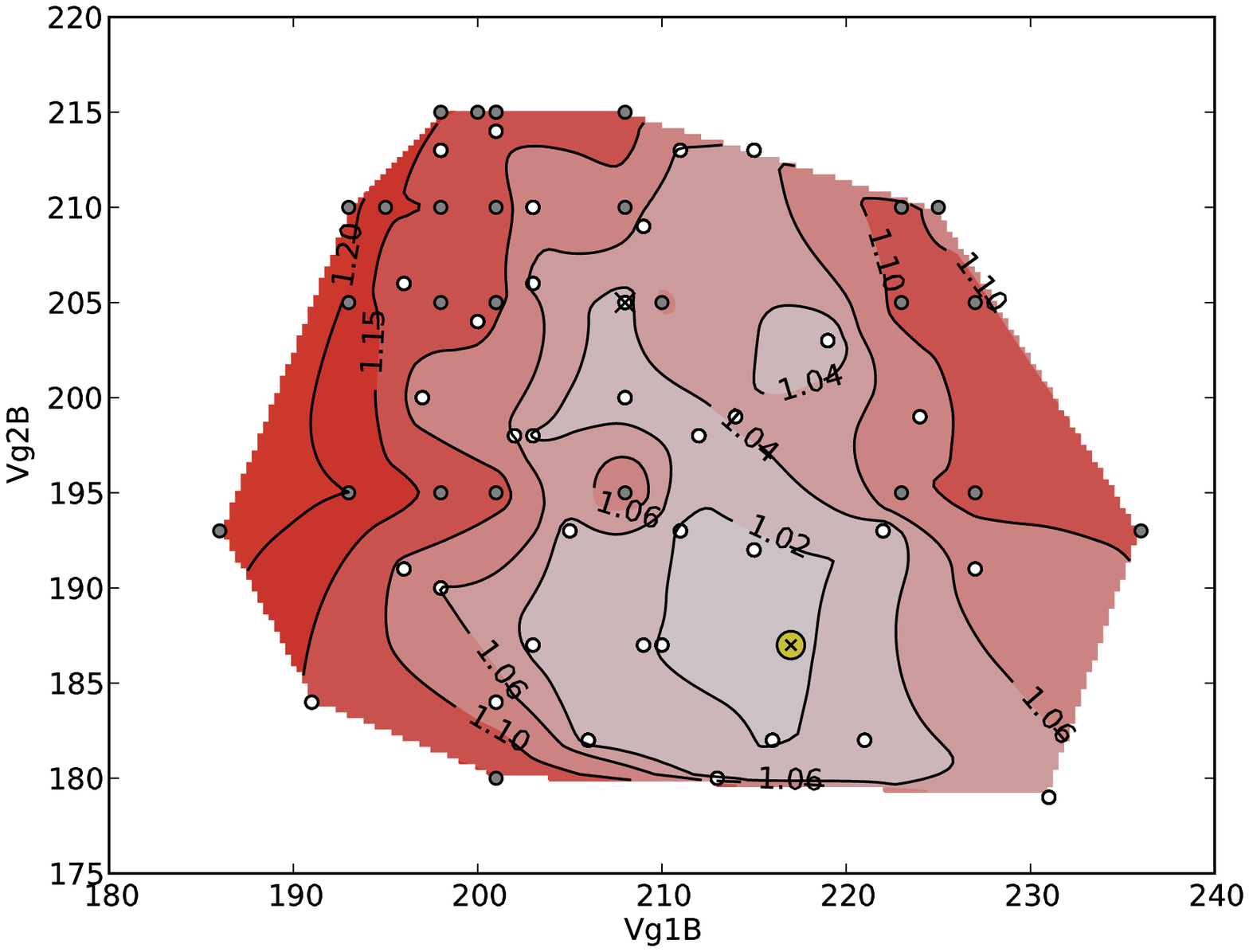}\\
            
            \textbf{LFI-19}\\
     \vskip 0.5 cm             
            \includegraphics[width=6.9cm]{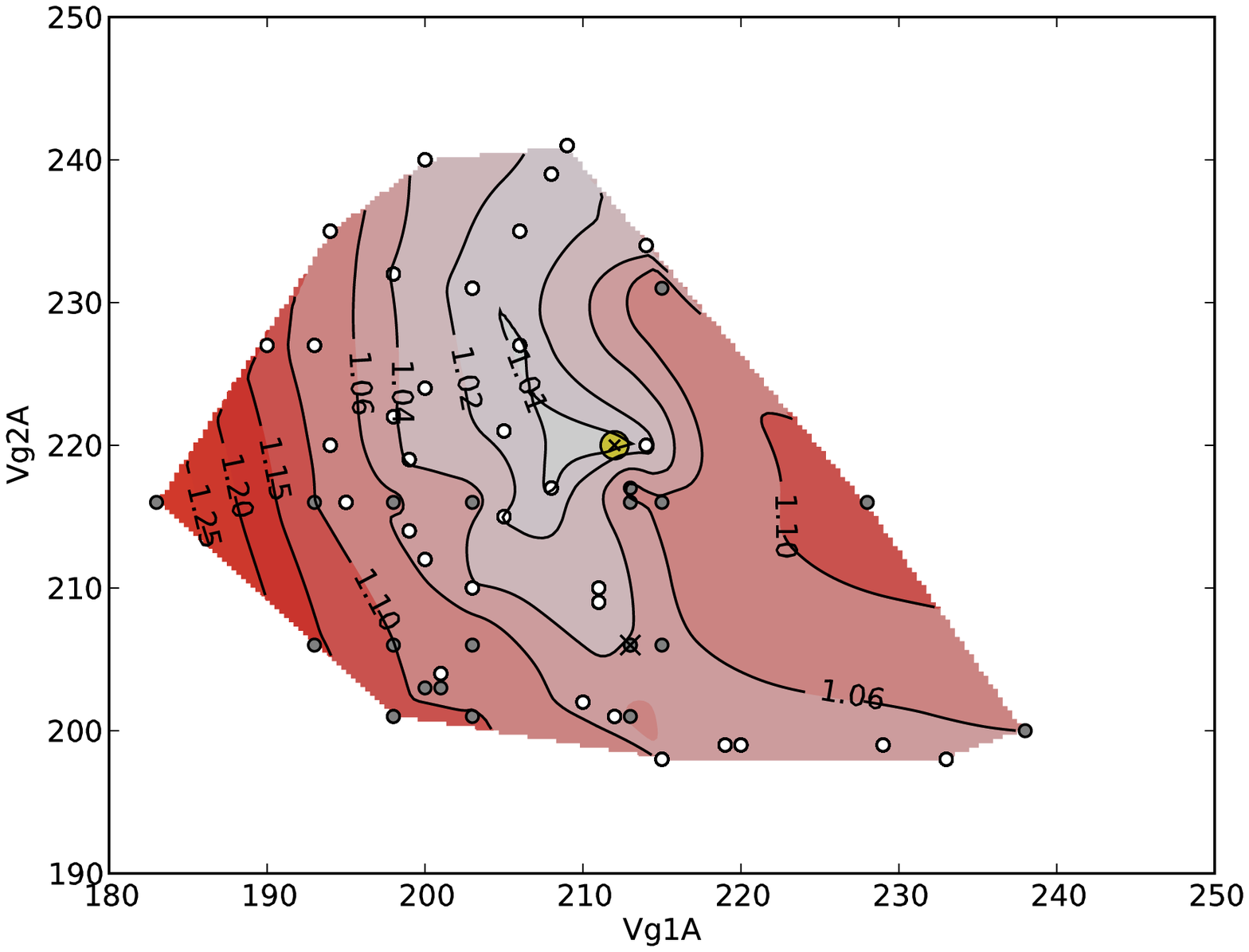}
            \includegraphics[width=6.9cm]{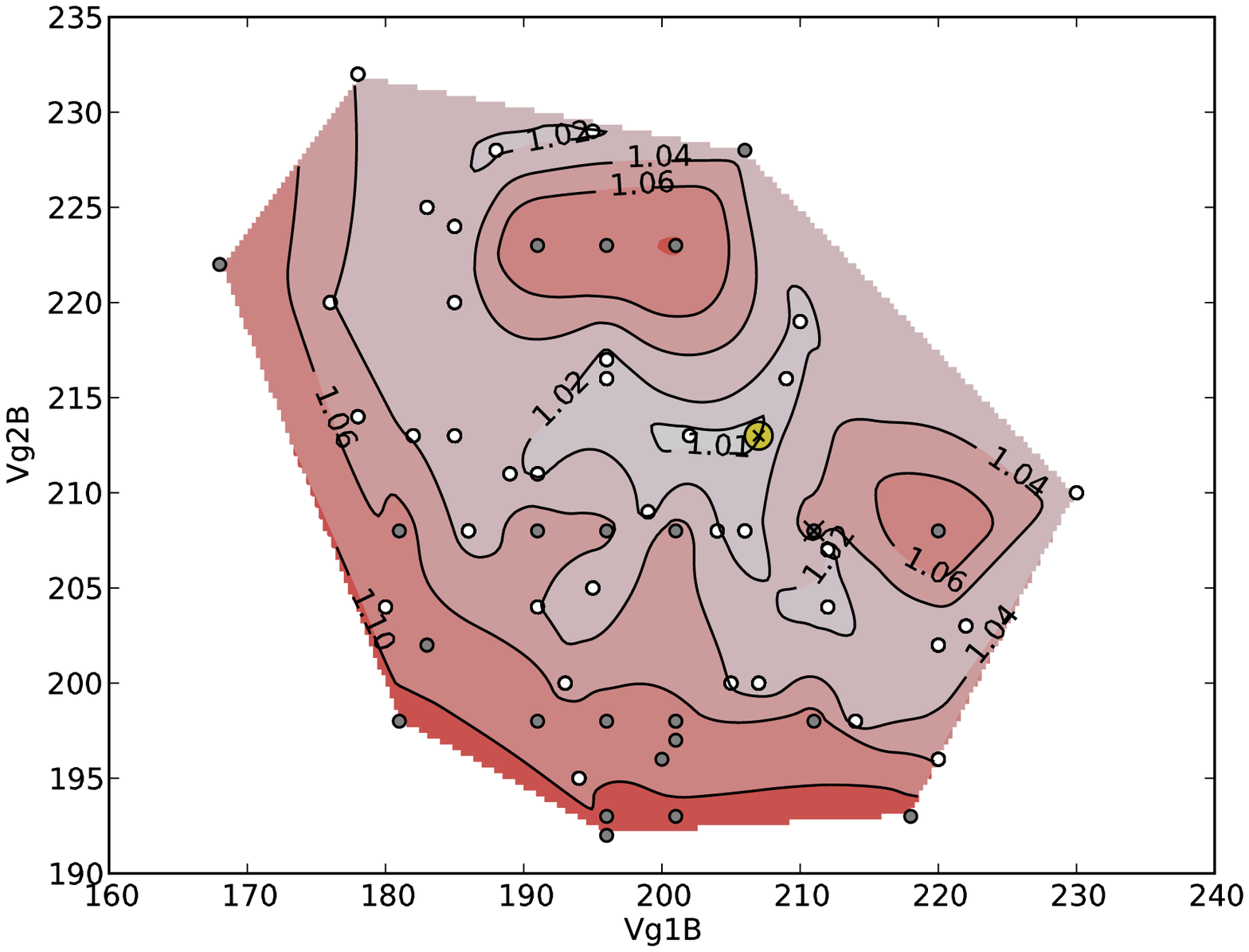}\\
            \includegraphics[width=6.9cm]{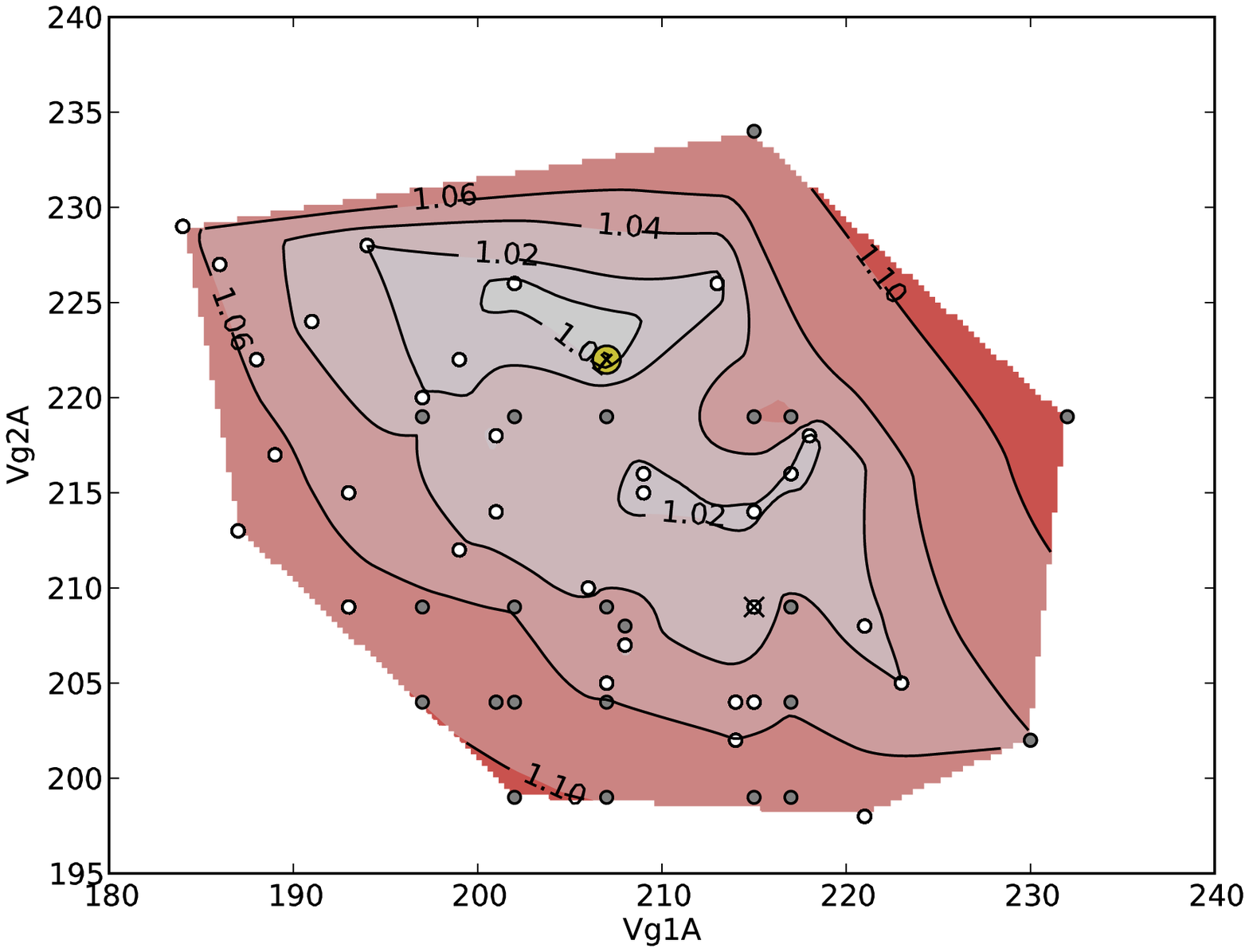}
            \includegraphics[width=6.9cm]{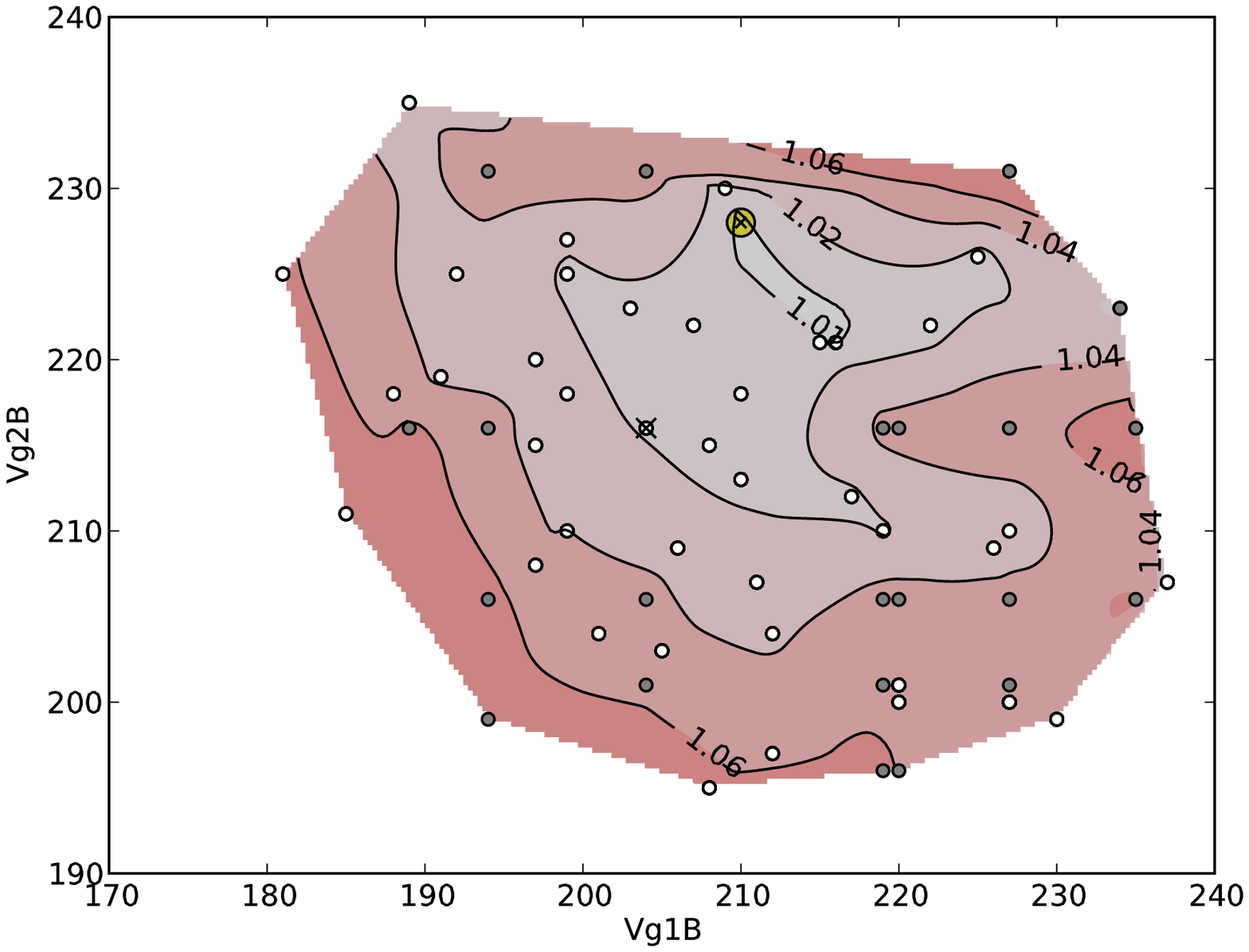}\\
                       \end{center}
    
        \label{fig_HYM_tun_18-19}
                    \end{figure}
                    
   \begin{figure}[htb]
        \begin{center}
                		\textbf{LFI-20}\\
    \vskip 0.5 cm  
  		       \includegraphics[width=6.9cm]{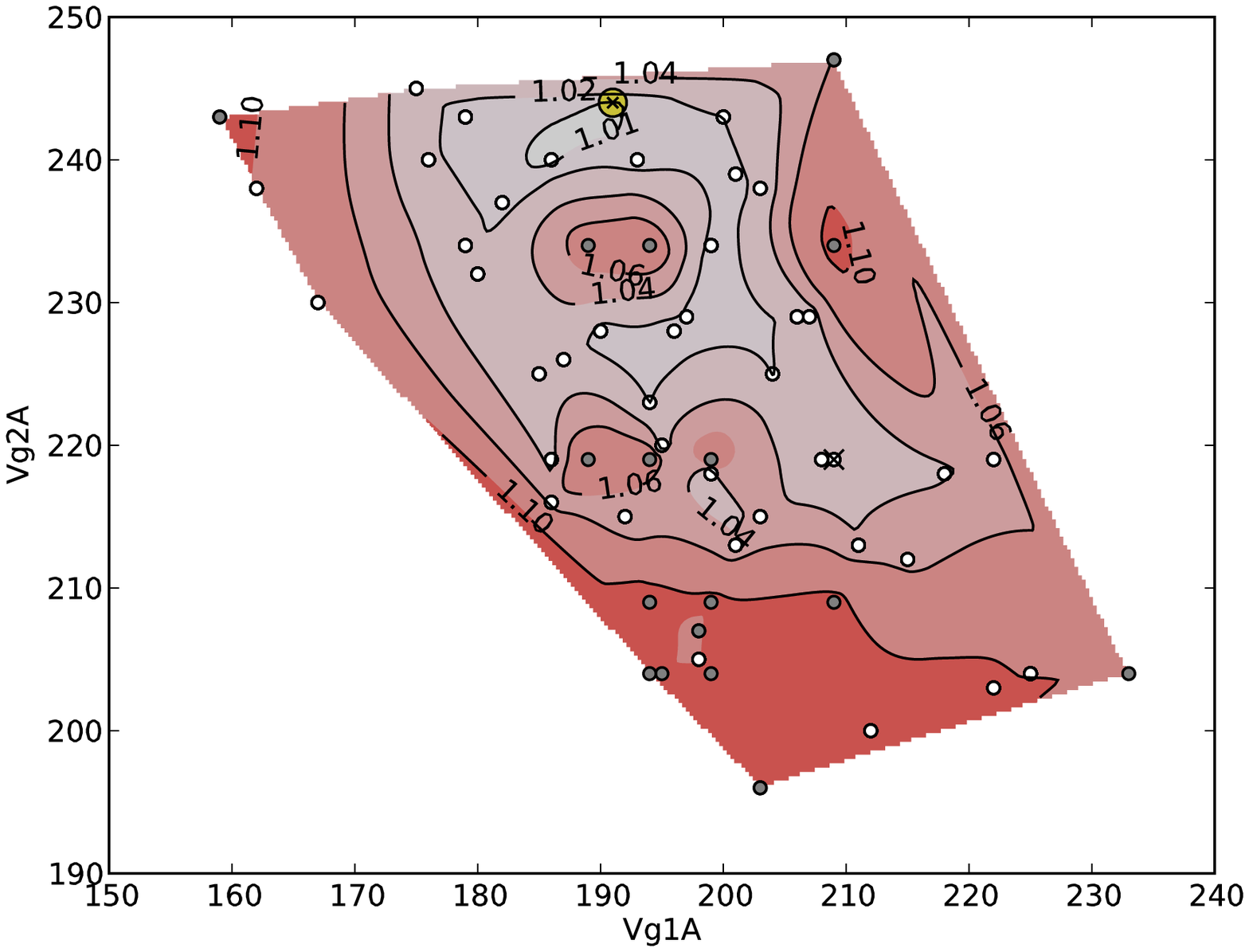} 
            \includegraphics[width=6.9cm]{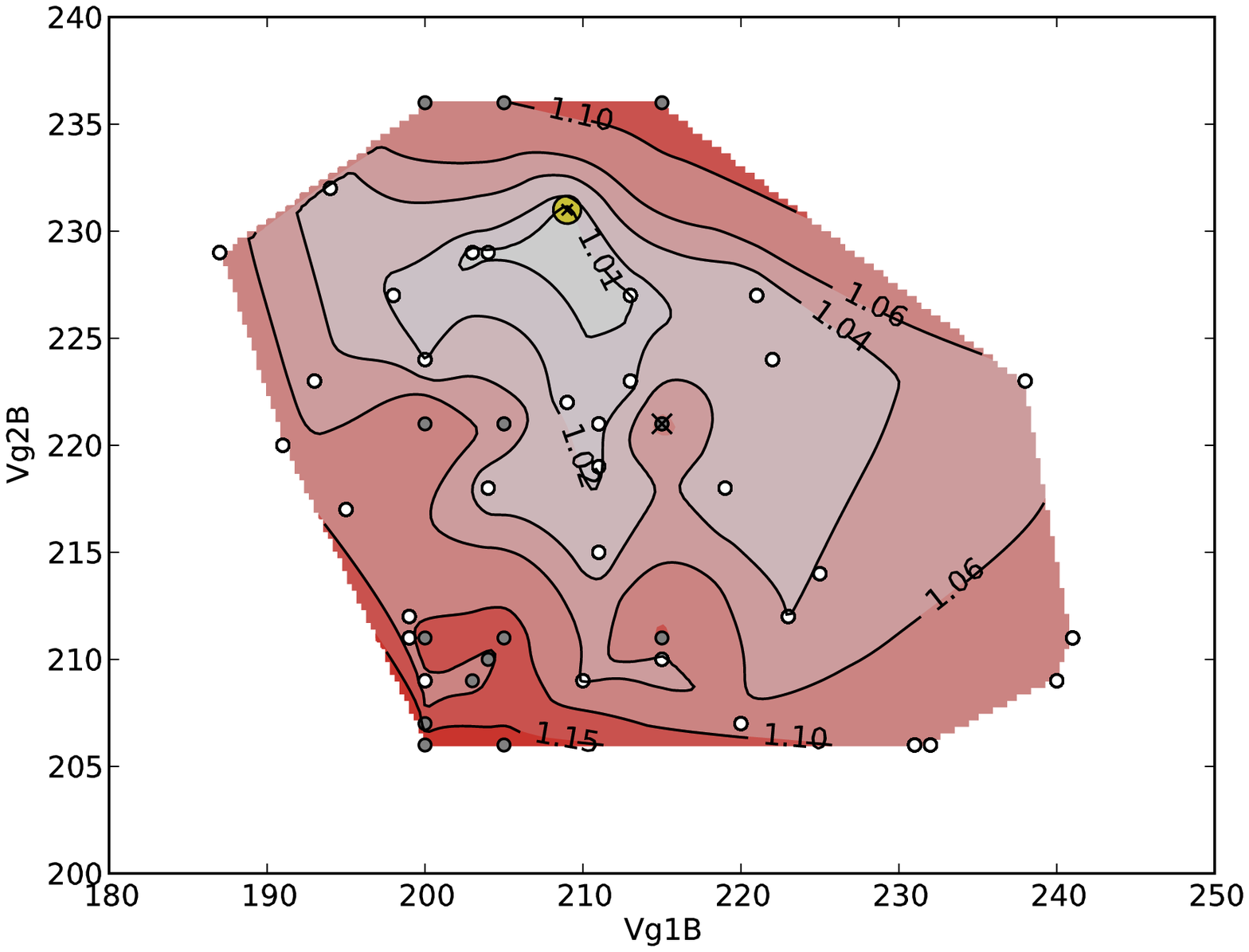}\\
            \includegraphics[width=6.9cm]{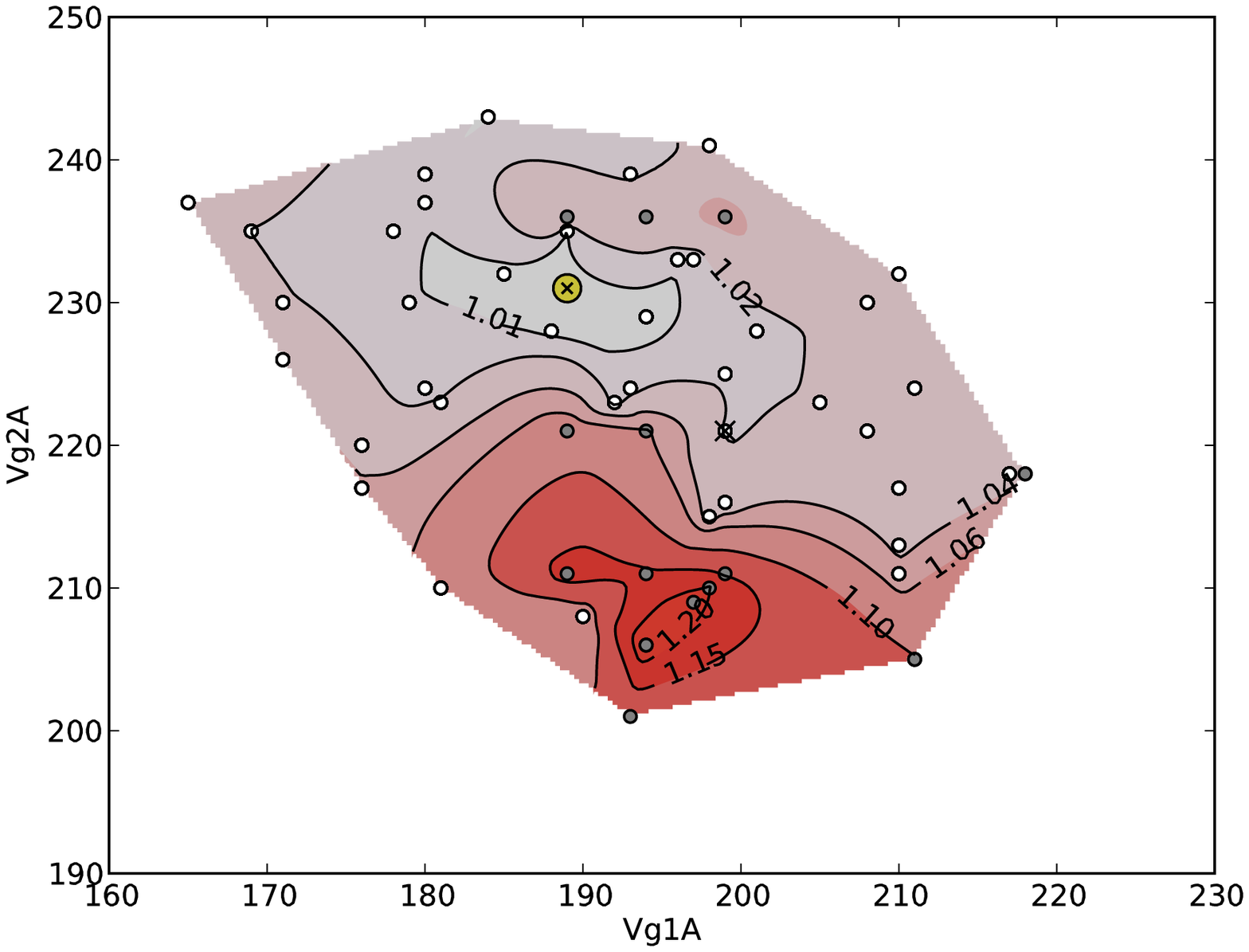}
            \includegraphics[width=6.9cm]{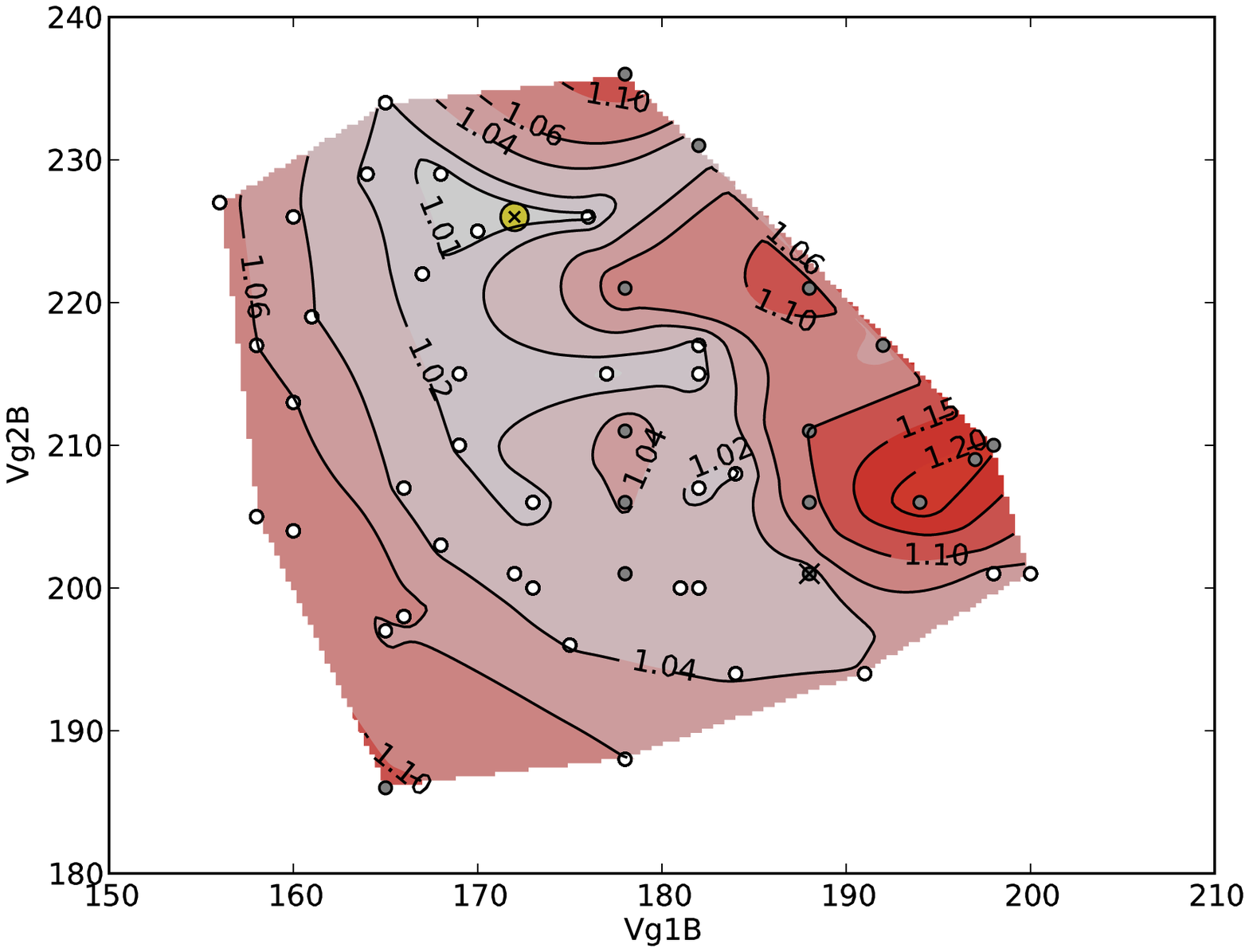}\\
            
            \textbf{LFI-21}\\
     \vskip 0.5 cm             
            \includegraphics[width=6.9cm]{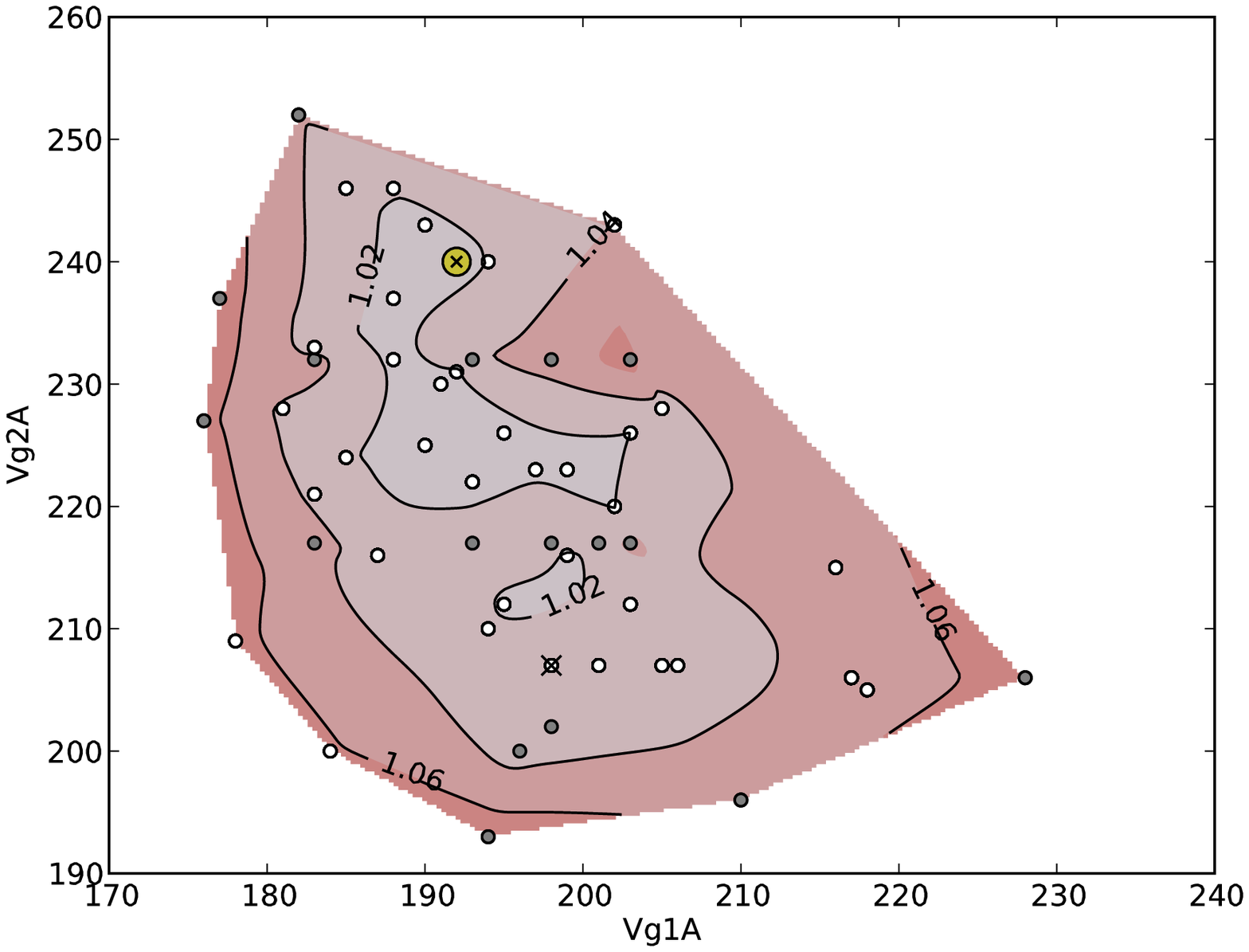}
            \includegraphics[width=6.9cm]{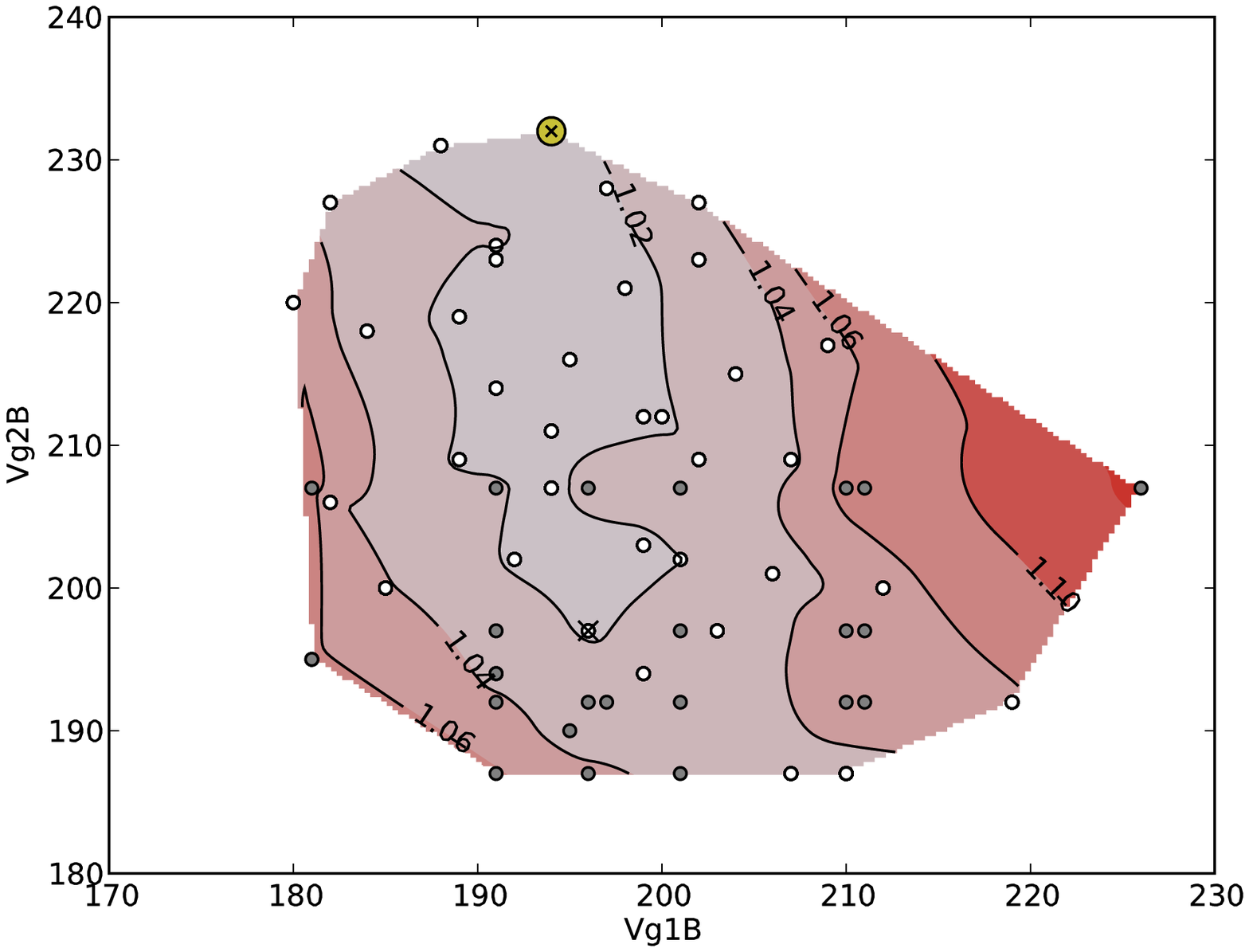}\\
            \includegraphics[width=6.9cm]{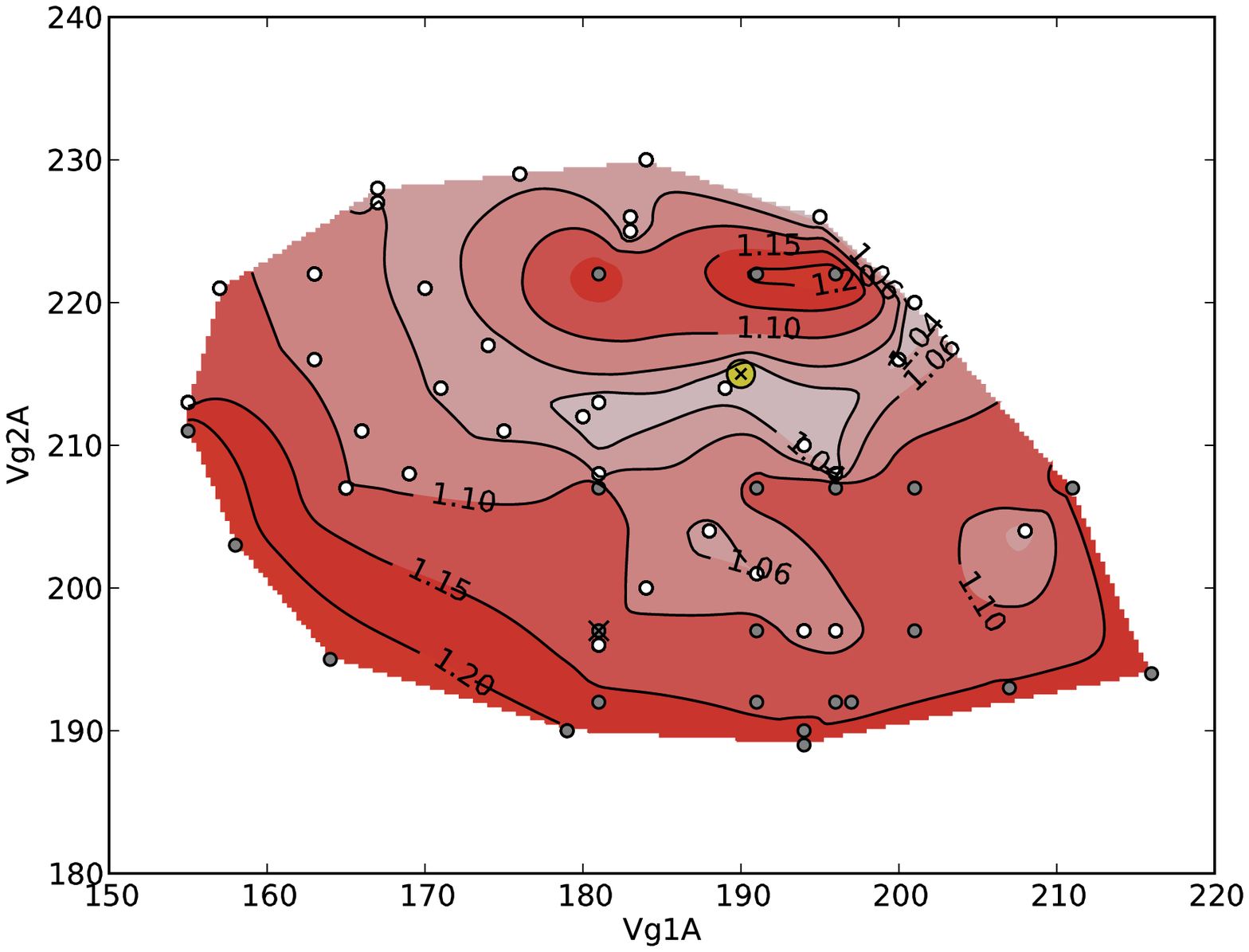}
            \includegraphics[width=6.9cm]{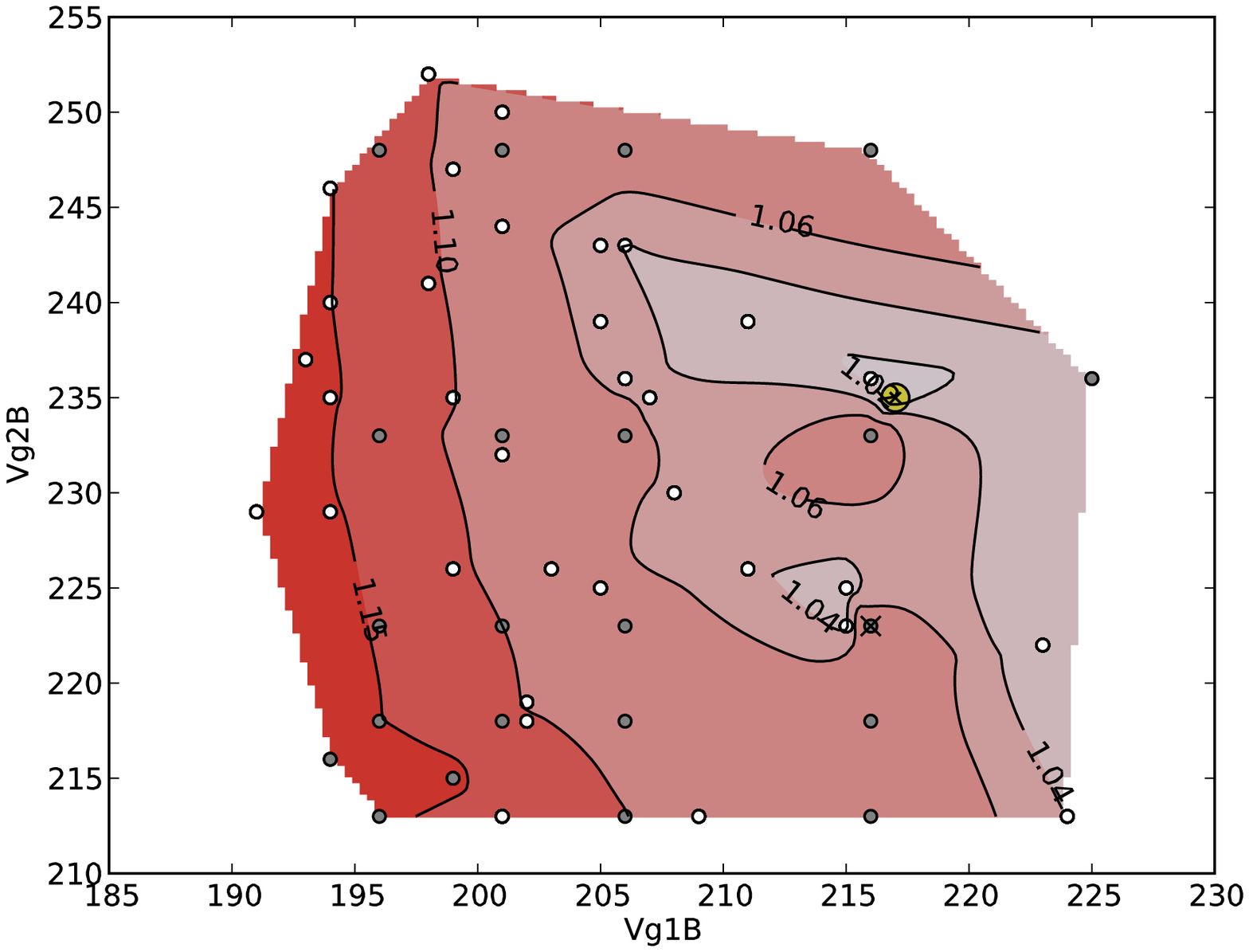}\\
                       \end{center}
            
        \label{fig_HYM_tun_20-21}
                    \end{figure}       
 
   \begin{figure}[htb]
        \begin{center}
                		\textbf{LFI-22}\\
    \vskip 0.5 cm  
  		       \includegraphics[width=6.9cm]{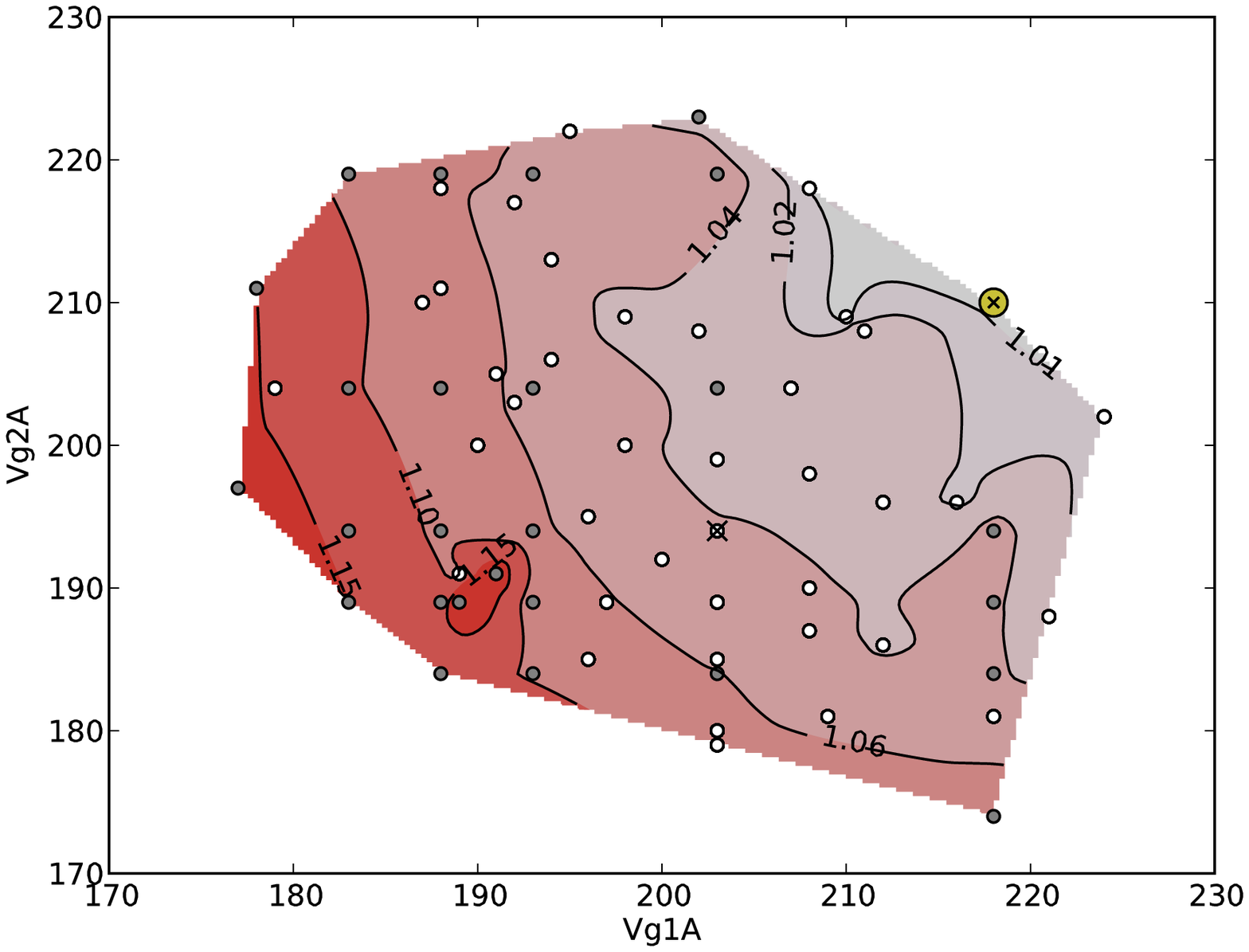} 
            \includegraphics[width=6.9cm]{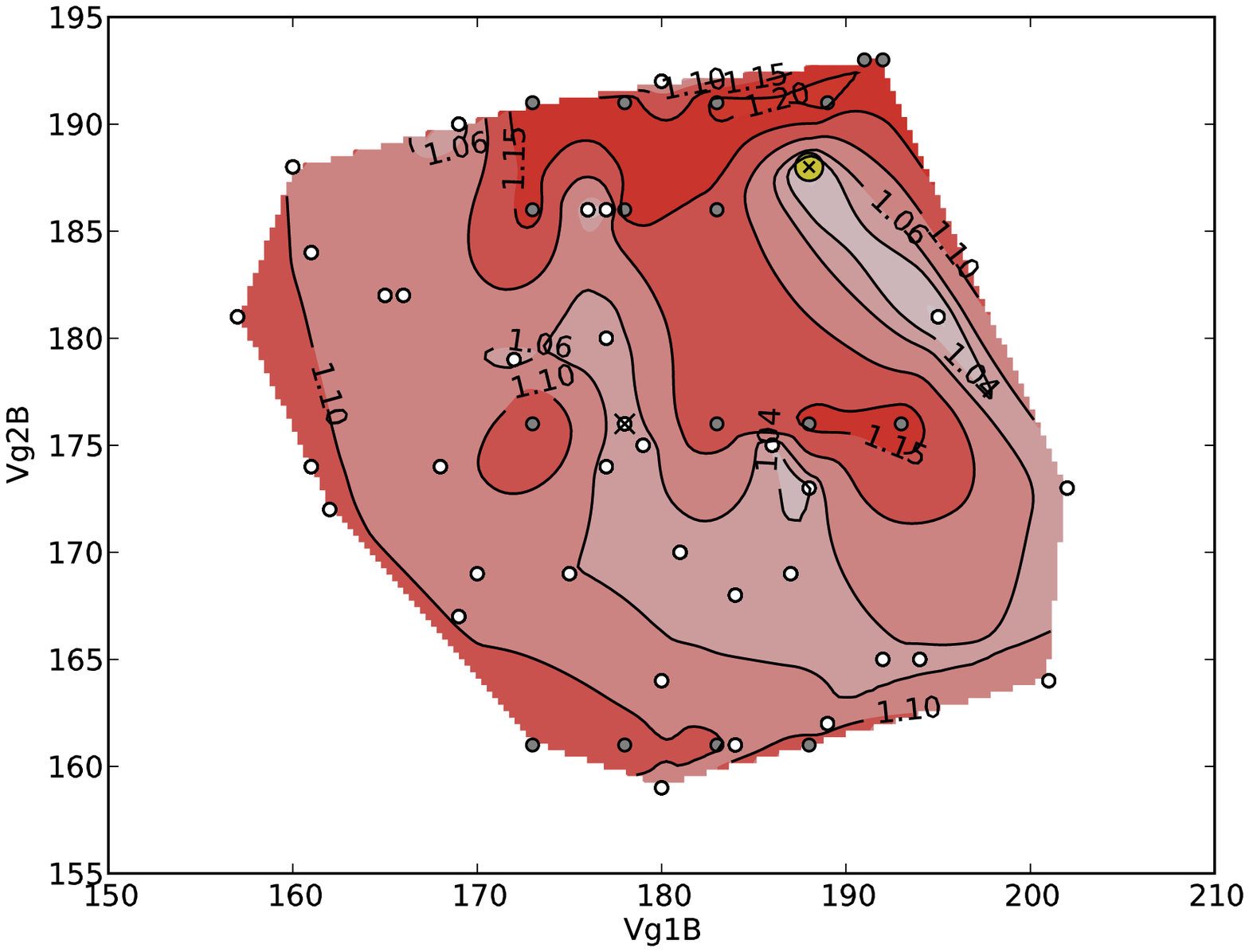}\\
            \includegraphics[width=6.9cm]{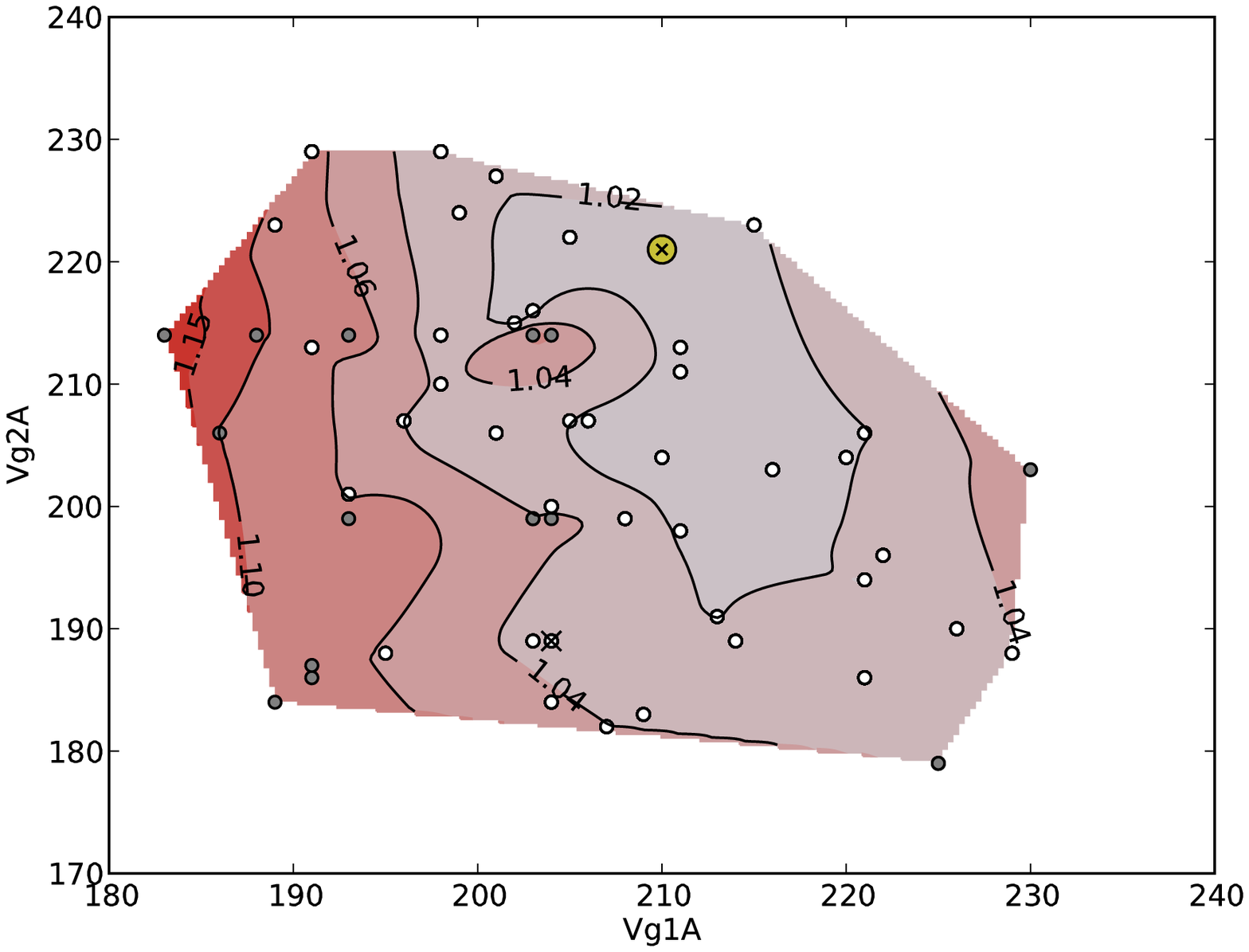}
            \includegraphics[width=6.9cm]{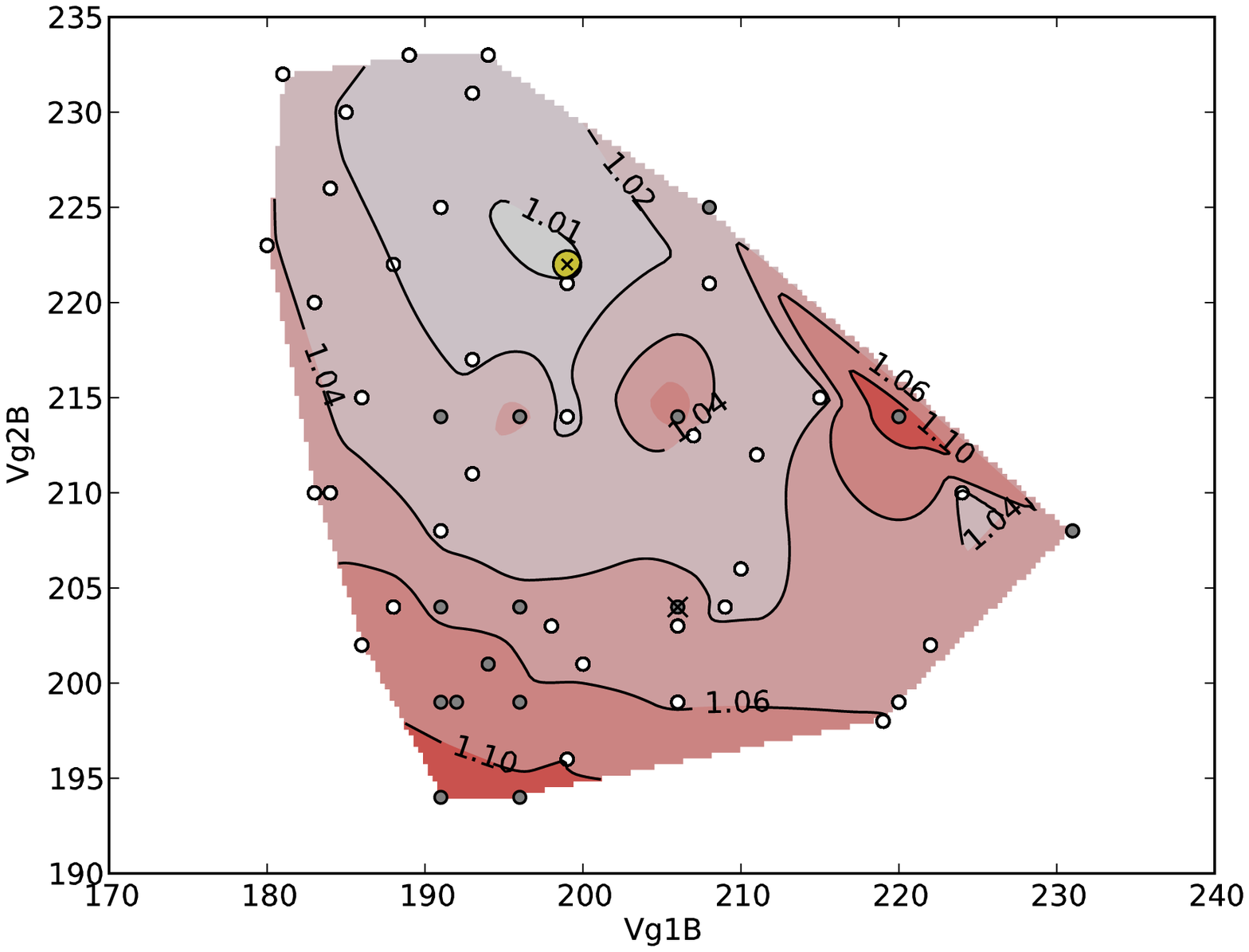}\\
            
            \textbf{LFI-23}\\
     \vskip 0.5 cm             
            \includegraphics[width=6.9cm]{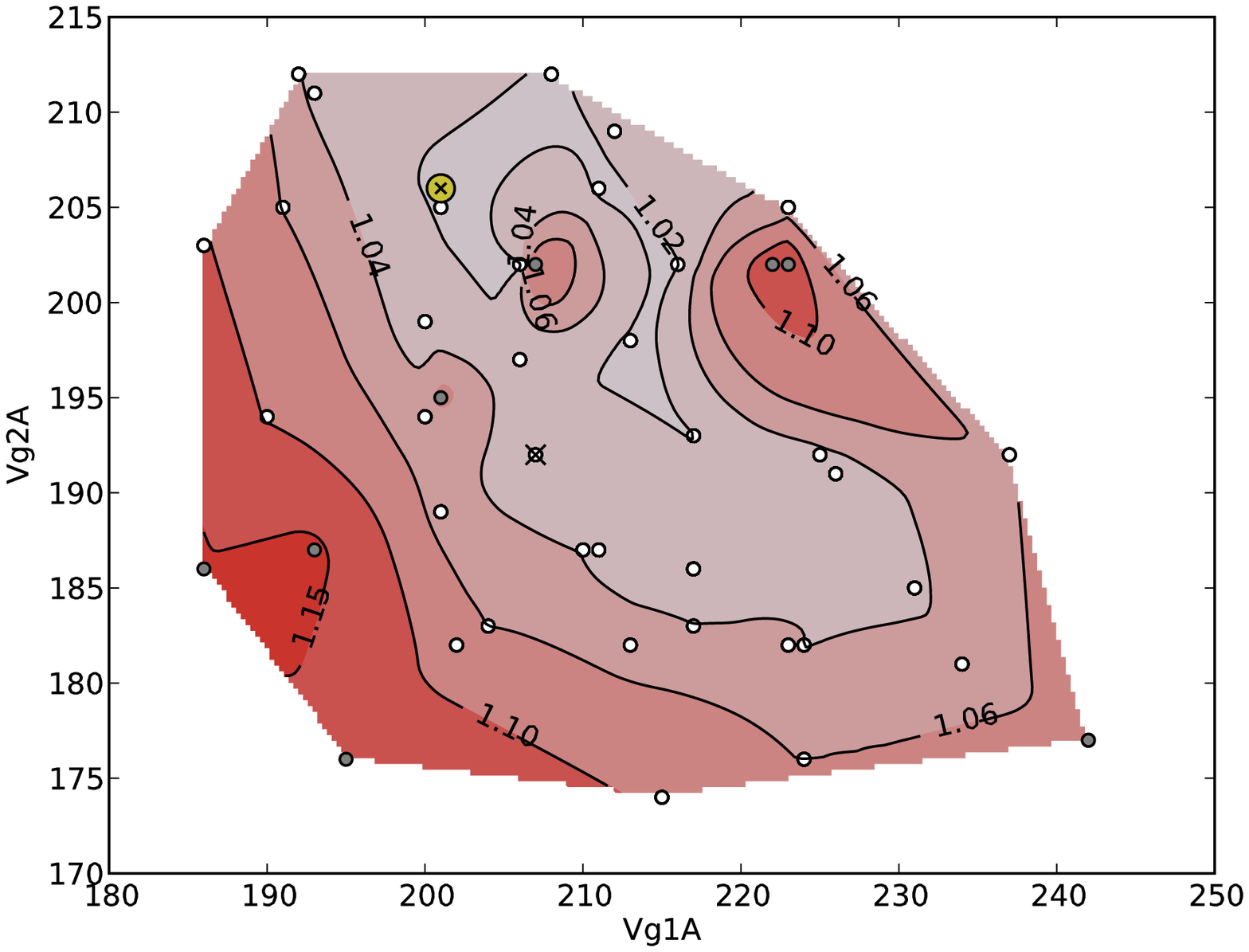}
            \includegraphics[width=6.9cm]{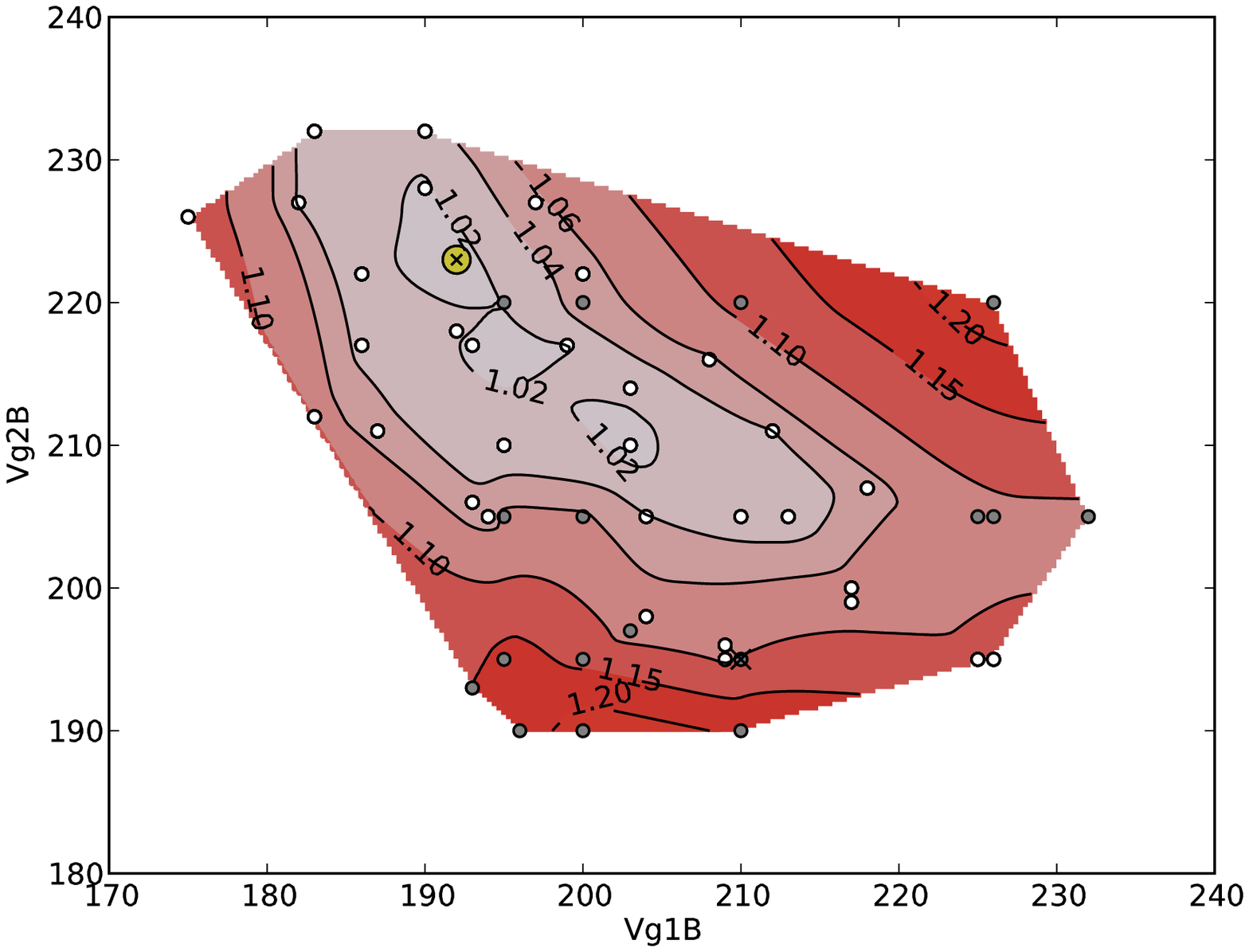}\\
            \includegraphics[width=6.9cm]{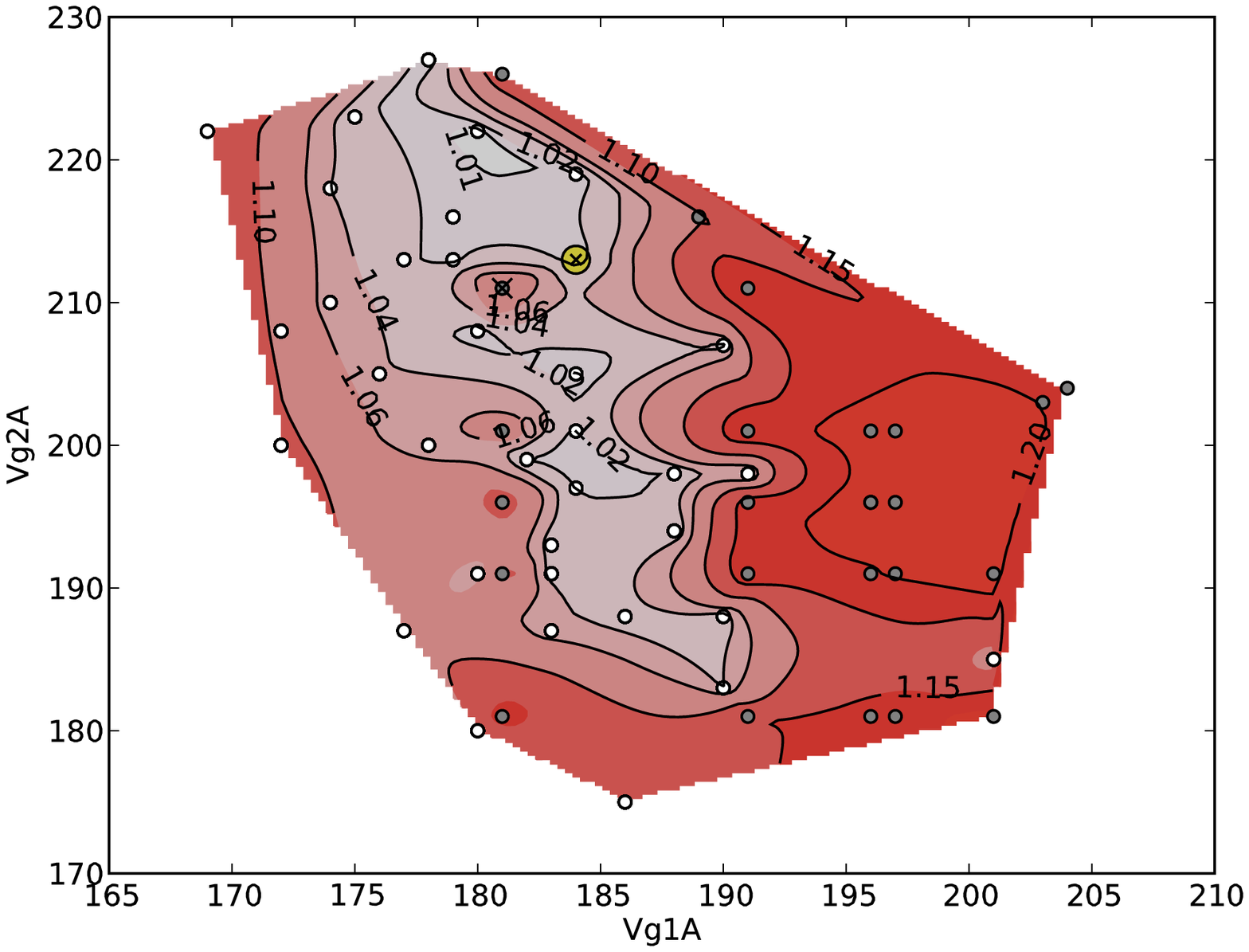}
            \includegraphics[width=6.9cm]{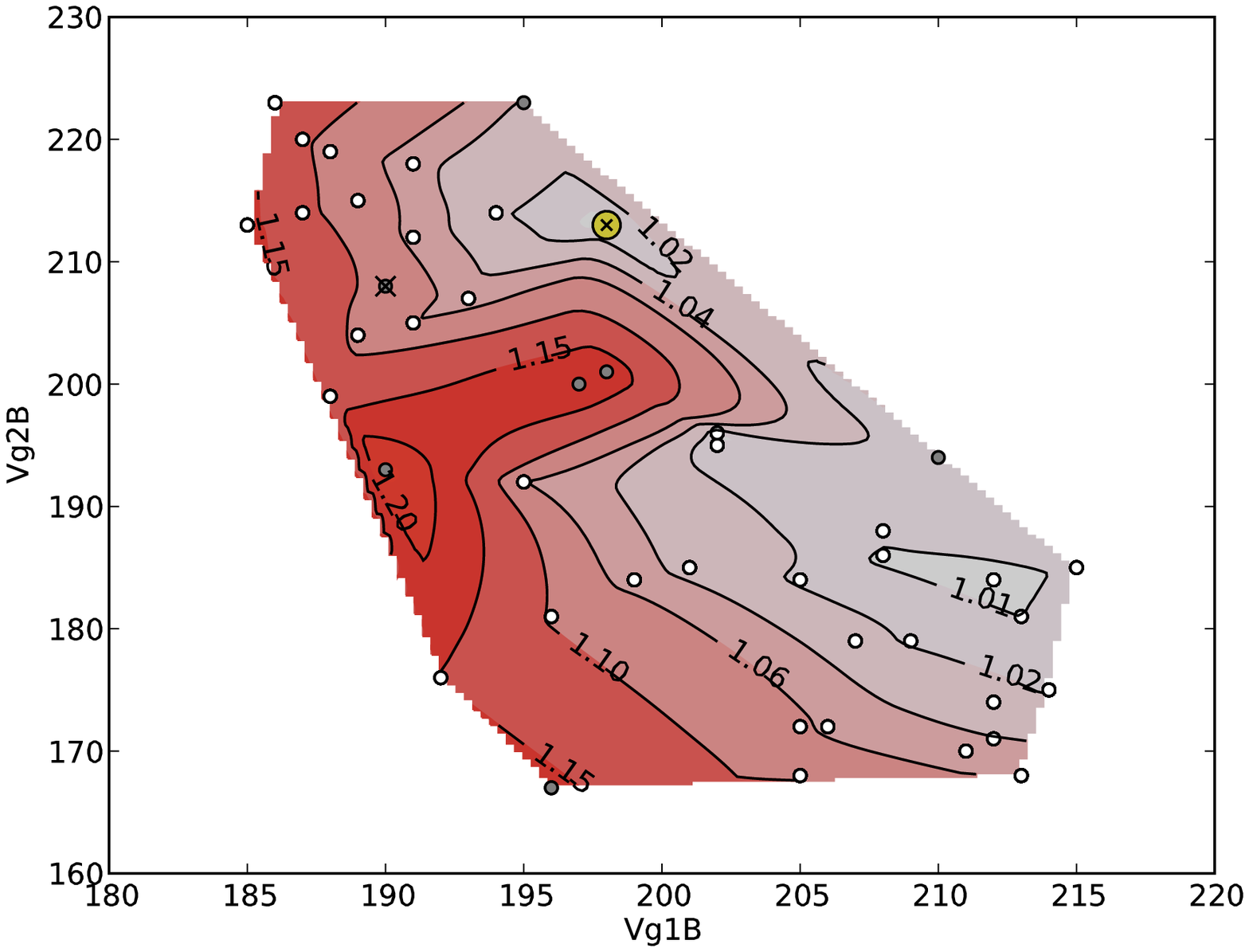}\\
                       \end{center}
            
        \label{fig_HYM_tun_22-23}
                    \end{figure}

   \begin{figure}[htb]
        \begin{center}
                		\textbf{LFI-24}\\
    \vskip 0.5 cm  
  		       \includegraphics[width=6.9cm]{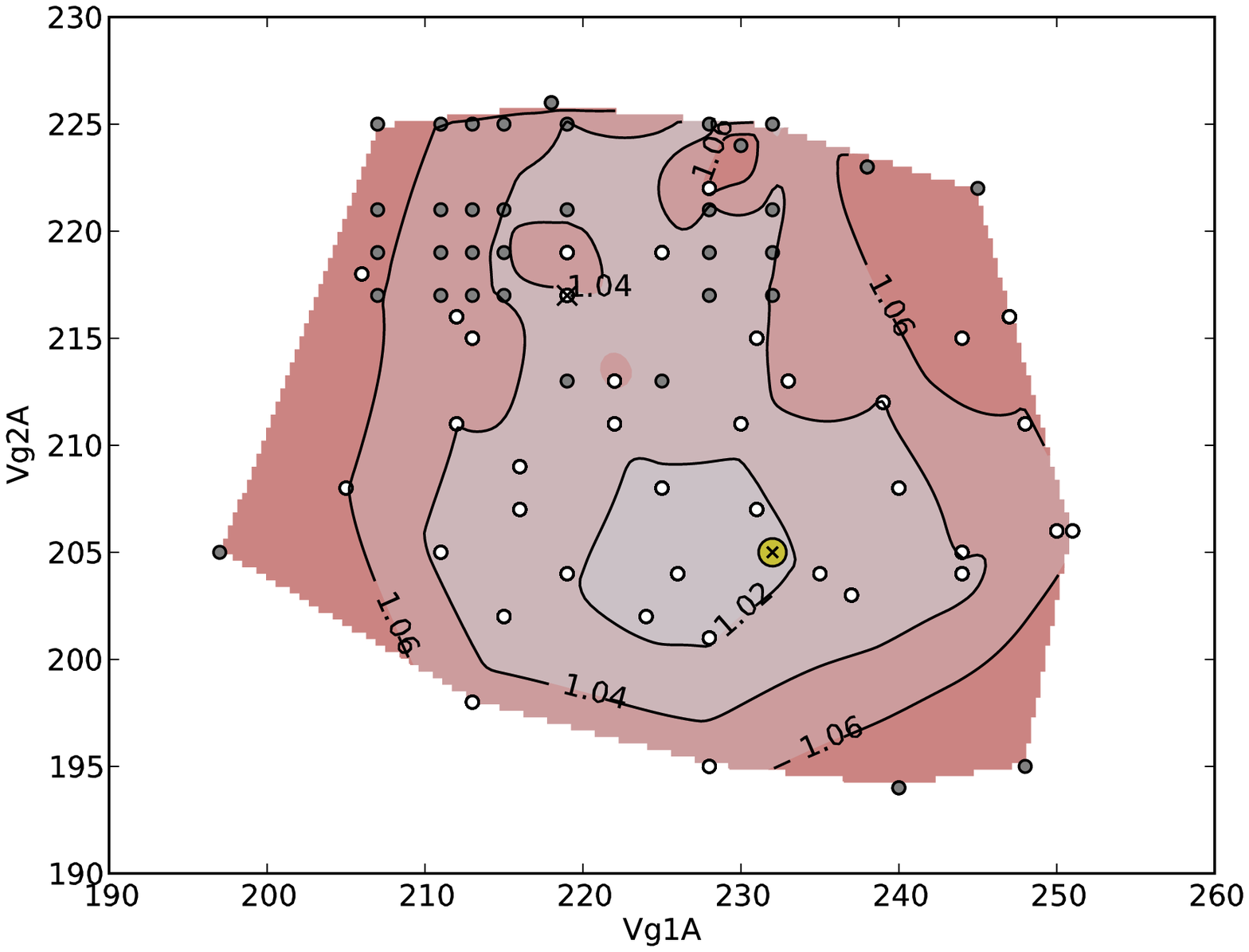} 
            \includegraphics[width=6.9cm]{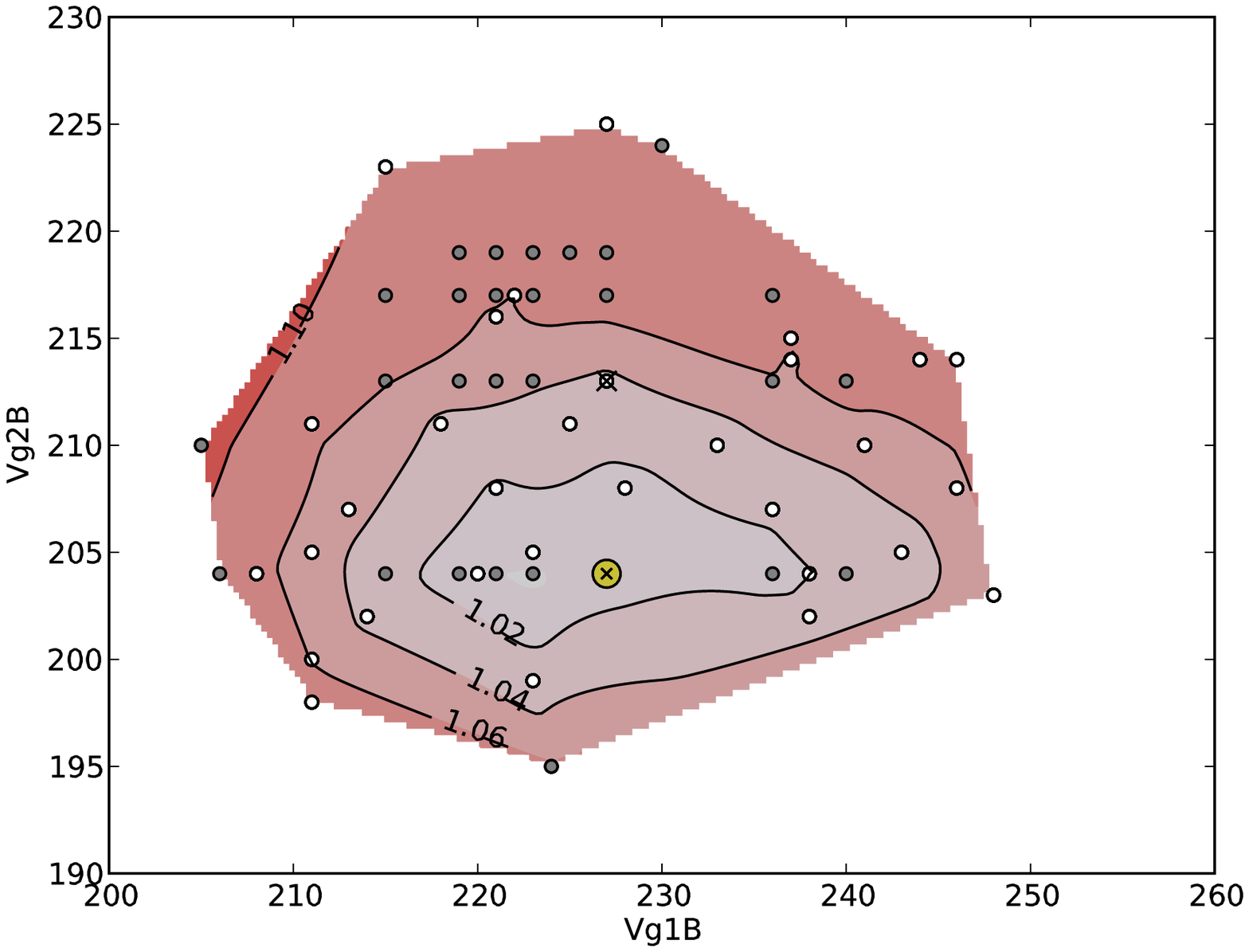}\\
            \includegraphics[width=6.9cm]{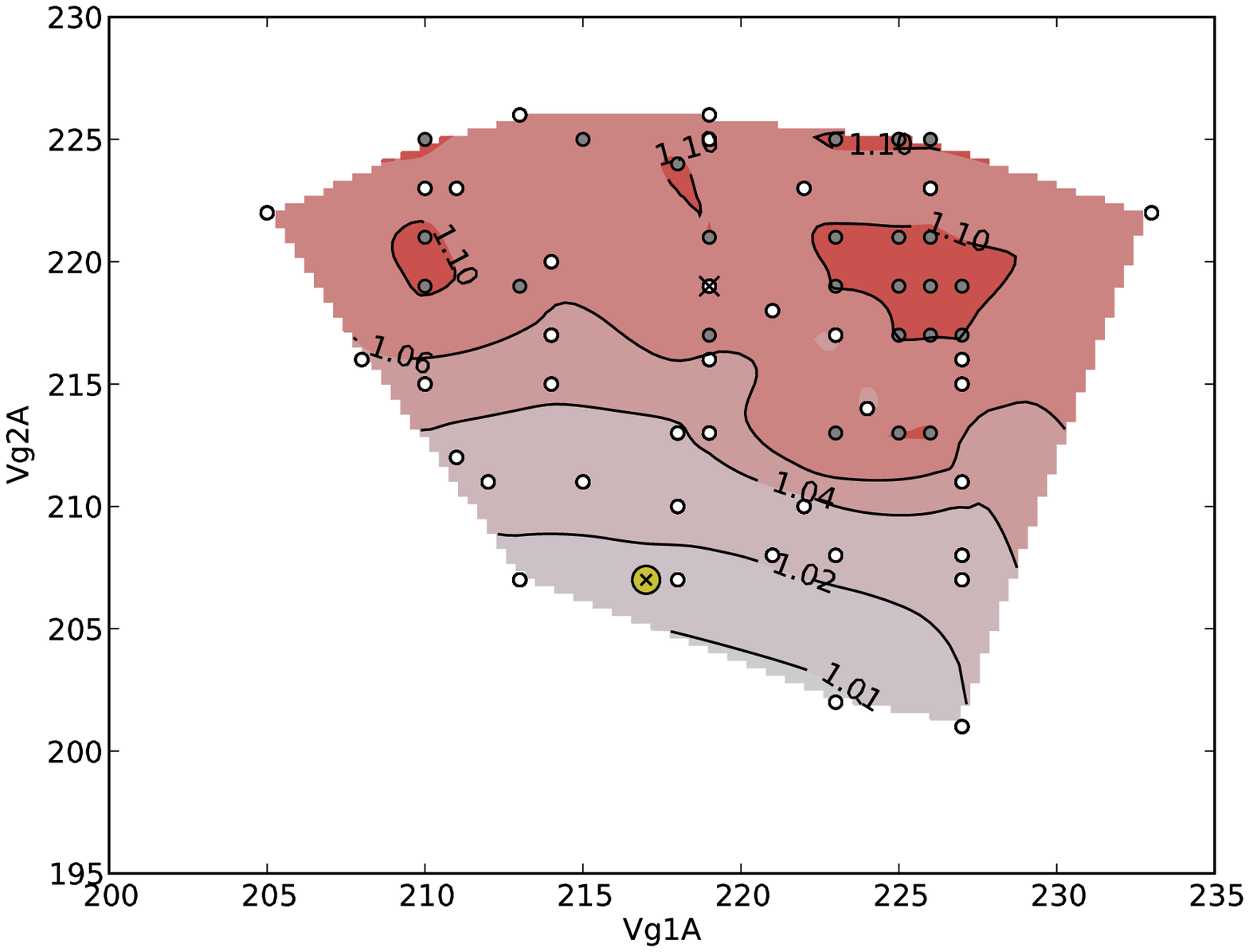}
            \includegraphics[width=6.9cm]{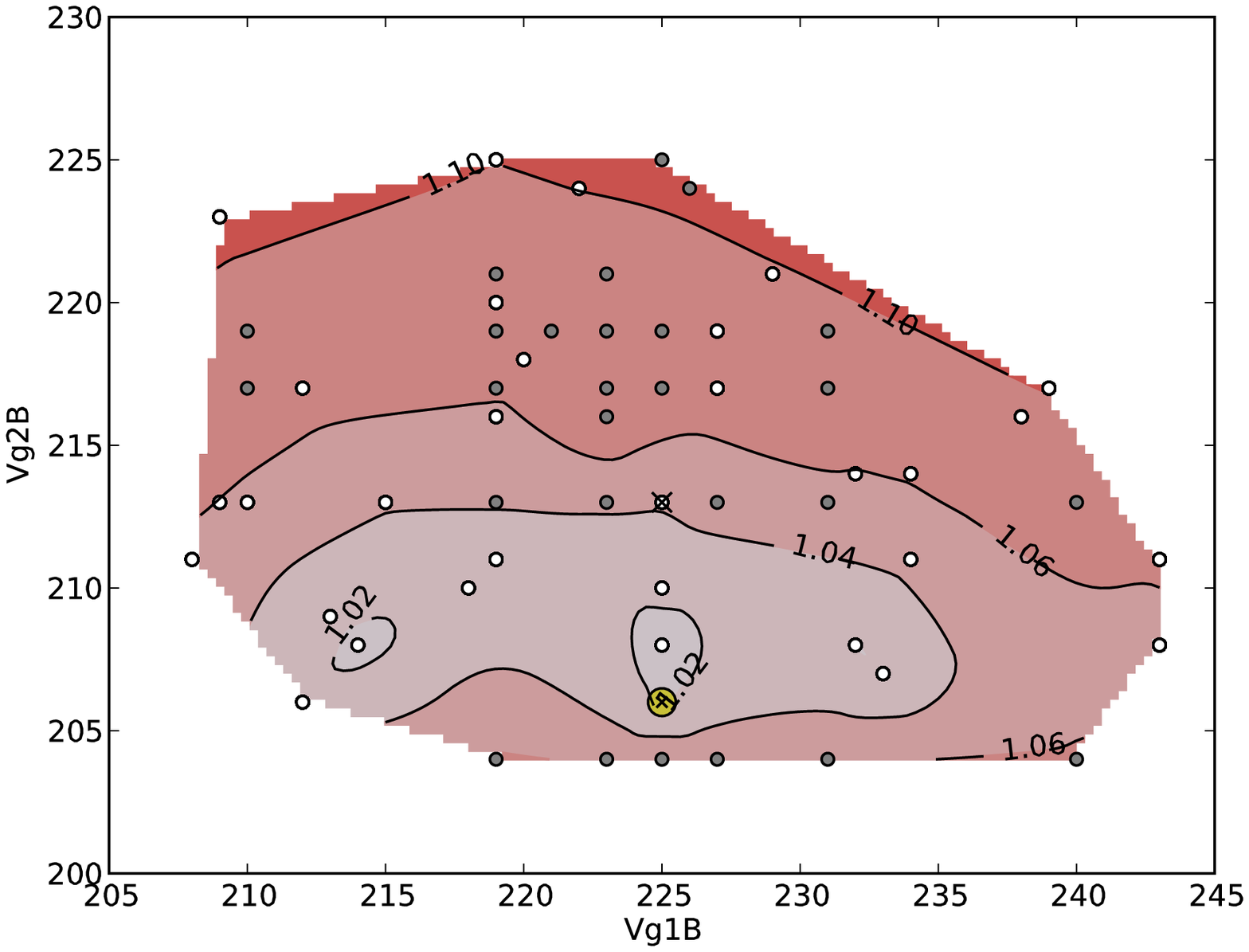}\\
            
            \textbf{LFI-25}\\
     \vskip 0.5 cm             
            \includegraphics[width=6.9cm]{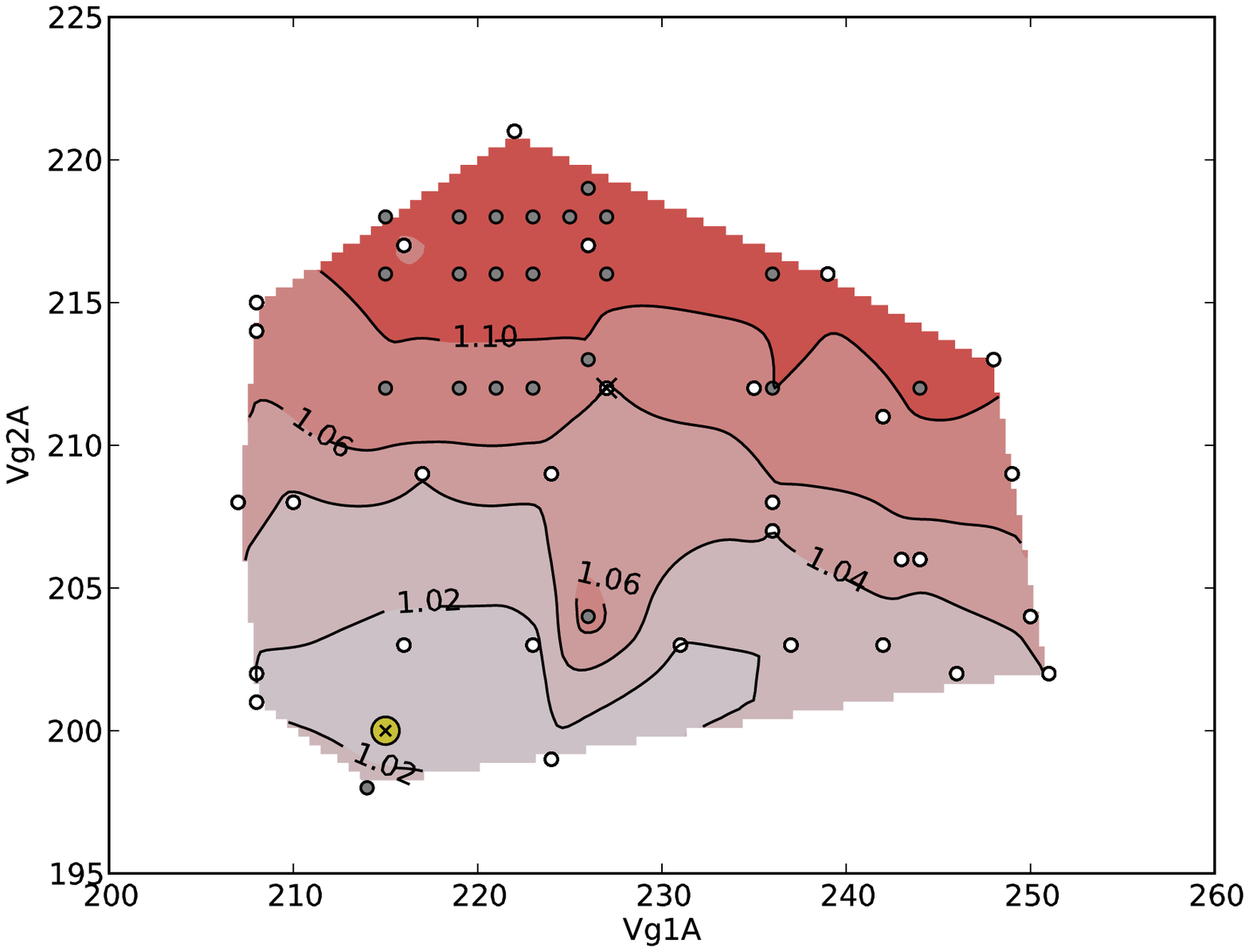}
            \includegraphics[width=6.9cm]{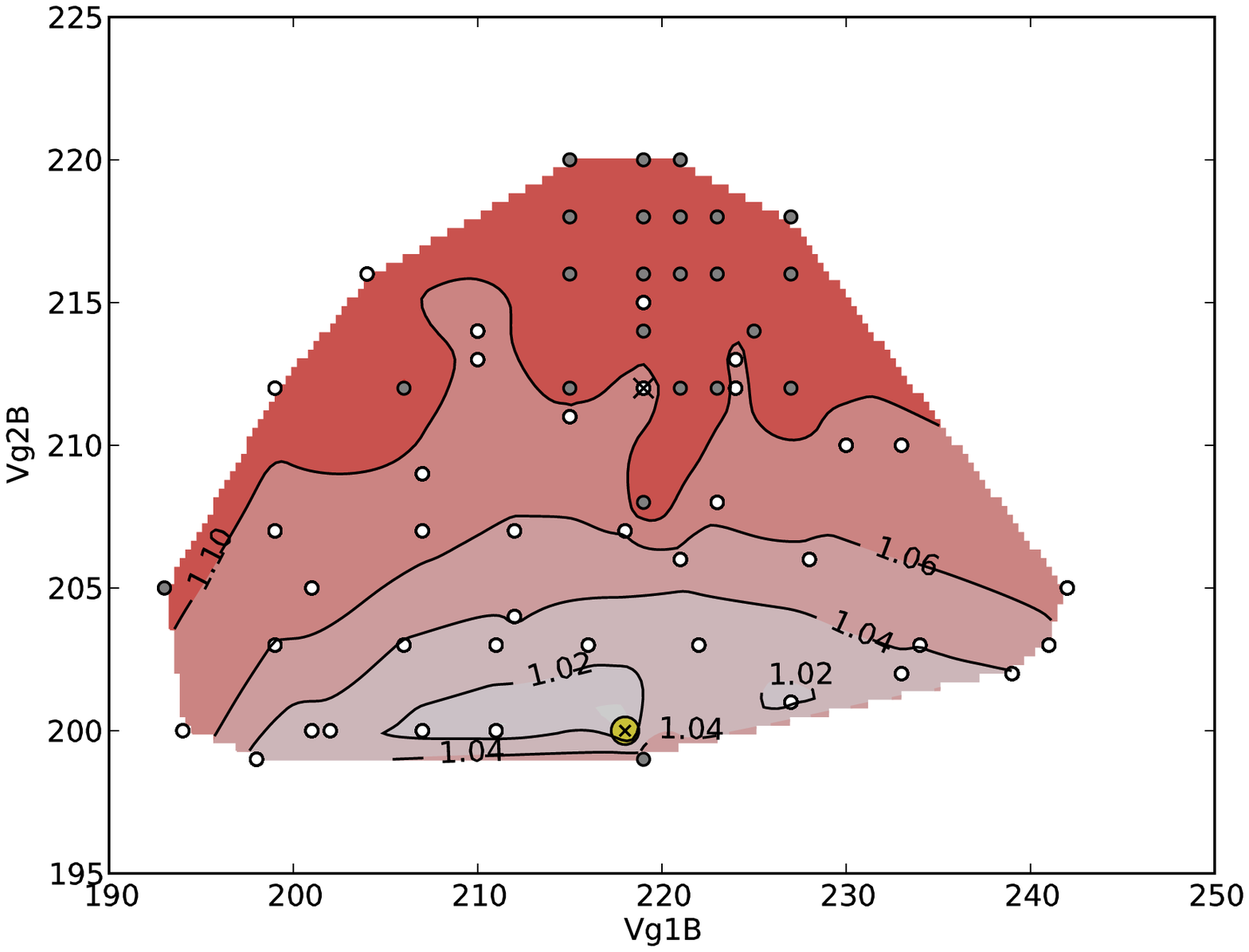}\\
            \includegraphics[width=6.9cm]{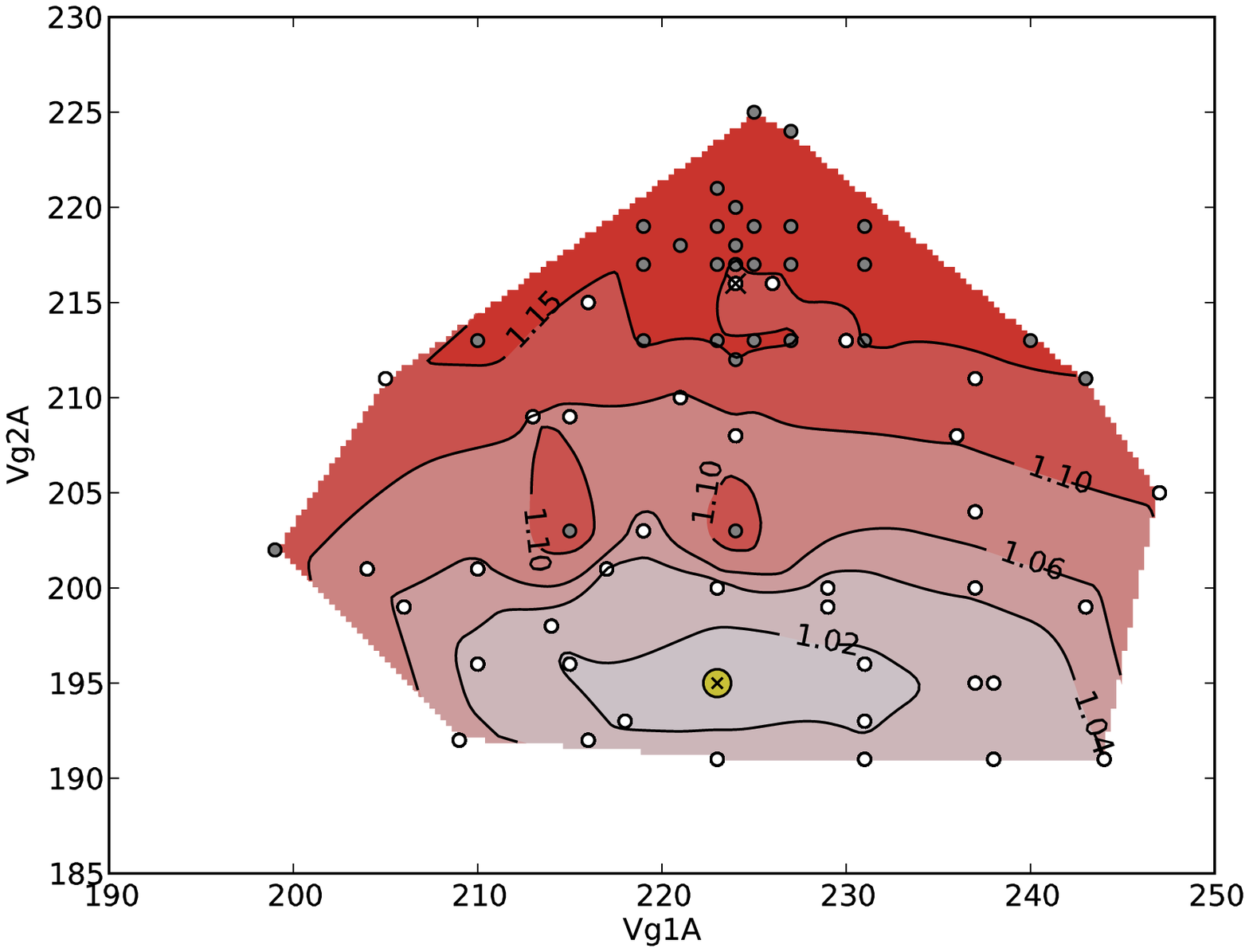}
            \includegraphics[width=6.9cm]{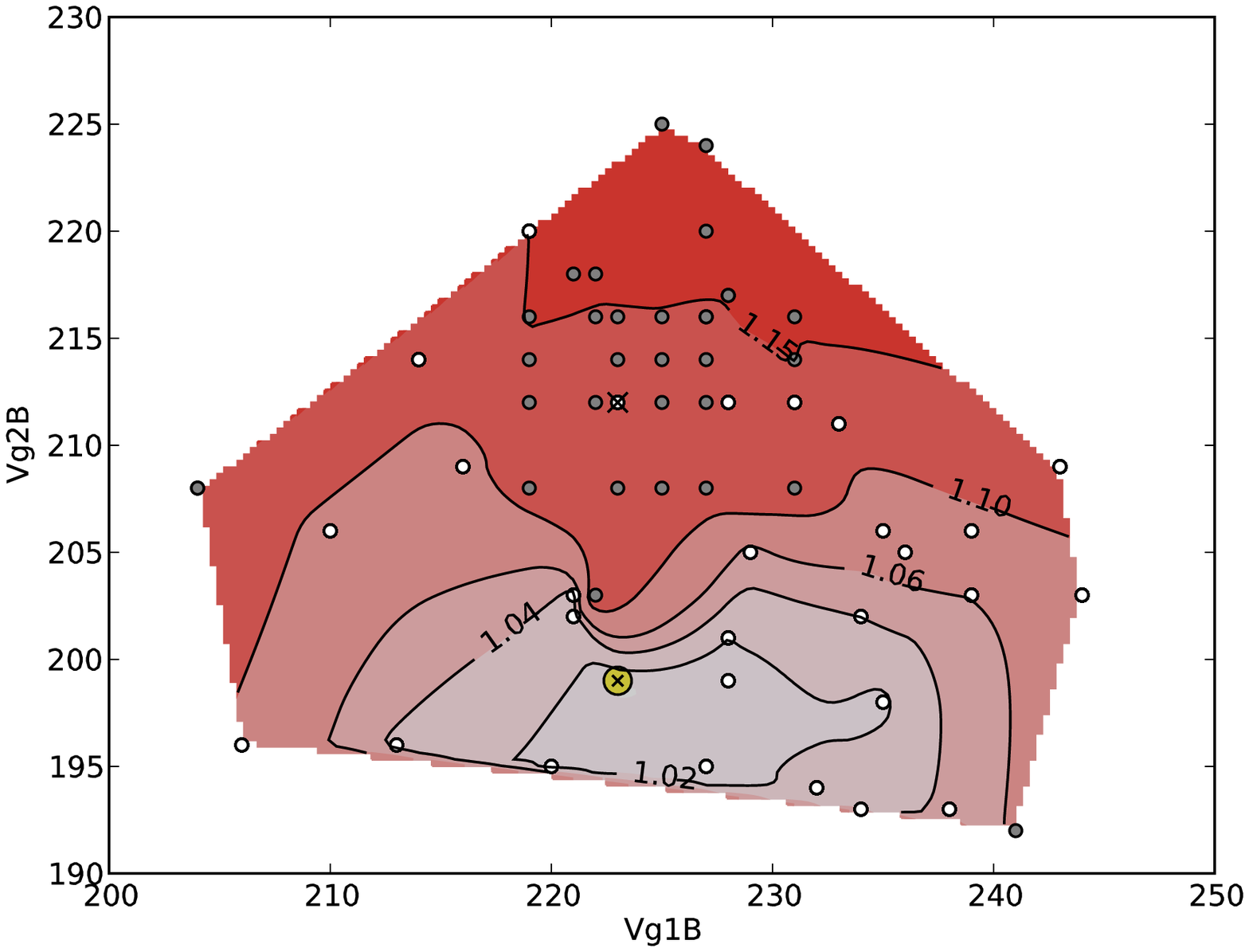}\\
                       \end{center}
            
        \label{fig_HYM_tun_24-25}
                    \end{figure}
 
   \begin{figure}[htb]

        \begin{center}
                		\textbf{LFI-26}\\
    \vskip 0.5 cm  
  		       \includegraphics[width=6.9cm]{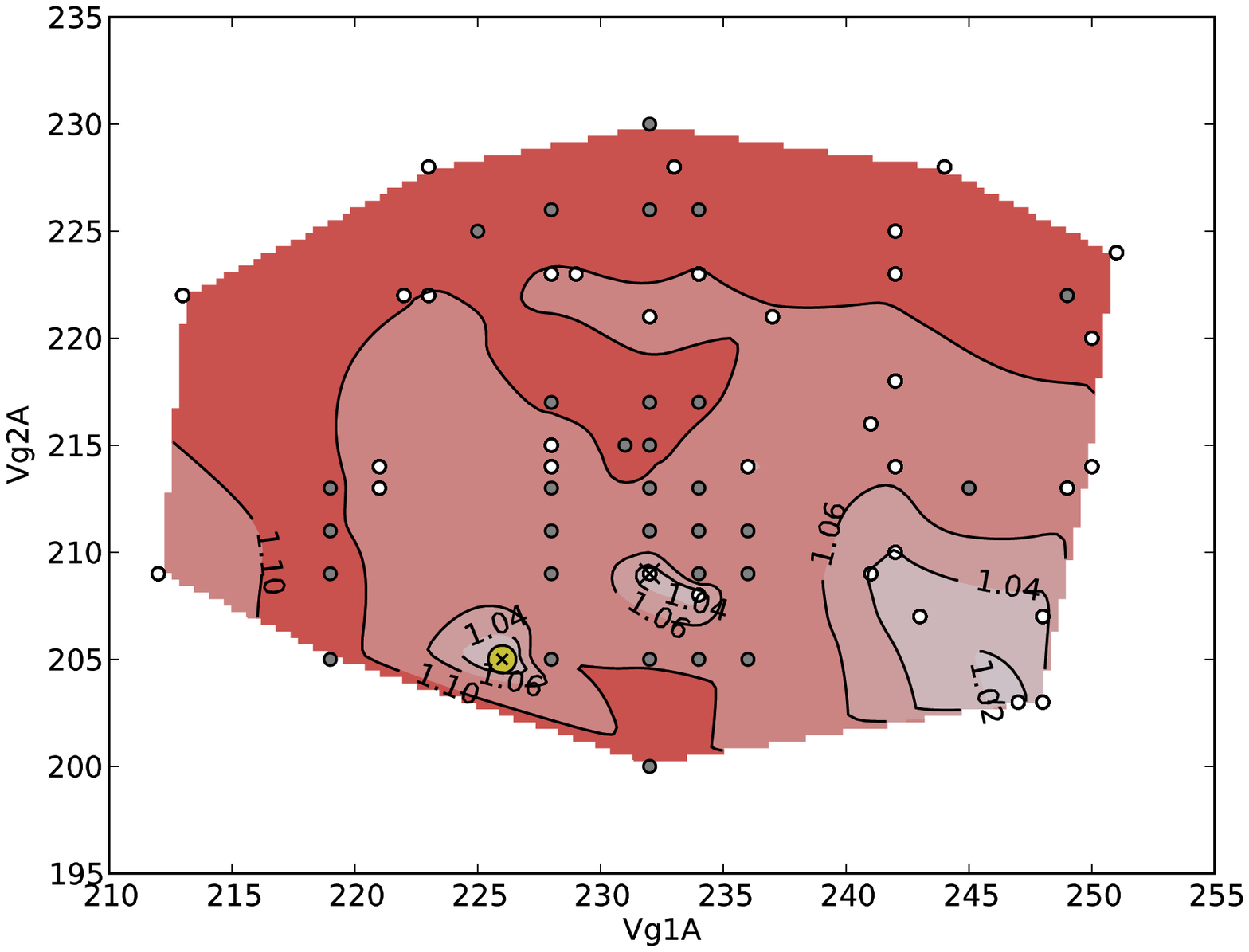} 
            \includegraphics[width=6.9cm]{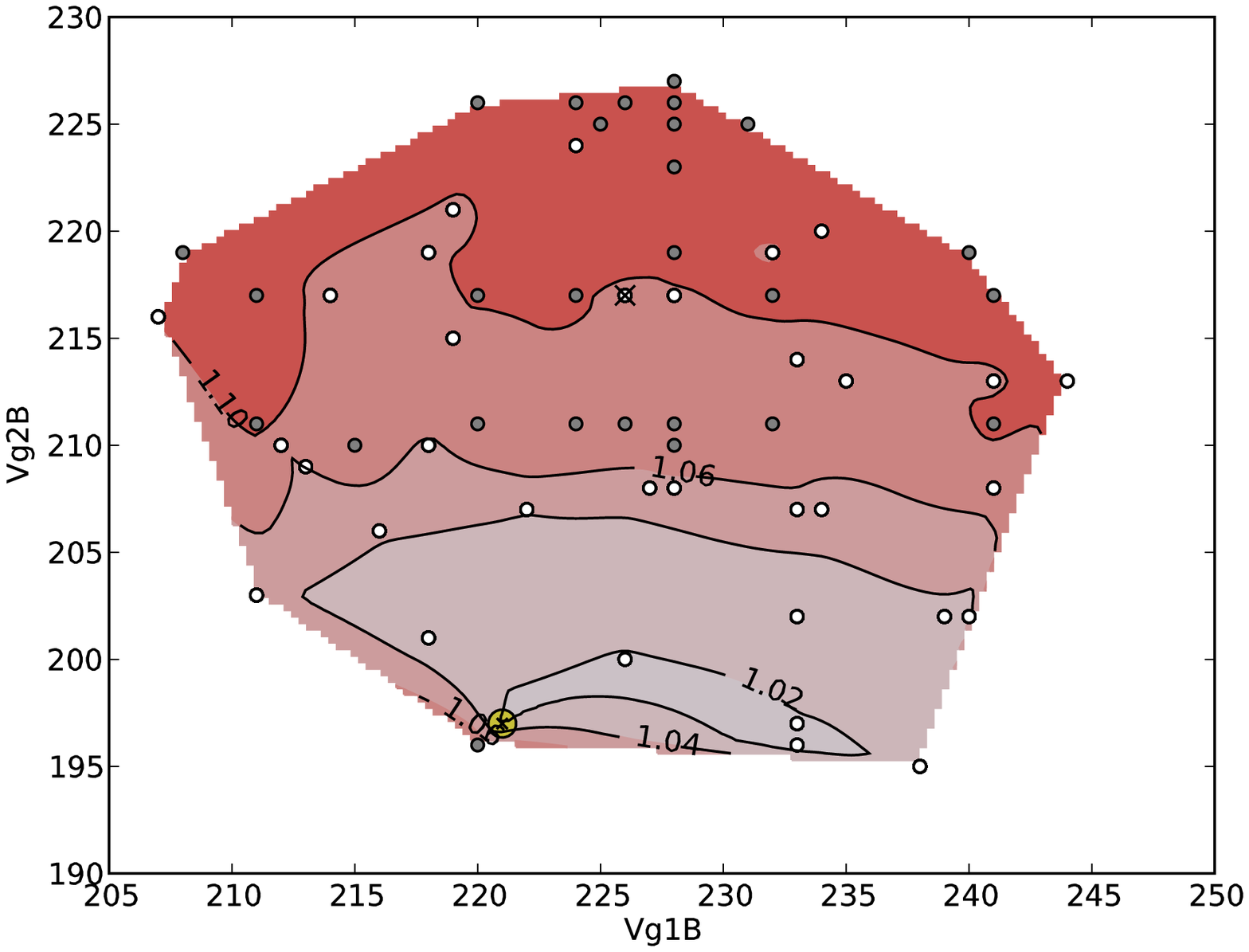}\\
            \includegraphics[width=6.9cm]{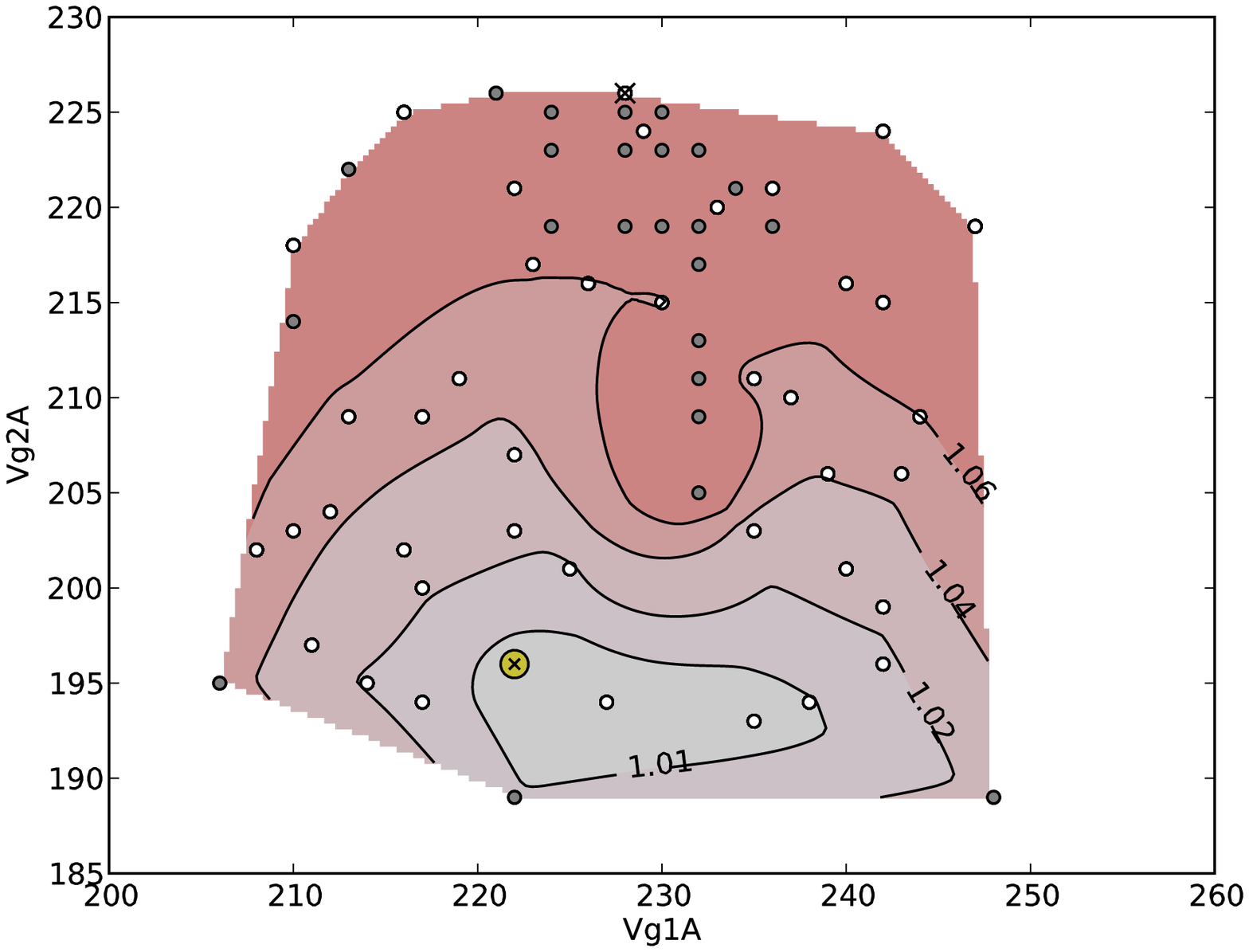}
            \includegraphics[width=6.9cm]{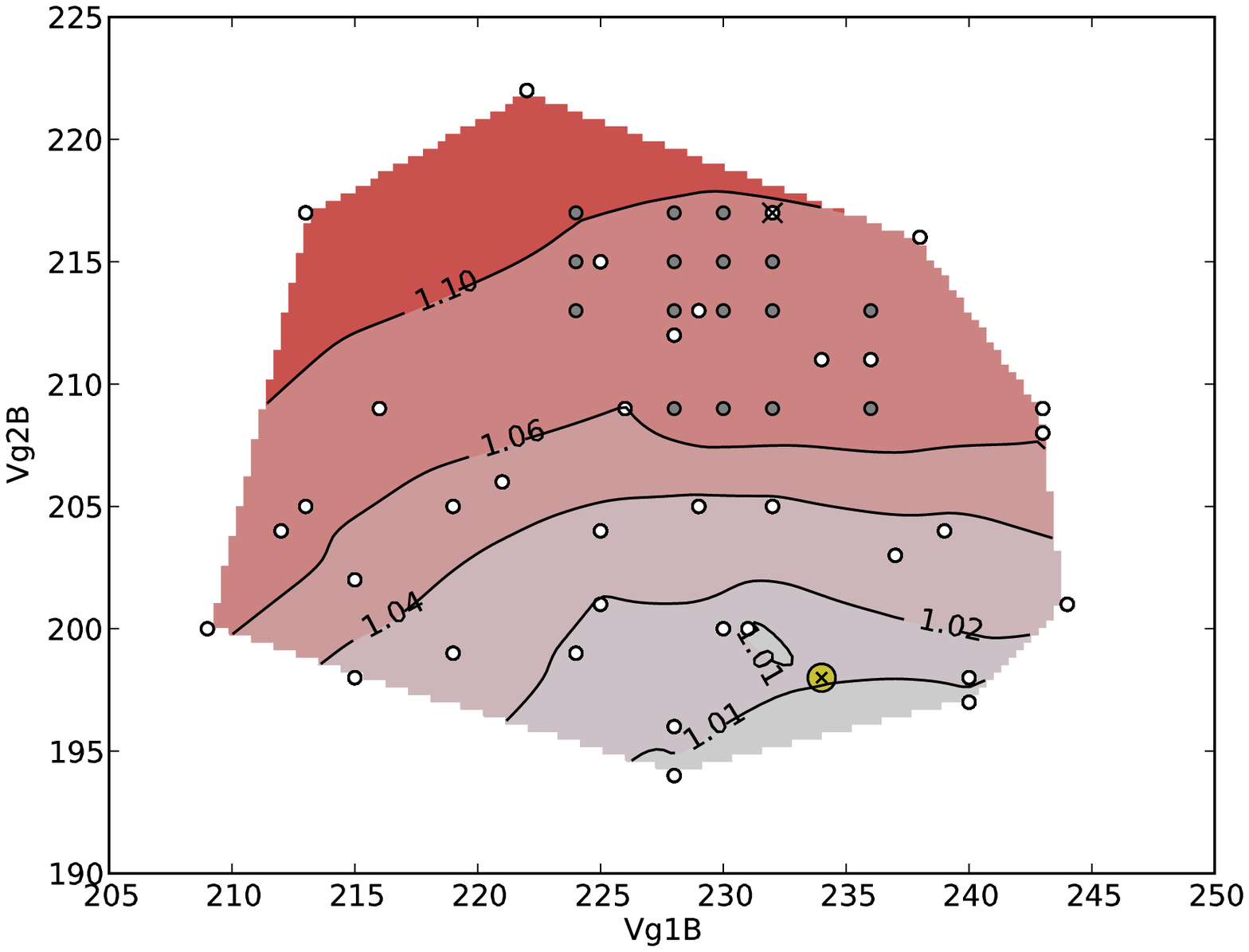}\\
            
            \textbf{LFI-27}\\
     \vskip 0.5 cm             
            \includegraphics[width=6.9cm]{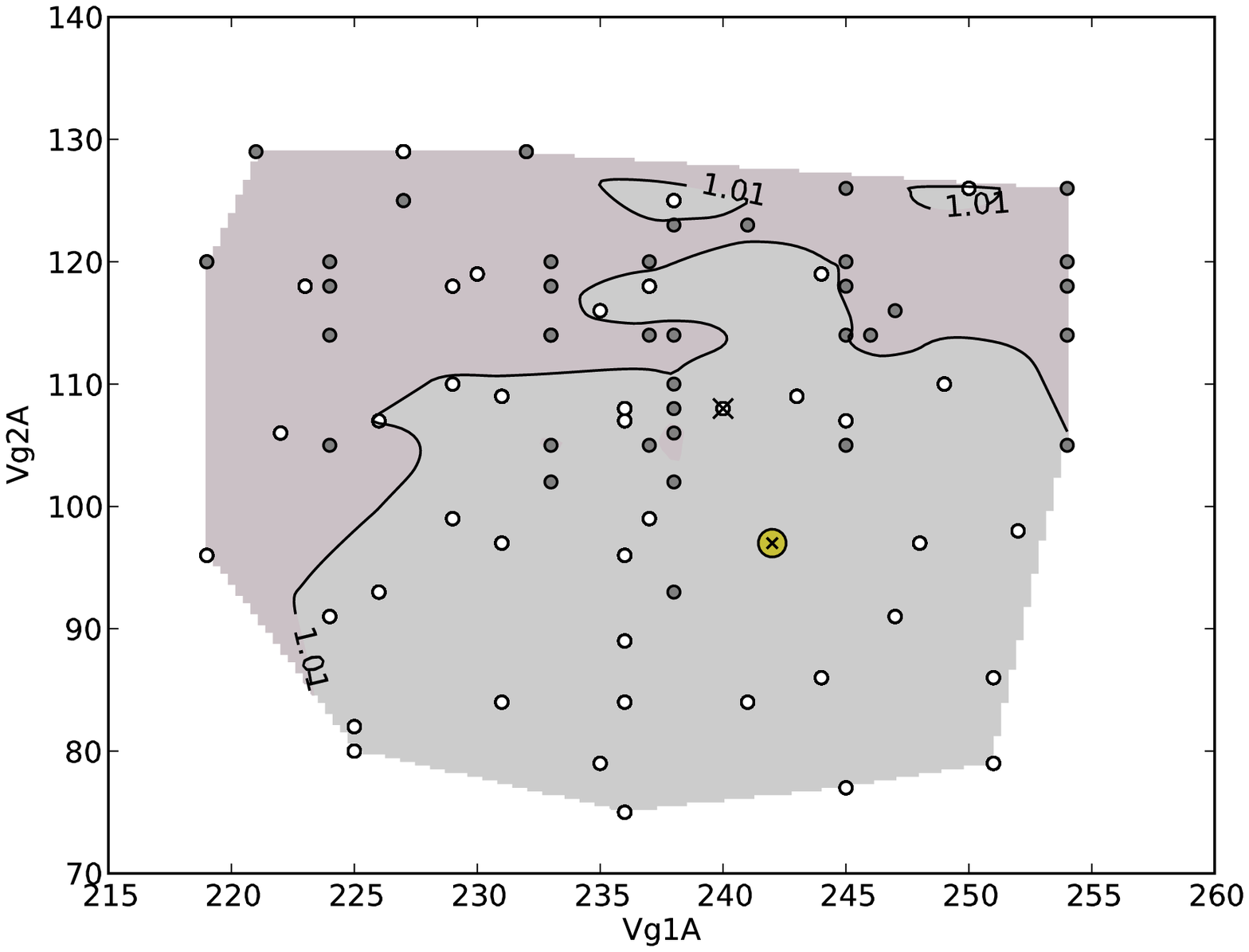}
            \includegraphics[width=6.9cm]{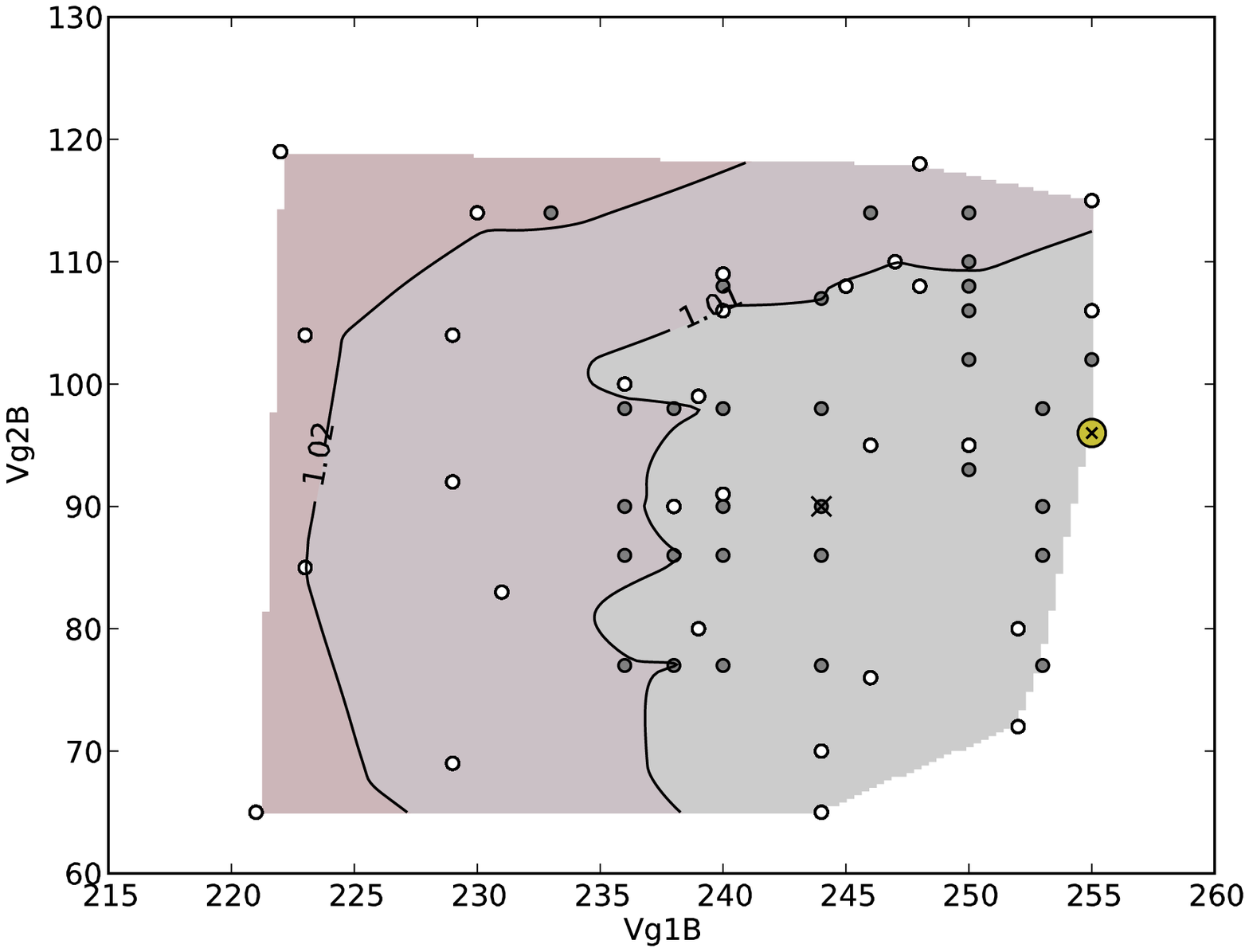}\\
            \includegraphics[width=6.9cm]{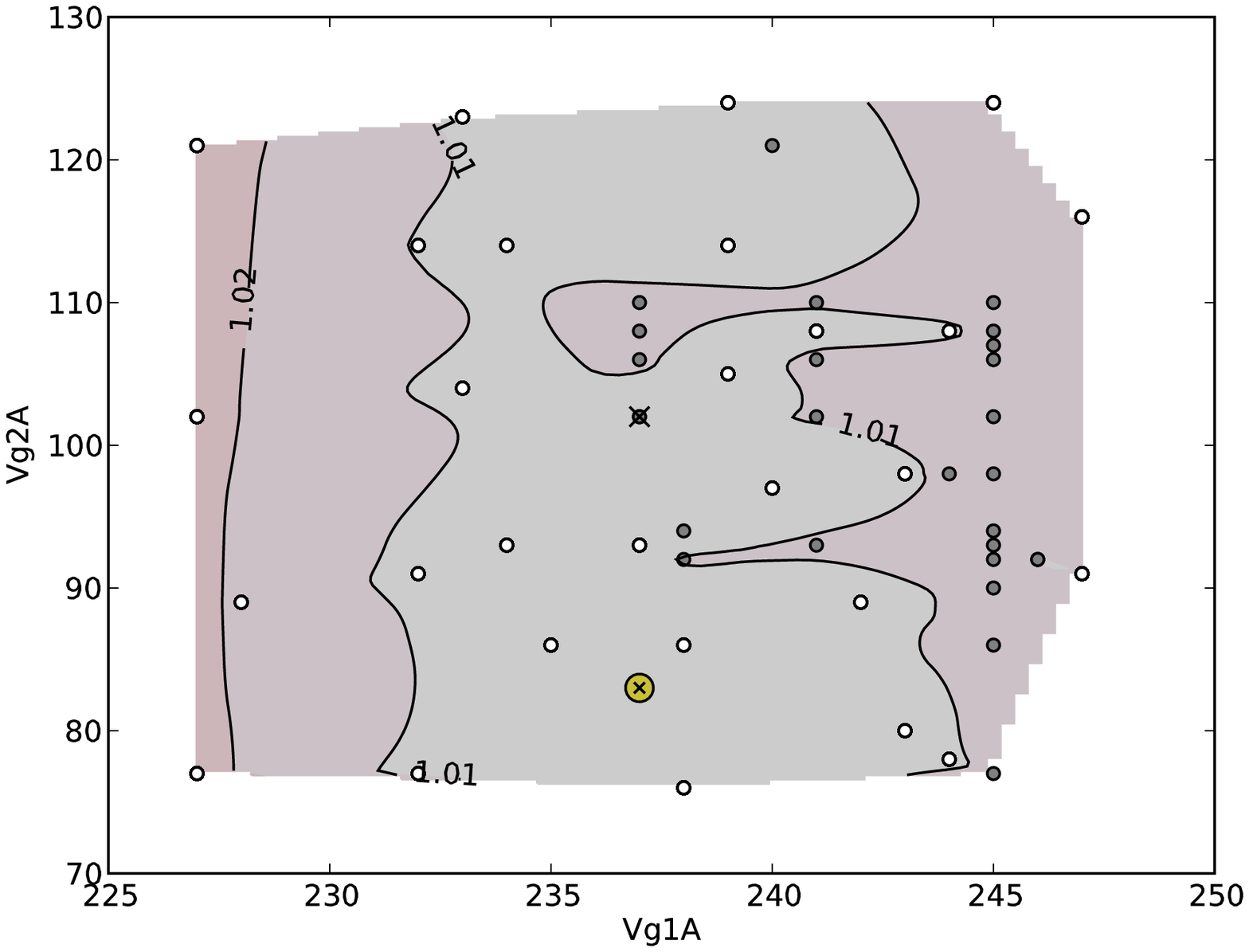}
            \includegraphics[width=6.9cm]{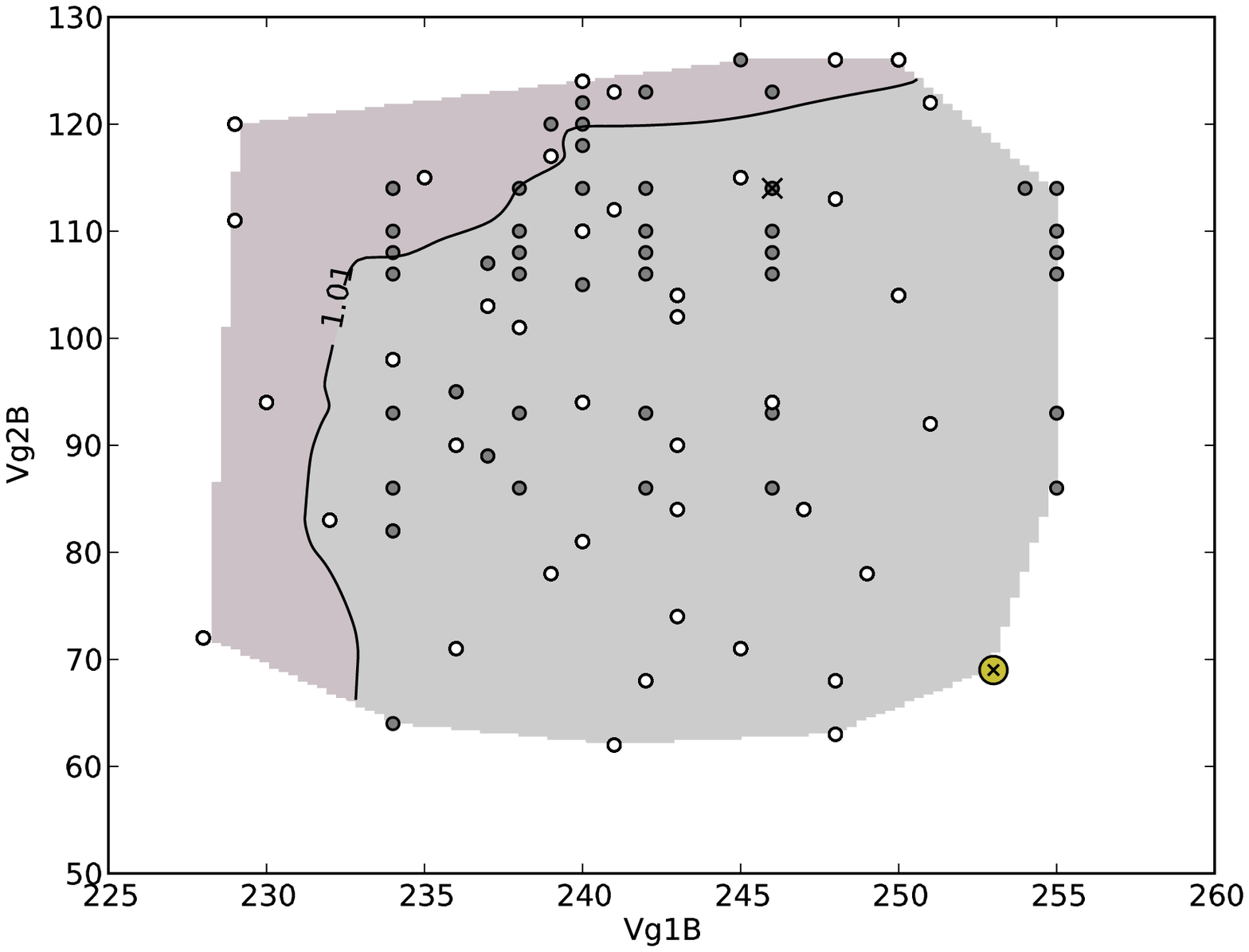}\\
                       \end{center}
            
        \label{fig_HYM_tun_26-27}
                    \end{figure}

\begin{figure}[htb]

        \begin{center}
                		\textbf{LFI-28}\\
    \vskip 0.5 cm  
  		       \includegraphics[width=6.9cm]{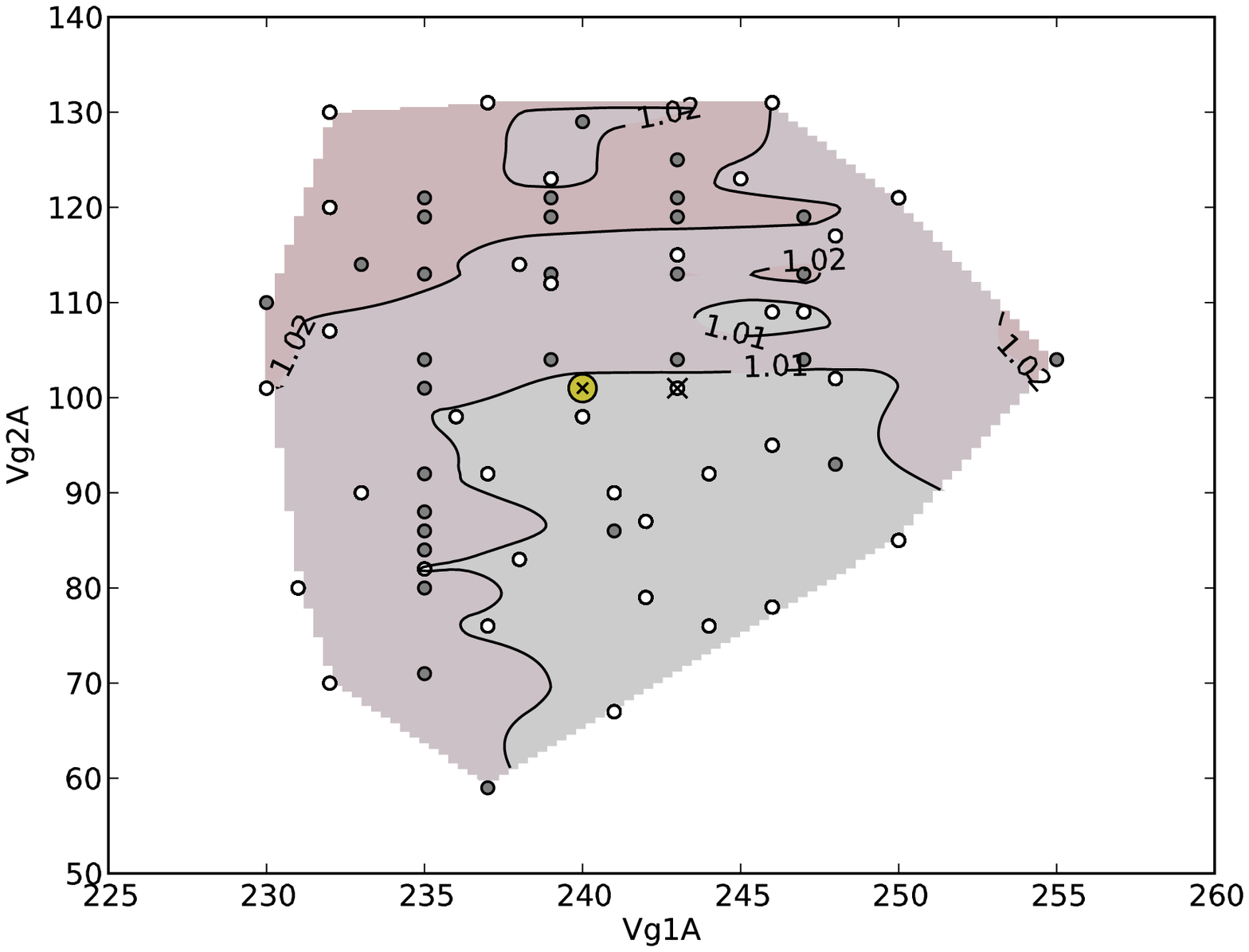} 
            \includegraphics[width=6.9cm]{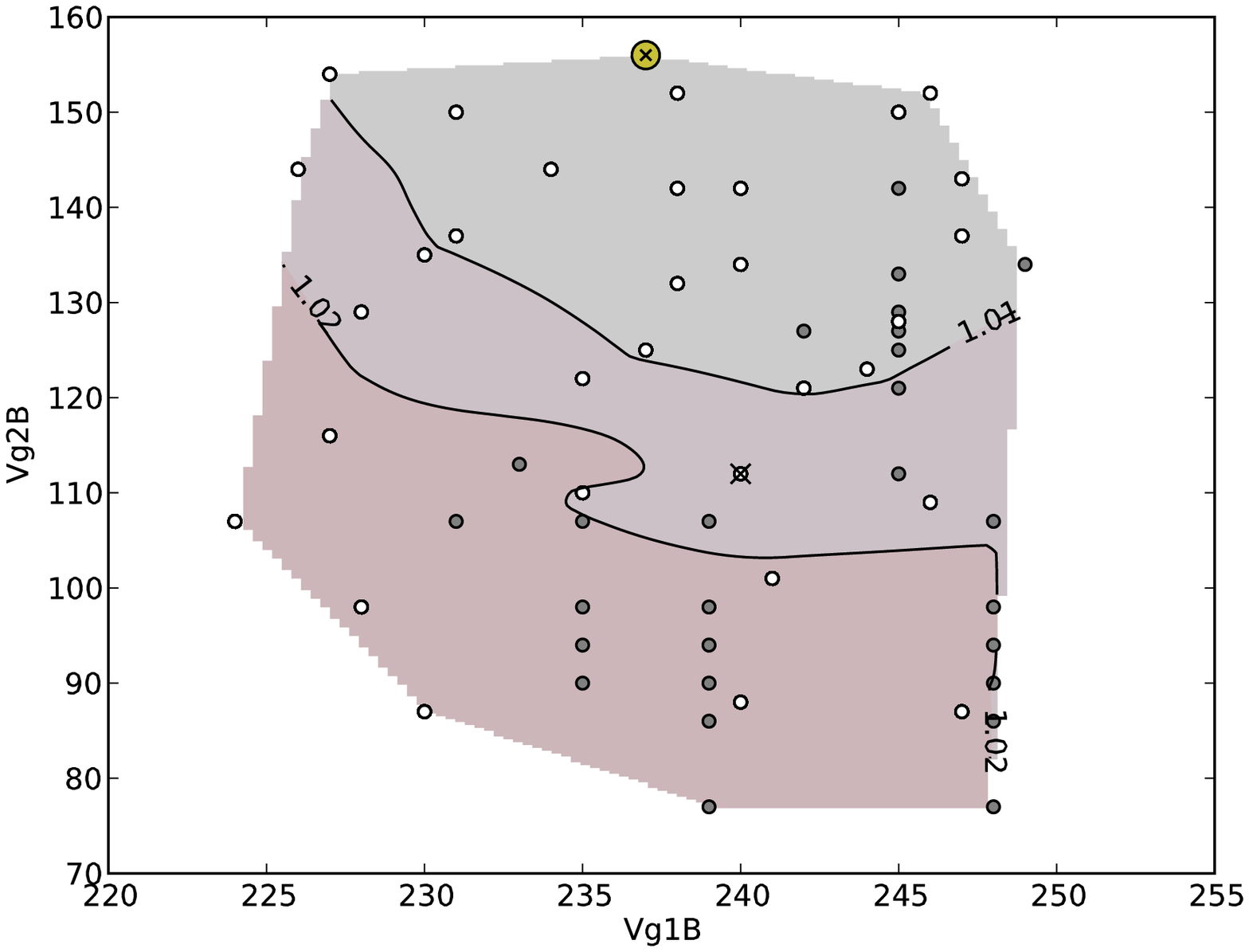}\\
            \includegraphics[width=6.9cm]{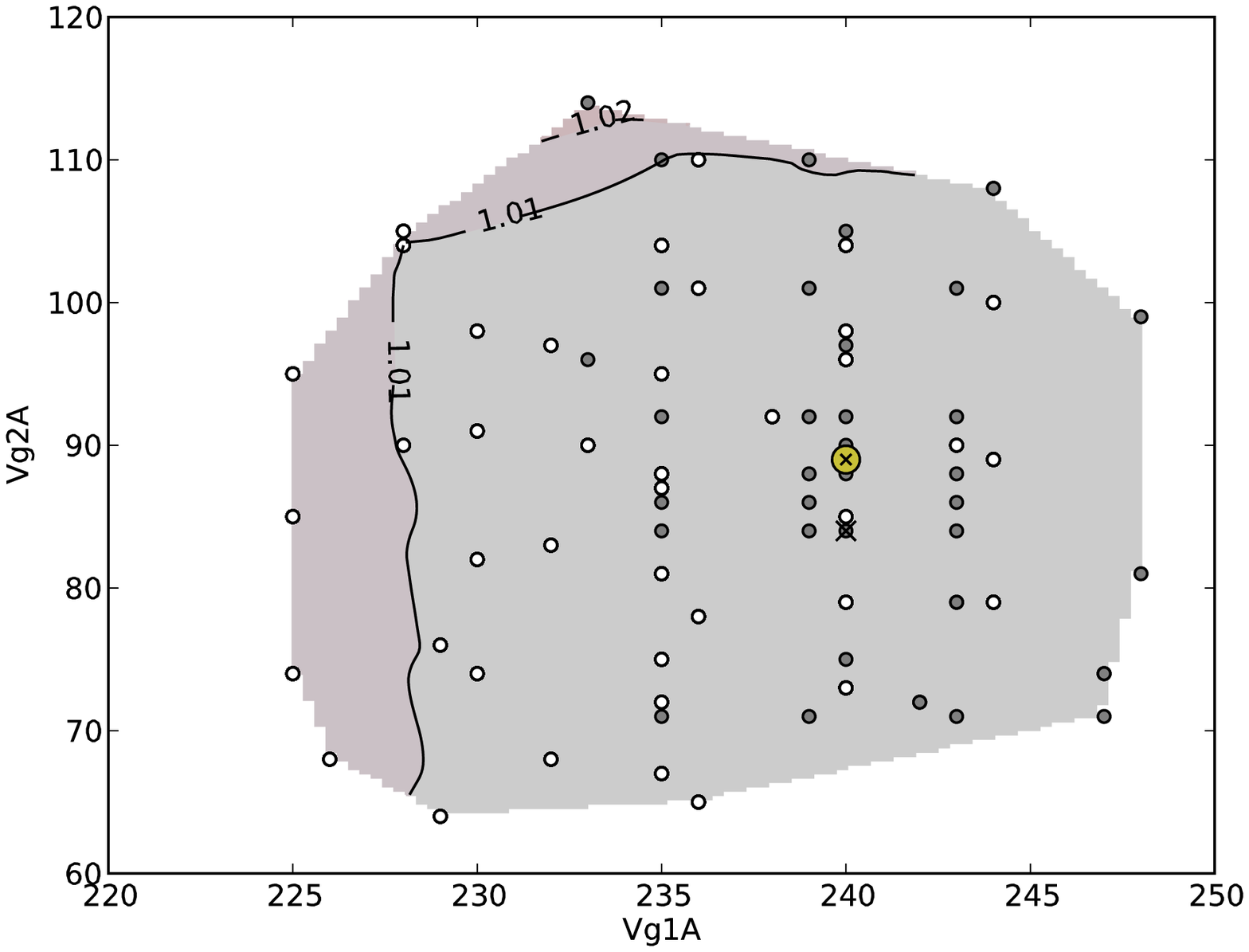}
            \includegraphics[width=6.9cm]{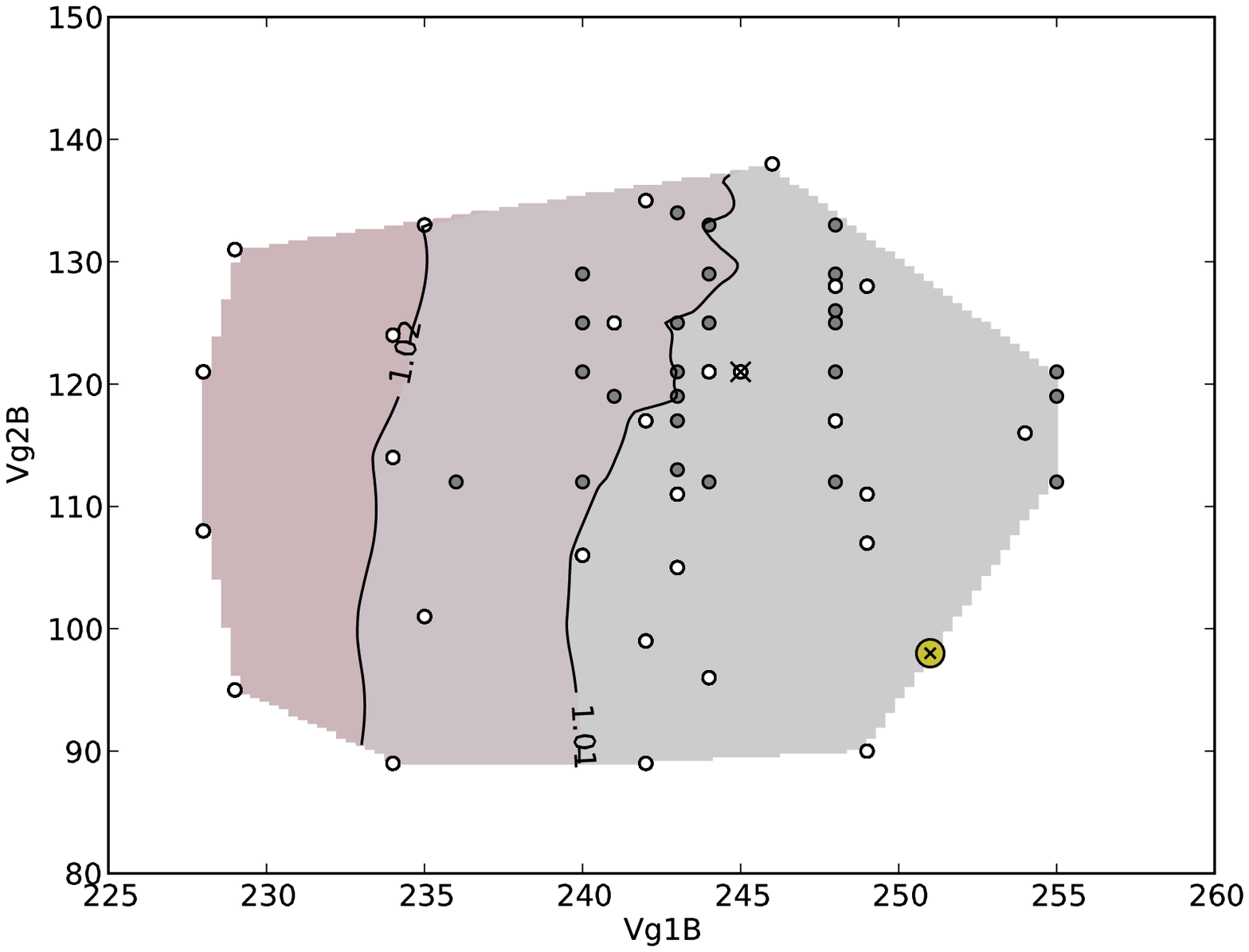}\\
                       \end{center}
                       
               \caption{Hypermatrix tuning noise temperature maps as a function of $V_{\rm g1}$, on x-axis, and $V_{\rm g2}$, on y-axis. Colour scale ranges from red (highest noise temperatures) to light-grey (lowest noise temperatures), normalized to the lowest noise temperature.The best condensed noise temperature is enhanced by a yellow circule; system level test  default by a black cross. Contours represents level of constant noise temperature. For each RCA, top panels: \texttt{M1} (left), \texttt{M2} (right); bottom panels: \texttt{S1} (left), \texttt{S2} (right). Note that the bias units are DEC units that are used to set the voltage in the DAE: see Table~4 for a rough conversion into physical units.} 
               
        \label{fig_HYM_tun}
    \end{figure}
    \clearpage 

%% file: a07_tuning_nonlin_plots.tex
\section{Hypermatrix tuning: non linear plots}
\label{app_HYM_nonlin_plots}

\begin{figure}[htb]
        \begin{center}              
            \textbf{LFI-26M}\\
            \includegraphics[width=7.0cm]{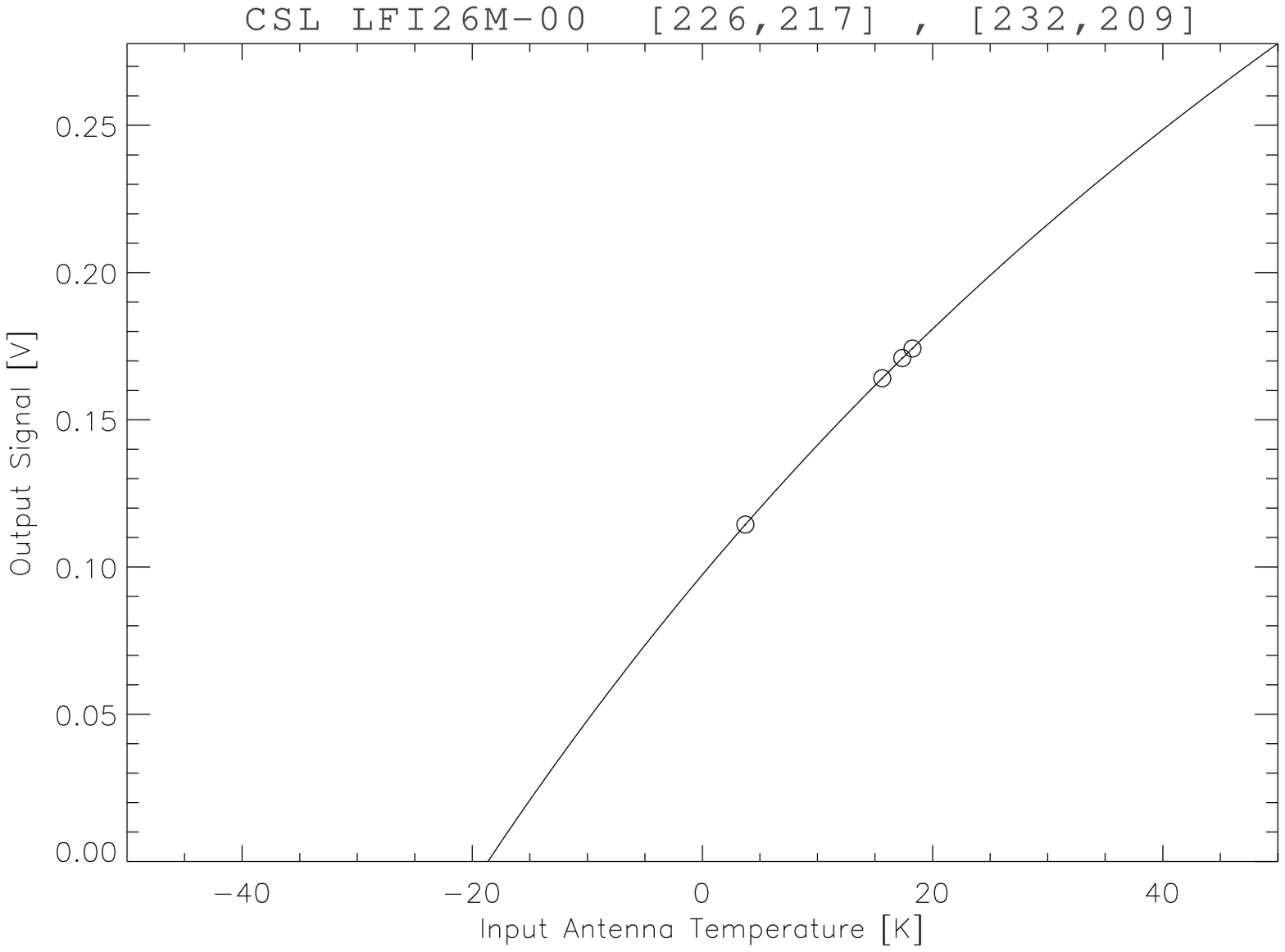}
            \includegraphics[width=7.0cm]{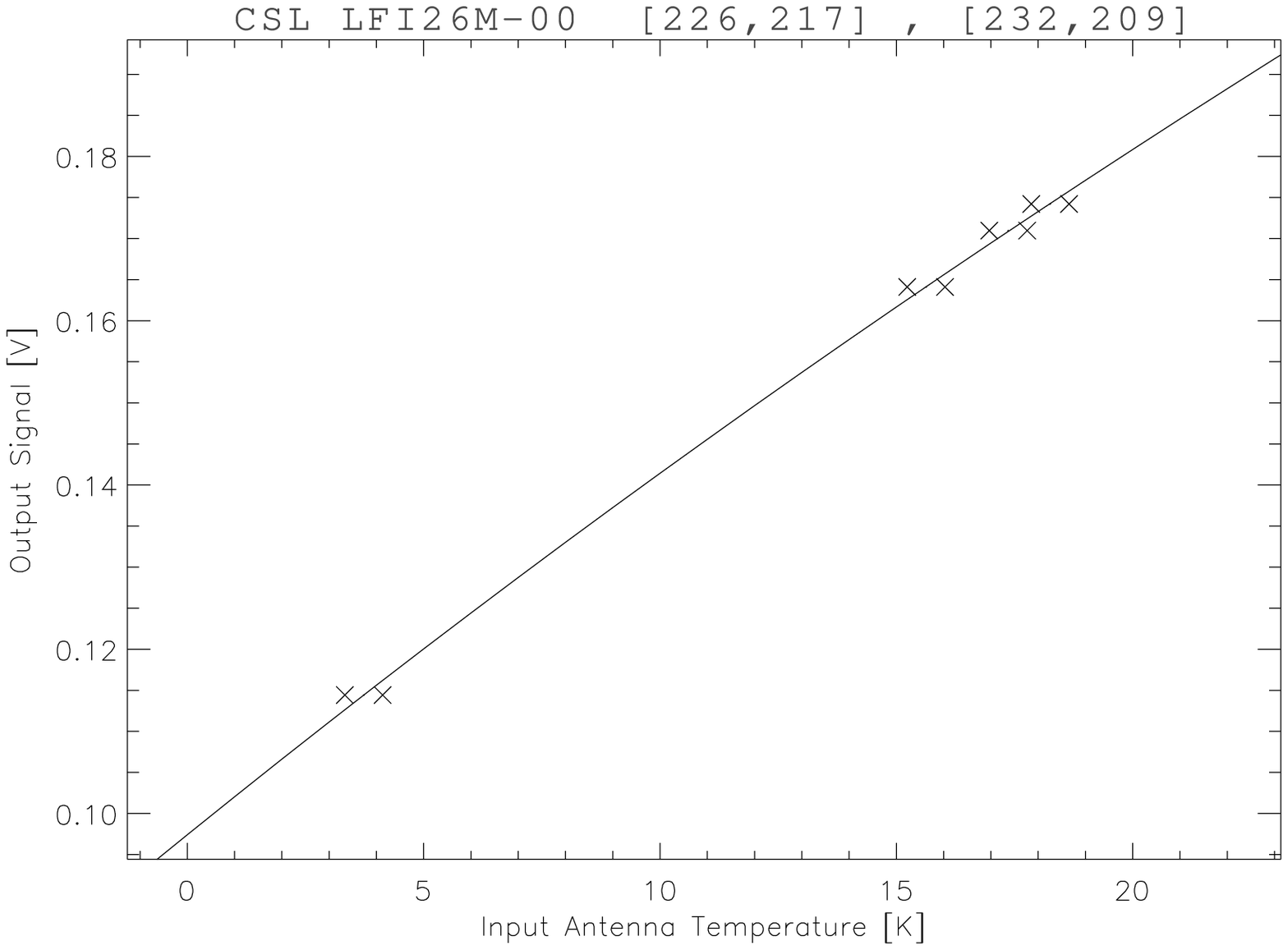}
            \includegraphics[width=7.0cm]{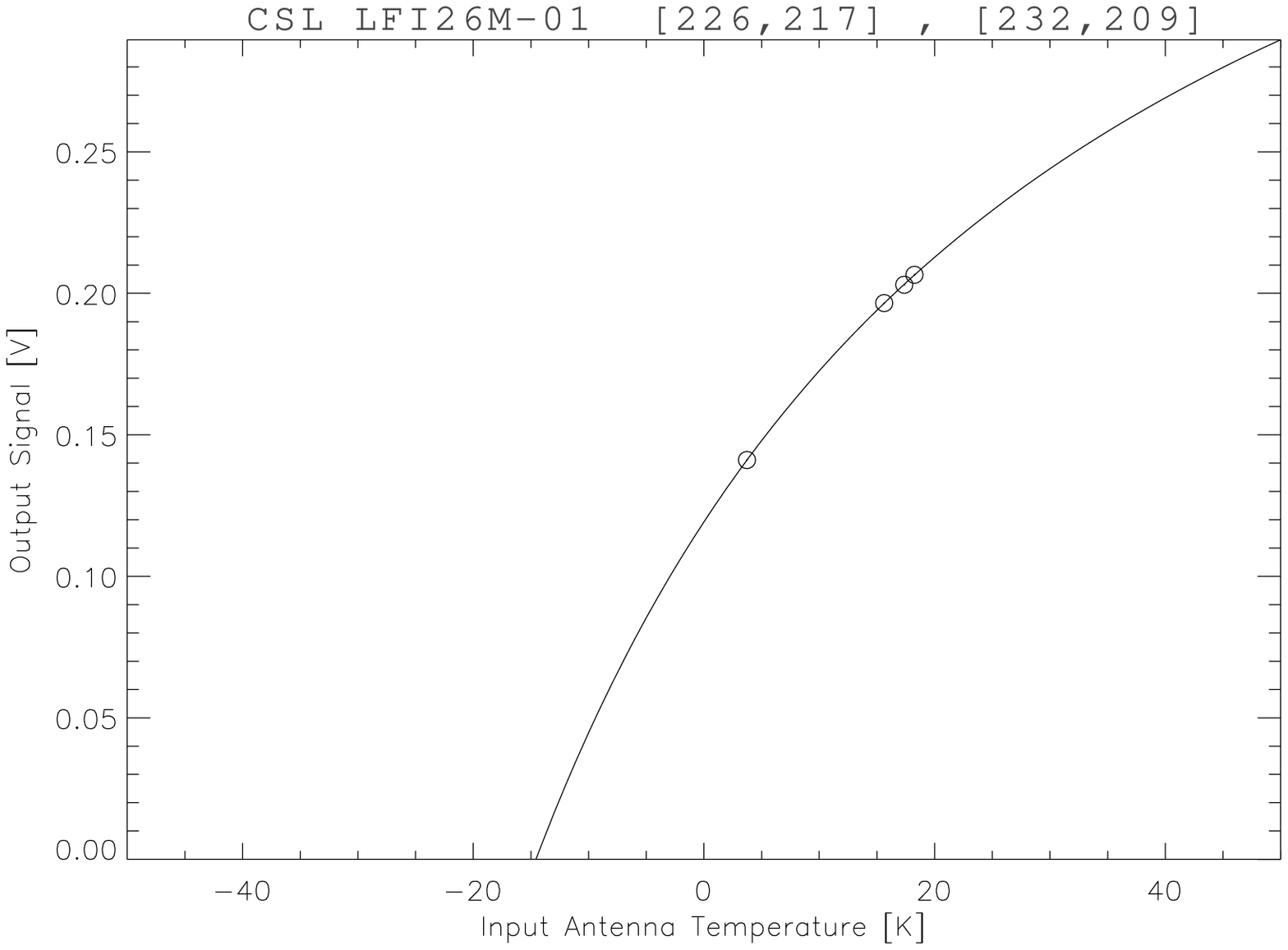}
            \includegraphics[width=7.0cm]{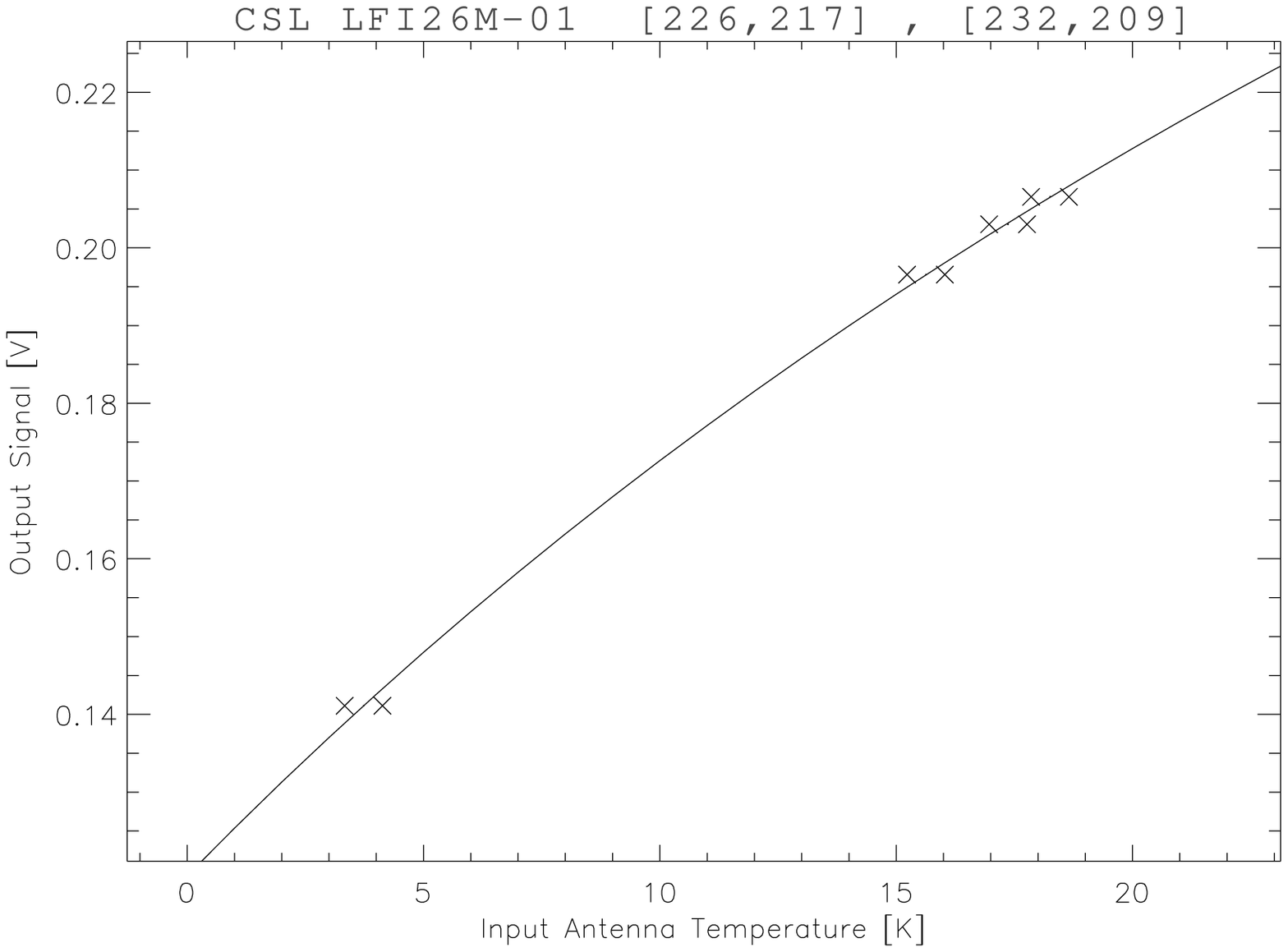}\\
            
            \textbf{LFI-26S}\\
            \includegraphics[width=7.0cm]{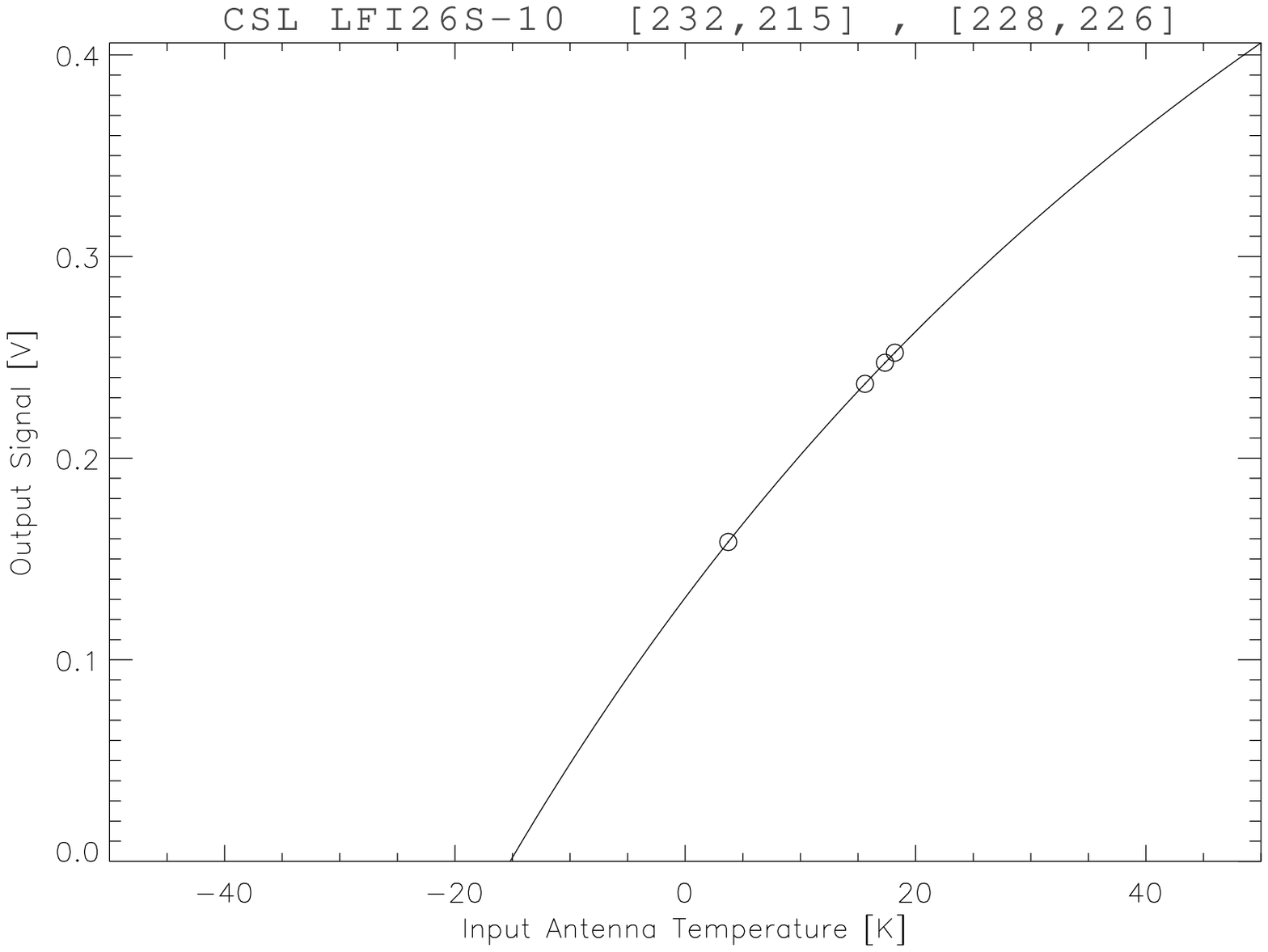}
            \includegraphics[width=7.0cm]{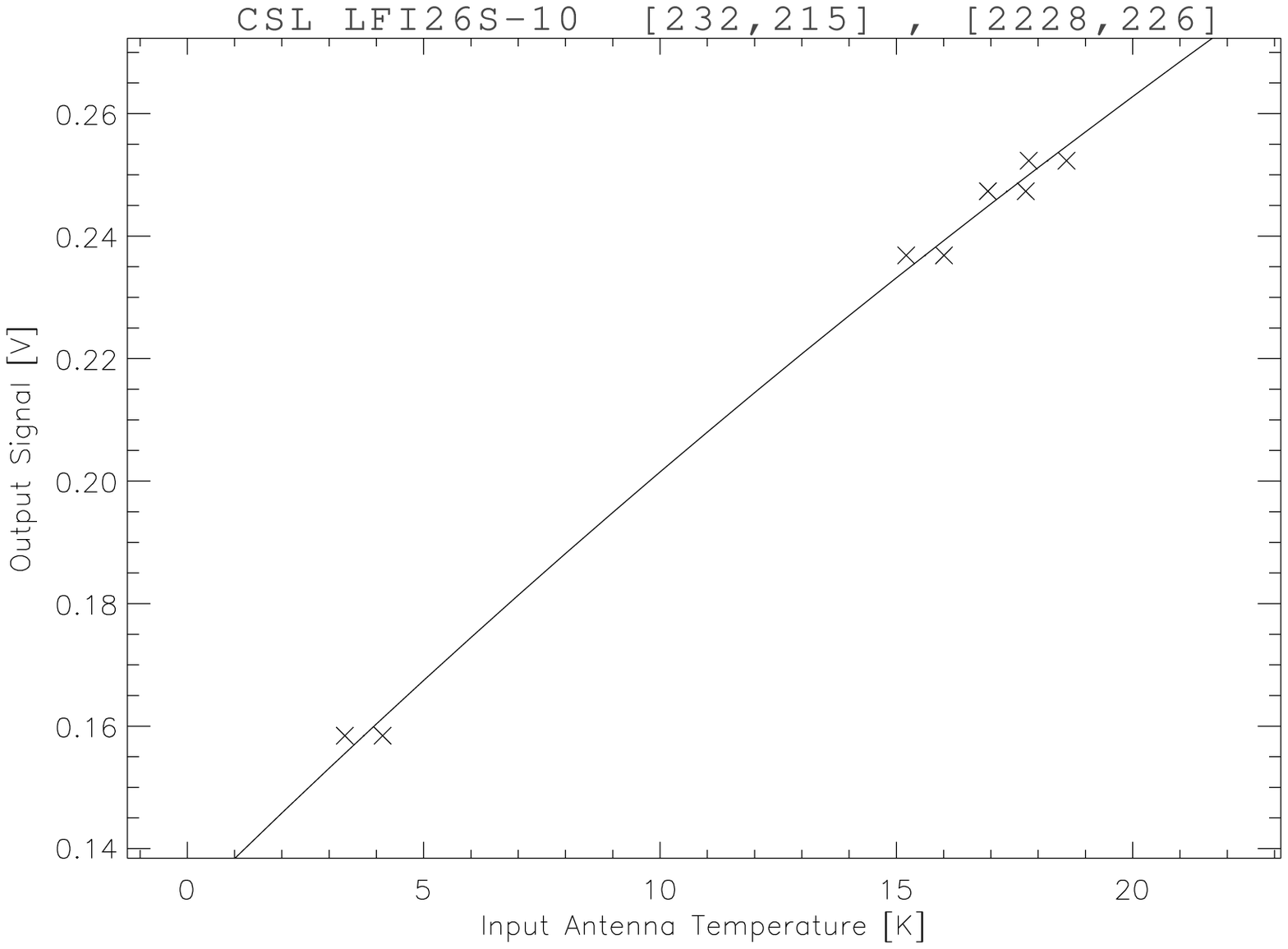}
            \includegraphics[width=7.0cm]{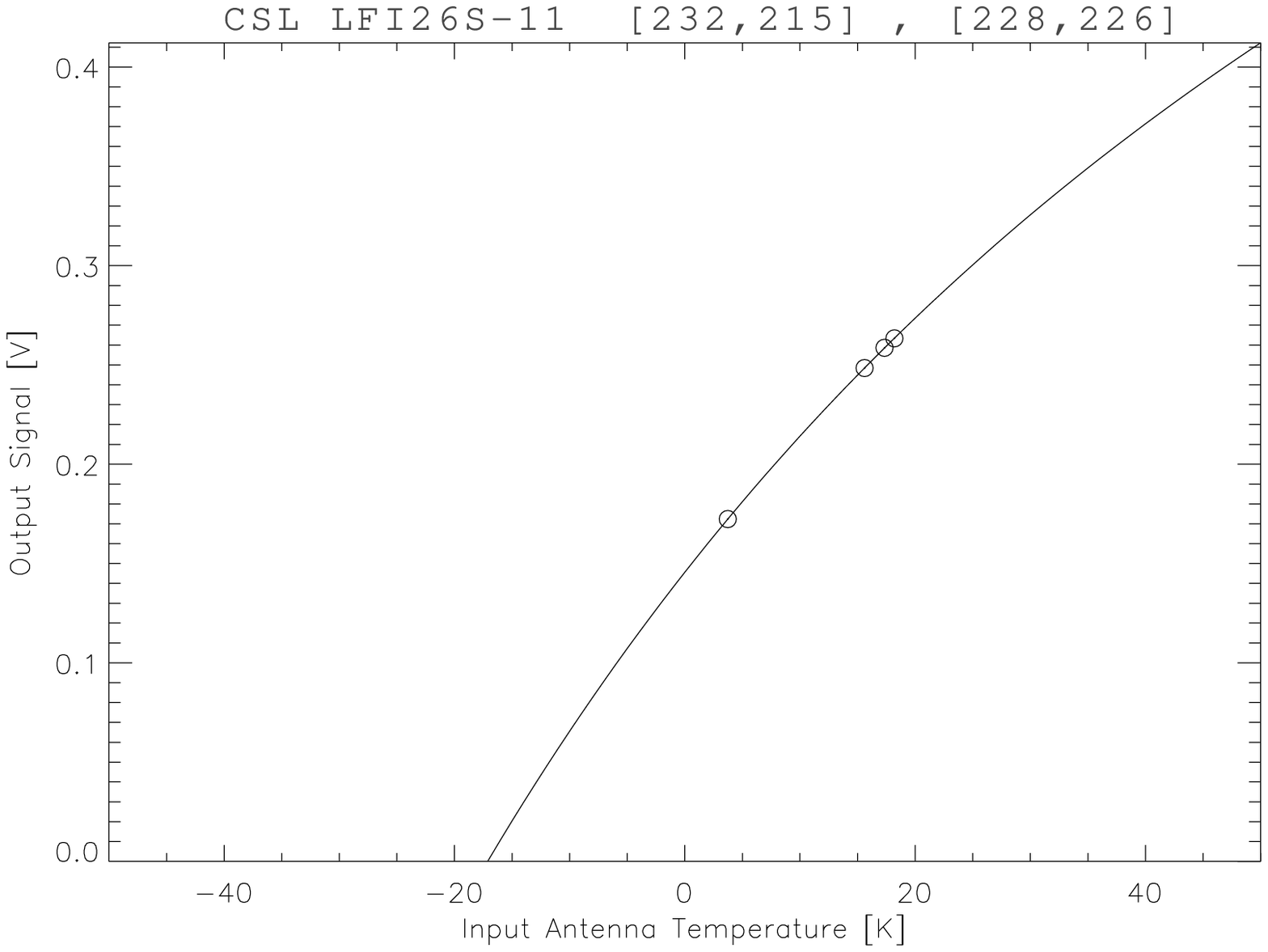}
            \includegraphics[width=7.0cm]{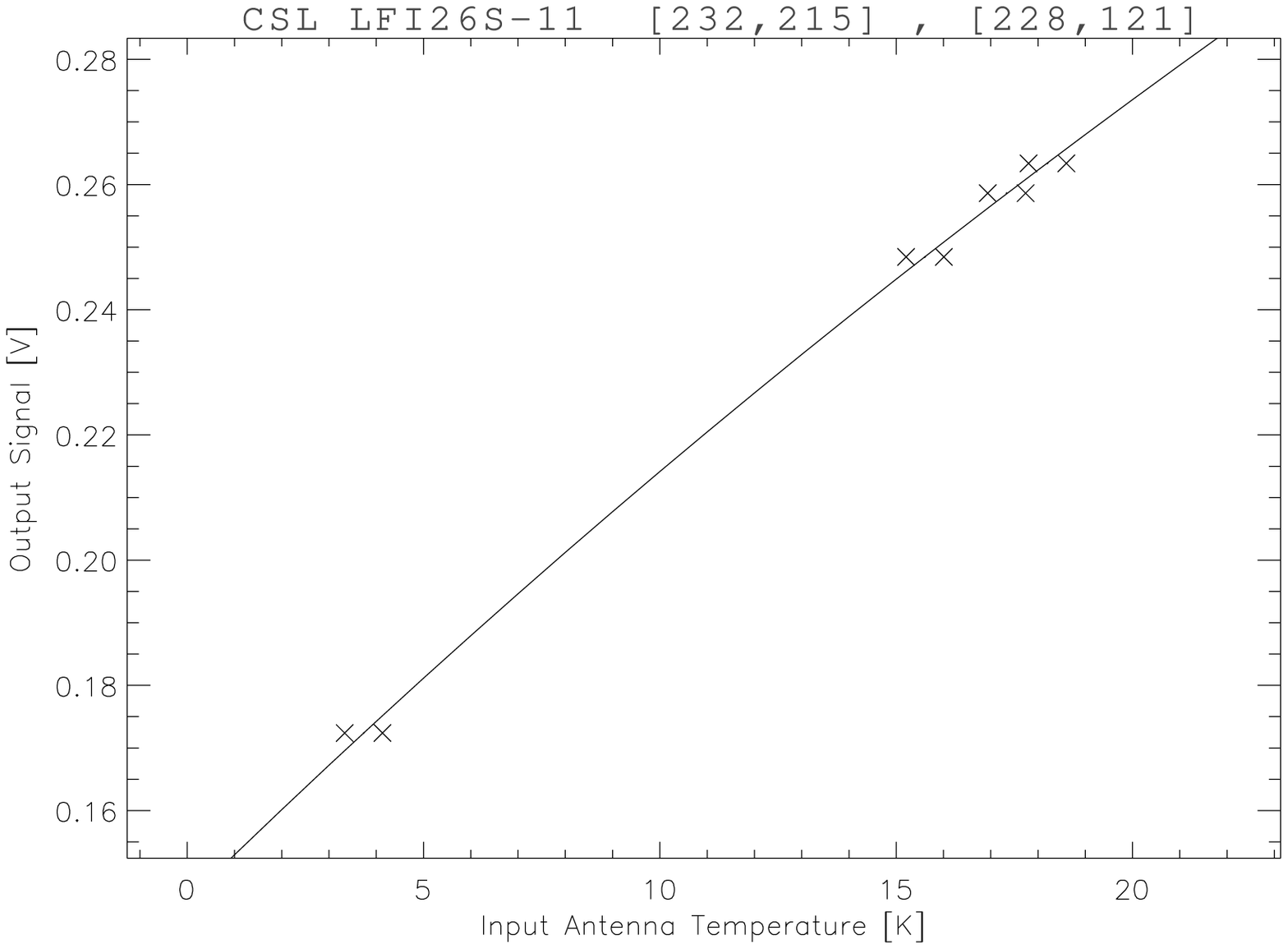}\\
            
          \end{center}
            
        \caption{Non linear response plots for 44~GHz \texttt{LFI-26} channel.}% Each detector (\texttt{M-00}, \texttt{M-01}, \texttt{S-10}, \texttt{S-11}) is represented from left to right: the voltage measured at \texttt{Diode-1} and \texttt{Diode-2} (y-axis) versus the extrapolated input temperature (x-axis); a zoom on the same plot  accounting also for uncertainties related to the knowledge of the 4~K reference load temperature and to the BEU thermal drift. }
        \label{fig_HYM_tun_nonlin_44}
            \end{figure}

\begin{figure}[htb]
        \begin{center}
        
        \textbf{LFI-27M}\\
            \includegraphics[width=7.0cm]{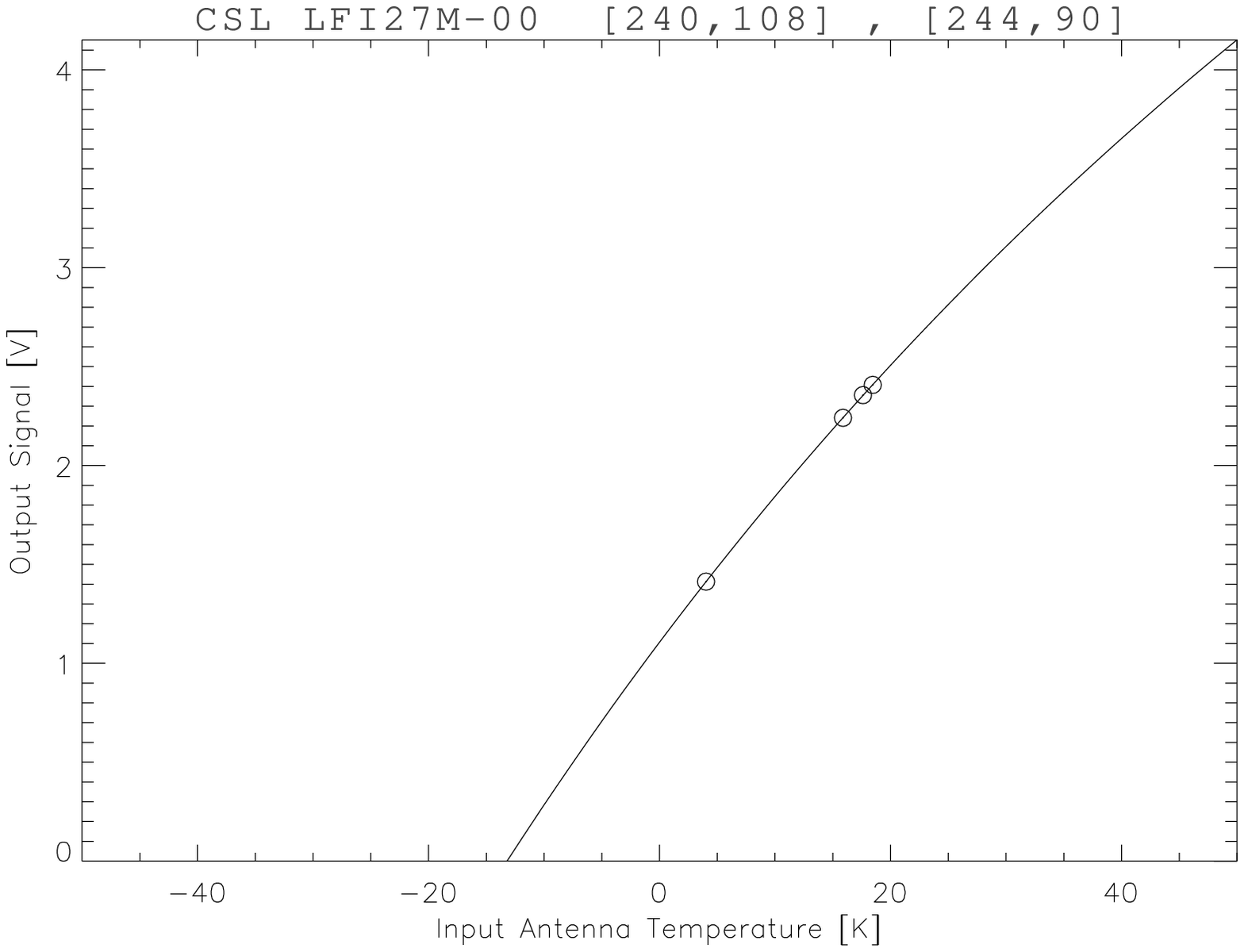}
            \includegraphics[width=7.0cm]{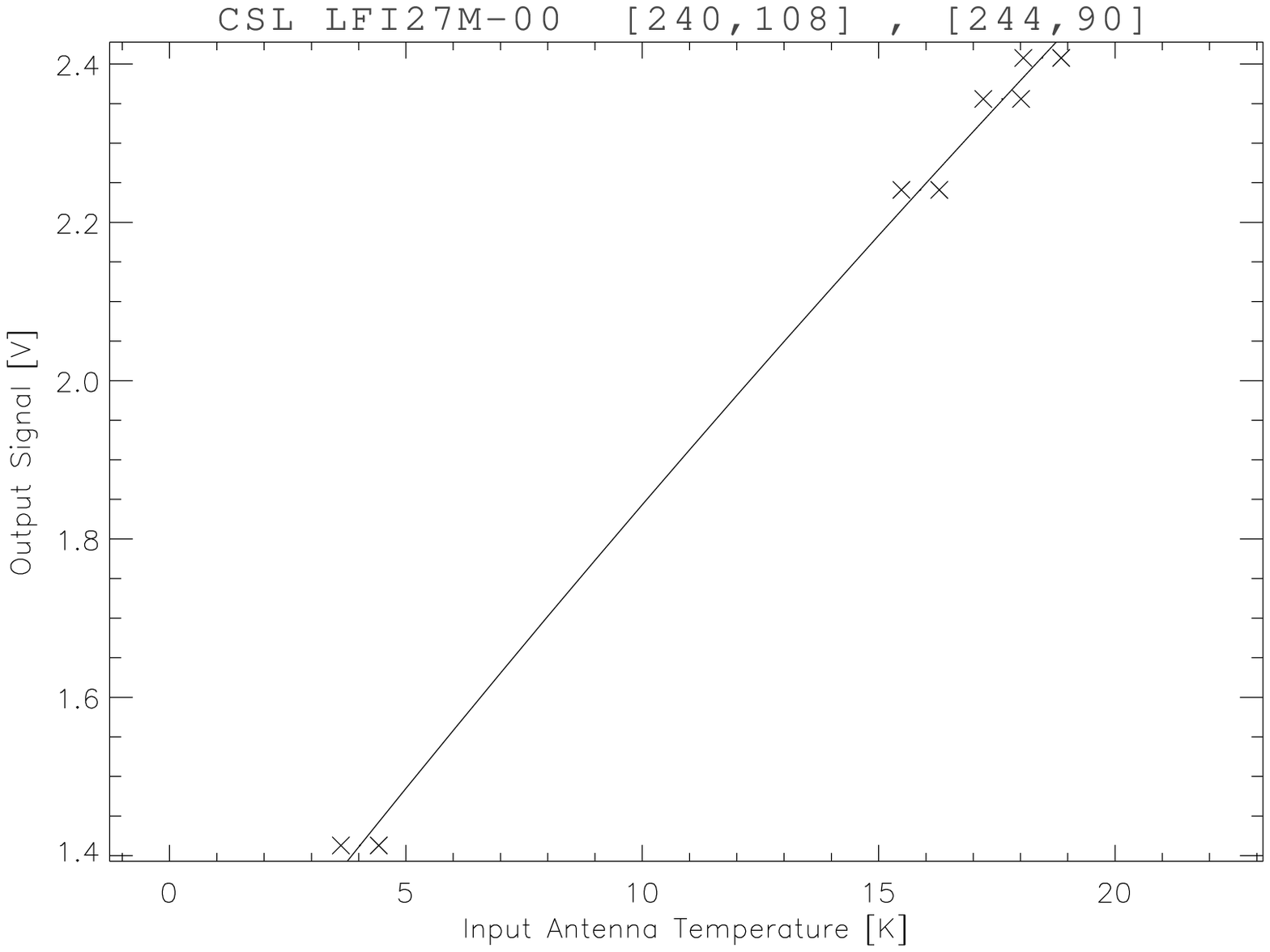}\\
            \includegraphics[width=7.0cm]{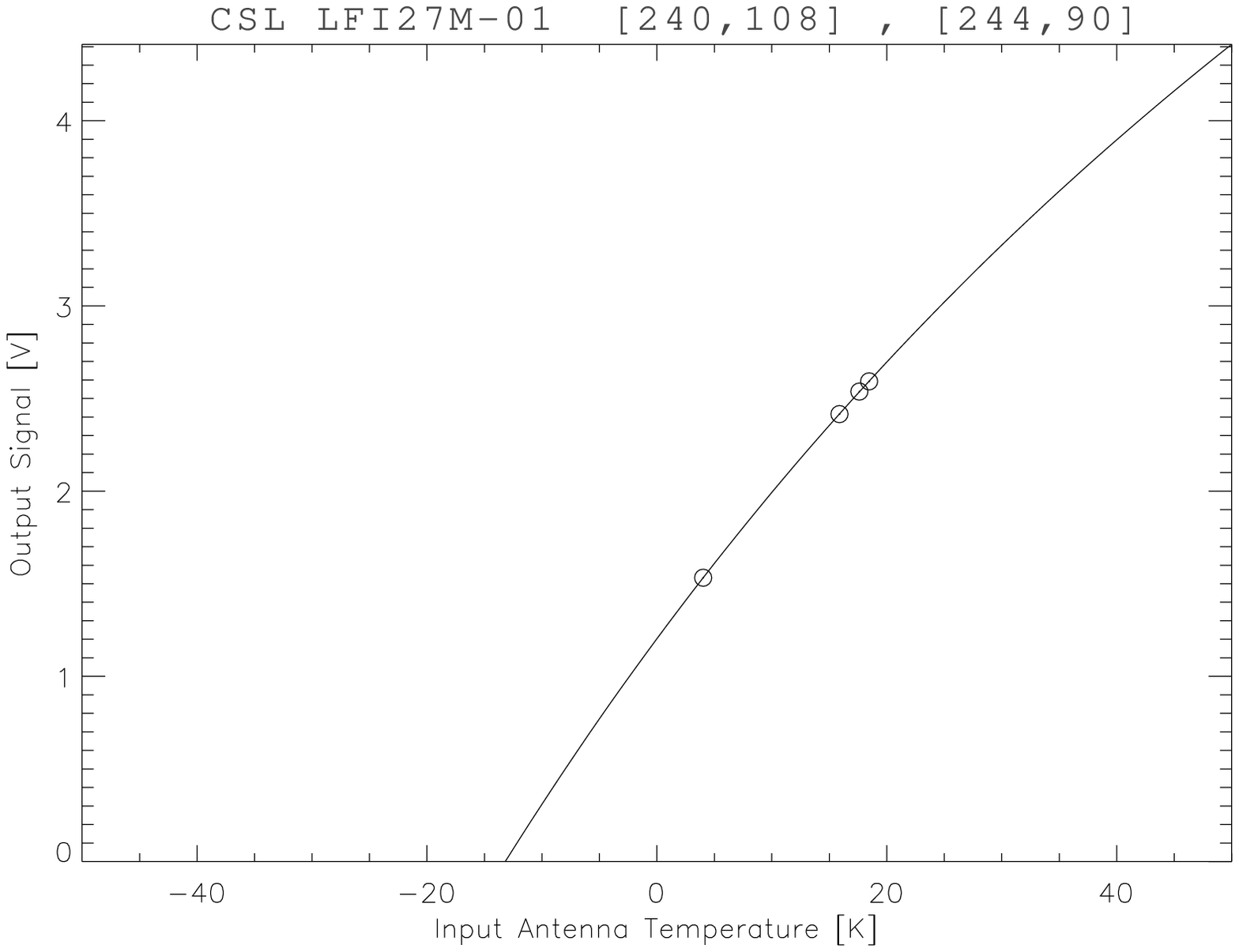}
            \includegraphics[width=7.0cm]{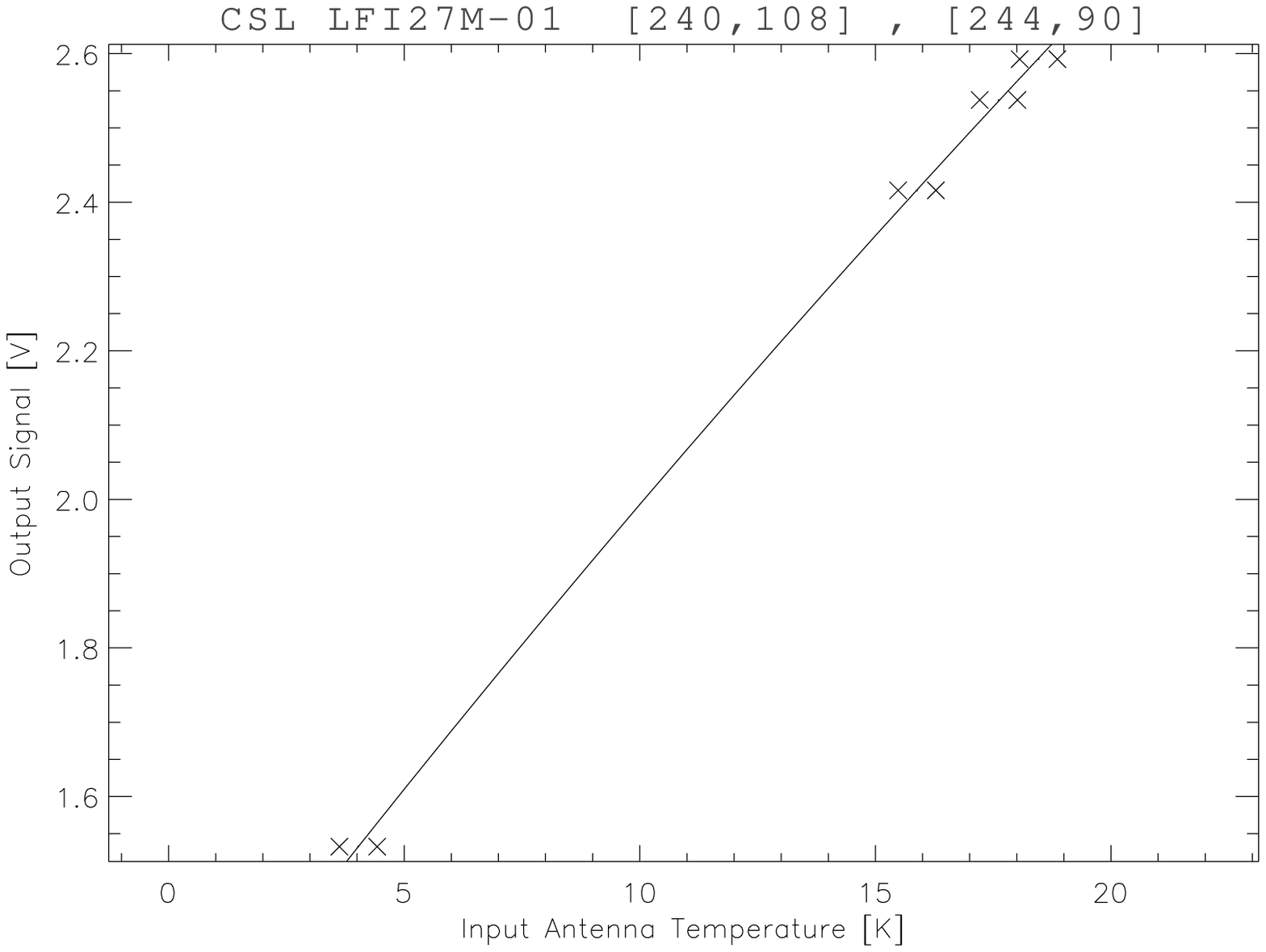}\\
            \textbf{LFI-27S}\\
            \includegraphics[width=7.0cm]{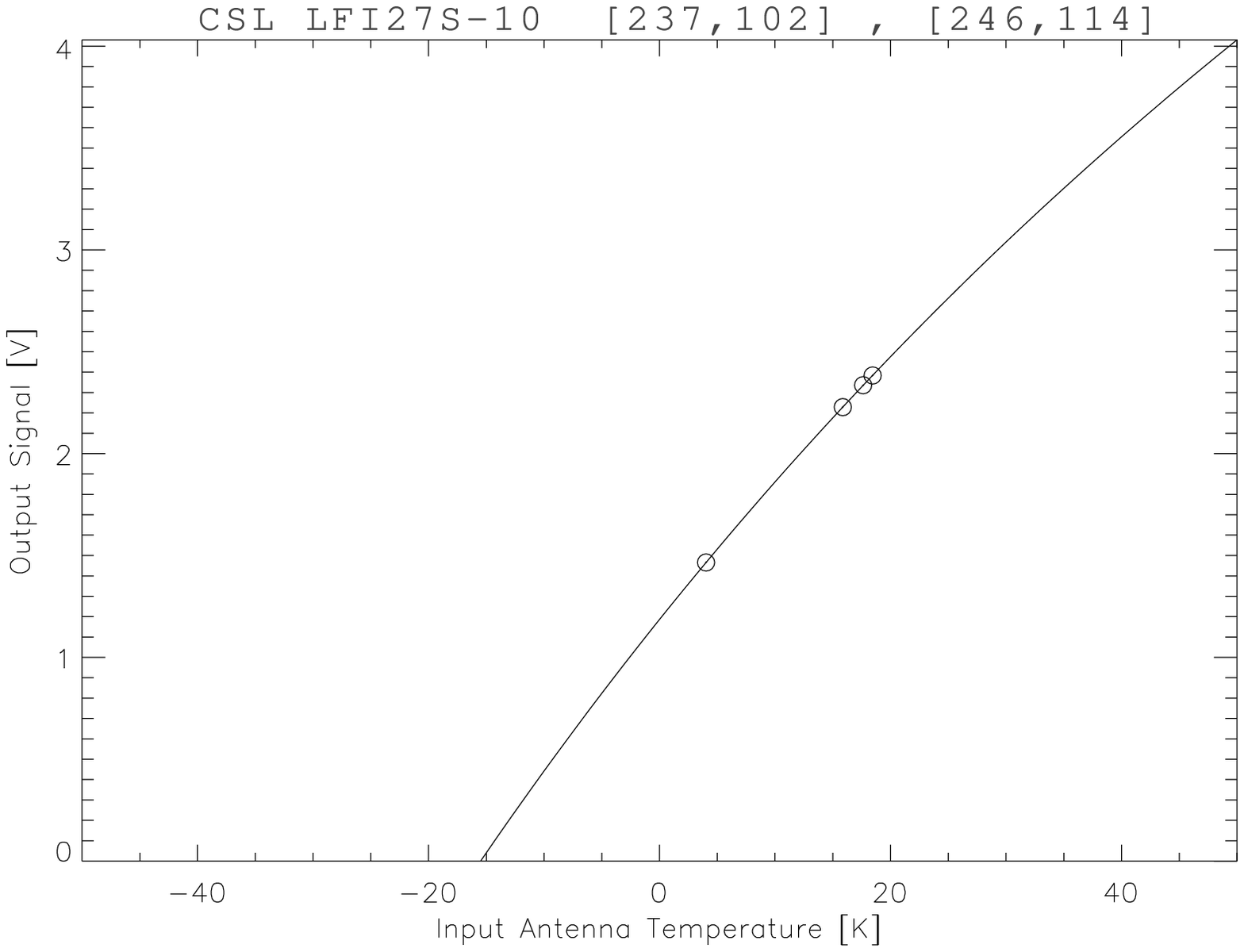}
            \includegraphics[width=7.0cm]{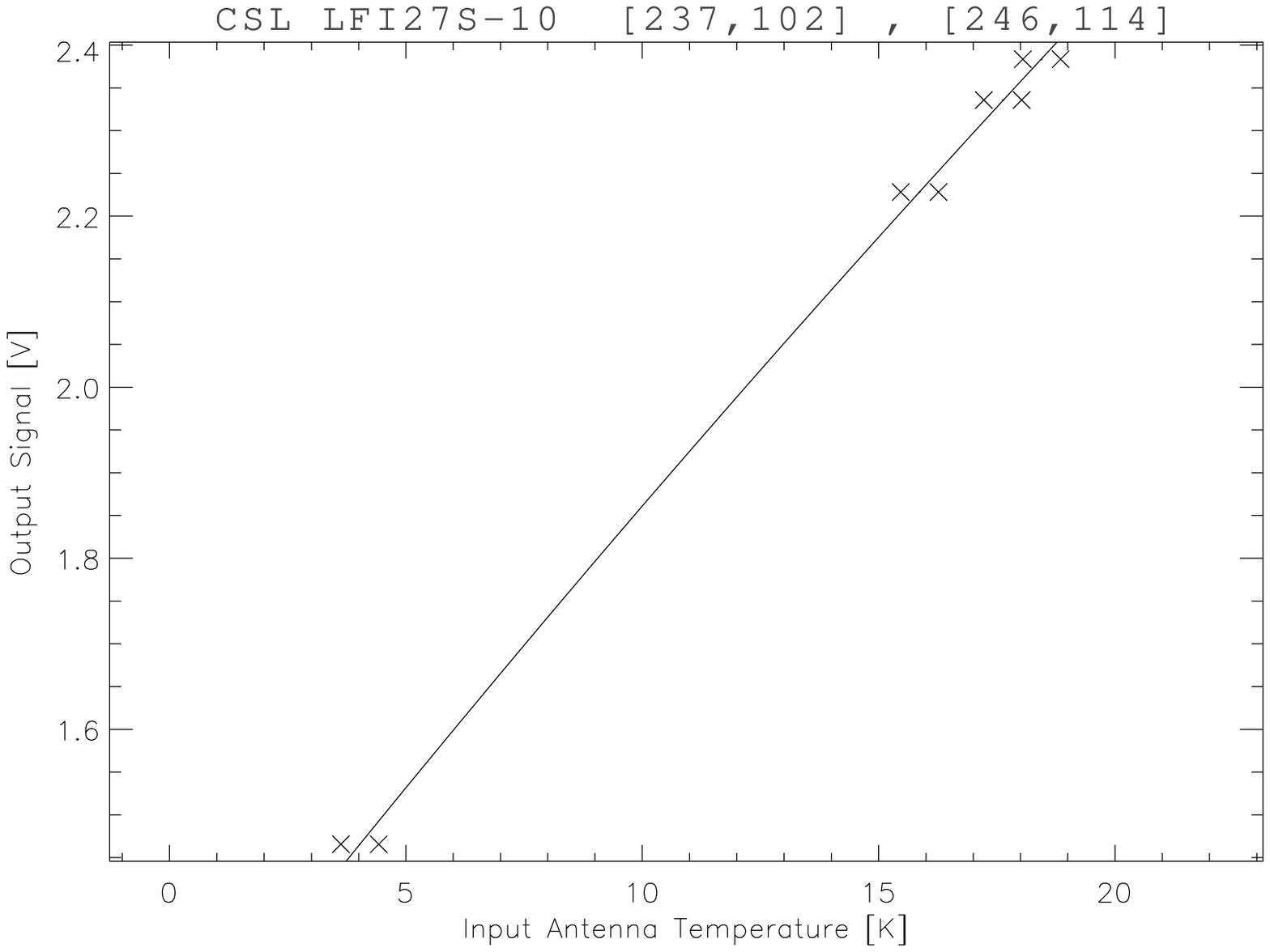}\\
            \includegraphics[width=7.0cm]{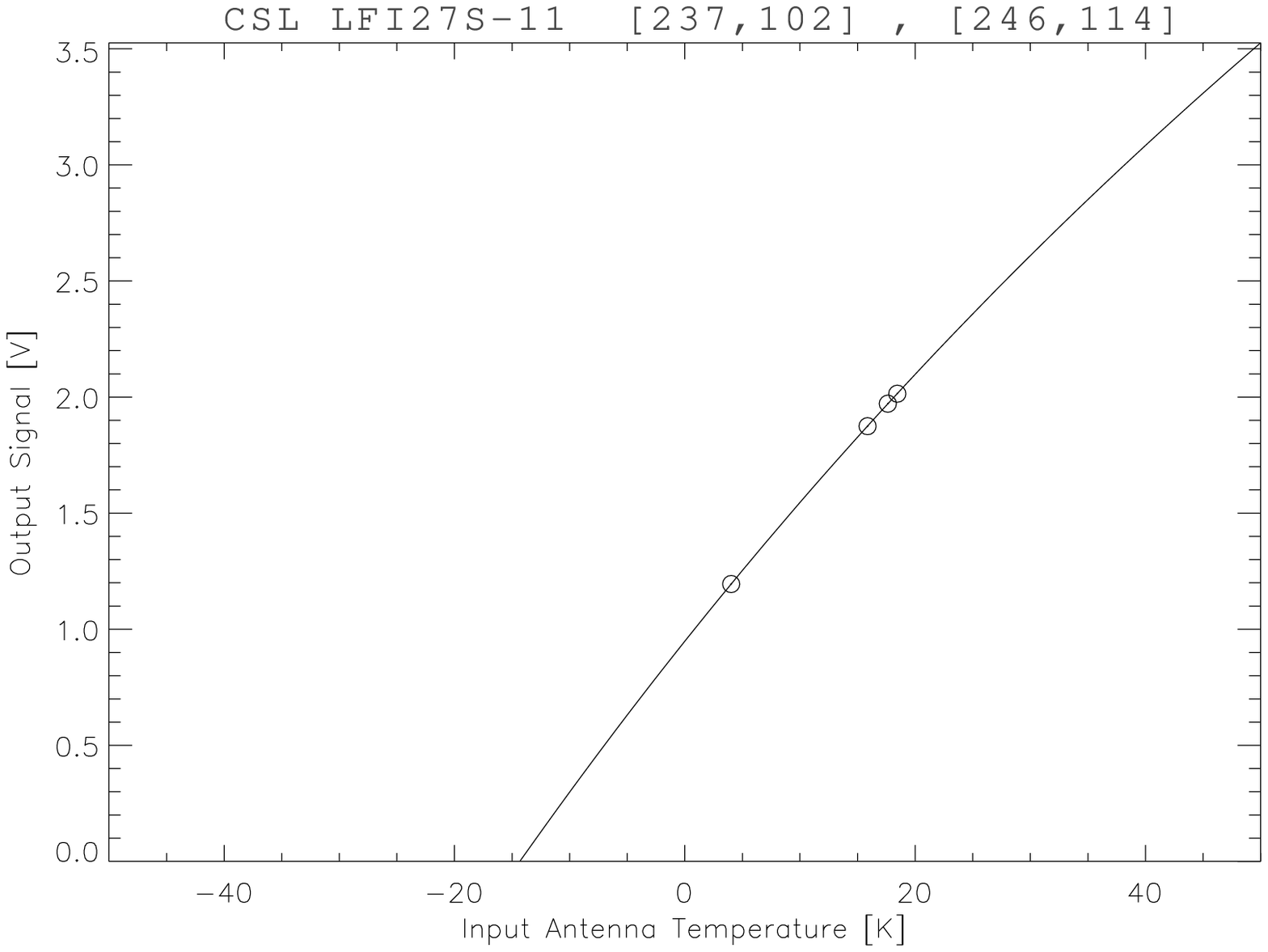}
            \includegraphics[width=7.0cm]{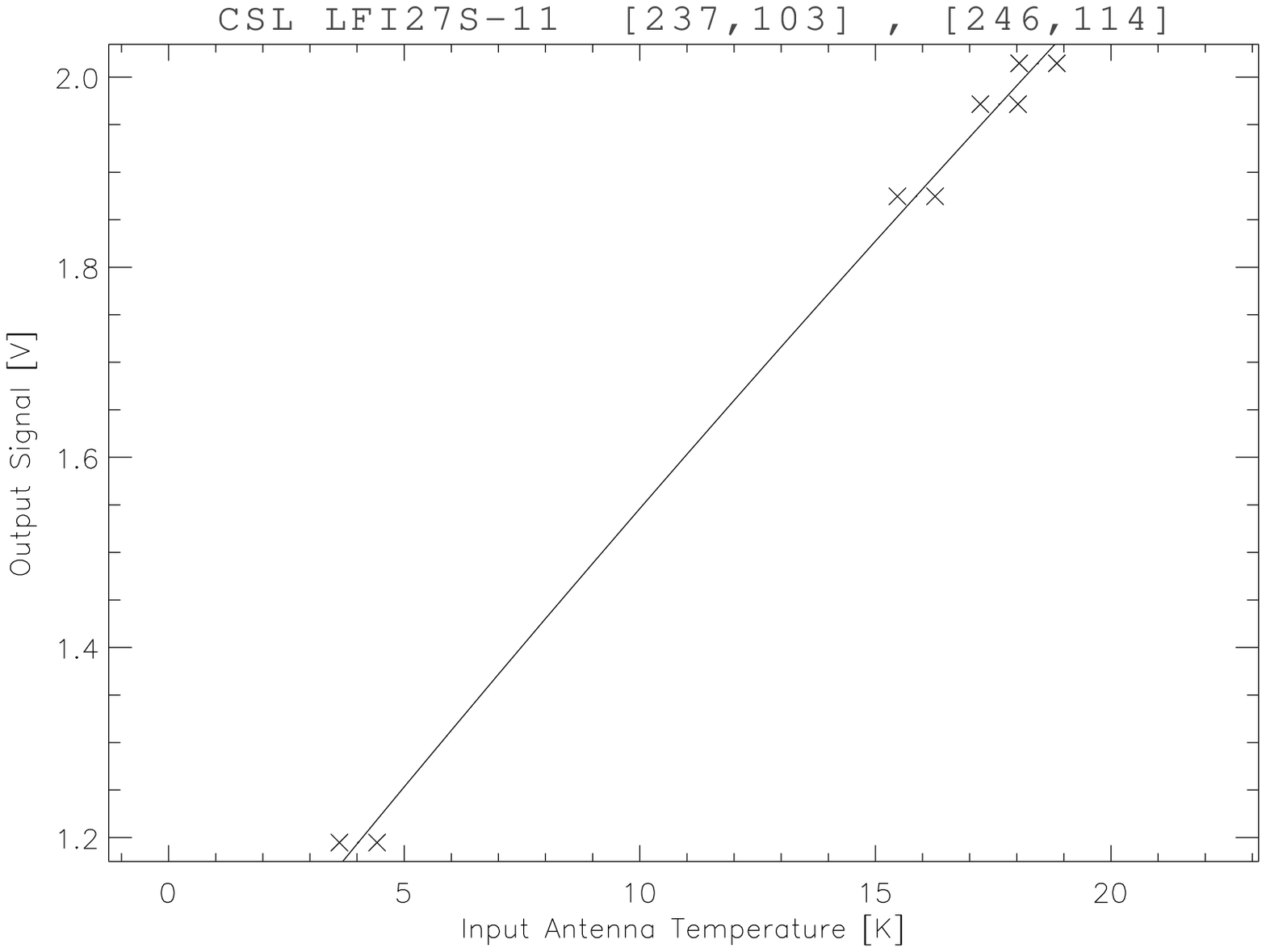}\\
        \end{center}
            
        \caption{Non linear response plots for 30~GHz \texttt{LFI-27} channel. }
        \label{fig_HYM_tun_nonlin_30-27}
    \end{figure}

    In Figures from ~\ref{fig_HYM_tun_nonlin_30-27} to \ref{fig_HYM_tun_nonlin_44} on each row two plots are showed for each detector (\texttt{M-00}, \texttt{M-01}, \texttt{S-10}, \texttt{S-11})  of the 30~GHz and 44~GHz Channels, each displaying the output voltage as a function of the extrapolated input antenna temperature: the panel on the right being a zoom of the one on the left. Data are corrected for the Back-End thermal drift; solid lines are obtained from the non linear gain model. \\
    The pair of crosses in the zoomed plots represent the minimum and maximum of the estimated 4~K temperatures. This uncertainty is driven by the knowledge of the 4~K reference load temperature and to the BEU thermal drift. 
     \begin{figure}[htb]
        \begin{center}
                 
             	\textbf{LFI-24M}\\
            \includegraphics[width=7.0cm]{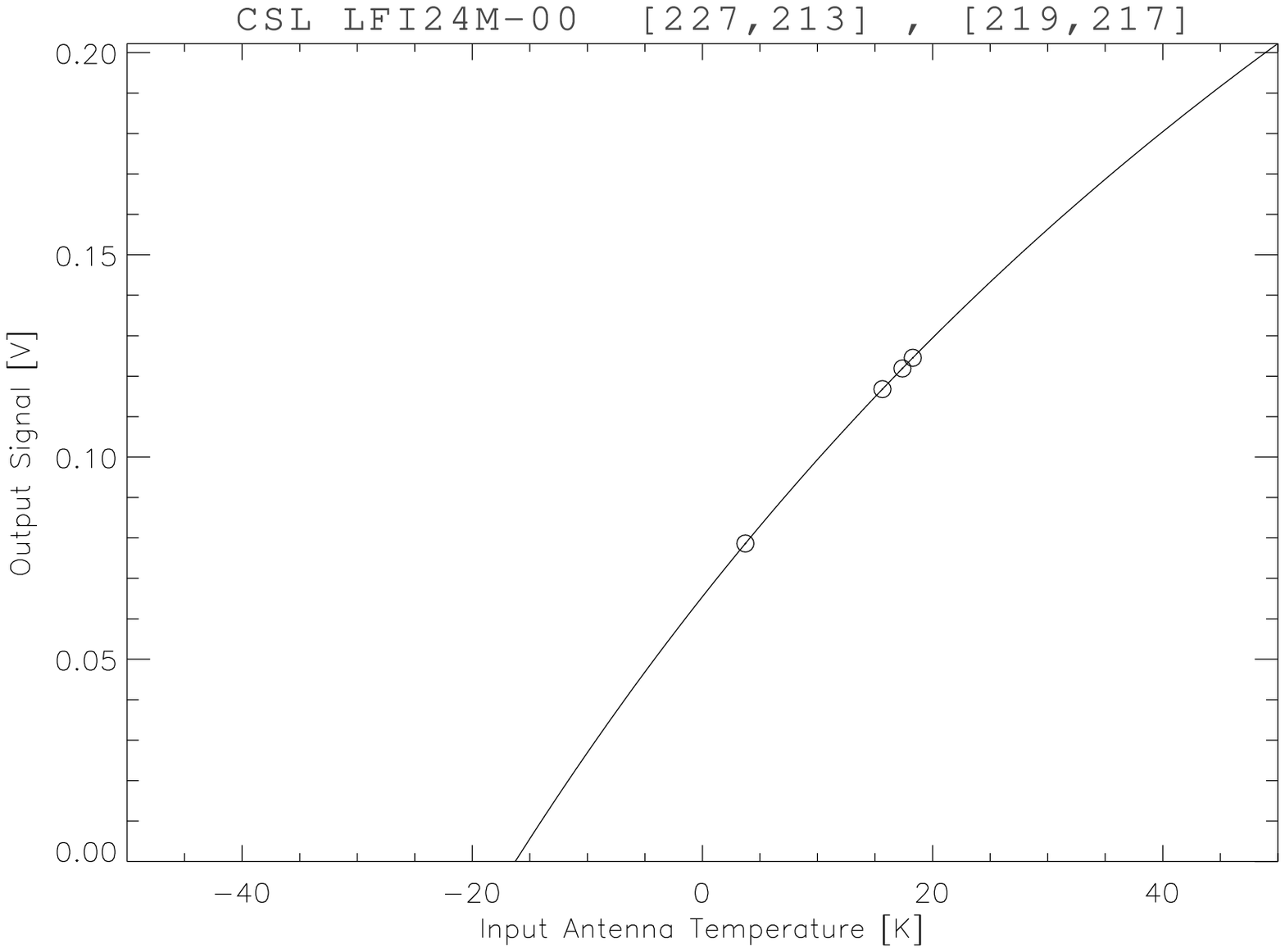}
            \includegraphics[width=7.0cm]{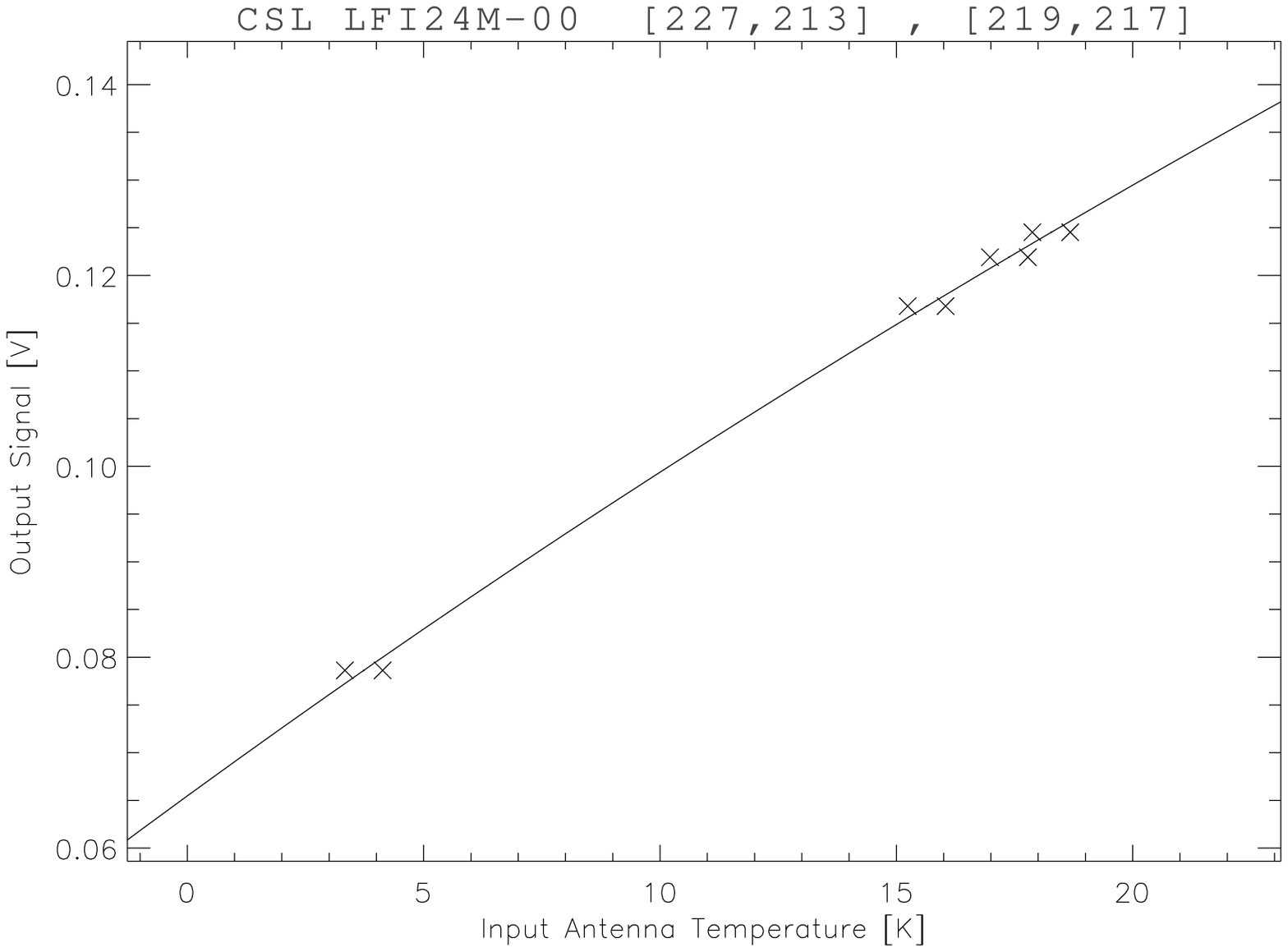}
            \includegraphics[width=7.0cm]{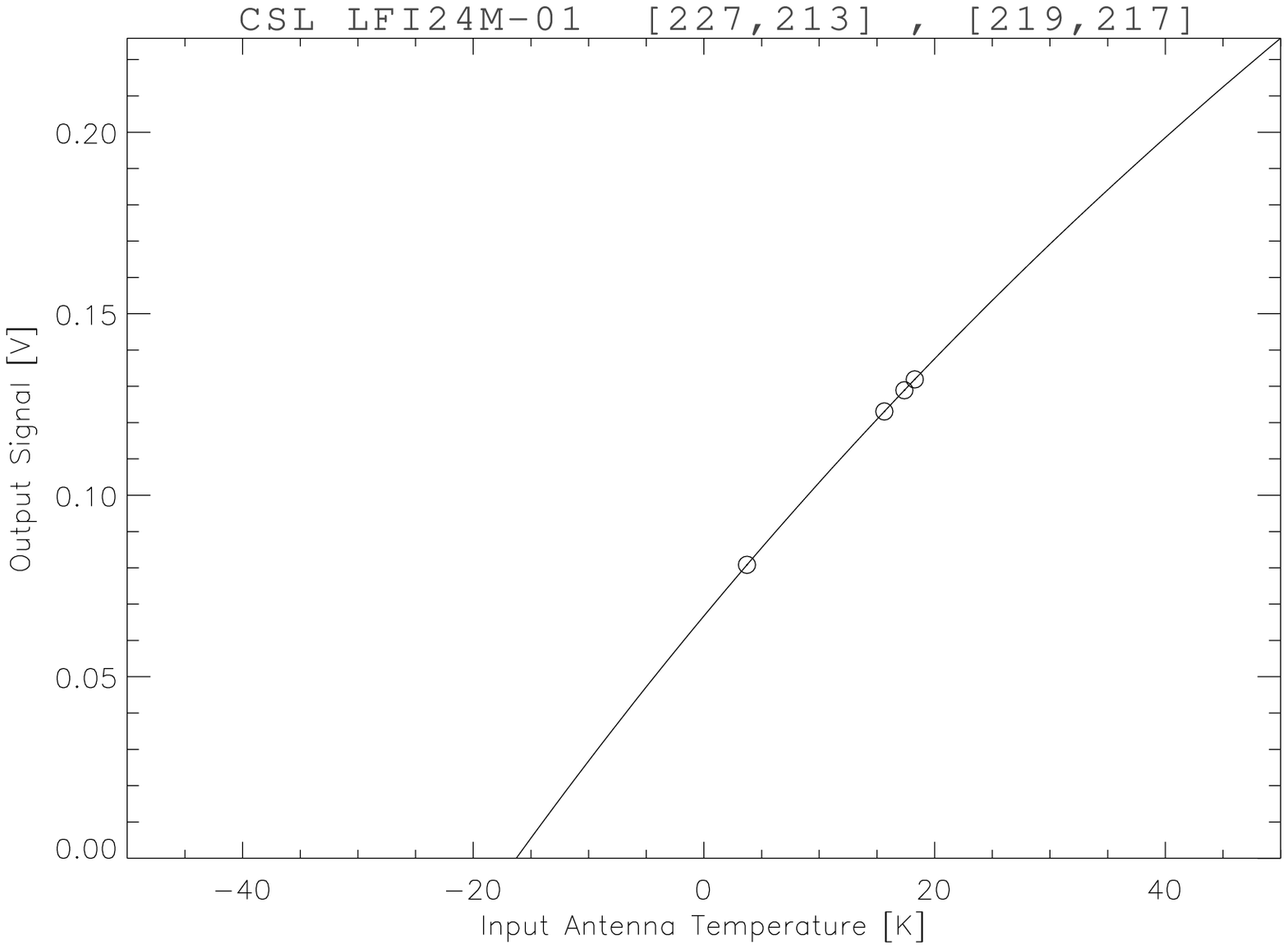}
            \includegraphics[width=7.0cm]{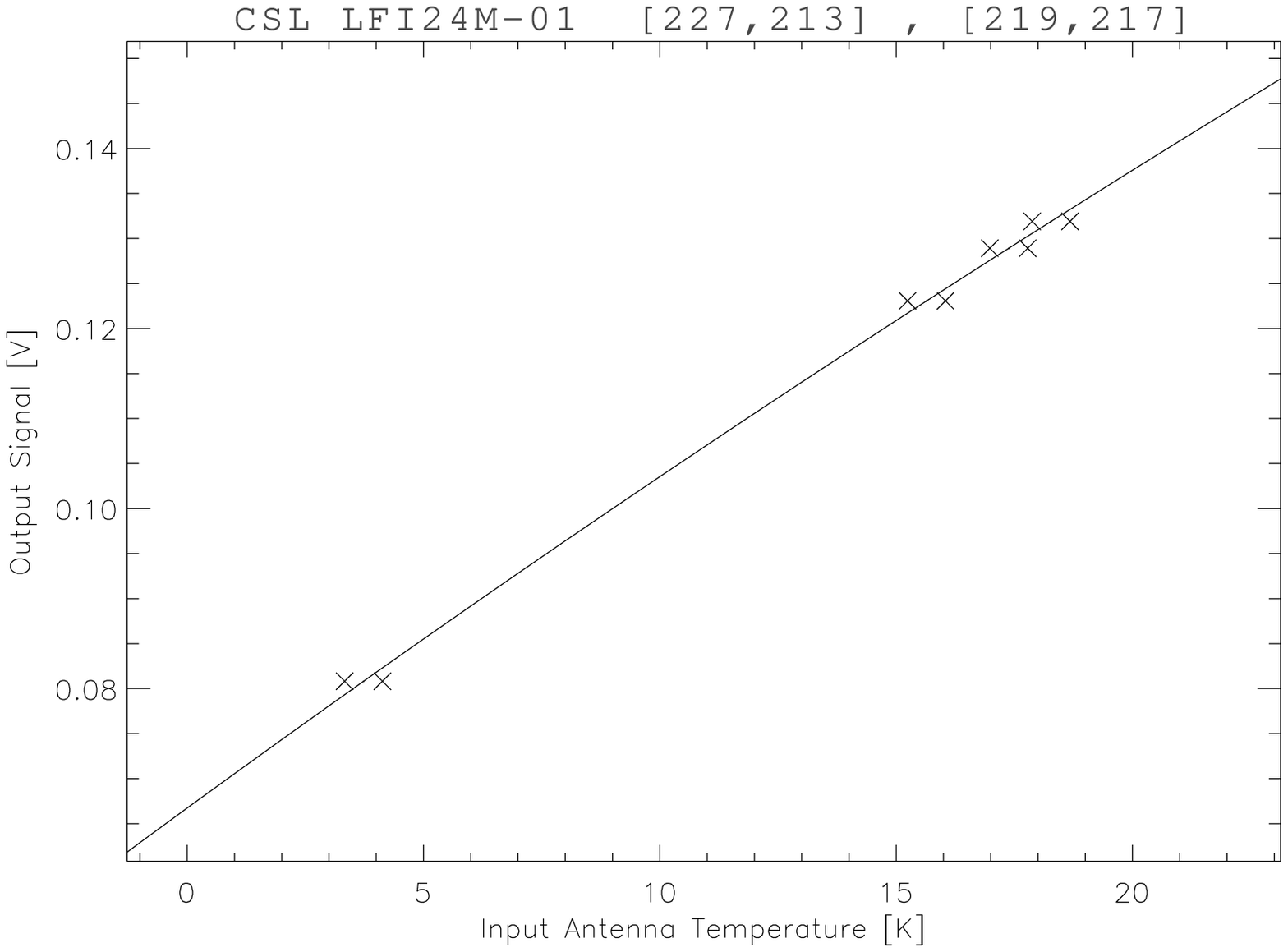}\\
             
            \textbf{LFI-24S}\\
            \includegraphics[width=7.0cm]{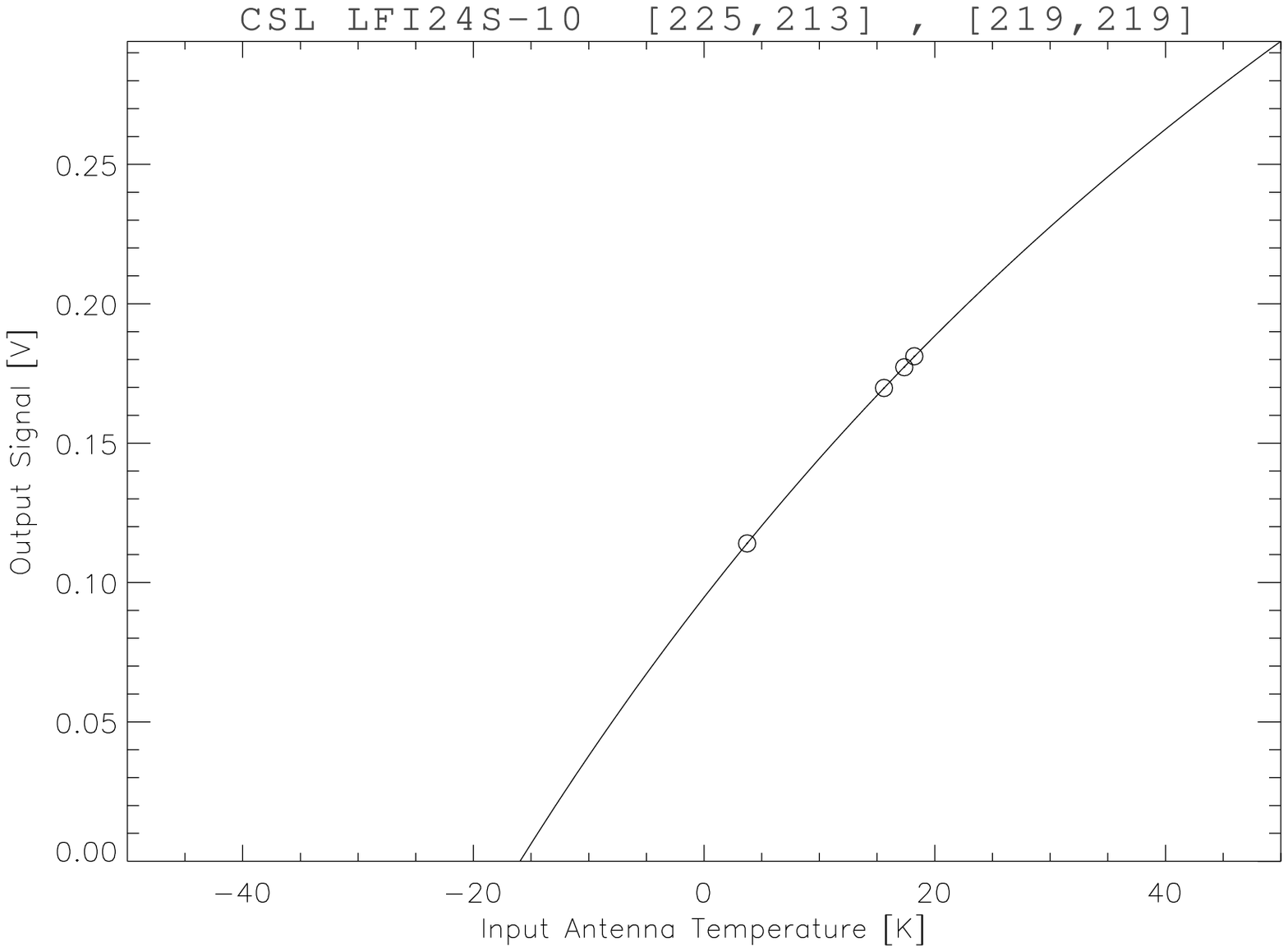}
            \includegraphics[width=7.0cm]{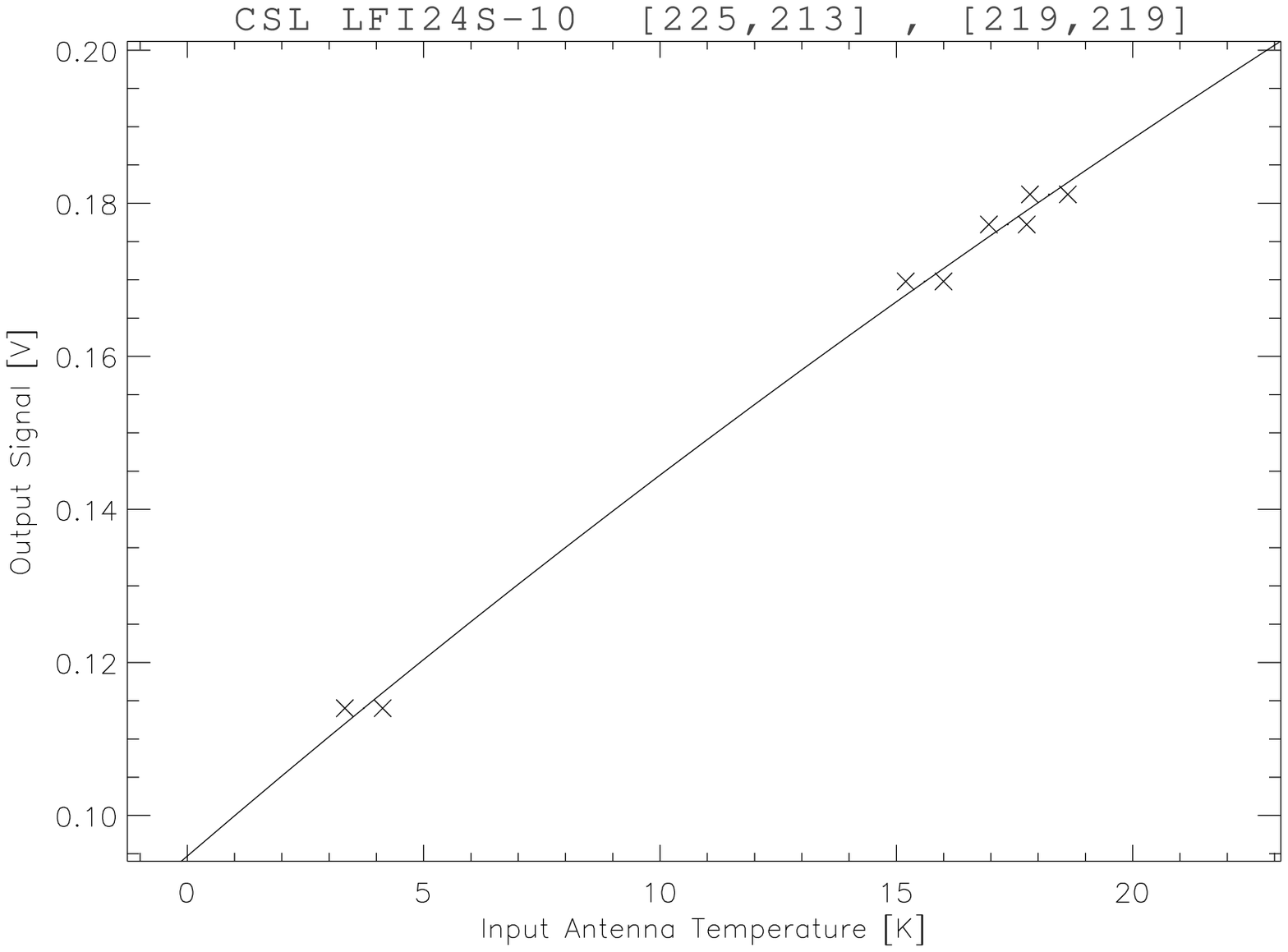}
            \includegraphics[width=7.0cm]{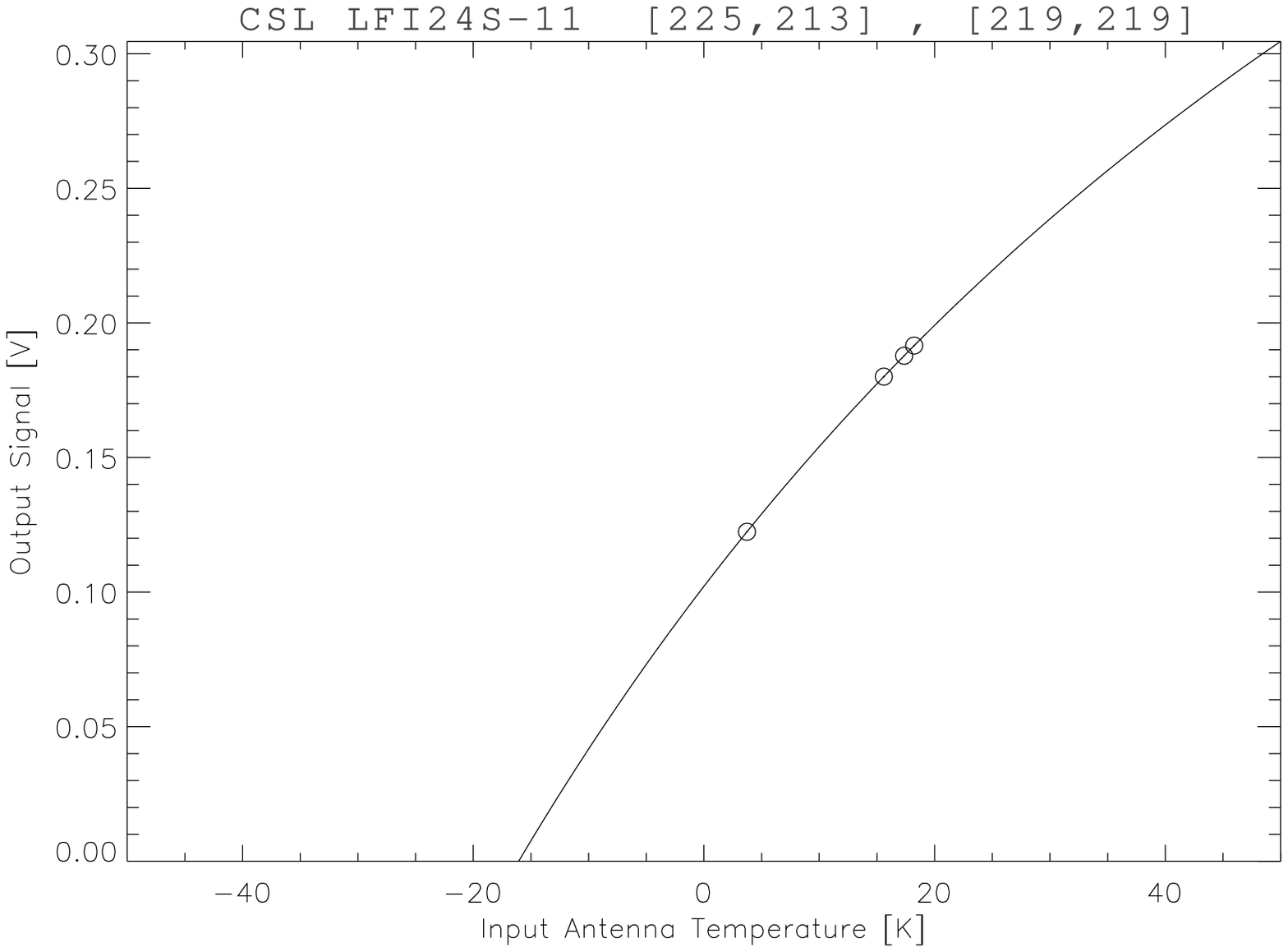}
            \includegraphics[width=7.0cm]{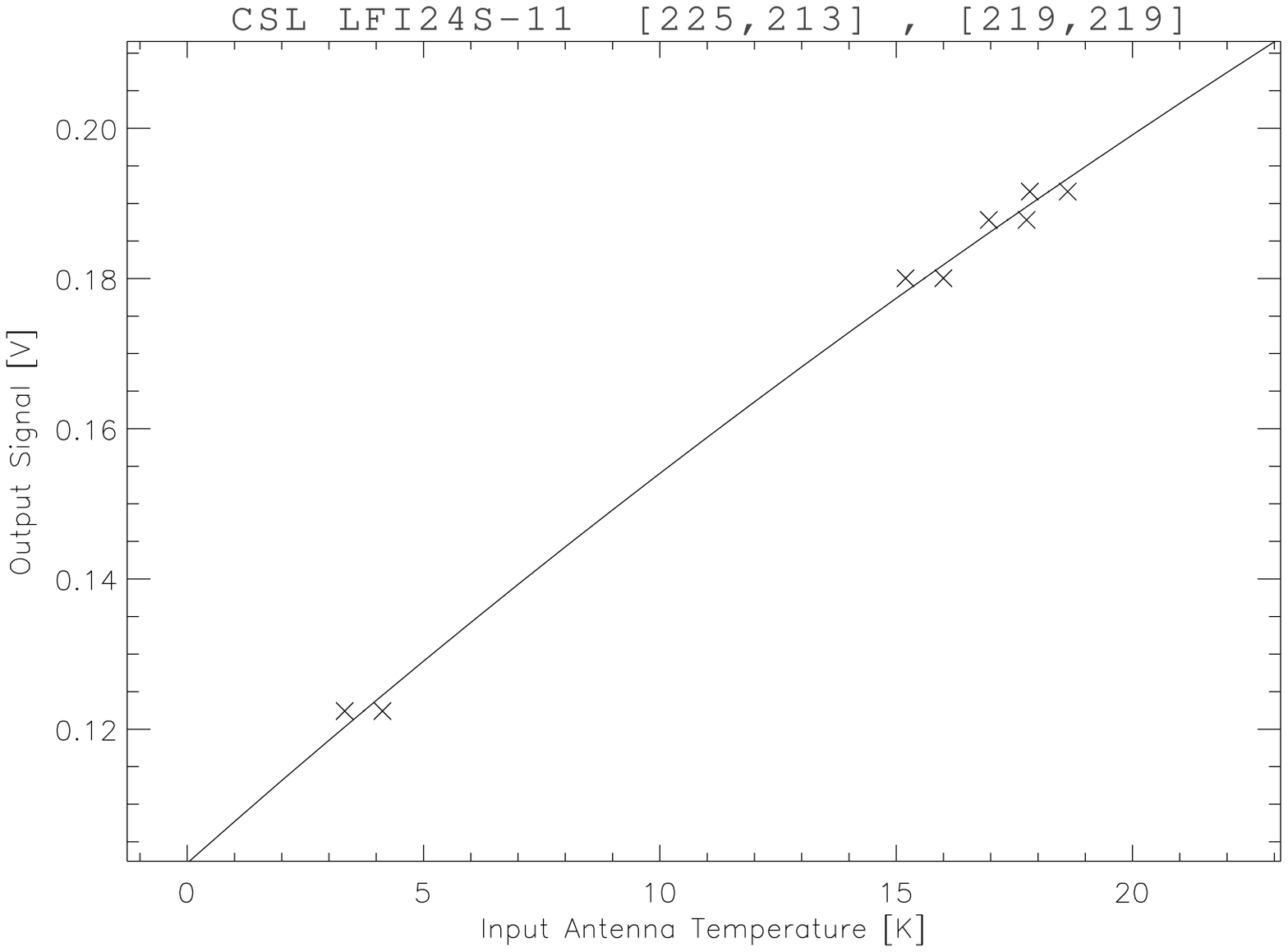}\\
          \end{center}
            
        \caption{Non linear response plots for 44~GHz \texttt{LFI-24} channel.}% Each detector (\texttt{M-00}, \texttt{M-01}, \texttt{S-10}, \texttt{S-11}) is represented from left to right: the voltage measured at \texttt{Diode-1} and \texttt{Diode-2} (y-axis) versus the extrapolated input temperature (x-axis); a zoom on the same plot  accounting also for uncertainties related to the knowledge of the 4~K reference load temperature and to the BEU thermal drift. }
        \label{fig_HYM_tun_nonlin_44-24}
                    \end{figure}
      \begin{figure}[htb]
        \begin{center}   
                         
            \textbf{LFI-25M}\\
            \includegraphics[width=7.0cm]{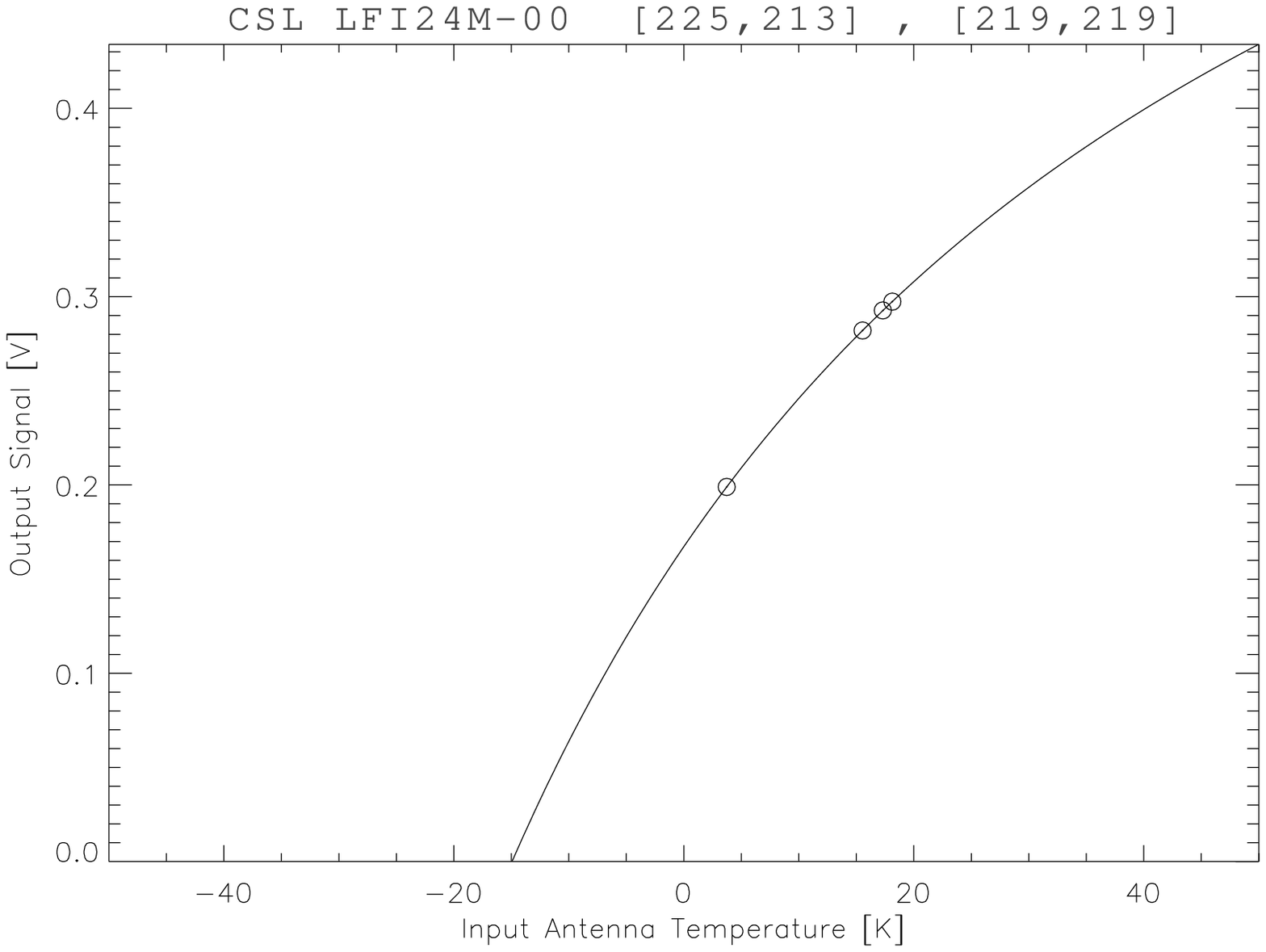}
            \includegraphics[width=7.0cm]{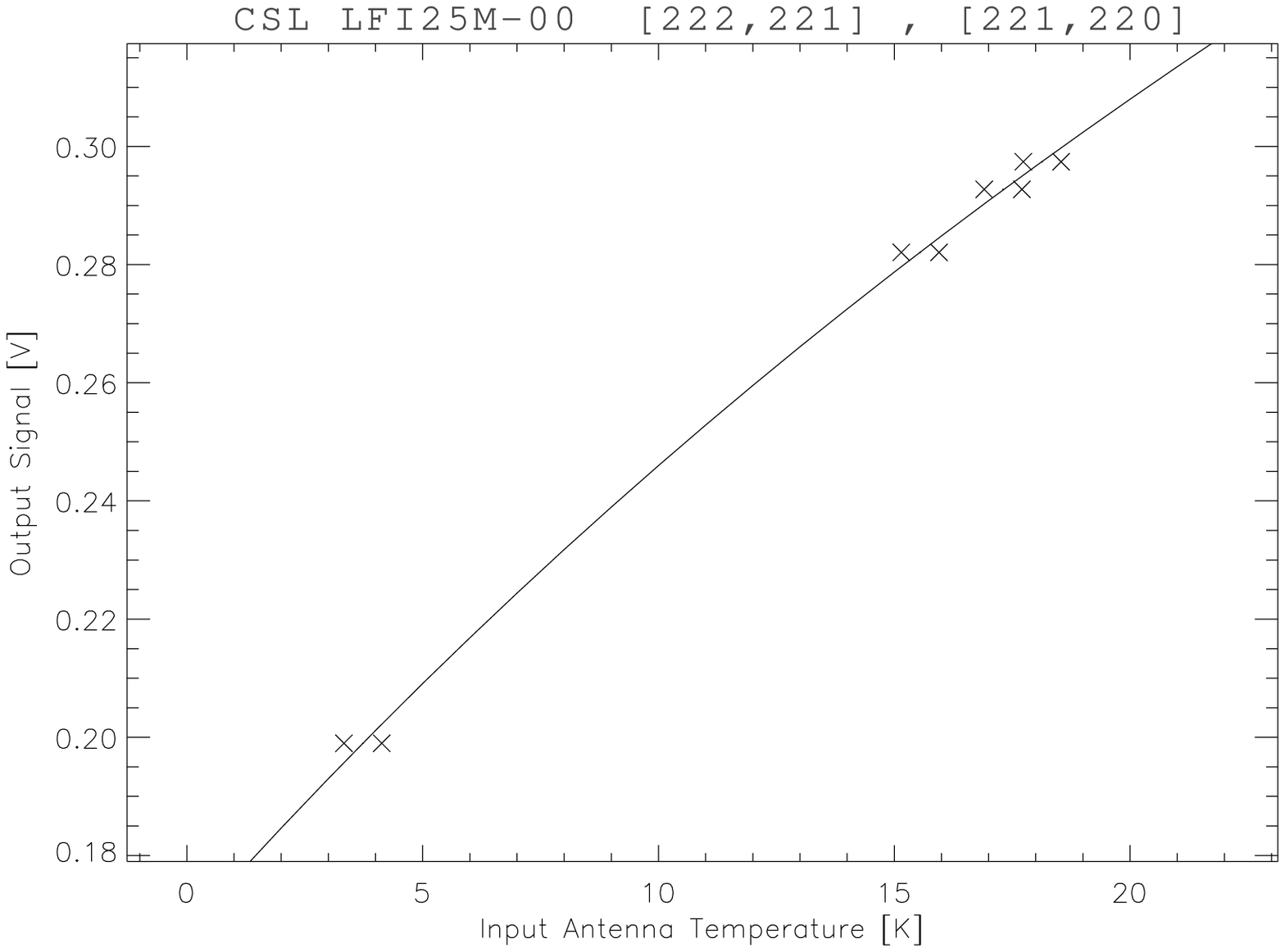}
            \includegraphics[width=7.0cm]{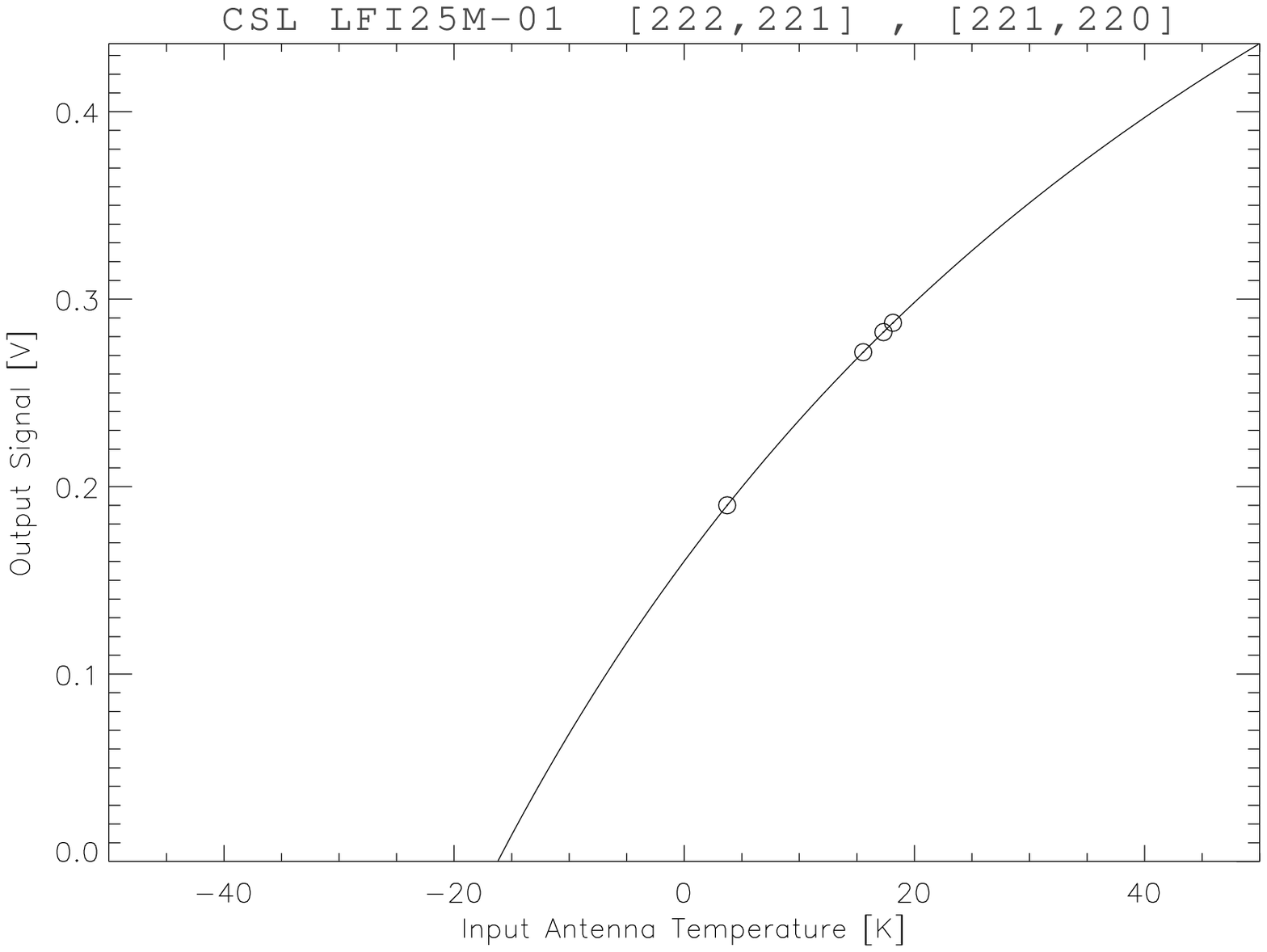}
            \includegraphics[width=7.0cm]{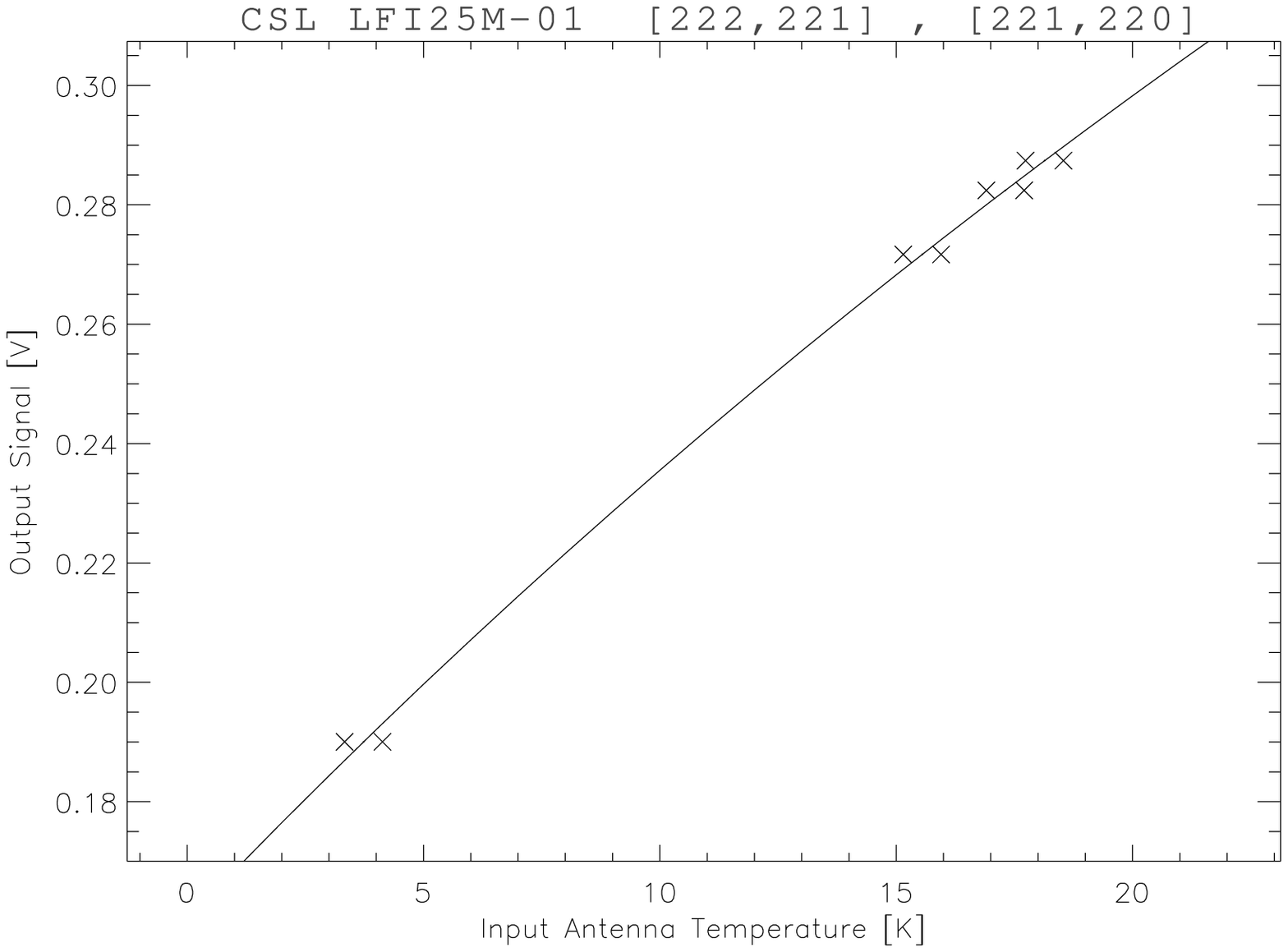}\\
            
            \textbf{LFI-25S}\\
            \includegraphics[width=7.0cm]{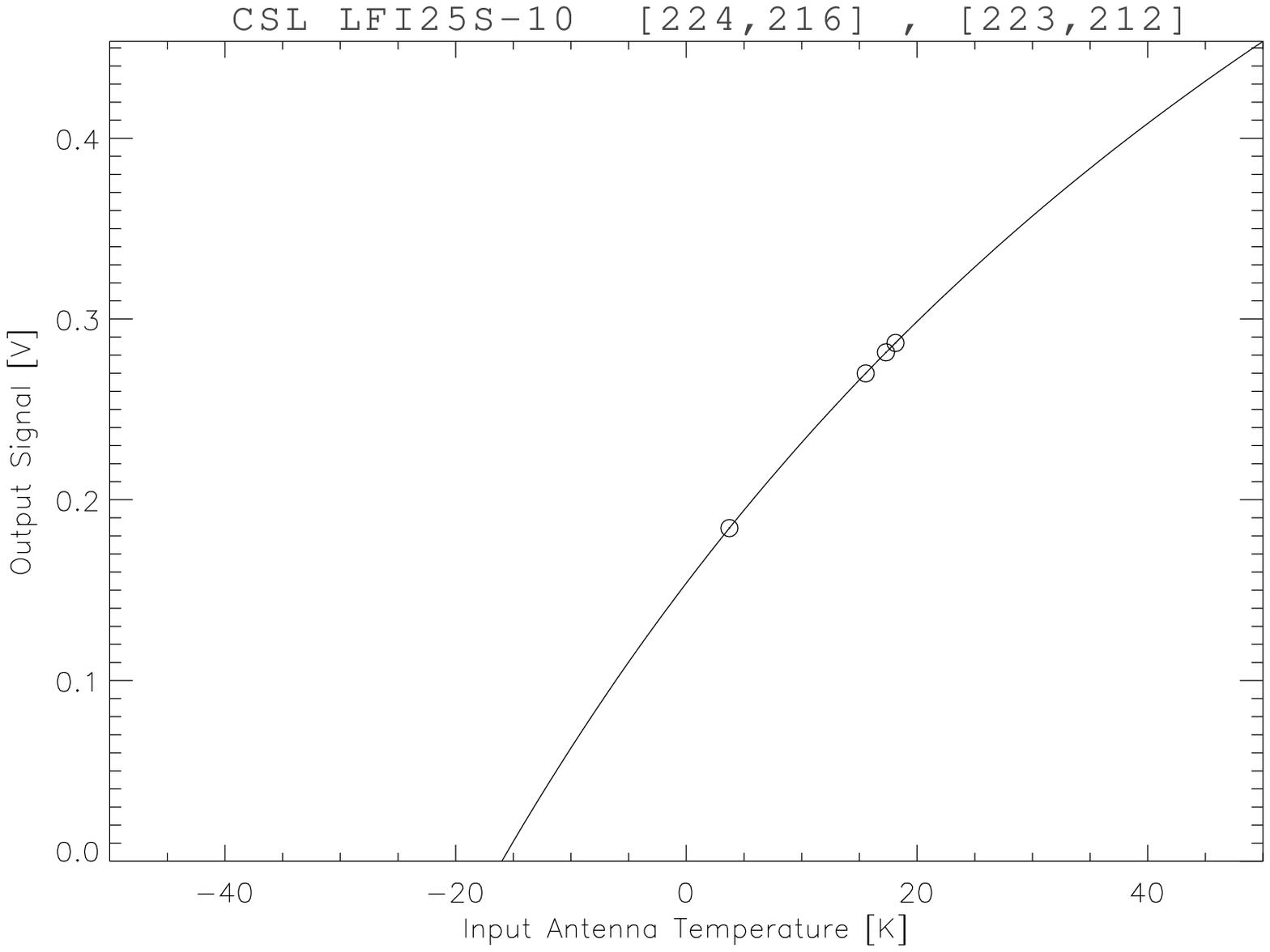}
            \includegraphics[width=7.0cm]{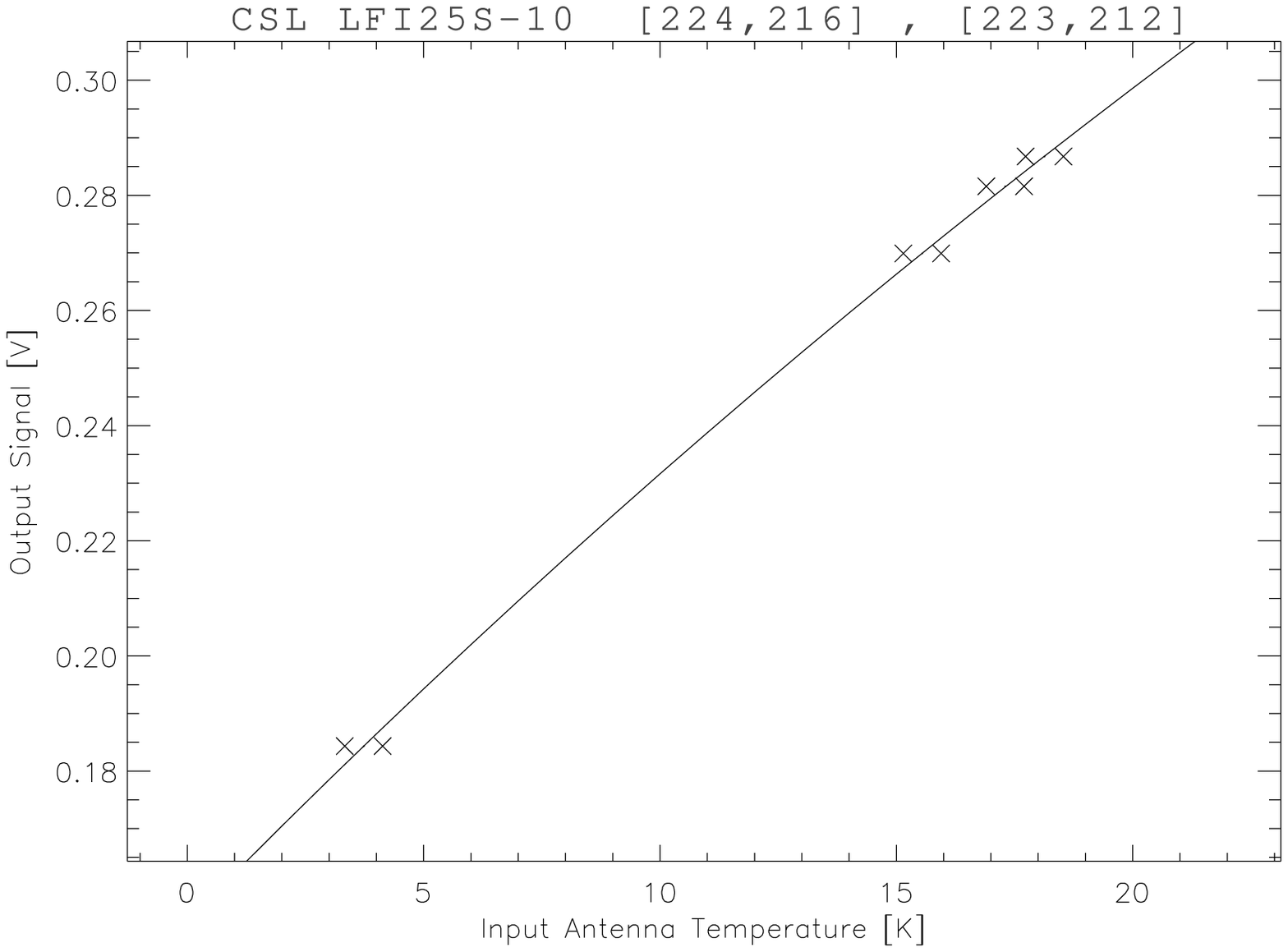}
            \includegraphics[width=7.0cm]{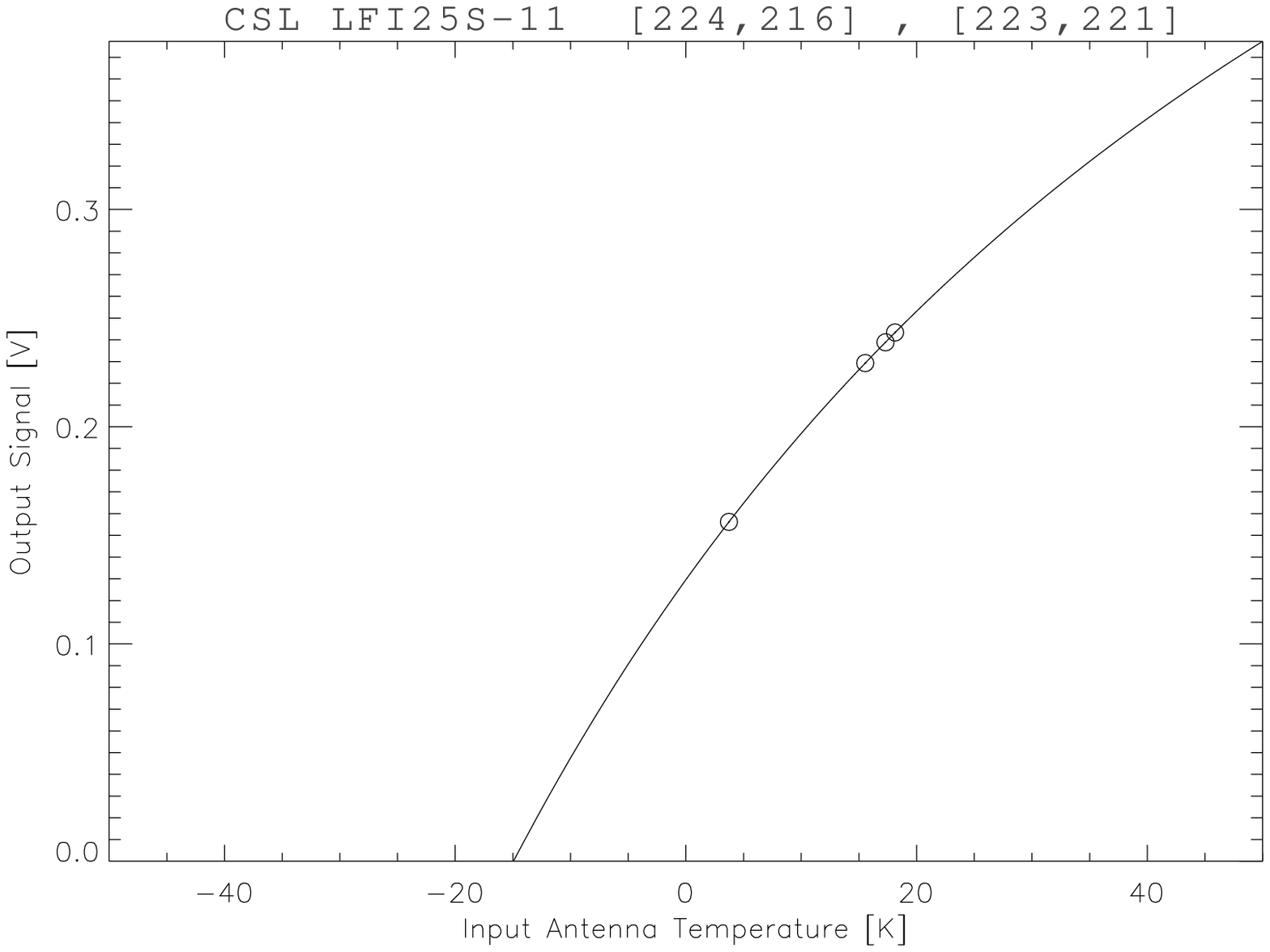}
            \includegraphics[width=7.0cm]{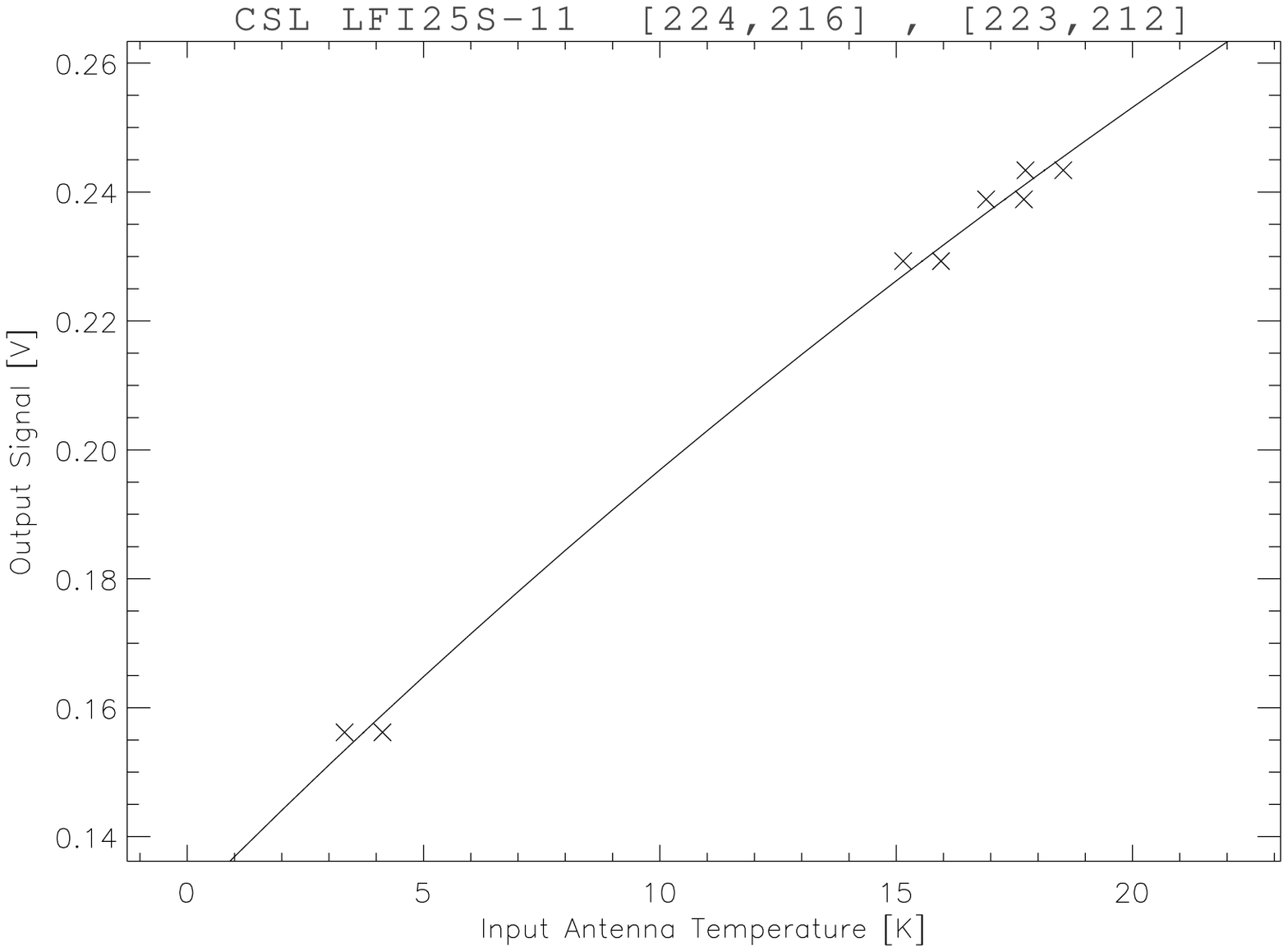}\\
          \end{center}
            
        \caption{Non linear response plots for 44~GHz \texttt{LFI-25} channel.}% Each detector (\texttt{M-00}, \texttt{M-01}, \texttt{S-10}, \texttt{S-11}) is represented from left to right: the voltage measured at \texttt{Diode-1} and \texttt{Diode-2} (y-axis) versus the extrapolated input temperature (x-axis); a zoom on the same plot  accounting also for uncertainties related to the knowledge of the 4~K reference load temperature and to the BEU thermal drift. }
        \label{fig_HYM_tun_nonlin_44-25}
            \end{figure}
 
     \begin{figure}[htb]
        \begin{center}     
            	\textbf{LFI-28M}\\
            \includegraphics[width=7.0cm]{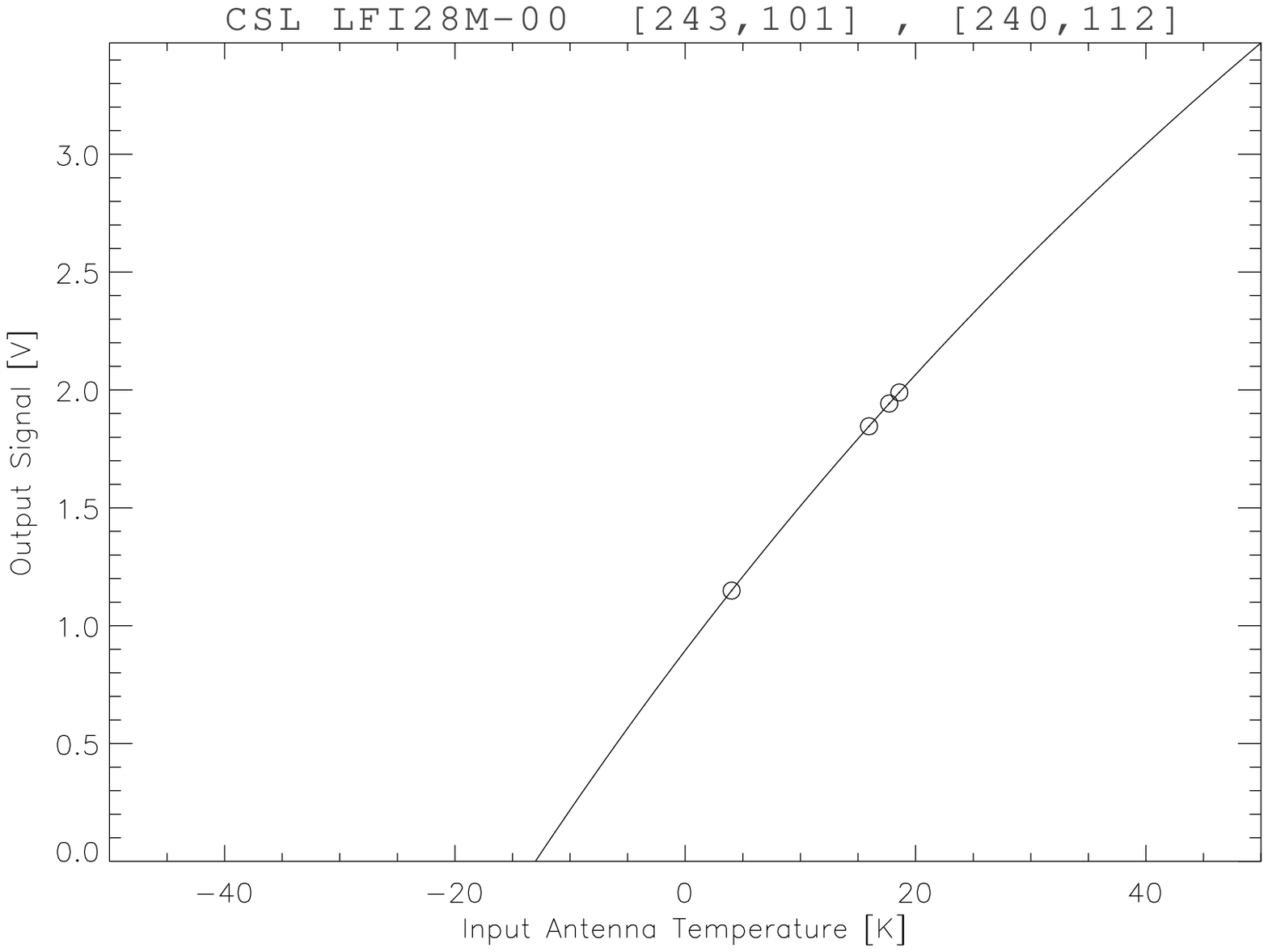}
            \includegraphics[width=7.0cm]{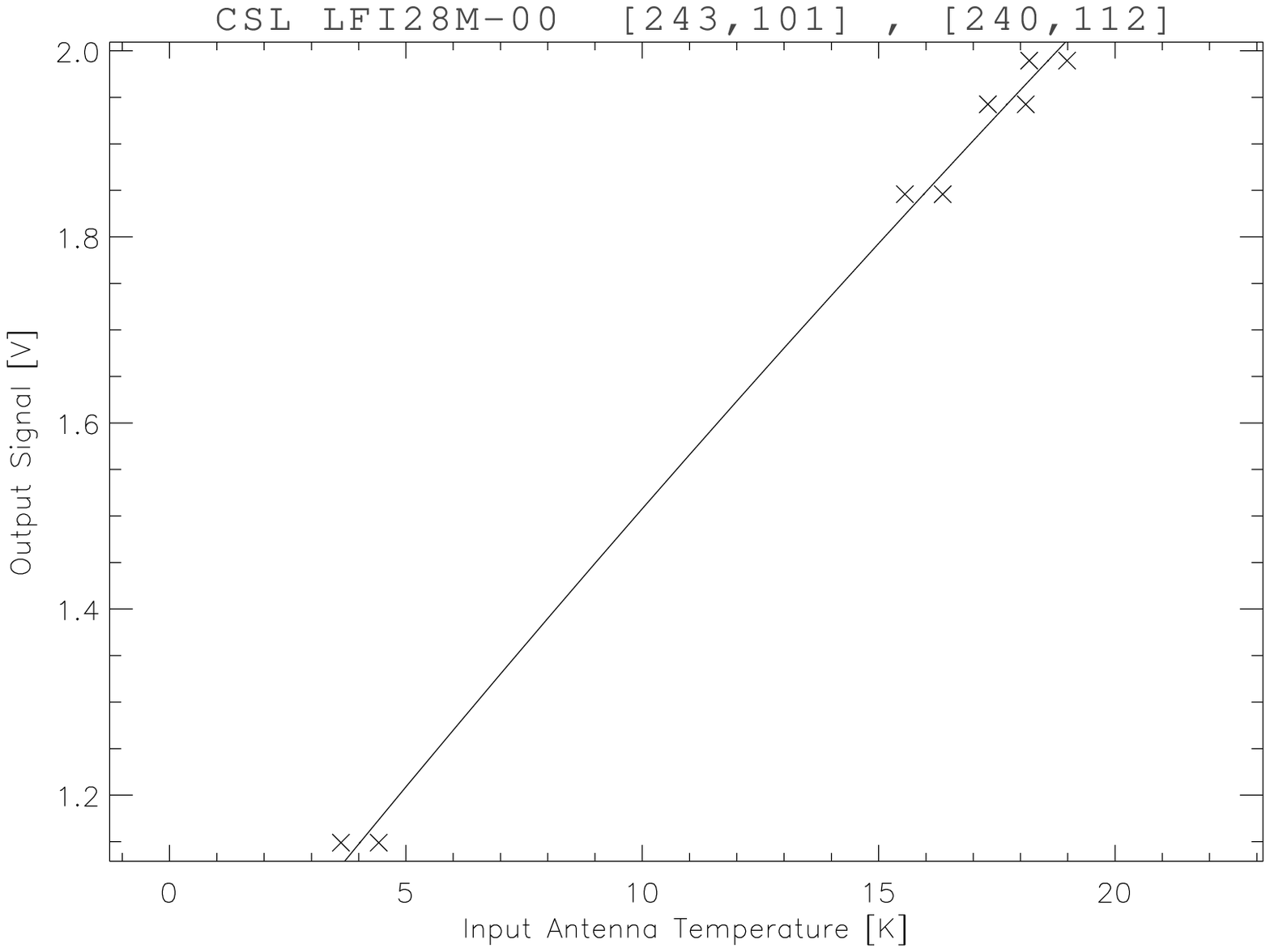}\\
            \includegraphics[width=7.0cm]{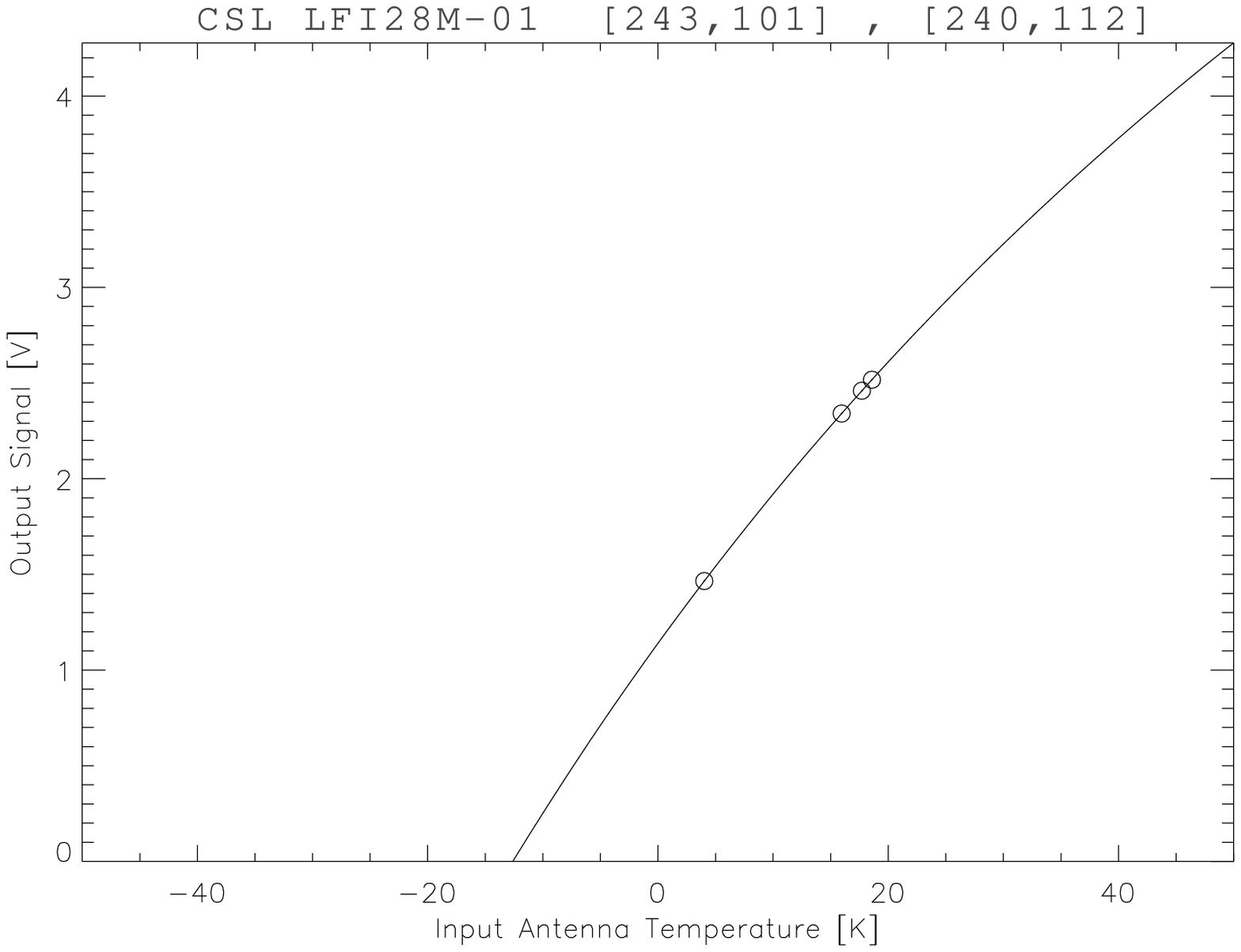}
            \includegraphics[width=7.0cm]{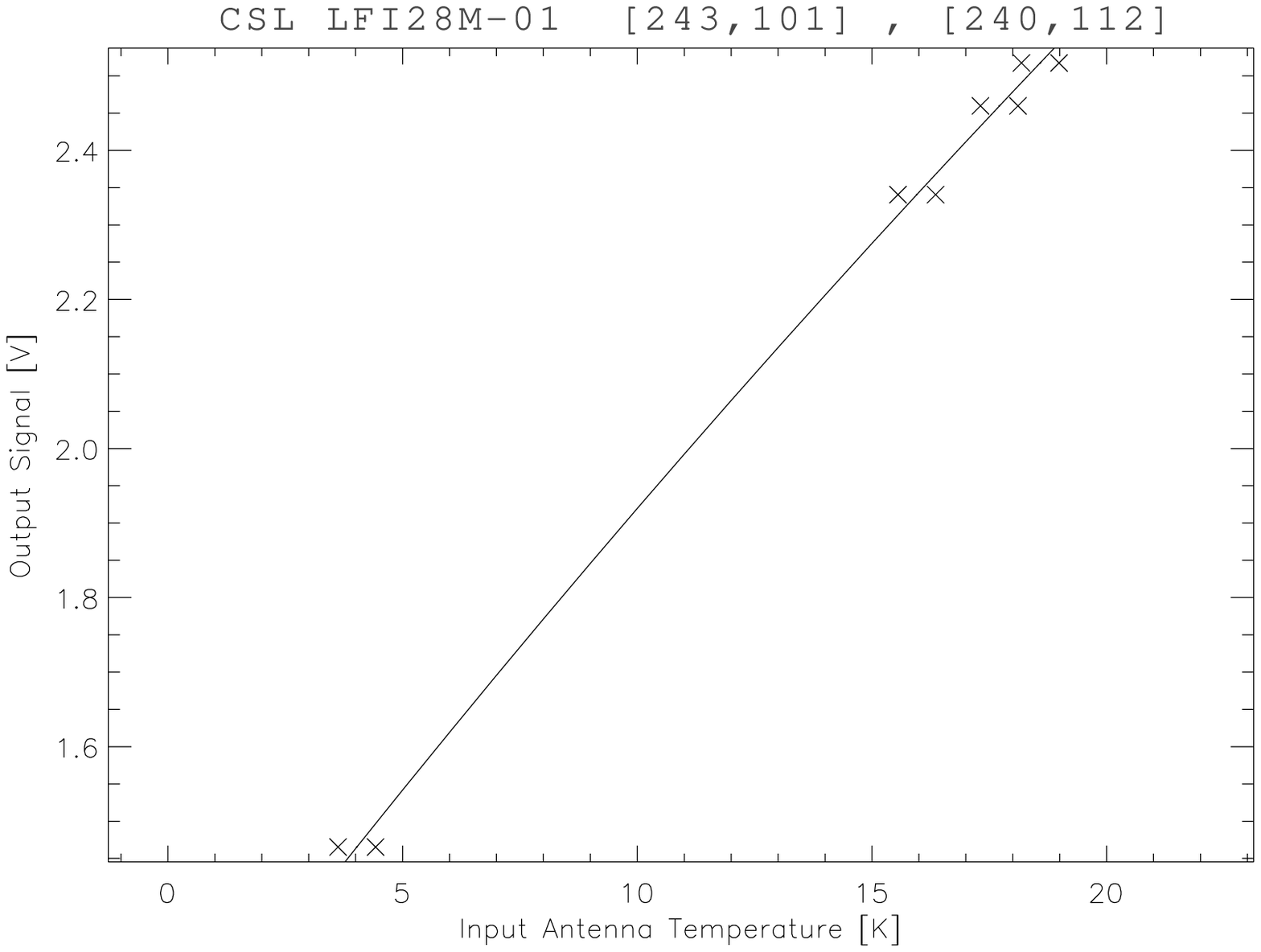}\\
   
            \textbf{LFI-28S}\\
            \includegraphics[width=7.0cm]{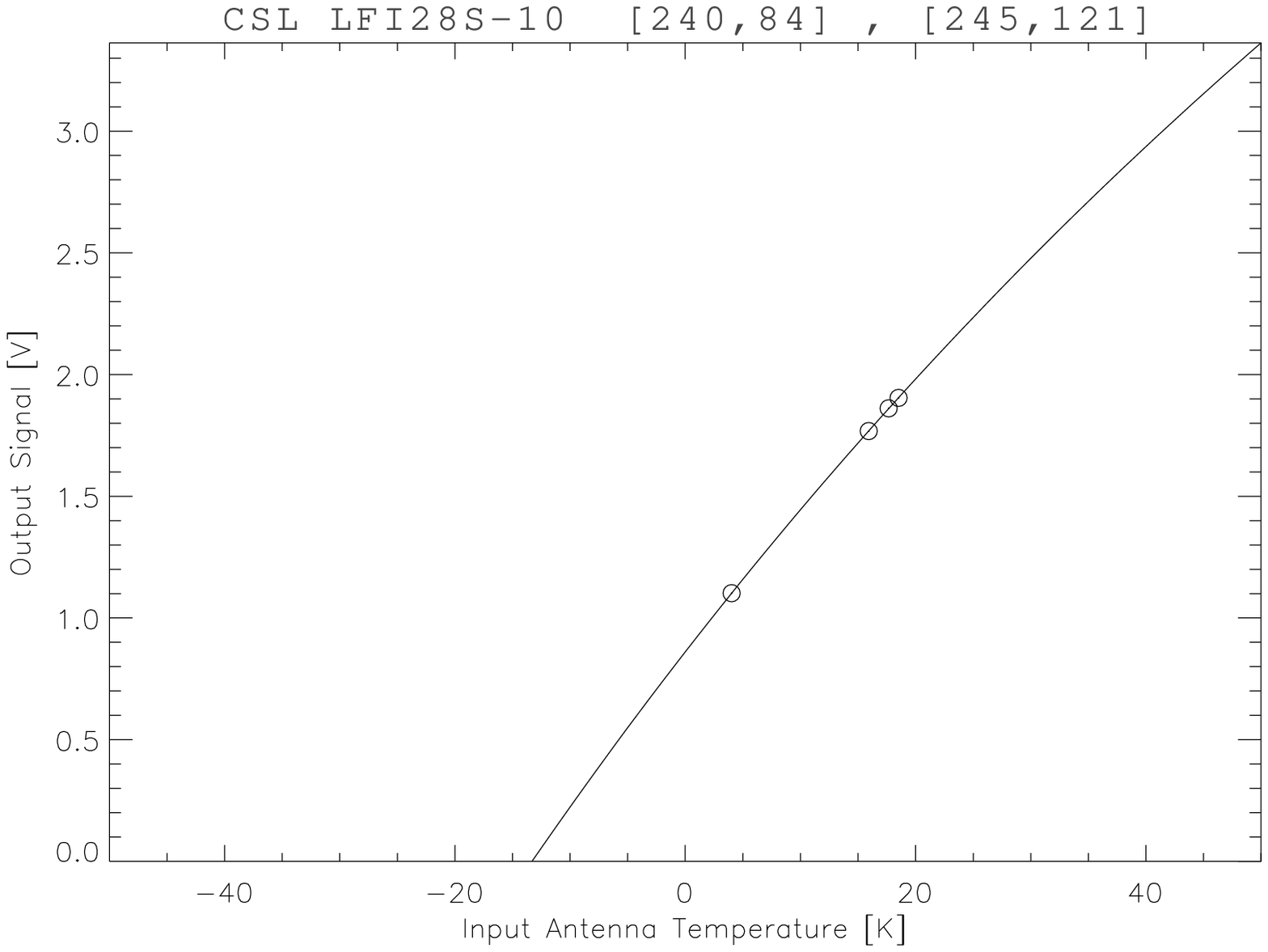}
            \includegraphics[width=7.0cm]{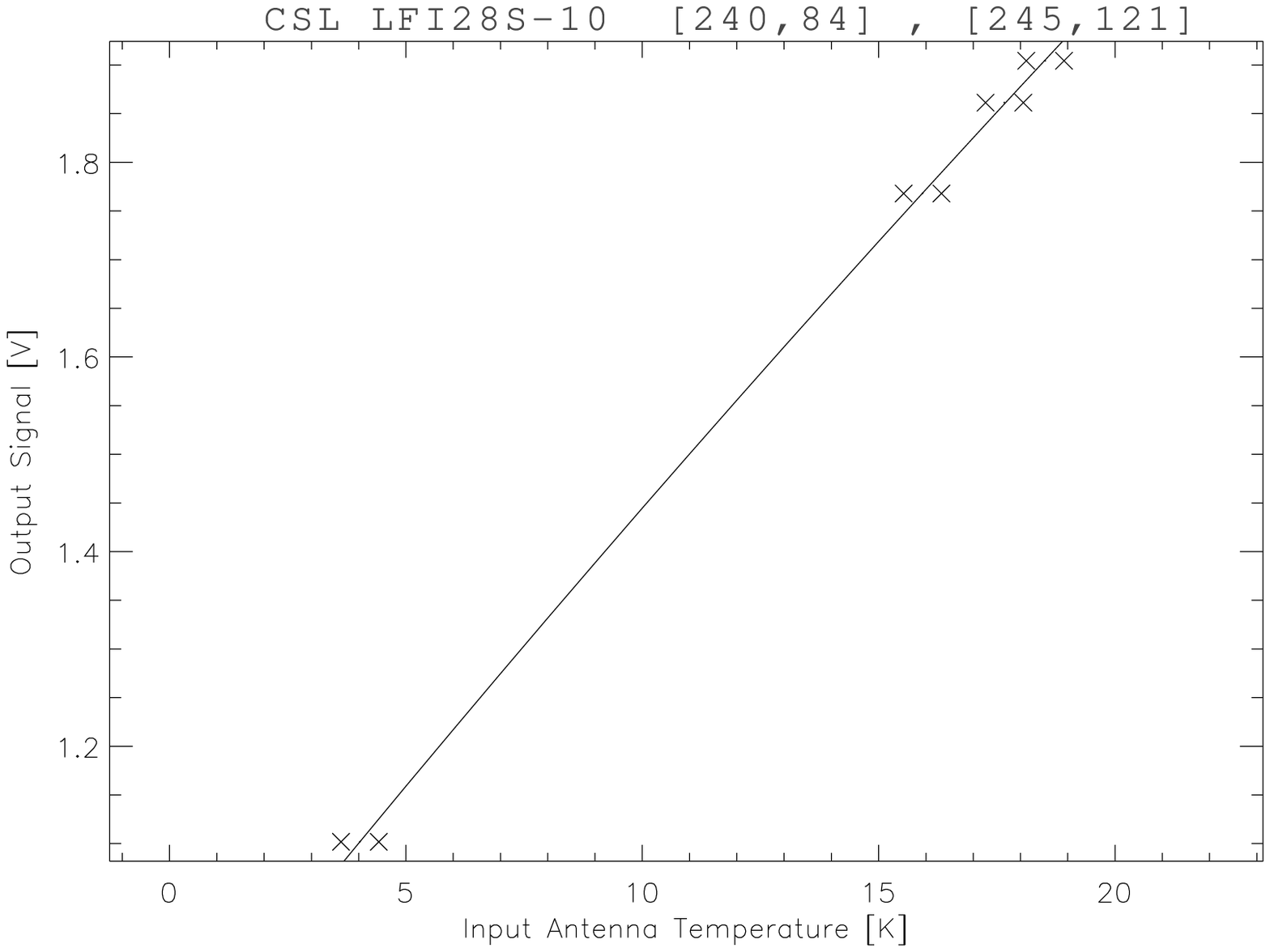}\\
            \includegraphics[width=7.0cm]{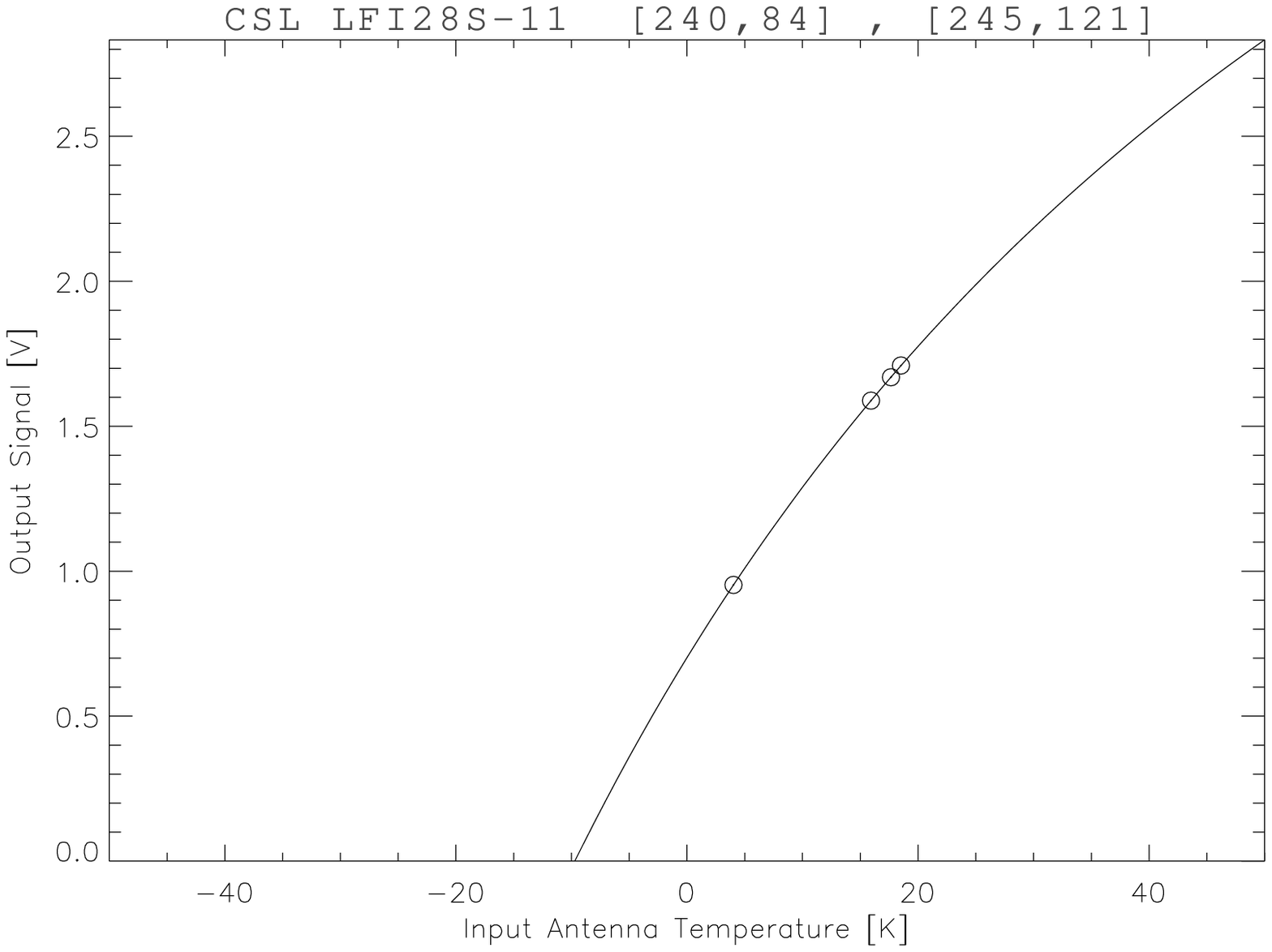}
            \includegraphics[width=7.0cm]{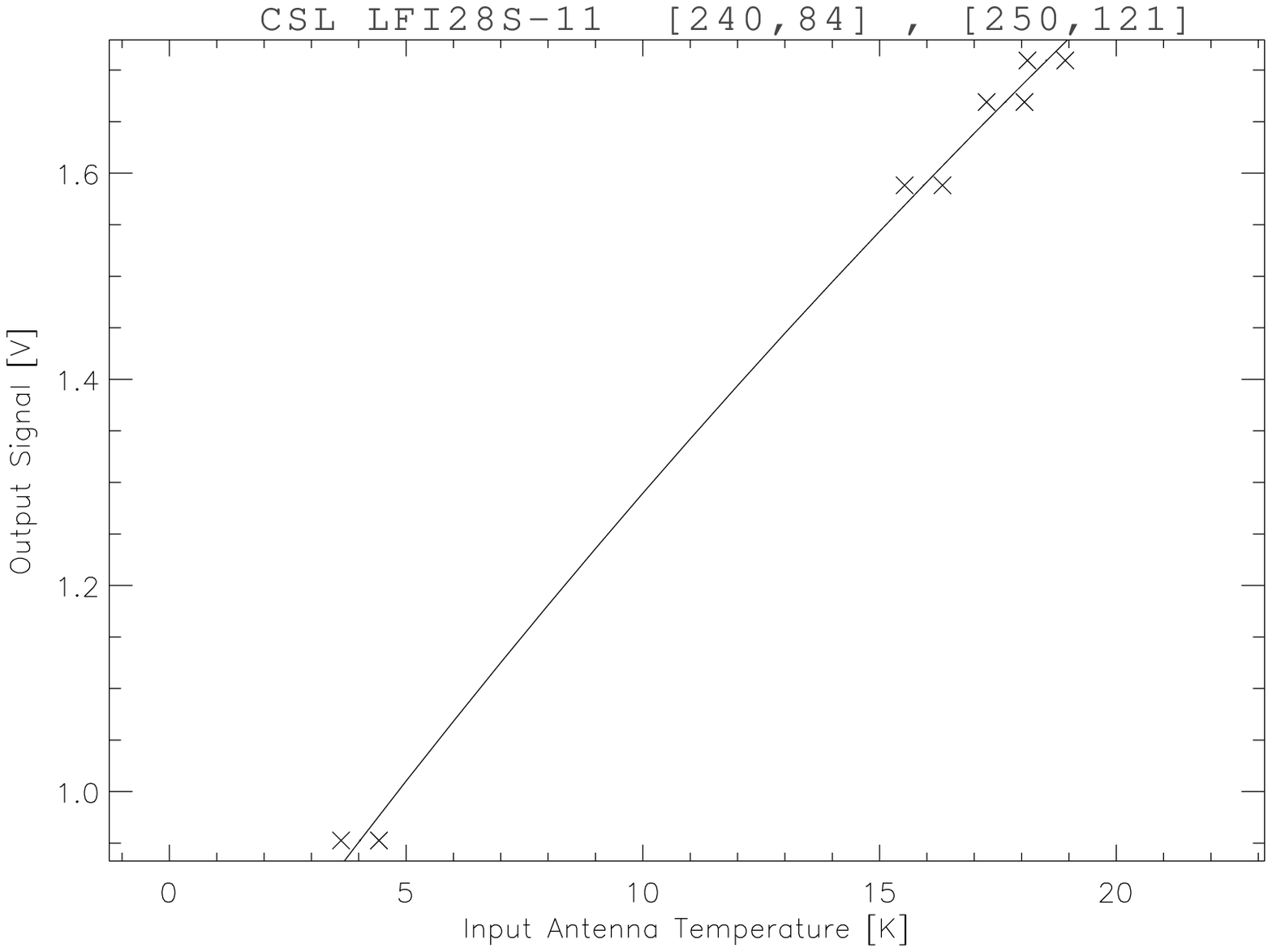}\\
        \end{center}
            
        \caption{Non linear response plots for 30~GHz \texttt{LFI-28} channel.}% Each detector (\texttt{M-00}, \texttt{M-01}, \texttt{S-10}, \texttt{S-11}) is represented from left to right: the voltage measured at \texttt{Diode-1} and \texttt{Diode-2} (y-axis) versus the extrapolated input temperature (x-axis); a zoom on the same plot  accounting also for uncertainties related to the knowledge of the 4~K reference load temperature and to the BEU thermal drift. }
        \label{fig_HYM_tun_nonlin_30}
    \end{figure}
            \clearpage 

%% file: a08_acronyms.tex
%\section{Appendix Acronym list}
\section{Acronym list}
\label{app_acronyms}

\begin{table}[h!]
    \begin{center}
        \caption{\label{tab_acronyms} Acronym list. }
        \vspace{.2cm}
        \begin{tabular}{l l }
            \hline
            \hline
            Acronym &  \\
            \hline
		ACA   	& Amplifiers Chain Assembly \\
		ADC   	& Analogue-to-Digital Converters \\
		ADU   	& Analogue-to-Digital Units \\
		ASD   	& Amplitude Spectral Density \\
		BEM   	& Back-End Module \\
		BEU   	& Back-End Unit \\
		CMB		& Cosmic Microwave Background \\
		CPV		& Calibration and Performance Verification \\
		CSL   	& Centre Spatiale de Li\'ege \\
		DAE   	& Data Acquisition Electronics \\
		DTCP   	& Daily Tele-Communication Period  \\
		FEM   	& Front-End Module \\
		FPU   	& Focal Plane Unit \\
		HFI		& High Frequency Instrument \\
		LFI		& Low Frequency Instrument \\
		LNA   	& Low Noise Amplifier \\
		OMT   	& Ortho-Mode Transducer \\
		PID   	& Proportional Integral Derivative \\
		PSD   	& Power Spectral Density \\
		RAA   	& Radiometer Array Assembly \\
		RCA   	& Radiometer Chain Assembly \\
		REBA   	& Radiometer Electronics Box Assembly \\
		TSA   	& Temperature Stabilisation Assembly \\
            \hline
        \end{tabular}
    \end{center}
\end{table}

\clearpage

%% file: lfi_cpv-June-2013-05-16.bbl
\hyphenation{Post-Script Sprin-ger}
\begin{thebibliography}{}

\bibitem{tauber2010}
J.~A.\,{Tauber}, N.\,{Mandolesi}, {J.-L.}\,{Puget} et~al.\, {Planck pre-launch
  status: The Planck mission}.\,\textit{\aap}, 520:A1, 2010.

\bibitem{mandolesi2010}
N.\,{Mandolesi}, M.\,{Bersanelli}, R.~C.\,{Butler} et~al.\, Planck pre-launch
  status: The Planck-LFI programme.\,\textit{\aap}, 520:A3, 2010.

\bibitem{lamarre2010}
J.~M.\,{Lamarre}, J.~L.\,{Puget} and al..\, {Planck pre-launch status: The HFI
  instrument, from specification to actual performance}.\,\textit{\aap},
  520:A9, 2010.

\bibitem{tauber2010a}
J.~A.\,{Tauber}, H.~U.\,{Norgaard-Nielsen}, P.~A.~R.\,{Ade} et~al.\, {Planck
  pre-launch status: The optical system}.\,\textit{\aap}, 520:A2, 2010.

\bibitem{2002_villa_planck_telescope}
F.\,{Villa}, M.\,{Bersanelli}, C.\,{Burigana} et~al.\,The Planck Telescope.\,In
  \textit{Experimental Cosmology at Millimetre Wavelengths}, 616:224--228,
  2002.

\bibitem{2005_dupac_planck_scanning_strategy}
X.\,{Dupac} and J.\,{Tauber}.\, Scanning strategy for mapping the Cosmic
  Microwave Background anisotropies with Planck.\,\textit{\aap}, 430:363--371,
  2005.

\bibitem{2006_maris_planck_scanning_strategy}
M.\,{Maris}, M.\,{Bersanelli}, C.\,{Burigana} et~al.\, The Flexible Planck
  Scanning Strategy.\,\textit{Memorie della Societ{\`a} Astronomica Italiana
  Supplement}, 9:460, 2006.

\bibitem{mennella2011}
A.\,{Mennella}, M.\,{Bersanelli}, R.~C.\,{Butler} et~al.\, {Planck early
  results. III. First assessment of the Low Frequency Instrument in-flight
  performance}.\,\textit{\aap}, 536:A3, 2011.

\bibitem{planck2011-1.5}
{Planck HFI Core Team}.\, {Planck early results. IV. First assessment of the
  High Frequency Instrument in-flight performance}.\,\textit{\aap}, 536:A4,
  2011.

\bibitem{bersanelli2010}
M.\,{Bersanelli}, N.\,{Mandolesi}, R.~C.\,{Butler} et~al.\, Planck pre-launch
  status: Design and description of the Low Frequency
  Instrument.\,\textit{\aap}, 520:A4, 2010.

\bibitem{d'arcangelo2009a}
O.\,{D'Arcangelo}, L.\,{Figini}, A.\,{Simonetto} et~al.\, The Planck-LFI flight
  model composite waveguides.\,\textit{Journal of Instrumentation}, 4:2007,
  2009.

\bibitem{valenziano2009}
L.\,{Valenziano}, F.\,{Cuttaia}, A.\,{De Rosa} et~al.\, Planck-LFI: design and
  performance of the 4 Kelvin Reference Load Unit.\,\textit{Journal of
  Instrumentation}, 4:2006, 2009.

\bibitem{herreros2009}
J.~M.\,{Herreros}, M.~F.\,{G{\'o}mez}, R.\,{Rebolo} et~al.\, The Planck-LFI
  Radiometer Electronics Box Assembly.\,\textit{Journal of Instrumentation},
  4:2008, 2009.

\bibitem{maris2009}
M.\,{Maris}, M.\,{Tomasi}, S.\,{Galeotta} et~al.\, Optimization of Planck-LFI
  on-board data handling.\,\textit{Journal of Instrumentation}, 4:2018, 2009.

\bibitem{seiffert2002}
M.\,{Seiffert}, A.\,{Mennella}, C.\,{Burigana} et~al.\, 1/f noise and other
  systematic effects in the Planck-LFI radiometers.\,\textit{\aap},
  391:1185--1197, 2002.

\bibitem{mennella2003}
A.\,{Mennella}, M.\,{Bersanelli}, M.\,{Seiffert} et~al.\, Offset balancing in
  pseudo-correlation radiometers for CMB measurements.\,\textit{\aap},
  410:1089--1100, 2003.

\bibitem{zacchei2011}
A.\,{Zacchei}, D.\,{Maino}, C.\,{Baccigalupi} et~al.\, {Planck early results.
  V. The Low Frequency Instrument data processing}.\,\textit{\aap}, 536:A5,
  2011.

\bibitem{davis2009}
R.~J.\,{Davis}, A.\,{Wilkinson}, R.~D.\,{Davies} et~al.\, Design, development
  and verification of the 30 and 44 GHz front-end modules for the Planck Low
  Frequency Instrument.\,\textit{Journal of Instrumentation}, 4:2002, 2009.

\bibitem{artal2009}
E.\,{Artal}, B.\,{Aja}, M.~L.\,{de la Fuente} et~al.\, LFI 30 and 44 GHz
  receivers Back-End Modules.\,\textit{Journal of Instrumentation}, 4:2003,
  2009.

\bibitem{varis2009}
J.\,{Varis}, N.~J.\,{Hughes}, M.\,{Laaninen} et~al.\, Design, development, and
  verification of the Planck Low Frequency Instrument 70 GHz Front-End and
  Back-End Modules.\,\textit{Journal of Instrumentation}, 4:2001, 2009.

\bibitem{villa2010}
F.\,{Villa}, L.\,{Terenzi}, M.\,{Sandri} et~al.\, Planck pre-launch status:
  Calibration of the Low Frequency Instrument flight model
  radiometers.\,\textit{\aap}, 520:A6, 2010.

\bibitem{mennella2010}
A.\,{Mennella}, M.\,{Bersanelli}, R.~C.\,{Butler} et~al.\, Planck pre-launch
  status: Low Frequency Instrument calibration and expected scientific
  performance.\,\textit{\aap}, 520:A5, 2010.

\bibitem{planck2011-1.3}
{Planck Collaboration}.\, {Planck early results. II. The thermal performance of
  Planck}.\,\textit{\aap}, 536:A2, 2011.

\bibitem{planck2011-1.1}
{Planck Collaboration}.\, {Planck early results. I. The Planck
  mission}.\,\textit{\aap}, 536:A1, 2011.

\bibitem{meinhold2009}
P.\,{Meinhold}, R.\,{Leonardi}, B.\,{Aja} et~al.\, Noise properties of the
  Planck-LFI receivers.\,\textit{Journal of Instrumentation}, 4:2009, 2009.

\bibitem{cuttaia2009}
F.\,{Cuttaia}, A.\,{Mennella}, L.\,{Stringhetti} et~al.\, Planck-LFI
  radiometers tuning.\,\textit{Journal of Instrumentation}, 4:2013, 2009.

\bibitem{mennella2009-2}
A.\,{Mennella}, F.\,{Villa}, L.\,{Terenzi} et~al.\, {The linearity response of
  the Planck-LFI flight model receivers}.\,\textit{Journal of Instrumentation},
  4:2011, 2009.

\bibitem{poutanen2004}
T.\,{Poutanen}, D.\,{Maino}, H.\,{Kurki-Suonio}, E.\,{Keih{\"a}nen} and
  E.\,{Hivon}.\, {Cosmic microwave background power spectrum estimation with
  the destriping technique}.\,\textit{\mnras}, 353:43-58, 2004.

\bibitem{tomasi2009}
M.\,{Tomasi}, A.\,{Mennella}, S.\,{Galeotta} et~al.\, Off-line radiometric
  analysis of Planck-LFI data.\,\textit{Journal of Instrumentation}, 4:2020,
  2009.

\bibitem{terenzi2009b}
L.\,{Terenzi}, M.~J.\,{Salmon}, A.\,{Colin} et~al.\, Thermal susceptibility of
  the Planck-LFI receivers.\,\textit{Journal of Instrumentation}, 4:2012-+,
  2009.

\end{thebibliography}
